\documentclass[mnsc,nonblindrev]{informs3mod} 

\OneAndAHalfSpacedXI

\usepackage{natbib}
 \bibpunct[, ]{(}{)}{,}{a}{}{,}%
 \def\bibfont{\small}%
\usepackage{rotating}
\usepackage{graphicx,array,multirow,color,framed}
\usepackage{fancyhdr}
\usepackage{float}
\usepackage{rcs}
\usepackage{caption}
\usepackage{verbatim} 
\usepackage[FIGTOPCAP]{subfigure}
\usepackage[dvips]{epsfig}     
\usepackage{epstopdf}
\usepackage{afterpage}
\usepackage{gensymb}
\usepackage{tablefootnote}
\usepackage{longtable}
\usepackage{hyperref}
\usepackage[utf8]{inputenc}
\usepackage[T1]{fontenc}
\newcommand*\rot{\rotatebox{90}}
\usepackage{capt-of}
\usepackage{booktabs,threeparttable}
\hypersetup{hidelinks}
\TheoremsNumberedThrough     
\ECRepeatTheorems

\EquationsNumberedThrough    

\MANUSCRIPTNO{}
\begin{document}



\RUNTITLE{The impact of Climate on Economic and Financial Cycles}

\TITLE{The impact of Climate on Economic and Financial Cycles: A Markov-switching Panel Approach}

\ARTICLEAUTHORS{%
\AUTHOR{Monica Billio, Roberto Casarin}
\AFF{Department of Economics, University Ca' Foscari of Venice, Fondamenta San Giobbe 873, 30121, Venice, Italy,
\EMAIL{billio@unive.it, r.casarin@unive.it}} 
\AUTHOR{Enrica De Cian, Malcolm Mistry}
\AFF{Centro Euro-Mediterraneo sui Cambiamenti Climatici (CMCC), 30175 Venice, Italy, \EMAIL{enrica.decian@unive.it, malcolm.mistry@unive.it}}
\AUTHOR{Anthony Osuntuyi}
\AFF{Department of Mathematics, Obafemi Awolowo University, 220005, Ile-Ife, Nigeria, \EMAIL{ayokunlet@yahoo.com}}
} 

\ABSTRACT{%
This paper examines the impact of climate shocks on 13 European economies analysing jointly business and financial cycles, in different phases and disentangling the effects for different sector channels. A Bayesian Panel Markov-switching framework is proposed to jointly estimate the impact of extreme weather events on the economies as well as the interaction between business and financial cycles. Results from the empirical analysis suggest that extreme weather events impact asymmetrically across the different phases of the economy and heterogeneously across the EU countries. Moreover, we highlight how the manufacturing output, a component of the industrial production index, constitutes the main channel through which climate shocks impact the EU economies. 
}%


\KEYWORDS{Bayesian inference, climate shocks, financial cycle, business cycle, Markov-switching, Multi-country Panel} 
\MSCCLASS{Primary 62F15; 60J20; Secondary 62P20; 91B64; 91B76}

\maketitle

%

\section{Introduction}
\label{sec:intro}

In the last few years, several emerging markets and advanced economies have experienced one form of contraction or another in their economic activities. This reality, among others, is of great concern to policymakers and other stakeholders who continually seek a better understanding of the mechanisms driving the economy in order to either mitigate or better manage the potential adverse socioeconomic impacts arising from such economic realizations. Given this, the investigation of several shocks and disturbances (e.g. technological shocks, policy shocks, economic shocks, etc.) as potential drivers of business cycles have been conducted. Recent studies (\cite{Acevedo2018},\cite{Gallic2020}, \cite{Pfleiderer2019}), however, show that extreme weather events have intensified with longer-lasting spells causing significant damage especially to health, agriculture, economy and the ecosystems (\cite{Ciscar2011}, \cite{Dell2012}). Similarly, the fallout of the 2008-09 global financial crisis highlights how events in the financial market can impact the real economy. 

In this paper, we limit attention to these two growing sources of macroeconomic fluctuations: financial shocks and climate shocks. We analyse the impact of climate shocks (measured as monthly temperature and precipitation variability) on short term fluctuations in the economic and financial activities of a set of European Union (EU) countries over 1981 -2016. Since temperature and precipitation are the two main weather attributes for characterizing changes in climate condition and any dramatic variations in these factors may likely cause cascading effects on community, ecosystem properties, and industrial production processes (see \cite{Arent2014} for a useful summary of the impact of climate change on key economic sector and services), three contrasting Climate Extreme Indices (CEIs) (\cite{Mistry2019}) that account for (i) high temperatures (ii) drought and (iii) very heavy rainfall events are considered in this study (see Section \ref{DataDescription} for details). 

We measure high temperature by considering the number of consecutive days per month for which the temperature rises above a given temperature threshold ($25\degree$C). As in \cite{Pfleiderer2019}, the purpose of the persistence concept introduced here is not to identify changes in heat-wave conditions per se, but rather to investigate changes in the distribution of the warm days spells on economic and financial activities. Following the recommendation by the Lincoln Declaration on Drought Indices (LDDI) and the subsequent approval by the World Meteorological Organization (WMO), we measure drought using the Standardized Precipitation Index (SPI) \cite{Mckee1993}. As opposed to other complex measures of drought, the SPI is relatively easy to construct, requires only monthly precipitation and can be compared across regions with obvious different climates. Lastly, we measure heavy rainfall events by computing, on a monthly basis, the number of days with rainfall exceeding $20 mm$. 

Our empirical approach involves estimating a Panel Markov Switching (PMS) model that jointly describes the cyclical behaviour of the EU economy at country-specific levels and an aggregate level while taking the interaction with the aggregate financial cycle and climate shocks into consideration. Dependence, among the different units (i.e., countries) in the panel, is captured by endogenous time-varying transition matrices. Precisely, we propose a convenient linear parametrization of the transition probabilities with a capacity of identifying the role played by both a global business factor (expressed by the proportion of chains in a given regime) and a local factor (expressed in terms of the state of the aggregate financial index). For the inferential exercise, we adopt a Bayesian inference procedure that allows us to deal with latent variable estimation through a data augmentation framework. We approximate the posterior distribution of the parameters and the latent MS chains by an efficient Gibbs sampling scheme (see, for example, \cite{fruhwirth2008}).


Our results from the empirical exercise are consistent with the literature on the relationship between financial and business cycles. While the estimated degree of synchronization between financial and business cycle does not appear to be significantly different when climate shocks are accounted for or not in the PMS model, we observe that ignoring climate shocks often result to a low degree of pairwise synchronization
among the business cycle. Further analysis suggests that climate shocks have an uneven impact across the different phases of the economy and heterogeneously across the EU countries. In particular, we observe, except for Greece, that the economies of countries within Southern Europe (the region with the highest exposure to a long spell of summer days) are negatively impacted by their exposure to a lengthy spell of summer days. The economies of France, Belgium, and Denmark on the other hand respond asymmetrically (positively during a recession and negatively during expansion) to high consecutive days of summer. Also, relative to the non-drought-like climatic condition, moderate to extreme dryness seem to contribute negatively to the IPI growth of most of the countries in Northern Europe. Besides, the economies of Ireland and Finland tend to be favoured by drought during the recession but hindered during expansion. Lastly, from the model selection exercise conducted, drought and precipitation indices appear to be the two most relevant climate shocks to consider when assessing the co-movement between business and financial cycle. Overall, France seems to be the only economy that is resilient to all the climate shocks during the recession. To gain a better understanding of the observed heterogeneous and asymmetric impact of extreme weather events on industrial production, we examine the impact of weather extremes on the indices of the manufacturing sector. Our deduction from this exercise indicates that the impact of climate shocks on the economy is mostly felt through the manufacturing sector.

The paper is organized as follows. Section \ref{sec:PMS model} introduces the PMS model and discusses inferential procedure adopted. In Section \ref{EmpriccalResusult} we provide an empirical application of our proposed model to the industrial production indices for a selected set of European countries and a composite financial index sampled at the monthly frequency from February 1981 to December 2016. The implications of climate shocks on the economic variable is also presented in this section. Section \ref{sec:conclude} concludes the paper.
\section{Literature review}
\label{sec:litreview}
Several attempts have been made in the economic literature to evaluate the potential impact of climate change on the economy (\cite{Burkeetal2015}, \cite{Dafermosetal2018}, \cite{Delletal2014}, \cite{DifenbaughBurke2019}), with particular attention paid to its effect on the agricultural system (\cite{SchlenkerRobert2009}, \cite{BurkeEmerick2016}, \cite{IizumiRamankutty2016}, \cite{Gallic2020}). Most of these studies have focused on analyzing the impact of weather on economic activities by either using a theoretical model to assess its long-term effects (\cite{Nordhaus1991}) or by centering attention on both short and long-term effects using empirical analysis ( \cite{Buckleetal2007}; \cite{Kamberetal2013} and \cite{DeCianetal2019}). There have however been strong objections (see \cite{Pindyck2017}) to the use of theoretical models largely for their lack of empirical foundation. In the instance that these models provide useful arguments for rationalizing the long-term implications of climate, they become immaterial for short-run analysis at a business cycle frequency. Contrary to this, empirical models have the capacity of providing important quantitative insights on the short-term transmission channels. Identifying the channel(s) of impact on the economy of the short term dimension of weather plays an important role in decision making as policymakers are required to regularly manage short term adverse weather events having significant implications (e.g. recessions, food security, currency depreciation, etc.). Recently \cite{Gallic2020} examined the role of drought conditions on agricultural production and their implications for macroeconomic fluctuations in New Zealand. The authors document that weather (drought) shocks account for more than a third of the Gross Domestic Product (GDP) and agricultural output fluctuations in New Zealand. However, to the best of our knowledge, there is no study of the impact of extreme weather events on the fluctuations in the economy through industrial production.

While a sizeable number of research activities focused on the effects of industrial production processes on global warming exists in the literature, there have been limited studies on the impact of climate change and variability on industrial production. \cite{Colmer2020} quantified the effectiveness of labour reallocation in mitigating the economic consequences of weather-driven agricultural productivity shocks. \cite{Sudarshan2014} provides empirical evidence suggesting that increase in high temperatures will impact India's non-agricultural sectors, especially manufacturing, with an attendant negative influence on economic output through their impact on worker productivity. \cite{Palutikof1983} shows that seasonal extremes, in the form of severe winter and drought summer, significantly impact Great Britain's industrial output. While previous literature has attributed aggregate losses in GDP to agriculture, \cite{Hsiang2010} argues that even in Central American countries agriculture represents a small share of GDP and that non-agricultural sectors are also heavily affected by temperature changes as well as extreme conditions experiences during cyclones. The findings by the SYKE team at the Finnish Environment Institute on the impact of climate change on industrial production\footnote{\url{https://ilmasto-opas.fi/en/ilmastonmuutos/vaikutukset/-/artikkeli/79840ec2-4723-442b-a6b3-5a2ebc46f6da/teollisuus.html\#cli_authors}} suggest that climate change has no significant direct effect on Finland's other industrial sector (i.e. forest, food, and construction industries). However, the difficulty in catering for the occurrence of extreme weather phenomena arising from changes in the climatic conditions are still left unresolved. Further research is therefore recommended to assess the impacts of climate change on the Finnish chemical, mining, metal, and electronics industry (raw materials or processes).

Our choice of the CEIs not only differs from other studies that employ deviations of annual temperature and precipitation from the climatological mean\footnote{An exception is the SPI which by design represents the abnormal precipitation deficit in relation to the long-term average conditions for a region, see \cite{Mckee1993} for further details} (referred to as 'long-term climate anomalies', e.g. \cite{Kahnetal2019}), but are also better representative of the physical and socio-economic impacts in various sectors (see \cite{Zhangetal2011}; \cite{GeoffreyPart2016}; \cite{Mysiaketal2018} and \cite{Mistry2019}), and are therefore considered to play a more prominent role in our modelling strategy focusing on economic and business cycles. While conventionally speaking, climate shocks refer to changes in environmental conditions over longer time-spans (e.g. increase in average regional/global temperatures; rise in sea-levels, typically over at least 30 year periods), the environmental shocks represented by the CEIs in our study are a blend of both shorter frequency events and departure from long-term conditions. Moreover, during the near-recent 35-years of our study period (1981-2016), the CEIs have been exacerbated by the acceleration in global warming \cite{seneviratne2012}. We therefore refer to these CEIs in our study as 'climate shocks' in lieu of 'weather shocks', but consider them to be indicators of extreme weather events or variations in climate due to both natural and anthropogenic causes. It is also important to emphasize that while reverse causality by way of expansion in industrial output and other anthropogenic activities are documented to be the dominant cause of the observed warming since the mid-20th century (e.g. see \cite{IPCC2014}), our study focuses on the transmission channel of: climate-change --$>$ variations in extreme weather events --$>$ impacts on the economy\footnote{Put another way, since we investigate business cycle fluctuations that are short term and not the long term trends, we can expect an impact of CEIs on the business cycles and not vice-versa because there is a lag of at least 50/80 years between economic activity, Green House Gas (GHG) emissions and changes in the distribution of extreme events}.

In the study of the relationship between business and financial indices, some of the readily asked questions relate to: the identification of the differences and common features of business and financial cycles; the extent of synchronization between both cycles; the identification of economic mechanisms that can explain the characteristics and differences of financial cycles relative to business cycles. Empirically, it has been reported in the literature (see \cite{Hubrichetal2013}, \cite{Borio2014}, among others) that financial cycles have a longer duration, typically between 10 and 30 years, than a typical business cycle (7 and 11 years). This observation, generally, suggests that in a stable macroeconomic environment, financial in-balance could go undetected. Other empirically observed features include but are not limited to: higher amplitude in financial cycle relative to the volatility in business indicator; greater symmetry in financial cycle phases than business cycle; persistence in financial downturn and short duration in recession. 

Four approaches may be identified in the literature for studying the relationship between cycles: the turning point analysis (see \cite{Harding2002}, \cite{Claessensetal2011}); frequency-based filters (see \cite{Hiebertetal2018}, and \cite{Aikmanetal2015}, \cite{Galatietal2016}); model-based filters (see \cite{RunstlerVlekke2017}); and spectral analysis techniques (see \cite{Schuleretal2017}). In this paper, attention will be given to the model-based approach for modelling and assessing the degree of synchronization between business and financial cycles. More specifically, following the report by \cite{Borio2014} (that financial cycles depend on financial regimes -liberalized market, controlled market- and the state of the business cycle -expansion or recession-), and further evidence by \cite{Claessensetal2011} (that business cycles are significantly dependent on financial cycle during financial crisis), we propose a Markov-switching model in our study of the synchronization of business and financial cycles. 

Markov-switching (MS) models have been extensively used in macroeconomics and finance to extract the different phases or regimes in a market environment. In the early days of its development, attention was given to either its application to a univariate series, see \cite{Hamilton1989}, \cite{FOR:FOR1148}, \cite{Bil11} and \cite{Bil12}, or to a small set of series under the assumption of an homogeneous transition probability describing the behaviour of the MS chain. There has, however, been growing interest to extend MS models to multivariate settings because of the very important role played by large databases to forecasting. See, for illustration, \cite{banbura2010large} in Vector Autoregressive (VAR) framework, \cite{Stock2014}, for turning point application, and \cite{CasarinGrassiRavazzoloVanDijk2015} for forecast combinations. In view of this, a dynamic panel model with variable-specific interacting Markov-switching processes is proposed in this paper.

In the study of business cycle of large panel of countries, the potential of Panel Markov-switching (PMS) models have often been leveraged on for analyzing the co-movement of country-specific business cycles. PMS model has largely been favoured in the literature because of its capability not only in characterizing the units-specific cycles, but also of its ability to showcase the importance of aggregate business cycle and time-invariant components of the cycle transition. Inline with this, \cite{kose2003international, kose2008understanding} document the common dynamic properties of the world business cycle fluctuations employing a Bayesian dynamic latent factor model. Their results suggest that regional components play only a minor role in explaining cycle fluctuations. However, recent studies suggest that the world component is not enough to explain the business cycle synchronization. \cite{francis2012endogenously} find that when the regional component is defined differently from simple geography, its effect becomes more important. \cite{aastveit2015drives, aastveit2016world} explicitly introduce regional factors into a global dynamic factor model. They find that both the global and the regional factors are relevant in explaining the business cycle variation. \cite{Leiva2014} proposed a new model that combines several bi-variate Markov-switching models and synchronization in order to create a link of inter-dependencies business cycles. However, the model cannot assess the importance of global and regional components in the cycle's fluctuations. 

\section{A panel Markov-switching model}
\label{sec:PMS model}
\subsection{The model}
\label{subsec:MSGARCH model}
\noindent Let $ y_{it}\in\mathbb{R} $, $t=1,\ldots,T$  be a sequence of monthly observations of industrial production growth rates for $i=1,\ldots,N$ countries. Also, let $ x_t\in\mathbb{R}$, $t=1,\ldots,T$ be a sequence of monthly observations of a common financial variable. In the following PMS model specification we assume that each economic variable, $y_{it}$, $i=1,2,\dots, N$, follow a conditionally linear and Gaussian process with mean and variance driven by 2-state Markov chain processes, $\lbrace S_{y,it}\rbrace$, $i=1,2,\dots, N$, with values in $\lbrace 1,2\rbrace $ and with time-varying transition probability matrices, $P_{it}$. Similarly, we assume that the common financial variable, $x_{t}$, follows a conditionally linear and Gaussian process with mean and variance driven by 2-state Markov chain process $\lbrace S_{x,t}\rbrace$ with values in $\lbrace 1,2\rbrace$ and with time-varying transition 
probability matrices, $P_{t}$. {\textcolor{black}{Regimes 1 and 2 respectively denotes recession and expansion.}} 

The measurement equations are written as:
\begin{eqnarray}
y_{it}&=&\sum\limits_{k=1,2}\mathbb{I}_{\{k\}}(S_{y,it})\left[ \Psi_{ik}'z^{BC}_{it}+\sigma_{ik}\varepsilon_{it}\right],\quad \varepsilon_{it} \sim \mathcal{N}(0,1),~ i=1, 2,\dots, N,~ t=1, 2,\dots, T 
\label{eq1:Measurement1}\\
x_{t}&=&\sum\limits_{k=1,2}\mathbb{I}_{\{k\}}(S_{x,t})\left[ \Phi_{k}'z^{FC}_{t}+\tau_{k}\eta_{t}\right],\quad\quad \eta_{t} \sim \mathcal{N}(0,1),~ t=1, 2,\dots, T  
\label{eq1:Measurement2}
\end{eqnarray}
where $(z^{BC}_{it})'=\left(1,z^{BC}_{i,2t},\ldots,z^{BC}_{i,mt}\right)$, and  $(z^{FC}_{t})'=\left(1,z^{FC}_{2t},\ldots,z^{FC}_{t}\right)$ are, respectively, vectors of the country specific and aggregate climate shocks and $\Psi_{ik}$ and $\sigma_{ik}$ are country-and-regime specific parameters, and $\Phi_{k}$ and $\tau_{k}$ are the common financial-and-regime specific parameters. The symbol $\mathbb{I}_{E}(X)$ is the indicator function which takes value $1$ if $X \in E$ and $0$ otherwise.  

We assume that the time-varying transition probabilities of each country-specific cycle and the financial cycle do not only depend on their own past values but involve also the past regimes of other countries and financial cycle chains in the panel. That is, the time varying transition matrices $P_{it}$ of the $i$-th country and $P_{t}$ of the  financial variable, respectively, have $l$-th row and $k$-th column elements $P_{it,lk}$ and $P_{t,lk}$ defined as:

\begin{eqnarray}
P_{it+1,lk}&=&\mathbb{P}\left( S_{y,it+1}=l \vert S_{y,it}=k,S_{y,-i,t},S_{x,t}\right)
\label{eq3:TransitionKernel1}\\
P_{t+1,lk}&=&\mathbb{P}\left( S_{x,t+1}=l \vert S_{x,t}=k,S_{y,t}\right)
\label{eq3:TransitionKernel2}
\end{eqnarray}
representing the conditional probabilities that unit $i$ and the financial variable, respectively, moves to the regime $l\in \lbrace 1, 2 \rbrace$ at time $t+1$, $S_{y,t}=(S_{y,1,t},\ldots,S_{y,N,t})'$ is the set of country business cycles and $S_{y,-i,t}=(S_{y,1,t},\ldots,,S_{y,i-1,t}, S_{y,i+1,t}, \ldots, S_{y,N,t})'$. The multivariate logistic transformation is widely used in the literature for parametrizing an endogenous time varying transition model. However, such specification implies a non-linear transformation of the parameters which makes inference more difficult. It is then convenient to use a linear parametrization of the transition matrices:
\begin{eqnarray}
P_{it+1,lk}&=&\alpha_{i} p_{i,lk}+\beta_{i} (S_{x,t}-1)+\gamma_{i} m_{k}\left( S_{y,t} \right),\, i=1,\dots, N \quad\quad \hbox{(business cycles transition)}
\label{eq3:transitionProb1}\\
P_{t+1,lk}&=&\alpha_{f} p_{f,lk}+\beta_{f} (S_{x,t}-1)+\gamma_{f} m_{k}\left( S_{y,t} \right), \,\,\,\,\quad \quad\quad\quad\quad  \hbox{(financial cycle transition)}
\label{eq3:transitionProb2}  
\end{eqnarray}
with $0<\alpha_{i} \leq 1$, $0\leq \beta_{i} <1$, $0\leq \gamma_{i} <1$, $0<\alpha_{f} \leq 1$, $0\leq \beta_{f} <1$, $0\leq \gamma_{f} <1$, $\alpha_{i}+\beta_{i}+\gamma_{i}=1$, $\alpha_{f}+\beta_{f}+\gamma_{f}=1$, $i=1,\dots, N$,  interactions parameters and $p_{i,lk}$ and $p_{f,lk}$  transition probability parameters such that $p_{i,l1}+ p_{i,l2}=1$, $ p_{f,l1} + p_{f,l2}=1$.  The time-invariant component of the transition probability, $p_{i,lk}$ detects the fixed component of country $i$'s  business cycle, while $p_{f,lk}$ detects that of the aggregated financial cycle. The time-varying component is driven by two factors: a global business cycle measure $m_{k}(S_{y,t})$ and a common financial cycle $S_{x,t}$. The global interaction factor $m_{k}\left( S_{y,t} \right)$ is given  by the proportion of chains in regime $k$ at time $t$ that is:
\begin{align}
m_{k}\left( S_{y,t} \right)&=\frac{1}{N}\sum \limits_{j=1}^{N} \mathbb{I}_{\{k\}}(S_{y,j,t})  &  \label{eq4:mtk}
\end{align}

The second term on the right-hand side of Equation \ref{eq3:transitionProb1} represents the contribution of the aggregate financial cycle on the country-specific business cycles while the third term represents the aggregate global business cycle, measured by the interactions among the business cycles of the participating units, on the country-specific business cycle. On the other hand, the second term in Equation \ref{eq3:transitionProb2} is the reinforcement effect on the financial cycle, while the third term provides information on the effect of the global business cycle on the financial cycle. Relative to the expected value of the time varying transition probabilities of the country-specific economic cycles when the aggregate financial cycle is in state 1 (recession), the parameters $\beta_{i}$'s and $\beta_{f}$, respectively, measures the contribution of the expansionary (state 2) phase of the aggregate financial cycle on the country-specific cycles and the financial cycle. The parameters $\gamma_{f}$ and $\gamma_{i}$'s allows us to measure, respectively, the sensitivity of the aggregate financial cycle and the country specific-business cycle to the fluctuation in the aggregated EU business cycle.

As regards inference, we build on the Bayesian setting and the Markov Chain Monte Carlo (MCMC) sampling for posterior approximation as in \cite{Billio2013} and \cite{Casarinetal2018}.
\subsection{Bayesian inference}
\label{sec:BayesInfe}
\subsubsection{Likelihood function and prior distributions}

\noindent Let $\boldsymbol{\theta}=(\Psi, \sigma, \Phi, \tau, vec(P_{1})',\ldots,vec(P_{N})',vec(P)',\alpha_{1},\dots,\alpha_{N}, \beta_{1},\dots, \beta_{N}, \gamma_{1},\dots, \gamma_{N}, \alpha_{f}, \beta_{f}, \gamma_{f} )$ 
be the vector of parameters with
$$
\begin{aligned}
&\Psi=(\Psi_{11}',\Psi_{12}',\ldots, \Psi_{i1}',\Psi_{i2}',\ldots,\Psi_{N1}',\Psi_{N2}'),\\
& \sigma = (\sigma_{11},\sigma_{12},\ldots,\sigma_{i1},\sigma_{i2},\ldots,\sigma_{N1},\sigma_{N2}),\\
&\Phi = (\Phi_{1}',\Phi_{2}'),\quad\quad \tau = (\tau_{1},\tau_{2}),\quad\quad \forall i=1,\ldots, N\\
\end{aligned}
$$
Let us define the allocation variables ${\xi_{yk,it}}=\mathbb{I}_{\{k\}}(S_{y,it})$, and ${\xi_{xk,t}}= \mathbb{I}_{\{k\}}(S_{x,t})$ to indicate the regime $k$ that the observation $y_{it}$ and $x_{t}$ belong to respectively and denote with $v_{1:T}= (v_{1},\dots,v_{T})$ the collection overtime of a variable $v_{t}$, $t=1,\dots,T$. By using the sequential factorization of the likelihood, the complete-data likelihood of the PMS model writes as follows:
\begin{flalign}
\begin{split}
&\mathcal{L} (Y_{1:T},X_{1:T}, S_{y,1:T}, S_{x,1:T} \mid \boldsymbol{\theta}, Z_{1:T})\\
&=\prod\limits_{t=1}^{T}\prod\limits_{l=1}^{2}\prod\limits_{k=1}^{2}
\left[\left(\prod\limits_{i=1}^{N}f_{y}(y_{it} | z_{t},S_{y,it},\boldsymbol{\theta})^{\xi_{yl,it}}P_{it,lk}^{\xi_{yl,it}\xi_{yk,it-1}}\right)
f_{x}(x_{t} | z_{t},S_{x,t},\boldsymbol{\theta})^{\xi_{xl,t}}P_{t,lk}^{\xi_{xl,t}\xi_{xk,t-1}}\right] \\ 
&=\prod\limits_{t=1}^{T}\prod\limits_{l=1}^{2}\prod\limits_{k=1}^{2}
\left[\left(
\prod\limits_{i=1}^{N}(2\pi\sigma_{il}^2)^{-\frac{{\xi_{yl,it}}}{2}}\exp\left({{-\frac{{\xi_{yl,it}}}{2\sigma_{il}^2}}}(y_{it}-\Psi_{il}'z^{BC}_{it})^2\right)(\alpha_{i} p_{i,lk}+\beta_{i} (S_{x,t}-1)+\gamma_{i} m_{t,k})^{\xi_{yl,it}\xi_{yk,it-1}}\right)\right.\\
&\quad\quad\quad\quad\quad\quad\left.\times(2\pi\tau_{l}^2)^{-\frac{{\xi_{xl,t}}}{2}}\exp\left({{-\frac{{\xi_{xl,t}}}{2\tau_{l}^2}}}(x_{t}-\Phi_{l}'z^{FC}_{t})^2\right)
 (\alpha_{f} p_{f,lk}+\beta_{f} (S_{x,t}-1)+\gamma_{f} m_{t,k})^{\xi_{xl,t}\xi_{xk,t-1}}\right]
\label{eqn:likelihood2} 
\end{split} &
\end{flalign}

Given that a Bayesian model specification is not complete without the priors, we make the following distributional assumptions about the priors for the unit specific and financial variable parameters:
\begin{flalign}
\Psi_{il} &\sim \mathcal{MN}(m_{y,il},\Sigma_{y,il}) \label{eqn:mupriorY} \\
\Phi_{l} &\sim
\mathcal{MN}(m_{x,l},\Sigma_{x,l}) \label{eqn:mupriorX} \\
\sigma_{il}^2 &\sim \mathcal{IG}(\alpha_{y,il},\beta_{y,il})  \label{eqn:sigmapriorY} \\
\tau_{l}^2 &\sim \mathcal{IG}(\alpha_{x,l},\beta_{x,l})  \label{eqn:sigmapriorX} \\
(p_{i,l1},p_{i,l2})&\sim \mathcal{D}ir(\delta_{y,i1},\delta_{y,i2}) \label{eqn:transprob}\\
(p_{f,l1},p_{f,l2})&\sim \mathcal{D}ir(\delta_{x,1},\delta_{x,2}) \label{eqn:transprobf}\\
(\alpha_{i}, \beta_{i}, \gamma_{i}) &\sim \mathcal{D}ir(\varphi_{y,i1},\varphi_{y,i2},\varphi_{y,i3}) \label{eqn:alpha}\\
(\alpha_{f}, \beta_{f}, \gamma_{f}) &\sim \mathcal{D}ir(\varphi_{x,1},\varphi_{x,2}, \varphi_{x,3}) \label{eqn:alphaf}
\end{flalign}

with $l=1,2$ and $i=1,\ldots,N$, where $\mathcal{IG}(\alpha,\beta)$ denote the inverse gamma distribution with parameters: $\alpha$ and $\beta$ and $\mathcal{D}ir(\delta_{1},\ldots,\delta_{K})$ the $K$ 
dimensional Dirichlet distribution with parameters: $\delta_{1},\ldots,\delta_{K}$. 

A commonly observed issue while dealing with Bayesian analysis involving Markov-switching processes, is the non-identifiability of the parameters due to label switching. The identification issue relates to the invariant property of the posterior distribution of the Markov-switching parameters when the labels of the parameters are permuted. As a consequence, an identical set of marginal posterior distributions is obtained for each switching component of the parameters. In addition, both the MCMC simulation of the posterior distribution of the switching parameters and the interpretation of the switching parameters are affected. See among others \cite{Celeux1998},  \cite{Fruhwirth2001}, and \cite{Fruhwirth2006} for discussion on label switching and non-identifiability issues as it affects MCMC based Bayesian inference. Several alternative approach are available in the literature for handling this identification issue. We however follow the commonly used approach in the macroeconomics literature by imposing a restriction on the intercept of each of the measurement equations i.e.  $\Psi_{i01} \leq \Psi_{i02}$ for all $i$ and $\Phi_{01} \leq \Phi_{02}$. This approach is adopted because of its ease and natural interpretation of the different states (e.g. recession and expansion) of the cycles.

\subsection{Posterior simulation}
\noindent The joint posterior distribution of the parameters $\boldsymbol{\theta}$ and the latent variables $S_{x}$ and $S_{y}$ is:
\begin{flalign}
\pi(\boldsymbol{\theta}, S_{y}, S_{x}  \mid Y, X, Z) &\propto \mathcal{L} (Y,X, S_{y},S_{x} \mid \boldsymbol{\theta}, Z_{1:T} ) \pi(\boldsymbol{\theta}),
\end{flalign}
where 
$$
\pi(\boldsymbol{\theta}) = \pi(\sigma)\pi(\Psi)\pi(\tau)\pi(\Phi)\prod\limits_{i=1}^{N}\pi(\alpha_{i},\beta_{i},\gamma_{i})\pi(\alpha_{f},\beta_{f},\gamma_{f})
$$
is the joint prior distribution. The derivation of the distributions and details of the sampling method are provided in Appendix \ref{Appendix2}. We develop a sampling algorithm based on the full conditional posterior distributions. The Gibbs sampler iterates as follows:

\begin{minipage}[t]{1\textwidth}
\begin{enumerate}
\item Draw $S_{y,i1:T}$ and $S_{x}$ respectively from $f(S_{y,i1:T}\mid Y_{1:T}, X_{1:T}, Z_{1:T},\boldsymbol{\theta}, S_{y,-i,1:T},S_{x,1:T})$ and \\
$f(S_{x}\mid Y_{1:T}, X_{1:T} ,Z_{1:T}, \boldsymbol{\theta}, S_{y,1:T})$, $i=1,\ldots,N$ . 
\item Draw $(\alpha_{i}, \beta_{i}, \gamma_{i})$, from $f((\alpha_{i}, \beta_{i}, \gamma_{i}) \mid Y_{1:T}, X_{1:T}, Z_{1:T}, S_{1:T},(p_{l1},p_{l2}))$, $i=1,\ldots,N$\\ where $S_{1:T}=(S_{y,1:T},S_{x,1:T})$.
\item Draw $(\alpha_{f}, \beta_{f}, \gamma_{f})$, from $f((\alpha_{f}, \beta_{f}, \gamma_{f})\mid Y_{1:T}, X_{1:T}, Z_{1:T}, S_{1:T},(p_{f,l1},p_{f,l2}))$.
\item Draw $(p_{i,l1},p_{i,l2})$ from $f((p_{i,l1},p_{i,l2}) \mid Y_{1:T}, X_{1:T}, Z_{1:T}, S_{1:T},(\alpha_{1},\dots,\alpha_{1N}, \beta_{1},\dots. \beta_{N}, \gamma_{1},\dots, \gamma_{N}))$,\\  $i=1,\ldots,N$, $l=1,2$.
\item Draw $(p_{f,l1},p_{f,l2})$ from $f((p_{f,l1},p_{f,l2}) \mid Y_{1:T}, X_{1:T}, Z_{1:T}, S_{1:T},(\alpha_{f}, \beta_{f},\gamma_{f}))$,  $i=1,\ldots,N$, $l=1,2$.
\item Draw $\Psi_{il}$ and $\Phi_{l}$, respectively from  $f(\Psi_{il} \mid Y_{1:T}, X_{1:T}, Z_{1:T}, S_{y,i1:T},\sigma_{il})$, and\\ $f(\Phi_{l} \mid Y_{1:T}, X_{1:T}, Z_{1:T}, S_{x,1:T},\tau_{l})$, $i=1,\ldots,N$, $l=1,2$. 
\item Draw $\sigma_{il}$ and $\tau_{l}$, respectively, from $f(\sigma_{il}\mid Y_{1:T}, X_{1:T}, Z_{1:T}, S_{y,i1:T},\Psi_{il})$  and \\
$f(\tau_{l} \mid Y_{1:T}, X_{1:T}, Z_{1:T}, S_{x,1:T},\Phi_{l})$ $i=1,\ldots,N$, $l=1,2$. 
\end{enumerate}
\end{minipage}	

\medskip 
Samples of the state trajectory corresponding to each of the unit-specific business cycle and financial cycle are drawn by Forward Filter Backward Sampling (FFBS) scheme. See \cite{Fruhwirth2006} for details of the FFBS.
We note here, as presented in Appendix \ref{Appendix2}, that the conditional posterior distributions of the interaction parameters, $(\alpha_{i}, \beta_{i}, \gamma_{i})$, $i=1,\dots, N$ and $(\alpha_{f}, \beta_{f}, \gamma_{f})$, are directly proportional to a polynomial of three (3) variables of degree $T+\varphi_{y,i1}-1$ in $\alpha$, $T+\varphi_{y,i2}-1$ in $\beta$, and  $T+\varphi_{y,i3}-1$ in $\gamma$. It may therefore be deduced that the conditional posterior distribution of the interaction parameters is a mixture of Dirichlet distributions. However, the weights and parameters of the mixture are not readily available or easily obtained. In-view of this, Metropolis Hastings (MH) strategies are implemented for generating proposals for the interaction parameters. More precisely, independent samples of the interacting parameters are generated from a mixture of three (3) equally weighted Dirichlet distributions with parameters $(T+\varphi_{y,i1},\varphi_{y,i2},\varphi_{y,i3})$, $(\varphi_{y,i1},T+\varphi_{y,i2},\varphi_{y,i3})$, and $(\varphi_{y,i1},\varphi_{y,i2},T+\varphi_{y,i3})$.

\section{Empirical results}
\label{EmpriccalResusult} 
\subsection{Data sources and description}
\label{DataDescription} 
We measure the business cycle of each of the 13 EU countries\footnote{Austria (AU), Belgium (BE), Denmark (DE), Finland (FI), France (FR), Germany (GE), Greece (GR), Ireland (IR), Italy (IT), Luxembourg (LU), Netherlands (NE), Portugal (PO), and Spain (SP)} using a seasonally adjusted IPI, sampled at the monthly frequency from February 1981 to December 2016 from the EUROSTAT/OECD databases. To measure the financial cycle, we follow the strategy described in \cite{TheConfBoard2011} and construct a composite financial index based on the three variables proposed by \cite{Borio2014} and \cite{Drehmannetal2012}: Credit to private sector in real terms, Credit-to-GDP-ratio and aggregate real property price measure reconstructed using single data on EU members countries released by FED. This composite index serve as handy summary measure of the behavior of the cyclical indicator and it tends to smooth out some of the volatility of individual series. The use of composite index is consistent with the traditional view of the business cycle developed by \cite{BurnsMitchell1946}. In addition, composite financial indicator is a relevant component used by authorities, such as the European Central Bank (ECB) or Bank for International Settlements (BIS), in monitoring EU stability, and usually the decisions made by some of these authorities can have a strong impact on the overall stability of the EU system, whereas industrial production decisions are not centralized. Since these series are all seasonally adjusted and available only at quarterly frequency, we implement the \cite{ChowLin1971} procedure to transform the data into monthly frequency. This procedure is undertaken to ensure that the sampling frequency of the financial index is in conformity with that of the economic index used for measuring the business cycle. Further description of sources and constructions of these data is provided in Appendix \ref{DataConstruction}.

\medskip
\noindent As regards to the climatic conditions, the three CEIs were selected for assessing the impact of climate shocks on the IPI growth and aggregate financial index growth of selected EU countries. We include three CEIs (Table \ref{ClimateIndexDescrip}) from a broader suite of CEIs defined and developed by the Expert Team on Climate Change Detection and Indices 
(ETCCDI)\footnote{\url{{https://www.wcrp-climate.org/data-etccdi}}}, and made available by the European Climate Assessment \& Dataset (ECA\&D) gridded observational dataset (E-OBS, version 19.0e)\footnote{Data accessed from \url{http://surfobs.climate.copernicus.eu/dataaccess/access_eobs_indices.php} on 12 Nov 2019} at a high-spatial resolution of $0.1\degree \times 0.1\degree $ ($\sim 11\times 11$ km at the equator).The three CEIs utilized in our study are assembled using daily maximum temperature (TX, Units: $\degree$C) and daily total precipitation (RR Units: $mm$), and are aggregated to the individual country level using country shape files from EUROSTAT\footnote{\url{http://ec.europa.eu/eurostat/web/gisco/geodata/reference-data/administrative-units-statistical-units.} The CEIs are aggregated from the gridded to country level using the extract function of the R raster package (\cite{HijmansEtten2014}). }. The monthly average of these extremes are used as proxy for the assessing their impact of the extremes on the financial index. 

\begin{table}[h]
\begin{center}
\begin{tabular}{ccccc}
\hline
CEI          & CEI  & Description & Units & Time scale\\
(Short Name) & (Long Name) &   &   &  \\
\hline
CSU  & Consecutive  & Maximum number of consecutive & Days & Monthly\\
     & Summer Days  & days when $TX\geq  25\degree C$ &   &  \\
SPI\tablefootnote{SPI is a probability index based on precipitation. It is designed to be a spatially invariant indicator of drought. SPI on 6-monthly timescale refers to precipitation in the previous 6-month period (with positive and negative values indicating wet and dry conditions respectively). We utilize monthly values falling under the following two intervals (i) [ less than -0.5 classified as drought-like conditions ] and (ii) [ greater than or equal to -0.5, classified as non-drought like conditions]. }  & Standardized & Measure of drought  & Unitless & 6-monthly\\
      & Precipitation Index & specified as a precipitation deficit &  & \\
r20mm & Very heavy  & Number of days when $RR\geq 20 mm$ & Days & Monthly\\
      & precipitation days &    &  &\\
\hline
\end{tabular}
\end{center}
\caption{Description of three CEIs used in our study. For a full description of CEIs and their potential applications in sectoral impacts, readers are guided to \cite{Mistry2019}.}\label{ClimateIndexDescrip}
\end{table}


While the CSU and r20mm follow the conventional definition in our study, the SPI on 6 monthly scales used here is transformed slightly for ease of interpretation. SPI which is computed by standardizing the deviation of observed monthly precipitation from the long-run mean, can take both negative- (representing dry period) and positive- (indicating wetness) values. Here we define a dichotomous variable to differentiate between the two conditions: SPI values less than -0.5 are tagged drought-like conditions and labeled 1, while values greater than or equal to -0.5 are tagged non-drought like conditions and labeled 0.

\begin{figure}[h!]
\begin{center}
\includegraphics[scale=1.0]{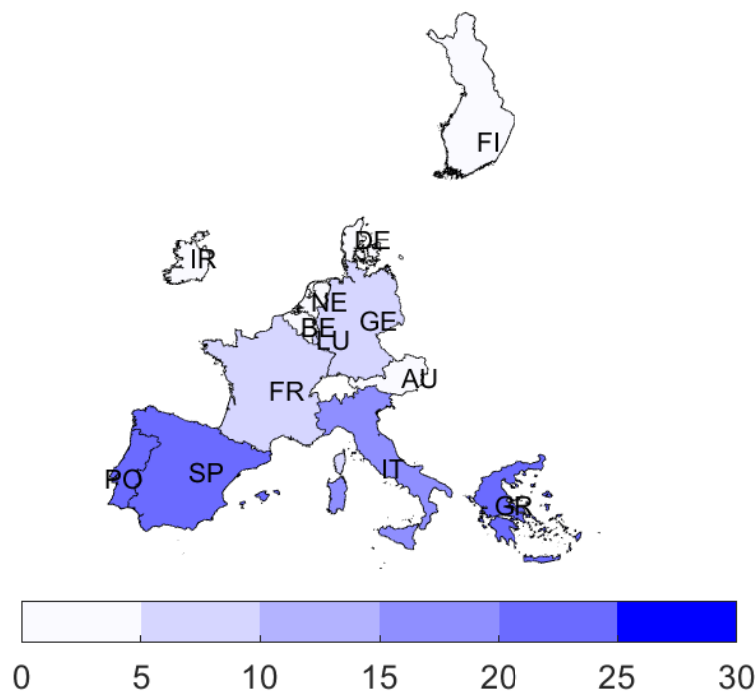}
\end{center} 
		\caption{Average temperature (Hot days (CSU, $T>25\degree C$)) in July over the period  1981 to  2016.}\label{PrelimCheck}		
\end{figure}

\medskip
\noindent In Appendix \ref{SummaryStats} a summary statistics of the data used in this study is presented. The average monthly industrial production growth rate of the EU countries under study falls between $0.038\%$ and $0.75\%$ with Ireland and Italy respectively having the highest and lowest monthly variation in their growth rates (see Table \ref{tab:Descrip_Stat}). In addition, Luxembourg and France are the only two nations with negatively skewed monthly growth rates. The variability in Ireland's monthly growth rate may be attributed to the sudden large  growth rate observed towards the end the sample period (see Figure \ref{PLOTIPI}). France, Germany and Italy tend to have a low and relatively stable monthly growth rate (see Figure \ref{PLOTIPI}). Greece, Italy, Portugal and Spain are countries with the highest spell of summer days (over 20 days) while Ireland experienced the least spell (mostly less than 5 days) of summer days (see Figure \ref{PLOTCSU}). As a visual illustration of the distribution of extreme weather events across the eurozone countries, Figure \ref{PrelimCheck}(a) displays the average CSU during the month of July. All the EU countries under study experienced a fair share of moderate to exceptionally dry weather condition (Figure \ref{PLOTSPI}). In Figure \ref{PLOTr20mm}, Austria and Portugal are noted to have experienced a relatively higher monthly frequency of heavy rainfall than others. 

In the following section, we apply our model (Equations \ref{eq1:Measurement2} - \ref{eq3:transitionProb2}) to a selected set of EU countries business cycles and the aggregate financial cycle for assessing the relevance of the climate shocks and financial shocks in explaining the synchronization between business and financial cycle.
We perform 20000 Gibbs iterations after a burn-in period of 2000 samples. However, to reduce the very high positive serial correlation often associated with samples produced by Gibbs sampler, we consider every 10th draw after the burn-in period. This approach reduces the effective sample size upon which the following results are based to 2000 samples.

\subsection{Business and financial cycles extraction}

\begin{figure}[h!]
        \centering
        \subfigure[Without climate shocks]{
                \includegraphics[width=0.4\textwidth]{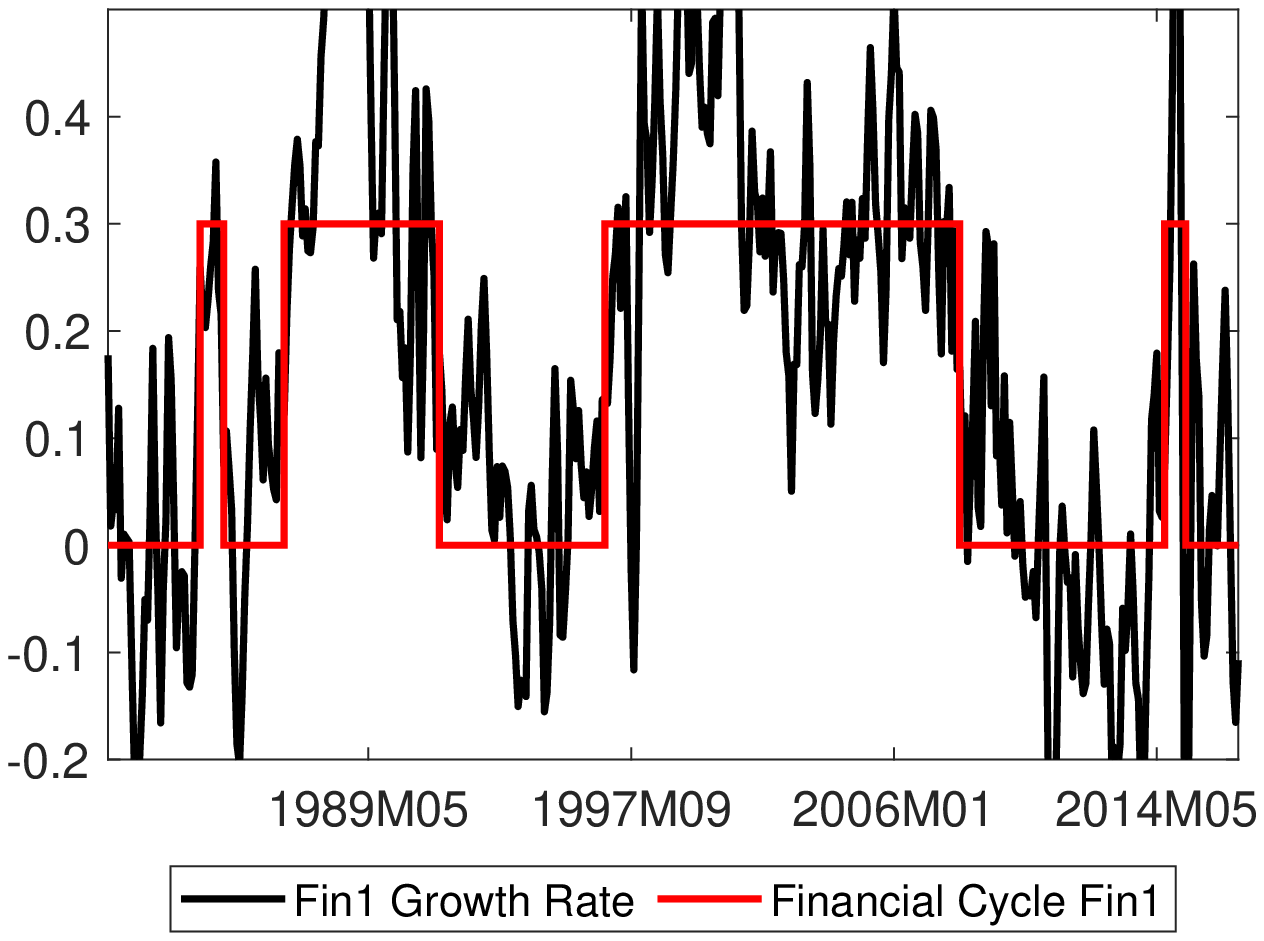}
                \label{fig:FinCycleWITHOUTCI}}
        ~
        \subfigure[With climate shocks]{
                \includegraphics[width=0.4\textwidth]{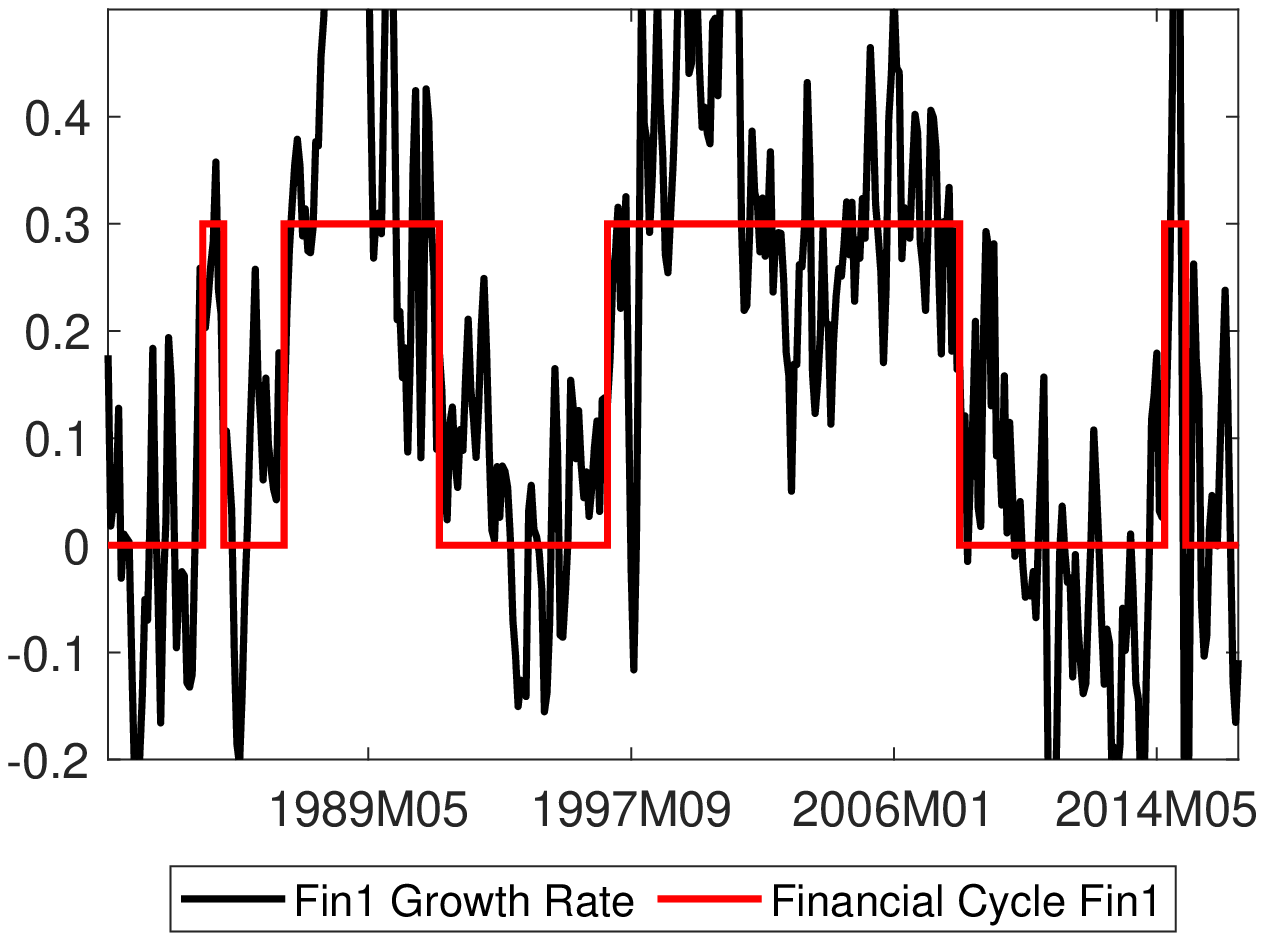}
                \label{fig:FinCycleWITHCI}}                   
		\caption{Financial index growth rates (black solid) and the estimated financial cycle $\hat{S}_{x,t}$ (step-wise red) for the PMS model in Equations \ref{eq1:Measurement2} - \ref{eq3:transitionProb2}. Figure \ref{fig:FinCycleWITHOUTCI} displays the result of the PMS model in the absence of climate shocks while Figure \ref{fig:FinCycleWITHCI} reports the result of the PMS model subject to climate extreme indices  CSU, SPI and r20mm capturing, respectively, consecutive hot days , droughts, and number of days with heavy rainfall. 
		}\label{FinCycle}
\end{figure}

\noindent Figure \ref{FinCycle} shows the growth rate in the financial index Fin1 (black solid line) together with the estimated financial cycle $\hat{S}_{x,t}$ (red stepwise line) under two different specifications of the PMS model. Figure \ref{fig:FinCycleWITHOUTCI} showcases the estimate of the financial cycle when climate shocks are exempted from the measurement equations of the PMS model while Figure \ref{fig:FinCycleWITHCI} reports estimate of the Financial cycle when climate shocks are taken into consideration in the PMS model. There appears to be no difference in the dynamics of the estimated financial cycle obtained under the two specifications of the PMS model. Furthermore, consistent with findings of studies that use similar methodology to measure the financial cycle (\cite{Drehmannetal2012},  \cite{ECB2014}) and reports on financial crisis episodes described by \cite{Dupreyetal2016} and \cite {Watts2016}, the following episodes of financial crisis in Euro Area: 1981-1983 Equity and currency crisis (France, Germany, Italy); 1992-1995 Banking, equity and currency crisis (France, Germany, Italy, Spain); 2008-2013 Banking and equity crisis (France, Germany, Italy, Spain, UK); 2015-2016 Banking crisis in Italy and Spain may be identified in Figure \ref{FinCycle}. 


\begin{figure}[h!]
        \centering
        \subfigure[Without climate shocks]{
                \includegraphics[width=0.4\textwidth]{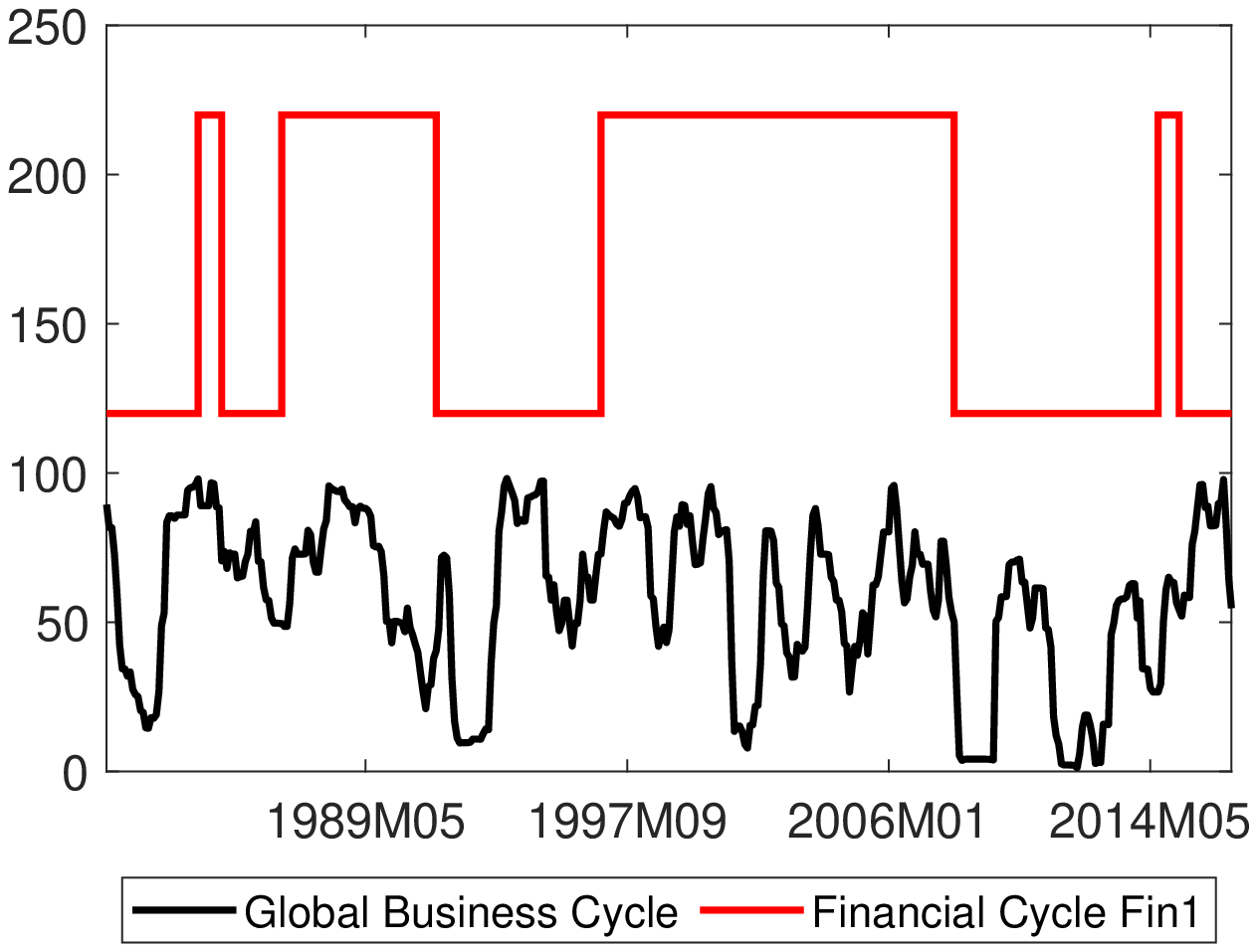}
                \label{fig:HidSwitch2WITHOUTCI}}
        ~
        \subfigure[With climate shocks]{
                \includegraphics[width=0.4\textwidth]{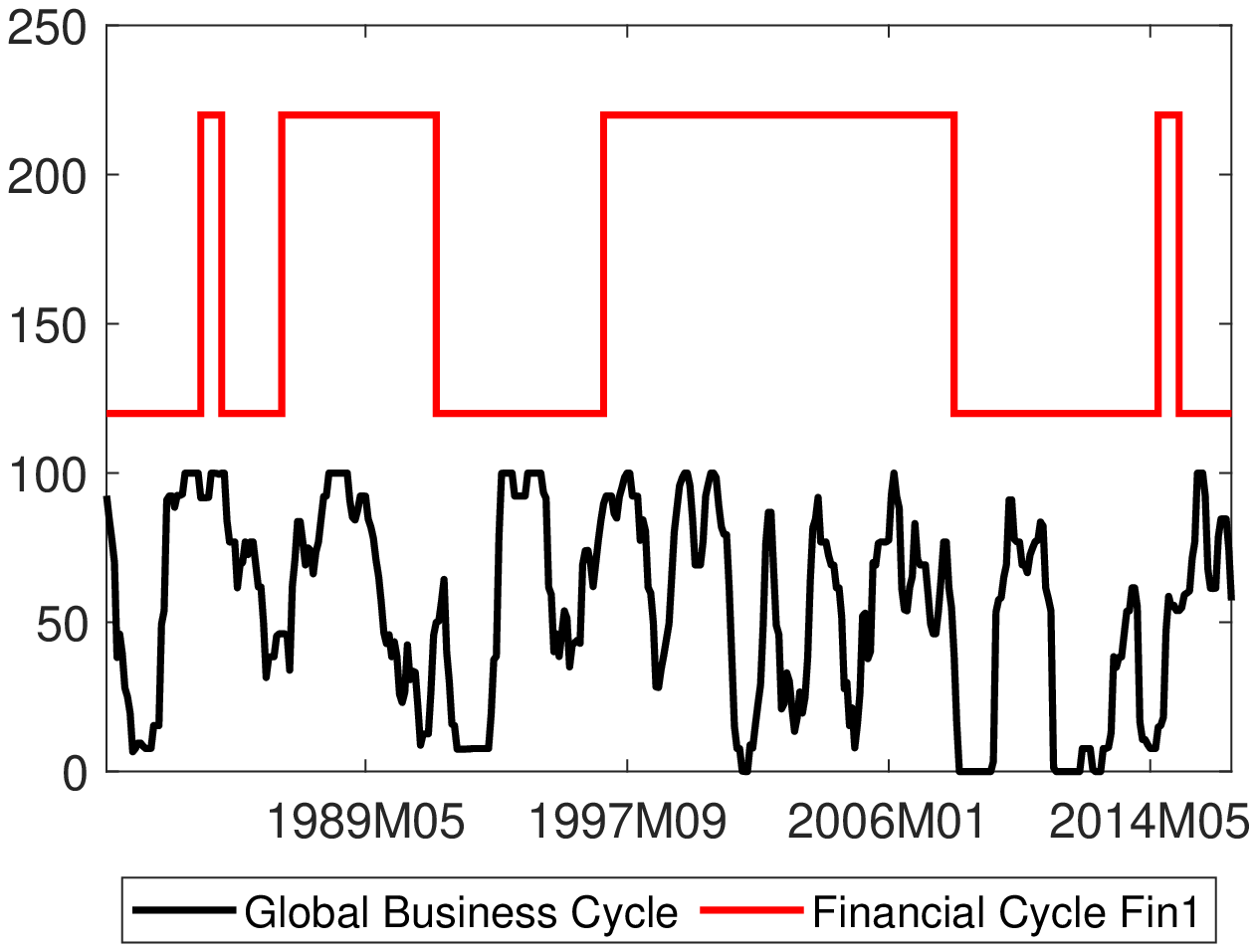}
                \label{fig:HidSwitch2WITHCI}}                   
		\caption{Estimated global common business cycles ($m_{2}(\hat{S}_{y,t})$, black line) and financial cycle ($\hat{S}_{x,t}$ red line) for PMS model. The Figures \ref{fig:HidSwitch2WITHOUTCI} and \ref{fig:HidSwitch2WITHCI}, respectively, display the results from the PMS model when climate shocks are not considered and when they are. The global interaction factor $m_{k}\left( S_{y,t} \right)$ is given by the proportion of chains in regime $k$ at time $t$ and computed using Equation \ref{eq4:mtk}. Sample period: February 1981 to December 2016 (month on month).}\label{Global}
\end{figure}

\medskip
\noindent Figure \ref{Global}, on the other hand, shows the extracted global business cycle $m_{2}(\hat{S}_{y,t})$ (black line), which represent the proportion of countries in regime 2 (expansion). The same figure exhibits the estimated financial cycle $\hat{S}_{x,t}$ (red line). Figures \ref{fig:HidSwitch2WITHOUTCI} and \ref{fig:HidSwitch2WITHCI} respectively display the extracted global business cycles obtained from the PMS model in the absence of climate index and that for which climate shocks are accounted for. The peaks and troughs of the global business cycle obtained under the PMS model subject to climate shocks appear to be slightly larger when compared to that obtained from the restricted PMS model. This difference may be attributed to the presence of Climate shocks.

Turning to the degree of synchronization between the aggregated financial cycle and the business cycle of the country $i$, we follow \cite{Harding2002} by computing the Concordance Index (CI) using 
\begin{equation}\label{eq:CI}
CI_{i} =\frac{1}{T} \left\{\sum_{t=1}^{T}(S_{x,t}-1) \cdot (S_{y,it}-1)+ \sum_{t=1}^{T} (2-S_{x,t})\cdot (2-S_{y,it})\right\}
\end{equation}
where $S_{y,it}$ indicates the business cycle series and $S_{x,t}$ the financial cycle series. The ${CI_i}$ takes value between zero and one. If the ${CI_i}$ equals one (zero), cyclical movements are perfectly synchronized (anti-cyclical). This statistics provides an avenue to examine the impact, if there exists any, of climate shocks on the synchronization of the cycles. By applying the pairwise concordance statistics given above, we obtain the synchronization results in Table \ref{tab:ConcodancePMS}. We note that the data above the diagonal in Table \ref{tab:ConcodancePMS} represents the 
result of the synchronization between the business and financial cycle extracted from the climate shocks free PMS model while the data below the diagonal reports estimates of the comovements of the business and financial cycle using PMS model subject to all the climate shocks. Almost all the countries exhibit a level of synchronization above 0.5 with each other and the financial cycle. The absolute difference in the degree of synchronization between the business cycle and financial cycle under the different specifications of the PMS model is less than 0.03. This observation suggests that climate shocks mildly impacts on the synchronization between business and financial cycle. We however
note from this table that by accounting for climate shocks, the estimated degree of pairwise synchronization between the business cycles are mostly larger than
those obtained by neglecting climate shocks. In particular, the marked increase in the estimated degree of pairwise synchronization between the business cycle of France and Denmark (0.23 units) and that of Denmark and Luxembourg (0.19 units) may be associated to the presence of shocks in the PMS model specification.

\begin{table}[h!]
\centering
\begin{threeparttable}
\captionsetup{font=small,labelfont=bf,labelsep=period}
  \caption{Pairwise concordance statistics between the business and financial cycles estimated using the PMS model}
\begin{scriptsize}  
    \begin{tabular}{c|rrrrrrrrrrrrr|r}
          & \multicolumn{1}{l}{AU} & \multicolumn{1}{l}{BE} & \multicolumn{1}{l}{DE} & \multicolumn{1}{l}{FI} & \multicolumn{1}{l}{FR} & \multicolumn{1}{l}{GE} & \multicolumn{1}{l}{GR} & \multicolumn{1}{l}{IR} & \multicolumn{1}{l}{IT} & \multicolumn{1}{l}{LU} & \multicolumn{1}{l}{NE} & \multicolumn{1}{l}{PO} & \multicolumn{1}{l}{SP} & \multicolumn{1}{l}{FIN1} \\
\hline              
    AU    & 1.00  & 0.82  & 0.61  & 0.71  & 0.65  & 0.82  & 0.70  & 0.61  & 0.78  & 0.63  & 0.68  & 0.64  & 0.79  & 0.64 \\
    BE    & 0.80  & 1.00  & 0.61  & 0.75  & 0.66  & 0.80  & 0.64  & 0.63  & 0.81  & 0.63  & 0.67  & 0.65  & 0.75  & 0.58 \\
    DE    & 0.69  & 0.66  & 1.00  & 0.59  & 0.47  & 0.61  & 0.65  & 0.69  & 0.62  & 0.49  & 0.65  & 0.59  & 0.51  & 0.50 \\
    FI    & 0.70  & 0.74  & 0.65  & 1.00  & 0.57  & 0.67  & 0.63  & 0.62  & 0.74  & 0.57  & 0.72  & 0.62  & 0.68  & 0.52 \\
    FR    & 0.78  & 0.81  & 0.70  & 0.77  & 1.00  & 0.67  & 0.59  & 0.54  & 0.66  & 0.76  & 0.57  & 0.59  & 0.79  & 0.58 \\
    GE    & 0.80  & 0.82  & 0.67  & 0.69  & 0.76  & 1.00  & 0.69  & 0.58  & 0.76  & 0.59  & 0.65  & 0.61  & 0.74  & 0.65 \\
    GR    & 0.71  & 0.68  & 0.65  & 0.65  & 0.71  & 0.70  & 1.00  & 0.67  & 0.62  & 0.56  & 0.64  & 0.59  & 0.64  & 0.58 \\
    IR    & 0.66  & 0.63  & 0.70  & 0.62  & 0.67  & 0.61  & 0.68  & 1.00  & 0.64  & 0.51  & 0.63  & 0.64  & 0.59  & 0.61 \\
    IT    & 0.73  & 0.81  & 0.69  & 0.72  & 0.79  & 0.78  & 0.65  & 0.66  & 1.00  & 0.68  & 0.70  & 0.65  & 0.77  & 0.58 \\
    LU    & 0.73  & 0.77  & 0.68  & 0.65  & 0.75  & 0.72  & 0.67  & 0.64  & 0.76  & 1.00  & 0.48  & 0.57  & 0.73  & 0.55 \\
    NE    & 0.68  & 0.64  & 0.63  & 0.69  & 0.70  & 0.66  & 0.63  & 0.60  & 0.71  & 0.60  & 1.00  & 0.62  & 0.63  & 0.55 \\
    PO    & 0.63  & 0.68  & 0.67  & 0.62  & 0.70  & 0.65  & 0.63  & 0.65  & 0.71  & 0.67  & 0.60  & 1.00  & 0.69  & 0.51 \\
    SP    & 0.72  & 0.74  & 0.64  & 0.71  & 0.77  & 0.72  & 0.70  & 0.68  & 0.82  & 0.73  & 0.67  & 0.68  & 1.00  & 0.61 \\
\hline    
    FIN1  & 0.66  & 0.58  & 0.48  & 0.53  & 0.59  & 0.64  & 0.58  & 0.60  & 0.56  & 0.52  & 0.54  & 0.51  & 0.58  & 1.00 \\
\hline    
    \end{tabular}%
\begin{tablenotes}
       \item NOTE: The matrix is not symmetric because the data below and above the diagonal, respectively, refer to estimates of the business
       cycle synchronization when the PMS model account for climate shocks and when they are ignored. 
   \end{tablenotes}
   \end{scriptsize}\label{tab:ConcodancePMS}%
\end{threeparttable}    
\end{table}%

\begin{figure}[h!]
\begin{center}
\begin{tabular}{|c|c|c|c|}
\hline
 &Idiosyncharctic ($\alpha$) & Financial cylcle ($\beta$) & Global ($\gamma$)\\
 \hline
\rot{without climate indicator}&
\includegraphics[scale=0.4]{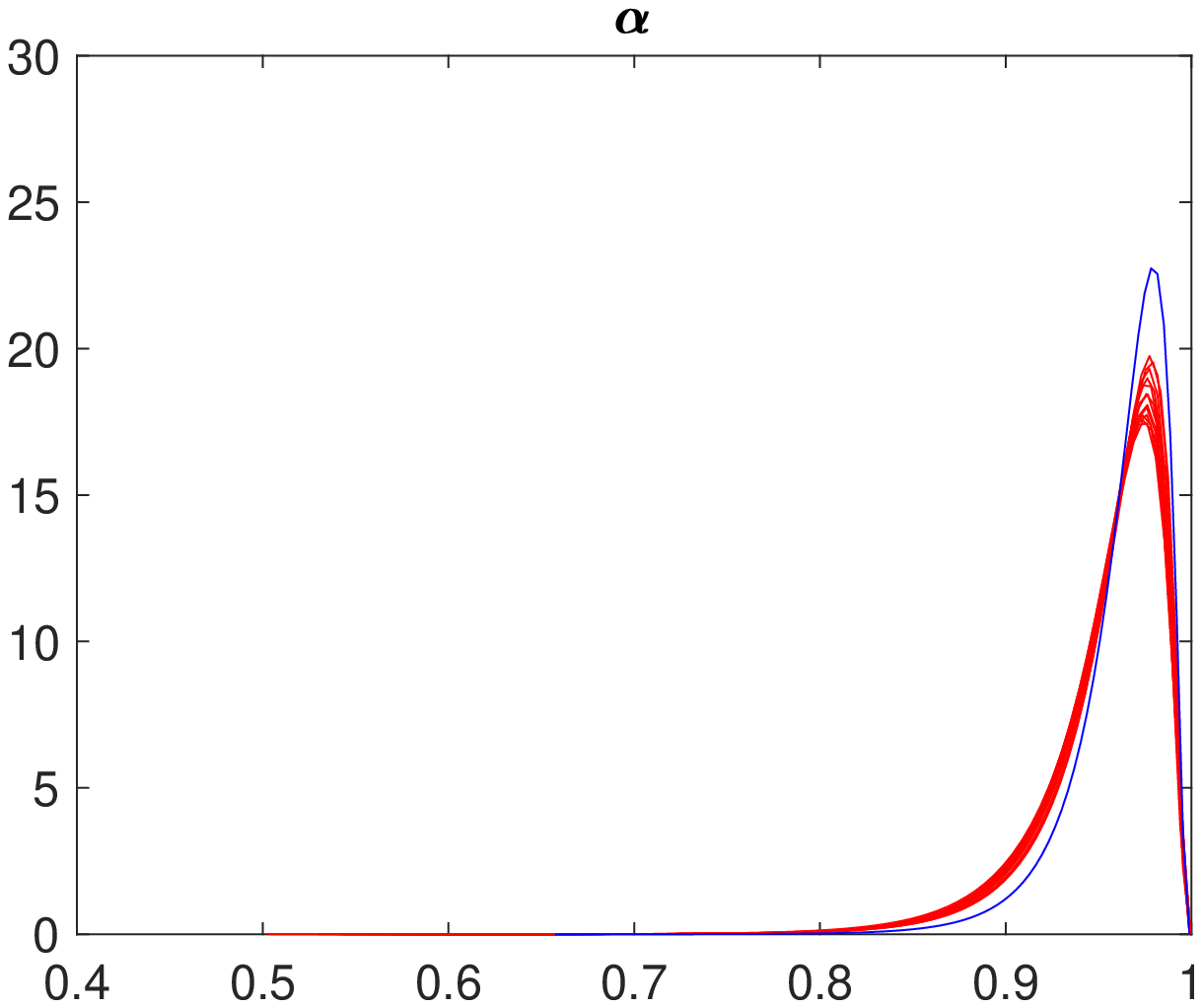}&\includegraphics[scale=0.4]{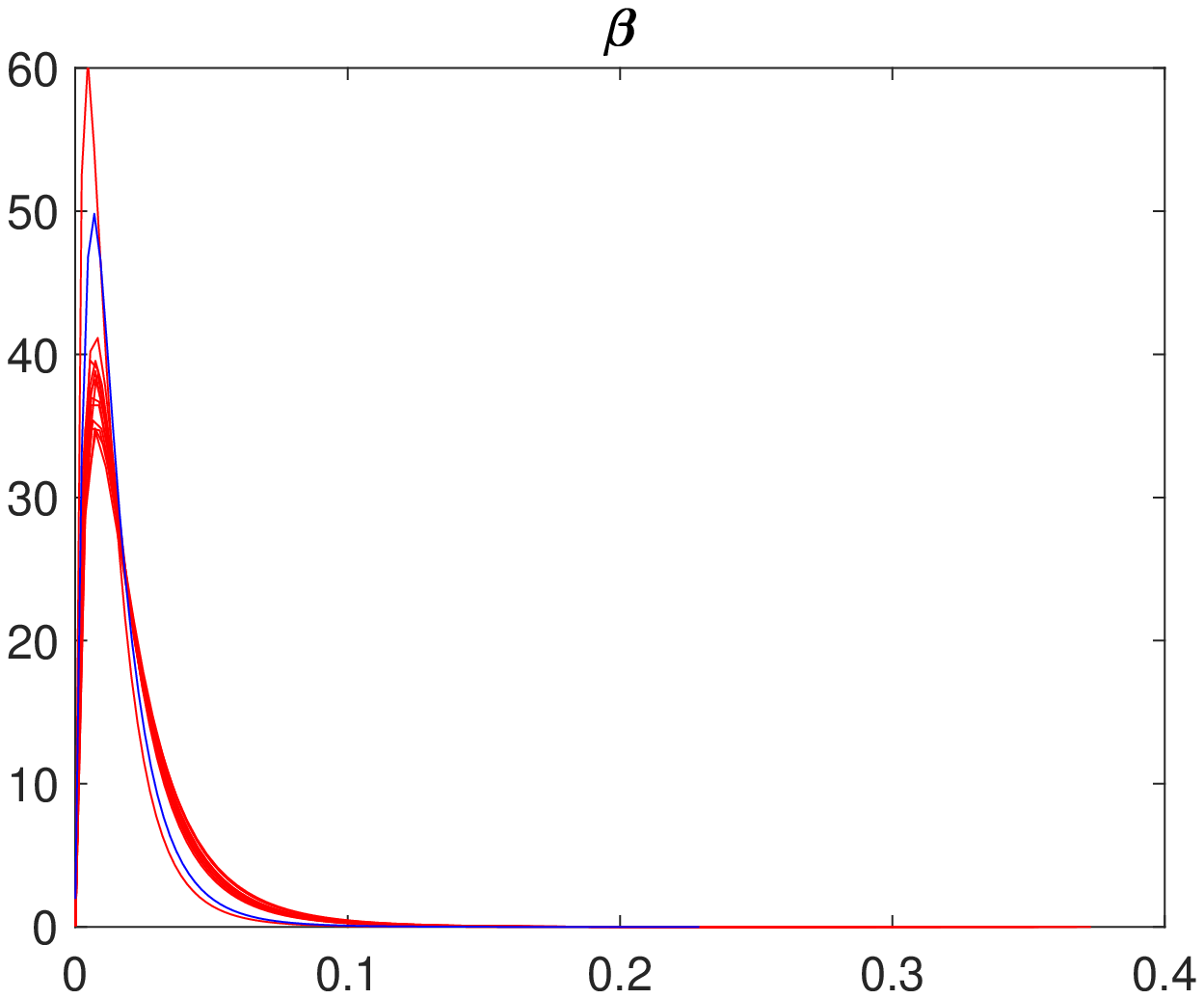}&\includegraphics[scale=0.4]{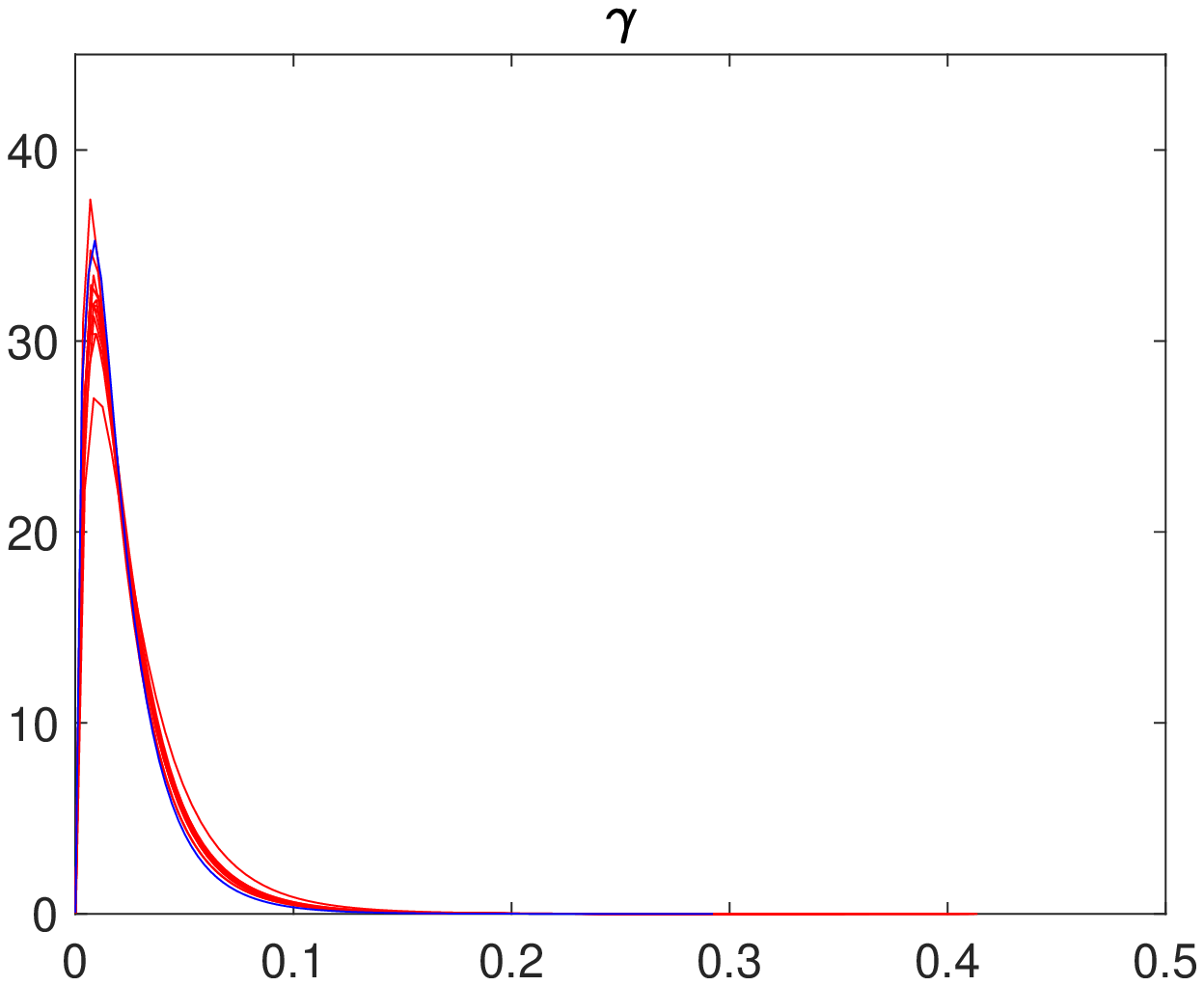}\vspace{5pt}\\
\hline
\rot{with ALL climate indicators}&
\includegraphics[scale=0.4]{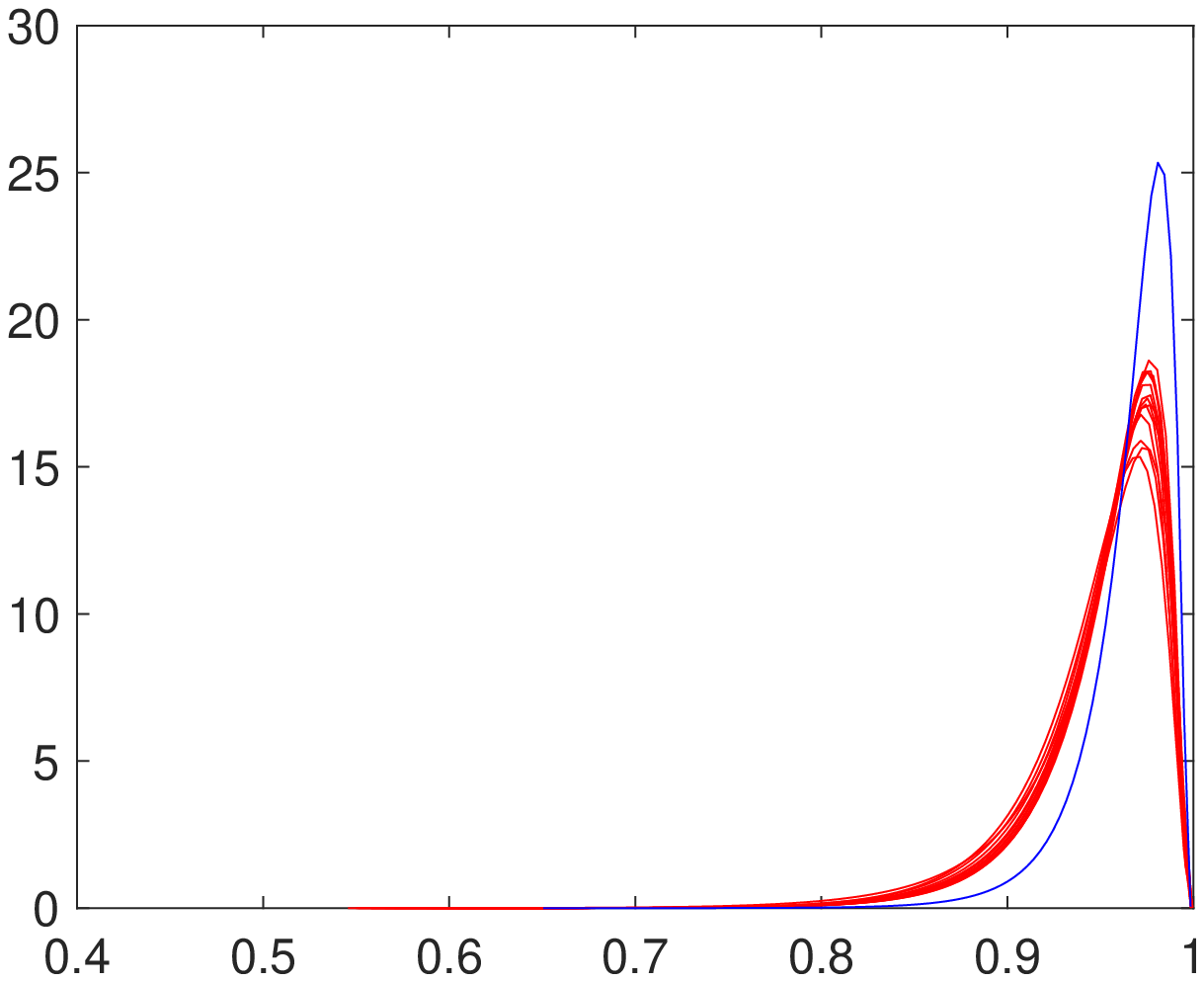}&\includegraphics[scale=0.4]{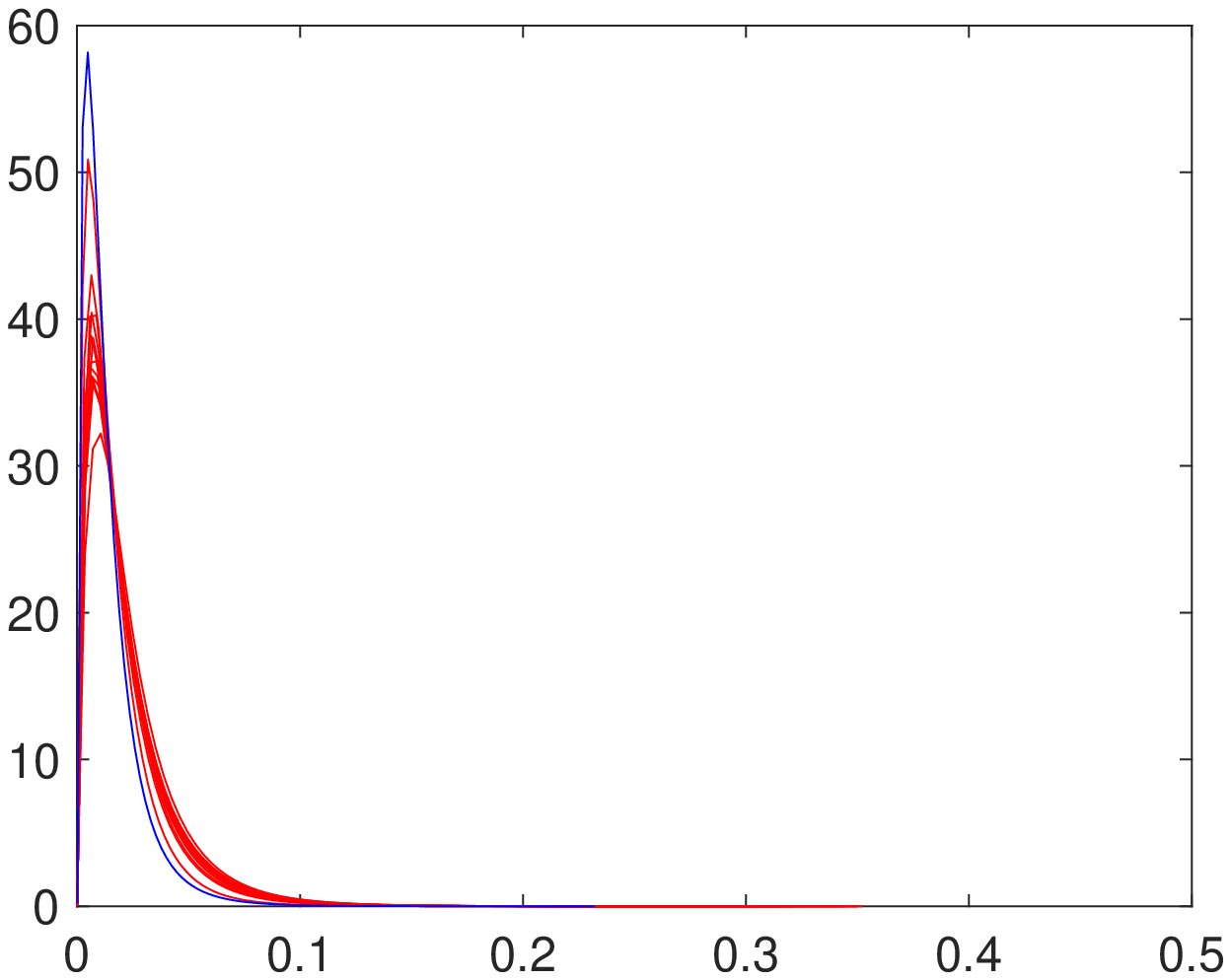}&\includegraphics[scale=0.4]{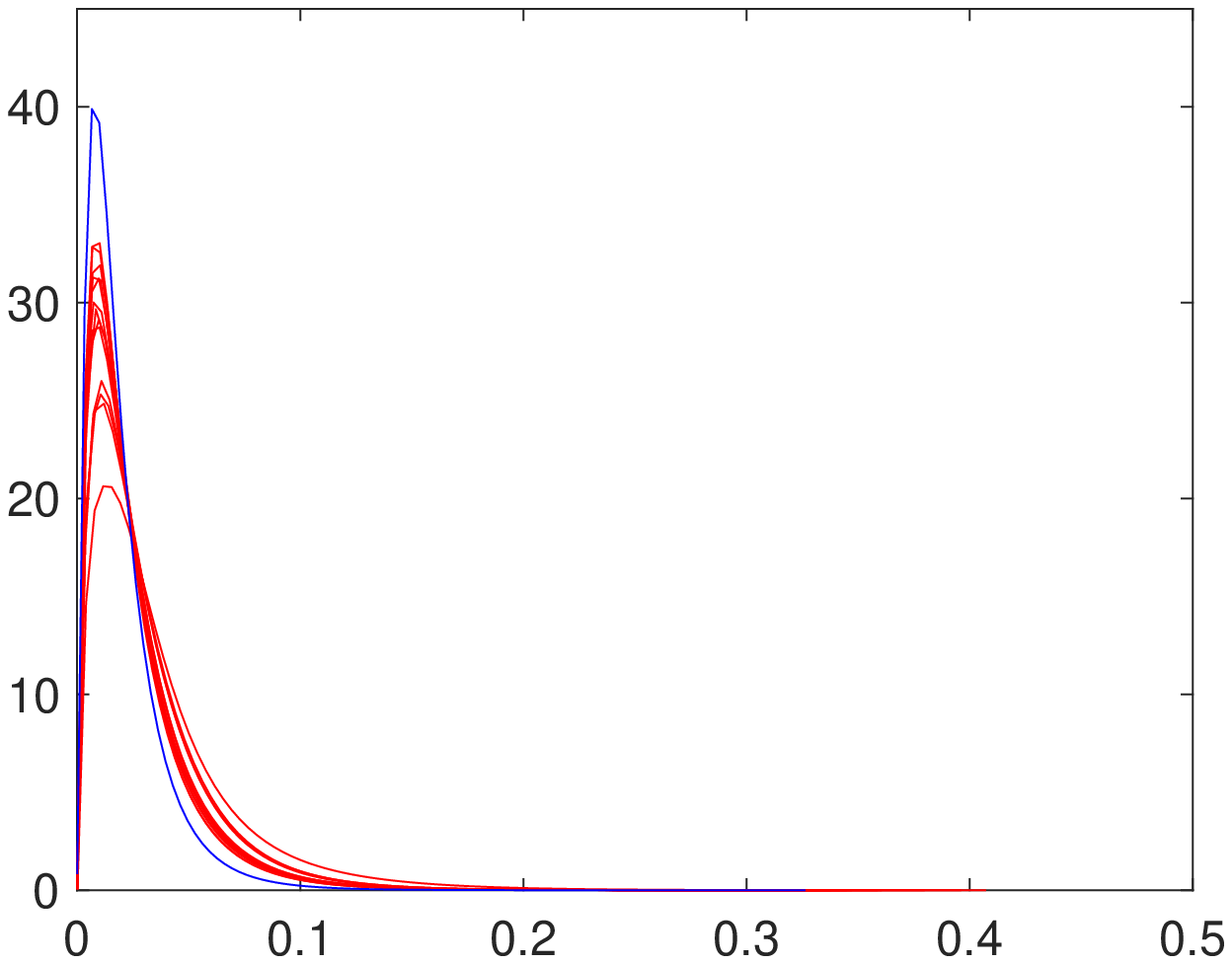}\vspace{5pt}\\
\hline 
\end{tabular}
\end{center}
\caption{ Posterior density for the interaction parameters for both the business cycles (red lines ) and the financial cycle (blue line) under the PMS model. $\alpha$ (column 1), $\beta$ (column 2) and $\gamma$ (column 3) respectively captures the distribution of the idiosyncratic effect of the fixed transition probability, financial cycle effect and the global cycle effect on time varying transition probabilities. The two panes on the other hand represents different specifications of the PMS model. }\label{interactionNPooled}
\end{figure}

\begin{figure}[h!]
\begin{center}
\begin{tabular}{|c|c|c|c|}
\hline
 &Idiosyncratic ($\alpha$) & Financial cylcle $\beta$ & Global $\gamma$\\
 \hline
\rot{without climate indicator}&
\includegraphics[scale=0.35]{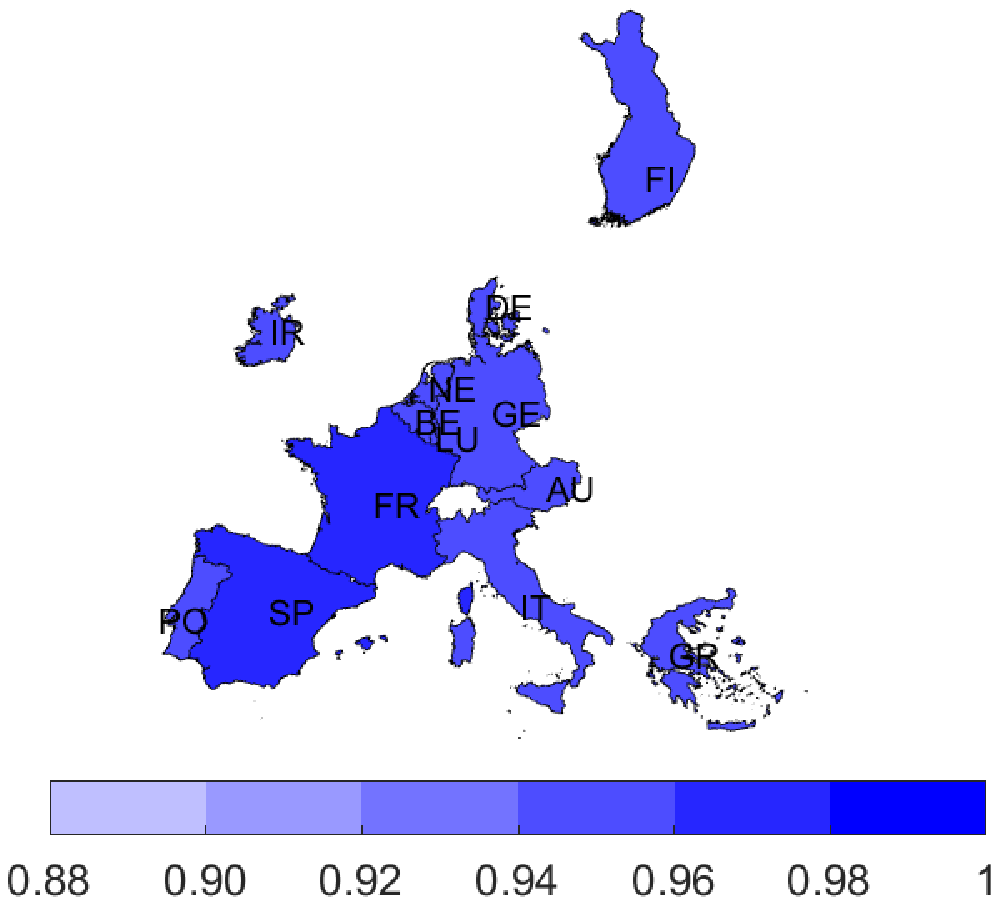}&\includegraphics[scale=0.35]{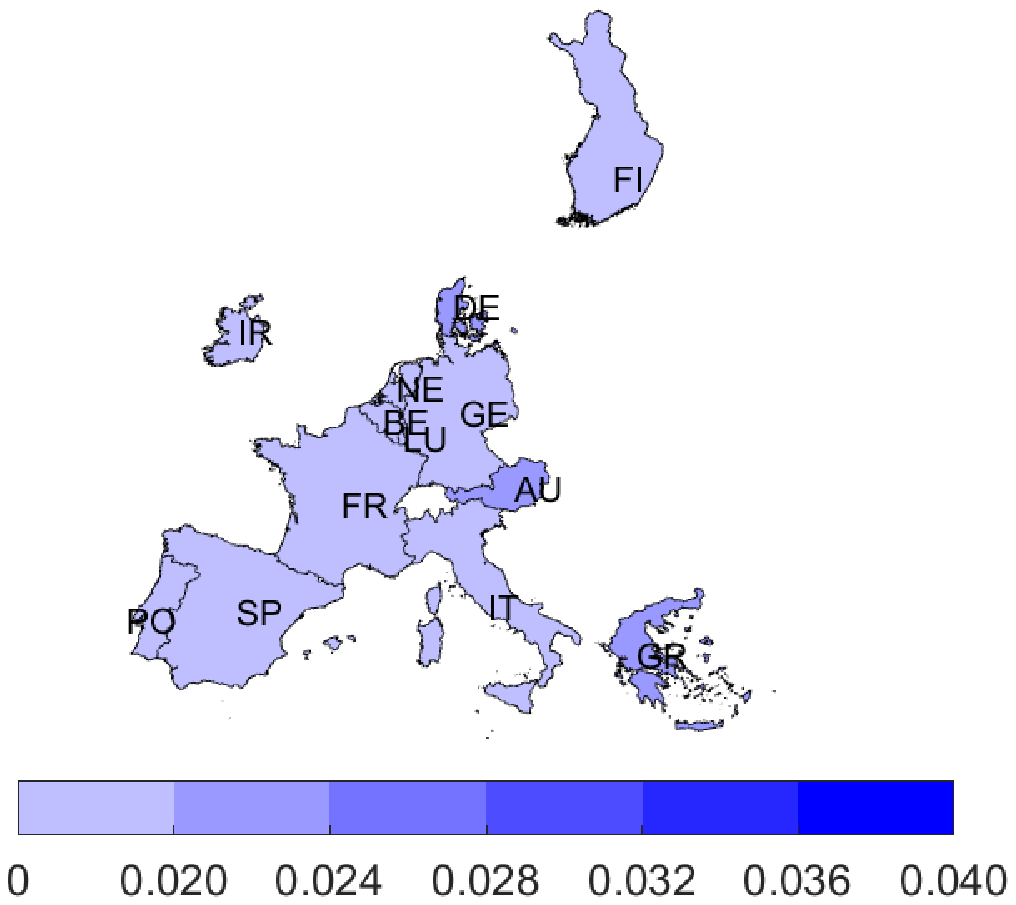}&\includegraphics[scale=0.35]{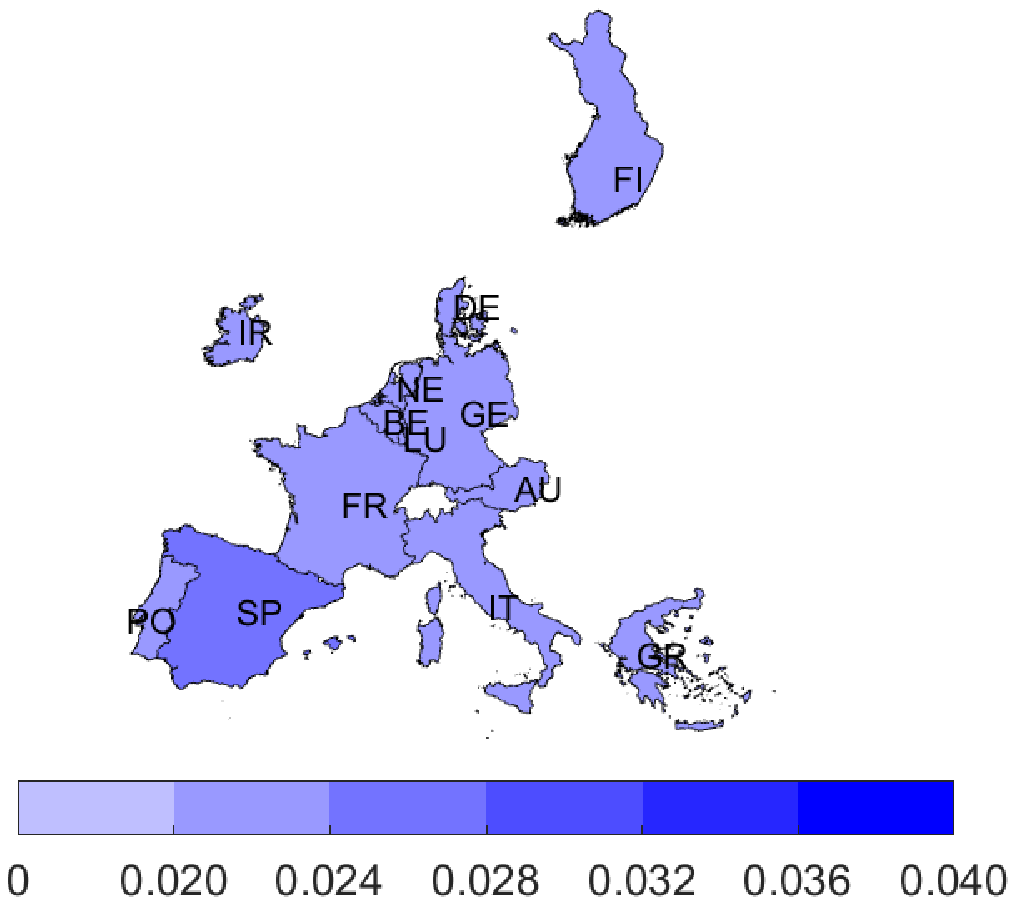}\vspace{5pt}\\
\hline
\rot{with ALL climate indicators}&
\includegraphics[scale=0.35]{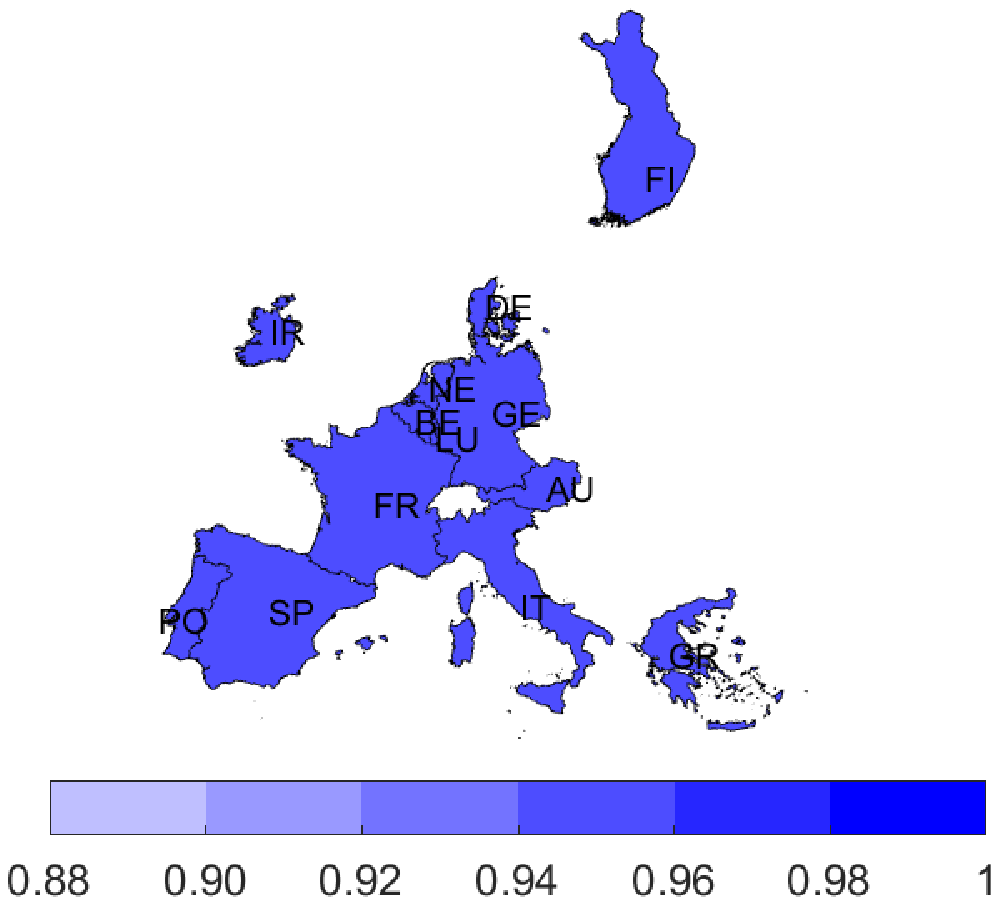}&\includegraphics[scale=0.35]{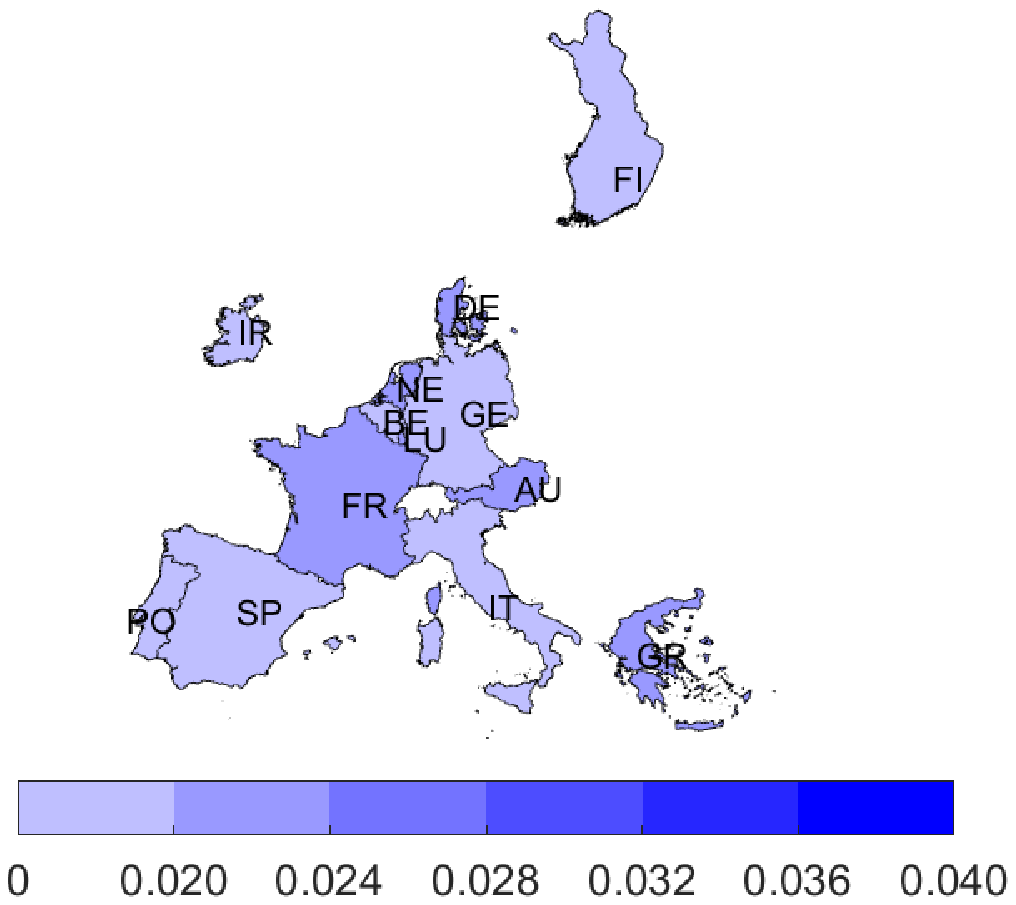}&\includegraphics[scale=0.35]{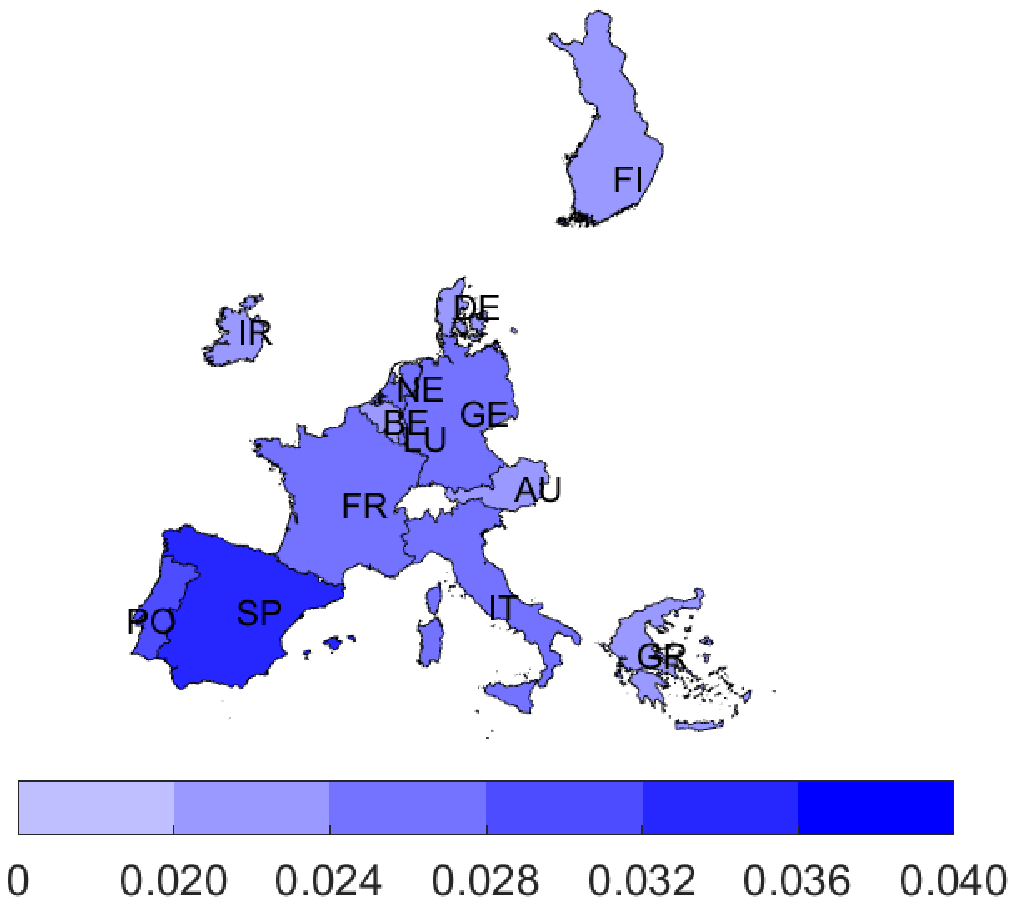}\vspace{5pt}\\
\hline 
\end{tabular}
\end{center}
\caption{Study area with patches representing the contribution of the interaction parameters (idiosyncratic ($\alpha$), financial cycle ($\beta$) and global business cycle ($\gamma$)) under different specifications of the PMS model (rows) on the dynamics of the business cycle of each of the EU countries.}\label{INteractionPar}
\end{figure}

\medskip
\noindent Figure \ref{interactionNPooled} displays the posterior distributions of the constant transition parameter $\alpha$, the interaction parameters $\beta$ between country-specific business cycles and the financial cycle and the global cycle coefficient $\gamma$ under two different specifications of our PMS model. Correspondingly, Tables \ref{InterParaWithClimateShocks} and  \ref{InterParaWithoutClimateShocks} reports their posterior means and 95\% credible intervals (in brackets) while Figure \ref{INteractionPar} gives a pictorial representation of the posterior mean of the interaction parameters. From Tables \ref{InterParaWithClimateShocks} and \ref{InterParaWithoutClimateShocks}, we observe that the mean persistence of the states $S_{y,it}=1$ and $S_{y,it}=2$, that is, the probability of being in recession (expansion) conditional upon being in recession (expansion) in the previous period are mostly above 0.90 across the countries with the aggregate financial cycle exhibiting higher probability (0.97) of staying either in recession or expansion. 

\medskip
\noindent The posterior mean of the idiosyncratic parameter, $\alpha_{i}$, is approximately 0.95 across the countries under study, providing support for time-variations in the transition probabilities of the business cycles i.e. the dynamics of both the financial cycle and the global business cycle are relevant in determining the state of the economy (see Equation \ref{eq3:transitionProb2}). The PMS model with climate shocks appears to capture a slightly higher impact of the global business cycle, $m_{k}(S_{y,t})$, on each country business cycle when compared to the result obtained from the climate shocks free PMS model.(See Tables \ref{InterParaWithClimateShocks}, \ref{InterParaWithoutClimateShocks} and Figure \ref{INteractionPar}). 
This observation however suggests that the interaction between business cycles is not sufficient to explain the variation in the dynamics of country-specific cycle.  
In addition, relative to the financial cycle being in state 1 (recession), the expansionary phase (state 2) of the financial cycle contributes between $0.01$ and $0.03$ in predicting the next phase of the economy (see \ref{eq3:transitionProb2}). This observation suggests that the financial cycle plays a relevant role in determining the phases of the business cycle. 

\begin{table}[h!] 
\caption{Posterior means and 95\% credible intervals (in brackets) of the interaction parameters of the PMS model in the presence of climate shocks} 
\begin{scriptsize}
\begin{center}
    \begin{tabular}{r|ccccc}
    \hline
    \multicolumn{1}{l|}{i} & $\hat{p}_{i,11}$ & $\hat{p}_{i,22}$ & $\alpha_{i}$ & $\beta_{i}$ & $\gamma_{i}$  \\
    \hline 
    \multicolumn{1}{l|}{AU} & 0.9011 & 0.9188 & 0.9557 & 0.021 & 0.0233  \\
          & (0.893,0.9356) & (0.912,0.9400) & (0.9147,0.9844) & (0.0042,0.0491)&  (0.004,0.0547)  \\
    \multicolumn{1}{l|}{BE} & 0.9179 & 0.9211 & 0.9569 & 0.0193&  0.0237 \\
          & (0.914,0.9427) & (0.9096,0.9574) & (0.9181,0.9861) & (0.0036,0.0451)& (0.0041,0.0555)  \\
    \multicolumn{1}{l|}{DE} & 0.9030 & 0.9031 & 0.9561 & 0.0203& 0.0236 \\
          & (0.8928,0.9349) & (0.8847,0.9259) & (0.9192,0.9853) & (0.0036,0.0476)&  (0.0046,0.0544) \\
    \multicolumn{1}{l|}{FI} & 0.9046 & 0.9173 & 0.9580 & 0.0198  & 0.0222\\
          & (0.8946,0.9417) & (0.9018,0.9571) & (0.9183,0.9851) & (0.0037,0.0451)  & (0.0043,0.0511)\\
    \multicolumn{1}{l|}{FR} & 0.9111 & 0.9061 & 0.9553  & 0.0202 & 0.0244 \\
          & (0.8688,0.9177) & (0.8921,0.9470) & (0.916,0.9838) & (0.0034,0.0465) & (0.0043,0.0556)\\
    \multicolumn{1}{l|}{GE} & 0.9116 & 0.9202 & 0.9571 & 0.0187 &  0.0242 \\
          & (0.884,0.9263) & (0.9155,0.9464) & (0.9201,0.9847) &  (0.0033,0.0434) & (0.0042,0.0551) \\
    \multicolumn{1}{l|}{GR} & 0.9033 & 0.9113 & 0.9578 & 0.0201  & 0.0221  \\
          & (0.8914,0.9409) & (0.9016,0.9441)& (0.9218,0.9848) & (0.0038,0.0470)  & (0.0041,0.0520) \\
    \multicolumn{1}{l|}{IR} & 0.9238 & 0.9194 & 0.9586 & 0.0186 & 0.0228 \\
          & (0.9060,0.9583) & (0.9095,0.9438) & (0.9233,0.9849)  & (0.0035,0.0428) & (0.0041,0.0541) \\
    \multicolumn{1}{l|}{IT} & 0.9171 & 0.9298 & 0.9582 & 0.0172 & 0.0246 \\
          & (0.9112,0.9489) & (0.9142,0.9674) & (0.9225,0.9852) &(0.0032,0.0398) & (0.0046,0.0565)\\
    \multicolumn{1}{l|}{LU} & 0.9112 & 0.9136 &0.9523 & 0.0200 & 0.0277\\
          & (0.8825,0.9241) & (0.9089,0.9418)& (0.9124,0.9829) & (0.0040,0.0460) & (0.0051,0.0623)\\
    \multicolumn{1}{l|}{NE} & 0.9095 & 0.9024 & 0.9507 & 0.0219 &  0.0273\\
          & (0.8761,0.9209) & (0.8643,0.9135) & (0.913,0.9819) & (0.0042,0.0494)&  (0.0052,0.0603) \\
    \multicolumn{1}{l|}{PO} & 0.9175 & 0.9208 & 0.9537 & 0.0182 &  0.0281 \\
          & (0.8934,0.9365) & (0.9140,0.9511) & (0.9145,0.9822) &(0.0036,0.0423)&  (0.0052,0.0601) \\
    \multicolumn{1}{l|}{SP} & 0.9130 & 0.9347 & 0.9511 & 0.0144 & 0.0345 \\
          & (0.9009,0.9518) & (0.9047,0.9747) & (0.8844,0.9851) & (0.0028,0.0335)&  (0.0065,0.0920) \\
    \hline
          & $\hat{p}_{f,11}$ & $\hat{p}_{f,22}$ & $\alpha_{f}$& $\beta_{f}$ & $\gamma_{f}$\\ 
    \hline    
    \multicolumn{1}{l|}{FIN1} & 0.9684 & 0.9678 & 0.9694 & 0.0126 & 0.018 \\
     
          & (0.9631,0.9873) & (0.9342,0.9864) & (0.9428,0.9883) & (0.0023,0.0298) & (0.0037,0.0399)\\
   \hline
    \end{tabular}%
    \end{center}     
    \end{scriptsize}   
    \label{InterParaWithClimateShocks}
\end{table}%

\begin{table}[h!] 
\caption{Posterior means and 95\% credible intervals (in brackets) of the interaction parameters of the PMS model  in the absence of climate shocks} 
\begin{scriptsize}
\begin{center} 
    \begin{tabular}{r|ccccc}
     \hline
    \multicolumn{1}{l|}{i} & $\hat{p}_{i,11}$ & $\hat{p}_{i,22}$ & $\alpha_{1}$ & $\beta_{i}$ & $\gamma_{i}$ \\
    \hline 
    \multicolumn{1}{l|}{AU} & 0.8907 & 0.9210 & 0.9558 & 0.0207 & 0.0234  \\
          & (0.8710,0.9239) & (0.9131,0.9542) & (0.9144,0.9841) & (0.0039,0.0482) & (0.0042,0.0553)  \\
     \hline
    \multicolumn{1}{l|}{BE} & 0.9011 & 0.9181 & 0.9590 & 0.0188 & 0.0222 \\
          & (0.8899,0.9323) & (0.9037,0.956) & (0.9233,0.9859) &(0.0032,0.0432) & (0.0043,0.0517)  \\
     \hline
    \multicolumn{1}{l|}{DE} & 0.9167 & 0.8861 & 0.9564 & 0.0212 & 0.0225  \\
          & (0.9031,0.9477) & (0.853,0.9144) & (0.9179,0.9850) & (0.0036,0.0498)& (0.0045,0.0527)  \\
     \hline
    \multicolumn{1}{l|}{FI} & 0.9062 & 0.9267 & 0.9578 & 0.0195 & 0.0227  \\
          & (0.8999,0.9401) & (0.9104,0.9651) & (0.9185,0.9856) & (0.0037,0.0459) & (0.0045,0.0545) \\
     \hline
    \multicolumn{1}{l|}{FR} & 0.8994 & 0.9261 & 0.9608 & 0.0183 & 0.0209  \\
          & (0.8474,0.9417) & (0.8804,0.9711) & (0.9265,0.9866) & (0.0033,0.0436) & (0.0041,0.0492) \\
     \hline
    \multicolumn{1}{l|}{GE} & 0.9136 & 0.9116 & 0.9575 & 0.0197& 0.0227\\
          & (0.8767,0.9198) & (0.9012,0.9487) & (0.9179,0.9851) & (0.0034,0.0442) & (0.0039,0.0525) \\
     \hline
    \multicolumn{1}{l|}{GR} & 0.9149 & 0.9142 & 0.9570 & 0.0209 & 0.0221 \\
          & (0.9000,0.9422) & (0.9040,0.9460) & (0.9216,0.9848) & (0.0041,0.0476) & (0.0042,0.0496)  \\
     \hline
    \multicolumn{1}{l|}{IR} & 0.9205 & 0.9116 & 0.9580 & 0.0194& 0.0226 \\
          & (0.9001,0.9577) & (0.9012,0.9460) & (0.9207,0.9844) & (0.0034,0.045) &  (0.0043,0.0543)  \\
     \hline
    \multicolumn{1}{l|}{IT} & 0.9200 & 0.9296 & 0.9591 & 0.0182 & 0.0228  \\
          & (0.9085,0.9405) & (0.9080,0.9699) & (0.9225,0.9847) & (0.0034,0.0416) & (0.0043,0.0526) \\
     \hline
    \multicolumn{1}{l|}{LU} & 0.9046 & 0.9057 & 0.9599 & 0.0189  & 0.0212 \\
          & (0.8425,0.9207) & (0.8495,0.9617) & (0.926,0.9858) & (0.0033,0.0442) & (0.004,0.0487)  \\
     \hline
    \multicolumn{1}{l|}{NE} & 0.9049 & 0.9133 & 0.9569 & 0.0198 & 0.0233 \\
          & (0.8918,0.9293) & (0.8952,0.9270) & (0.9191,0.9848) & (0.0038,0.0465) & (0.004,0.0532)  \\
     \hline
    \multicolumn{1}{l|}{PO} & 0.9168 & 0.9240 & 0.9588 & 0.0179  & 0.0232 \\
          & (0.9046,0.9489) & (0.9031,0.9627) & (0.9245,0.9848) & (0.0035,0.0427) & (0.0044,0.0521)  \\
     \hline
    \multicolumn{1}{l|}{SP} & 0.9242 & 0.9533 & 0.9608 & 0.0124 & 0.0268 \\
          & (0.8989,0.9472) & (0.9091,0.9877) & (0.9248,0.9861)& (0.0025,0.0288) & (0.0051,0.0601)  \\
    \hline
          & $\hat{p}_{f,11}$ & $\hat{p}_{f,22}$ & $\alpha_{f}$ & $\beta_{f}$ & $\gamma_{f}$  \\
    \hline
    \multicolumn{1}{l|}{FIN1} & 0.9715 & 0.9717 & 0.9661 & 0.0143 & 0.0197 \\
          & (0.9668,0.9873) & (0.9673,0.9865) & (0.9392,0.9873) & (0.0028,0.0318) & (0.0029,0.0448)  \\
    \hline
        \end{tabular}%
        \end{center}     
        \end{scriptsize}   
        \label{InterParaWithoutClimateShocks}
    \end{table}%
\medskip
\noindent We further investigate, in the following section, if aggregate financial shock is sufficient for driving business cycle synchronization by assessing the role played climate shocks as a mechanism driving industrial production growth. There has been no conclusive result in the literature on the mechanisms driving business cycle synchronization of EU member countries. \cite{RoseEngel2002}, for instance, finds a positive impact of currency unions on the business cycle synchronization of EU member countries. This stance has however been challenged by \cite{BaxterKouparitsas2005}. Weather, on the other hand, has been argued to play a potential role in influencing economic activity through various channels (\cite{Acevedo2018}). In-line with this, we analyze, in the following section, the potential contribution of extreme weather conditions in explaining the dynamics of the growth of IPI. The estimated business cycles, $\hat{S}_{y,t}=(\hat{S}_{y,1t},\ldots,\hat{S}_{y,13t})$ for the countries are reported in Figures \ref{FigCyc1NPooling1} - \ref{FigCyc1NPooling2} in Appendix \ref{Appendix3}. 

\subsection{Impact of climate change on industrial production}
\begin{figure}[h!]
        \centering
        \subfigure[Without climate shocks]{
                \includegraphics[width=0.4\textwidth]{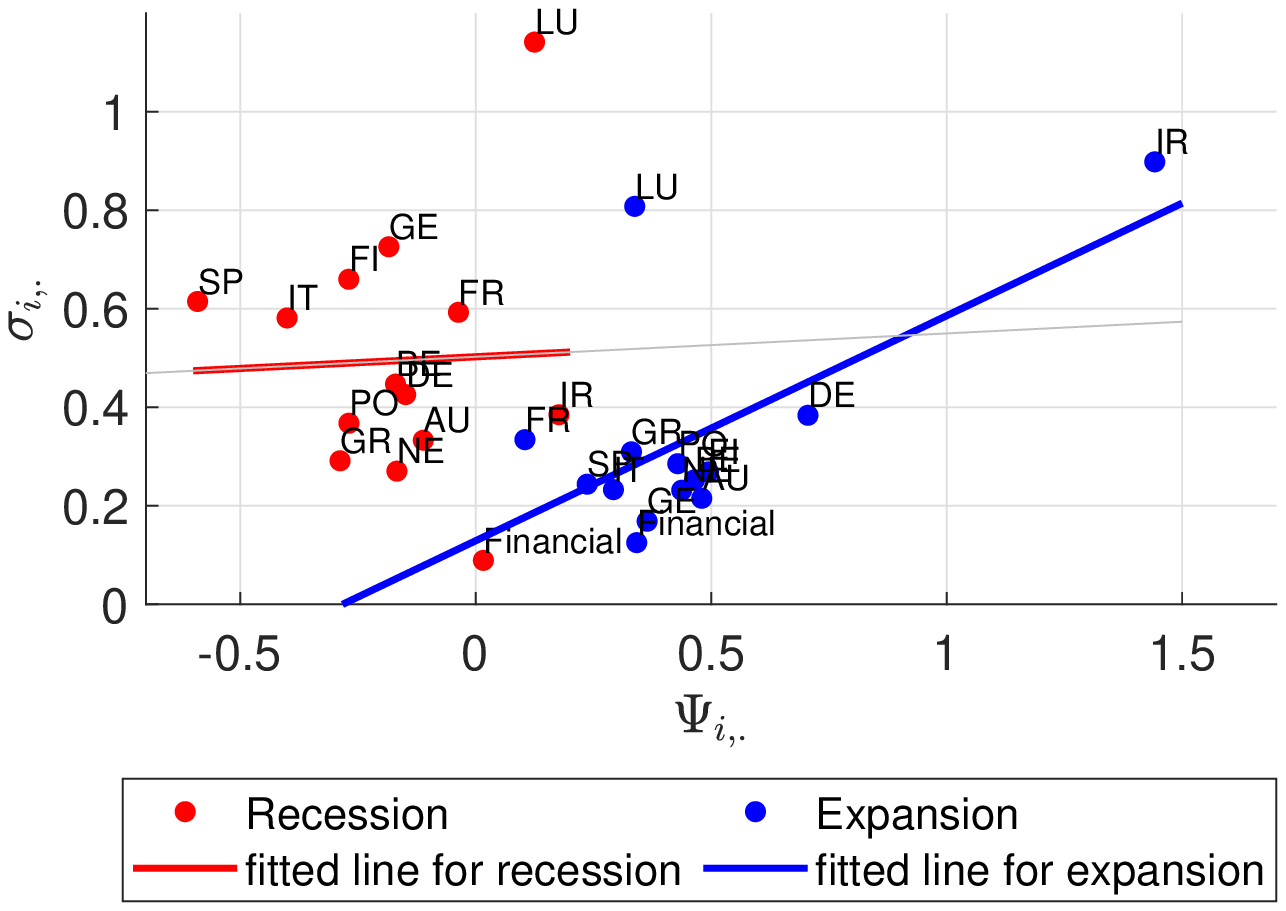}
                \label{fig:ParametersWITHOUTCI}}
        ~
        \subfigure[With climate shocks]{
                \includegraphics[width=0.4\textwidth]{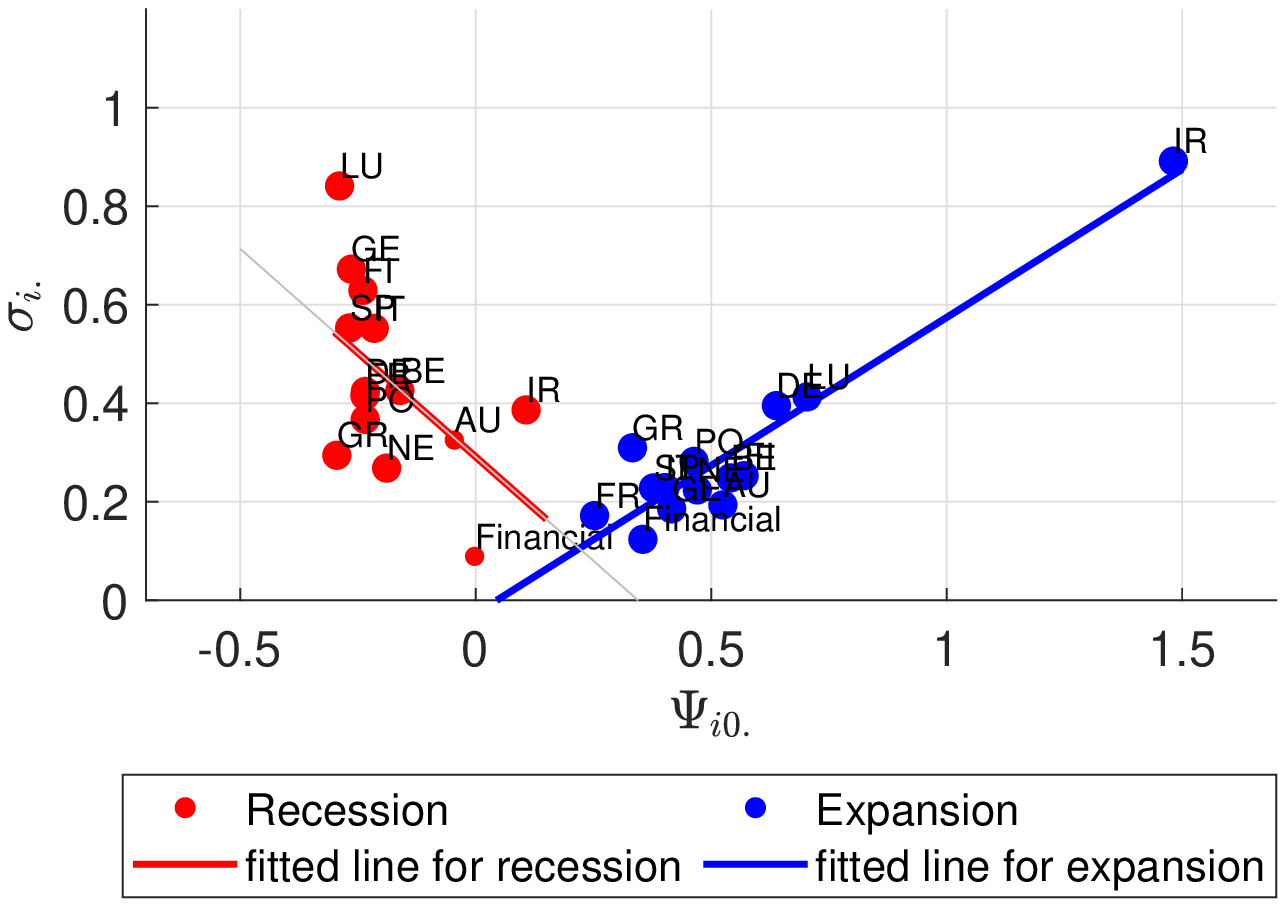}
                \label{fig:ParametersWITHCI}}                   
		\caption{Estimates of monthly state-level IPI and Financial index (Fin1) mean growth rates (horizontal axis) and volatilities (vertical axis) with the Bayesian Panel Markov-switching model. The horizontal axis represents the mean growth rates and the vertical axis the volatilities. Sample period: February 1981 to December 2016 (month on month). Figures \ref{fig:ParametersWITHOUTCI} and \ref{fig:ParametersWITHCI} respectively report the output of the PMS model in the absence and presence of climate shocks.}\label{Param}
\end{figure}

\begin{figure}[h!]
\begin{center}
\begin{tabular}{|c|c|c|}
\hline
 & Recession & Expansion\\
 \hline
\rot{Intercept (without climate indicator)}&
\includegraphics[scale=0.5]{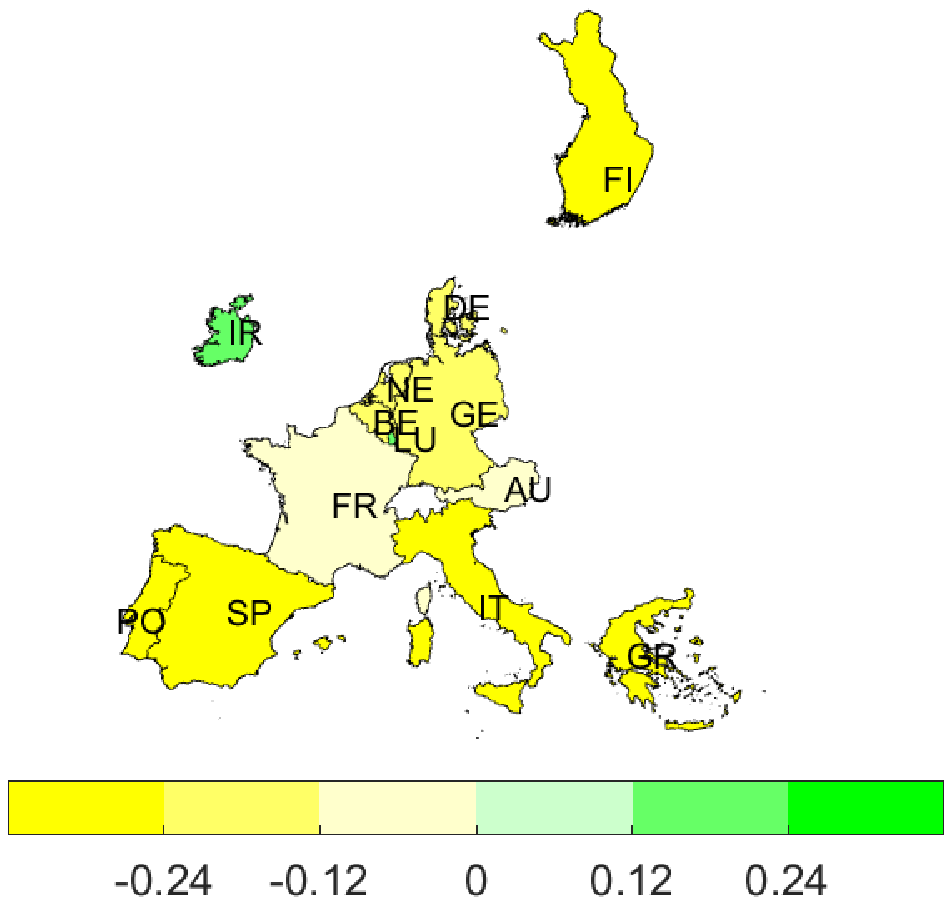}&\includegraphics[scale=0.5]{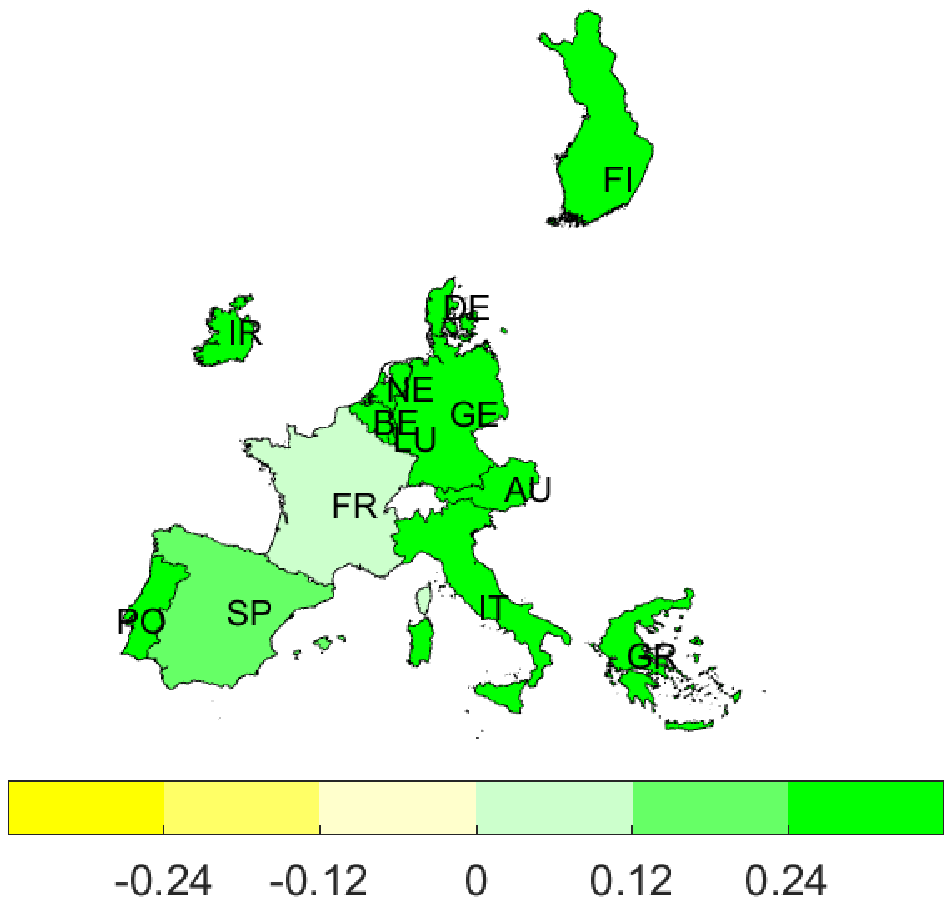}\vspace{5pt}\\
\hline
\rot{Intercept (with ALL climate indicators)}&
\includegraphics[scale=0.5]{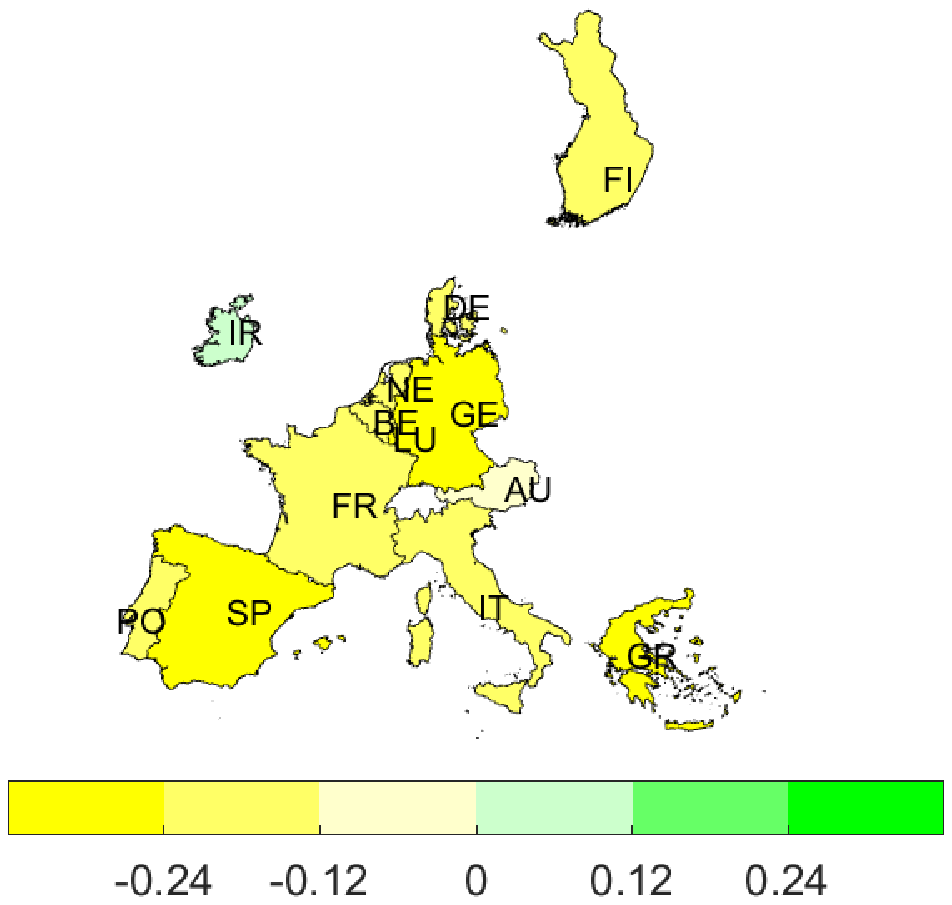}&\includegraphics[scale=0.5]{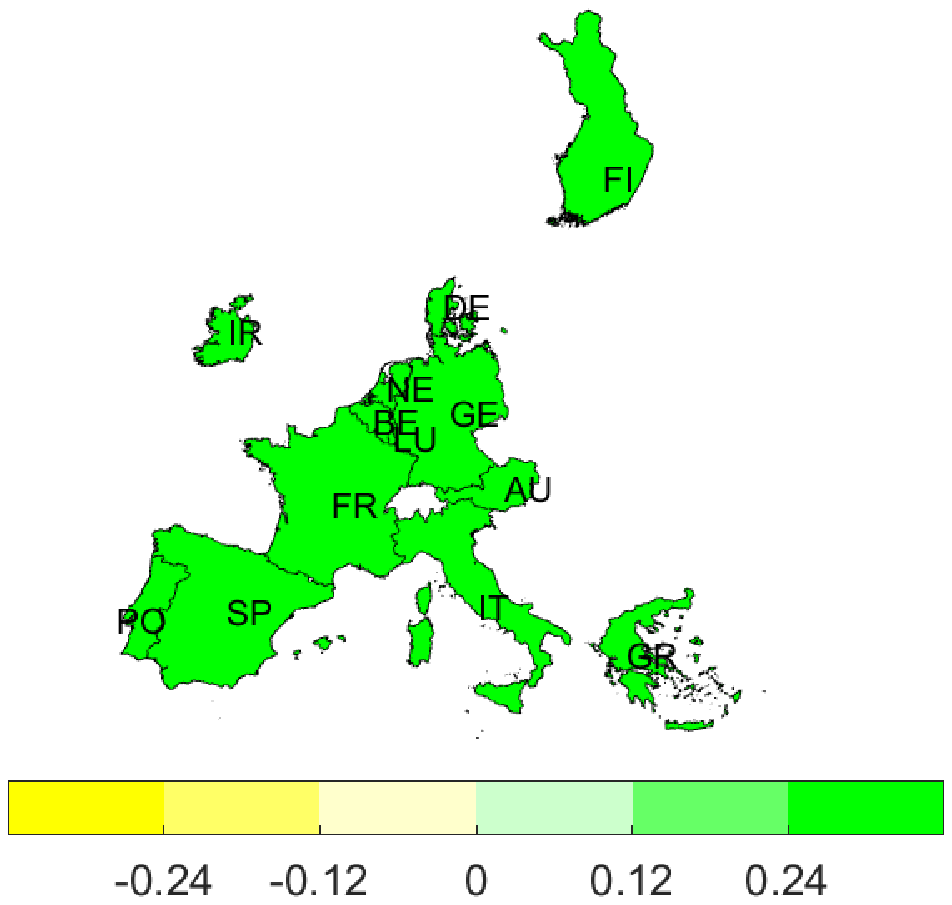}\vspace{5pt}\\
\hline 
\end{tabular}
\end{center}
\caption{Study area with yellow and green patches denoting negative and positive impact of climate shocks on IPI growth rate of the represented EU Countries under different phases economy (columns) and specifications of the PMS model (rows).
}\label{Map1intercept}
\end{figure}

\noindent Noting that the standardized precipitation index (SPI) measuring drought has been transformed into a dichotomous variable, our results here will be discussed in relative terms. The estimated IPI and financial mean growth rates as well as their volatilities are given in the scatter plots of Figure \ref{Param}, while Figure \ref{Map1intercept} presents a map of the study area with yellow and green patches representing the average IPI during the two phases of the economy. In both figures, we report the results obtained under two different specifications of the PMS model. We observe from Figure \ref{Param} that a clear separation of the two phases of the economy is less obvious by neglecting climate shocks in the PMS model specification. In addition, in a non-drought like climatic condition an inverse relationship exists between the growth rates and the volatities during recession, while a positive relationship is observed between the growth rates and the volatilities during the expansionary phase of the economy ie. during recession a fall in industrial production is accompanied by an increase in volatility whereas during expansion increase in industrial production is associated with increase in volatility. This observation suggests that insights into the mechanisms driving the IPI growth rate of EU countries may be understood by accounting for climate shocks. Irrespective of the PMS model specification adopted, Ireland is observed to have experienced the highest positive average growth rate and at the same time the most volatile economy during the expansionary phase of the economy. 

\begin{figure}[h!]
\begin{center}
\begin{tabular}{|c|c|c|}
\hline
 & Recession & Expansion\\
\hline
\rot{Temperature (CSU)}&
\includegraphics[scale=0.5]{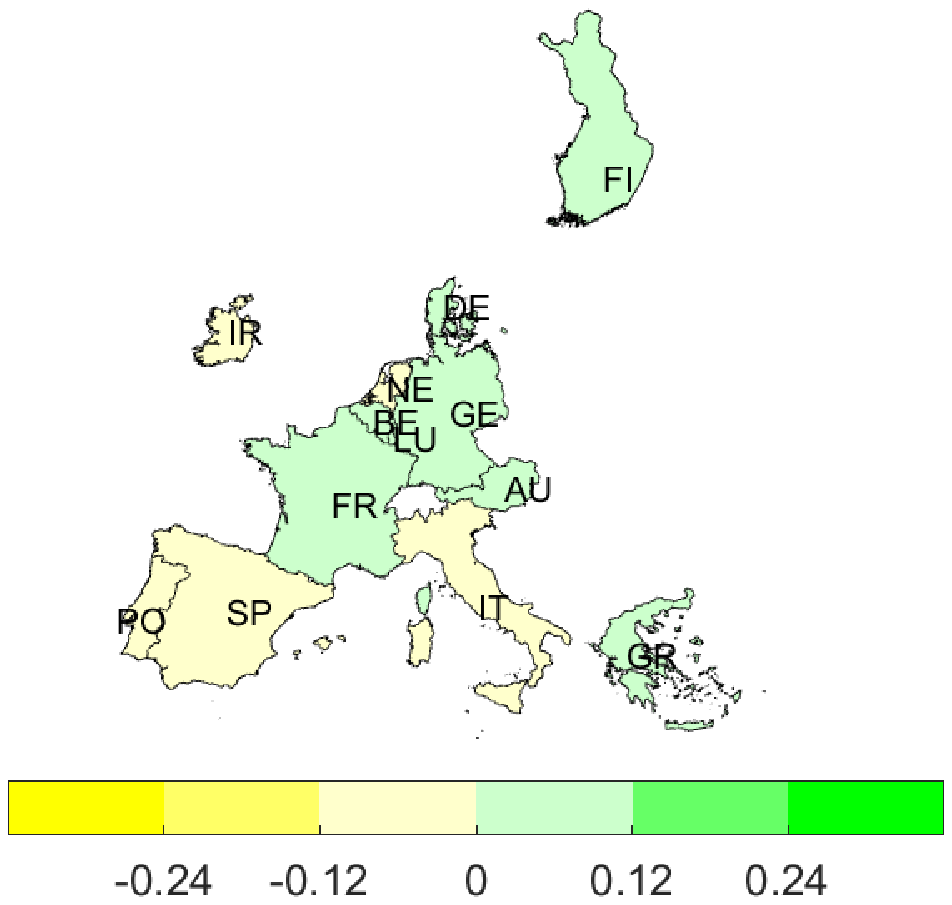}&\includegraphics[scale=0.5]{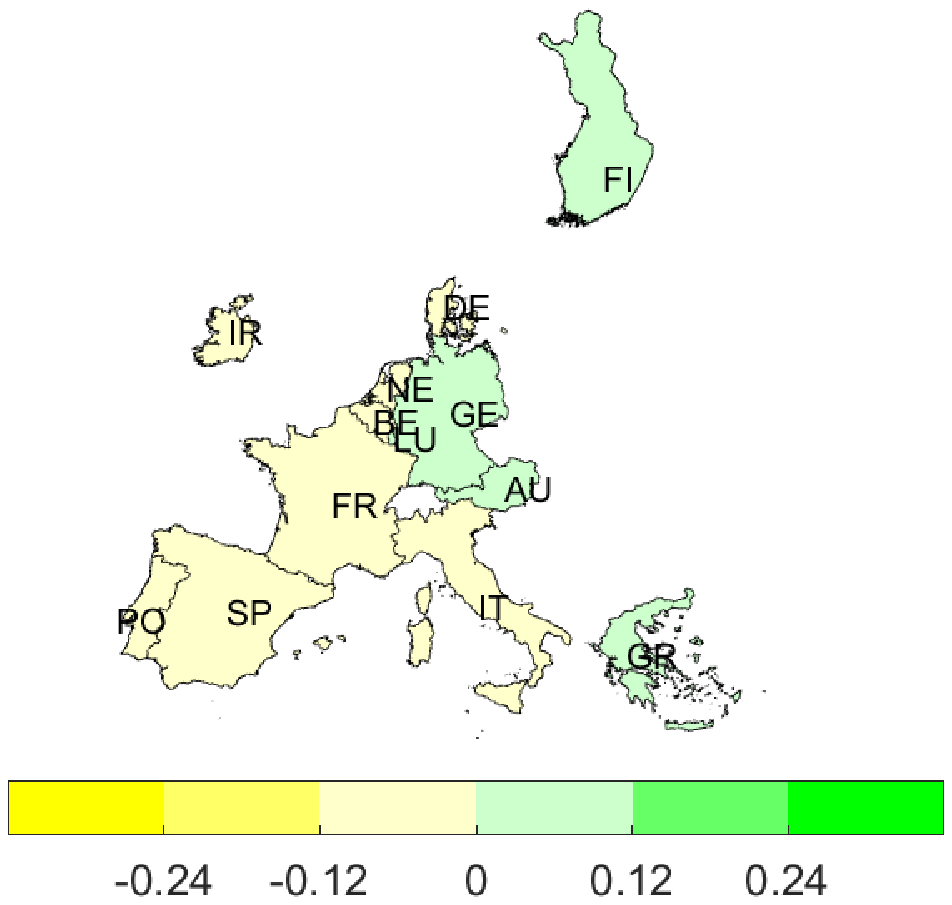}\vspace{5pt}\\
\hline
\rot{Drought (SPI)}&
\includegraphics[scale=0.5]{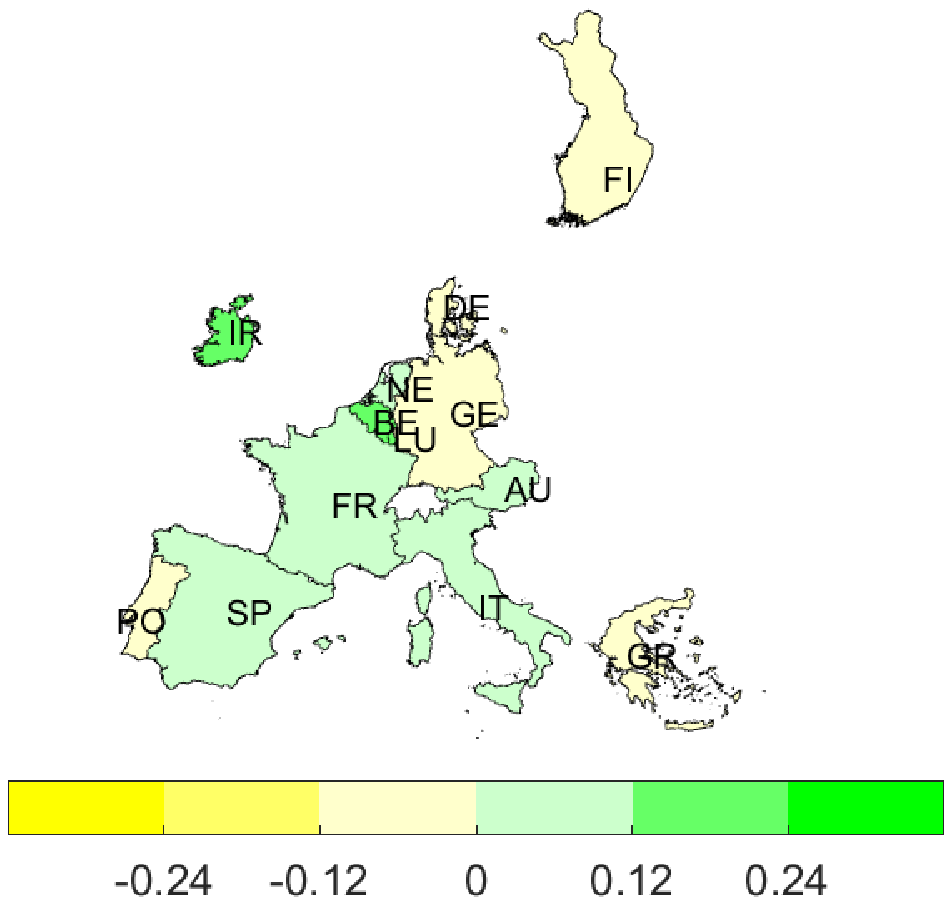}&\includegraphics[scale=0.5]{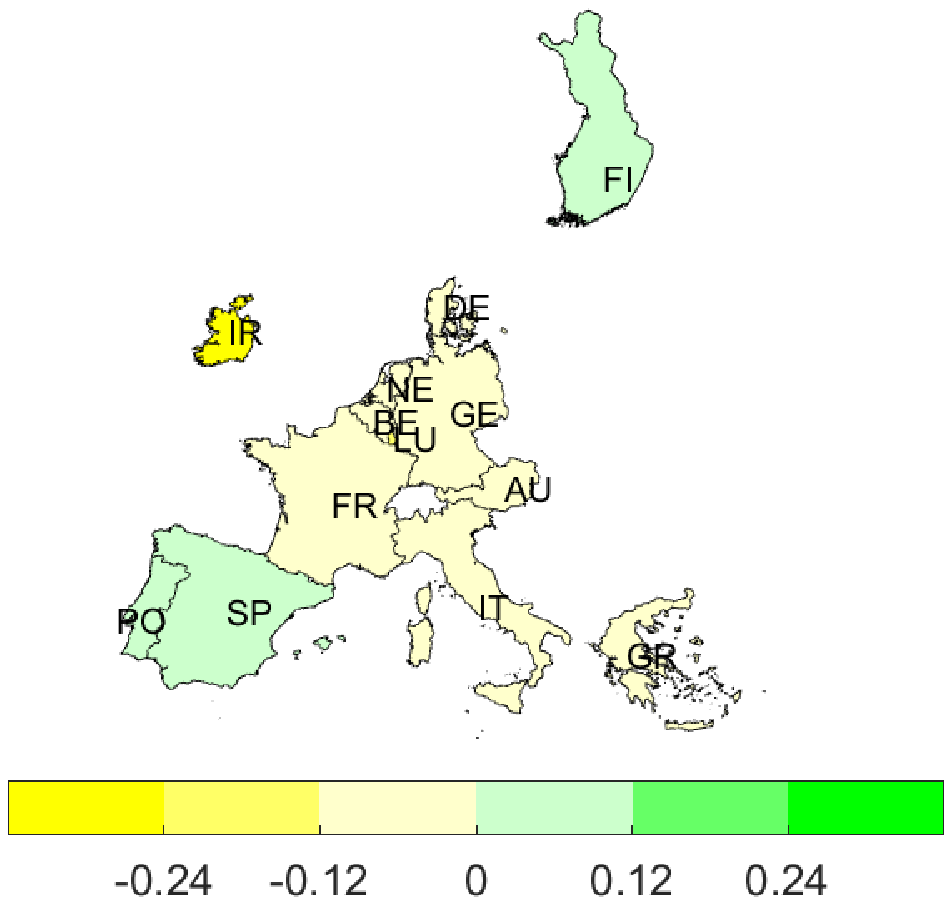}\\
\hline
\rot{Precipitation (r20mm)}&
\includegraphics[scale=0.5]{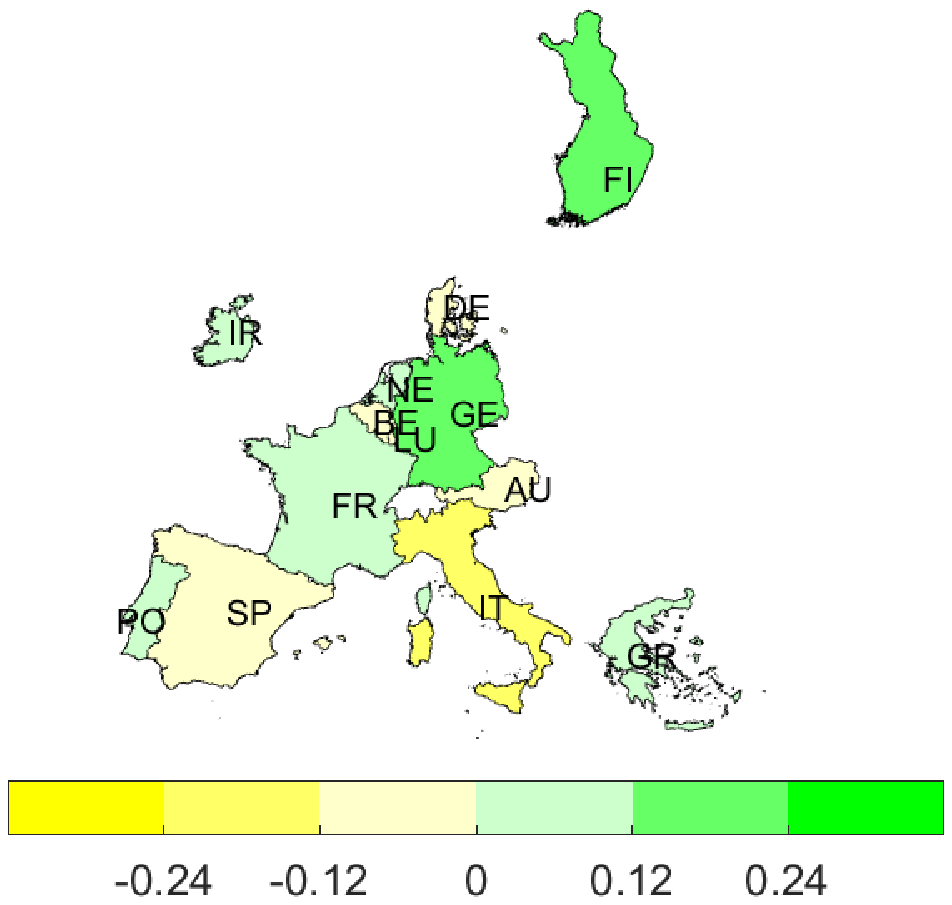}&\includegraphics[scale=0.5]{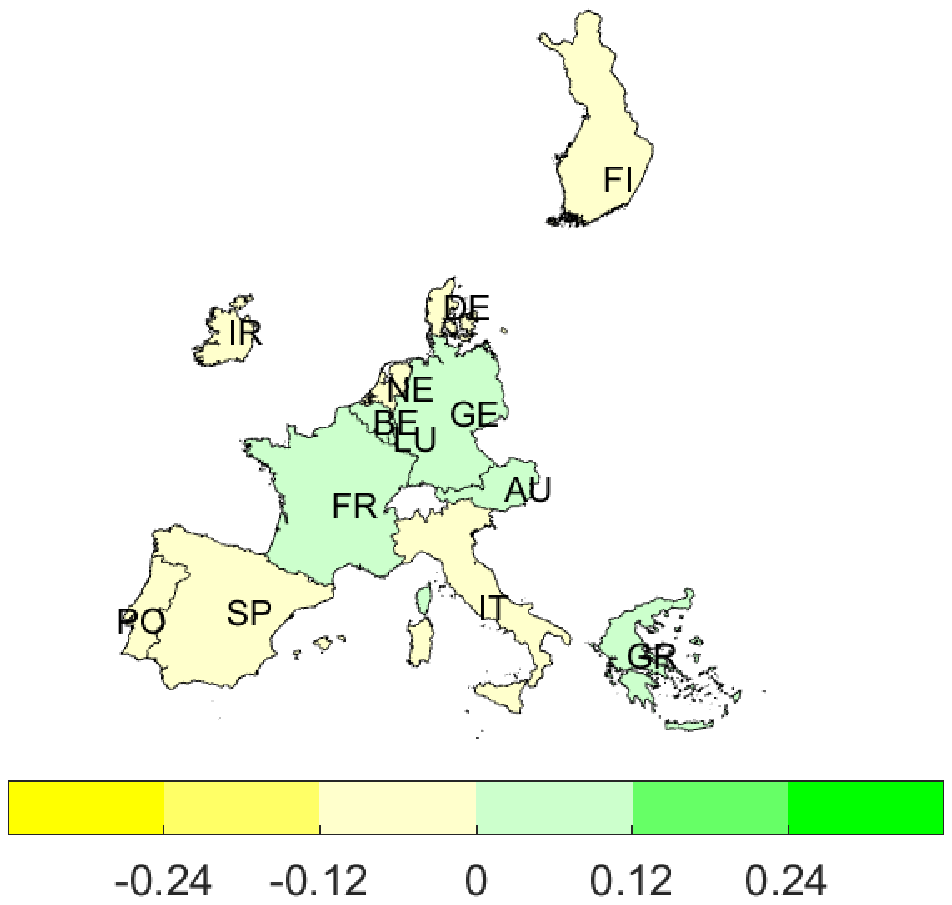}\\
\hline
\end{tabular}
\end{center}
\caption{Study area with the impact of climate shocks (indicated by different colours) on the growth rate on the represented EU Countries during recession and expansion. Yellow and green patches respectively denote negative and positive impact. The SPI index is transformed into two contrasting weather scenarios using a dummy variable. We set the dummy variable to one if the SPI is less than -0.5 (dry situation) and zero otherwise.}\label{Map1}
\end{figure}

\medskip
\noindent Figure \ref{Map1} displays estimates of the coefficients capturing the regime specific impact of the changes in extreme weather event on country specific industrial production indices. In this figure, yellow and green patches respectively denote negative and positive impact of climate shocks on IPI growth under different phases of the economy and specifications of the PMS model. While economic activities of countries within Southern Europe (region with the highest exposure to long spell of summer days), with the exception of Greece, appear to be hampered by extreme temperature, we observe asymmetric effect (positive during recession and negative during expansion) of lengthy spell of consecutive summer days on the economies of France, Belgium and Denmark.

\medskip 
\noindent Relative to the non-drought-like climatic condition, moderate to extreme dryness seem to contribute negatively to the IPI growth of most of the countries in Northern Europe. The economies of Ireland and Finland tend to be favoured by drought during recession but hindered during expansion. Spain seems to be favoured by drought and thus compensates for the negative impact of extreme temperature on its economy. 
In addition, drought tend to favour growth by reducing (by about 0.08) the adverse impact of recession on Netherlands while Portugal's economy (historically, the country with the highest exposure to moderate to extreme dryness) seem to enjoy a boost (by about 0.09) from drought during expansion. Italy and Luxembourg's growths are negatively impacted by drought during expansion. An increase in the frequency of heavy precipitation days per month seems to negatively affect Italy's average industrial production growth with a reduction of -0.059 during expansion and -0.18 during recession. On the contrary, increasing frequency of high precipitation per month tend to contribute to the industrial production growth indices in Germany (0.15), Netherlands (0.07) and Portugal (0.06) during recession and to Greece (0.07) during expansion.  

\medskip
\noindent The posterior means and 95\% credible intervals (in brackets) for the parameters of the measurement equations of our  PMS models are reported in Tables
\ref{tab:WithClimateShock} and \ref{tab:withoutCL}.

\subsubsection{Channel of impact by sector}

\medskip
\noindent Climate shocks can influence economic activity through various channels. The most obvious one is agricultural output, given that temperature and precipitation are direct inputs in crop production. However, studies show evidence of broader impacts, including labour productivity, mortality, health, and conflict. In this section, we examine the impact of extreme weather on sectoral outputs of IPI namely: mining and quarrying; manufacturing; and utilities (ie. electricity, gas, steam and air conditioning ). However, due to data availability issue, we focus on the impact of climate shocks on the manufacturing sector only by estimating the PMS model in Equation \ref{eq1:Measurement1} and \ref{eq1:Measurement2} but with the growth of the manufacturing sector as our outcome of interest. The manufacturing sector accounts for about 90\% of the entire IPI. 

\medskip
\noindent Before we proceed further, we observe that data on Italy's  manufacturing sector is only available from January 1990. We therefore reconstruct the missing  observation by fitting a simple linear regression of Italy's manufacturing data on its corresponding IPI for the period January 1990 to December 2016. The fitted regression model together with Italy's IPI for the January 1981 to December 1989 were then used to estimate the missing observations. The result of the estimation of the PMS model reported in Table \ref{tab:ManPMS} and Figure \ref{fig:MANunconstrainedR20MMAsymCheckALL} suggests that unlike the aggregate IPI, France, Portugal and Germany's manufacturing sector do not seem to be responsive to the increase in the spell of summer days. Similarly, Belgium and Luxembourg's manufacturing are not sensitive to drought. In addition, high precipitation does not significantly impact the manufacturing sector of the Netherlands. 
Apart from these few cases, the result in Table \ref{tab:ManPMS} generally suggests that most of our deductions on the impact of climate shocks on the IPI EU countries affects the manufacturing sector.  

\subsection{Model selection}
\noindent In this section, we compare different alternative specifications of the PMS model to see which covariates are significant (exclusion of covariates). In this case, we consider situations when the measurement equations are: (i) independent of the climate change indices;  (ii) dependent on only one the climate index at a time, and; (iii) dependent simultaneously on all the selected climate change indices. Corresponding, we have Models 1-5 for PMS model. 

\medskip
\noindent A pairwise comparison of all the competing models $\{M_{j}: j=1,2,\dots,J\}$ will be carried out by evaluating the Bayes factor (BF)
$$
B(j,j') = \dfrac{f(Y_{1:T}, X_{1:T}|Z_{1:T},M_{j})}{f(Y_{1:T}, X_{1:T}|Z_{1:T},M_{j'})}
$$
between model $M_{j}$ and $M_{j'}$, for all pairs of models based on the full sample.

\medskip
\noindent Noting that under model $M_{j}$, the posterior distribution of $(\theta, S_{y,1:T}, S_{x,1:T})$ given the observations can be written as
$$
P(\theta,S_{y,1:T},S_{x,1:T}| Y_{1:T},X_{1:T},Z_{1:T},M_{j}) =\dfrac{f( Y_{1:T},X_{1:T}|\theta,S_{y,1:T},S_{x,1:T},Z_{1:T},M_{j} ) P( \theta, S_{y,1:T},S_{x,1:T} |M_{j}   ) }{f(Y_{1:T},X_{1:T} | Z_{1:T},M_{j}  )   }
$$
where $f( Y_{1:T},X_{1:T}|\theta,S_{y,1:T},S_{x,1:T},Z_{1:T},M_{j} )$ is the joint density function of the observations conditional on the parameter set, $f( Y_{1:T},X_{1:T}|,Z_{1:T},M_{j} )$ is the marginal joint density function of the observations and $P( \theta, S_{y,1:T},S_{x,1:T} |M_{j})$ is the prior distribution  of $(\theta, S_{y,1:T}, S_{x,1:T})$. Furthermore, the logarithm of the ratio between the posterior distribution of the parameters of a pair of competing models is
\begin{equation}
\begin{aligned}
&\log\left( \dfrac{P(\theta,S_{y,1:T},S_{x,1:T}| Y_{1:T},X_{1:T},Z_{1:T},M_{j})}{P(\theta,S_{y,1:T},S_{x,1:T}| Y_{1:T},X_{1:T},Z_{1:T},M_{k})}\right) = \log\left(\dfrac{f(Y_{1:T},X_{1:T} | Z_{1:T},M_{k}  ) }{ f(Y_{1:T},X_{1:T} | Z_{1:T},M_{j}  )  }\right)\\
&+\log\left(\dfrac{f( Y_{1:T},X_{1:T}|\theta,S_{y,1:T},S_{x,1:T},Z_{1:T},M_{j} )}{f( Y_{1:T},X_{1:T}|\theta,S_{y,1:T},S_{x,1:T},Z_{1:T},M_{k} )}  \right) 
+ \log\left( \dfrac{P( \theta, S_{y,1:T},S_{x,1:T} |M_{j}   ) }{P( \theta, S_{y,1:T},S_{x,1:T} |M_{k}   ) }   \right)
\end{aligned} \label{LogRatio}
\end{equation}
By setting 
$$
p_{M_{j}} = \dfrac{P(\theta,S_{y,1:T},S_{x,1:T}| Y_{1:T},X_{1:T},Z_{1:T},M_{j})}{P(\theta,S_{y,1:T},S_{x,1:T}| Y_{1:T},X_{1:T},Z_{1:T},M_{j}) + P(\theta,S_{y,1:T},S_{x,1:T}| Y_{1:T},X_{1:T},Z_{1:T},M_{k})}
$$
and assuming that the prior distribution of the augmented parameter set, $(\theta, S_{y,1:T},S_{x,1:T})$, is equal under the two competing models, then Equation \ref{LogRatio} may be written as
\begin{equation} 
 \begin{aligned} 
&\log{\left( \dfrac{p_{M_{j}}}{1- p_{M_{j}}} \right)  } \\
&= \log{B(k,j)}   +  \log{f( Y_{1:T},X_{1:T}|\theta,S_{y,1:T},S_{x,1:T},Z_{1:T},M_{j} )} - \log{f( Y_{1:T},X_{1:T}|\theta,S_{y,1:T},S_{x,1:T},Z_{1:T},M_{k} )}.
\end{aligned}\label{LogOdds}
\end{equation} 
Equation \ref{LogOdds} can be thought of as a logistic regression. Hence, the problem of estimating the Bayes factor can be recast as a reverse logistic regression (see \cite{Geyer1994} and \cite{CameronPettitt2014}). That is, given the log-likelihoods corresponding to each MCMC draws from the posterior distribution under $M_{j}$, the
\begin{equation}
\log{\left( \dfrac{p_{M}}{1- p_{M}} \right)  }= \alpha + \beta \log{f( Y_{1:T},X_{1:T}|\theta,S_{y,1:T},S_{x,1:T},Z_{1:T},M )}
\label{ReverseLogistic}
\end{equation}
where $M\in \{M_{j},M_{j'}\}$. 

\begin{table}[h!]
\begin{center}
\caption{Pairwise estimate of the log-Bayes factor, $\log(BF(j', j))$, (in multiples of $10^3$) in favor of $M_{j'}$ over $M_{j}$ for models.}\label{tLogBayesFactor}
\begin{tabular}{cccccc}
\cline{2 -6}
          &    \multicolumn{5}{c}{Unconstrained PMS}\\
$\log B(j',j)$ &   & CSU    & SPI & r20mm  & ALL \\
\hline
$M_{j'}$ & Model1 & Model2 & Model3 & Model4 & Model5\\
\hline
Model1  &      0  & -0.4461 &  -2.7454  & -3.1959  & -0.2163\\
Model2  & 0.4461  &     0   &  -5.1733  & -6.5194  &  4.6588\\
Model3  & 2.7454  &  5.1733 &       0   &  0.0980  &  1.9526\\
Model4  & 3.1959  &  6.5194 &  -0.0980  &       0  &  2.0426\\
Model5  & 0.2163  & -4.6588 &  -1.9526  & -2.0426  &       0\\
\hline   
\end{tabular}
\end{center}
{\footnotesize{Note: The results in this Table is computed by estimating the value of $\alpha$ (log-Bayes factor) in the logistic regression given in Equation (\ref{ReverseLogistic}).}} 
\end{table}

\medskip
\noindent From Table \ref{tLogBayesFactor}, the PMS model with climate shocks perform better than Model 1 (PMS model without climate shocks). Furthermore, Model 3 (PMS model with drought index) and Model 4 (PMS model with rainfall index) appear to be consistently perform better than other specifications of the PMS model. This observation suggests drought and precipitation index are relevant climate shock indices to consider in the study of the economic and financial cycle. 
\section{Conclusion}
\label{sec:conclude}
\noindent This paper studies the co-movement of the business cycles and the aggregate financial cycle of a selected set of European Union countries by using a Panel Markov-Switching (PMS) model. An attempt at explaining the mechanism driving the euro-area economic cycle through the assessment of the potential impact of extreme weather conditions on industrial production is also undertaken. Our results suggests that extreme weather conditions, aggregate financial and global cycles are relevant in describing the dynamics of country-specific business cycles. Economies of EU countries are generally observed to be sensitive to changes in the spell length of summer days and drought during the expansionary phase of the economy while increasing frequency in heavy precipitation tend to affect the economies more during recession. In addition, climate shocks are observed to have uneven impact across the two states of the economy and heterogeneous impact across the EU countries. A further analysis in the channel of impact suggests that the impact of extreme weather events on the economy is mostly felt through the manufacturing sector of the economy. 

Our focus in this study has been limited to the impacts of climate shocks on the economies of EU countries, accounting for the interaction of extreme weather events with the economies. Thus as a further line of research, the study of the impact of climate shocks on EU economies through their influence on raw materials and processes originating from outside the EU may provide insight into the sources of variations in the economies of EU nations. 

\newpage
%
%
%
 \begin{APPENDICES}
 \end{APPENDICES}

\newpage
\ACKNOWLEDGMENT{Enrica De Cian and Malcolm Mistry gratefully acknowledge the funding received from the European Research Council (ERC) under the European
Union's Horizon 2020 research and innovation programme under grant agreement No 756194 (ENERGYA); Monica Billio, Roberto Casarin and Anthony Osuntuyi are grateful for the funding received from the EURO Horizon 2020 project on Energy efficient Mortgages Action Plan (EeMAP), grant agreement No 746205 and the VERA project on economic and financial cycles under the grant number No. 479/2018 Prot. 33287 -VII / 16 of 08.06.2018. We also wish to thank seminar participants at the Econometric Society Virtual World Congress, Milan, 2020..}


\bibliographystyle{informs2014} 
\bibliography{Pan,References} 

\ECSwitch


\ECHead{Appendices}
\tableofcontents
\begin{APPENDICES} 
 \section{Information on construction of composite financial cycle indicator}
 \label{DataConstruction}
 \begin{enumerate}
 \item {\emph{Credit to private sector.}} The time series is available on the BIS (Bank for International
 Settlements) Statistical portal website for the period 1999Q1-2016Q1 and in nominal terms.\footnote{\url{https://www.bis.org/statistics/totcredit.htm?m=6\%7C326.}} For the period 1981Q1-1998Q4 we reconstructed the data using a weighted average of the credit data for GER, FRA, ITA, SPA, BEL, NET, UK with the weights reflecting each country's share of GDP in the group. The reconstructed time series is then converted at monthly frequency using the \cite{ChowLin1971} procedure. The monthly series is transformed in real terms using the monthly CPI index (source: BIS\footnote{\url{https://www.bis.org/statistics/cp.htm?m=6\%7C348.}}) after having removed the seasonal component using X-12-ARIMA method.\footnote{{The code was run in Eviews using the automatic selecting procedure available. From a visual inspection, using TRAMO-SEATS does not alter the dynamics of the SA series.}} The period 1981M1-1998M12 for the Euro Area CPI is reconstructed using the CPI
 seasonally adjusted time series for GER, FRA, ITA, SPA, BEL, NET.
 \item {\emph{Credit-to-GDP-ratio.}} The time series is available on the BIS Statistical portal at quarterly frequency for the period  1999Q1-2016Q1.\footnote{\url{https://www.bis.org/statistics/totcredit.htm?m=6\%7C326.}} For the period 1981Q1-1998Q4 we reconstructed the data using a backward forecasting procedure using using a weighted
 average of the credit-to-GDP-ratio data for GER, FRA, ITA, SPA, BEL, NET, UK. The reconstructed time series is then converted at monthly frequency using the \cite{ChowLin1971} procedure.
 \item {\emph{Real property price index.}} Three different house price data are used:
      \begin{itemize}
      \item Aggregate EU data reconstructed using data available from FED-International House Price Database at quarterly frequency.\footnote{\url{http://www.dallasfed.org/institute/houseprice/}} The house price index in real terms for the Euro Area is computed as a weighted average of the house price growth indices for GER, FRA, ITA, SPA, BEL, NET, UK. The aggregation is based on GDP weights. All data are seasonally adjusted and expressed in real terms. The house price indices of above countries are converted at monthly frequency using the \cite{ChowLin1971} procedure.
      \item Aggregate EU data taken from Eurostat. Data is seasonally adjusted using X-13 and deflated using aggregate CPI. CPI for Euro Area is seasonally adjusted using X-13. Monthly conversion is implemented using \cite{ChowLin1971} procedure.
      \item  Aggregate EU data reconstructed using data available from BIS at quarterly frequency. The house price index in real terms for the Euro Area is computed as a weighted average of the house price growth indices for GER, FRA, ITA, SPA, BEL, NET, UK. The aggregation is based on GDP weights. All data are seasonally adjusted using X-13 and are expressed in real terms. The house price indices
      of above countries are converted at monthly frequency using the \cite{ChowLin1971} procedure.     
      \end{itemize}
 \end{enumerate}
 \newpage
 \clearpage 
 \section{Summary statistics}
 \label{SummaryStats}
 \begin{sidewaystable}[p]
 \setlength{\tabcolsep}{2pt}
 \begin{center}
   \caption{Descriptive statistics}
     \begin{tabular}{lrrrrrrrrrrrrrr}
           & \multicolumn{1}{c}{AU} & \multicolumn{1}{c}{BE} & \multicolumn{1}{c}{DE} & \multicolumn{1}{c}{FI} & \multicolumn{1}{c}{FR} & \multicolumn{1}{c}{GE} & \multicolumn{1}{c}{GR} & \multicolumn{1}{c}{IR} & \multicolumn{1}{c}{IT} & \multicolumn{1}{c}{LU} & \multicolumn{1}{c}{NE} & \multicolumn{1}{c}{PO} & \multicolumn{1}{c}{SP} & \multicolumn{1}{c}{FIN1} \\
 \hline          
     IPI   &       &       &       &       &       &       &       &       &       &       &       &       &       &  \\
 \hline              
     Mean  & 0.2560 & 0.2100 & 0.1868 & 0.1996 & 0.0625 & 0.1398 & 0.0386 & 0.7550 & 0.0356 & 0.2571 & 0.1403 & 0.1311 & 0.0711 & 0.0759 \\
     Std   & 1.8332 & 2.3772 & 3.3067 & 2.5375 & 1.3144 & 1.7167 & 3.0789 & 5.4963 & 1.5454 & 3.8528 & 2.6840 & 2.7114 & 1.7927 & 0.0972 \\
     Skewness & 0.3954 & 0.2285 & 0.3831 & 0.2076 & -0.1318 & 0.1329 & 0.7345 & 0.0502 & 0.0310 & -0.0297 & 0.2639 & 0.3540 & 0.1465 & 0.3861 \\
     Kurtosis & 1.2710 & 4.5441 & 3.8653 & 6.9877 & 0.6727 & 8.5348 & 6.8569 & 4.5452 & 0.9849 & 1.6774 & 4.2471 & 1.2520 & 3.2684 & 0.4287 \\
     max   & 7.0651 & 12.4194 & 15.7783 & 16.4445 & 4.0196 & 12.3112 & 21.7115 & 32.1652 & 6.1118 & 12.5760 & 15.1818 & 11.1819 & 9.2994 & 0.4888 \\
     min   & -5.8406 & -9.9855 & -15.3117 & -12.9707 & -5.0183 & -9.4637 & -13.1990 & -24.0269 & -4.4759 & -16.9189 & -10.3773 & -8.7307 & -6.7392 & -0.2196 \\
           &       &       &       &       &       &       &       &       &       &       &       &       &       &  \\
 \hline                    
     CSU   &       &       &       &       &       &       &       &       &       &       &       &       &       &  \\
 \hline              
     Mean  & 1.3028 & 1.2826 & 0.4882 & 0.4736 & 2.3550 & 1.6087 & 6.3723 & 0.1178 & 4.6107 & 1.4774 & 1.1756 & 6.7536 & 6.3192 &  \\
     Std   & 2.1428 & 2.1756 & 1.2256 & 1.1225 & 3.6091 & 2.6082 & 8.6569 & 0.5403 & 6.7987 & 2.6117 & 2.0299 & 8.4002 & 8.0993 &  \\
     Skewness & 2.0392 & 2.4368 & 3.9735 & 3.2364 & 2.0226 & 2.1376 & 1.1720 & 6.4918 & 1.4318 & 2.5567 & 2.5118 & 1.0235 & 1.0330 &  \\
     Kurtosis & 4.4571 & 8.5284 & 20.9902 & 12.2254 & 4.8443 & 5.7010 & -0.0122 & 45.7514 & 0.8383 & 8.8871 & 9.2100 & -0.3015 & -0.4008 &  \\
     max   & 13.4400 & 16.0500 & 11.2200 & 7.9000 & 20.2800 & 16.5500 & 29.5000 & 4.8700 & 26.2500 & 17.9200 & 15.7000 & 29.9700 & 26.7700 &  \\
     min   & 0.0000 & 0.0000 & 0.0000 & 0.0000 & 0.0000 & 0.0000 & 0.0000 & 0.0000 & 0.0000 & 0.0000 & 0.0000 & 0.0000 & 0.0000 &  \\
           &       &       &       &       &       &       &       &       &       &       &       &       &       &  \\
 \hline                    
     SPI   &       &       &       &       &       &       &       &       &       &       &       &       &       &  \\
 \hline              
     Mean  & 0.0378 & 0.1449 & 0.0953 & 0.2619 & 0.0271 & 0.0570 & -0.1082 & 0.0949 & -0.1017 & 0.1614 & 0.1783 & -0.1319 & -0.1101 &  \\
     Std   & 0.6666 & 0.7814 & 0.8448 & 0.6721 & 0.7103 & 0.6939 & 0.5881 & 0.8229 & 0.5970 & 0.8663 & 0.8447 & 0.8715 & 0.6958 &  \\
     Skewness & 0.0483 & -0.2315 & -0.6507 & -0.2790 & -0.0661 & -0.0071 & -0.0263 & 0.0493 & -0.3446 & -0.1045 & -0.4833 & -0.1202 & -0.1822 &  \\
     Kurtosis & -0.2176 & -0.2245 & 1.4928 & -0.1456 & -0.3719 & 0.0547 & 0.6377 & -0.0696 & 0.1285 & -0.0659 & 0.8899 & -0.3187 & -0.5633 &  \\
     max   & 2.1900 & 2.0000 & 2.0900 & 1.7400 & 1.9300 & 1.9000 & 2.0500 & 2.4400 & 1.4200 & 2.4600 & 2.1400 & 2.0600 & 1.3900 &  \\
     min   & -1.7400 & -2.2100 & -3.7300 & -1.8500 & -1.9200 & -2.1600 & -1.6600 & -2.1400 & -1.9200 & -2.2900 & -3.0600 & -2.7000 & -1.8300 &  \\
           &       &       &       &       &       &       &       &       &       &       &       &       &       &  \\
 \hline                    
     r20mm &       &       &       &       &       &       &       &       &       &       &       &       &       &  \\
 \hline              
     Mean  & 0.7672 & 0.2877 & 0.1364 & 0.1275 & 0.3567 & 0.3530 & 0.2718 & 0.5791 & 0.5540 & 0.4026 & 0.3196 & 0.5687 & 0.2641 &  \\
     Std   & 0.6657 & 0.4249 & 0.2521 & 0.1940 & 0.2946 & 0.3010 & 0.3669 & 0.5595 & 0.4354 & 0.6049 & 0.3816 & 0.8638 & 0.2847 &  \\
     Skewness & 1.2087 & 2.3590 & 2.7616 & 2.2500 & 1.4449 & 1.2677 & 1.9363 & 1.6066 & 1.4062 & 2.1017 & 1.8536 & 2.2725 & 1.7723 &  \\
     Kurtosis & 1.5461 & 6.3801 & 9.2356 & 6.1447 & 2.4451 & 1.8739 & 4.3536 & 3.5642 & 2.3306 & 5.1077 & 4.0128 & 5.7232 & 3.1147 &  \\
     max   & 3.6000 & 2.5000 & 1.6800 & 1.2400 & 1.6600 & 1.5900 & 2.1500 & 3.8300 & 2.5100 & 3.4100 & 2.3500 & 5.4100 & 1.3700 &  \\
     min   & 0.0000 & 0.0000 & 0.0000 & 0.0000 & 0.0000 & 0.0000 & 0.0000 & 0.0000 & 0.0000 & 0.0000 & 0.0000 & 0.0000 & 0.0000 &  \\
 \hline              
     \end{tabular}%
   \label{tab:Descrip_Stat}%
 \end{center}
\end{sidewaystable}
 
 \begin{figure}[h!]
         \centering
         \subfigure[AU]{                
                 \includegraphics[width=0.17\textwidth]{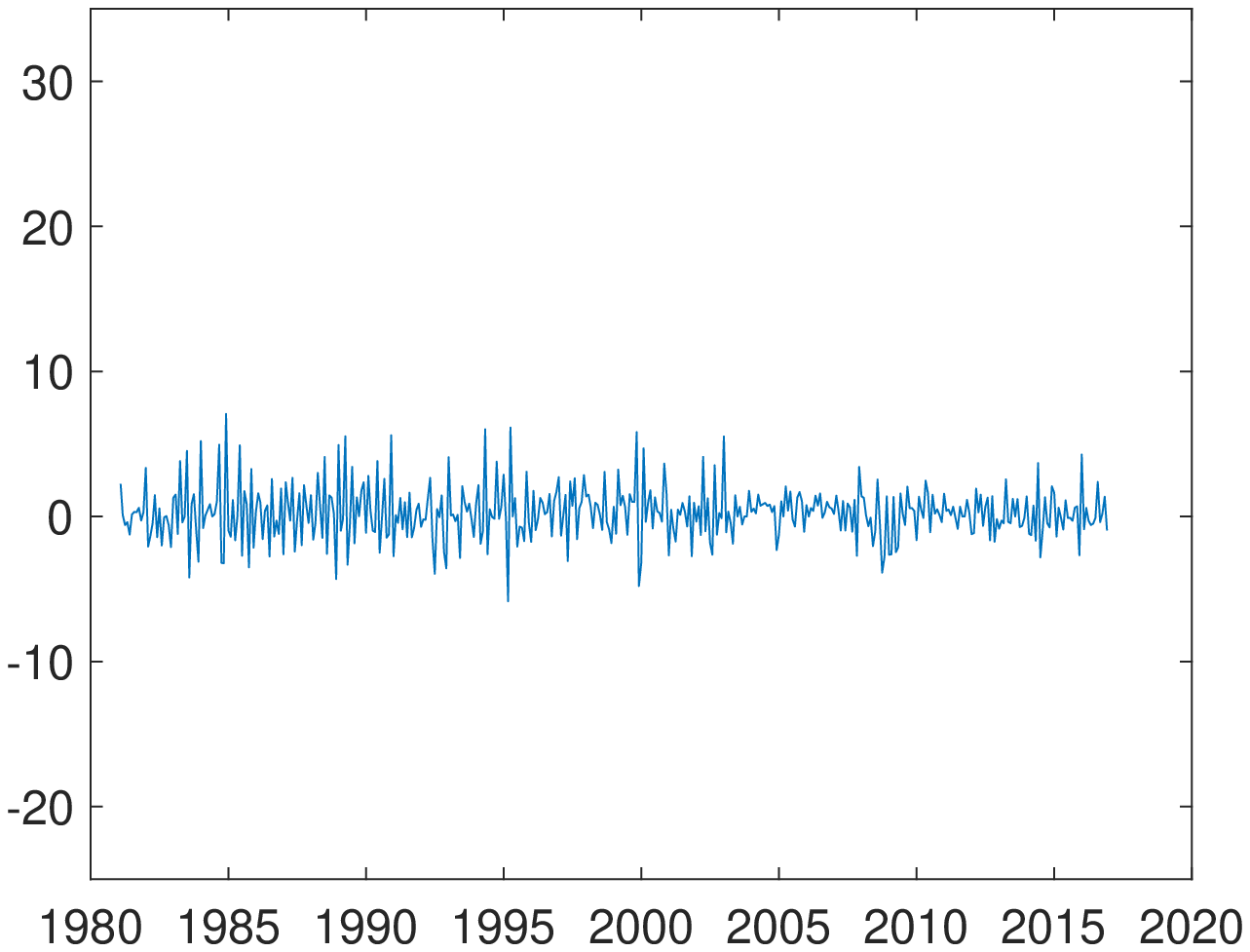}
                 \label{fig:plotIPIAU}}
         ~
         \subfigure[BE]{                
                 \includegraphics[width=0.17\textwidth]{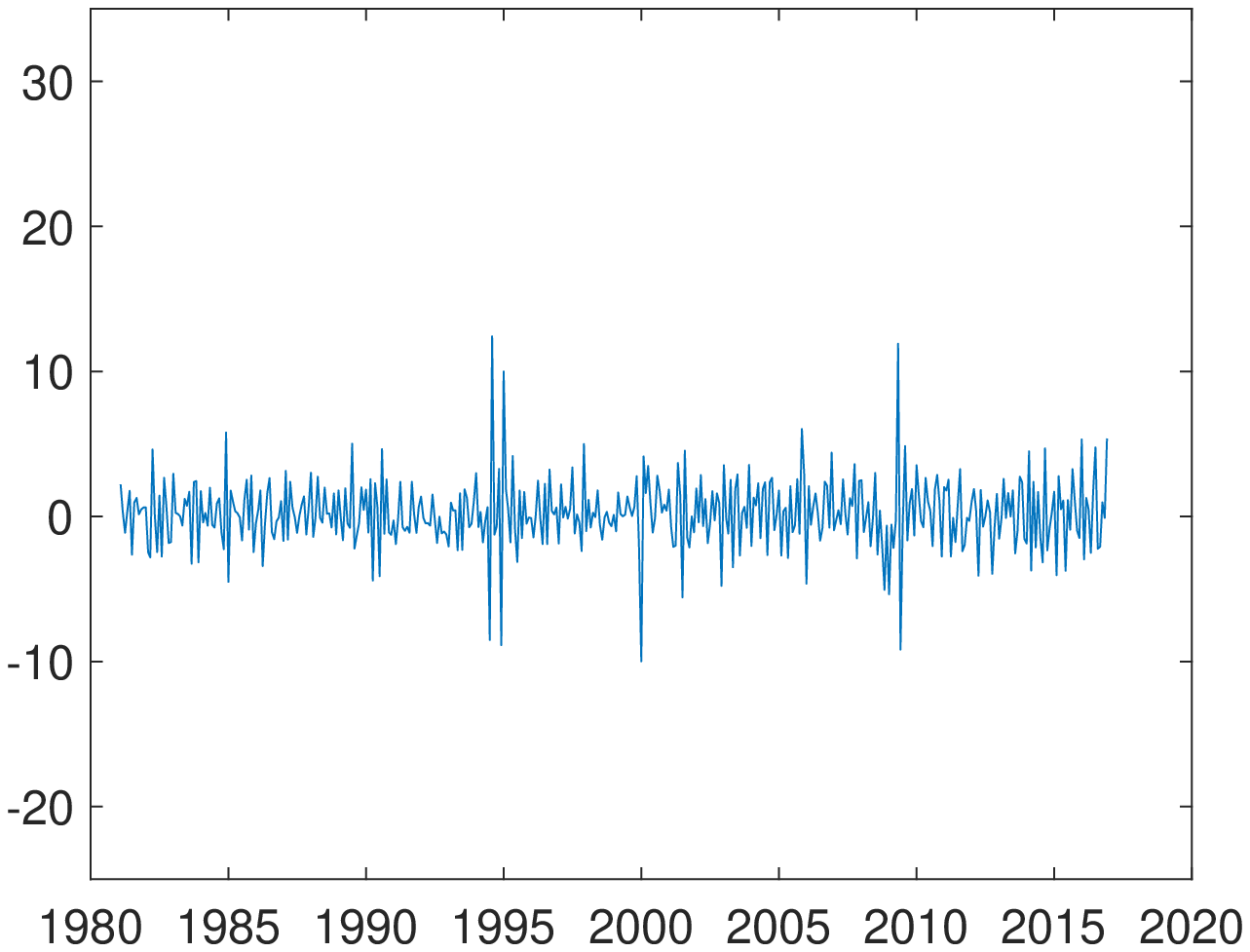}
                 \label{fig:plotIPIBE}}
         ~
         \subfigure[DE]{
                 \includegraphics[width=0.17\textwidth]{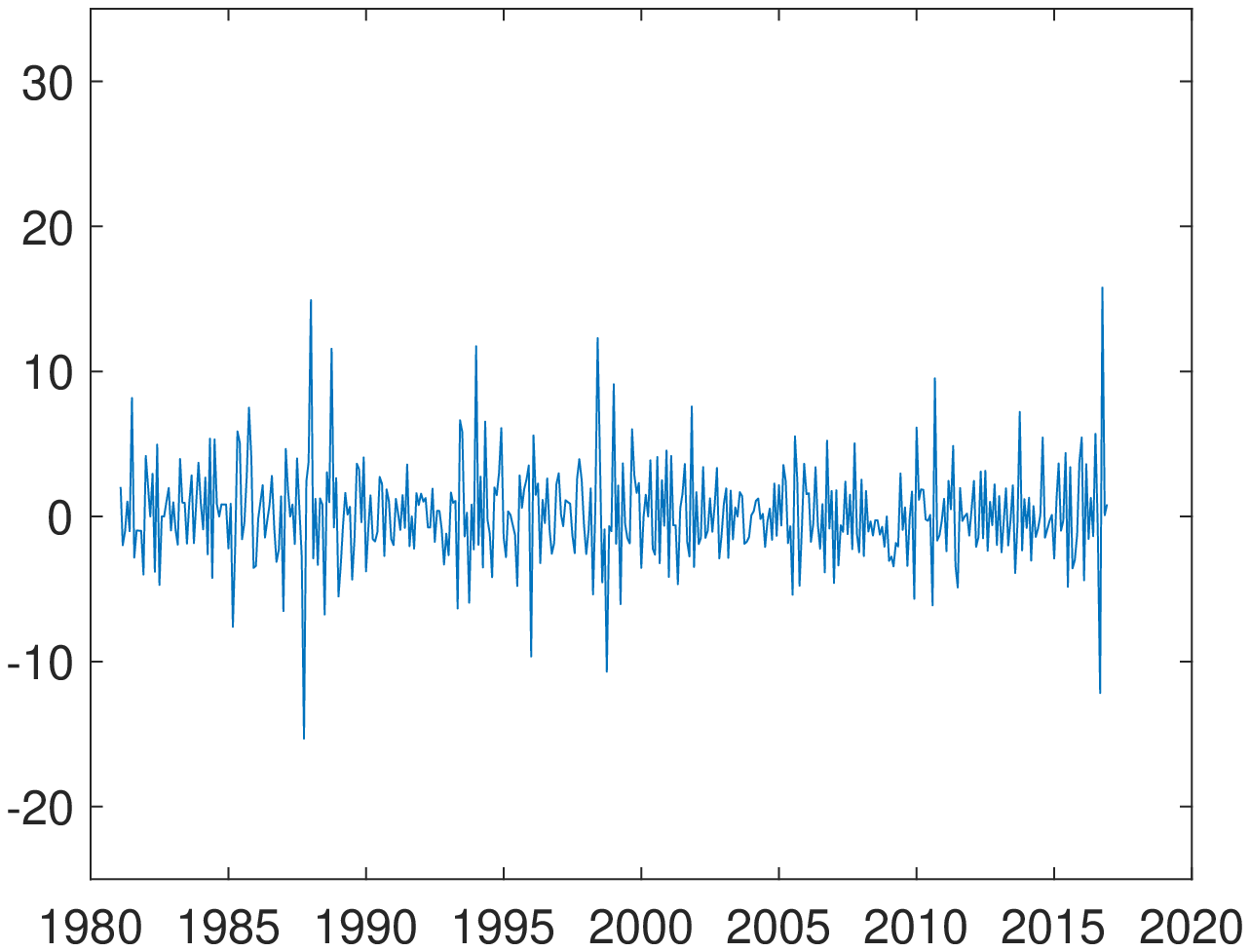}
                 \label{fig:plotIPIDE}}    
         ~
         \subfigure[FI]{                
                  \includegraphics[width=0.17\textwidth]{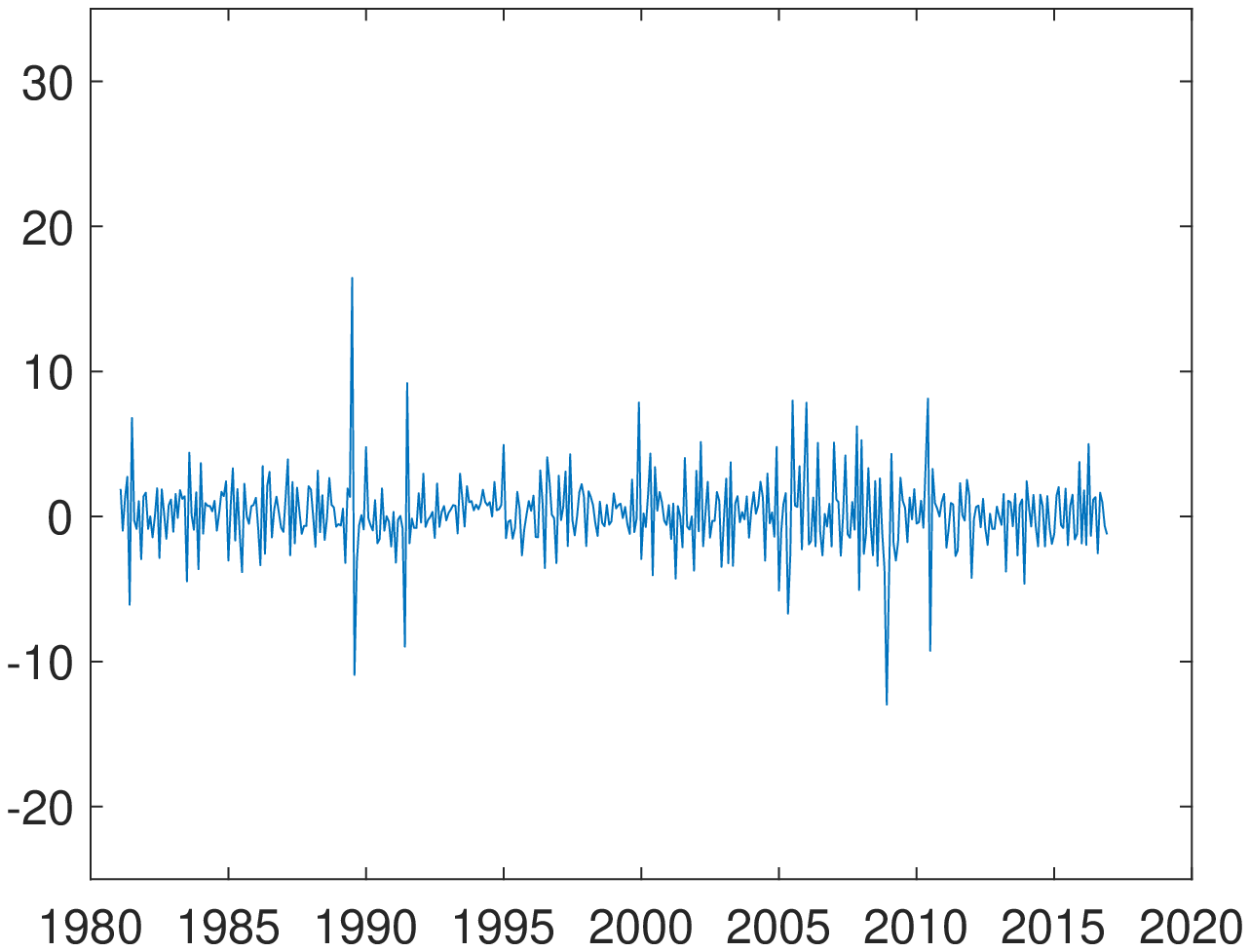}
                  \label{fig:plotIPIFI}}
         ~
         \subfigure[FR]{
                  \includegraphics[width=0.17\textwidth]{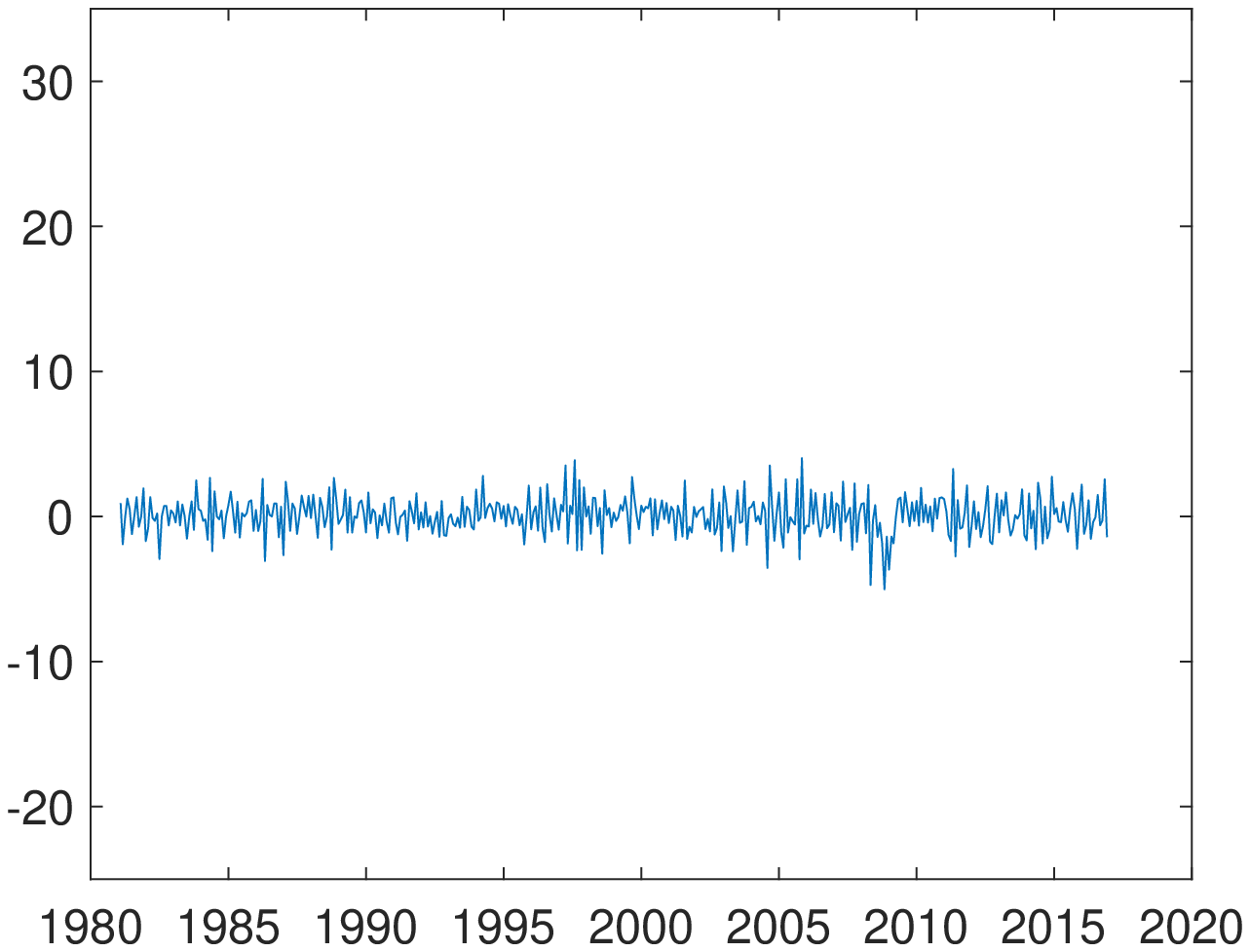}
                  \label{fig:plotIPIFR}} 
         ~
         \subfigure[GE]{                
                  \includegraphics[width=0.17\textwidth]{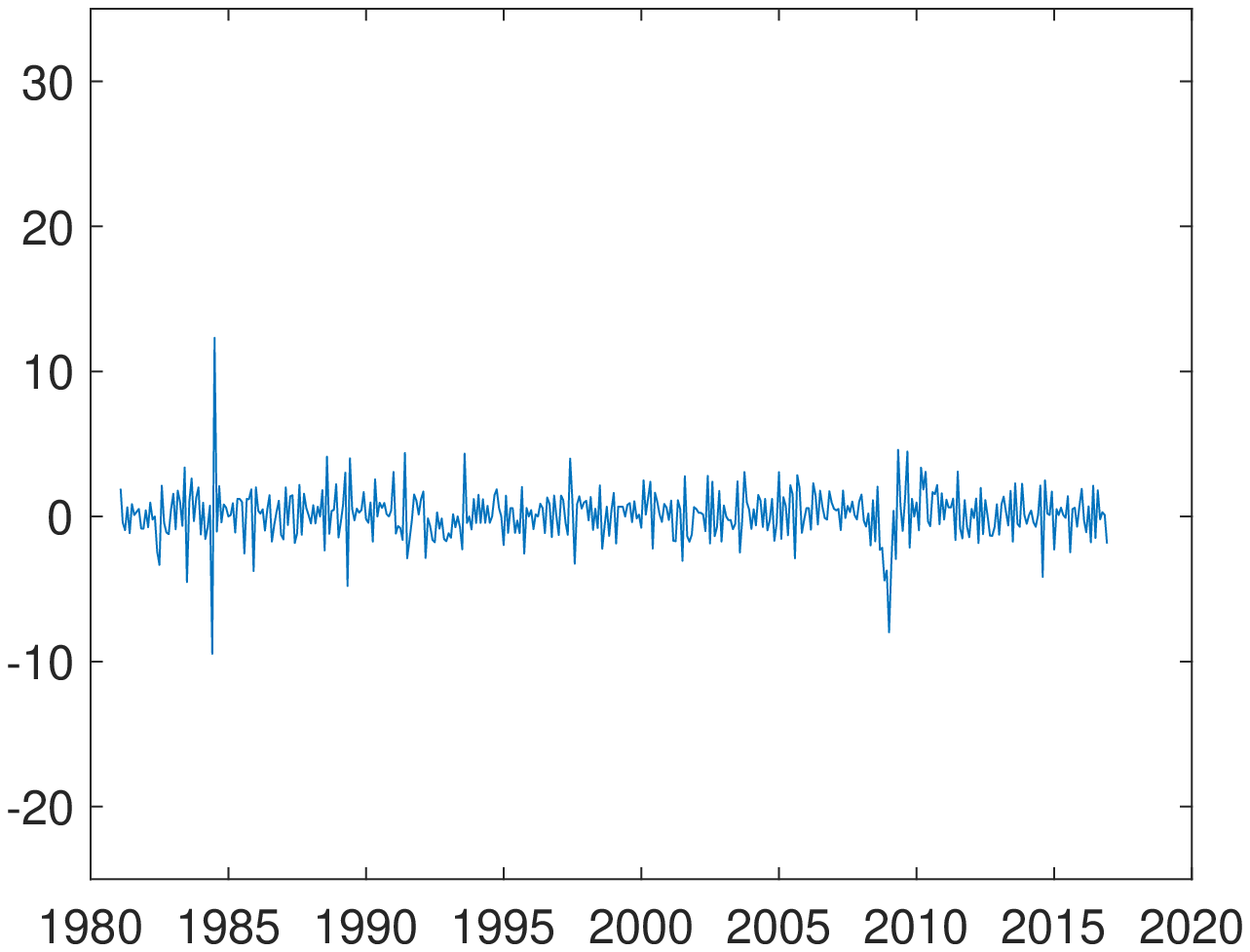}
                  \label{fig:plotIPIGE}}
         ~
         \subfigure[GR]{
                  \includegraphics[width=0.17\textwidth]{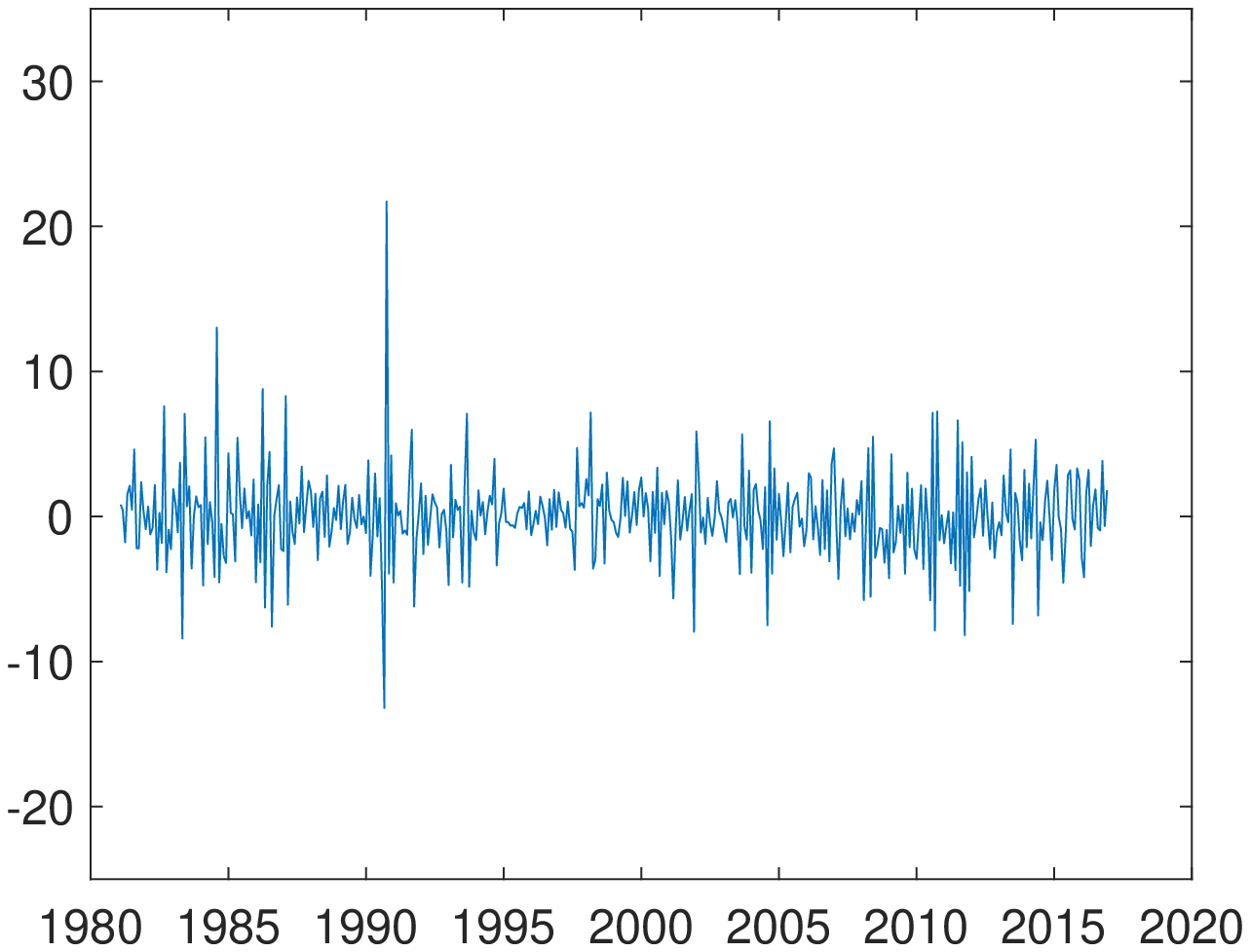}
                  \label{fig:plotIPIGR}} 
         ~
         \subfigure[IR]{                
                  \includegraphics[width=0.17\textwidth]{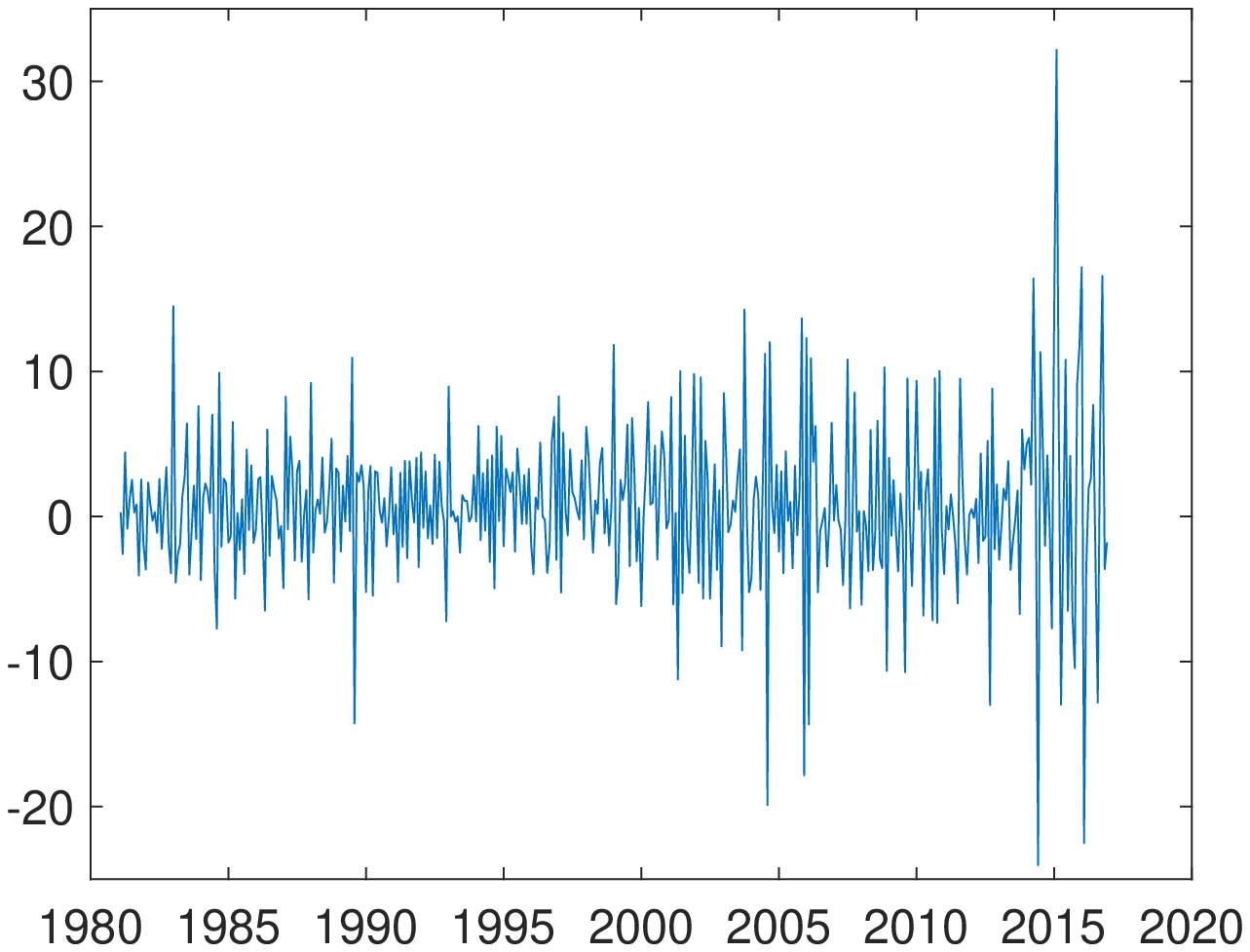}
                  \label{fig:plotIPIIR}}
         ~
         \subfigure[IT]{
                  \includegraphics[width=0.17\textwidth]{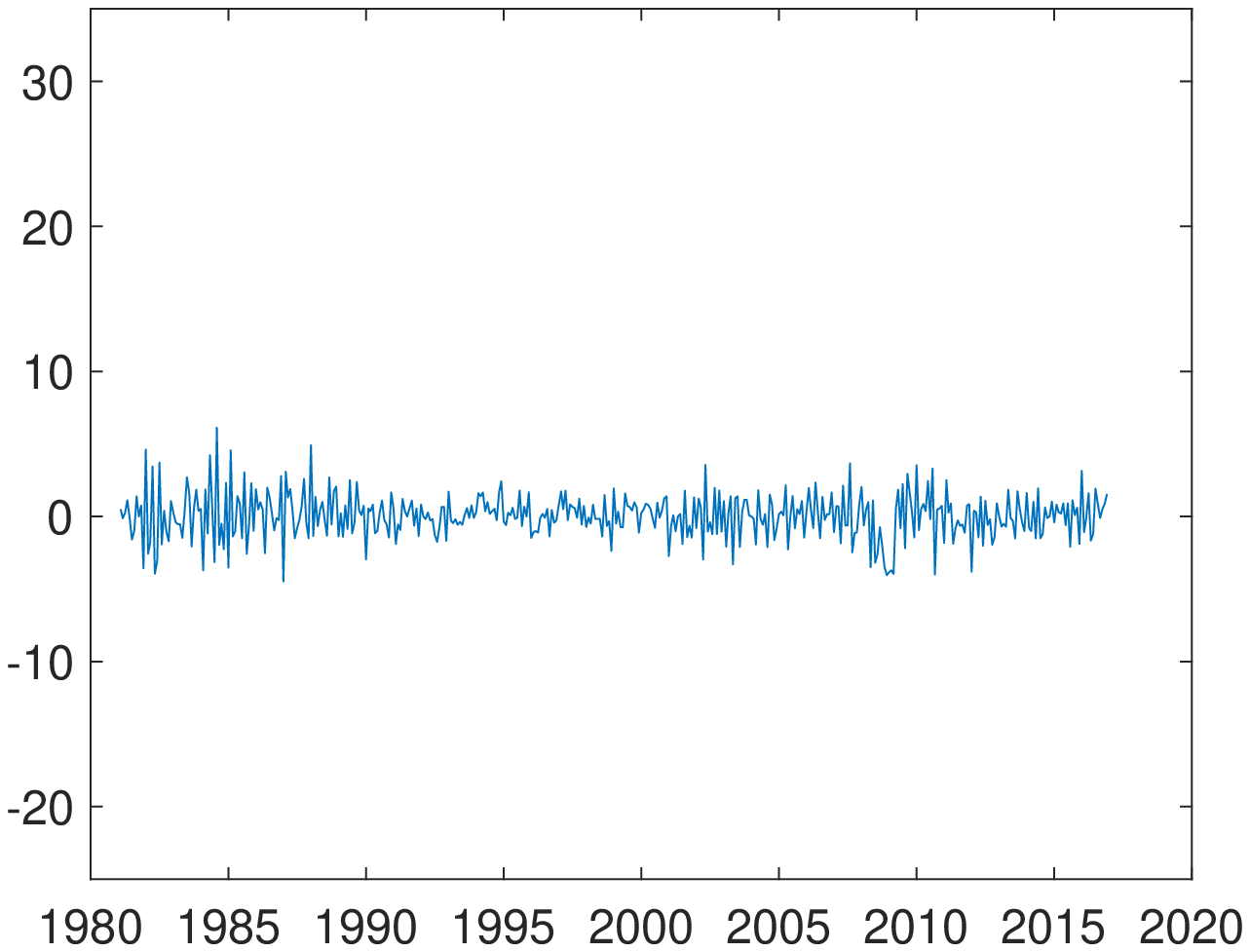}
                  \label{fig:plotIPIIT}}       
         ~
         \subfigure[LU]{                
                  \includegraphics[width=0.17\textwidth]{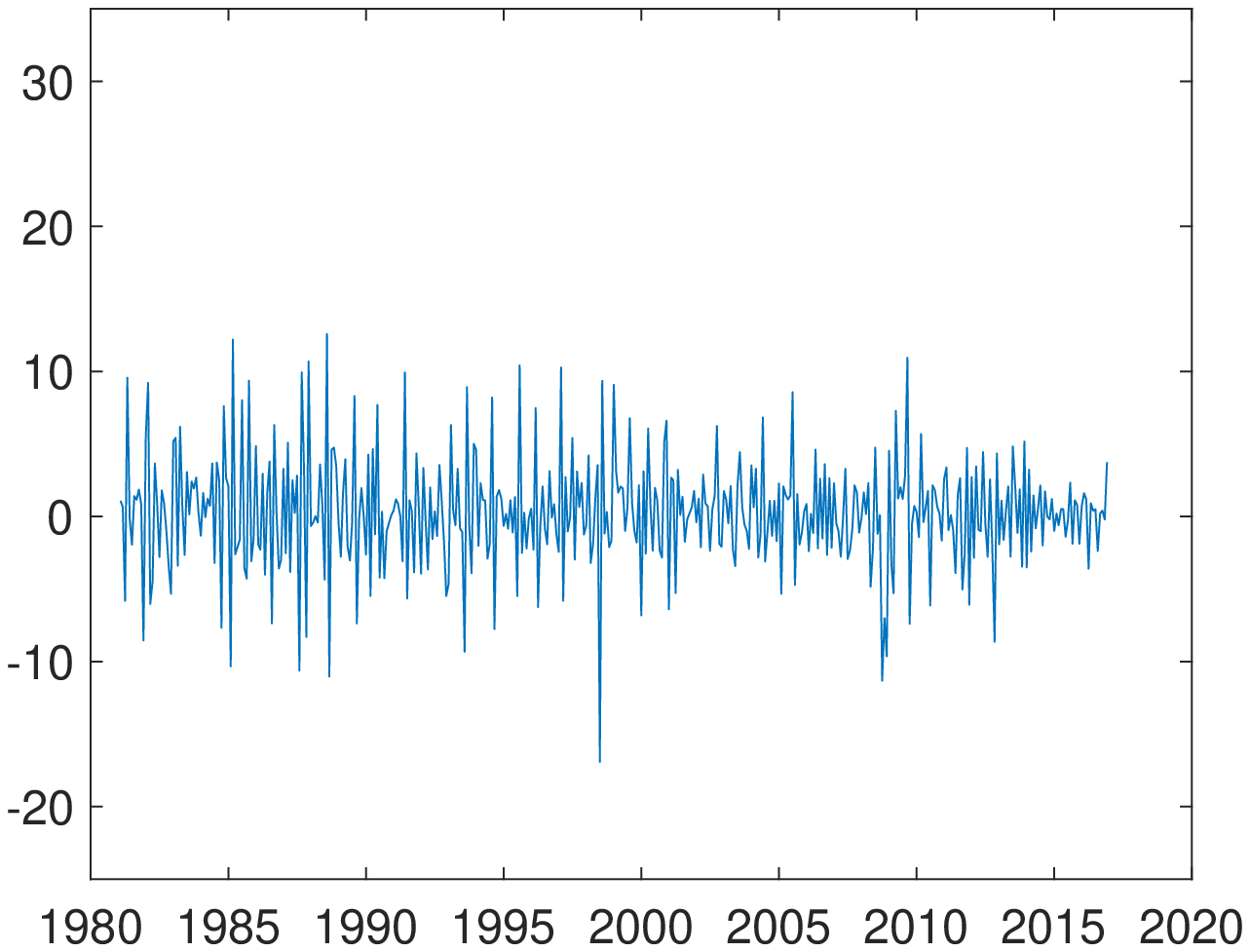}
                  \label{fig:plotIPILU}}
         ~
         \subfigure[NE]{
                  \includegraphics[width=0.17\textwidth]{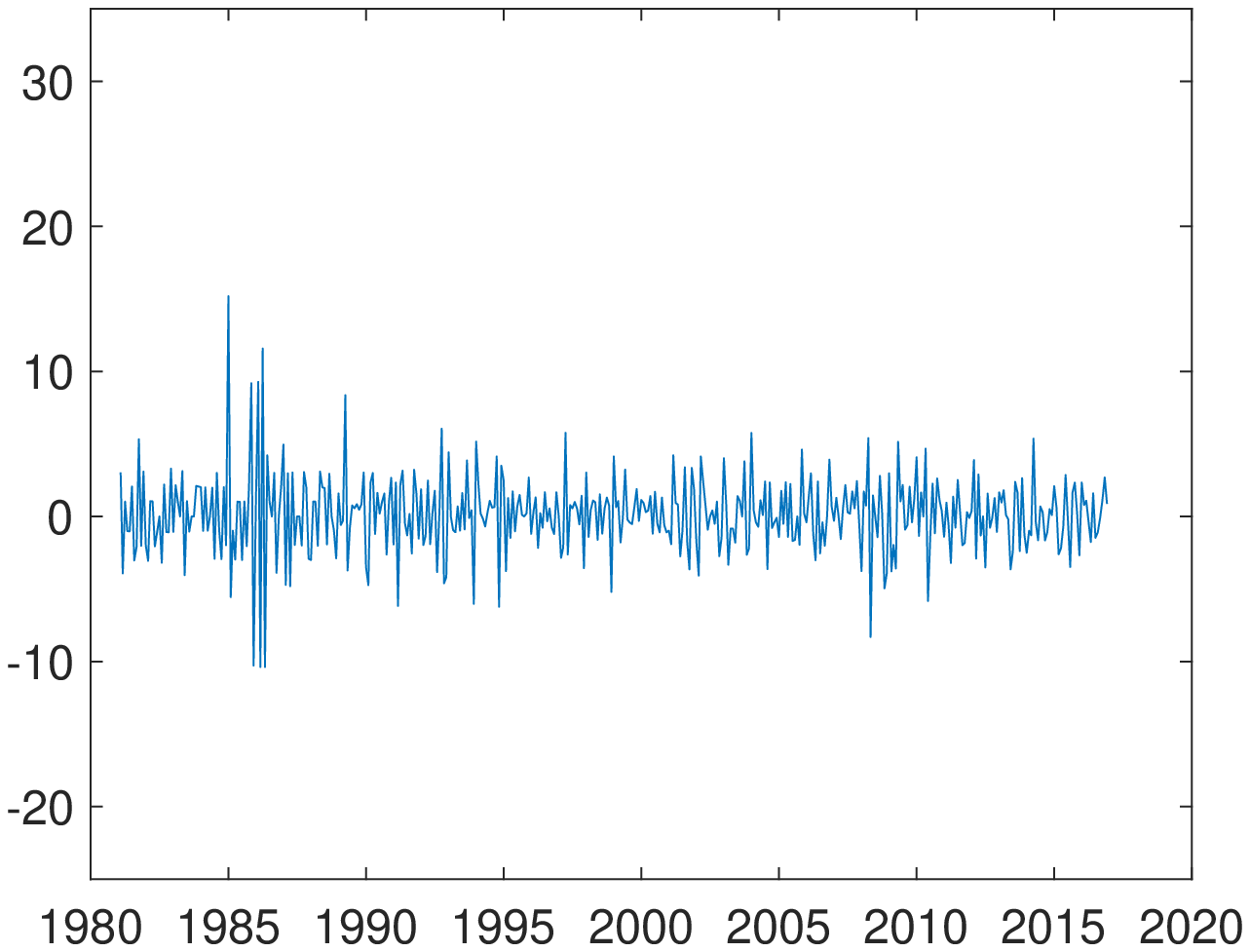}
                  \label{fig:plotIPINE}} 
         ~
         \subfigure[PO]{                
                  \includegraphics[width=0.17\textwidth]{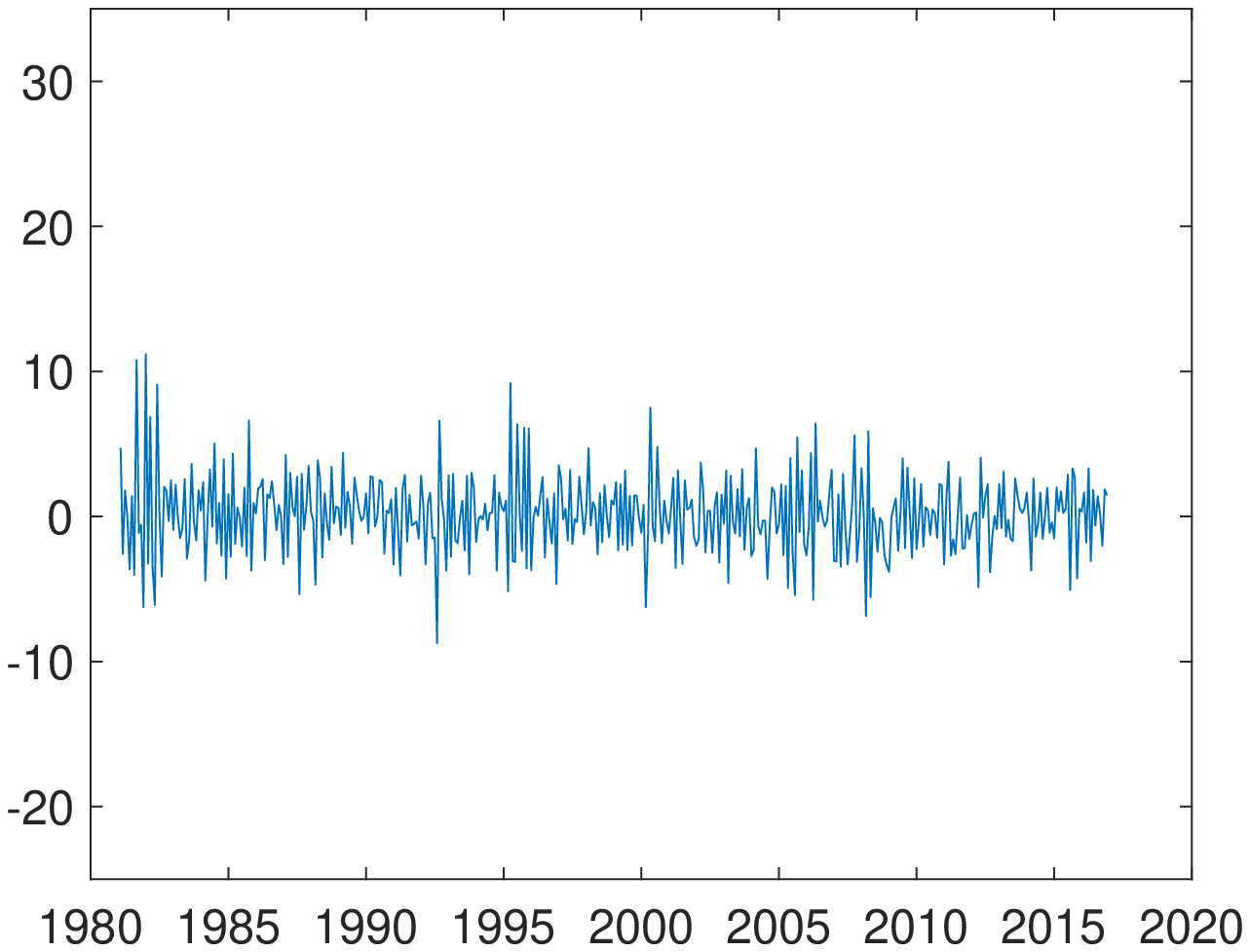}
                  \label{fig:plotIPIPO}}
         ~
         \subfigure[SP]{
                  \includegraphics[width=0.17\textwidth]{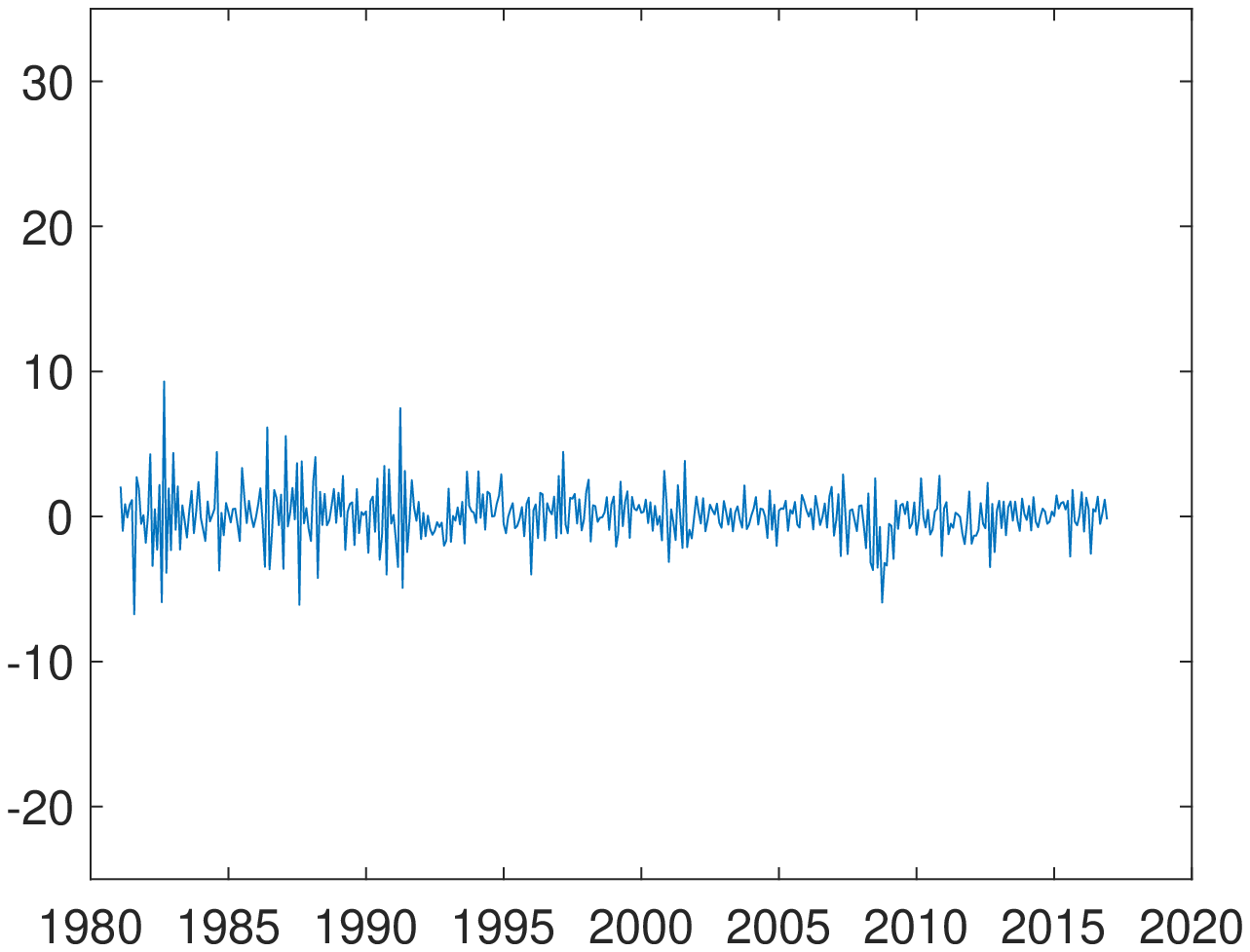}
                  \label{fig:plotIPISP}} 
         ~
         \subfigure[Fin]{
                  \includegraphics[width=0.17\textwidth]{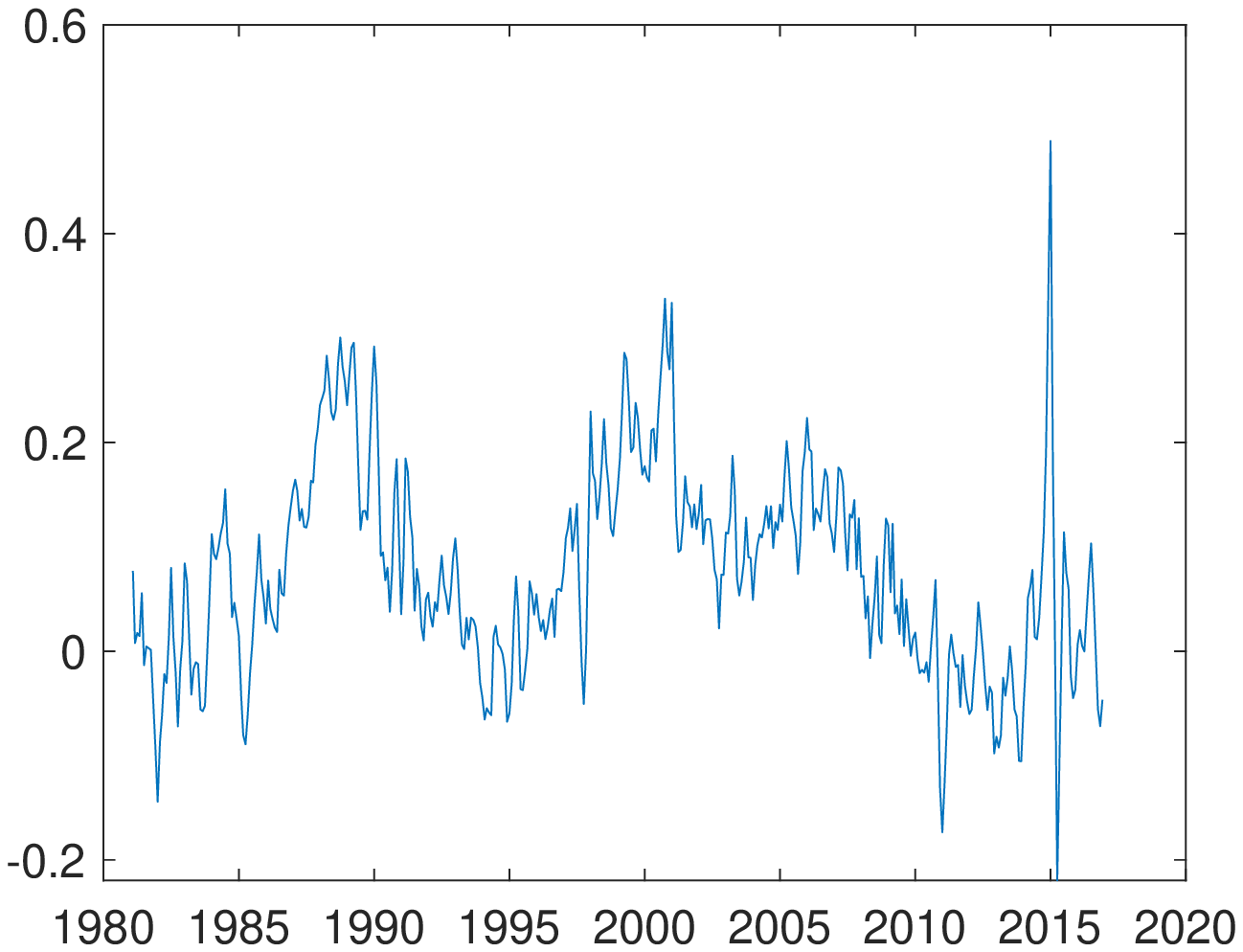}
                  \label{fig:plotIPIFin1}}                  
 		\caption{Time series plot of the percentage change on previous month's Industrial Production Indices (IPI) and aggregate (composite) Financial index for 13 EU countries for the sample period February 1981 to December 2016. }\label{PLOTIPI}
 \end{figure}
 
 \begin{figure}[h!]
         \centering
         \subfigure[AU]{                
                 \includegraphics[width=0.17\textwidth]{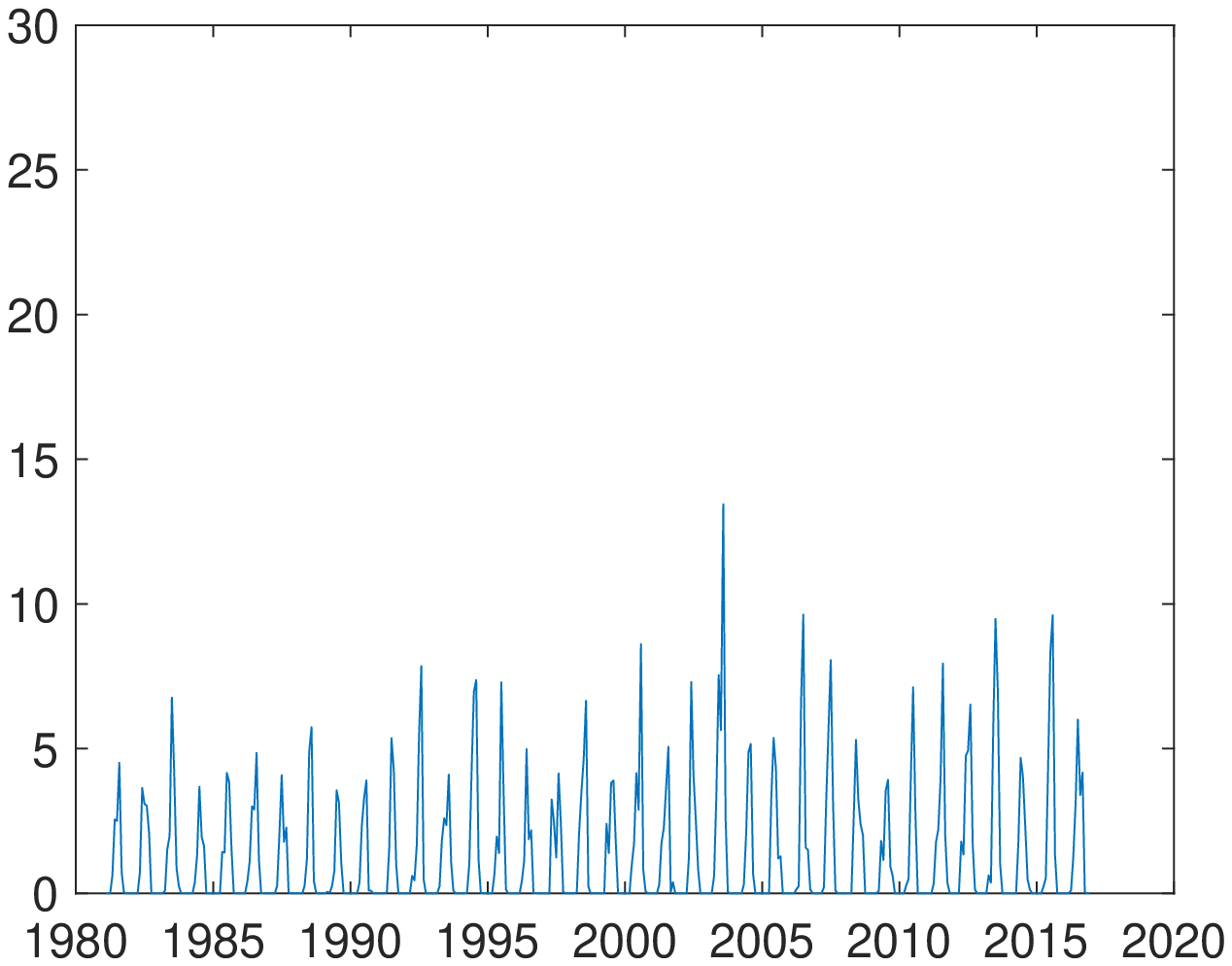}
                 \label{fig:plotCSUAU}}
         ~
         \subfigure[BE]{                
                 \includegraphics[width=0.17\textwidth]{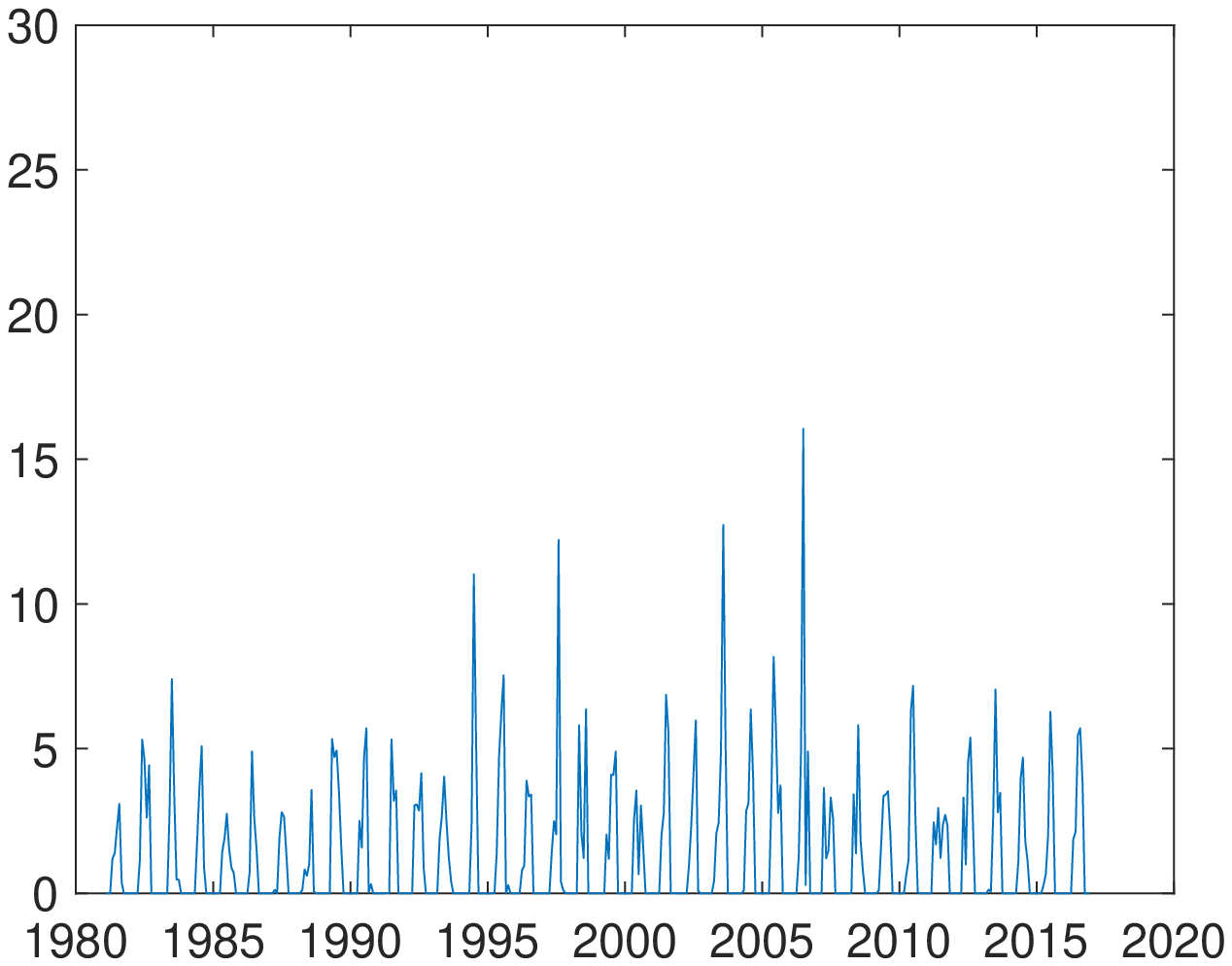}
                 \label{fig:plotCSUBE}}
         ~
         \subfigure[DE]{
                 \includegraphics[width=0.17\textwidth]{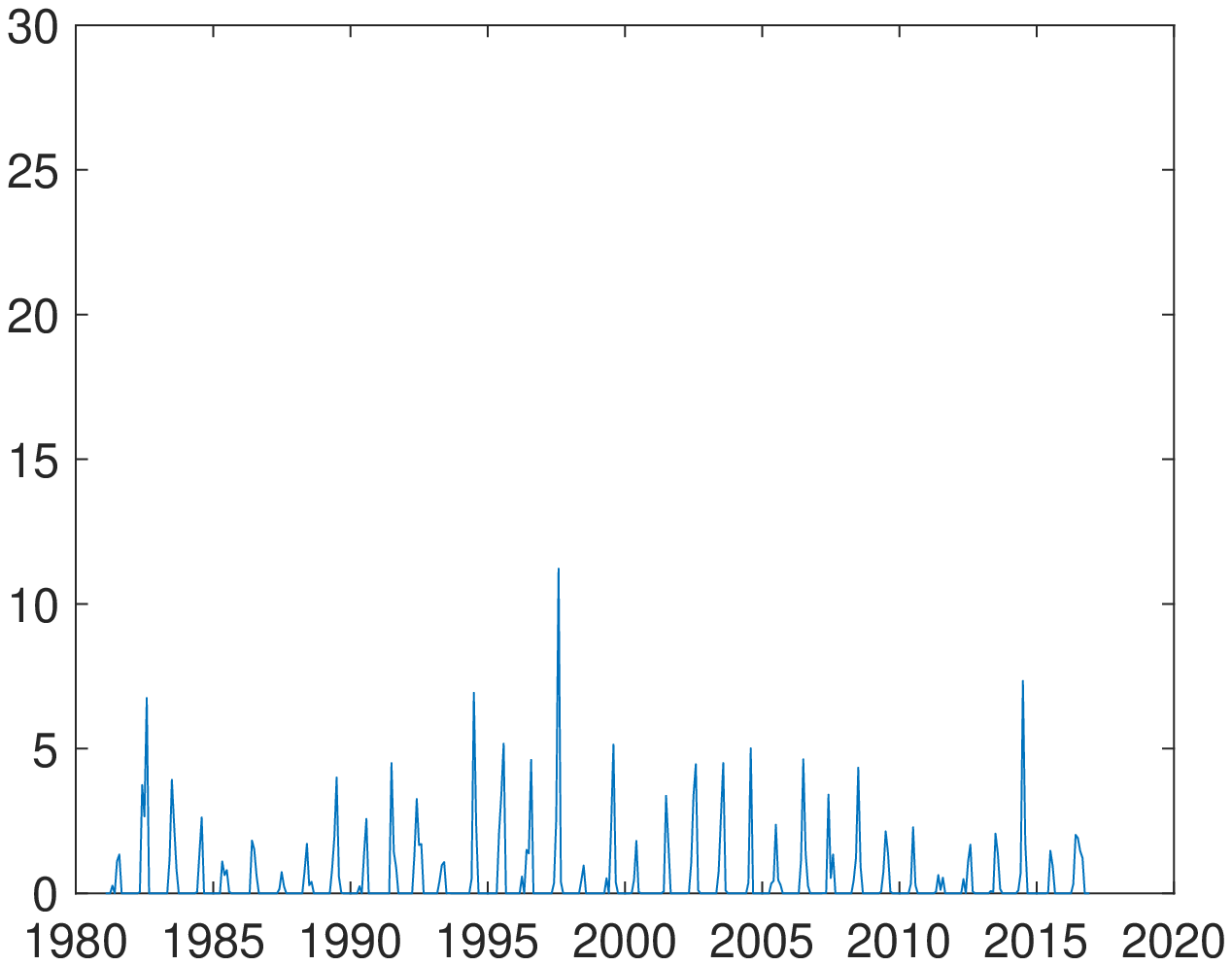}
                 \label{fig:plotCSUDE}}    
         ~
         \subfigure[FI]{                
                  \includegraphics[width=0.17\textwidth]{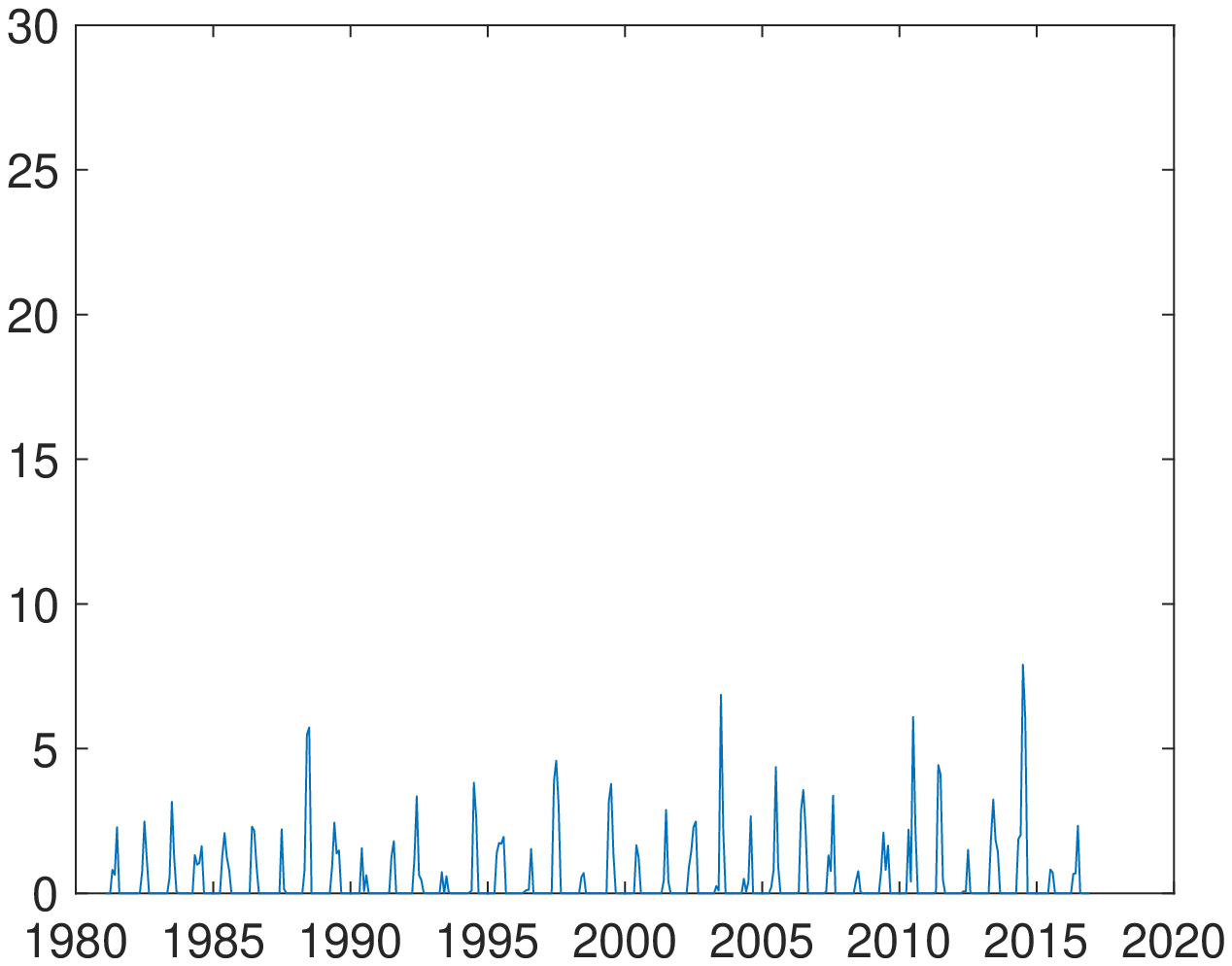}
                  \label{fig:plotCSUFI}}
         ~
         \subfigure[FR]{
                  \includegraphics[width=0.17\textwidth]{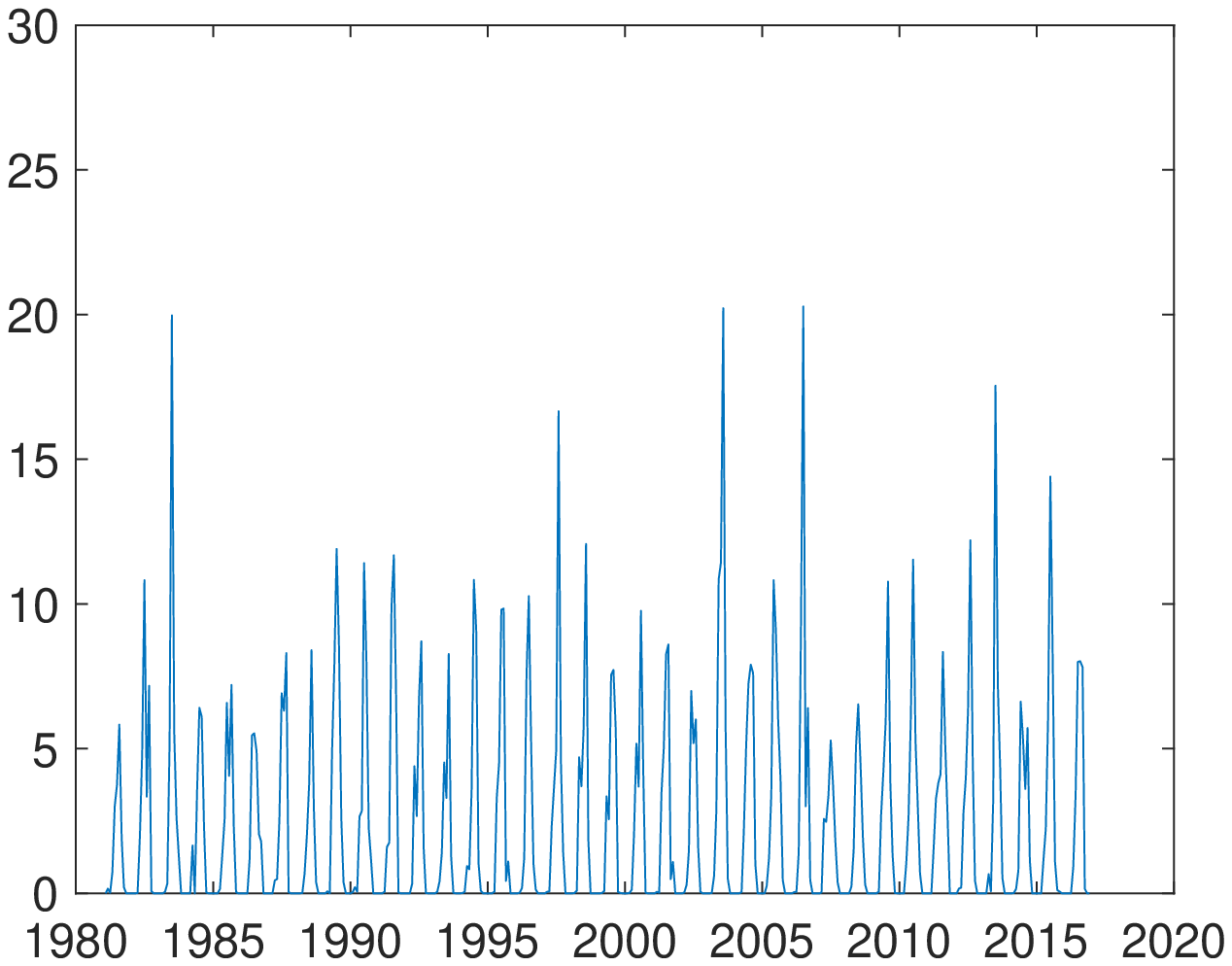}
                  \label{fig:plotCSUFR}} 
         ~
         \subfigure[GE]{                
                  \includegraphics[width=0.17\textwidth]{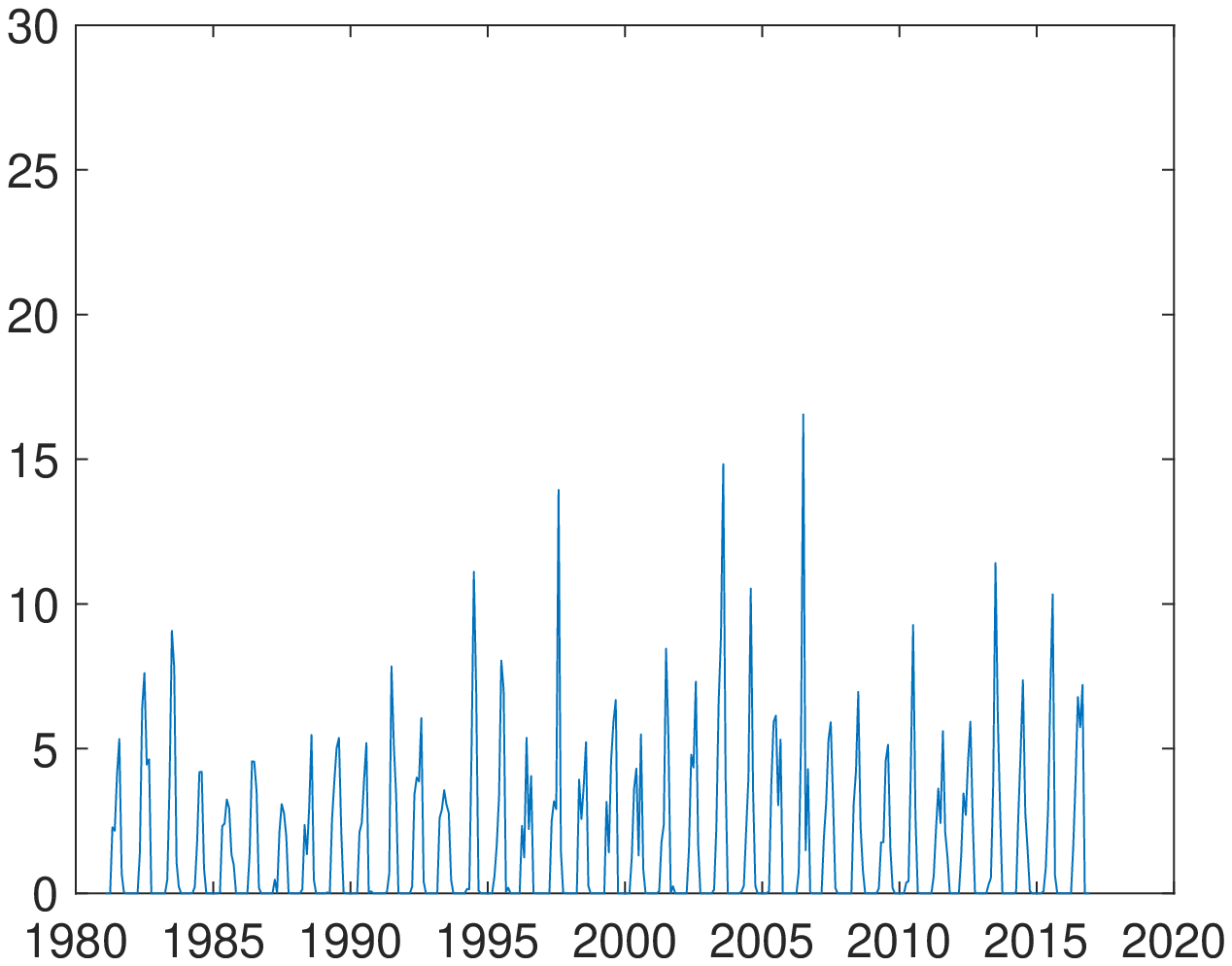}
                  \label{fig:plotCSUGE}}
         ~
         \subfigure[GR]{
                  \includegraphics[width=0.17\textwidth]{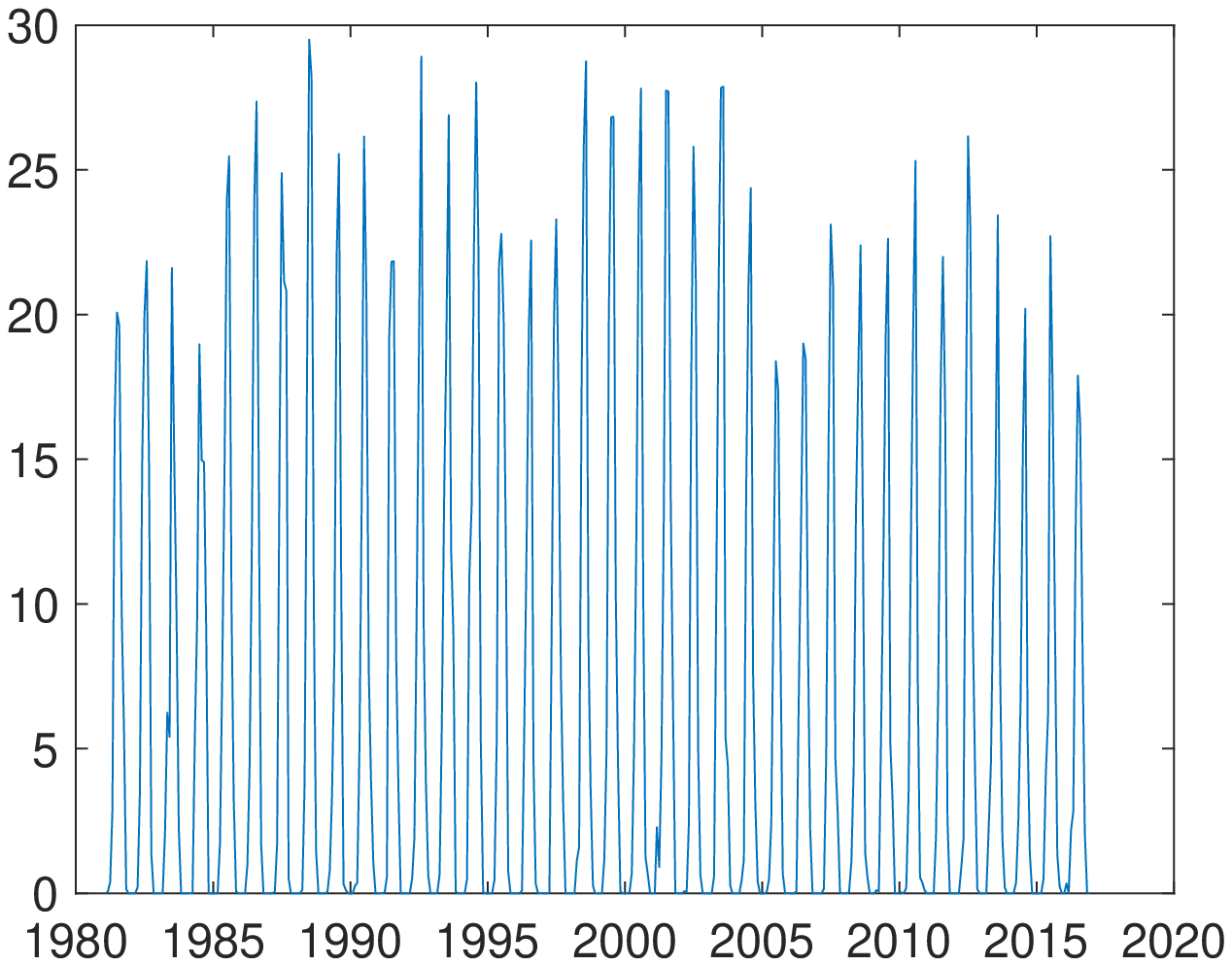}
                  \label{fig:plotCSUGR}} 
         ~
         \subfigure[IR]{                
                  \includegraphics[width=0.17\textwidth]{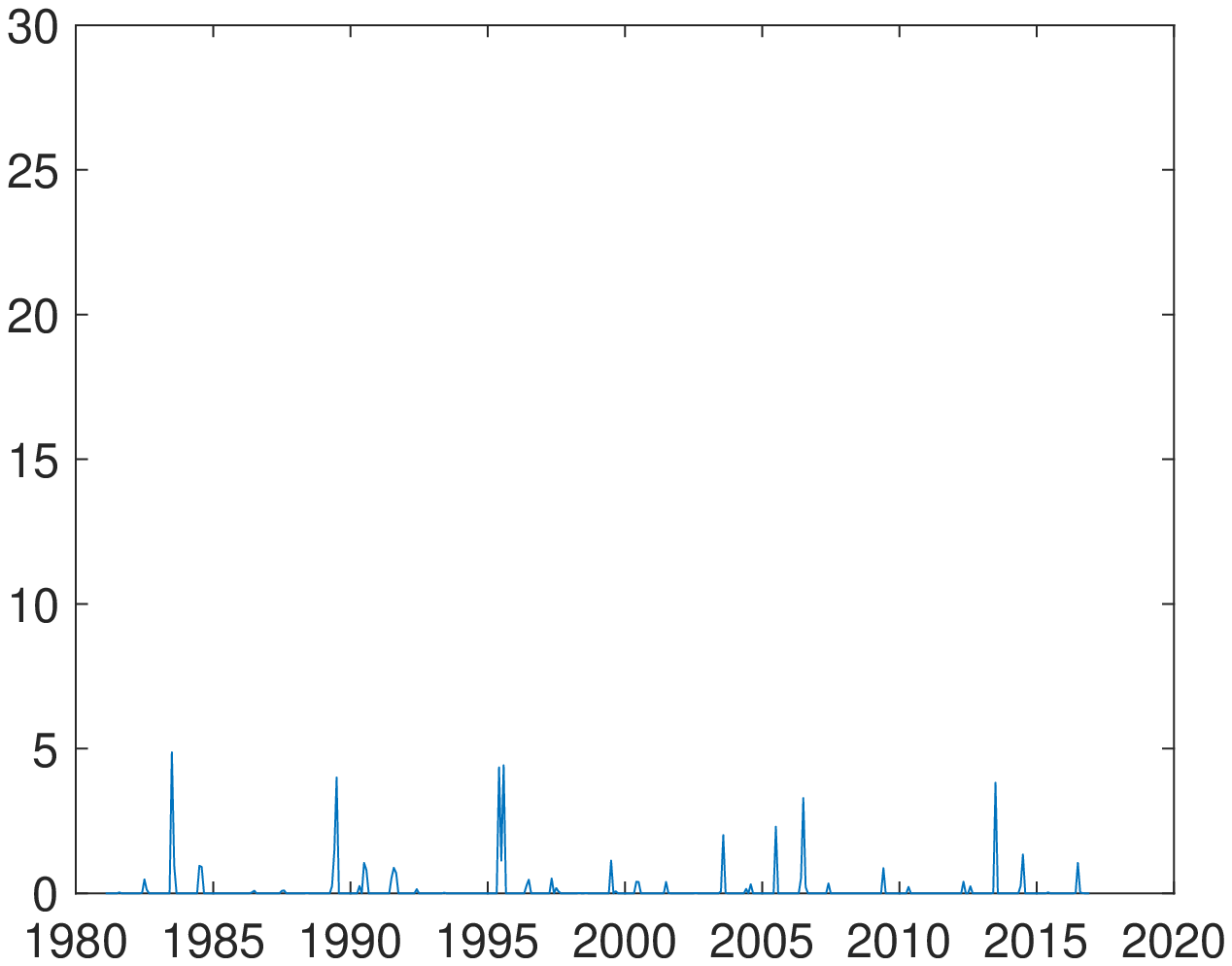}
                  \label{fig:plotCSUIR}}
         ~
         \subfigure[IT]{
                  \includegraphics[width=0.17\textwidth]{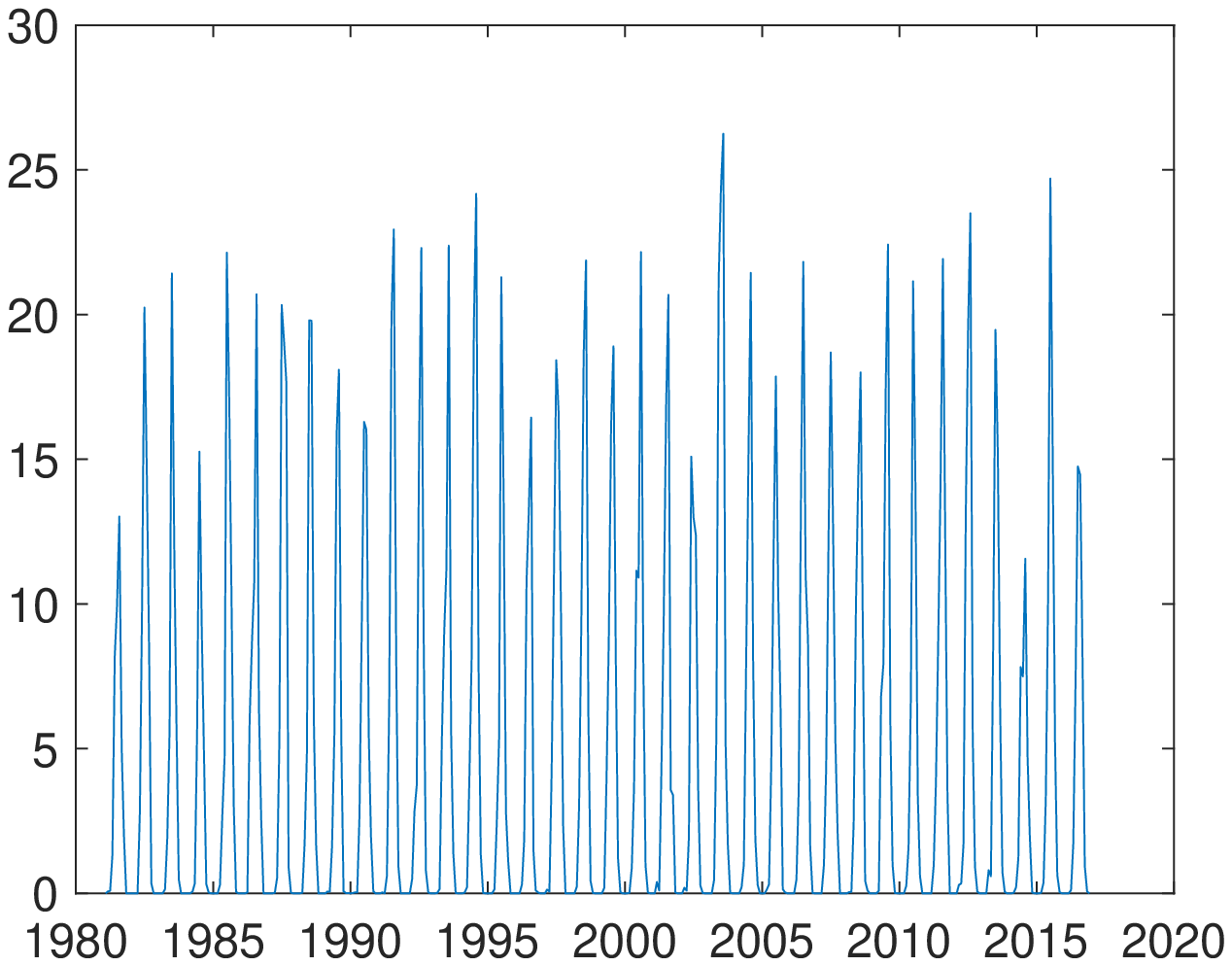}
                  \label{fig:plotCSUIT}}       
         ~
         \subfigure[LU]{                
                  \includegraphics[width=0.17\textwidth]{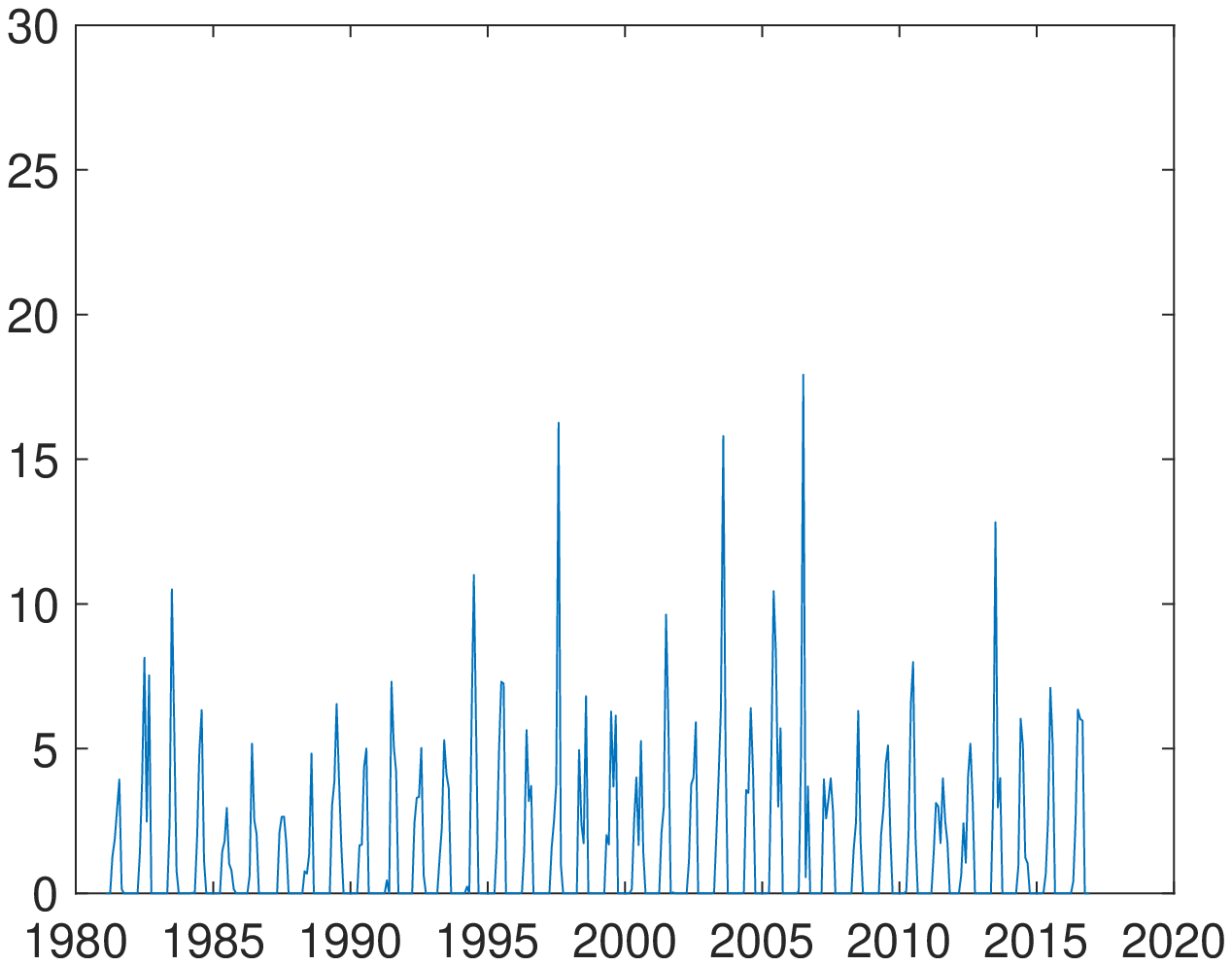}
                  \label{fig:plotCSULU}}
         ~
         \subfigure[NE]{
                  \includegraphics[width=0.17\textwidth]{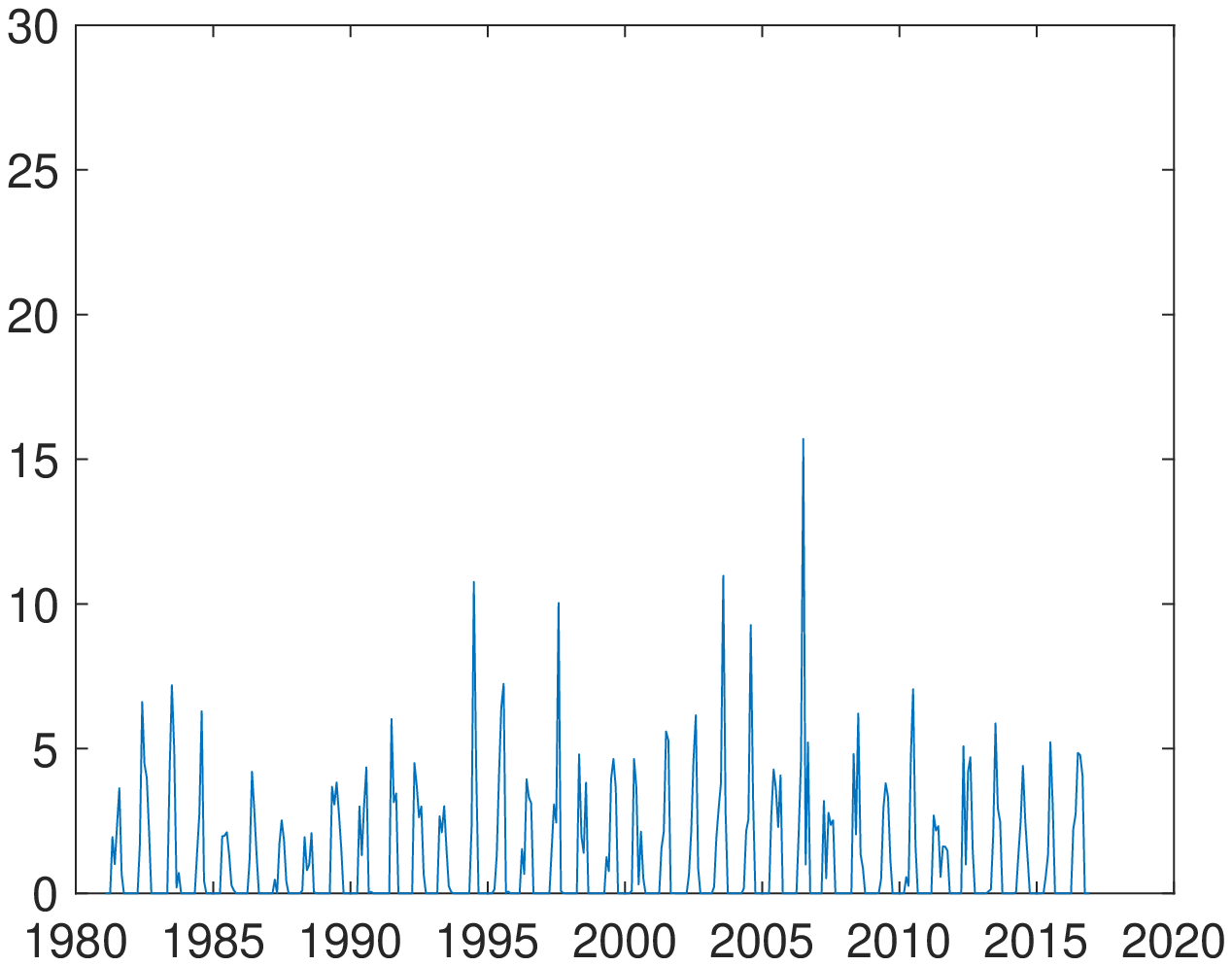}
                  \label{fig:plotCSUNE}} 
         ~
         \subfigure[PO]{                
                  \includegraphics[width=0.17\textwidth]{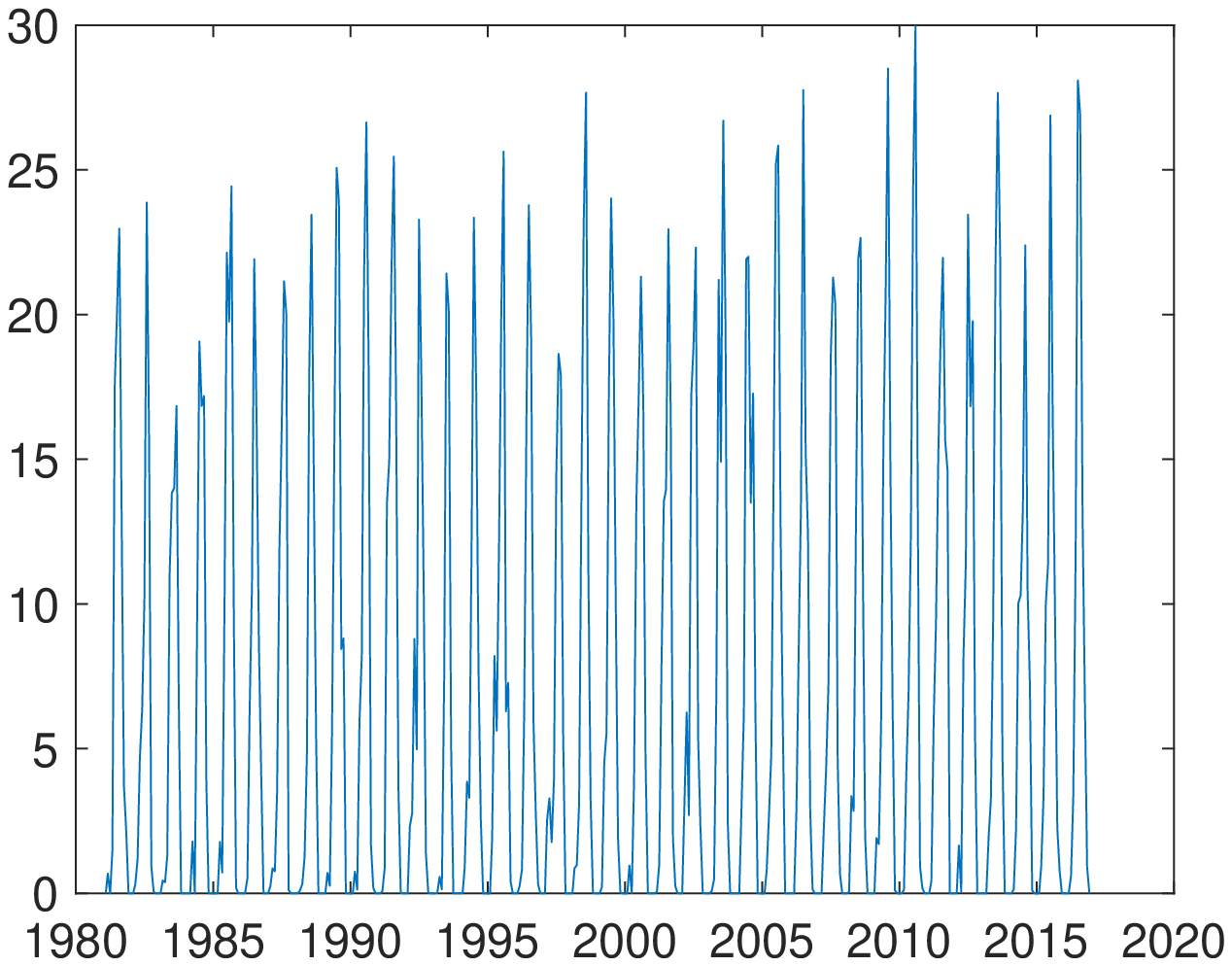}
                  \label{fig:plotCSUPO}}
         ~
         \subfigure[SP]{
                  \includegraphics[width=0.17\textwidth]{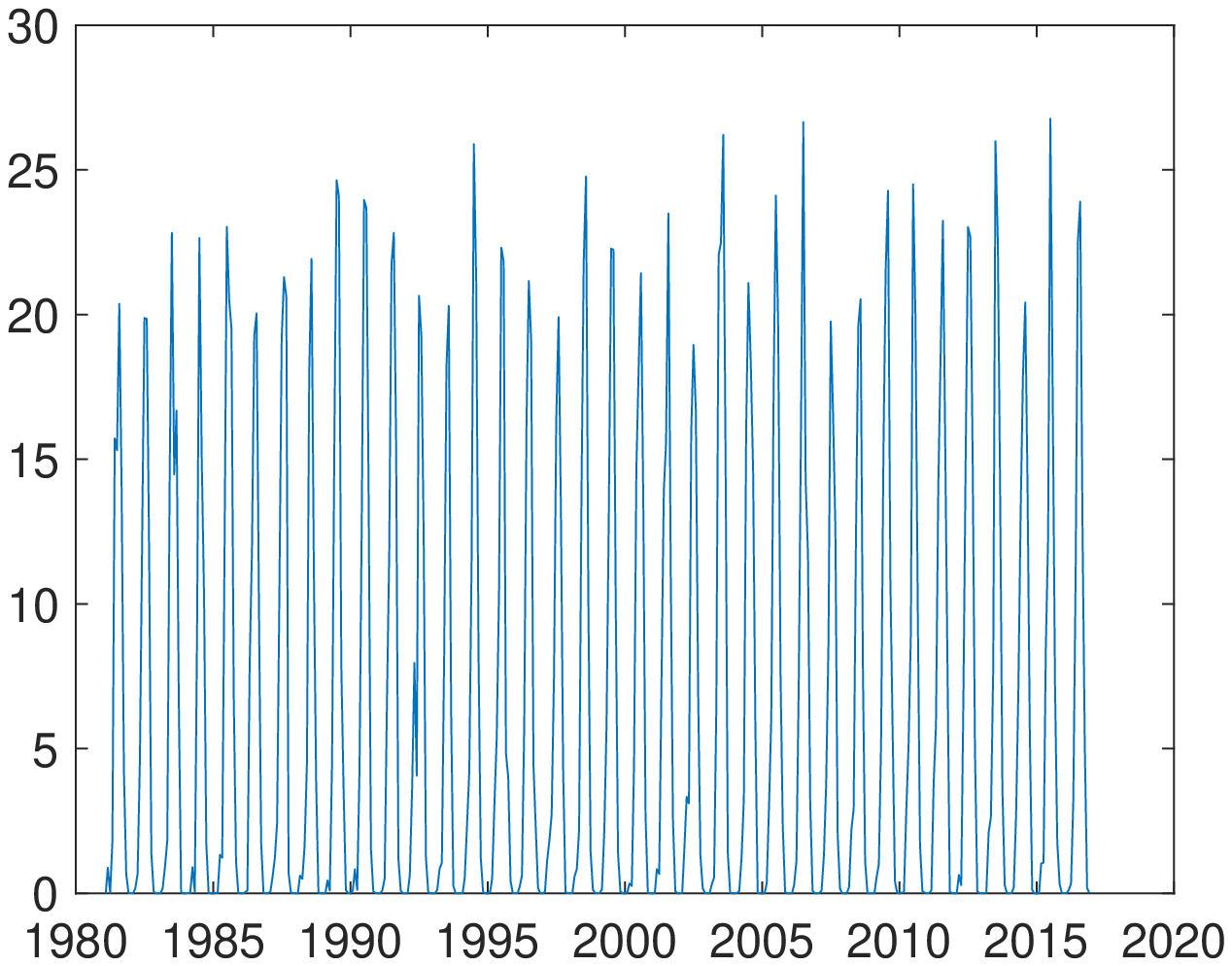}
                  \label{fig:plotCSUSP}} 
 		\caption{Time series plot of monthly spell of summer days (CSU) with daily temperature exceeding $25\degree$C for 13 EU countries for the sample period February 1981 to December 2016.}\label{PLOTCSU}
 \end{figure}
 
 \begin{figure}[h!]
         \centering
         \subfigure[AU]{                
                 \includegraphics[width=0.17\textwidth]{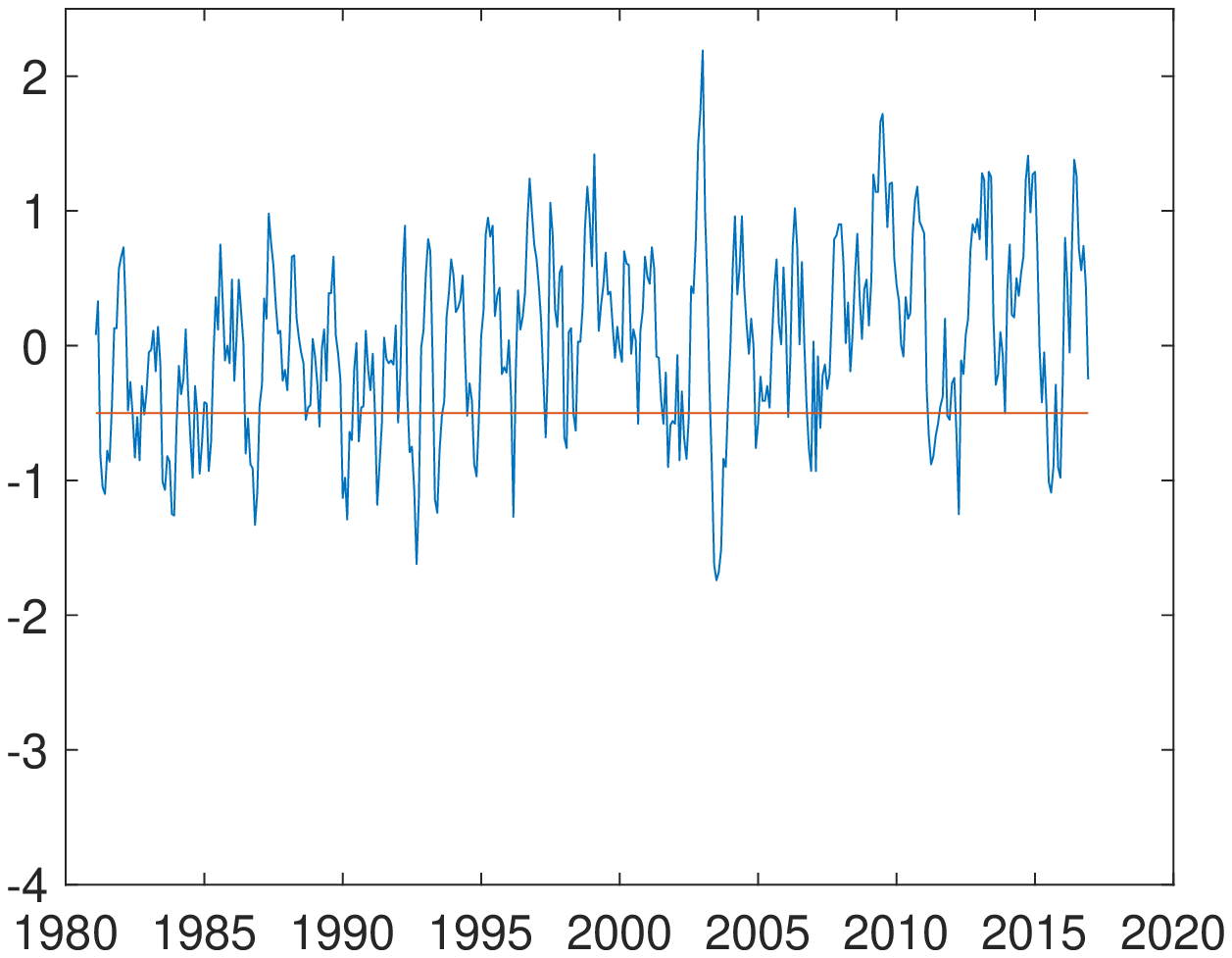}
                 \label{fig:plotSPIAU}}
         ~
         \subfigure[BE]{                
                 \includegraphics[width=0.17\textwidth]{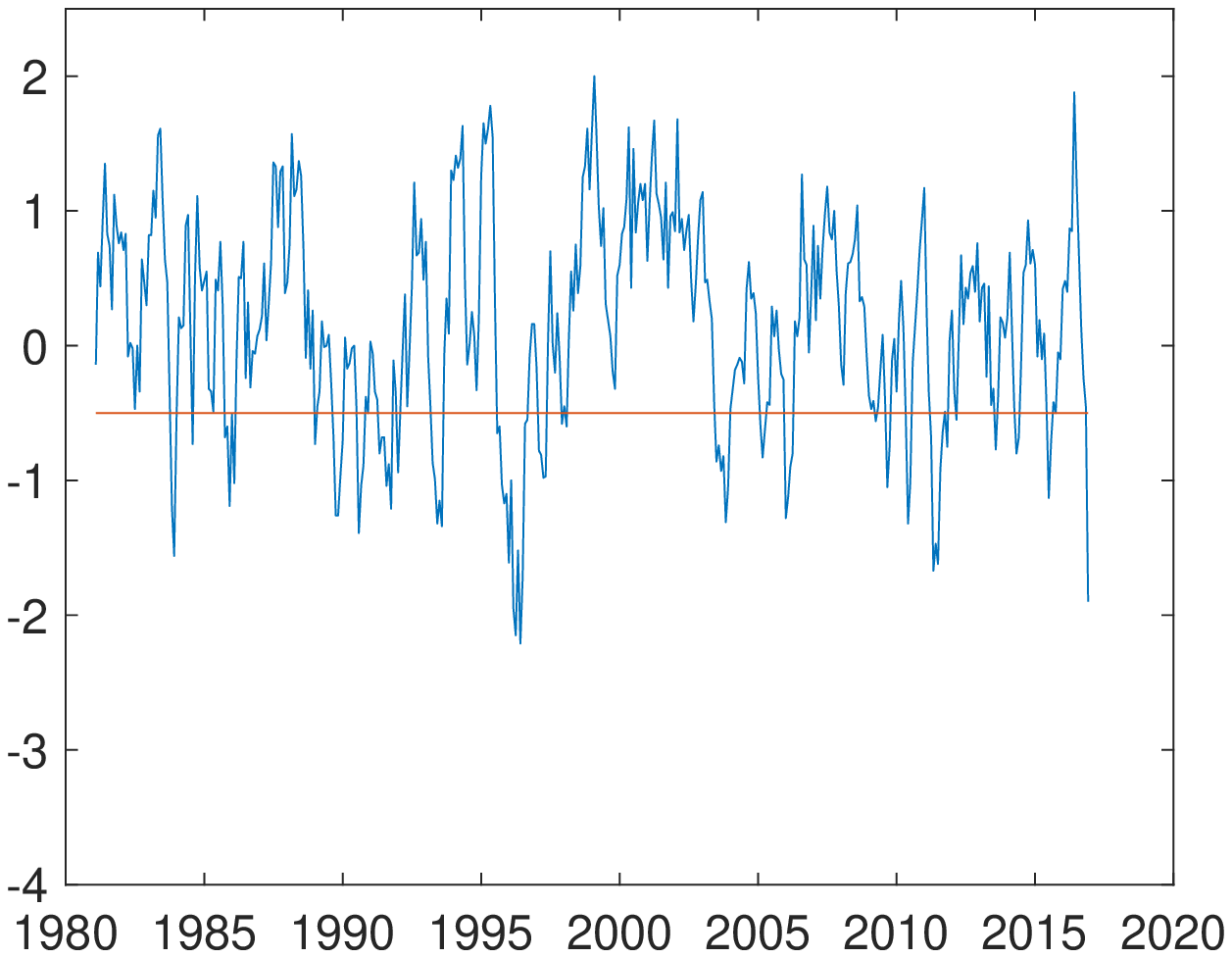}
                 \label{fig:plotSPIBE}}
         ~
         \subfigure[DE]{
                 \includegraphics[width=0.17\textwidth]{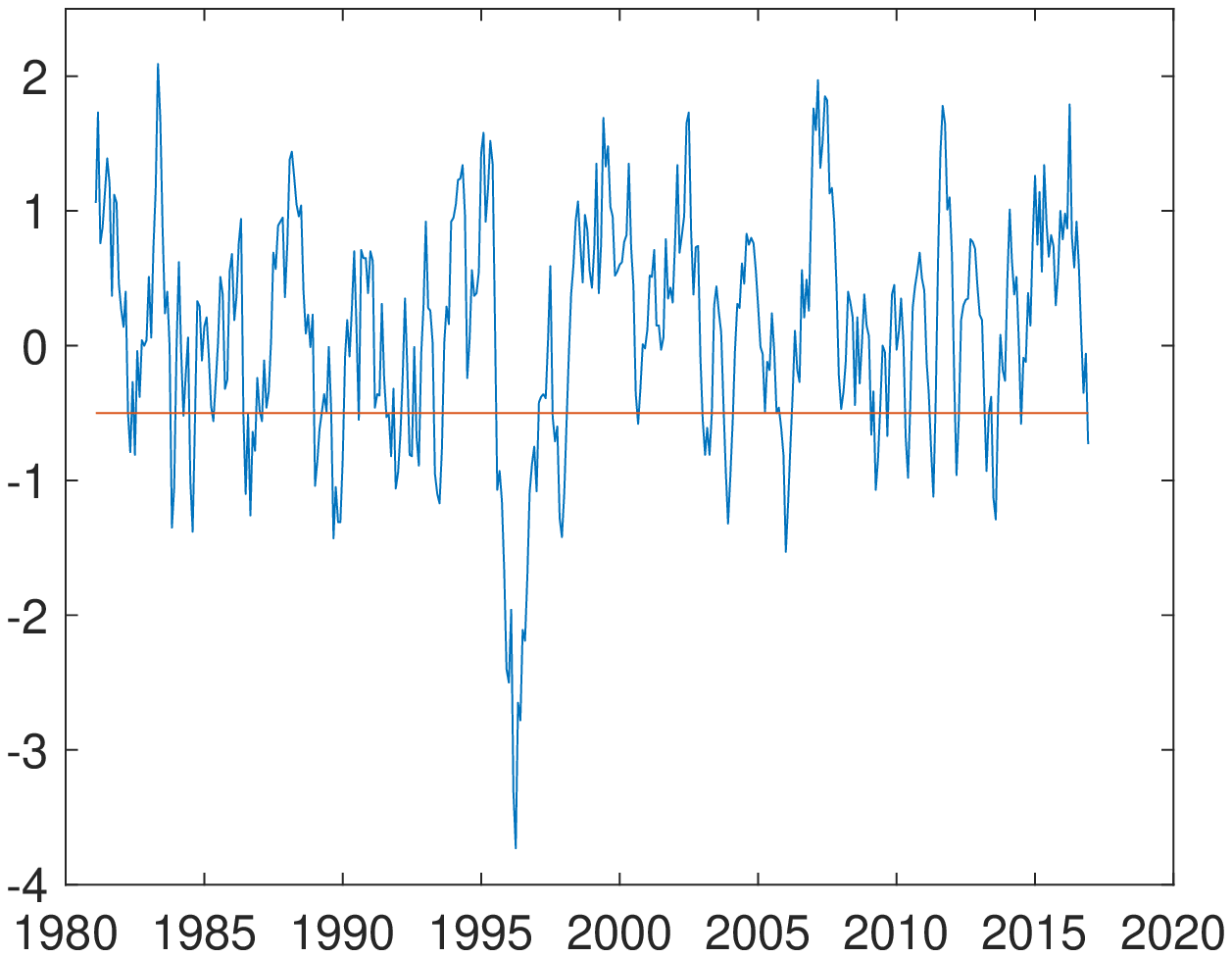}
                 \label{fig:plotSPIDE}}    
         ~
         \subfigure[FI]{                
                  \includegraphics[width=0.17\textwidth]{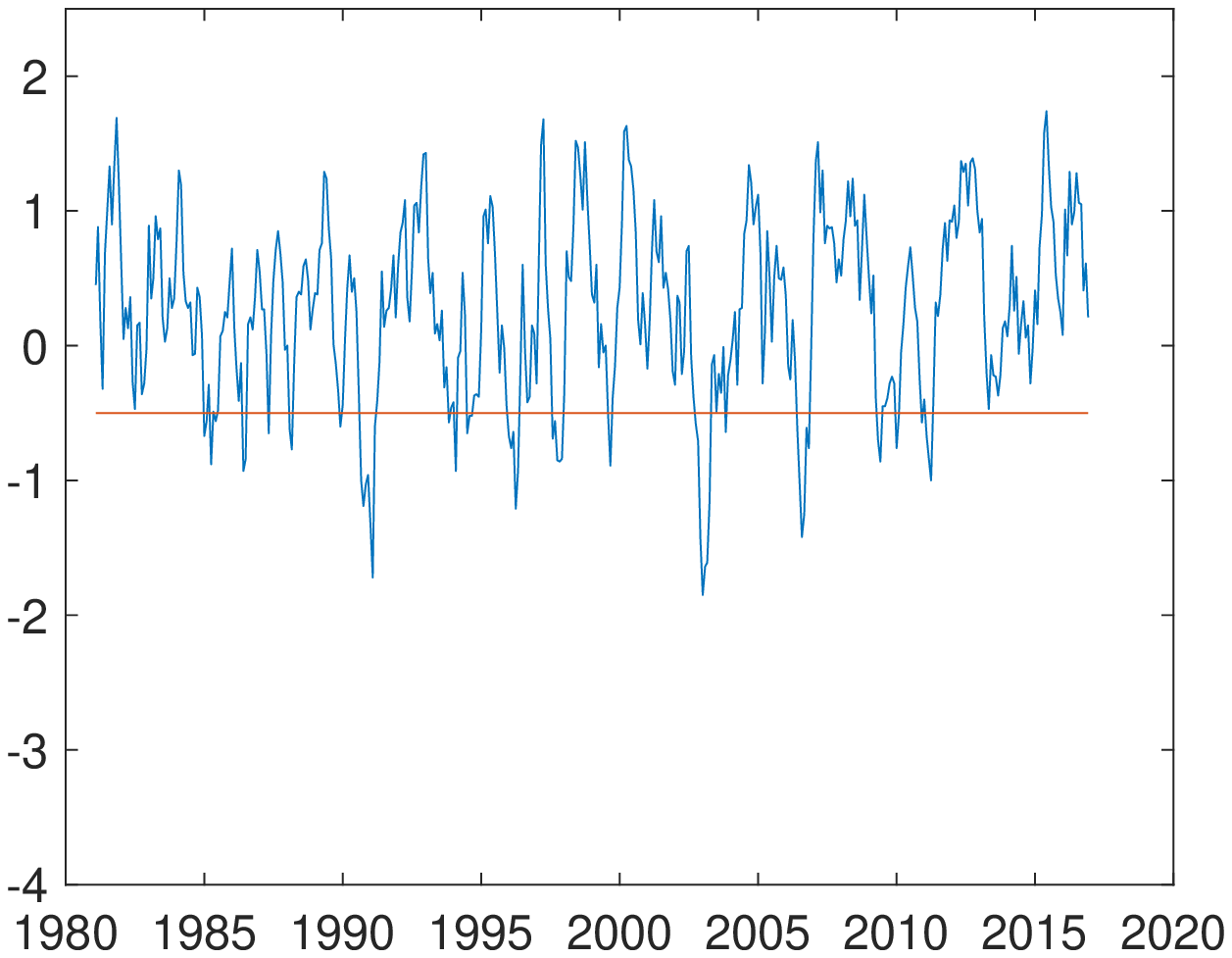}
                  \label{fig:plotSPIFI}}
         ~
         \subfigure[FR]{
                  \includegraphics[width=0.17\textwidth]{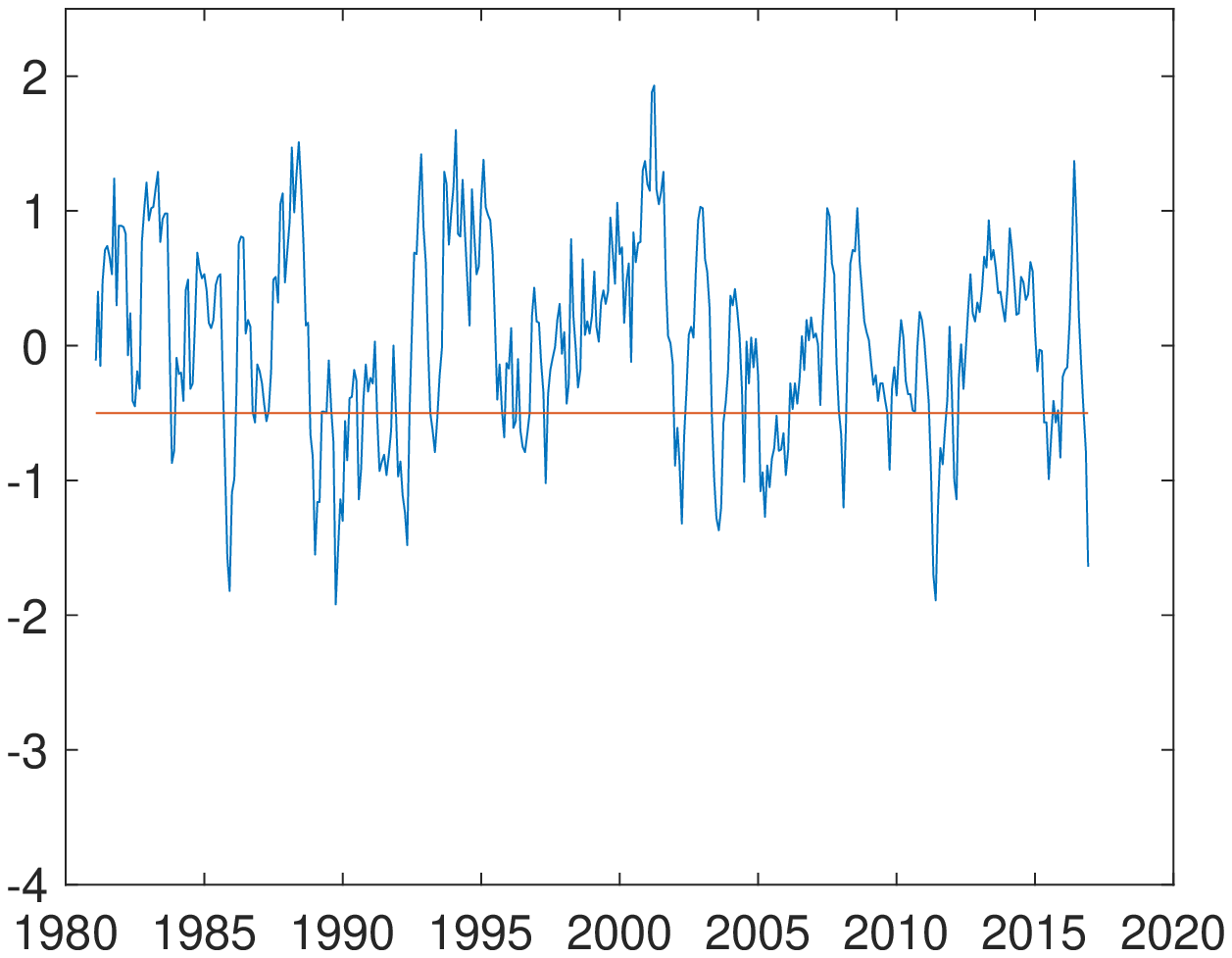}
                  \label{fig:plotSPIFR}} 
         ~
         \subfigure[GE]{                
                  \includegraphics[width=0.17\textwidth]{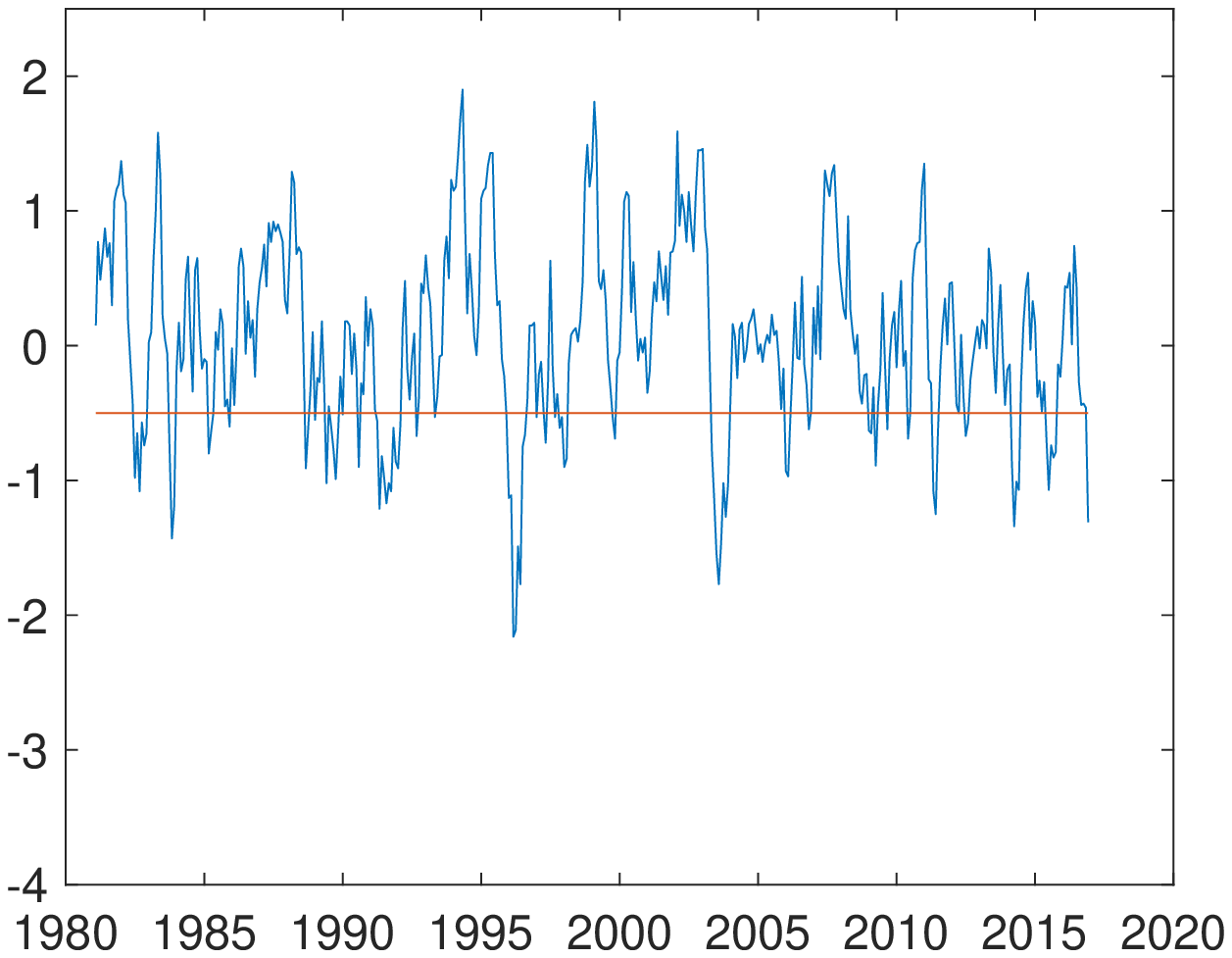}
                  \label{fig:plotSPIGE}}
         ~
         \subfigure[GR]{
                  \includegraphics[width=0.17\textwidth]{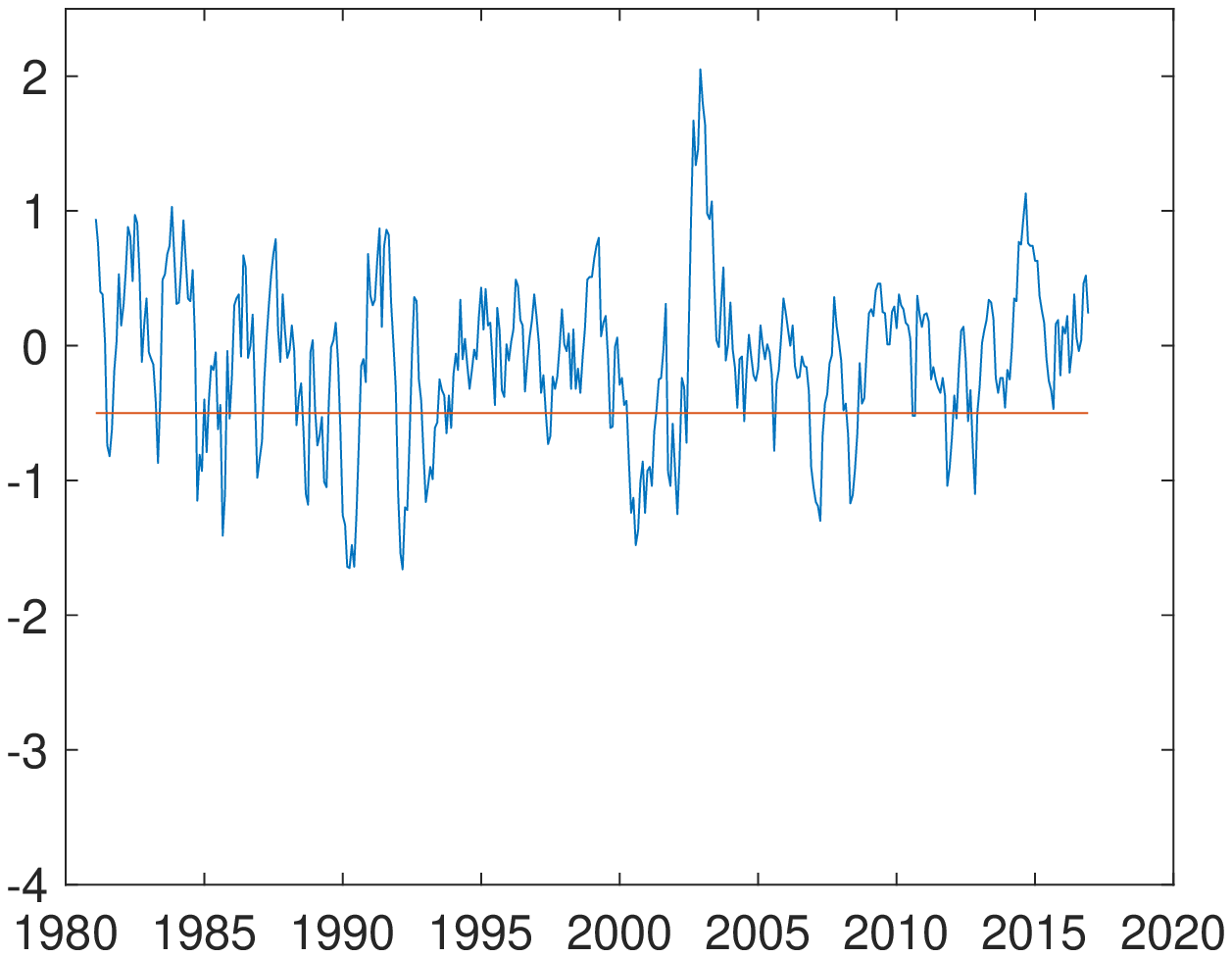}
                  \label{fig:plotSPIGR}} 
         ~
         \subfigure[IR]{                
                  \includegraphics[width=0.17\textwidth]{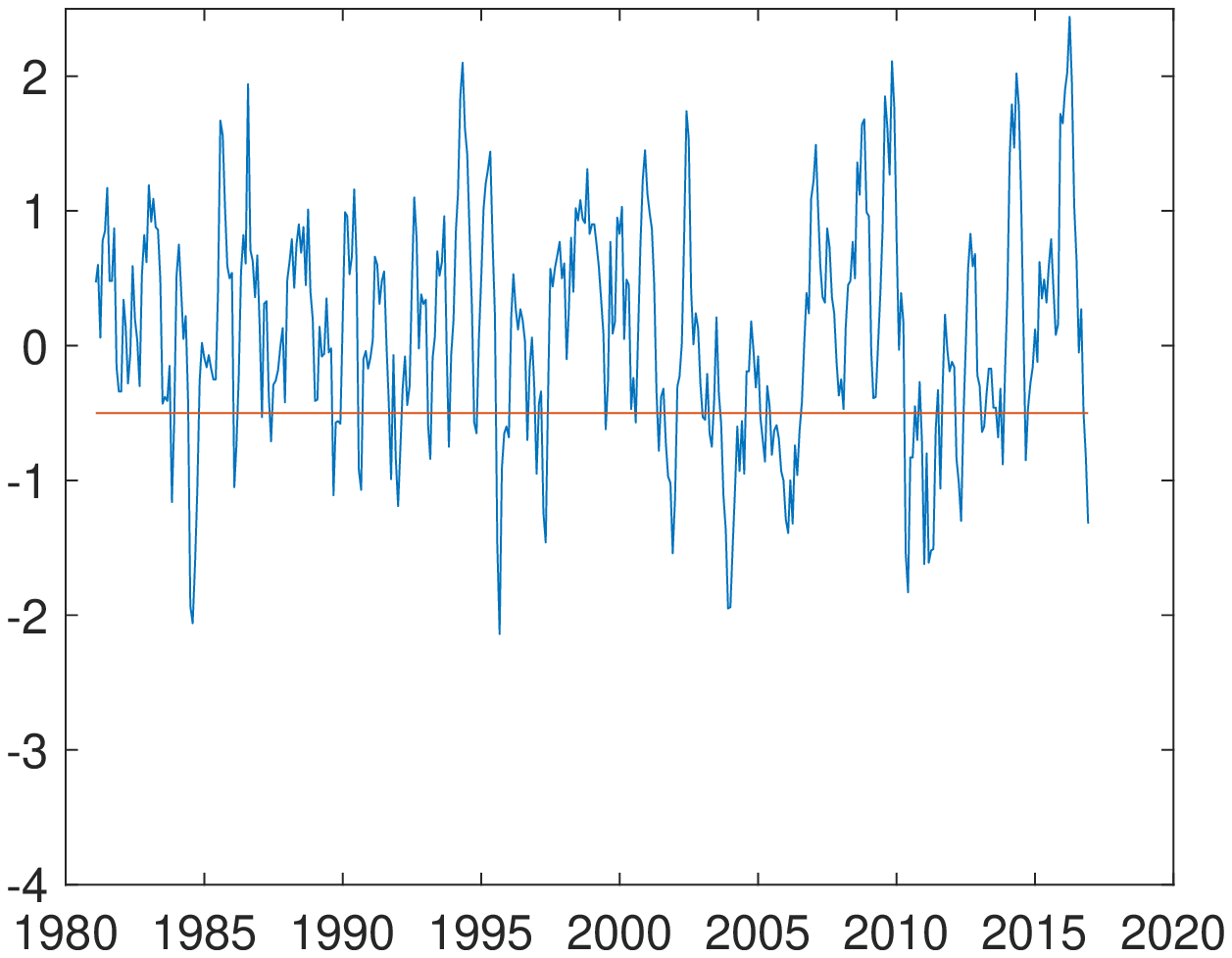}
                  \label{fig:plotSPIIR}}
         ~
         \subfigure[IT]{
                  \includegraphics[width=0.17\textwidth]{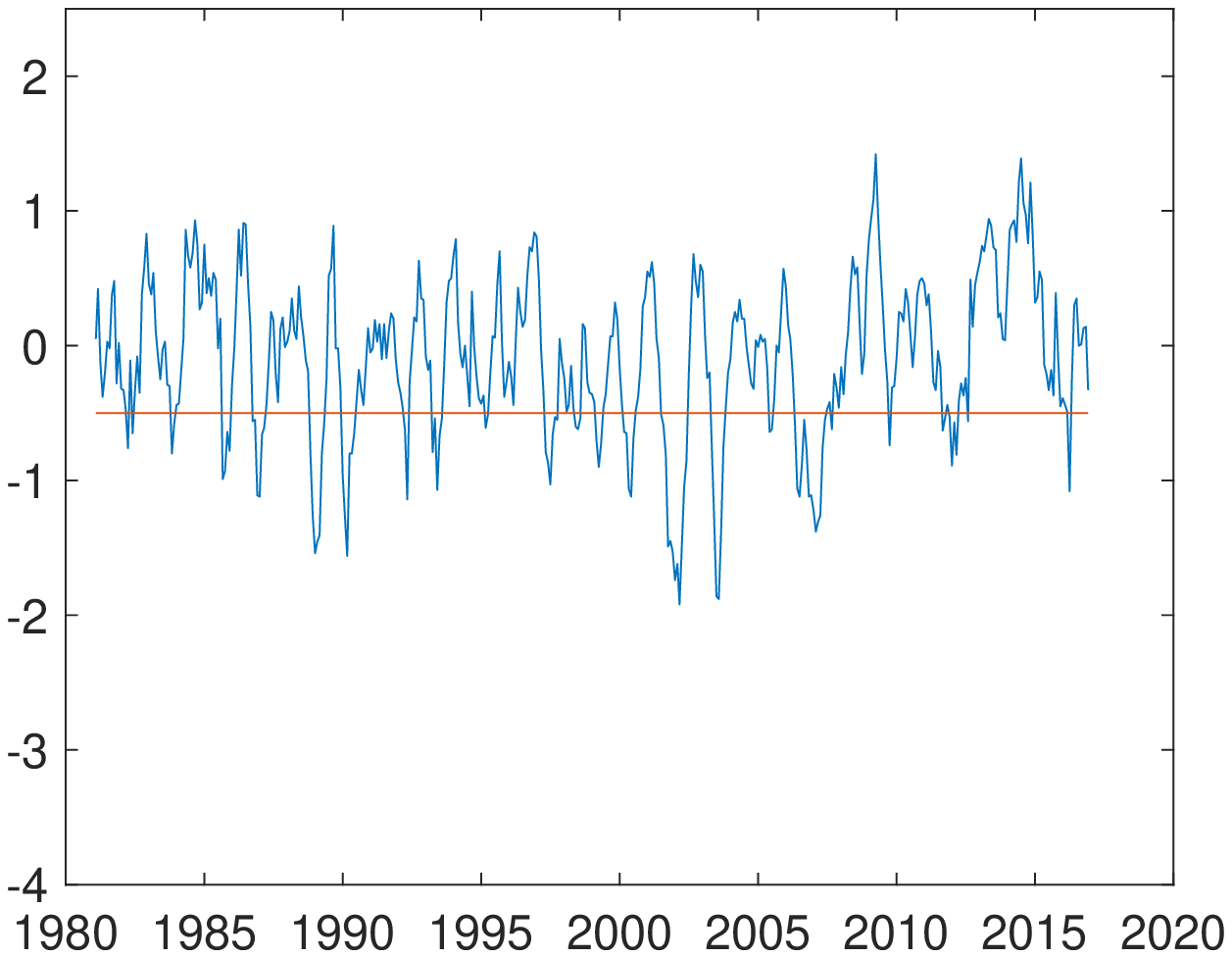}
                  \label{fig:plotSPIIT}}       
         ~
         \subfigure[LU]{                
                  \includegraphics[width=0.17\textwidth]{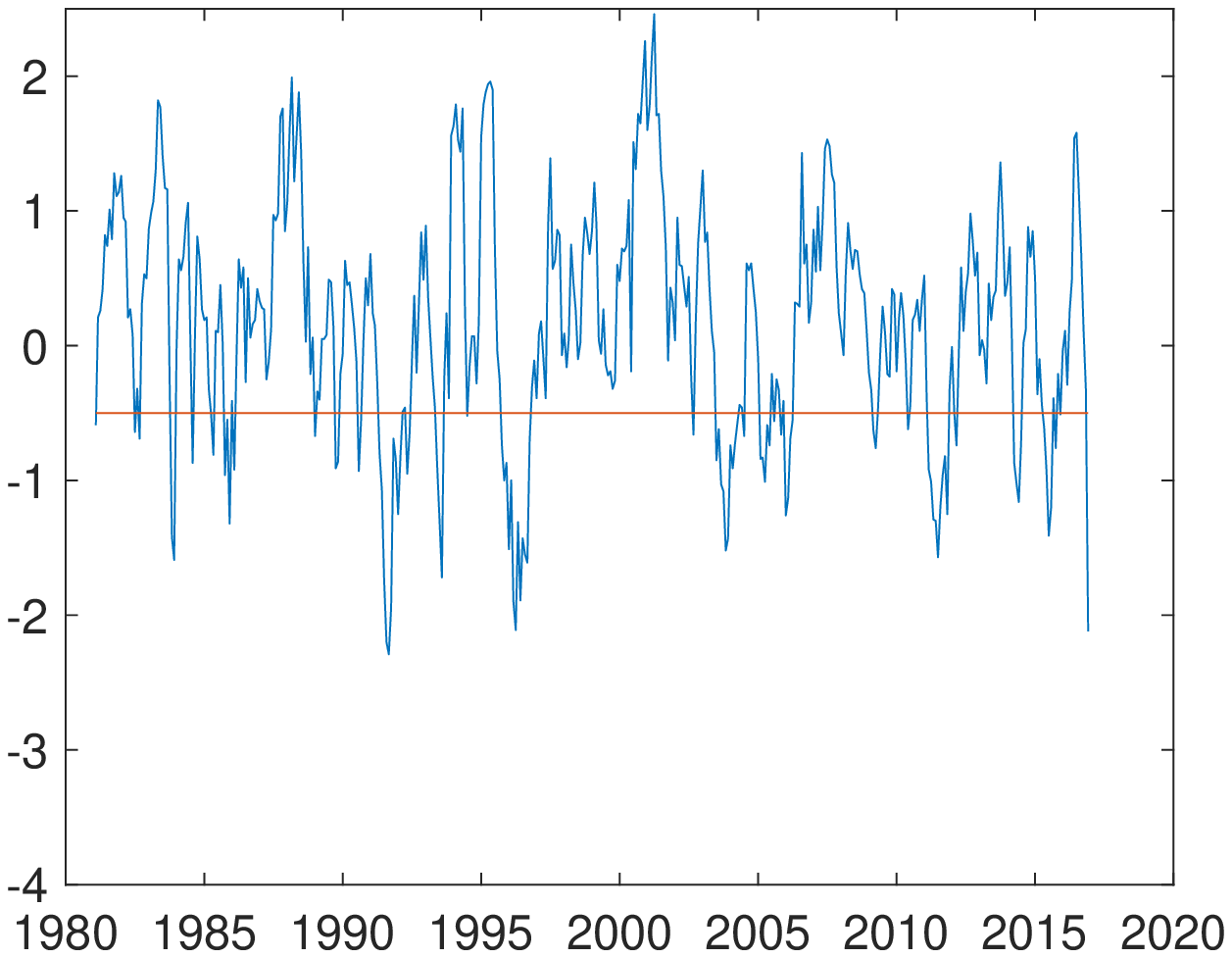}
                  \label{fig:plotSPILU}}
         ~
         \subfigure[NE]{
                  \includegraphics[width=0.17\textwidth]{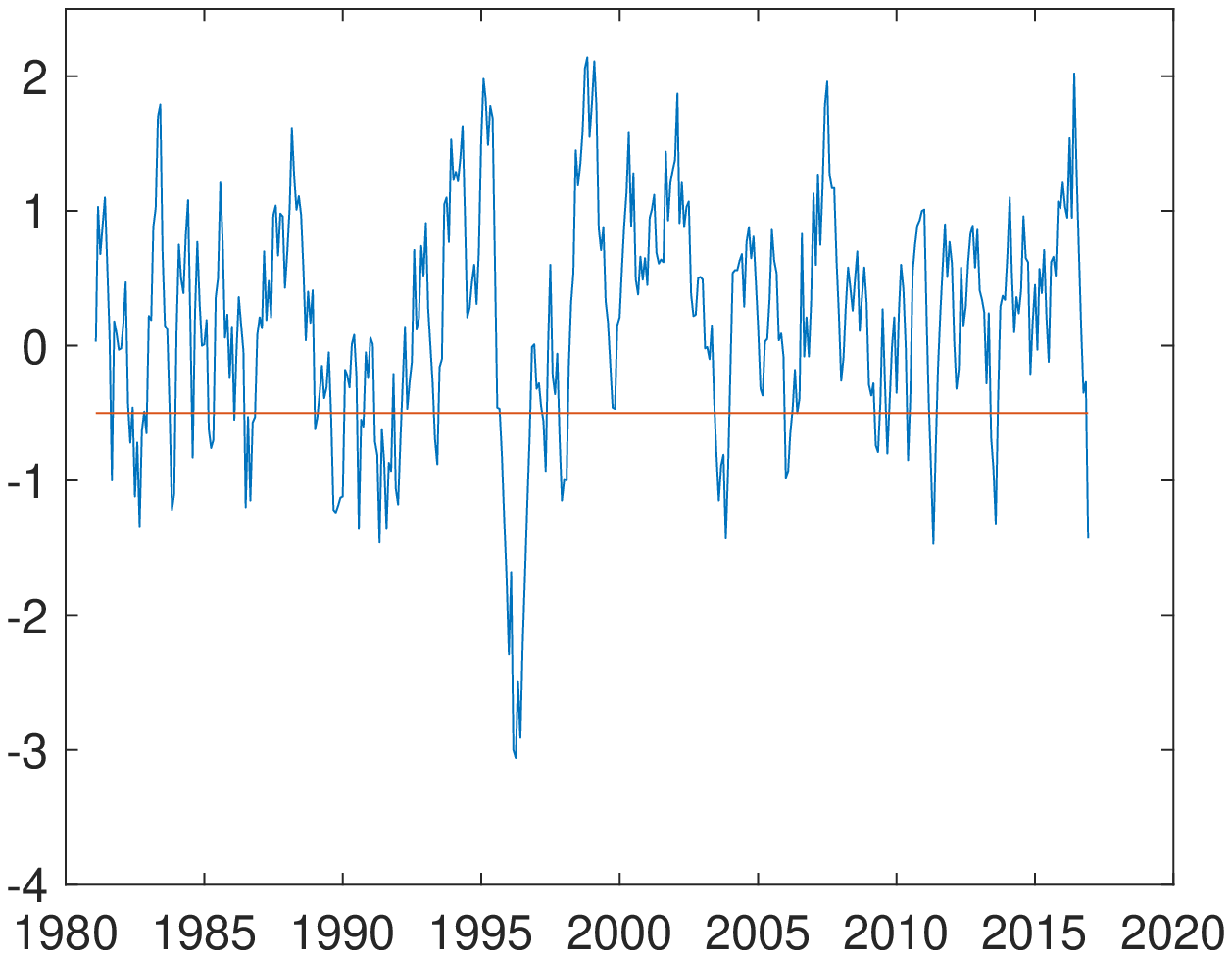}
                  \label{fig:plotSPINE}} 
         ~
         \subfigure[PO]{                
                  \includegraphics[width=0.17\textwidth]{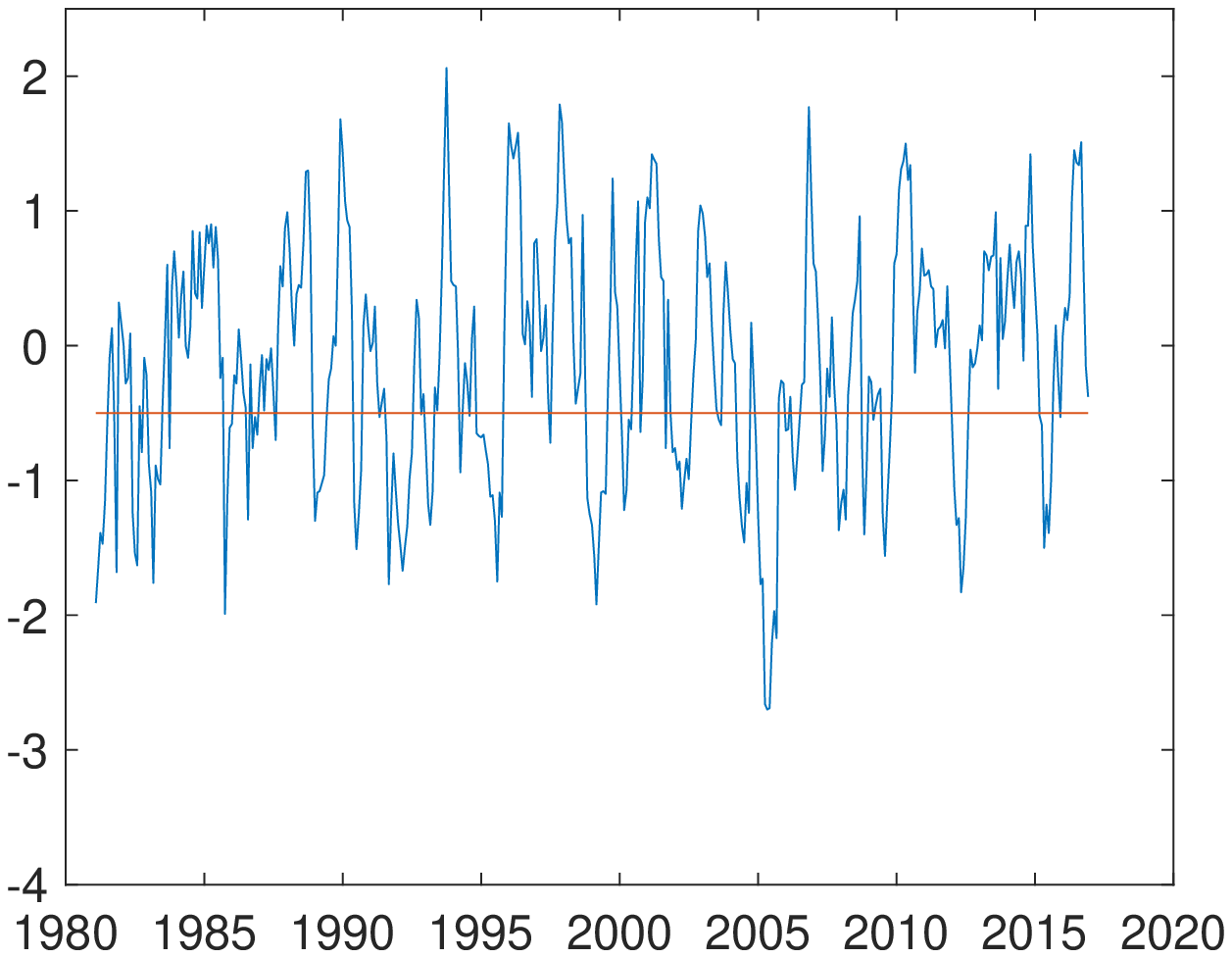}
                  \label{fig:plotSPIPO}}
         ~
         \subfigure[SP]{
                  \includegraphics[width=0.17\textwidth]{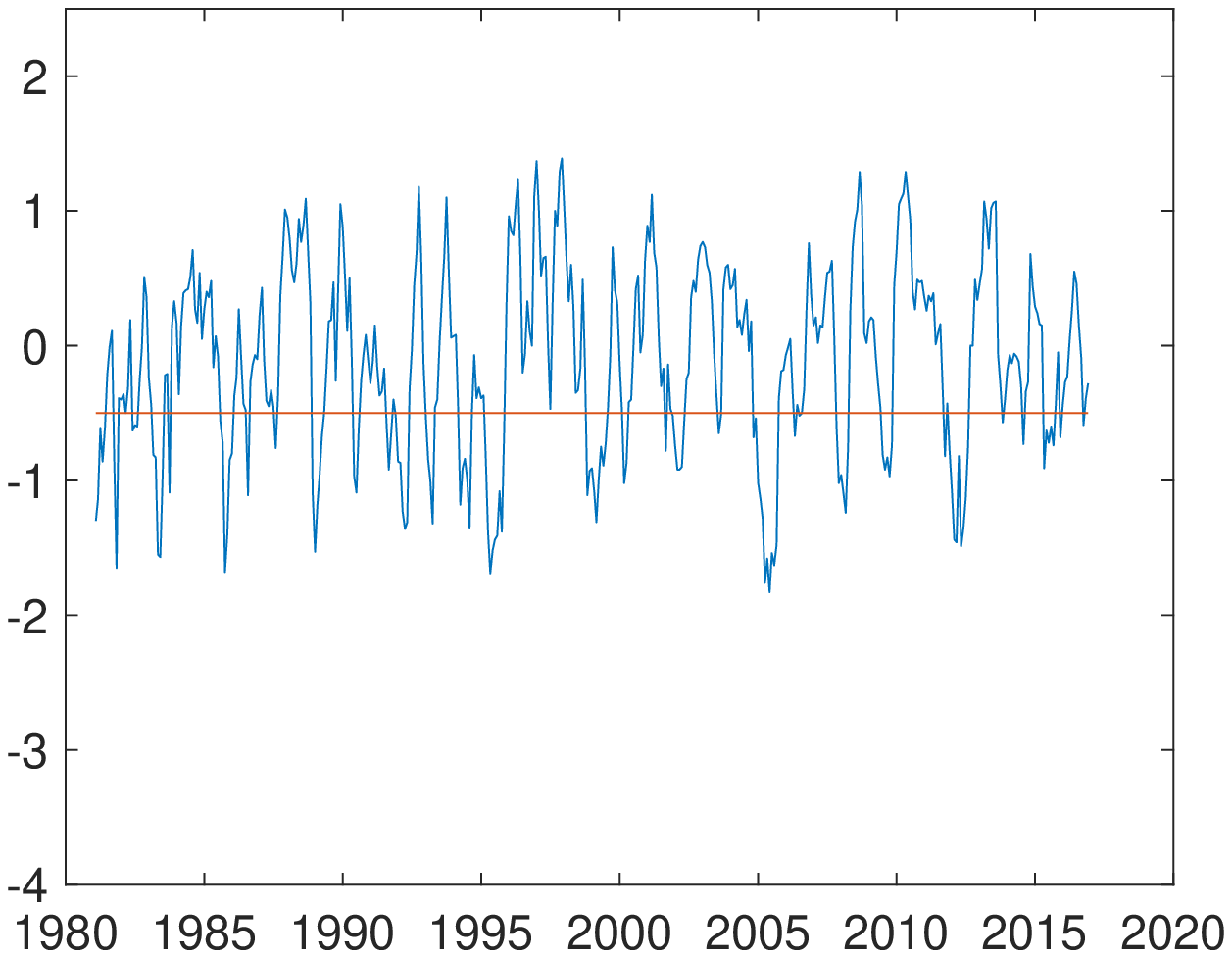}
                  \label{fig:plotSPISP}} 
 		\caption{Time series plot of the drought index measured using monthly standard precipitation index (SPI) for the accumulation period of 6-months for 13 EU countries for the period February 1981 to December 2016.}\label{PLOTSPI}
 \end{figure}

 \begin{figure}[h!]
         \centering
         \subfigure[AU]{                
                 \includegraphics[width=0.17\textwidth]{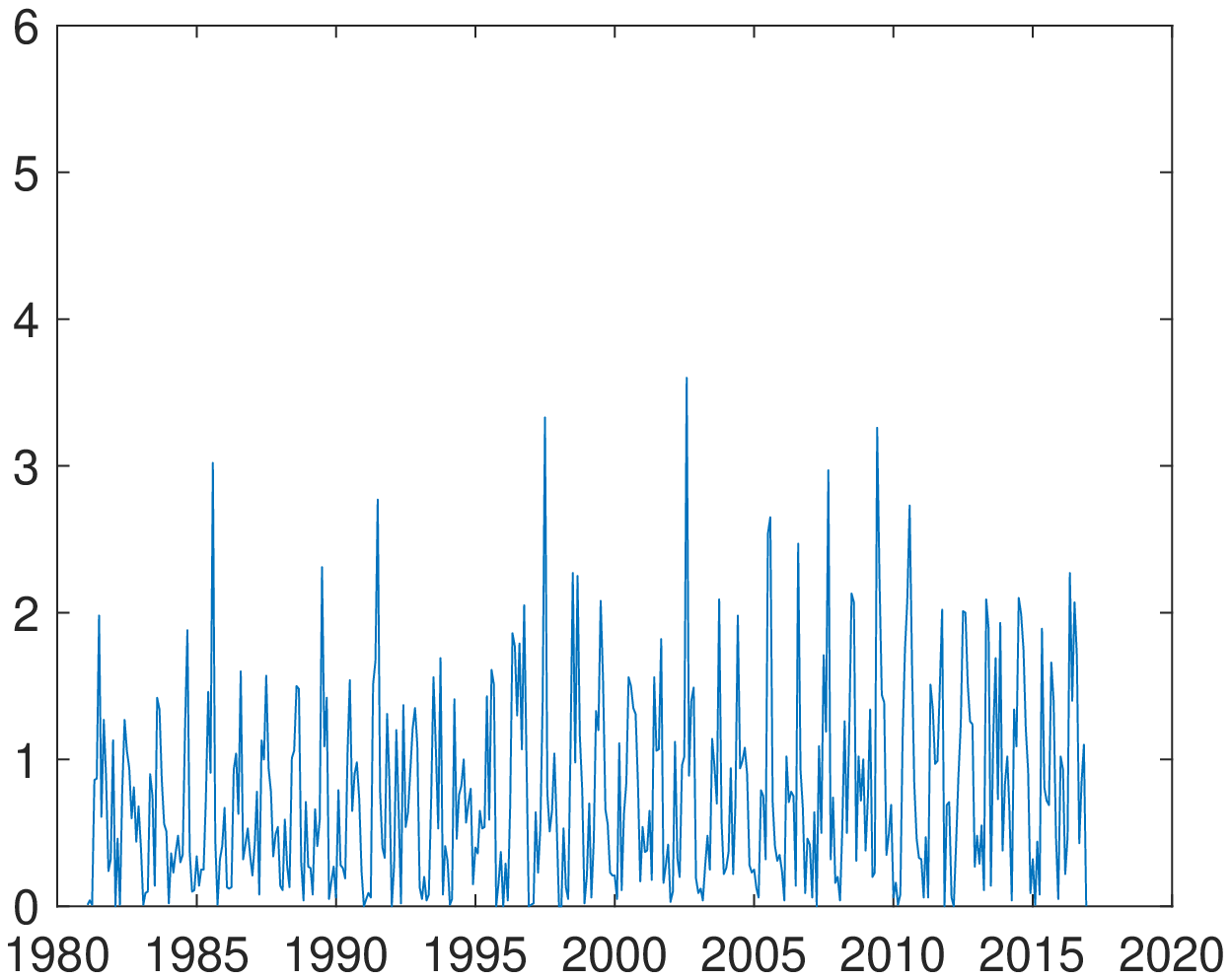}
                 \label{fig:plotr20mmAU}}
         ~
         \subfigure[BE]{                
                 \includegraphics[width=0.17\textwidth]{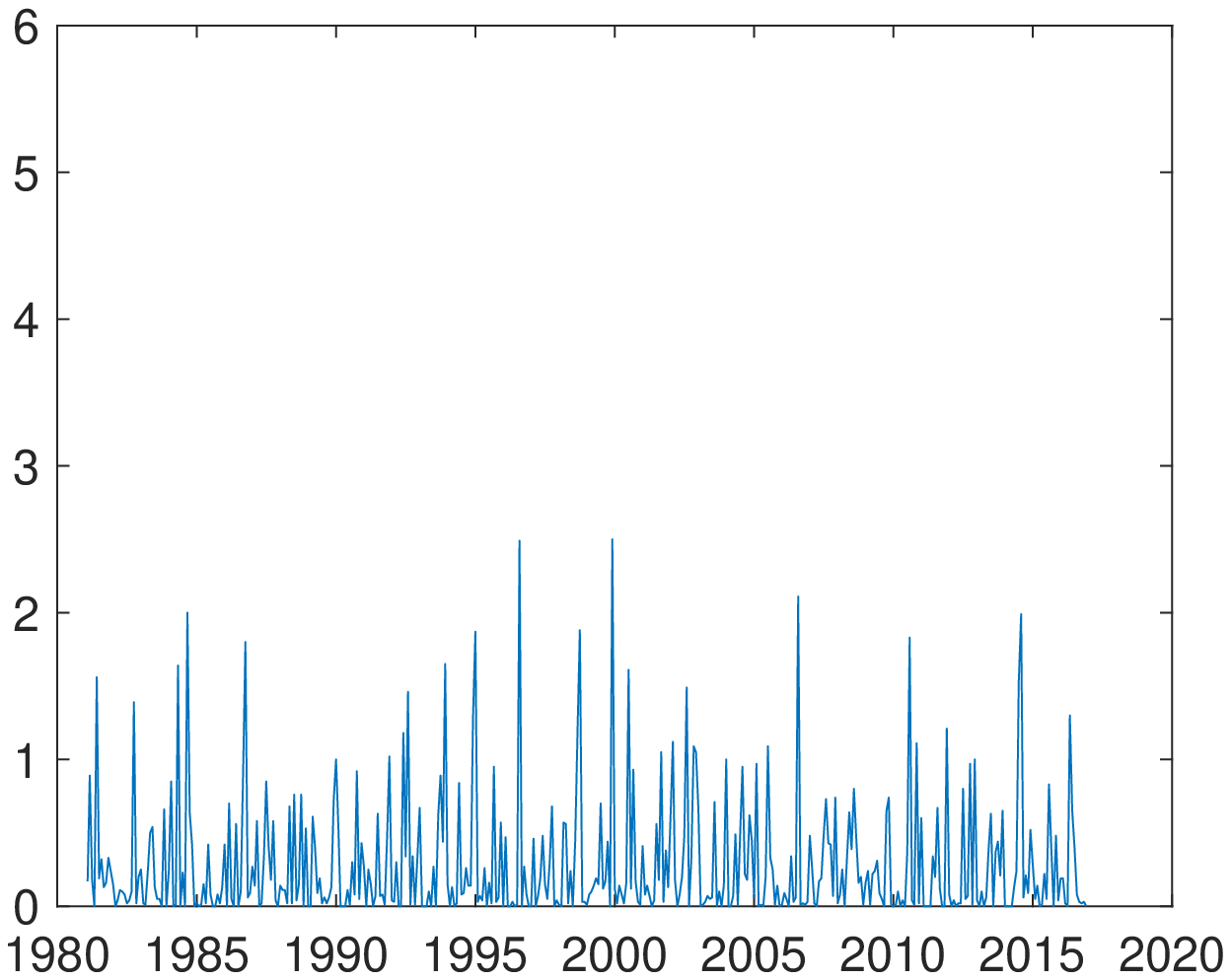}
                 \label{fig:plotr20mmBE}}
         ~
         \subfigure[DE]{
                 \includegraphics[width=0.17\textwidth]{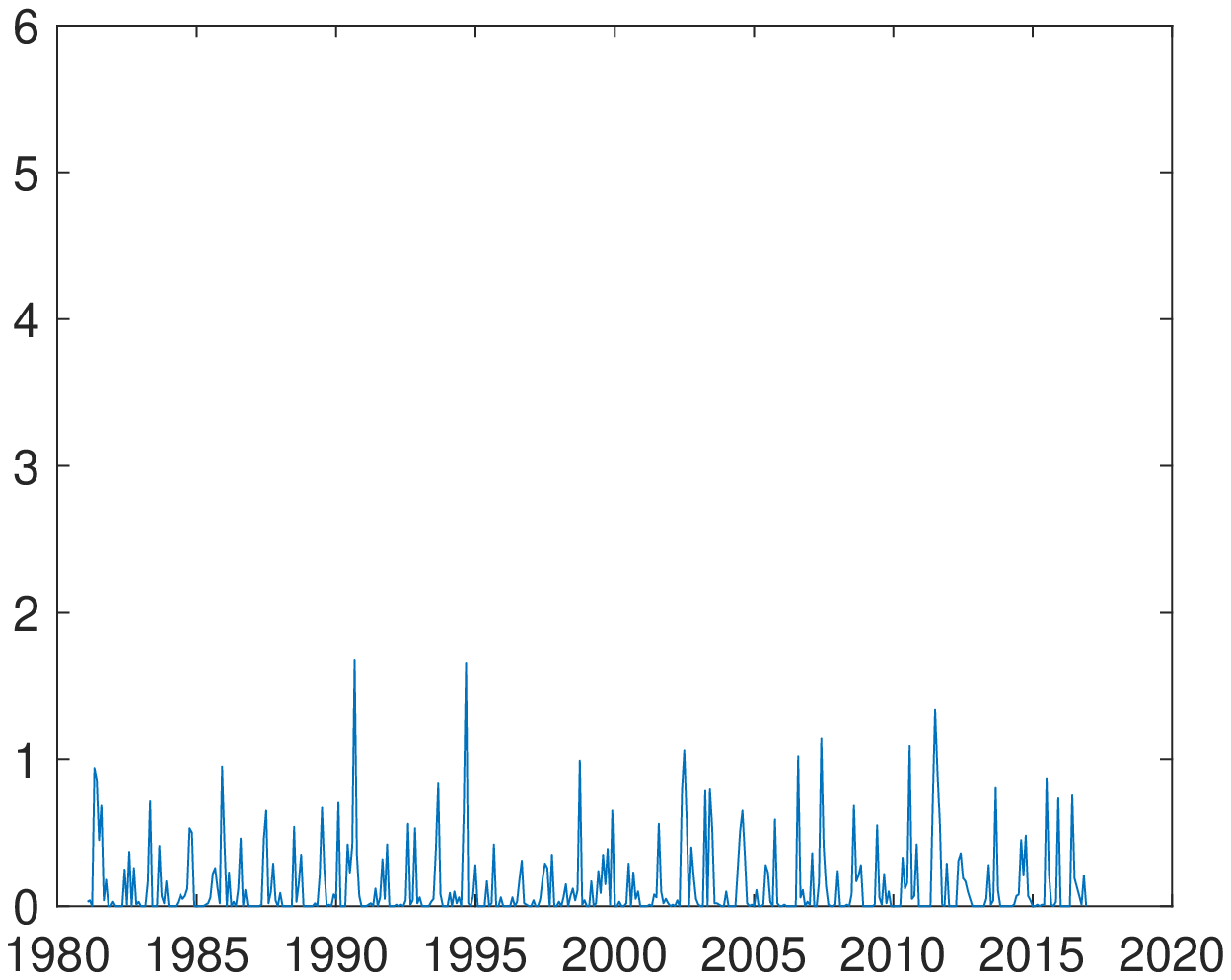}
                 \label{fig:plotr20mmDE}}    
         ~
         \subfigure[FI]{                
                  \includegraphics[width=0.17\textwidth]{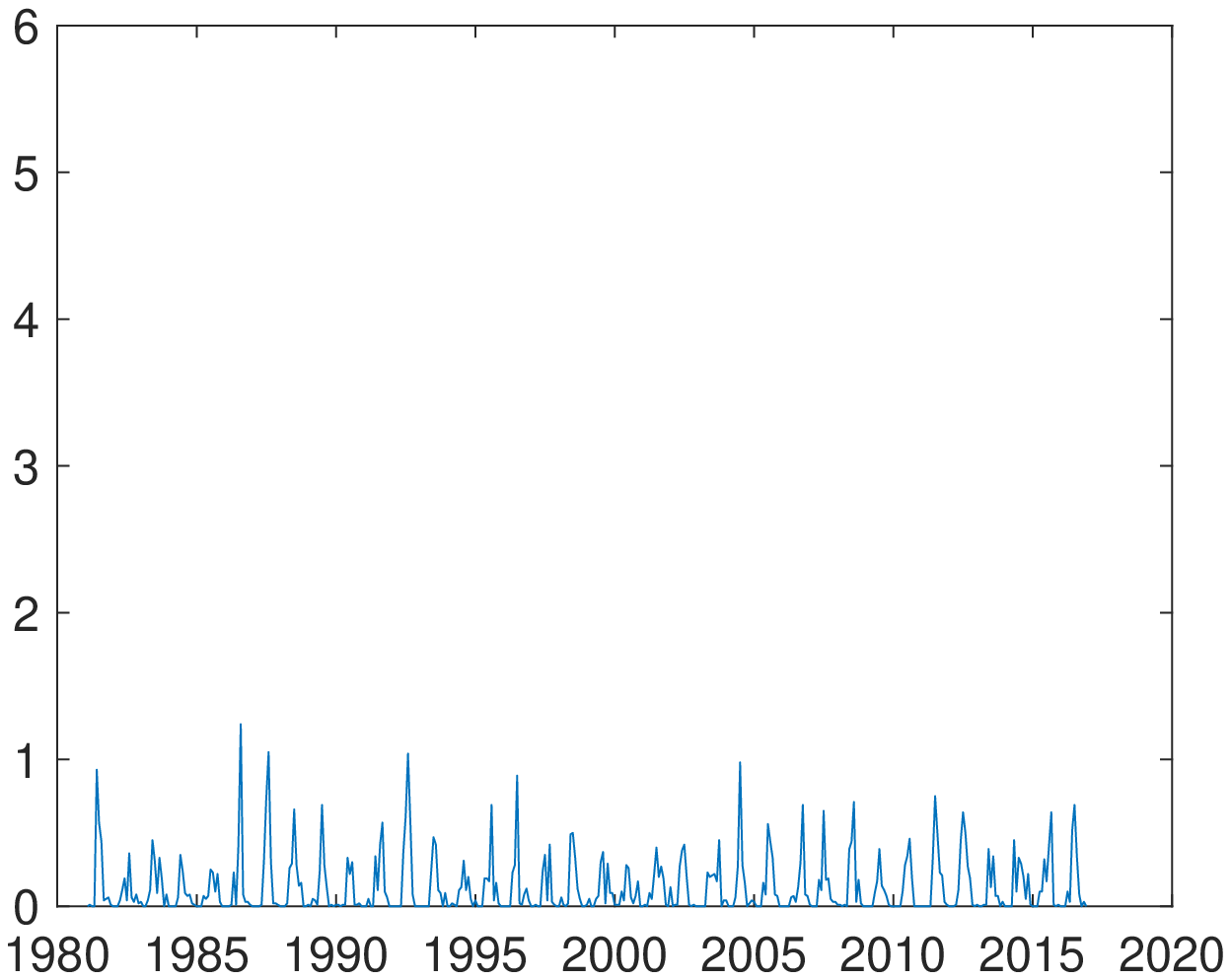}
                  \label{fig:plotr20mmFI}}
         ~
         \subfigure[FR]{
                  \includegraphics[width=0.17\textwidth]{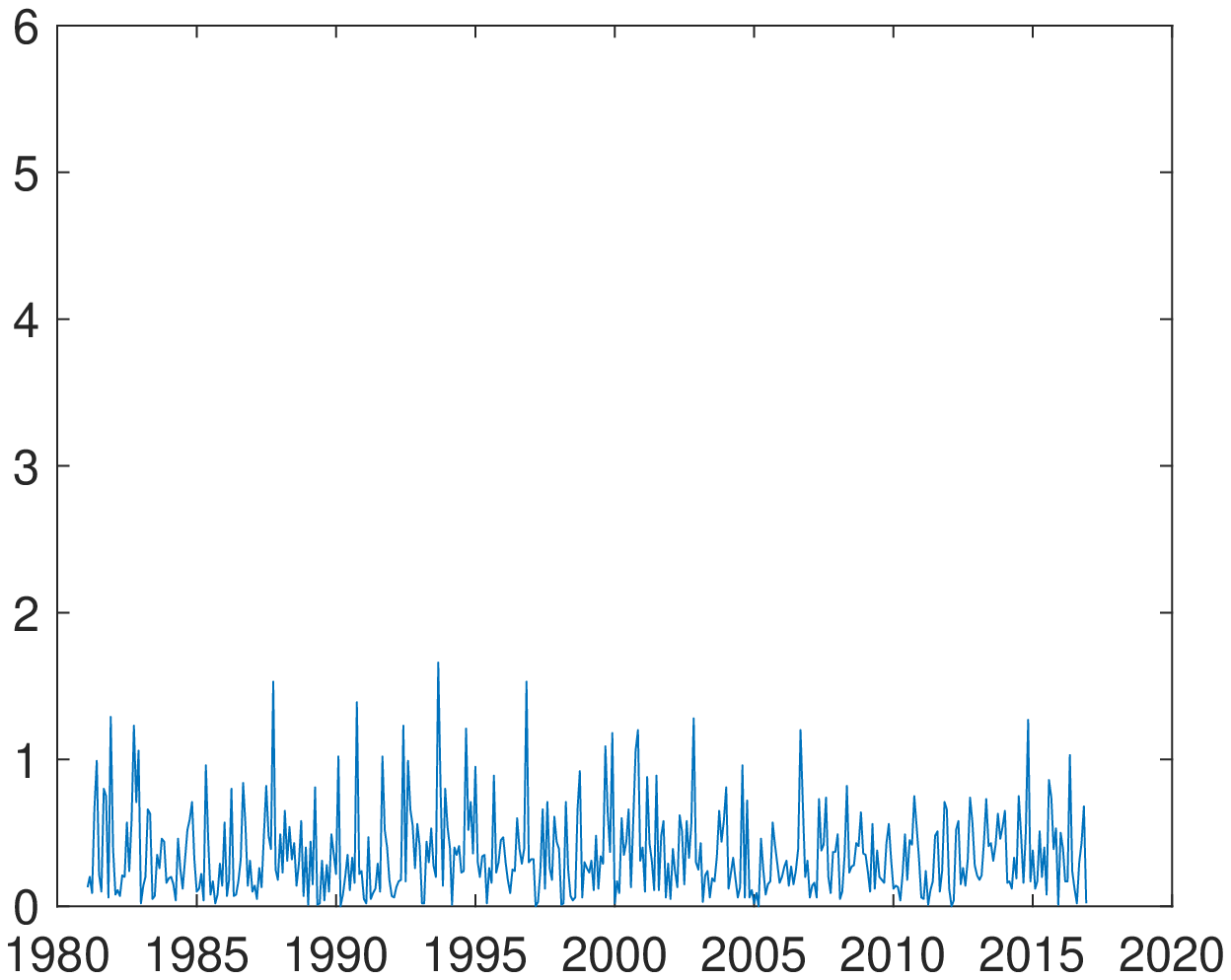}
                  \label{fig:plotr20mmFR}} 
         ~
         \subfigure[GE]{                
                  \includegraphics[width=0.17\textwidth]{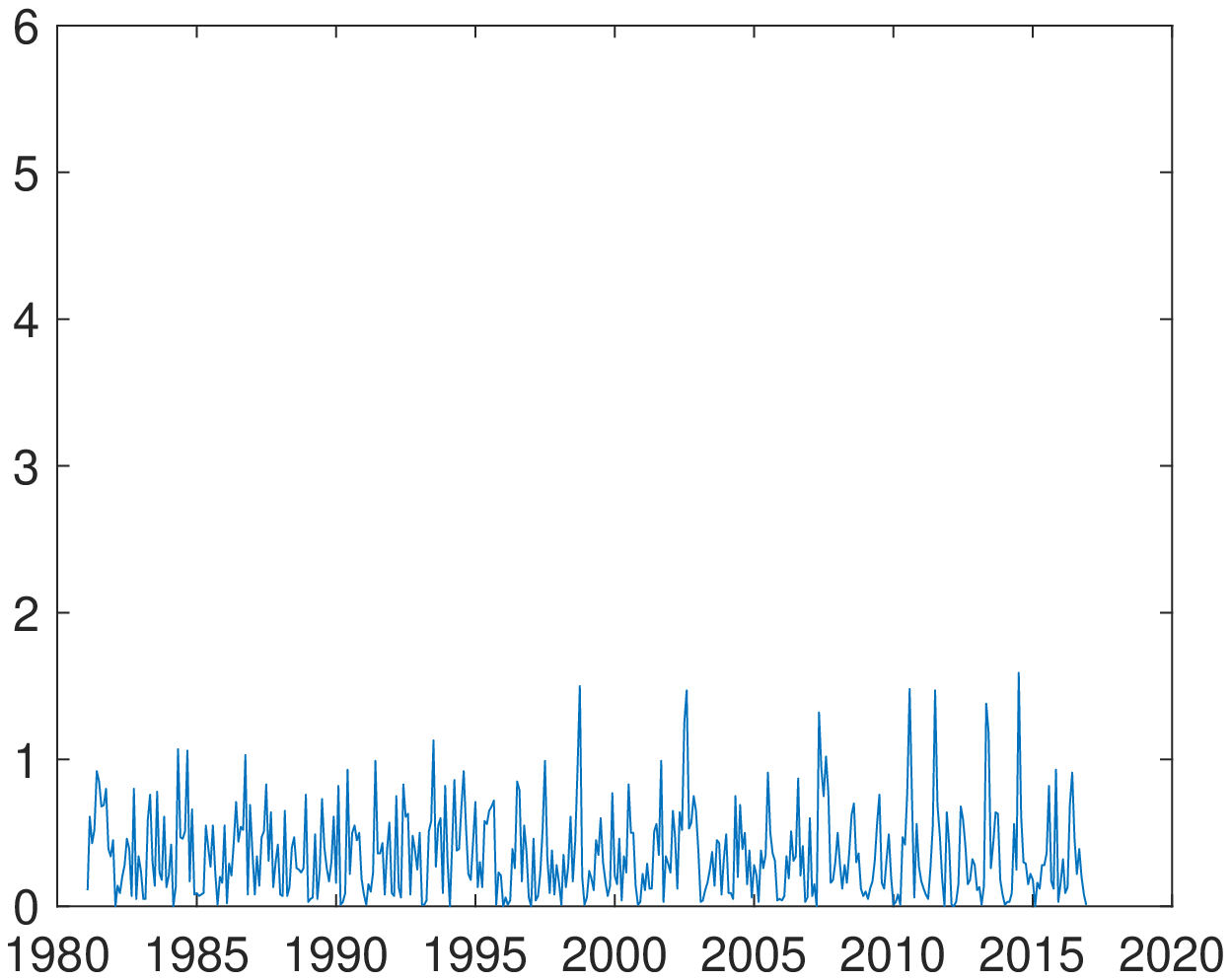}
                  \label{fig:plotr20mmGE}}
         ~
         \subfigure[GR]{
                  \includegraphics[width=0.17\textwidth]{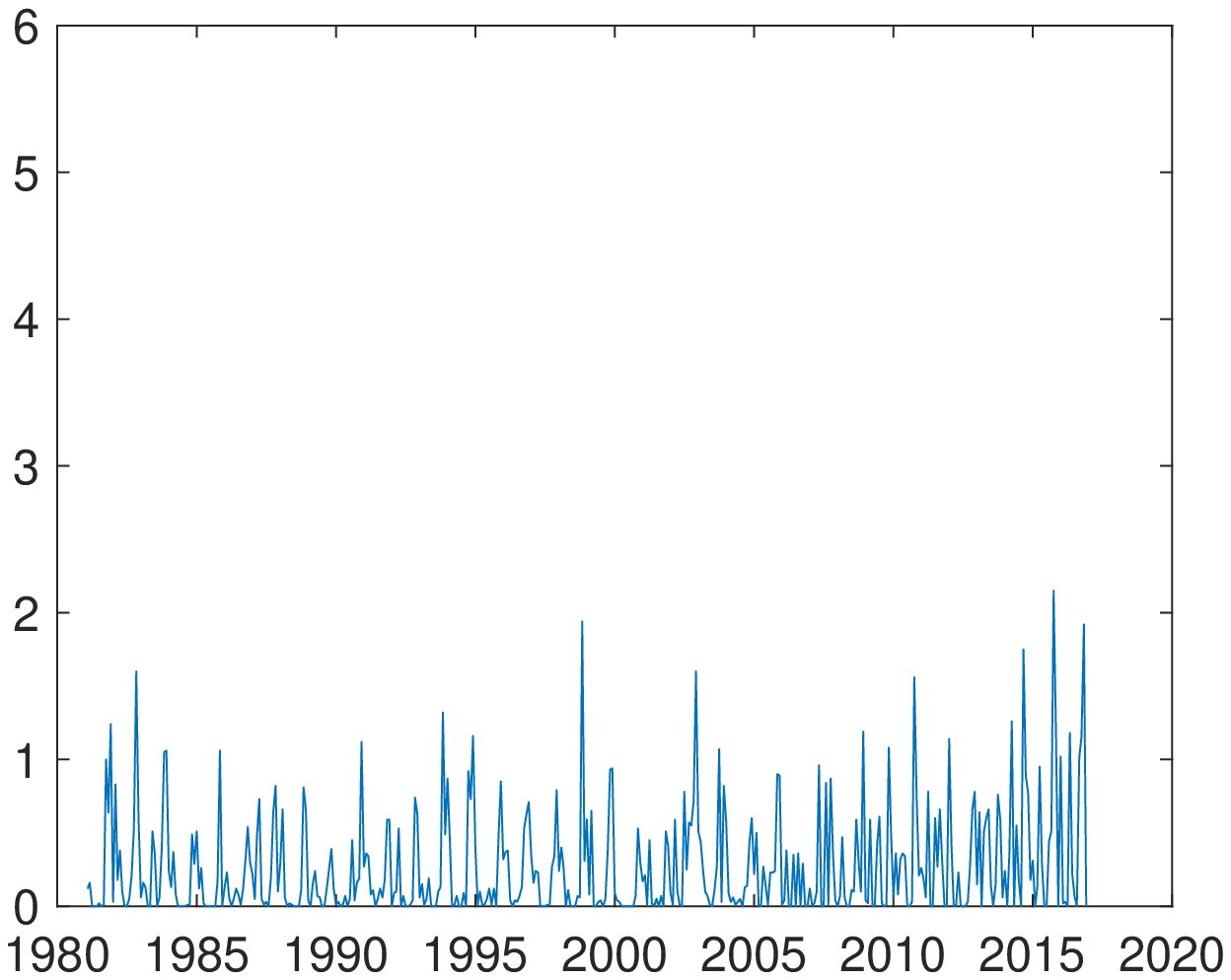}
                  \label{fig:plotr20mmGR}} 
         ~
         \subfigure[IR]{                
                  \includegraphics[width=0.17\textwidth]{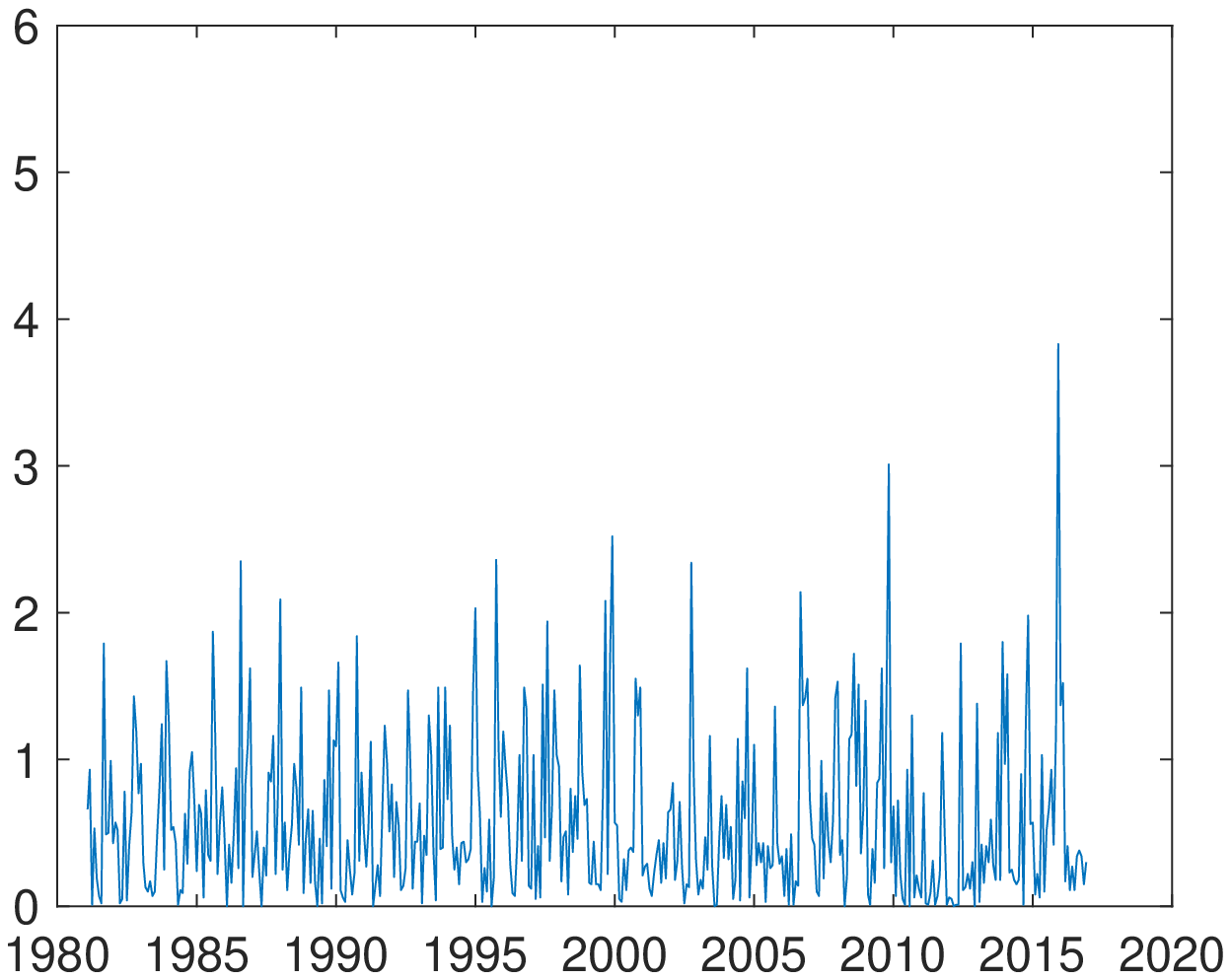}
                  \label{fig:plotr20mmIR}}
         ~
         \subfigure[IT]{
                  \includegraphics[width=0.17\textwidth]{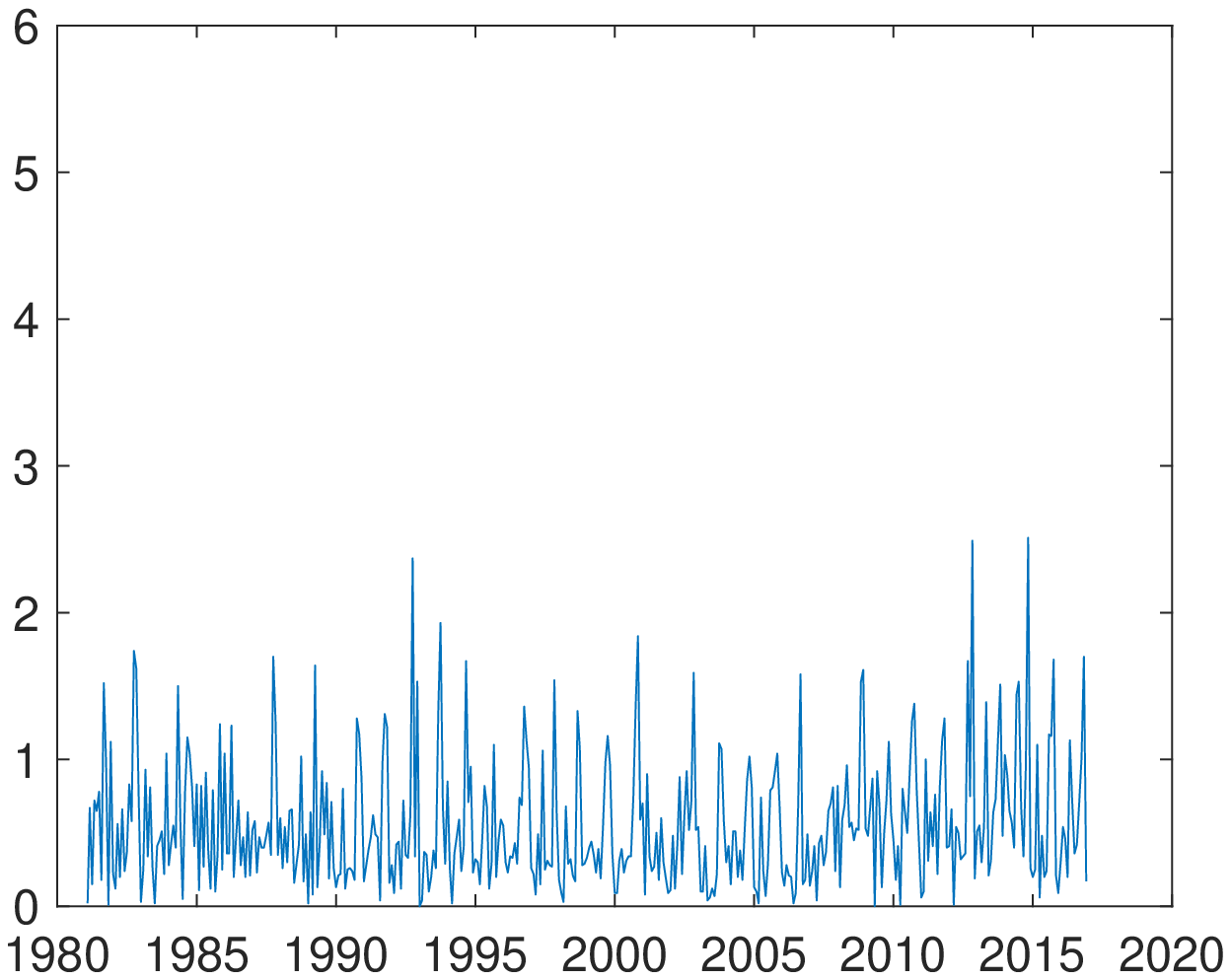}
                  \label{fig:plotr20mmIT}}       
         ~
         \subfigure[LU]{                
                  \includegraphics[width=0.17\textwidth]{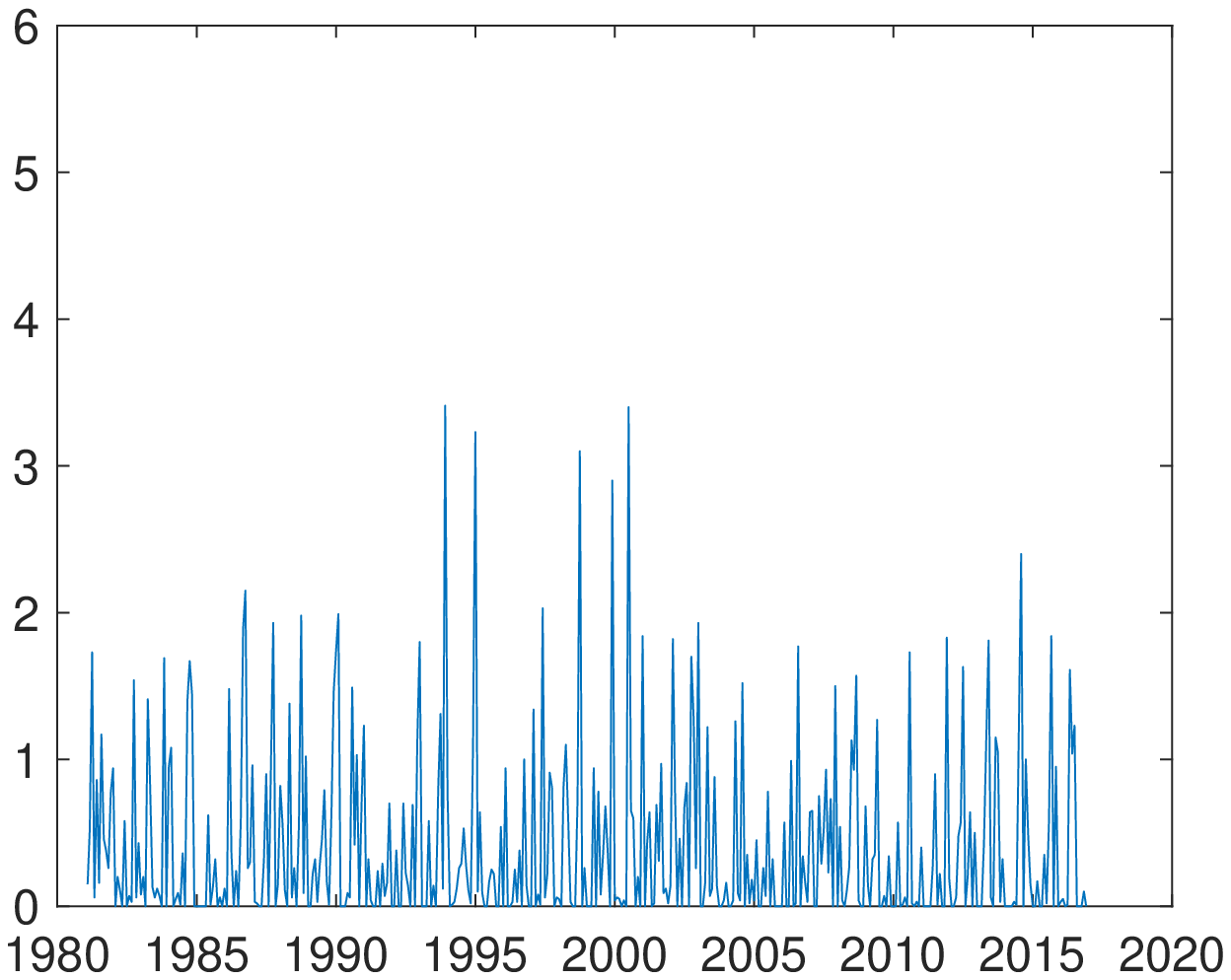}
                  \label{fig:plotr20mmLU}}
         ~
         \subfigure[NE]{
                  \includegraphics[width=0.17\textwidth]{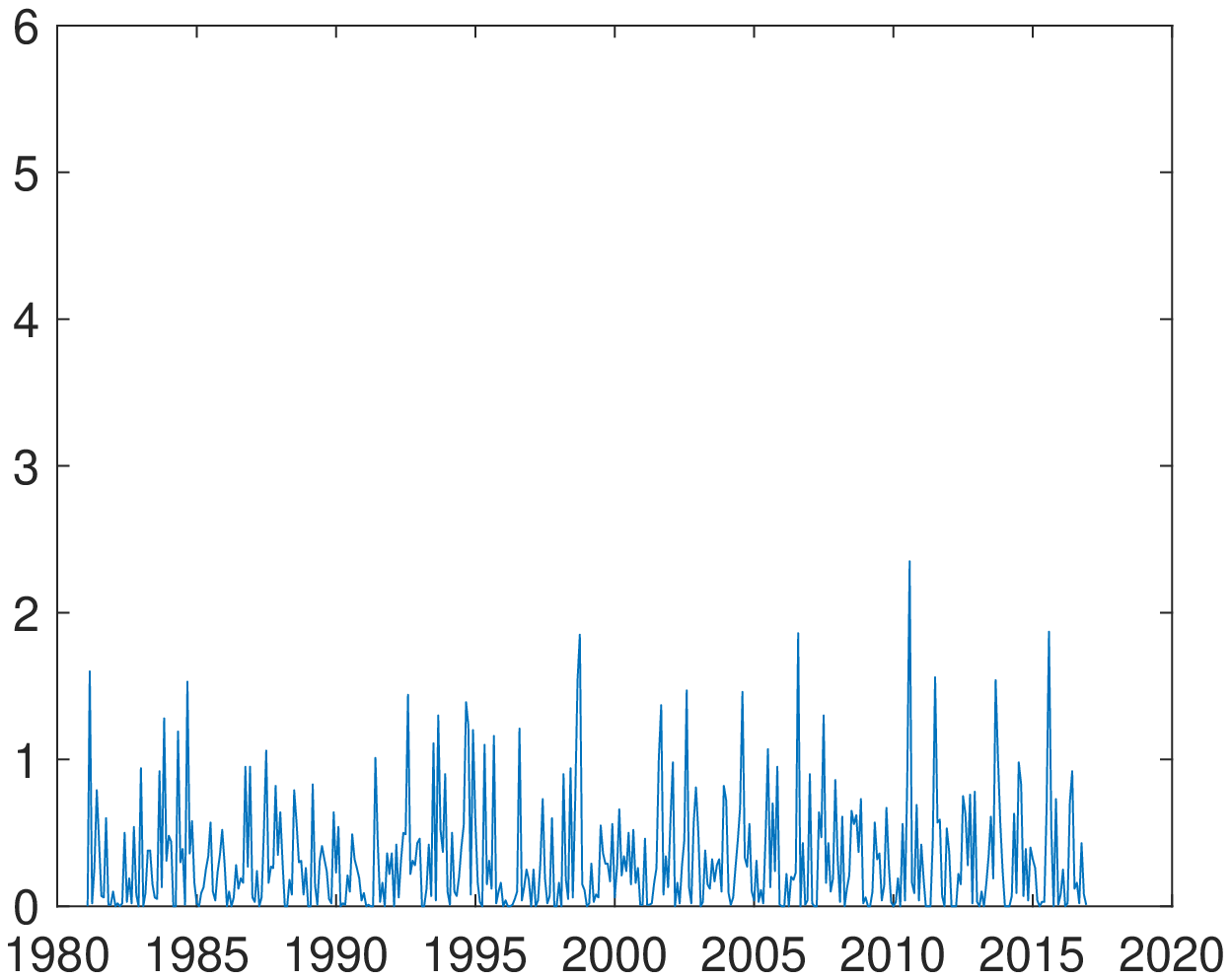}
                  \label{fig:plotr20mmNE}} 
         ~
         \subfigure[PO]{                
                  \includegraphics[width=0.17\textwidth]{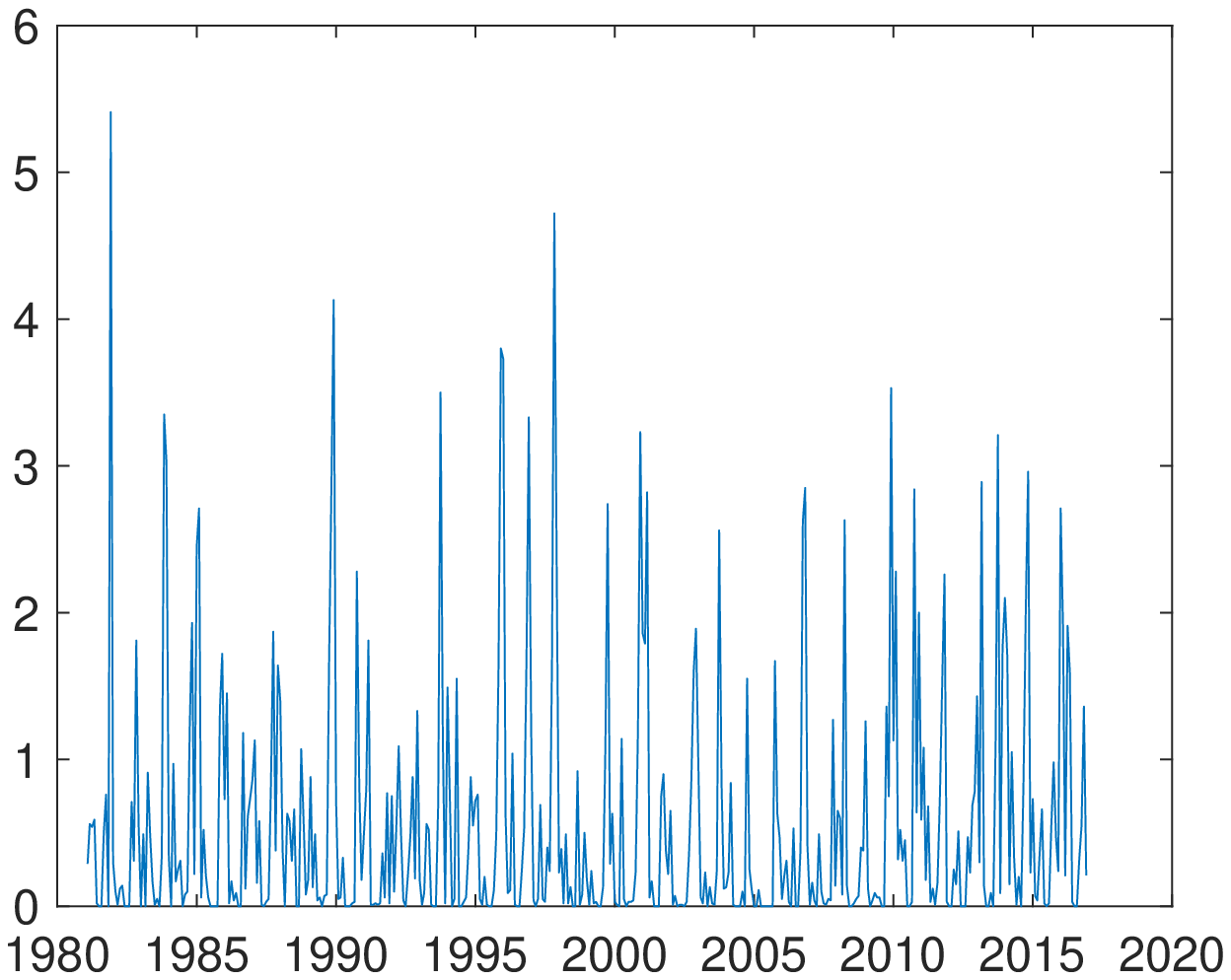}
                  \label{fig:plotr20mmPO}}
         ~
         \subfigure[SP]{
                  \includegraphics[width=0.17\textwidth]{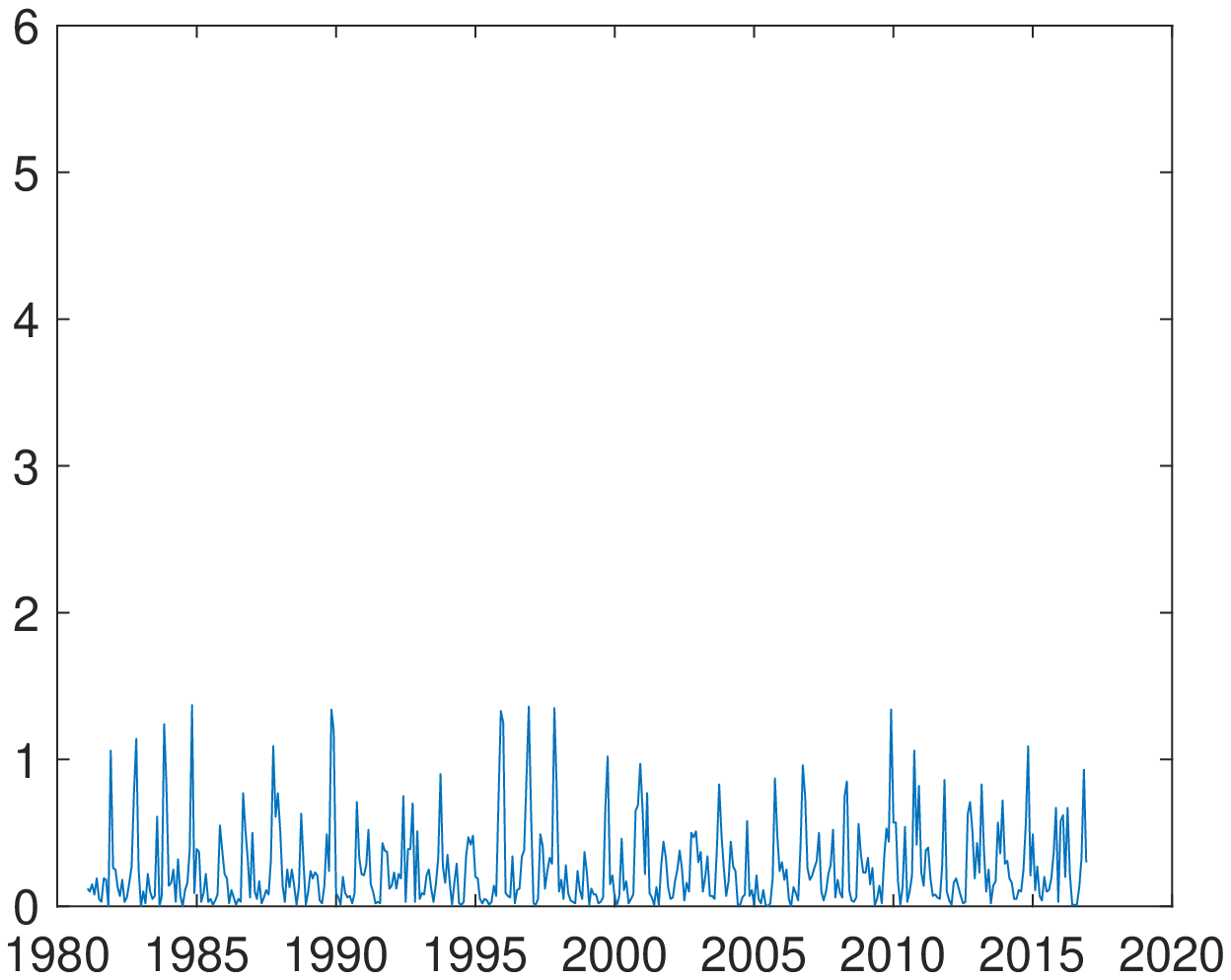}
                  \label{fig:plotr20mmSP}} 
 		\caption{Time series plot of the monthly frequency of daily heavy rainfall exceeding 20mm for 13 EU countries for the period February 1981 to December 2016.}\label{PLOTr20mm}
 \end{figure}
 \newpage
 \clearpage
 \section{Parameter conditional distributions}
 \label{Appendix2}
 \noindent In the following we derive the posterior distributions and give details of the Gibbs sampling scheme (see Section \ref{sec:BayesInfe}) implemented in this paper. 
 \subsection{Sampling the trajectories of the hidden variables, $S_{y}$ and $S_{x}$}
 \noindent We sample the state trajectory $S_{y,i1:T}$, for $i=1,\dots,N$ and $S_{x,1:T}$ corresponding to each of the unit-specific business cycle  and financial cycle from their respective
 joint posterior distribution\\ $f(S_{yi,1:T} | S_{y,-i,1:T},S_{x,1:T}, Y_{1:T}, X_{1:T}, Z_{1:T},\boldsymbol{\theta})$, $i=1,\dots, N$ and $f(S_{x,1:T} | S_{y,i,1:T}, Y_{1:T}, X_{1:T}, Z_{1:T},\boldsymbol{\theta})$ by the Forward Filter Backward Sampling (FFBS) algorithm (see  \cite{fruhwirth2008} for illustration). By means of dynamic factorization, the full conditional distribution of the unit specific hidden state associated with the business cycles is
 \begin{flalign}
 f(S_{y,i1:T} | S_{y,-i1:T}, S_{x,1:T}, Y_{1:T}, X_{1:T}, Z_{1:T}, \boldsymbol{\theta}) \nonumber\\
 = f(S_{y,iT} | S_{y,-i,1:T}, S_{x,,1:T}, Y_{1:T}, X_{1:T}, Z_{1:T}, \boldsymbol{\theta})\nonumber\\ f(S_{y,i,1:T-1} |S_{y,iT},S_{y,-i,1:T}, S_{x,,1:T}, Y_{1:T}, X_{1:T}, Z_{1:T}, \boldsymbol{\theta}) \nonumber \\
  = f(S_{y,iT} | S_{y,-i,1:T}, S_{x,,1:T}, Y_{1:T}, X_{1:T}, Z_{1:T},\boldsymbol{\theta} )\nonumber\\ 
  \times\prod\limits_{t=1}^{T-1} f(S_{y,i,t} | S_{y,i,t+1:T},S_{y,-i,1:T},S_{x,,1:T}, Y_{1:T}, X_{1:T}, Z_{1:T}, \boldsymbol{\theta}) \nonumber \\
  \propto f(S_{y,iT} | S_{y,-i,1:T}, S_{x,,1:T}, Y_{1:T}, X_{1:T}, Z_{1:T},\boldsymbol{\theta})\nonumber\\
 \times \prod\limits_{t=1}^{T-1} f(S_{y,i,t+1} | S_{y,i,t},S_{y,-i,t}, S_{x,t},\boldsymbol{\theta})f(S_{y,i,t} | S_{y,-i,1:T}, S_{x,,1:T}, Y_{1:T}, X_{1:T}, Z_{1:T}, \boldsymbol{\theta})  \nonumber 
 \end{flalign}
 \begin{flalign}
  \propto \left(f(S_{y,iT} | S_{y,-i,1:T}, S_{x,,1:T}, Y_{1:T}, X_{1:T}, Z_{1:T},\boldsymbol{\theta}) 
 \prod\limits_{t=1}^{T-1} f(S_{y,i,t} | S_{y,-i,1:t}, S_{x,,1:t}, Y_{1:T}, X_{1:T}, Z_{1:T}, \boldsymbol{\theta})\right)\nonumber \\
  \times \left(\prod\limits_{t=1}^{T-1} f(S_{y,i,t+1} | S_{y,i,t},S_{y,-i,t}, S_{x,t},\boldsymbol{\theta})\right)\nonumber
 \end{flalign}
 
 \medskip
 \noindent While, the joint posterior distribution of the hidden variable associated with the financial cycle may be factorized as 
 
 \begin{flalign}
 f(S_{x,1:T} | S_{y,1:T}, Y_{1:T}, X_{1:T}, Z_{1:T}, \boldsymbol{\theta}) \nonumber\\
  \propto \left(f(S_{x,T} | S_{y,1:T},  Y_{1:T}, X_{1:T}, Z_{1:T}, \boldsymbol{\theta}) 
 \prod\limits_{t=1}^{T-1} f(S_{x,t} | S_{y,1:t}, Y_{1:T}, X_{1:T}, Z_{1:T}, \boldsymbol{\theta})\right)\nonumber \\
  \times \left(\prod\limits_{t=1}^{T-1} f(S_{x,t+1} | S_{x,t},S_{y,t},\boldsymbol{\theta})\right)\nonumber
 \end{flalign}
 
 \subsection{Sampling $\boldsymbol{\theta}$}
 \noindent Sampling $\boldsymbol{\theta}$ from its full conditional distribution will be done by separating it into two sub-vectors: regime dependent parameters and the regime independent parameters. The regime dependent parameters are essentially the parameters of the measurement equations, while the regime independent parameters are the interaction parameters. 
 \subsubsection{Sampling parameters of the measurement equations}
 \noindent The parameters of the measurement equation consists of $\Psi_{il}$, $i=1,\dots,N$ (regime specific measure of the effect of climate condition on each country's IPI growth), $\Phi_{l}$ (regime specific measure of aggregate effect of climate condition on financial activity), $\sigma_{il}^{2}$, ($i=1,\dots, N$), (regime specific measure of the variability in the residual of each country's IPI growth), and $\tau_{l}^{2}$ (regime specific measure of the variability in the residual financial index) for $l=1,2$. The regime specific mean parameters,  $\Psi_{il}$, $i=1,\dots,N$ and $\Phi_{l}$ of the measurement equations will be sampled respectively from the multivariate distributions given by Equations \ref{posterior:phi} and \ref{posterior:psi}, while the regime specific variances
 parameters, $\sigma_{il}^{2}$, ($i=1,\dots, N$) and $\tau_{l}^{2}$, will sampled from the inverse gamma distributions given in Equations \ref{posterior:sigma} and \ref{posterior:tau}. 
 
 \begin{enumerate}
 
 \item[1.] Given that the prior distributions of the regime specific mean parameters, $\Psi_{il}$ (where $l=1,2$ and $i=1,\ldots,N$), of the measurement equation associated with each country's IPI growth are multivariate normally distributed (see Equation (\ref{eqn:mupriorY})), then their posterior conditional density functions are given as follows:
 
 \begin{flalign}
 f(\Psi_{il}|Y_{1:T},Z_{1:T},S_{y,1:T},\boldsymbol{\theta}_{-\Psi_{il}})\nonumber\\
 \propto
 \exp\left(-\frac{1}{2}\left(\Psi_{il}-m_{y,il}\right)'\Sigma_{y,il}^{-1}(\Psi_{il}-m_{y,il})\right)\prod\limits_{t\in \mathcal{T}_{y,i}}\exp\lbrace{-\frac{{\xi_{yl,it}}}{2\sigma^{2}_{il}}  \left( y_{it}-\Psi_{il}'z_{t}\right)^{2}}\rbrace \nonumber \\
 \propto \exp\lbrace{-\frac{1}{2}
 \left(\Psi_{i1}'[\Sigma_{y,il}^{-1}+\sum\limits_{t\in \mathcal{T}_{y,il}}z_{t}z_{t}'/\sigma_{il}]\Psi_{i1} -2 [m_{y,il}'\Sigma_{y,il}^{-1} + 
 \sum\limits_{t\in \mathcal{T}_{y,il}} y_{it}z_{t}'/\sigma_{il} ]\Psi_{il} \right)\rbrace} \nonumber \\
 \propto\,\mathcal{MN}(\overline{m}_{y,il},\overline{\Sigma}_{y,il}) \label{posterior:phi} 
 \end{flalign}
 with 
 $\overline{m}_{y,il}=\overline{\Sigma}_{y,il}\left(\sum\limits_{t\in \mathcal{T}_{y,il}} y_{t}z_{t}/\sigma_{il} + \Sigma_{y,il}^{-1}m_{y,il}\right)$ and
 $\overline{\Sigma}_{y,il}^{-1}= \Sigma_{y,il}^{-1} + \sum\limits_{t\in \mathcal{T}_{y,il}}z_{t}z_{t}'/\sigma_{il}$.
 \\ We defined $\mathcal{T}_{y,il} = \lbrace t=1,\ldots,T \vert S_{it} = l \rbrace $, $ T_{y,il} = card(\mathcal{T}_{y,il}) $,  $Y_{1:T}=\left(Y_1,\ldots,Y_T\right)$, $Z_{1:T}=\left(Z_1,\ldots,Z_T\right)$ and $S_{y,1:T}=\left(S_{y,1},\ldots,S_{y,T}\right)$. The notation $\boldsymbol{\theta}_{-r}$ indicates that element $r$ is excluded from the vector $\boldsymbol{\theta}$. 
 
 \item[2.] Following a similar argument as above and given that the prior distributions of $\Psi_{il}$ (where $l=1,2$) are multivariate normal distributed (see Equation (\ref{eqn:mupriorY})), we obtain their corresponding conditional posterior density functions as follows: 
 
 \begin{flalign}
 f(\Phi_{l}|X_{1:T},Z_{1:T},S_{x,1:T},\boldsymbol{\theta}_{-\Phi_{il}})\nonumber\\
 \propto \exp\lbrace{-\frac{1}{2}(\Phi_{l}'\Sigma_{x,l}^{-1}\Phi_{l}-2m_{x,l}'\Sigma_{x,l}^{-1}\Phi_{l})}\rbrace\exp\lbrace{-\sum\limits_{t\in \mathcal{T}_{x,l}}\frac{{1}}{2\tau^{2}_{l}}(\Phi_{l}'z_{t}z_{t}'\Phi_{l}-2x_{t}z_{t}'\Phi_{l})}\rbrace \nonumber  \\  
 \propto \exp\lbrace{-\frac{1}{2}
 \left(\Phi_{l}'[\Sigma_{x,l}^{-1}+\sum\limits_{t\in \mathcal{T}_{x,l}}z_{t}z_{t}'/\tau_{l}]\Phi_{l} -2 [m_{x,l}'\Sigma_{x,l}^{-1} + 
 \sum\limits_{t\in \mathcal{T}_{x,l}} x_{t}z_{t}'/\tau_{l} ]\Phi_{l} \right)\rbrace} \nonumber \\
 \propto\,\mathcal{MN}(\overline{m}_{x,l},\overline{\Sigma}_{x,l}) \label{posterior:psi} 
 \end{flalign}
 with 
 $\overline{m}_{x,l}=\overline{\Sigma}_{x,l}\left(\sum\limits_{t\in \mathcal{T}_{x,l}} x_{t}z_{t}/\tau_{l} + \Sigma_{x,l}^{-1}m_{x,l}\right)$, 
 $\overline{\Sigma}_{x,l}^{-1}= \Sigma_{x,l}^{-1} + \sum\limits_{t\in \mathcal{T}_{x,l}}z_{t}z_{t}'/\tau_{l}$, $\mathcal{T}_{x,l} = \lbrace t=1,\ldots,T \vert S_{x,t} = l \rbrace $, $ T_{x,l} = card(\mathcal{T}_{x,l}) $,  $X_{1:T}=\left(X_1,\ldots,X_T\right)$ and $S_{x,1:T}=\left(S_{x,1},\ldots,S_{x,T}\right)$. 
 
 \item[3.] As regards the variance parameters ($\sigma_{il}^{2}$, $i=1,\dots, N$, and $\tau_{l}^{2}$, $l=1,2$), we derive their conditional posterior distribution, under the assumption of inverse gamma distribution priors (see Equations (\ref{eqn:sigmapriorY}) and (\ref{eqn:sigmapriorX})), as follows
 
 \begin{eqnarray}{}
 f\left(\sigma_{il}|Y_{1:T},Z_{1:T},S_{y,1:T},\theta_{-\sigma_{il}}\right)  \propto \left(\frac{1}{\sigma_{il}^{2}}\right)^{(\alpha_{y,il}+1)}\exp\lbrace{-\frac{\beta_{y,il}}{\sigma_{il}^2}}\rbrace\prod\limits_{t\in \mathcal{T}_{y,il}}{\frac{1}{\sigma_{il}^{2}}}\nonumber\\ \times \exp\lbrace{-\sum\limits_{t\in \mathcal{T}_{y,il}}\frac{{1}}{2\sigma^{2}_{il}}(y_{it}-\Psi_{il}'z_{t})^{2}}\rbrace \nonumber \\
 \propto \left(\frac{1}{\sigma_{il}^{2}}\right)^{\left(\alpha_{y,il}+T_{y,il}+1\right)}\exp\lbrace{-\frac{1}{\sigma_{il}^{2}}\lbrace \beta_{y,il}+\sum\limits_{t\in \mathcal{T}_{y,il}}(y_{it}-\Psi_{il}'z_{t})^{2}\rbrace}\rbrace \nonumber \\
 \propto \mathcal{IG}\left(\alpha_{y,il}+T_{y,il},\beta_{y,il}+\sum\limits_{t\in \mathcal{T}_{y,il}}\left( y_{it}-\Psi_{il}'z_{t}\right)^{2}\right) \label{posterior:sigma}
 \end{eqnarray}  
 
 \item[4.] and 
 
 \begin{eqnarray}{}
 f\left(\tau_{l}|X_{1:T},Z_{1:T},S_{x,1:T},\theta_{-\tau_{l}}\right)  \propto \left(\frac{1}{\tau_{l}^{2}}\right)^{(\alpha_{x,l}+1)}\exp\lbrace{-\frac{\beta_{x,l}}{\tau_{l}^2}}\rbrace\prod\limits_{t\in \mathcal{T}_{x,l}}{\frac{1}{\tau_{l}^{2}}}\nonumber\\ \times \exp\lbrace{-\sum\limits_{t\in \mathcal{T}_{x,l}}\frac{{1}}{2\tau^{2}_{l}}(x_{t}-\Phi_{l}'z_{t})^{2}}\rbrace \nonumber \\
 \propto \left(\frac{1}{\tau_{l}^{2}}\right)^{\left(\alpha_{x,l}+T_{l}+1\right)}\exp\lbrace{-\frac{1}{\tau_{l}^{2}}\lbrace \beta_{x,l}+\sum\limits_{t\in \mathcal{T}_{x,l}}(x_{t}-\Phi_{l}'z_{t})^{2}\rbrace}\rbrace \nonumber \\
 \propto \mathcal{IG}\left(\alpha_{x,l}+T_{x,l},\beta_{x,l}+\sum\limits_{t\in \mathcal{T}_{x,l}}\left( x_{t}-\Phi_{l}'z_{t}\right)^{2}\right) \label{posterior:tau}
 \end{eqnarray}  
 \end{enumerate}
 
 \subsubsection{Sampling interaction parameters}
 \noindent The interaction parameters consists of the fixed transition probabilities ($P_{il,1:2}=(p_{i,l1},p_{i,l2})$, $i=1,\dots, N$, and $P_{fl,1:2}=(p_{f,l1},p_{f,l2})$, $l=1,2$), the idiosyncratic parameter ($\alpha_{i}$, $i=1,\dots, N$ and $\alpha_{f}$), the local (financial cycle) effect ($\beta_{i}$, $i=1,\dots, N$, and $\beta_{f}$) and the global effect ($\gamma_{i}$, $i=1,\dots, N$, and $\gamma_{f}$). We implement a Metropolis Hastings (MH) sampler for drawing each row of the transition probability matrix from the Beta distribution in Equations \ref{proposal:transitionBC} and \ref{proposal:transitionFC}, and the vector of other interaction parameters  ($\alpha_{i},\beta_{i}, \gamma_{i}$, $i=1,\dots, N$ and ($\alpha_{f},\beta_{f},\gamma_{f}$)) form a mixture of three component Dirichlet distributions in Equations \ref{proposal:interactionBC} and  \ref{proposal:interactionFC}.
 
 \begin{enumerate}
 \item[5.] Given that the prior distribution of each $l$-th row of the transition matrix $P_{il,1:2}=(p_{i,l1},p_{i,l2})$ follows a Beta distribution (See Equation \ref{eqn:transprob}), then the posterior conditional density function:
 is
 \begin{flalign}
 f\left(p_{il,1:2}|Y_{1:T}, X_{1:T},Z_{1:T},S_{y,1:T},S_{x,1:T},\theta_{-\left(p_{l,1:2}\right)}\right) \nonumber\\
 \propto \left(\prod\limits_{k=1}^{2}p_{lk}^{\left(\delta_{y,ik}-1\right)}\right)\prod\limits_{t=1}^{T}\prod\limits_{k=1}^{2}\left(\alpha_{i} p_{i,lk}+\beta_{i} (S_{x,t}-1)+\gamma_{i} m_{t,k}\right)^{\xi_{yk,it}\xi_{yl,it-1}} \nonumber \\
 \propto \left(\prod\limits_{k=1}^{2}p_{lk}^{\left(\delta_{y,ik}-1\right)}\right)\prod\limits_{k=0}^{1} \prod\limits_{t \in \mathcal{T}_{y,ilk}}\left(\alpha_{i} p_{i,lk}+\beta_{i} (S_{x,t}-1)+\gamma_{i} m_{t,k}\right) \nonumber \\
 \propto \prod\limits_{k=1}^{2}\left(p_{i,lk}^{\left(\delta_{y,ik}-1\right)}
 \left[ \prod\limits_{t \in \mathcal{T}_{y,ilk}}\left(\alpha_{i} p_{i,lk}+\beta_{i} (S_{x,t}-1)+\gamma_{i} m_{t,k}\right) \right]
  \right) \label{posterior:transprob} 
 \end{flalign}
 Since sampling directly from the non-standard posterior conditional distribution given in Equation  \ref{posterior:transprob} will pose a challenge in the implementation of the Gibbs iteration, we adopt a Metropolis Hastings (MH) sampling procedure. The Beta distribution in Equation \ref{proposal:transitionBC} obtained by setting $\alpha_{i}=1$ and $\beta_{i}=\gamma_{i}=0$ posterior distribution in Equation \ref{posterior:transprob} is used as the proposal distribution within the MH. We denote the proposal distribution by 
 \begin{flalign}
 g\left(p_{il,1:2}|S_{y,i1:T}\right) 
 \propto \prod\limits_{k=1}^{2} p_{i,lk}^{\left(\delta_{y,ik}-1\right)} p_{i,lk}^{T_{y,ilk}} \nonumber \\
 \propto Beta\left(\delta_{y,i1} +T_{y,il1}, \delta_{y,i2} +T_{y,il2} \right) \label{proposal:transitionBC}
 \end{flalign}
 
 \item[6.] Following a similar argument as above, we implement an MH sampling algorithm for drawing samples from the posterior distribution in Equation \ref{posterior:transitionFC} of each $l$-th row of the transition matrix $P_{fl,1:2}=(p_{f,l1},p_{f,l2})$. The Beta distribution in Equation \ref{proposal:transitionFC} derived by setting $\alpha_{f}=1$ and $\beta_{f}=\gamma_{f}=0$ in Equation \ref{posterior:transitionFC} is set as the proposal distribution for sampling the components of the transition probability matrix. 
 
 \medskip
 \noindent Given that the prior distribution of the $l$-th row of the transition matrix $P_{fl,1:2}=(p_{f,l1},p_{f,l2})$
 is Dirichlet distribution (see Equation (\ref{eqn:transprobf})) then the corresponding posterior conditional density function is:
 
 \begin{flalign}
 f\left(p_{fl,1:2}|Y_{1:T}, X_{1:T},Z_{1:T},S_{y,1:T},S_{x,1:T},\theta_{-\left(p_{fl,1:2}\right)}\right) \nonumber\\
 \propto \left(\prod\limits_{k=1}^{2}p_{f,lk}^{\left(\delta_{x,k}-1\right)}\right)
 \prod\limits_{t=1}^{T}\prod\limits_{k=1}^{2}\left(\alpha_{f} p_{f,lk}+\beta_{f} (S_{x,t}-1)+\gamma_{f} m_{t,k}\right)^{\xi_{xk,t}\xi_{xl,t-1}} \nonumber \\
 \propto \left(\prod\limits_{k=1}^{2}p_{f,lk}^{\left(\delta_{x,k}-1\right)}\right) 
 \prod\limits_{k=1}^{2} \prod\limits_{t \in \mathcal{T}_{x,lk}}\left(\alpha_{f} p_{f,lk}+\beta_{f} (S_{x,t}-1)+\gamma_{f} m_{t,k}\right) \nonumber \\ 
 \propto \prod\limits_{k=1}^{2}\left(p_{f,lk}^{\left(\delta_{x,k}-1\right)}
 \left[
 \prod\limits_{t \in \mathcal{T}_{x,lk}}\left(\alpha_{f} p_{f,lk}+\beta_{f} (S_{x,t}-1)+\gamma_{f} m_{t,k}\right) \right] \right) \label{posterior:transitionFC}
 \end{flalign}
 and the proposal distribution is denoted by 
 \begin{flalign}
 q\left(p_{fl,1:2}|S_{x,1:T},\boldsymbol{\theta}_{-\left(p_{fl,1:2}\right)}\right) \propto \prod\limits_{k=1}^{2}p_{f,lk}^{\left(\delta_{x,k}-1\right)}   p_{f,lk}^{T_{x,lk}} \nonumber\\
 \propto Beta(\delta_{x,1}+T_{x,l1}, \delta_{x,2}+T_{x,l2}) \label{proposal:transitionFC}
 \end{flalign} 
 
 \item[7.] As regards the vector of interaction parameters $(\alpha_{i}, \beta_{i}, \gamma_{i})$, assuming the Dirichlet prior distribution given in Equation (\ref{eqn:alpha}), the posterior density function is:
 \begin{flalign}
 f\left(\alpha_{i}, \beta_{i}, \gamma_{i}|Y_{1:T},X_{1:T},Z_{1:T}S_{y,1:T},S_{x,1:T},\theta_{-\left(\alpha_{i}, \beta_{i}, \gamma_{i} \right)}\right)\nonumber\\
 \propto \left(\alpha_{i}^{\varphi_{y,i1}-1}\beta_{i}^{\varphi_{y,i2}-1}\gamma_{i}^{\varphi_{y,i3}-1}\right)\prod\limits_{t=1}^{T}\prod\limits_{l=1}^{2}\prod\limits_{k=1}^{2}\left(\alpha_{i} p_{i,lk}+\beta_{i} (S_{x,t}-1)+\gamma_{i} m_{t,k}\right)^{\xi_{yk,it}\xi_{yl,it-1}} \nonumber \\
 \propto \left(\alpha_{i}^{\varphi_{y,i1}-1}\beta_{i}^{\varphi_{y,i2}-1}\gamma_{i}^{\varphi_{y,i3}-1}\right)\prod\limits_{k=1}^{2}\prod\limits_{l=1}^{2} \prod\limits_{t \in \mathcal{T}_{y,ilk}}\left(\alpha_{i} p_{i,lk}+\beta_{i} (S_{x,t}-1)+\gamma_{i} m_{t,k}\right) \nonumber
 \end{flalign}
 the above conditional posterior distributions of the interaction parameters, $(\alpha_{i}, \beta_{i}, \gamma_{i})$, $i=1,\dots, N$ are directly proportional to a polynomial of three (3) variables, of degree $T+\varphi_{y,i1}-1$ in $\alpha$, $T+\varphi_{y,i2}-1$ in $\beta$, and  $T+\varphi_{y,i3}-1$ in $\gamma$. It may therefore be deduced that the conditional posterior distribution of the interaction parameters is a mixture of Dirichlet distributions. However, the weights and parameters of the mixture are not readily available or easily obtained. This observation poses a difficulty for directly sampling from these distributions. In-view of this, MH samplers are implemented for drawing independent samples of the interacting parameters from the conditional posterior distributions.  More precisely, a mixture of three (3) equally weighted Dirichlet distributions with parameters $(T+\varphi_{y,i1},\varphi_{y,i2},\varphi_{y,i3})$, $(\varphi_{y,i1},T+\varphi_{y,i2},\varphi_{y,i3})$, and $(\varphi_{y,i1},\varphi_{y,i2},T+\varphi_{y,i3})$ denoted by 
 \begin{equation}
 \begin{aligned} 
 &q(\alpha_{i}, \beta_{i}, \gamma_{i})\\
 & = \dfrac{1}{3}\left(\mathcal{D}ir(T+\varphi_{y,i1},\varphi_{y,i2},\varphi_{y,i3}) +  \mathcal{D}ir (\varphi_{y,i1},T+\varphi_{y,i2},\varphi_{y,i3}) +\mathcal{D}ir(\varphi_{y,i1},\varphi_{y,i2},T+\varphi_{y,i3}) \right) ,
 \end{aligned}\label{proposal:interactionBC}
 \end{equation}
 is proposed for generating samples for the conditional posterior distribution of the interaction parameters. 
 
 \item[8.] Lastly, for the vector of interaction parameters $(\alpha_{f}, \beta_{f}, \gamma_{f})$, we combine the Dirichlet prior distribution in Equation \ref{eqn:alphaf} with the likelihood function to obtain the posterior distribution in Equation \ref{posterior:interactionFC}. 
 \begin{flalign}
 f\left(\alpha_{f},\beta_{f}, \gamma_{f}|Y_{1:T},X_{1:T},Z_{1:T}S_{y,1:T},S_{x,1:T},\theta_{-\left(\alpha_{f}, \beta_{f},\gamma_{f} \right)}\right)\nonumber\\
 \propto \left(\alpha_{f}^{\varphi_{x,1}-1}\beta_{f}^{\varphi_{x,2}-1}\gamma_{f}^{\varphi_{x,3}-1}\right)\prod\limits_{t=1}^{T}\prod\limits_{l=1}^{2}\prod\limits_{k=1}^{2}\left(\alpha_{f} p_{f,lk}+\beta_{f}(S_{x,t}-1)+\gamma_{f} m_{t,k}\right)^{\xi_{xk,t}\xi_{xl,t-1}} \nonumber \\
 \propto \left(\alpha_{f}^{\varphi_{x,1}-1}\beta_{f}^{\varphi_{x,2}-1}\gamma_{f}^{\varphi_{x,3}-1}\right)\prod\limits_{k=1}^{2}\prod\limits_{l=1}^{2} \prod\limits_{t \in \mathcal{T}_{x,lk}}\left(\alpha_{f} p_{f,lk}+\beta_{f}(S_{x,t}-1) + \gamma_{f} m_{t,k}\right).  \label{posterior:interactionFC}
 \end{flalign}
 We are however faced with a similar problem as described in item 8 above. In view of this, a mixture of Dirichlet distribution (see Equation \ref{proposal:interactionFC}) is proposed in the implementation of an MH sampling algorithm for drawing samples from the posterior distribution in Equation \ref{posterior:interactionFC} of the interacting parameters $(\alpha_{f}, \beta_{f}, \gamma_{f})$. 
 
 \begin{equation}
 \begin{aligned} 
 &q(\alpha_{f}, \beta_{f}, \gamma_{f})\\
 & = \dfrac{1}{3}\left(\mathcal{D}ir(T+\varphi_{x,1},\varphi_{x,2},\varphi_{x,3}) +  \mathcal{D}ir (\varphi_{x,1},T+\varphi_{x,2},\varphi_{x,3}) +\mathcal{D}ir(\varphi_{x,1},\varphi_{x,2},T+\varphi_{x,3}) \right) ,
 \end{aligned}\label{proposal:interactionFC}
 \end{equation}
 \end{enumerate}
 \newpage 
 \clearpage 
 \section{Tables}
 \begin{table}[htbp]
 \begin{scriptsize}
   \centering   
       \begin{tabular}{l|c|c|ccc|c}
   \hline        
       \multicolumn{1}{l|}{i} & \multicolumn{1}{l|}{Regime (k)} & \multicolumn{1}{c|}{Intercept(${\hat\Psi}_{i,0k}$)} &       \multicolumn{1}{c}{CSU(${\hat\Psi}_{i,1k}$)} &       \multicolumn{1}{c}{SPI(${\hat\Psi}_{i,2k}$)} &      \multicolumn{1}{c|}{r20mm (${\hat\Phi}_{i,3k}$)} &       \multicolumn{1}{c}{${\hat\sigma}_{i,k}$}  \\
   \hline 
     \multicolumn{1}{l|}{AU} & 1     & -0.0456 & 0.0122 & 0.0081 & -0.0183  & 0.3253 \\
           &       & (-0.0906,  0.0001) & (0.0002, 0.0245) & (-0.0522,0.0688) & (-0.0595, 0.0204) & (0.3065, 0.3447) \\
           & 2     & 0.5247  & 0.0127  & -0.0164   & 0.0050   & 0.1936   \\
           &       & (0.5008, 0.5494) & (0.0040, 0.0212) & (-0.0570, 0.0260) & (-0.0220, 0.0324) & (0.1824, 0.2053) \\
\hline            
     \multicolumn{1}{l|}{BE} & 1     & -0.1612  & 0.0126 & 0.1220   & -0.0337  & 0.4257  \\
           &       & (-0.2044, -0.1149) & (-0.0027, 0.0281) & (0.0443, 0.2034) & (-0.1031, 0.0346) & (0.3954, 0.4558) \\
           & 2     & 0.5413   & -0.0163    & -0.0894     & 0.0056     & 0.2485   \\
           &       & (0.5124, 0.5702) & (-0.0251, -0.0075) & (-0.1446, -0.0348) & (-0.0447, 0.0568) & (0.2328, 0.2649) \\
\hline            
     \multicolumn{1}{l|}{DE} & 1     & -0.2344   & 0.0509  & -0.0629  & -0.0446  & 0.4240 \\
           &       & (-0.2792, -0.1916) & (0.0227, 0.0781) & (-0.1485, 0.0218) & (-0.1783, 0.0871) & (0.4006, 0.4486) \\
           & 2     & 0.6384 &  -0.0185  & -0.0454  & -0.0667  & 0.3952 \\
           &       & (0.5989, 0.6792) & (-0.0450, 0.0091) & (-0.1190, 0.0277) & (-0.1951, 0.0649) & (0.3740, 0.4179) \\
\hline            
     \multicolumn{1}{l|}{FI} & 1     & -0.2395  & 0.0236  & -0.0439  & 0.1505  & 0.6291  \\
           &       & (-0.3022, -0.1779) & (-0.0228, 0.0668) & (-0.1755, 0.0887) & (-0.1127, 0.4116) & (0.5912, 0.6676) \\
           & 2     & 0.5699  & 0.0113  & 0.0442 & -0.1101  & 0.2530 \\
           &       & (0.5417, 0.5966) & (-0.0093, 0.0320) & (-0.0224, 0.1098) & (-0.2198, 0.0063) & (0.2365, 0.2705) \\
\hline            
     \multicolumn{1}{l|}{FR} & 1     & -0.2353 &  0.0081 &   0.1187  & 0.0257 & 0.4152  \\
           &       & (-0.3016, -0.1695) & (-0.0007, 0.0175) & (0.0354, 0.2008) & (-0.0858, 0.1409) & (0.3878, 0.4438) \\
           & 2     & 0.2523 & -0.0004 & -0.0647 & 0.0350 &  0.1720 \\
           &       & (0.2302, 0.2758) & (-0.0043, 0.0033) & (-0.0956, -0.0337) & (-0.0099, 0.0787) & (0.1633, 0.1812) \\
\hline            
     \multicolumn{1}{l|}{GE} & 1     & -0.2641  & 0.0097  & -0.0713 & 0.1519 & 0.6721  \\
           &       & (-0.3454, -0.1829) & (-0.0120,  0.0307) & (-0.1936, 0.0558) & (-0.0177, 0.3280) & (0.6316, 0.7157) \\
           & 2     & 0.4160 & 0.0068  & -0.0083 & 0.0313 & 0.1857  \\
           &       & (0.3877, 0.4447) & (0.0007, 0.0130) & (-0.0500, 0.0330) & (-0.0263, 0.0867) & (0.1711, 0.2018) \\
\hline            
     \multicolumn{1}{l|}{GR} & 1     & -0.2950   & 0.0021 & -0.0166  & 0.0379  & 0.2944 \\
           &       & (-0.3389, -0.2530) & (-0.0009, 0.0048) & (-0.0711, 0.0387) & (-0.0315, 0.1099) & (0.2778, 0.3116) \\
           & 2     & 0.3328  & 0.0008 & -0.0630  & 0.0656 & 0.3096  \\
           &       & (0.2919, 0.3763) & (-0.0023, 0.0037) & (-0.1297, 0.0046) & (-0.0038, 0.1362) & (0.2931, 0.3276) \\
           \hline 
     \multicolumn{1}{l|}{IR} & 1     & 0.1069 & -0.0070& 0.1203  & 0.0308  & 0.3864\\
           &       & (0.0597, 0.1539) & (-0.0836, 0.0710) & (0.0476, 0.1895) & (-0.0263, 0.0873) & (0.3649, 0.4079) \\
           & 2     & 1.4815 & -0.0747 & -0.2481 & -0.0072 & 0.8914  \\
           &       & (1.3625, 1.5990) & (-0.1876, 0.0374) & (-0.4189, -0.0808) & (-0.1314, 0.1185) & (0.8372, 0.9467) \\
\hline            
     \multicolumn{1}{l|}{IT} & 1     & -0.2156  & -0.0056  & 0.0576 & -0.1835  & 0.5521  \\
           &       & (-0.3044, -0.1282) & (-0.0124, 0.0014) & (-0.0601, 0.1724) & (-0.2889, -0.0785) & (0.5184, 0.5871) \\
           & 2     & 0.4020& -0.0019  & -0.0853 & -0.0593 & 0.2289 \\
           &       & (0.3667, 0.4379) & (-0.0044, 0.0006) & (-0.1270, -0.0439) & (-0.1016, -0.0169) & (0.2166, 0.2413) \\
\hline            
     \multicolumn{1}{l|}{LU} & 1     & -0.2893 & 0.0067 & 0.1456  & -0.0132   & 0.8411  \\
           &       & (-0.3902, -0.1861) & (-0.0200, 0.0358) & (-0.0035, 0.2945) & (-0.1208, 0.0909) & (0.6777, 0.9818) \\
           & 2     & 0.7036  & 0.0121 & -0.2358  & 0.0144 & 0.4129 \\
           &       & (0.6381, 0.7774) & (0.0005, 0.0231) & (-0.3477, -0.1002) & (-0.0427, 0.0707) & (0.3491, 0.4840) \\
\hline            
     \multicolumn{1}{l|}{NE} & 1     & -0.1893  & -0.0093  & 0.0862 & 0.0681 &  0.2683  \\
           &       & (-0.2214, -0.1561) & (-0.0199, 0.0015) & (0.0308, 0.1434) & (0.0152, 0.1227) & (0.2535, 0.2831) \\
           & 2     & 0.4704 &  -0.0168 & -0.0039 &  -0.0125    & 0.2240  \\
           &       & (0.4428, 0.4978) & (-0.0260, -0.0074) & (-0.0515, 0.0475) & (-0.0663 , 0.0423) & (0.2107, 0.2379) \\
\hline            
     \multicolumn{1}{l|}{PO} & 1     & -0.2346 & -0.0021 & -0.0571& 0.0597  & 0.3672 \\
           &       & (-0.2929, -0.1743) & (-0.0061, 0.0020) & (-0.1231, 0.0084) & (0.0124, 0.1055) & (0.3458, 0.3898) \\
           & 2     & 0.4636 & -0.0041  & 0.0986 & -0.0214 & 0.2820 \\
           &       & (0.4233, 0.5036) & (-0.0067, -0.0014) & (0.0505, 0.1457) & (-0.0465, 0.0030) & (0.2661, 0.2984) \\
\hline            
     \multicolumn{1}{l|}{SP} & 1     & -0.2679  & -0.0016 & 0.0454   & -0.0599 & 0.5529 \\
           &       & (-0.3626, -0.1769) & (-0.0076, 0.0045) & (-0.0529, 0.1439) & (-0.2281, 0.1192) & (0.5131, 0.5954) \\
           & 2     & 0.3775  & -0.0048  & 0.0066 & -0.0617  & 0.2281 \\
           &       & (0.3437, 0.4098) & (-0.0069, -0.0027) & (-0.0314, 0.0453) & (-0.1258, 0.0021) & (0.2164, 0.2395) \\
 \hline        
     \multicolumn{1}{l|}{} & \multicolumn{1}{l|}{} & \multicolumn{1}{c|}{Intercept(${\hat\Phi}_{0k}$)} &       \multicolumn{1}{c}{CSU(${\hat\Phi}_{1k}$)} &       \multicolumn{1}{c}{SPI(${\hat\Phi}_{2k}$)} &      \multicolumn{1}{c|}{r20mm (${\hat\Phi}_{3k}$)} &       \multicolumn{1}{c}{${\hat\tau}_{k}$}  \\
 \hline 
     \multicolumn{1}{l|}{FIN1} & 1     & -0.0025  & 0.0025   & 0.0443  & 0.0124  & 0.0892 \\
           &       & (-0.0194, 0.0144) & (0.0004, 0.0046) & (0.0228, 0.0656) & (-0.0211, 0.0474) & (0.0842, 0.0945) \\
           & 2     & 0.3548  & -0.0018 & 0.0000   & -0.0238  & 0.1238 \\
           &       & (0.3333, 0.3765) & (-0.0045, 0.0011) & (-0.0358, 0.0357) & (-0.0723, 0.0241) & (0.1167, 0.1313) \\
\hline            
     \end{tabular}%
   \end{scriptsize}   
\caption{Posterior means and 95\% credible intervals (in brackets) of the parameter of measurement equations of our PMS model subject to climate shocks.}\label{tab:WithClimateShock}%
 \end{table}%

\begin{table}[htbp]
\begin{scriptsize}
  \centering
       \begin{tabular}{l|c|cc}
   \hline        
       \multicolumn{1}{l|}{i} & \multicolumn{1}{l|}{Regime (k)} & \multicolumn{1}{c}{Intercept(${\hat\Psi}_{i,0k}$)} &              \multicolumn{1}{c}{${\hat\sigma}_{i,k}$}  \\
   \hline
    \multicolumn{1}{l|}{AU} & 1     & -0.1114  & 0.3330   \\
          &       & (-0.1587, -0.0617) & (0.3127, 0.3555) \\
          & 2     & 0.4799 & 0.3330  \\
          &       & (0.4545, 0.5071) & (0.3127, 0.3555) \\
\hline           
    \multicolumn{1}{l|}{BE} & 1     & -0.1704  & 0.4469  \\
          &       & (-0.2285 , -0.1115) & (0.4171, 0.4786) \\
          & 2     & 0.4646 & 0.4469  \\
          &       & (0.4356, 0.4916) & (0.4171, 0.4786) \\
\hline           
    \multicolumn{1}{l|}{DE} & 1     & -0.1488 & 0.4257  \\
          &       & (-0.1985, -0.0989) & (0.4034, 0.4480) \\
          & 2     & 0.7054 &  0.4257 \\
          &       & (0.6475, 0.7654) & (0.4034, 0.4480) \\
\hline           
    \multicolumn{1}{l|}{FI} & 1     & -0.2696  & 0.6598 \\
          &       & (-0.3555, -0.1877) & (0.6171, 0.7042) \\
          & 2     & 0.4941& 0.6598  \\
          &       & (0.4653, 0.5226) & (0.6171, 0.7042) \\
\hline           
    \multicolumn{1}{l|}{FR} & 1     & -0.0367 & 0.5927  \\
          &       & (-0.3175, 0.0781) & (0.1355, 0.7892) \\
          & 2     & 0.1044 &0.5927 \\
          &       & (0.0690, 0.1768) & (0.1355, 0.7892) \\
\hline           
    \multicolumn{1}{l|}{GE} & 1     & -0.1845  & 0.7257  \\
          &       & (-0.2802, -0.0916) & (0.6787, 0.7761) \\
          & 2     & 0.3639 &   0.7257  \\
          &       & (0.3426, 0.3849) & (0.6787, 0.7761) \\
\hline           
    \multicolumn{1}{l|}{GR} & 1     & -0.2881  & 0.2913  \\
          &       & (-0.3372, -0.2419) & (0.2744, 0.3095) \\
          & 2     & 0.3306 &    0.2913  \\
          &       & (0.2830, 0.3765) & (0.2744, 0.3095) \\
\hline           
    \multicolumn{1}{l|}{IR} & 1     & 0.1767 &  0.3849  \\
          &       & (0.1322, 0.2209) & (0.3633, 0.4068) \\
          & 2     & 1.4419 &    0.3849  \\
          &       & (1.3397, 1.5485) & (0.3633, 0.4068) \\
\hline           
    \multicolumn{1}{l|}{IT} & 1     & -0.4006 & 0.5812 \\
          &       & (-0.4779, -0.325)0 & (0.5429, 0.6233) \\
          & 2     & 0.2924     & 0.5812   \\
          &       & (0.2670, 0.3175) & (0.5429, 0.6233) \\
\hline           
    \multicolumn{1}{l|}{LU} & 1     & 0.1249  & 1.1412  \\
          &       & (-0.1642, 0.2827) & (0.3001, 1.4932) \\
          & 2     & 0.3376  & 1.1412   \\
          &       & (0.2553, 0.5187) & (0.3001, 1.4932) \\
\hline           
    \multicolumn{1}{l|}{NE} & 1     & -0.1674  & 0.2703  \\
          &       & (-0.2061, -0.1310) & (0.2544, 0.2862) \\
          & 2     & 0.4373 &    0.2703  \\
          &       & (0.4037, 0.4692) & (0.2544, 0.2862) \\
\hline           
    \multicolumn{1}{l|}{PO} & 1     & -0.2693   & 0.3676  \\
          &       & (-0.3258, -0.2184) & (0.3453, 0.3912) \\
          & 2     & 0.4289 &     0.3676  \\
          &       & (0.3926, 0.4624) & (0.3453, 0.3912) \\
\hline           
    \multicolumn{1}{l|}{SP} & 1     & -0.5907 & 0.6147\\
          &       & (-0.7305, -0.4470) & (0.5587, 0.6762) \\
          & 2     & 0.2366 &     0.6147  \\
          &       & (0.2095, 0.2651) & (0.5587, 0.6762) \\
\hline        
       \multicolumn{1}{l|}{i} & \multicolumn{1}{l|}{Regime (k)} & \multicolumn{1}{c}{Intercept(${\hat\Psi}_{f,0k}$)} &              \multicolumn{1}{c}{${\hat\sigma}_{f,k}$}  \\
   \hline
    \multicolumn{1}{l|}{FIN1} & 1     & 0.0162 & 0.0891  \\
          &       & (0.0061, 0.0261) & (0.0843, 0.0942) \\
          & 2     & 0.3418 & 0.0891   \\
          &       & (0.3271, 0.3557) & (0.0843, 0.0942) \\
\hline           
    \end{tabular}%
  \end{scriptsize}
\caption{Posterior means and 95\% credible intervals (in brackets) of the parameter of measurement equations of our PMS model in the absence of climate shocks.}\label{tab:withoutCL}%
\end{table}%

 \begin{table}[htbp]    
 \begin{scriptsize}
   \centering   
     \begin{tabular}{l|c|c|ccc|c}
 \hline        
     \multicolumn{1}{l|}{i} & \multicolumn{1}{l|}{Regime (k)} & \multicolumn{1}{c|}{Intercept(${\hat\Psi}_{i,0k}$)} &       \multicolumn{1}{c}{CSU(${\hat\Psi}_{i,1k}$)} &       \multicolumn{1}{c}{SPI(${\hat\Psi}_{i,2k}$)} &      \multicolumn{1}{c|}{r20mm (${\hat\Phi}_{i,3k}$)} &       \multicolumn{1}{c}{${\hat\sigma}_{i,k}$}  \\
 \hline           
     \multicolumn{1}{l|}{AU} & 1   & -0.1383 &  \textbf{0.0173} & 0.0345 &  0.0021 & 0.4444 \\
           &       & (-0.1994, -0.0762) & (0.0003, 0.0337) & (-0.0462, 0.1149) & (-0.0535, 0.0581) & (0.4203, 0.4704) \\
           & 2     & 0.6370  & \textbf{0.0130}   & \textbf{-0.0594}  & -0.0095 & 0.2595  \\
           &       & (0.6044, 0.6700)& (0.0016, 0.0246) & (-0.1108, -0.0049) & (-0.0448, 0.0256) & (0.2445, 0.2747) \\
 \hline                     
     \multicolumn{1}{l|}{BE} & 1   & -0.1513 & \textbf{0.0184} & 0.0795 & -0.0319 & 0.4968  \\
           &       & (-0.2040, -0.0995) & (0.0002, 0.0364) & (-0.0067, 0.1692) & (-0.1145, 0.0466) & (0.4696, 0.5274) \\
           & 2     & 0.6099  & \textbf{-0.0213} & \textbf{-0.0699}  & 0.0244  & 0.2770 \\
           &       & (0.5781, 0.6422) &( -0.0319, -0.0107) & (-0.1368, -0.0030) & (-0.0341, 0.0830) & (0.2610, 0.2943) \\
 \hline                     
     \multicolumn{1}{l|}{DE} & 1     & -0.1961 & \textbf{0.0397}  & \textbf{-0.1511} &        -0.0131& 0.5095  \\
           &       & (-0.2471, -0.1446) & (0.0075, 0.0740) & (-0.2555, -0.0495) & (-0.1594, 0.1366) & (0.4812, 0.5390) \\
           & 2     & 0.7071  & -0.0292   & \textbf{-0.1161}  & -0.0473 & 0.4662   \\
           &       & (0.6588, 0.7550) & (-0.0607, 0.0007) & (-0.1996, -0.0321) & (-0.1922, 0.0921) & (0.4400, 0.4918) \\
 \hline                     
     \multicolumn{1}{l|}{FI} & 1     & -0.2824 & \textbf{0.0467} & 0.0118 & 0.2261  & 0.6243 \\
           &       & (-0.3501, -0.2145) & (0.0013, 0.0934) &( -0.1299, 0.1563) & (-0.0478, 0.4998)& (0.5876, 0.6623) \\
           & 2     & 0.6112  & 0.0109  & 0.0554  & -0.0433  & 0.2755  \\
           &       & (0.5839, 0.6376) & (-0.0107, 0.0327) & (-0.0234, 0.1291) & (-0.1658, 0.0777) & (0.2593, 0.2920) \\
 \hline                     
     \multicolumn{1}{l|}{FR} & 1     & -0.2462  & 0.0084   & \textbf{0.1028}   & -0.0226 & 0.4512 \\
           &       & (-0.3146, -0.1784) & (-0.0021, 0.0186) & (0.0113, 0.1950) & (-0.1435, 0.0958) & (0.4187, 0.4844) \\
           & 2     & 0.2832 & -0.0024 & \textbf{-0.0813} & 0.0186  & 0.2034 \\
           &       & (0.2556, 0.3110) & (-0.0065, 0.0020) & (-0.1162, -0.0446) & (-0.0328, 0.0701) & (0.1919, 0.2145) \\
 \hline                     
     \multicolumn{1}{l|}{GE} & 1     & -0.2362 & 0.0067  & -0.1136  & 0.1822   & 0.6629  \\
           &       & (-0.3181, -0.1576) & (-0.0139, 0.0276) & (-0.2389, 0.0157) & (0.0129, 0.3466) & (0.6254, 0.7018) \\
           & 2     & 0.4190  & 0.0055 & -0.0182 & 0.0082   & 0.1431 \\
           &       & (0.3946, 0.4432) &( 0.0000, 0.0111) & (-0.0529, 0.0164) & (-0.044, 0.0597) & (0.1343, 0.1524) \\
 \hline                     
     \multicolumn{1}{l|}{GR} & 1     & -0.3368  & 0.0030  & -0.0196  & 0.0499  & 0.3050 \\
           &       & (-0.3769, -0.2943) & (0.0000, 0.0060) & (-0.0764, 0.0382) & (-0.0157, 0.1152) & (0.2878, 0.3225) \\
           & 2     & 0.3800 & -0.0026 & \textbf{-0.0969} & 0.0508  & 0.3103 \\
           &       & (0.3357, 0.4243) & (-0.0056, 0.0003) & (-0.1627, -0.0309) & (-0.0275, 0.1298) & (0.2932, 0.3280) \\
 \hline                     
     \multicolumn{1}{l|}{IR} & 1     & 0.1231  & 0.0328  & \textbf{0.1688}  & -0.0060 &  0.4084   \\
           &       & (0.0721, 0.1735) & (-0.0453, 0.1107) & (0.0925, 0.2434) & (-0.0655, 0.0543) & (0.3838, 0.4320) \\
           & 2     & 1.5562  & -0.0572  & -0.1767 & -0.0125   & 0.9821 \\
           &       & (1.4268, 1.6867) & (-0.1855, 0.0740) & (-0.3652, 0.0130) & (-0.1517, 0.1270) & (0.9268, 1.0402) \\
 \hline                     
     \multicolumn{1}{l|}{IT} & 1     & -0.2442 & -0.0061  & 0.0630 & \textbf{-0.1769} &       0.5632  \\
           &       & (-0.3354, -0.1537) & (-0.0129, 0.0013) & (-0.0520, 0.1835) & (-0.2846, -0.0684) & (0.5296, 0.5994) \\
           & 2     & 0.4371& \textbf{-0.0028}  & \textbf{-0.0762} & \textbf{-0.0813}        & 0.2256  \\
           &       & (0.4011, 0.4739) & (-0.0053, -0.0002) & (-0.1160, -0.0365) & (-0.1210, -0.0385) & (0.2140, 0.2377) \\
 \hline                     
     \multicolumn{1}{l|}{LU} & 1     & -0.2916  & 0.0225 & 0.1142  & -0.0278  & 0.7417 \\
           &       & (-0.3800, -0.2016) & (-0.0035, 0.0482) & (-0.0164, 0.2546) & (-0.1323, 0.0755) & (0.6815, 0.8229) \\
           & 2     & 0.7275 & \textbf{0.0208}  & -0.0909  & 0.0288 & 0.4084  \\
           &       & (0.6761, 0.7772) & (0.0084, 0.0338) & (-0.1811, 0.0022) & (-0.0239, 0.0824) & (0.3598, 0.4451) \\
 \hline                     
     \multicolumn{1}{l|}{NE} & 1     & -0.1231& 0.0010  & \textbf{0.0798}  & 0.0551  & 0.3486 \\
           &       & (-0.1619, -0.0854) & (-0.0145, 0.0166) & (0.0133, 0.1463) & (-0.0176, 0.1272) & (0.3058, 0.3742) \\
           & 2     & 0.4528  & \textbf{-0.0080} & -0.0206 & -0.0043  & 0.1921  \\
           &       & (0.4277, 0.4778) & (-0.0154, -0.0008) & (-0.0634, 0.0220) & (-0.0514, 0.0439) & (0.1769, 0.2219) \\
 \hline                     
     \multicolumn{1}{l|}{PO} & 1     & -0.2954  & -0.0013& -0.0564  & \textbf{0.0740} & 0.4084   \\
           &       & (-0.3679, -0.2224) & (-0.0059, 0.0035) & (-0.1281, 0.0175) & (0.0202, 0.1276) & (0.3832, 0.4343) \\
           & 2     & 0.4053  & -0.0007 & \textbf{0.0647}& 0.0132  & 0.2962 \\
           &       & (0.3663, 0.4440) & (-0.0033, 0.0020) & (0.0146, 0.1143) & (-0.0135, 0.0384) & (0.2804, 0.3127) \\
 \hline                     
     \multicolumn{1}{l|}{SP} & 1     & -0.2965  & -0.0044 & 0.0801 & 0.0197 & 0.5555 \\
           &       & (-0.3887, -0.2039) & (-0.0105, 0.0016) & (-0.0226, 0.1802) & (-0.1457, 0.1954) & (0.5151, 0.6016) \\
           & 2     & 0.4704 & \textbf{-0.0057}& \textbf{0.0689}  & \textbf{-0.1331} & 0.2839 \\
           &       & (0.4298, 0.5129) & (-0.0083, -0.0029) & (0.0218, 0.1138) & (-0.2149, -0.0525) & (0.2668, 0.3012) \\
 \hline           
     \multicolumn{1}{l|}{FIN1} & 1 & -0.0009& 0.0012  & \textbf{0.0193}  & 0.0039  & 0.0386   \\
           &       & (-0.0081, 0.0062) & (0.0003, 0.0020) & (0.0102, 0.0284) & (-0.0110, 0.0193) & (0.0365, 0.0408) \\
           & 2     & 0.1542 & -0.0007& -0.0005  & -0.0122  & 0.0536 \\
           &       & (0.1449, 0.1637) & (-0.0019, 0.0005) & (-0.0161, 0.0152) & (-0.0330, 0.0082) & (0.0505, 0.0569) \\
 \hline                     
     \end{tabular}%
 \end{scriptsize}    
 \caption{Posterior means and 95\% credible intervals (in brackets) of the parameter of measurement equations of our PMS model with the growth in the Manufacturing sector as the dependent variable.}
   \label{tab:ManPMS}%
 \end{table}%
\clearpage 
 
 \section{Business cycles and financial cycle Fin1}
 \label{Appendix3}

 \begin{figure}[h!]
 \centering
    \subfigure[None]{                
 \includegraphics[width=0.17\textwidth]{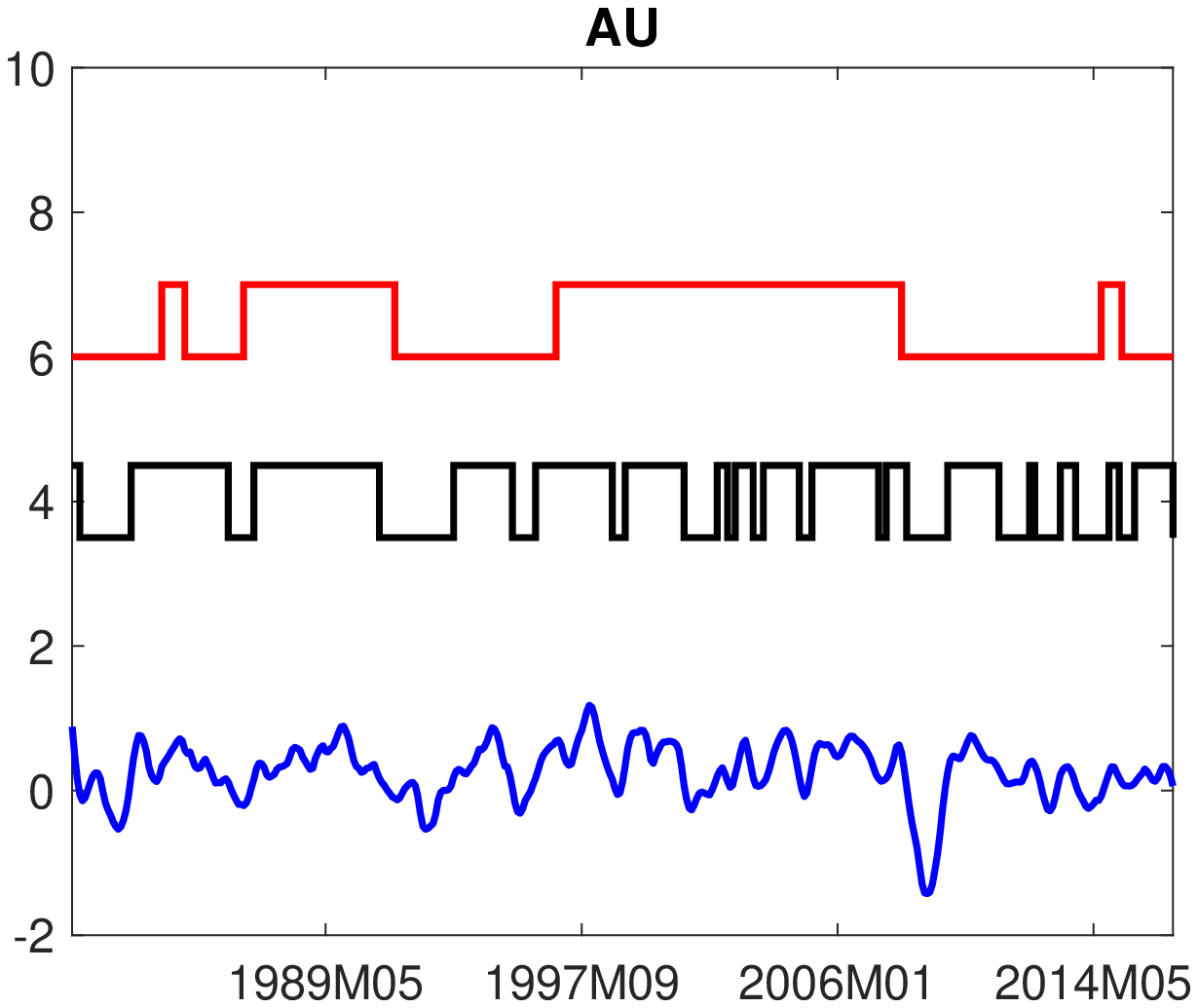}}
  ~
    \subfigure[Temperature only (CSU)]{                
 \includegraphics[width=0.17\textwidth]{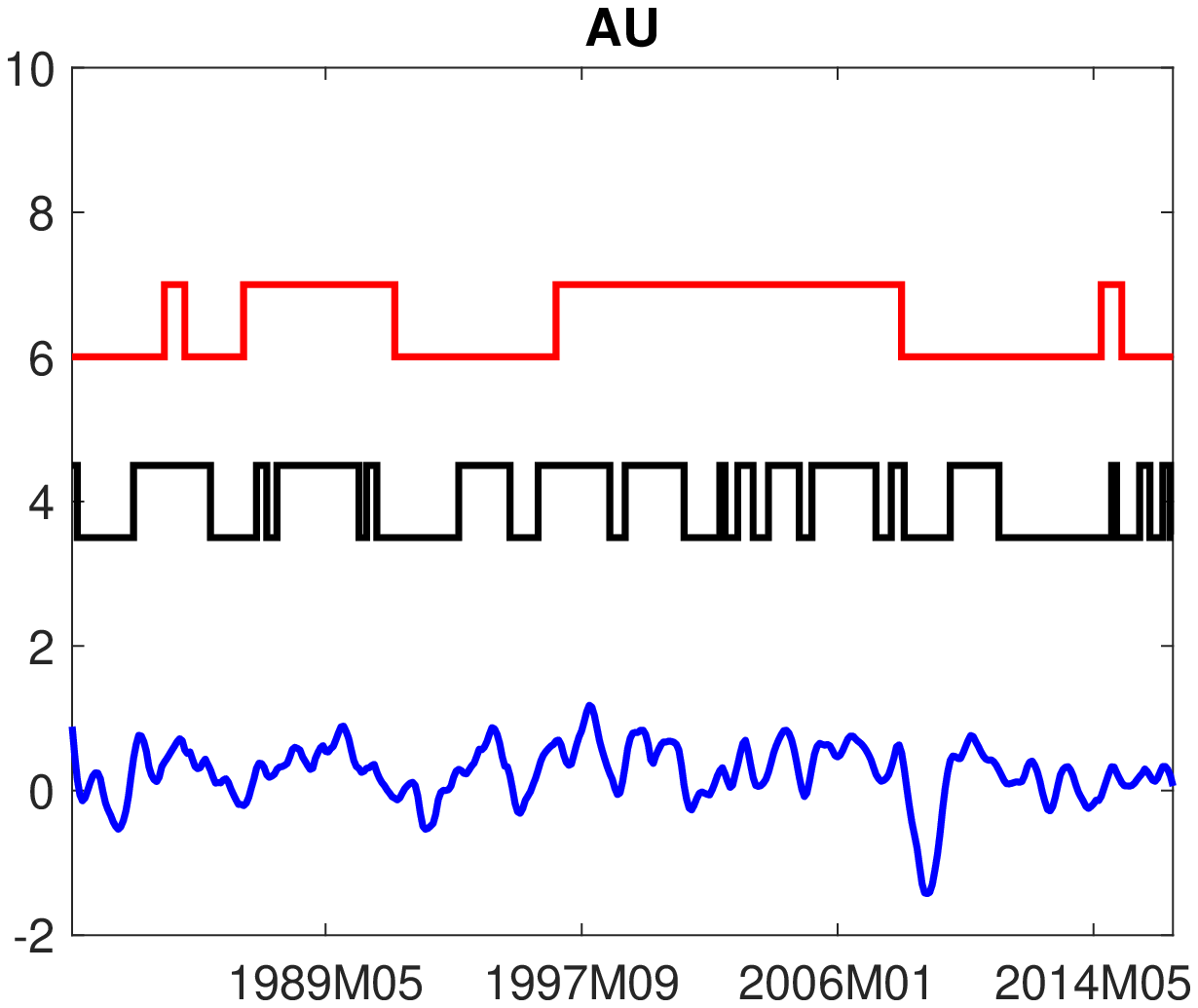}}
 ~  
    \subfigure[Drought only (SPI)]{                
 \includegraphics[width=0.17\textwidth]{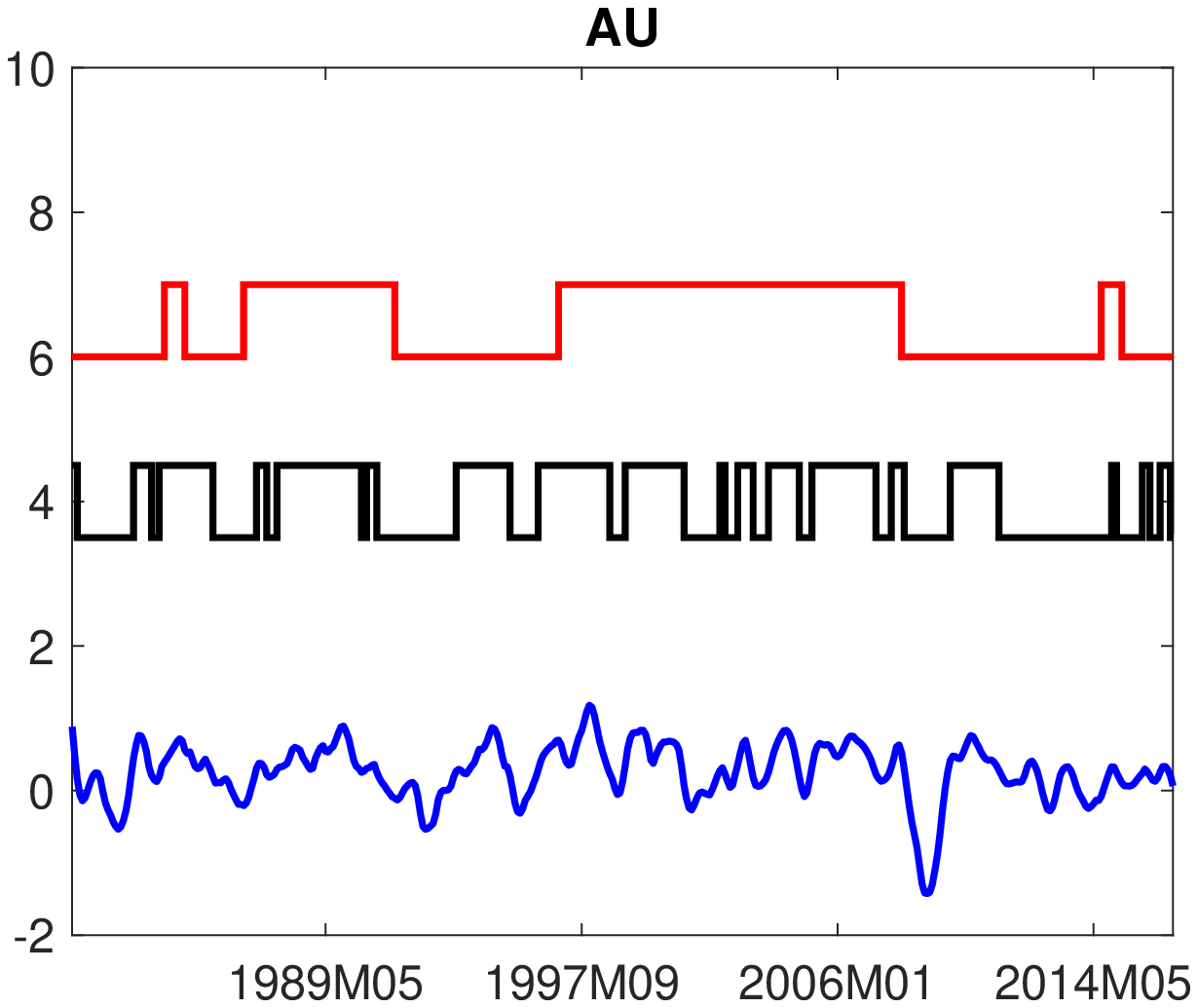}}
 ~ 
    \subfigure[Rainfall only (r20mm)]{                
 \includegraphics[width=0.17\textwidth]{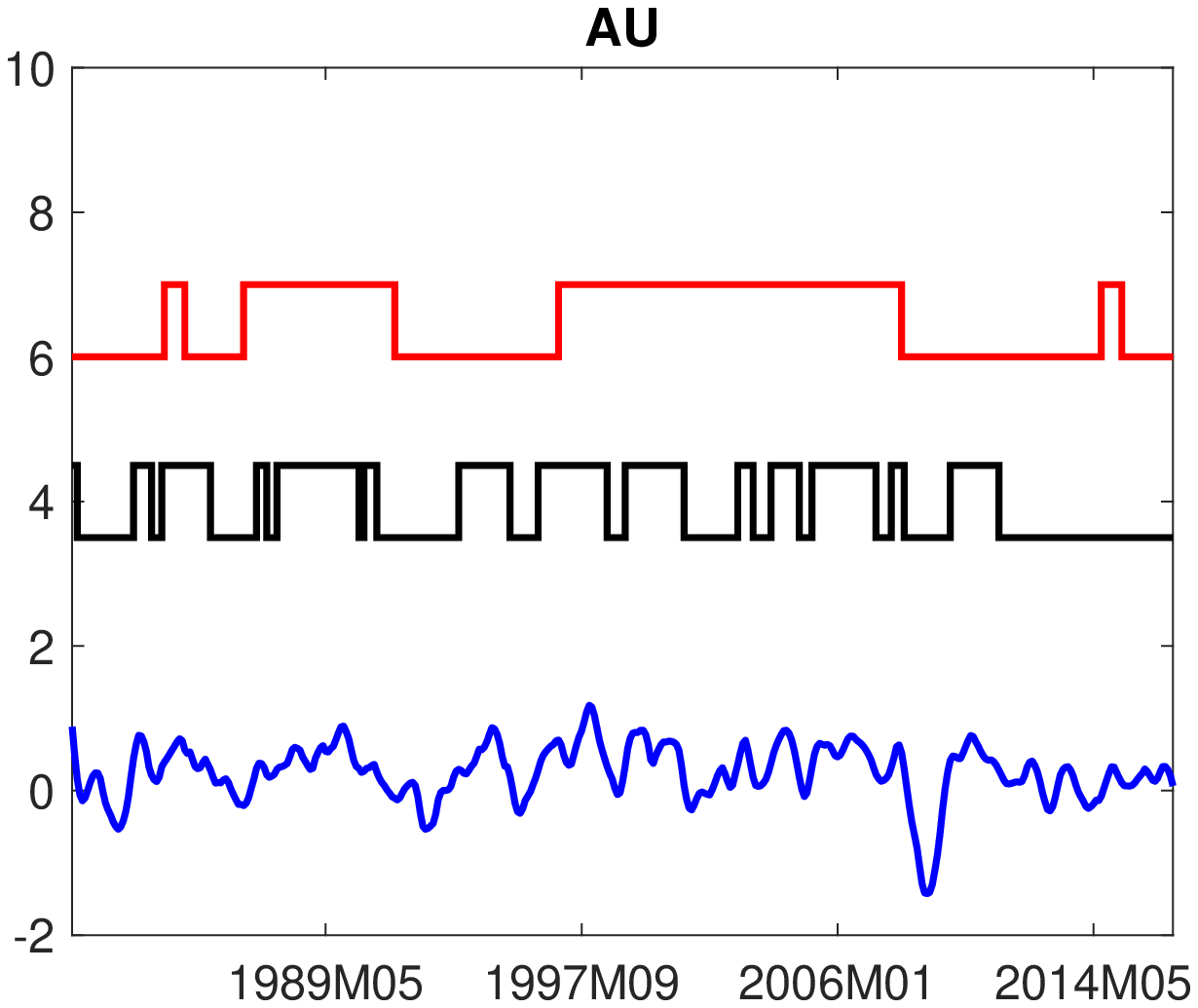}}
 ~ 
    \subfigure[All]{                
 \includegraphics[width=0.17\textwidth]{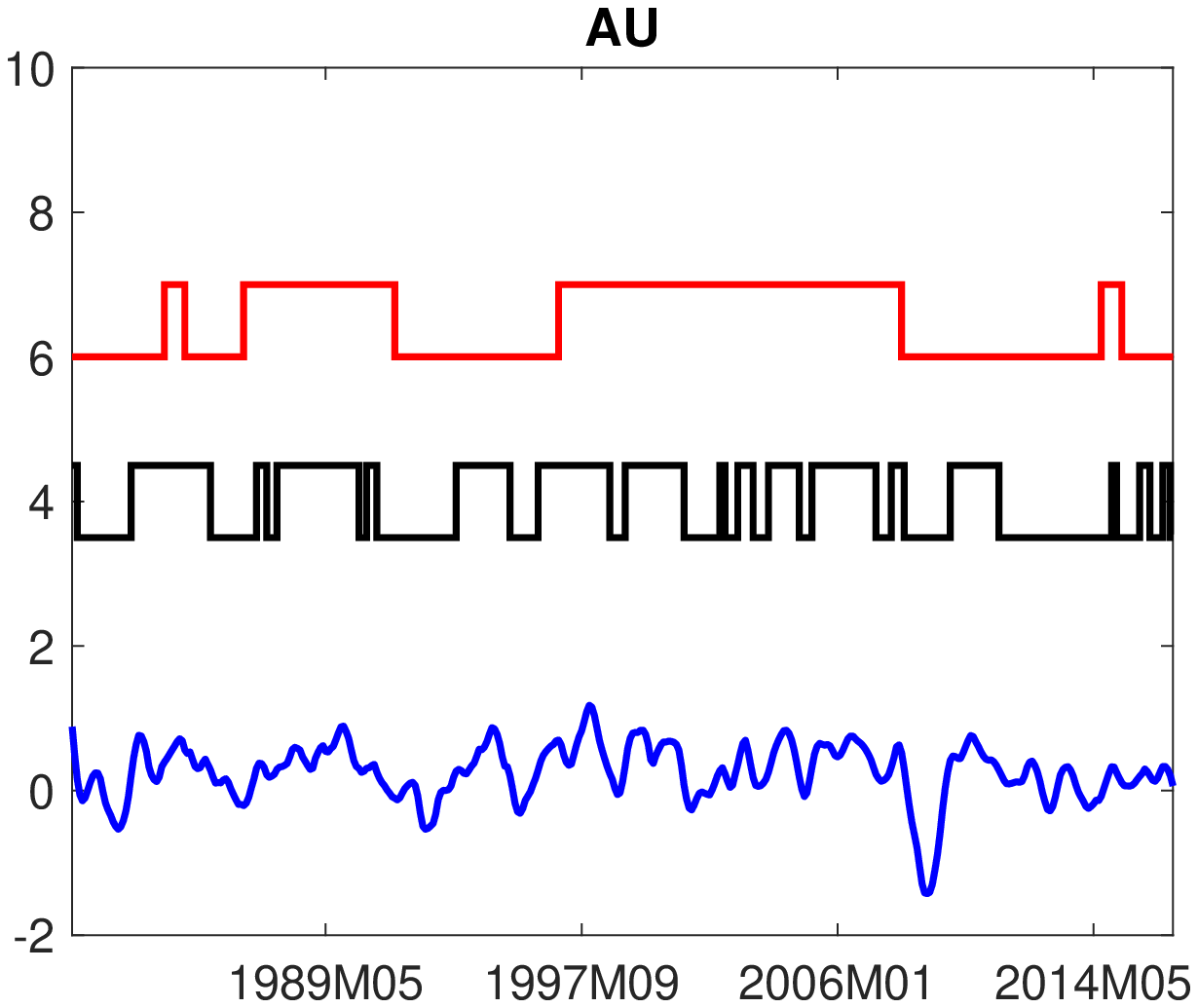}}
 ~ 
    \subfigure{                
 \includegraphics[width=0.17\textwidth]{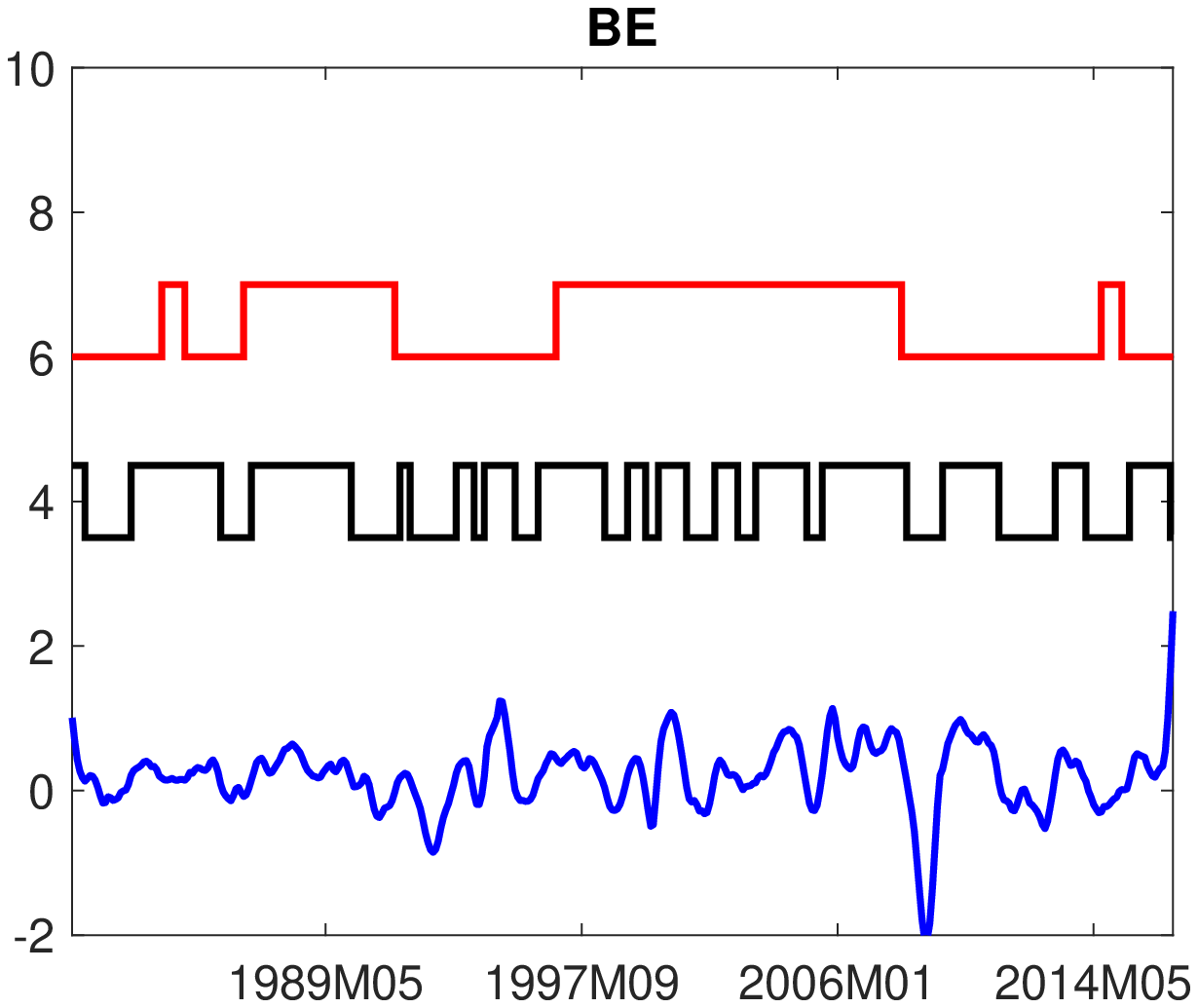}}
  ~
    \subfigure{                
 \includegraphics[width=0.17\textwidth]{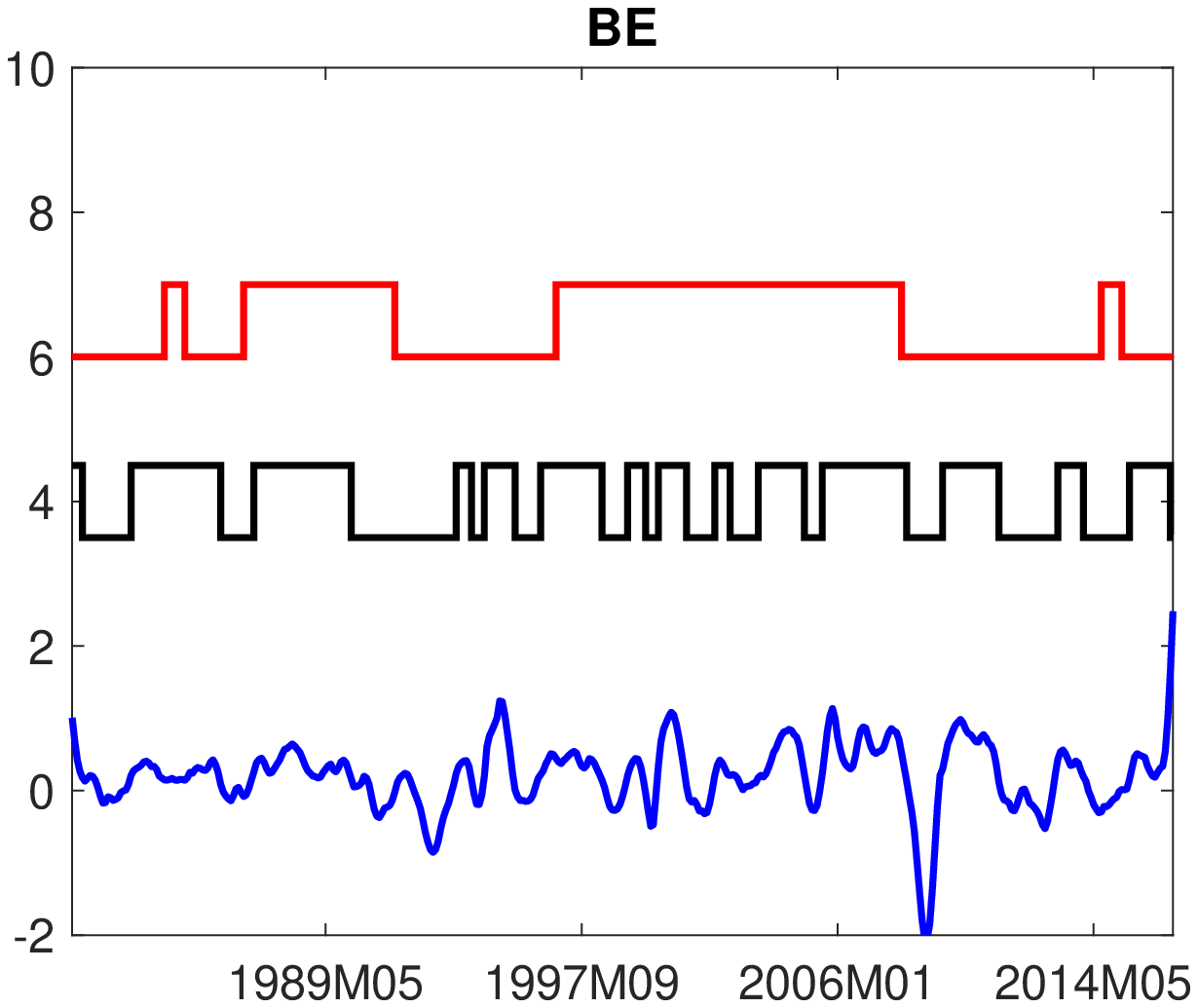}}
 ~  
    \subfigure{                
 \includegraphics[width=0.17\textwidth]{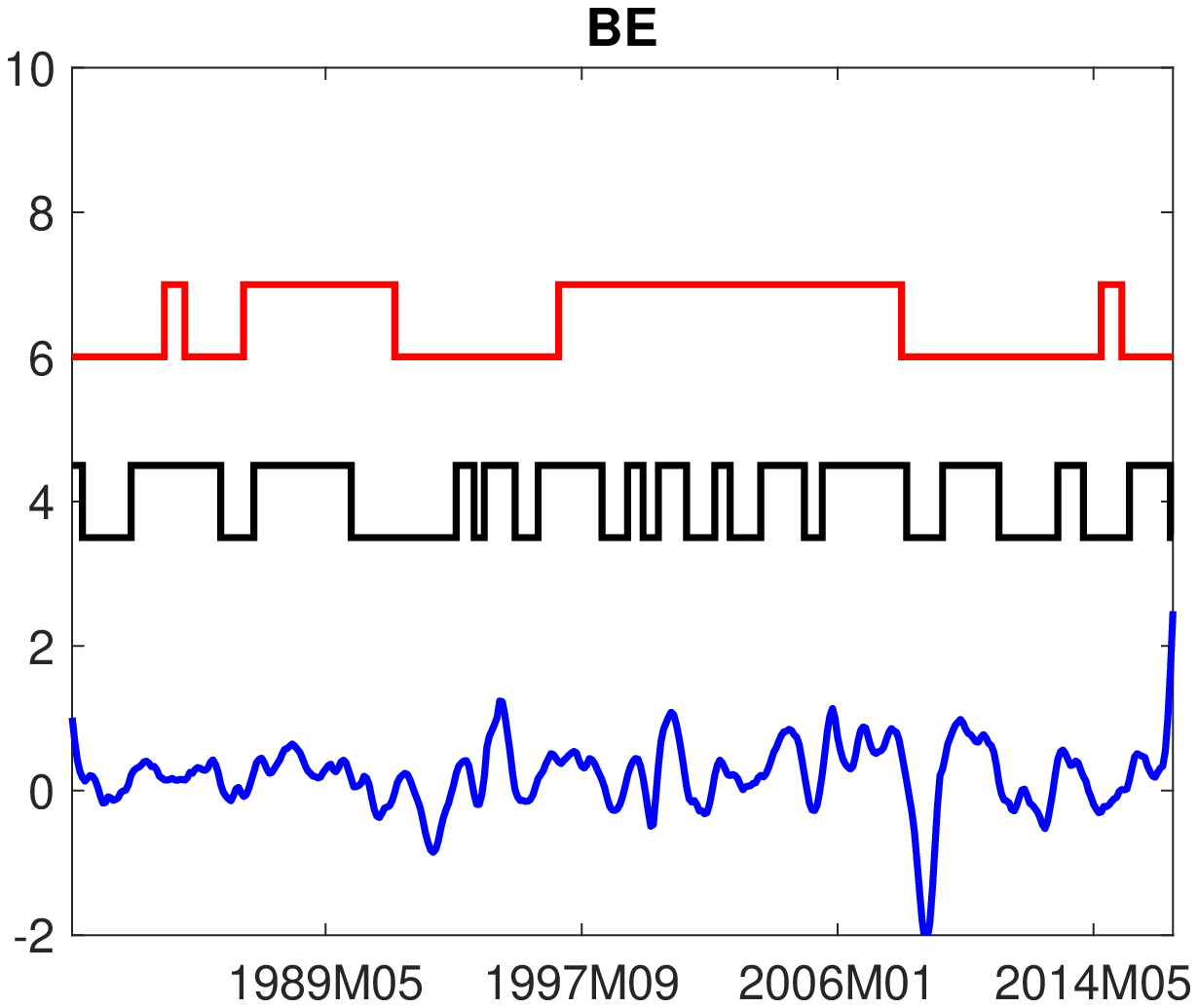}}
 ~ 
    \subfigure{                
 \includegraphics[width=0.17\textwidth]{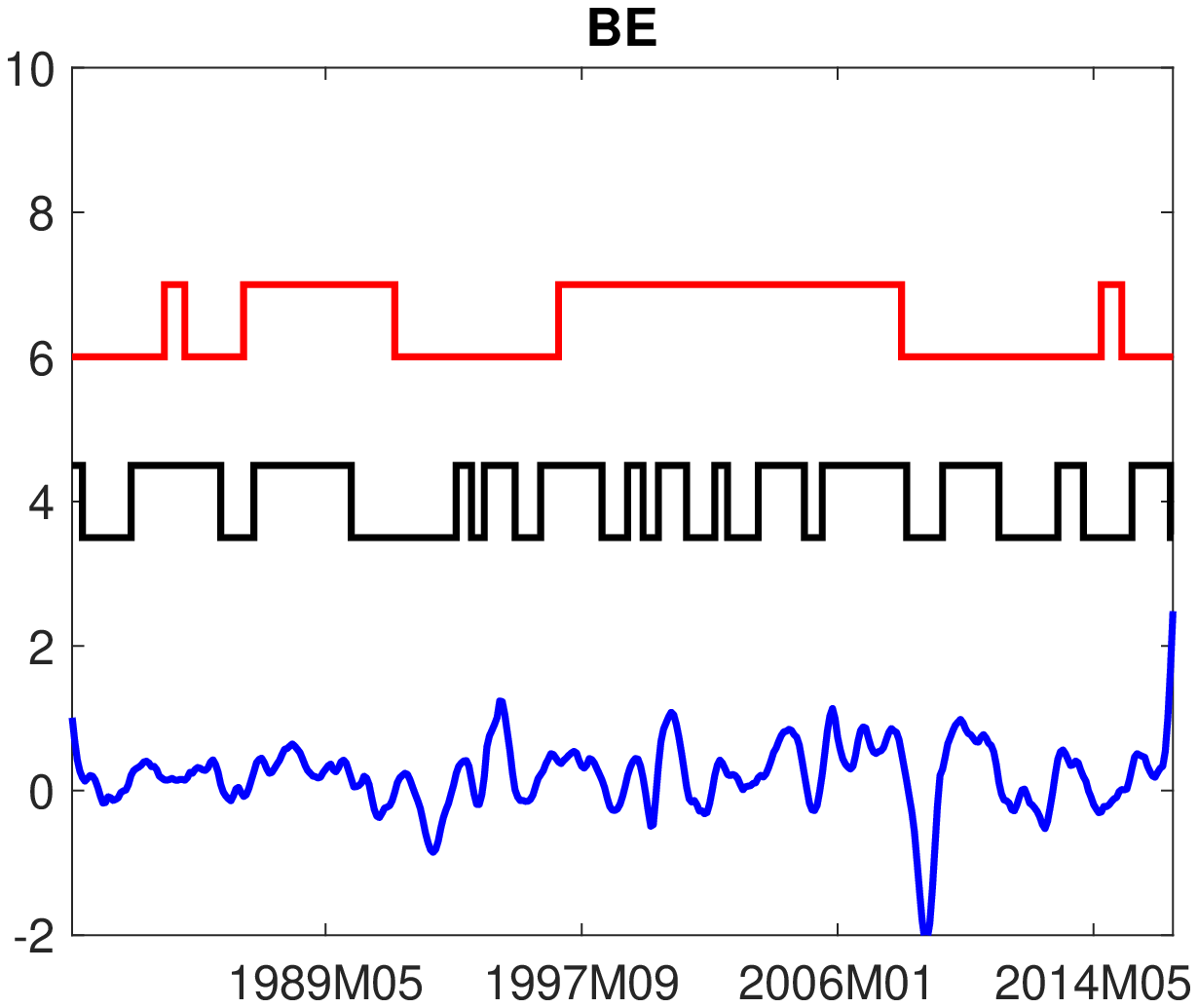}}
 ~ 
    \subfigure{                
 \includegraphics[width=0.17\textwidth]{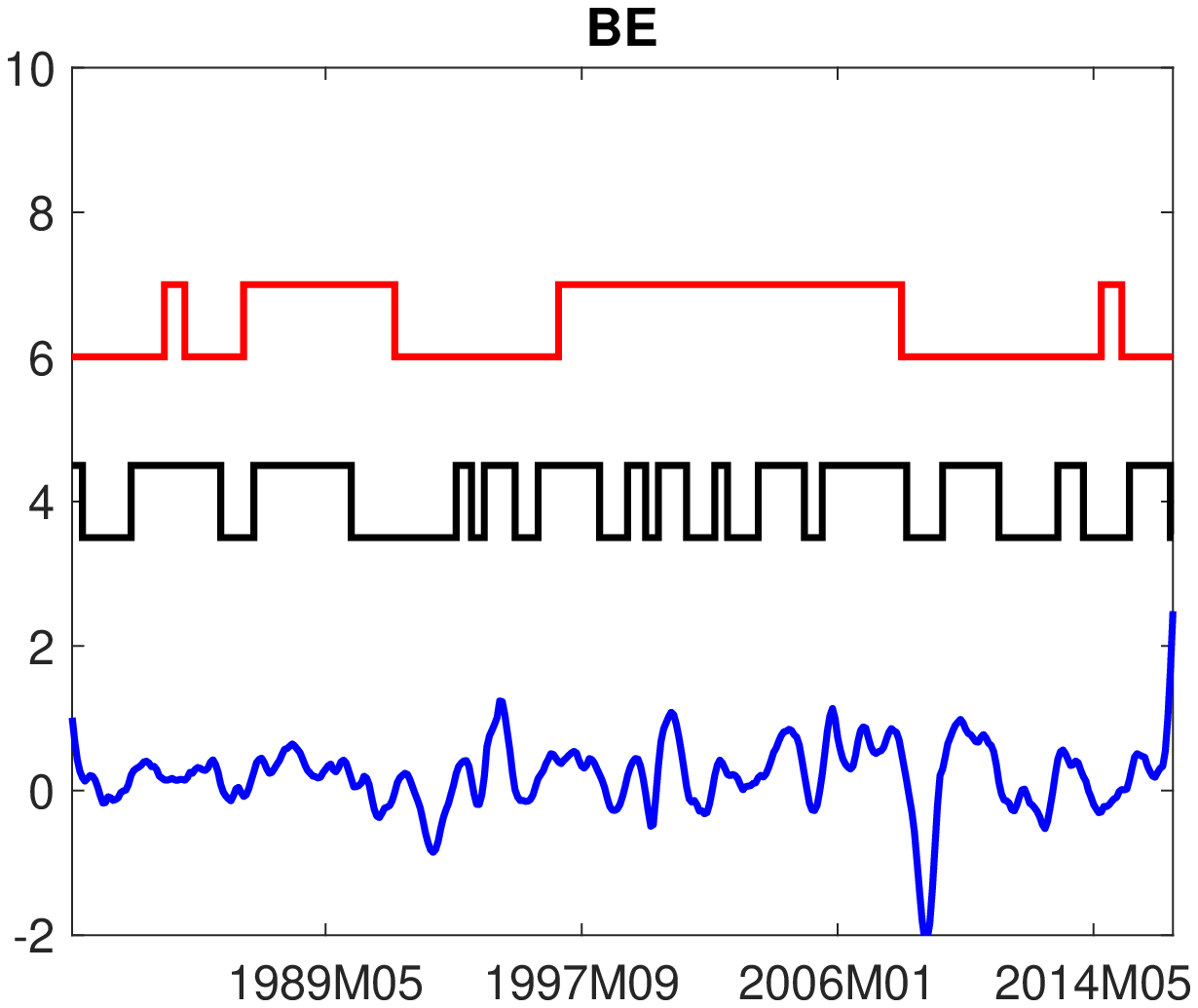}}
 ~ 
    \subfigure{                
 \includegraphics[width=0.17\textwidth]{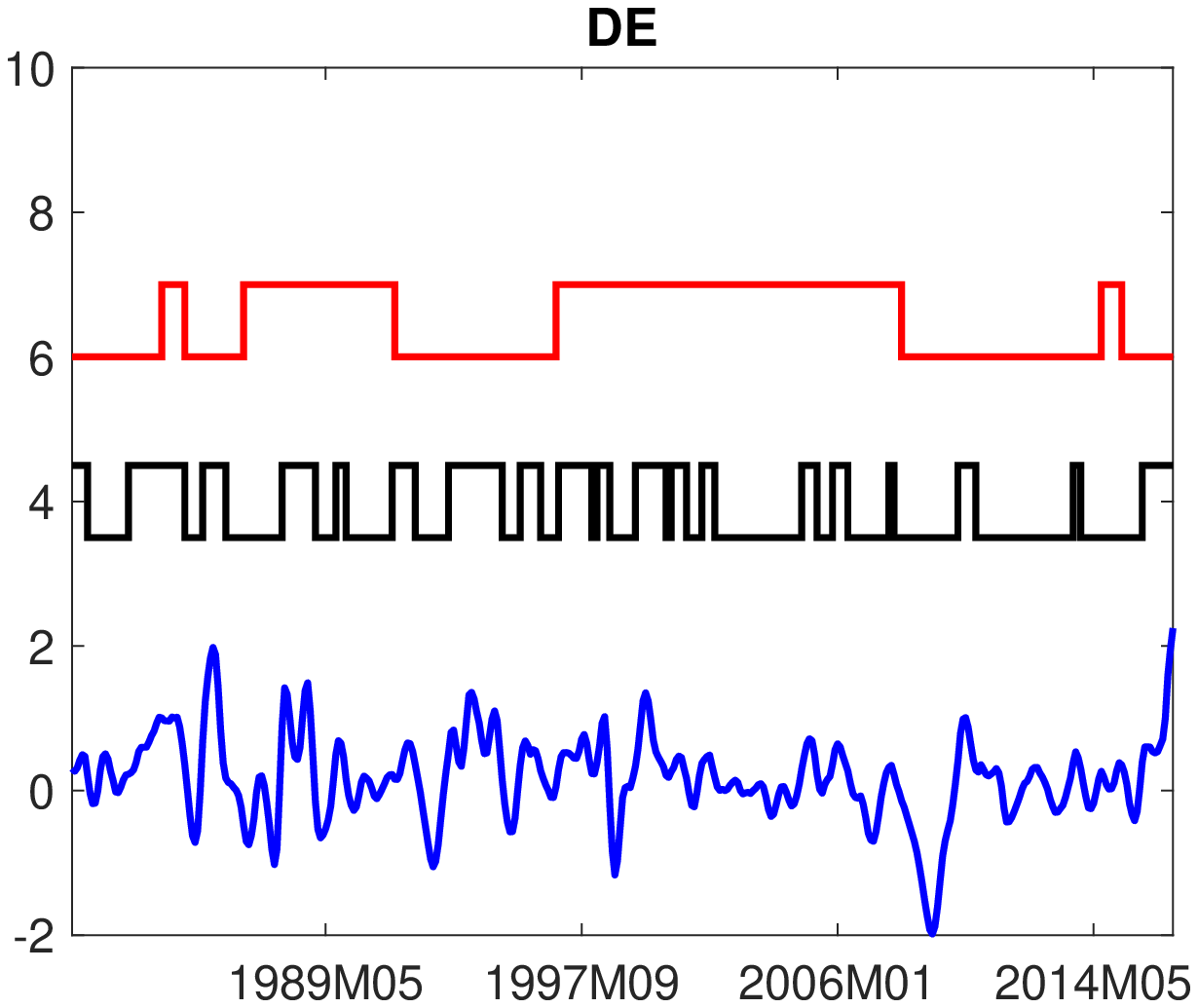}}
  ~
    \subfigure{                
 \includegraphics[width=0.17\textwidth]{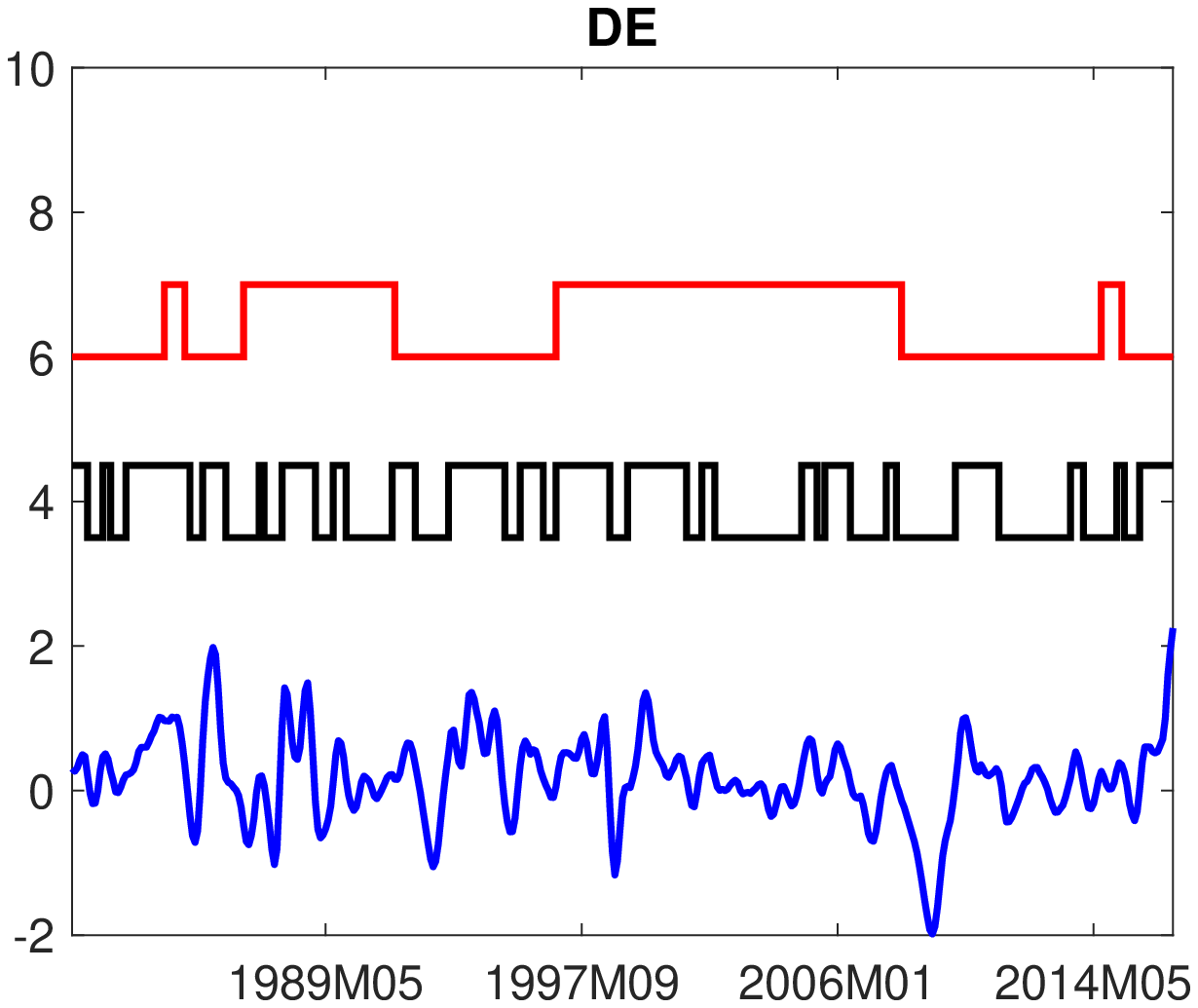}}
 ~  
    \subfigure{                
 \includegraphics[width=0.17\textwidth]{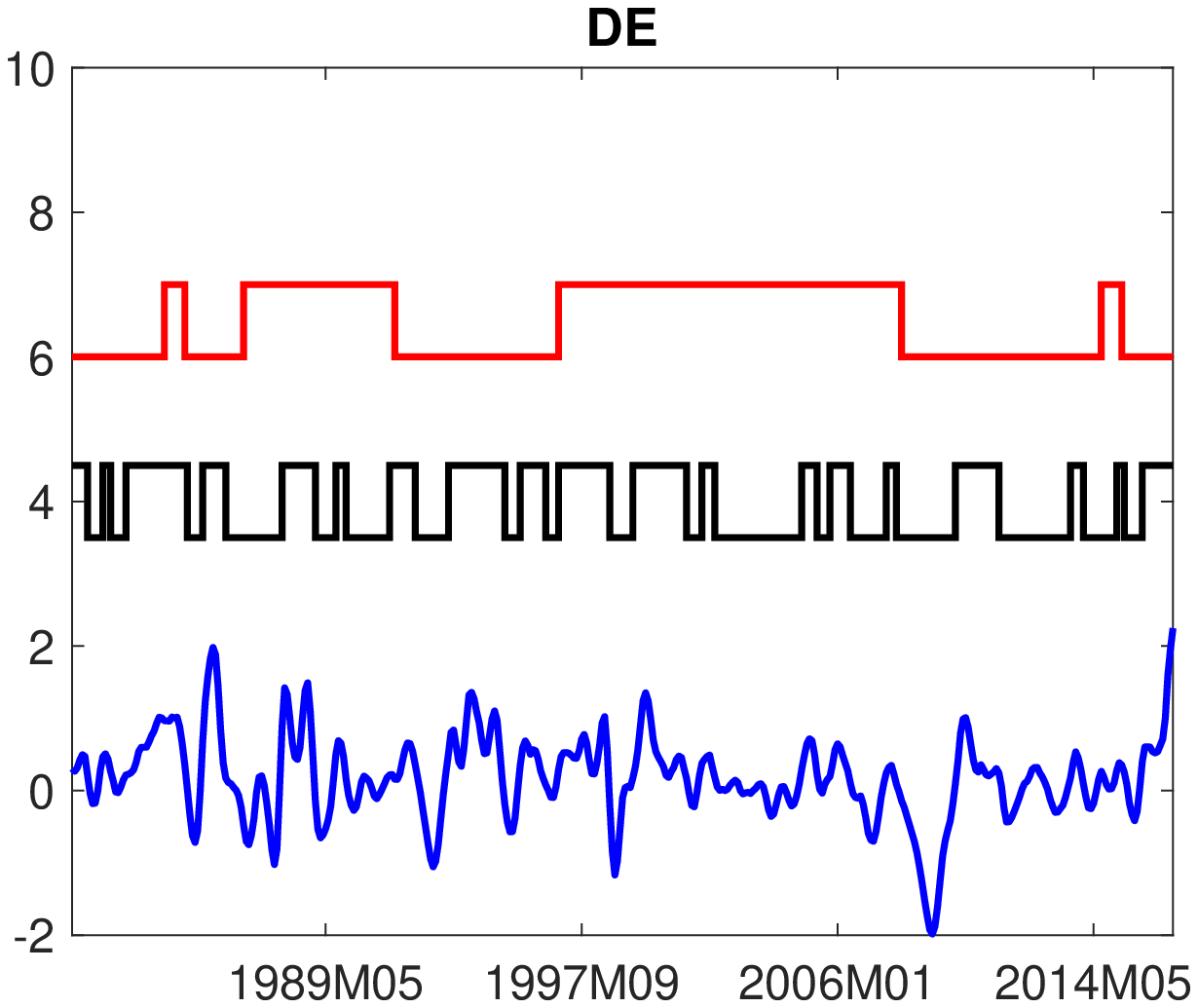}}
 ~ 
    \subfigure{                
 \includegraphics[width=0.17\textwidth]{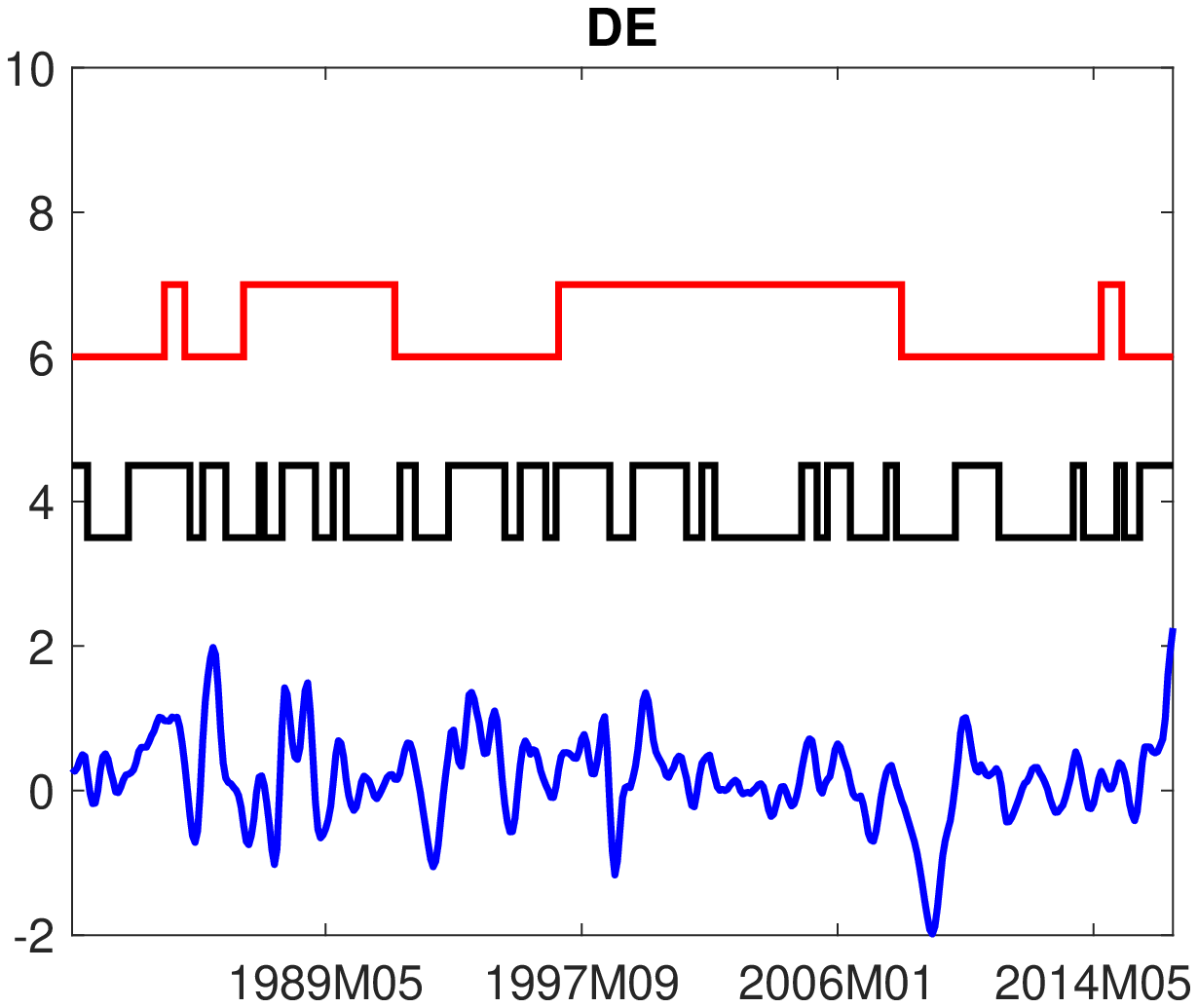}}
 ~ 
    \subfigure{                
 \includegraphics[width=0.17\textwidth]{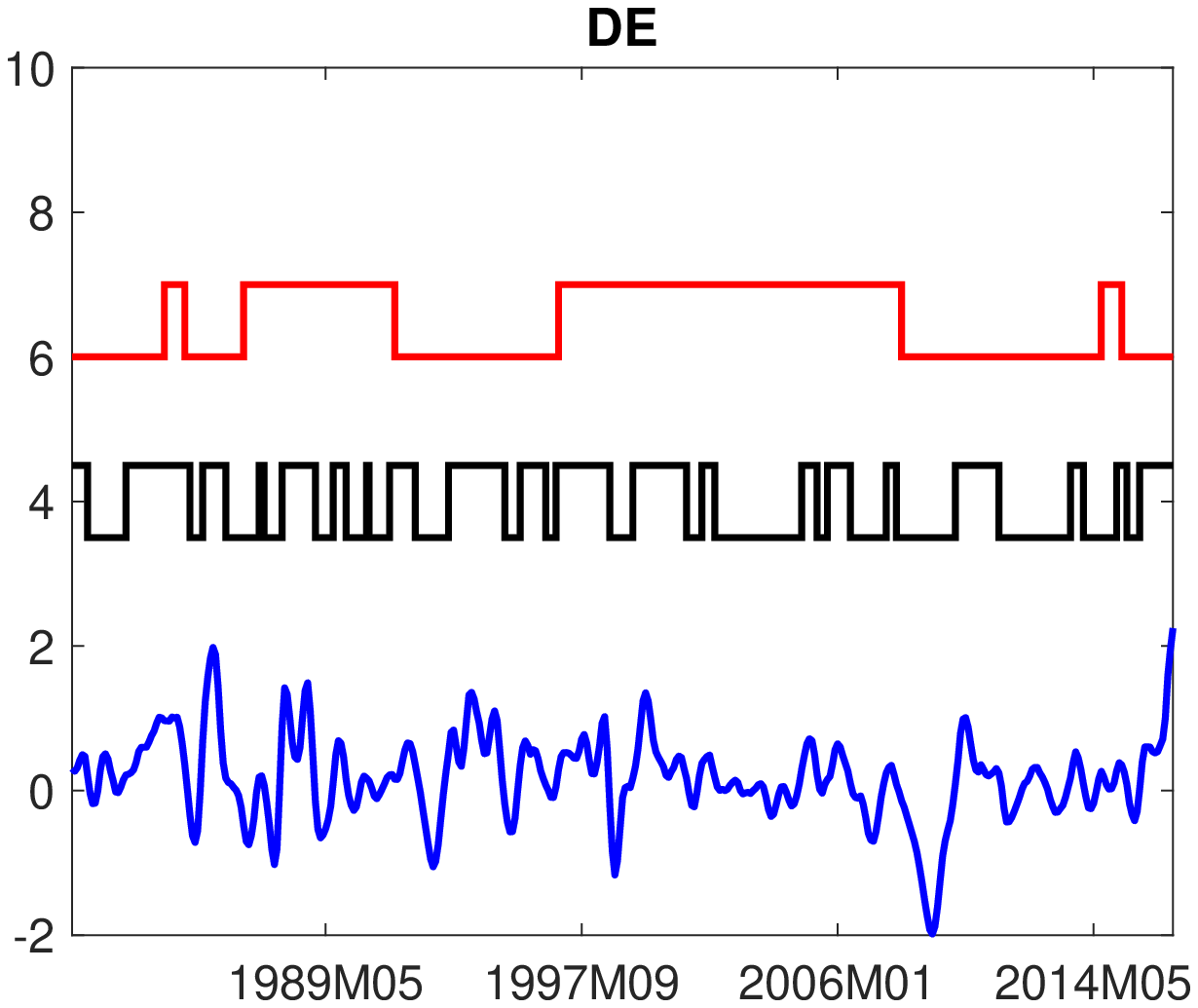}}
 ~ 
    \subfigure{                
 \includegraphics[width=0.17\textwidth]{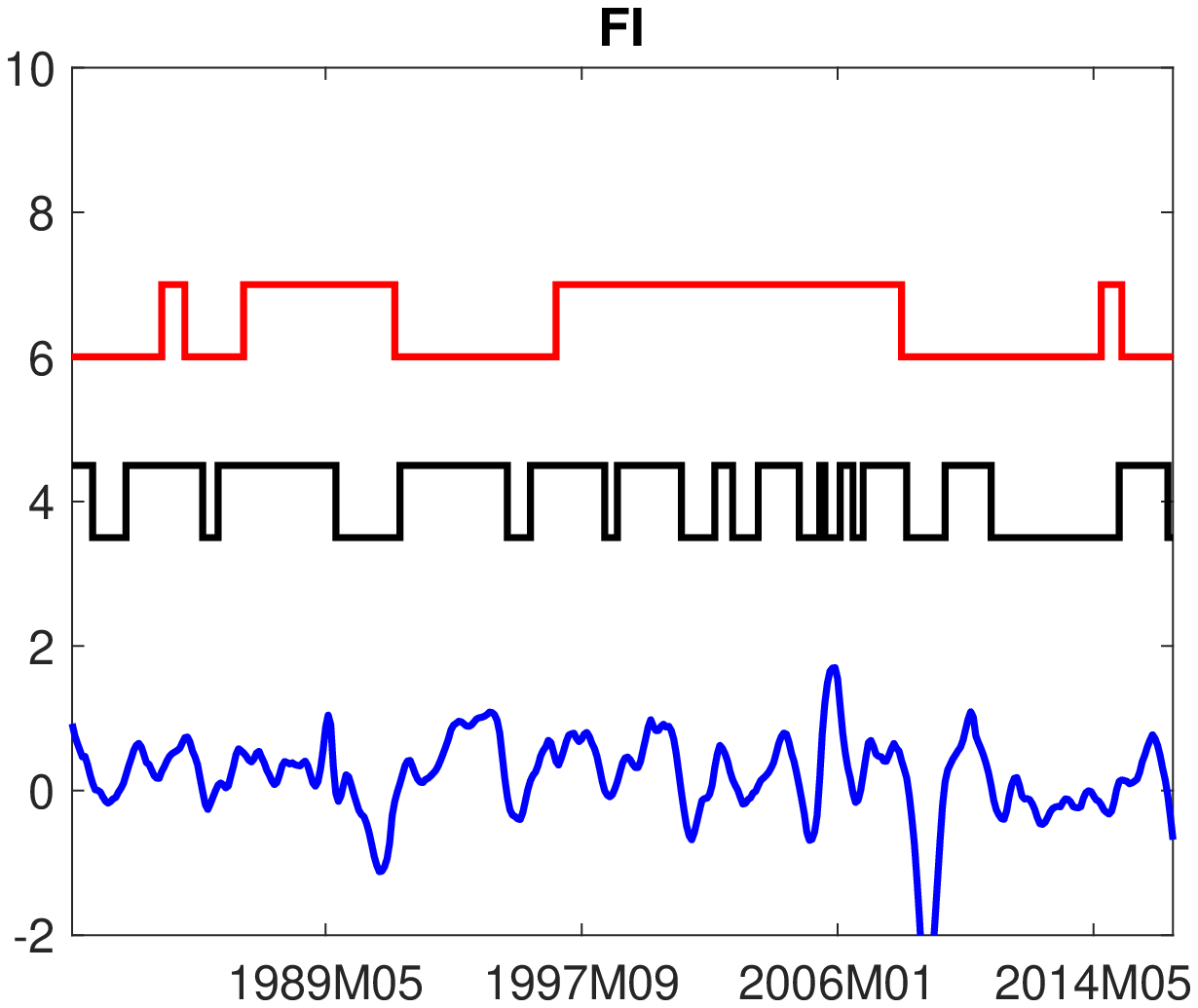}}
  ~
    \subfigure{                
 \includegraphics[width=0.17\textwidth]{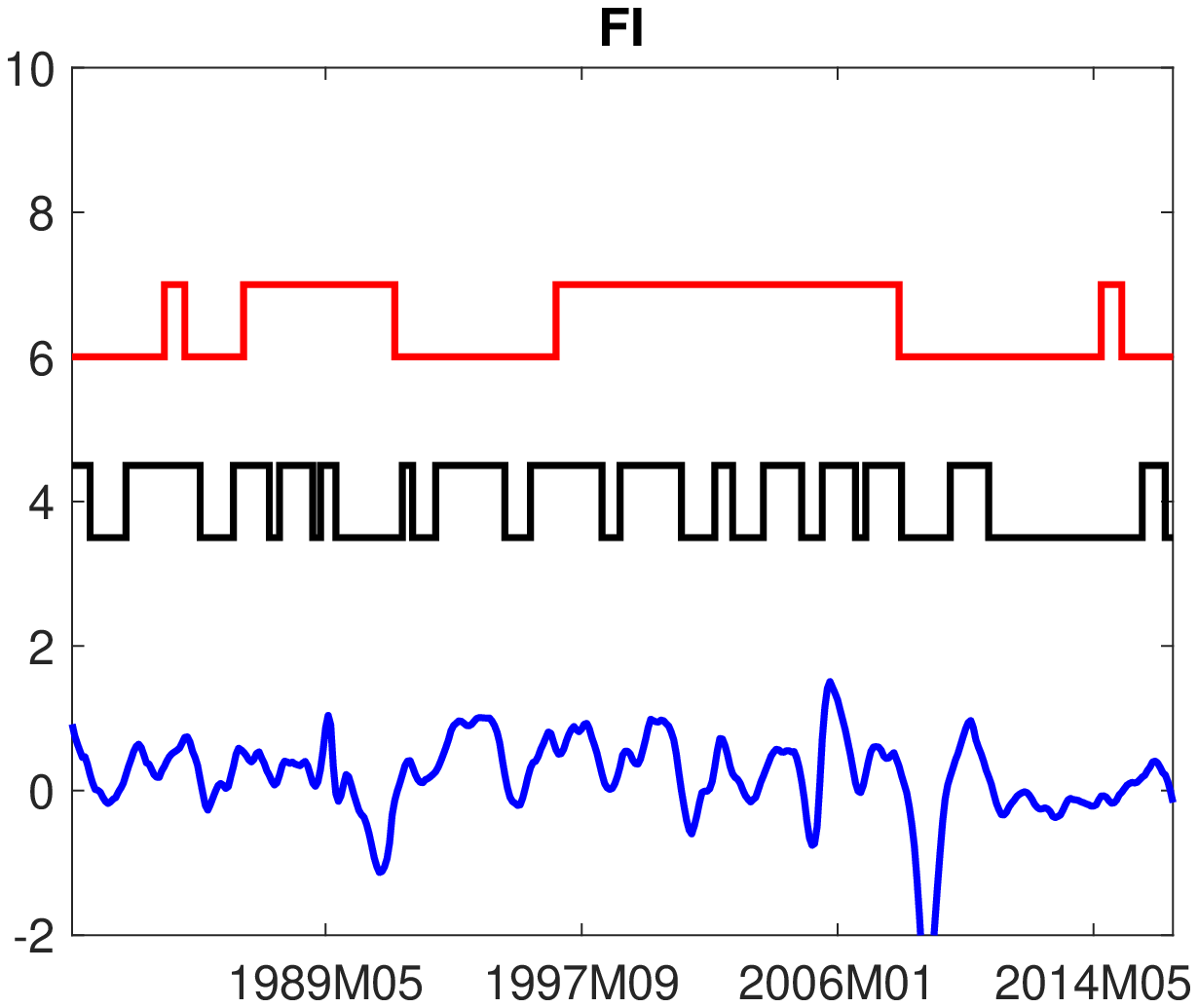}}
 ~  
    \subfigure{                
 \includegraphics[width=0.17\textwidth]{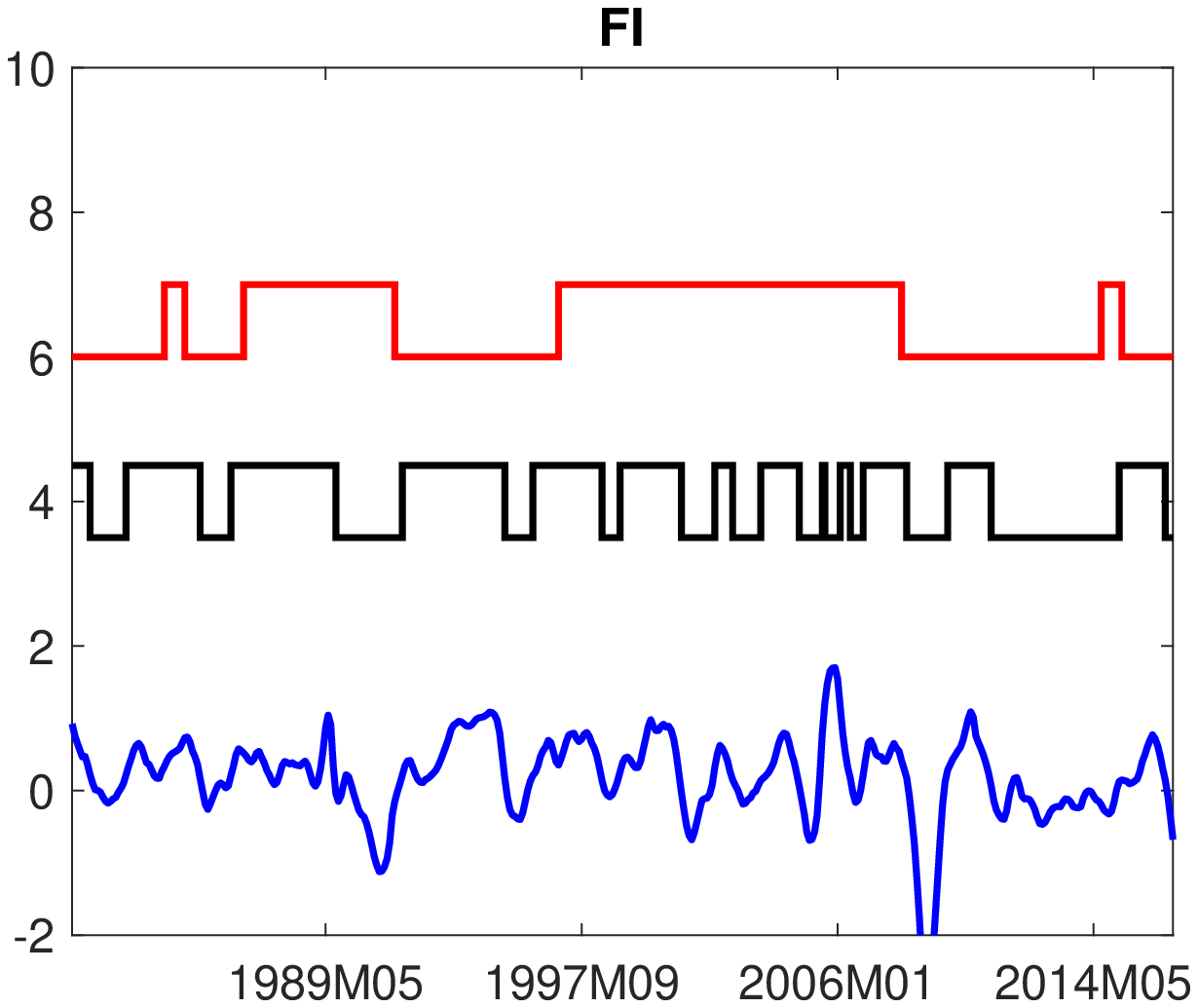}}
 ~ 
    \subfigure{                
 \includegraphics[width=0.17\textwidth]{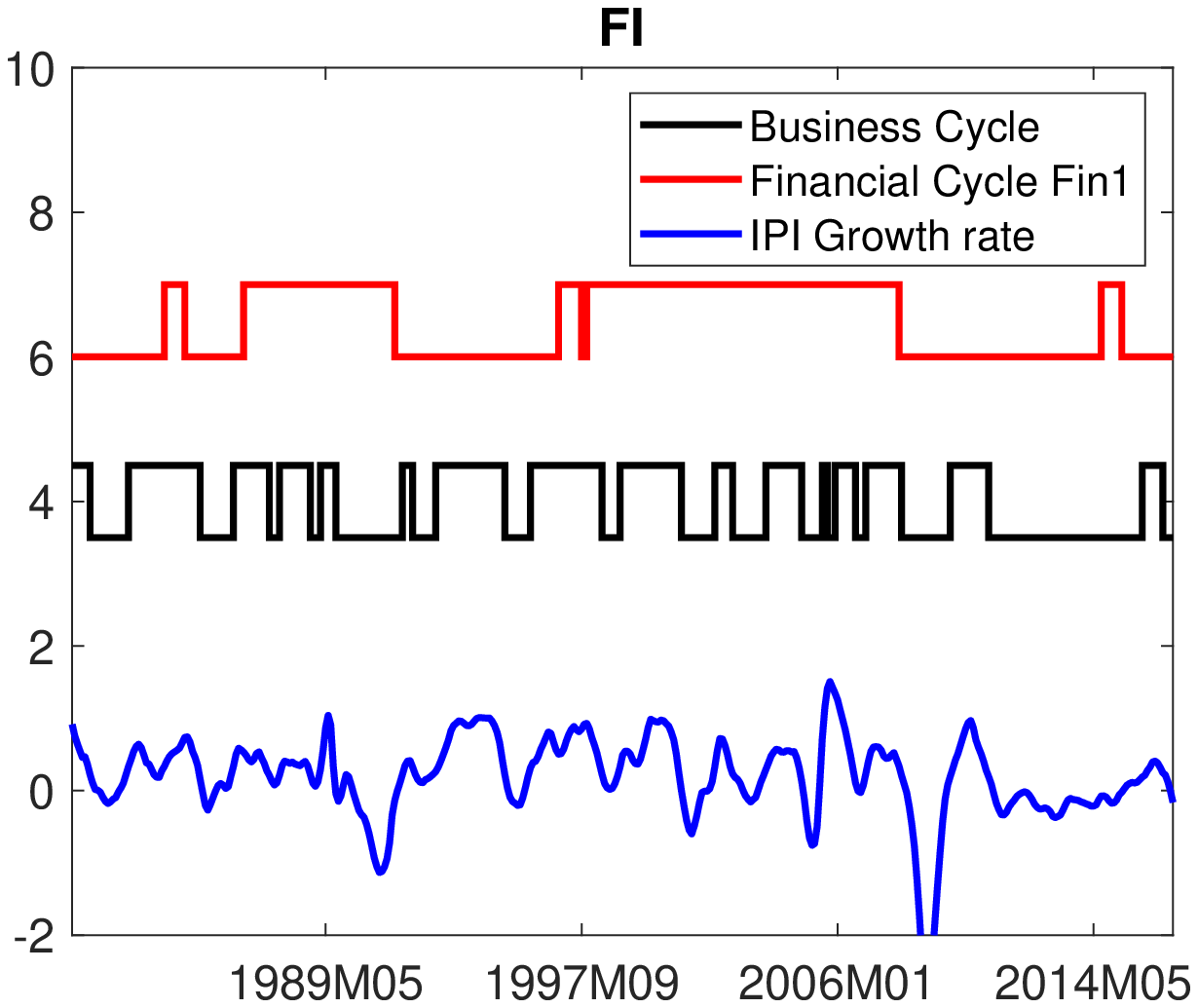}}
 ~ 
    \subfigure{                
 \includegraphics[width=0.17\textwidth]{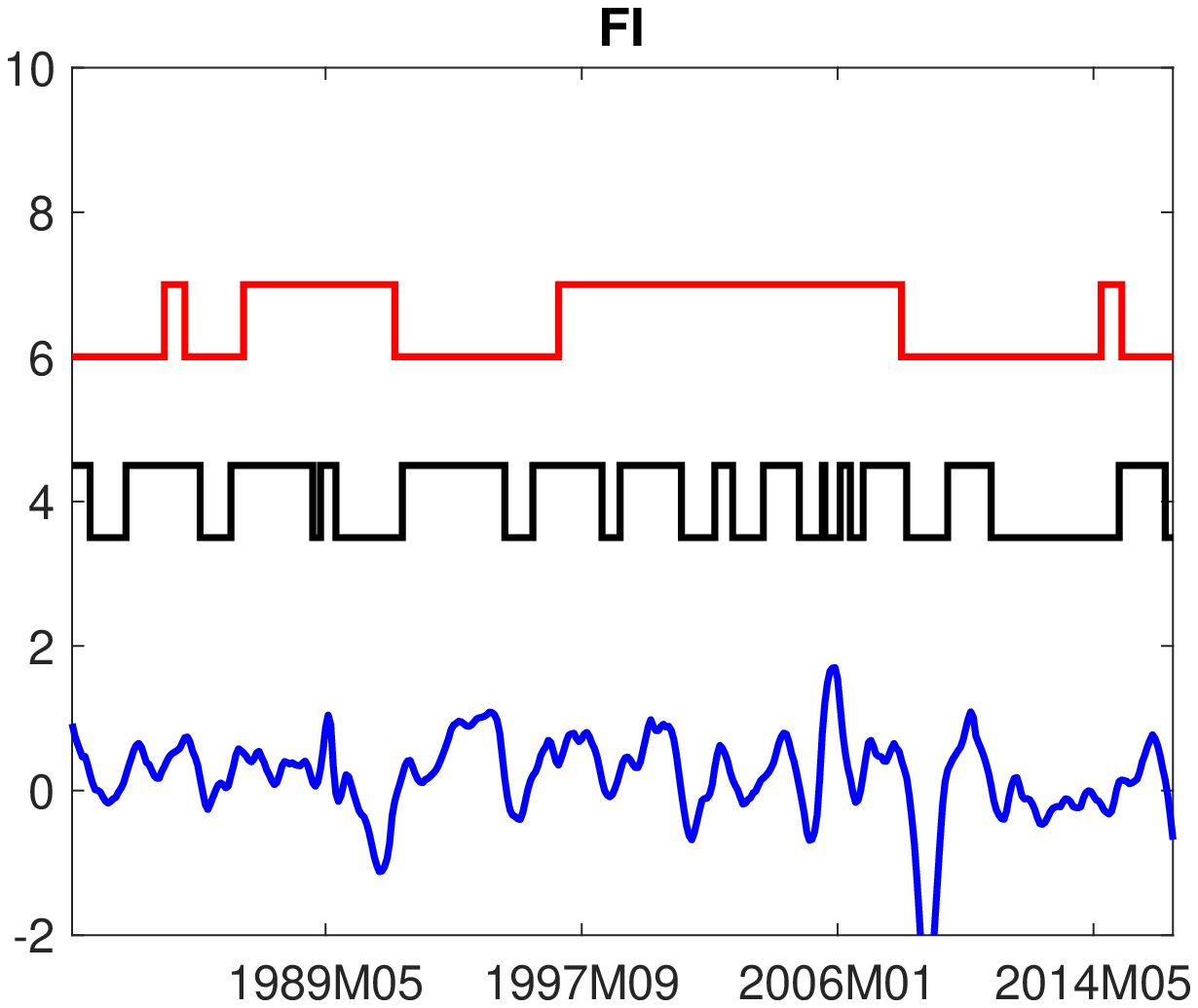}}
 ~ 
    \subfigure{                
 \includegraphics[width=0.17\textwidth]{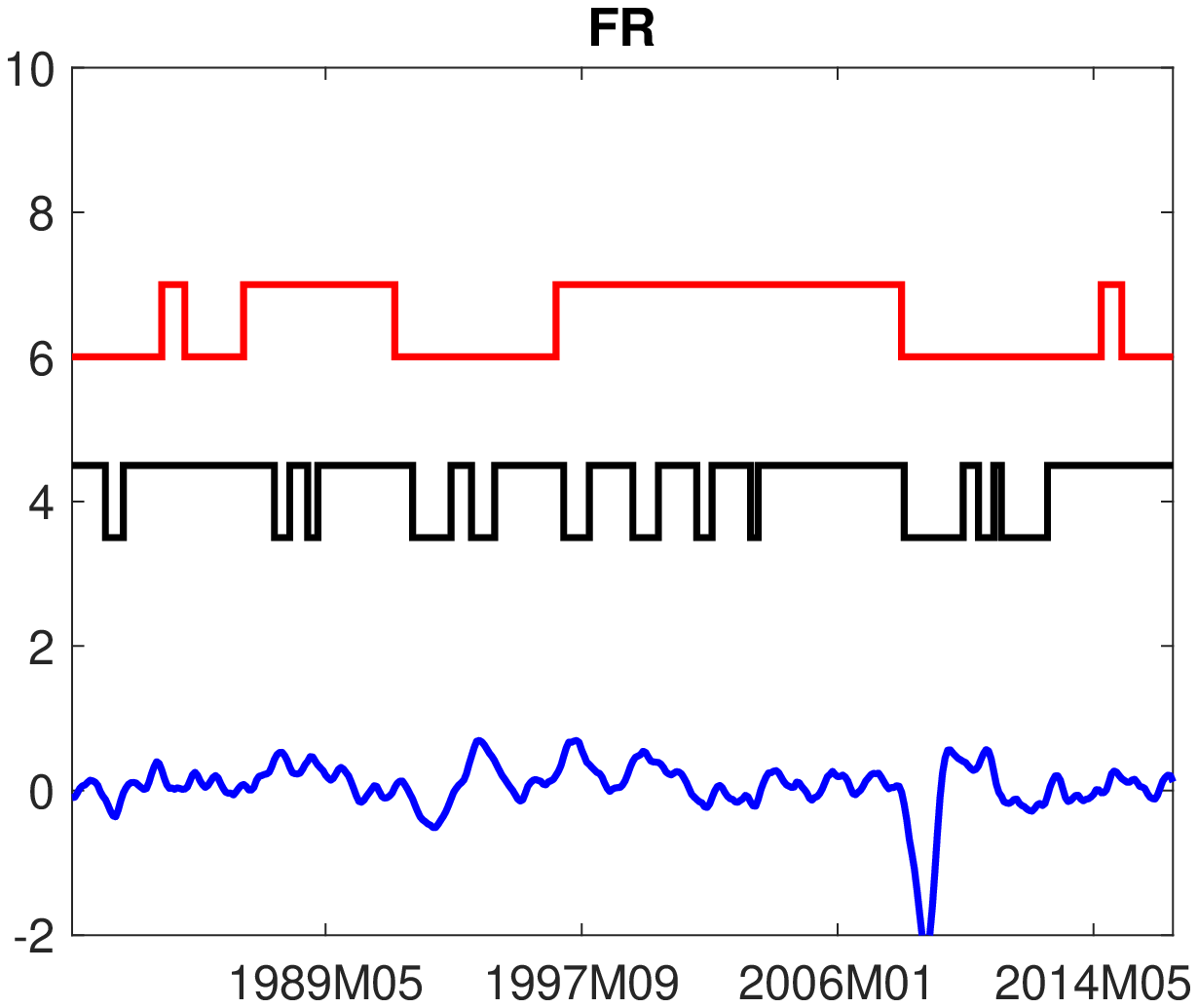}}
  ~
    \subfigure{                
 \includegraphics[width=0.17\textwidth]{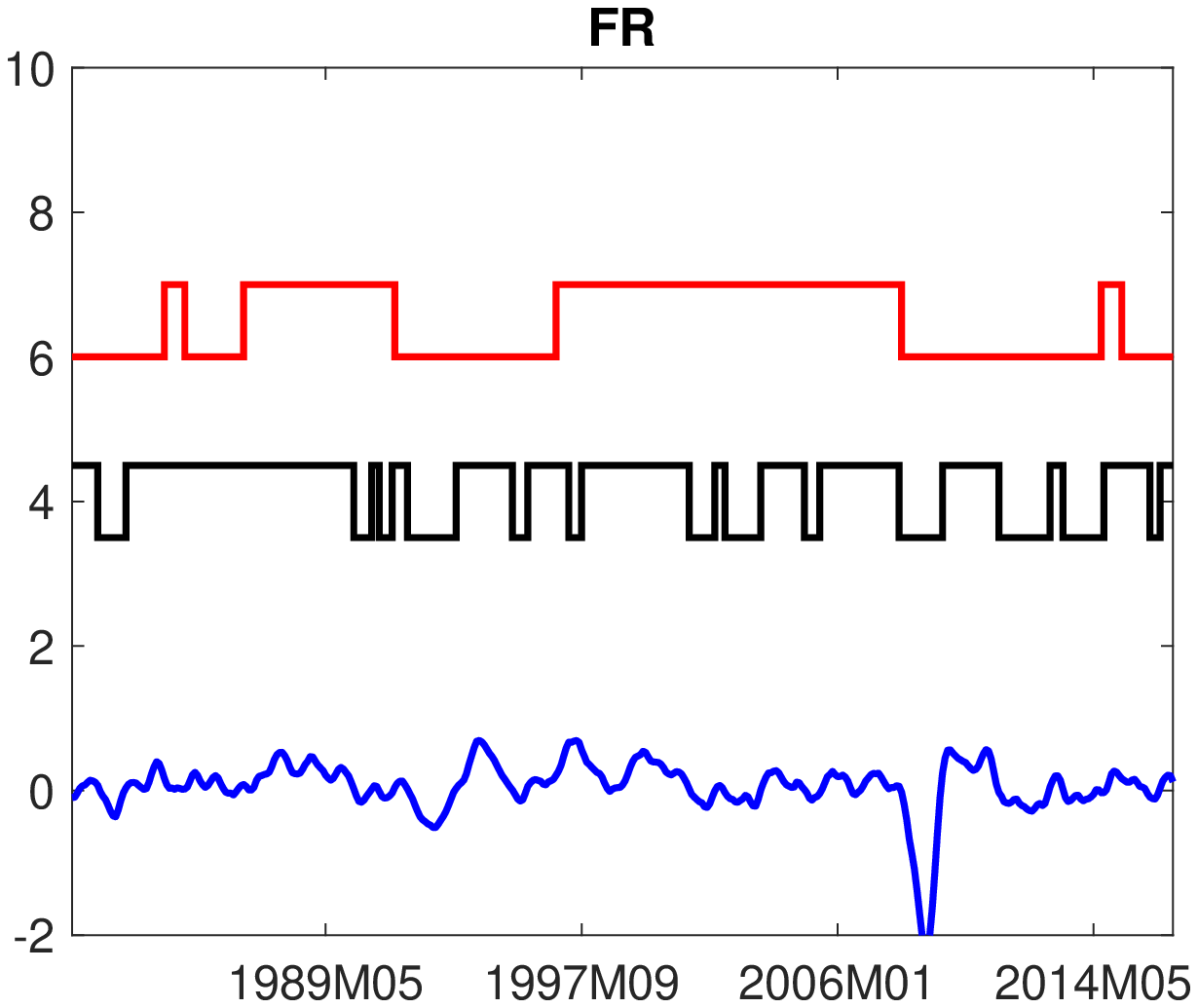}}
 ~  
    \subfigure{                
 \includegraphics[width=0.17\textwidth]{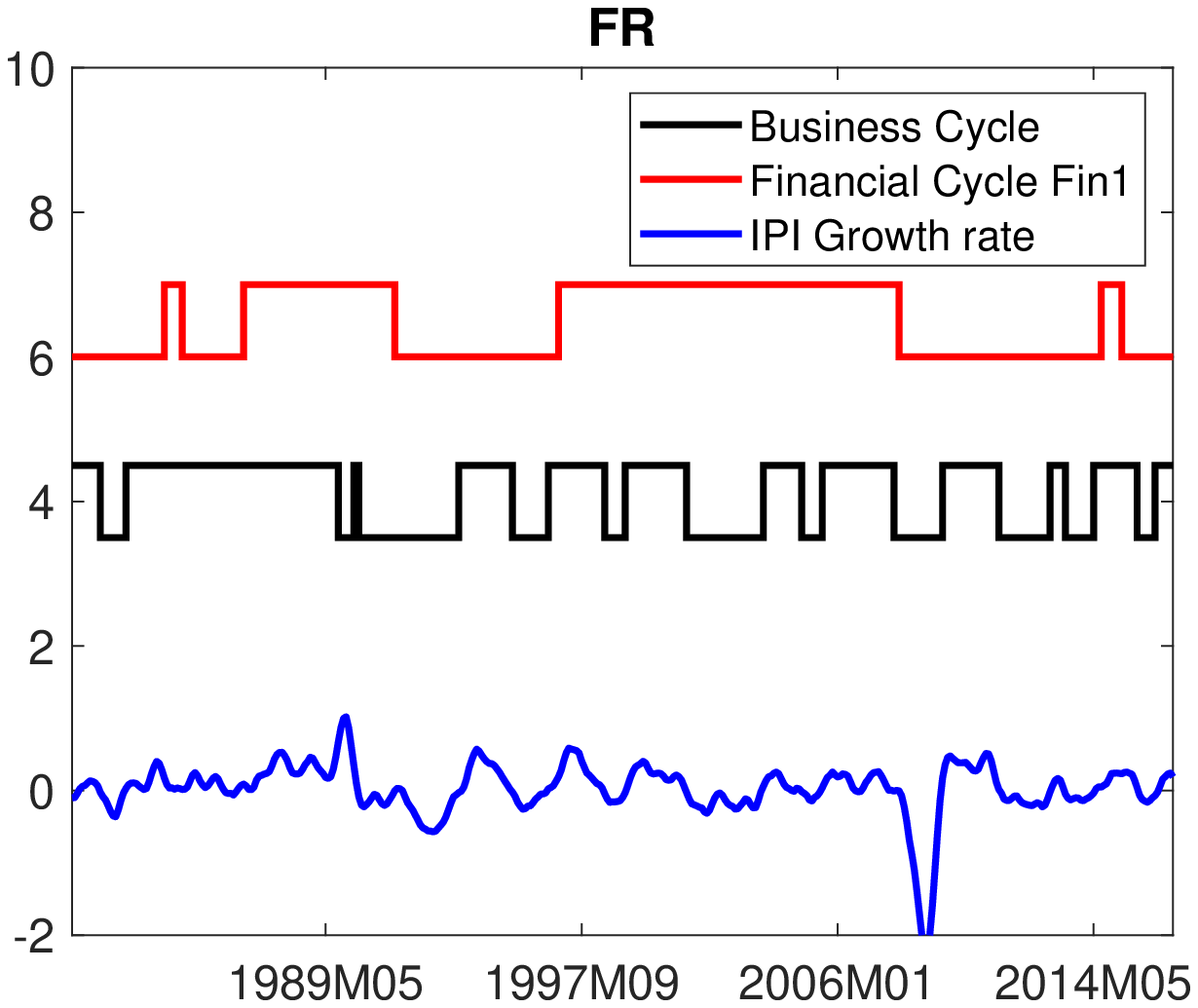}}
 ~ 
    \subfigure{                
 \includegraphics[width=0.17\textwidth]{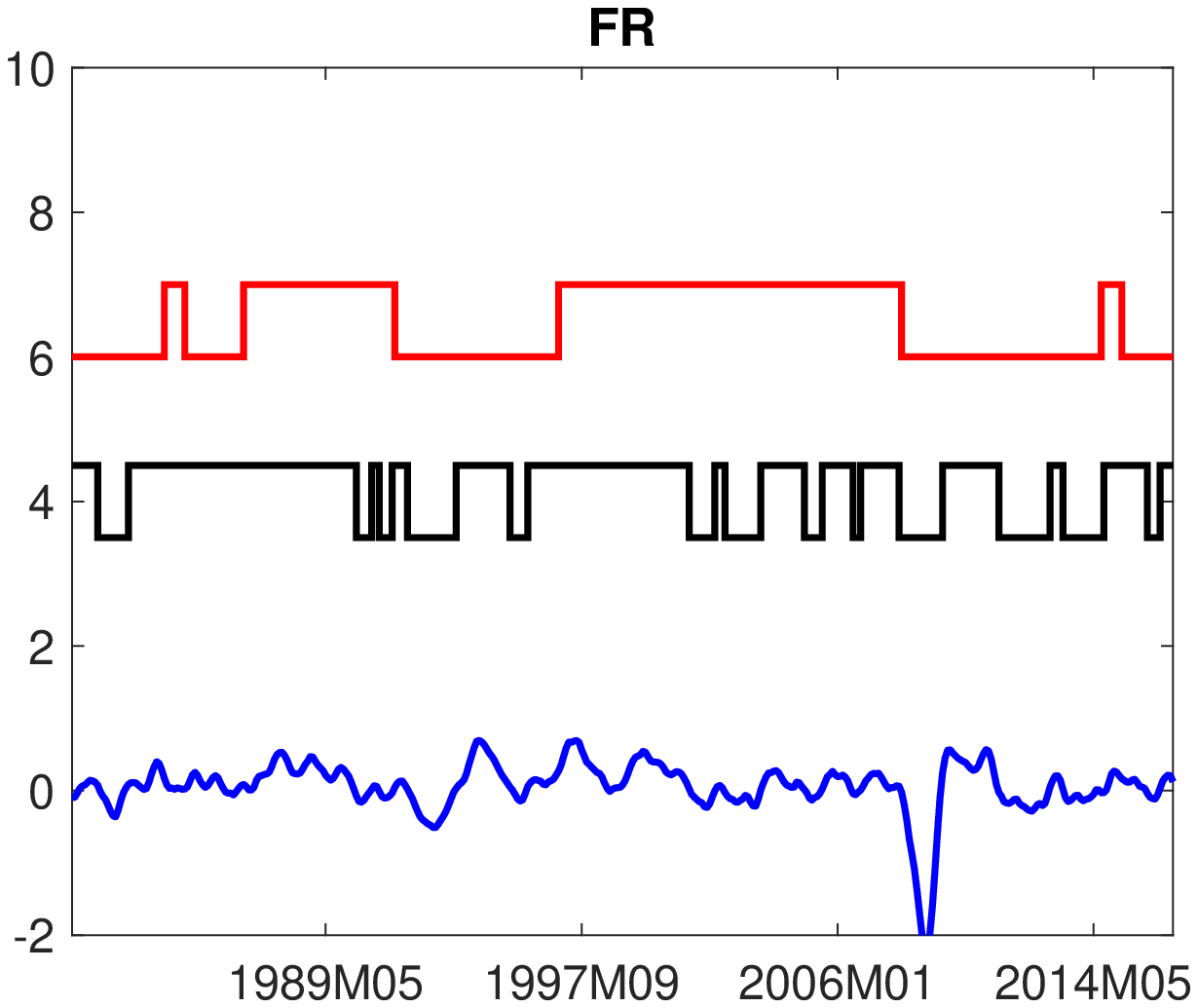}}
 ~ 
    \subfigure{                
 \includegraphics[width=0.17\textwidth]{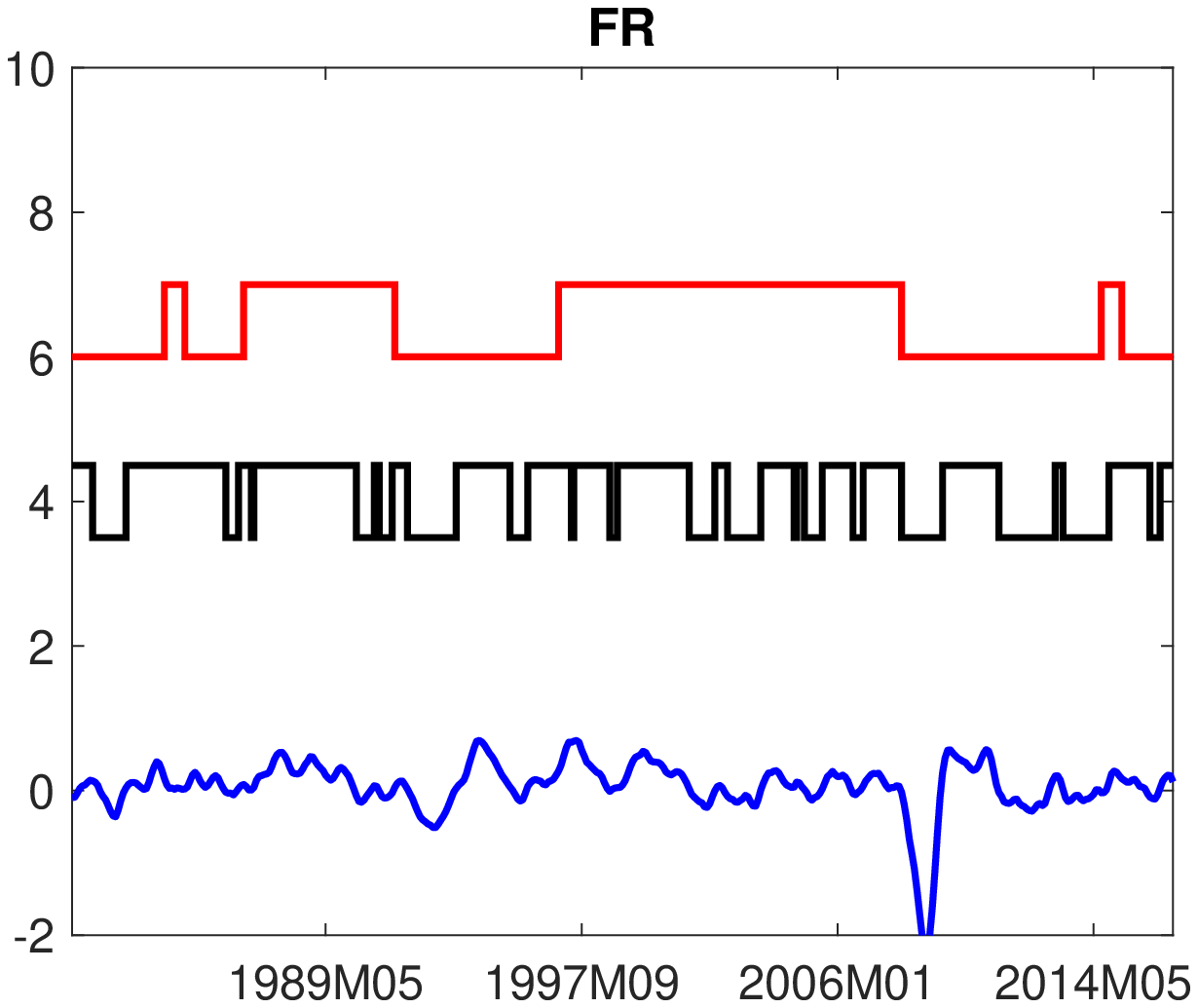}}
 ~ 
    \subfigure{                
 \includegraphics[width=0.17\textwidth]{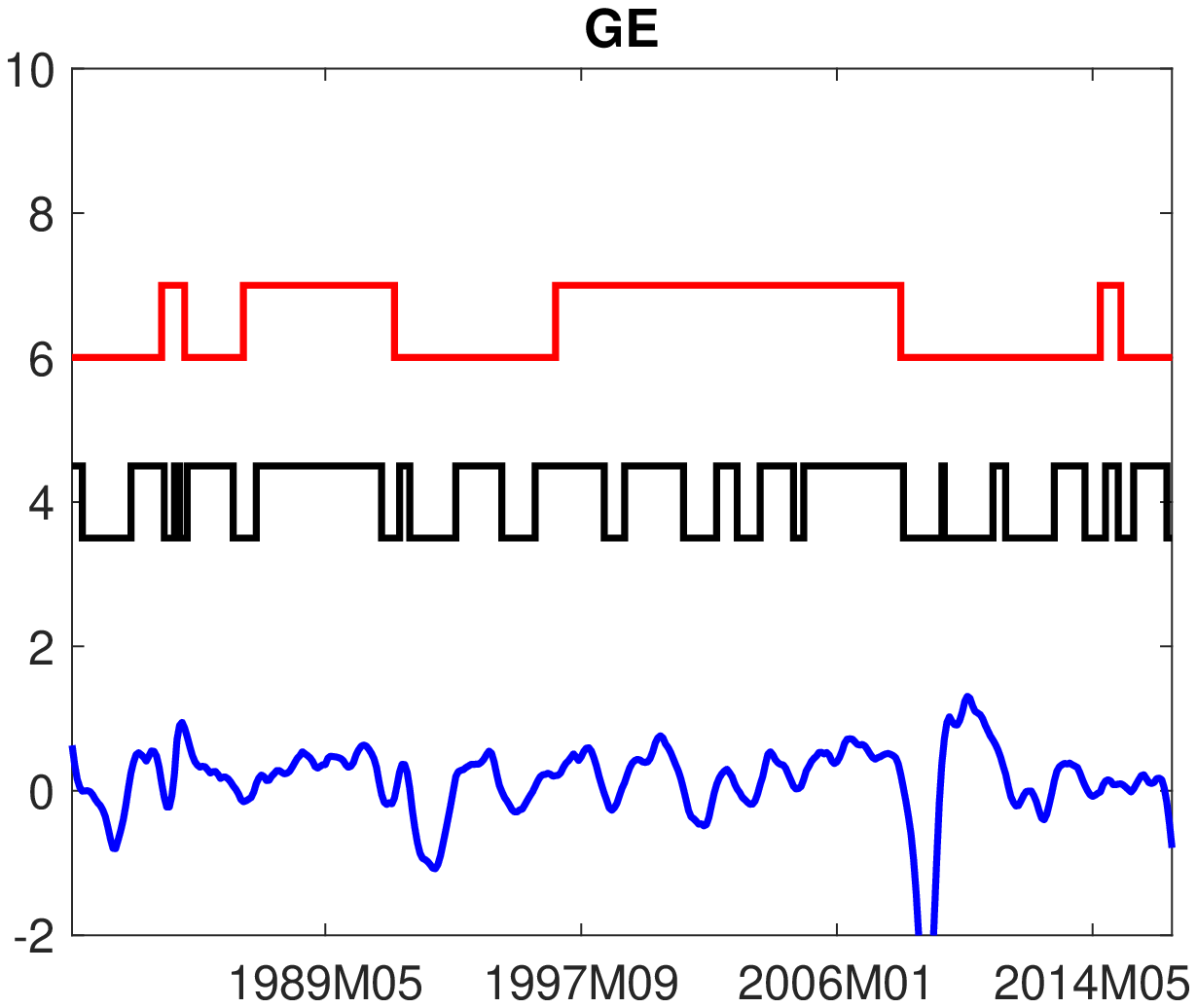}}
  ~
    \subfigure{                
 \includegraphics[width=0.17\textwidth]{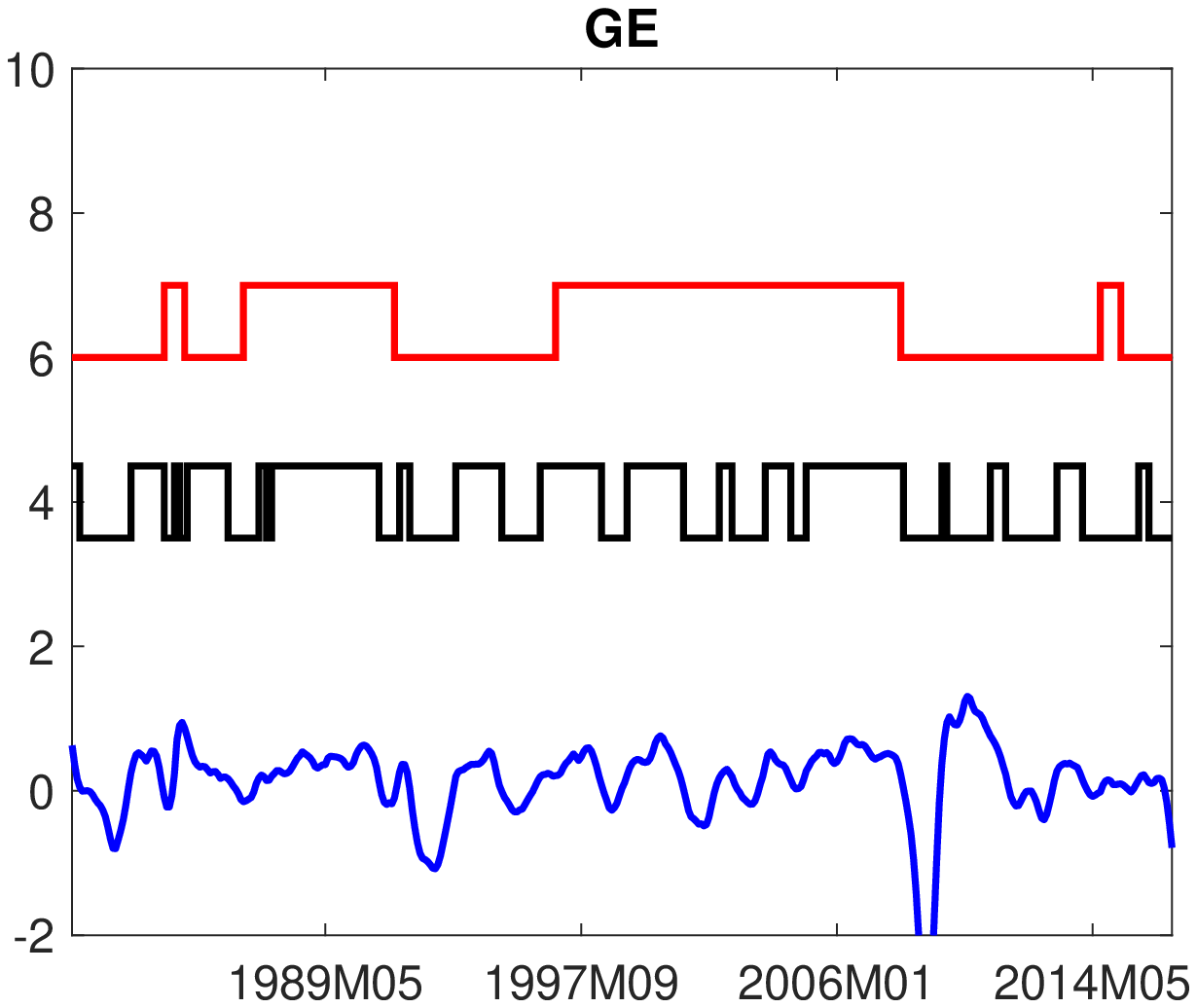}}
 ~  
    \subfigure{                
 \includegraphics[width=0.17\textwidth]{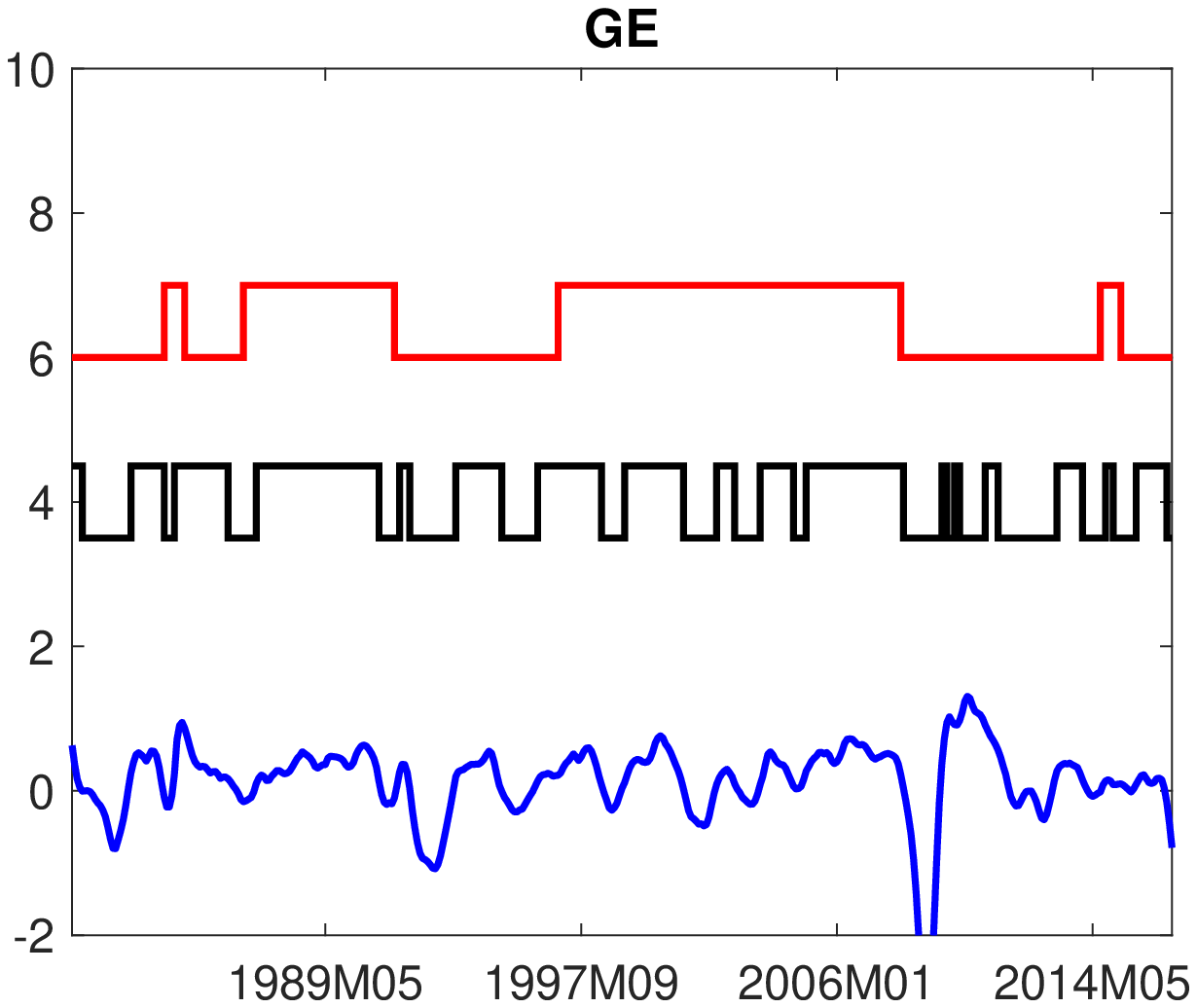}}
 ~ 
    \subfigure{                
 \includegraphics[width=0.17\textwidth]{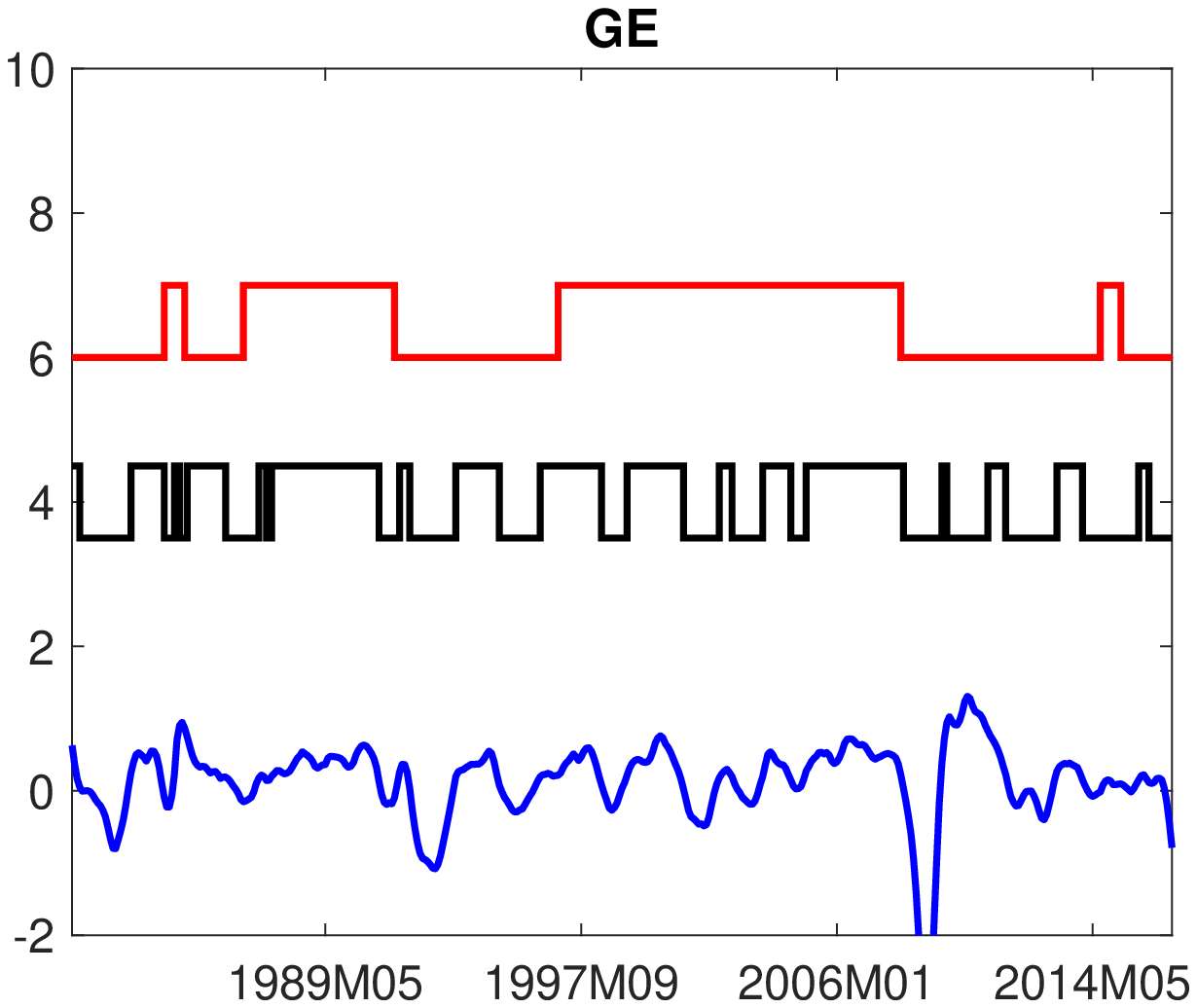}}
 ~ 
    \subfigure{                
 \includegraphics[width=0.17\textwidth]{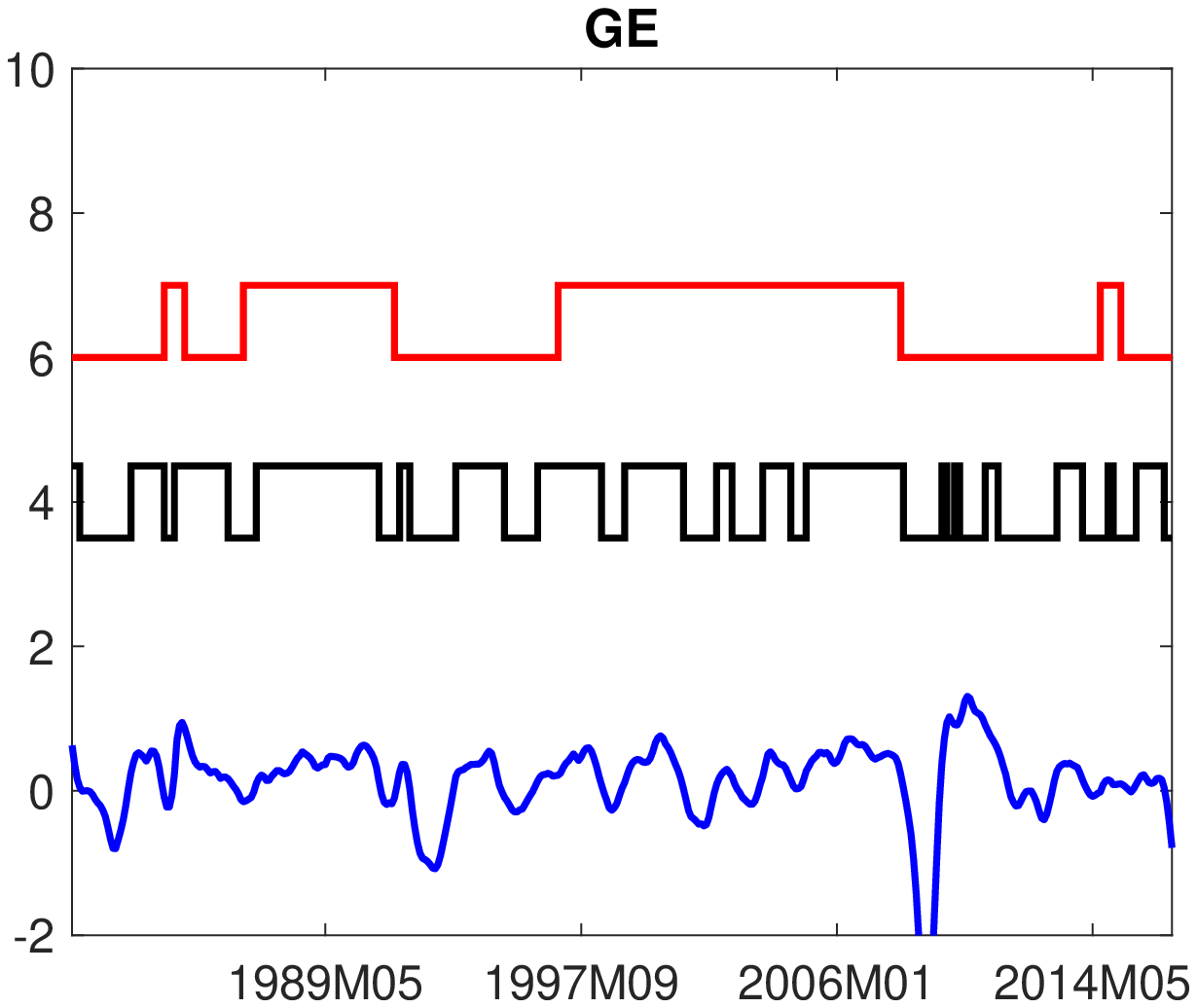}}
 ~ 
 \caption{Estimated cycles for different specifications of the unconstrained PMS models (columns) and countries (rows). The first column represent estimates of the climate conditions independent unconstrained PMS model, the second columns to the fourth column respectively correspond to the estimates of the unconstrained PMS model subject to temperature, drought and rainfall, while the last column represent estimates from unconstrained PMS taking all the climate risk indicators into consideration simultaneously.  In each plot country-specific business cycles (black line), financial cycle (red line) and country specific industrial production (IPI) growth (blue line).}\label{FigCyc1NPooling1}
 \end{figure}
 
 \begin{figure}[h!]
 \centering
    \subfigure[None ]{                
 \includegraphics[width=0.17\textwidth]{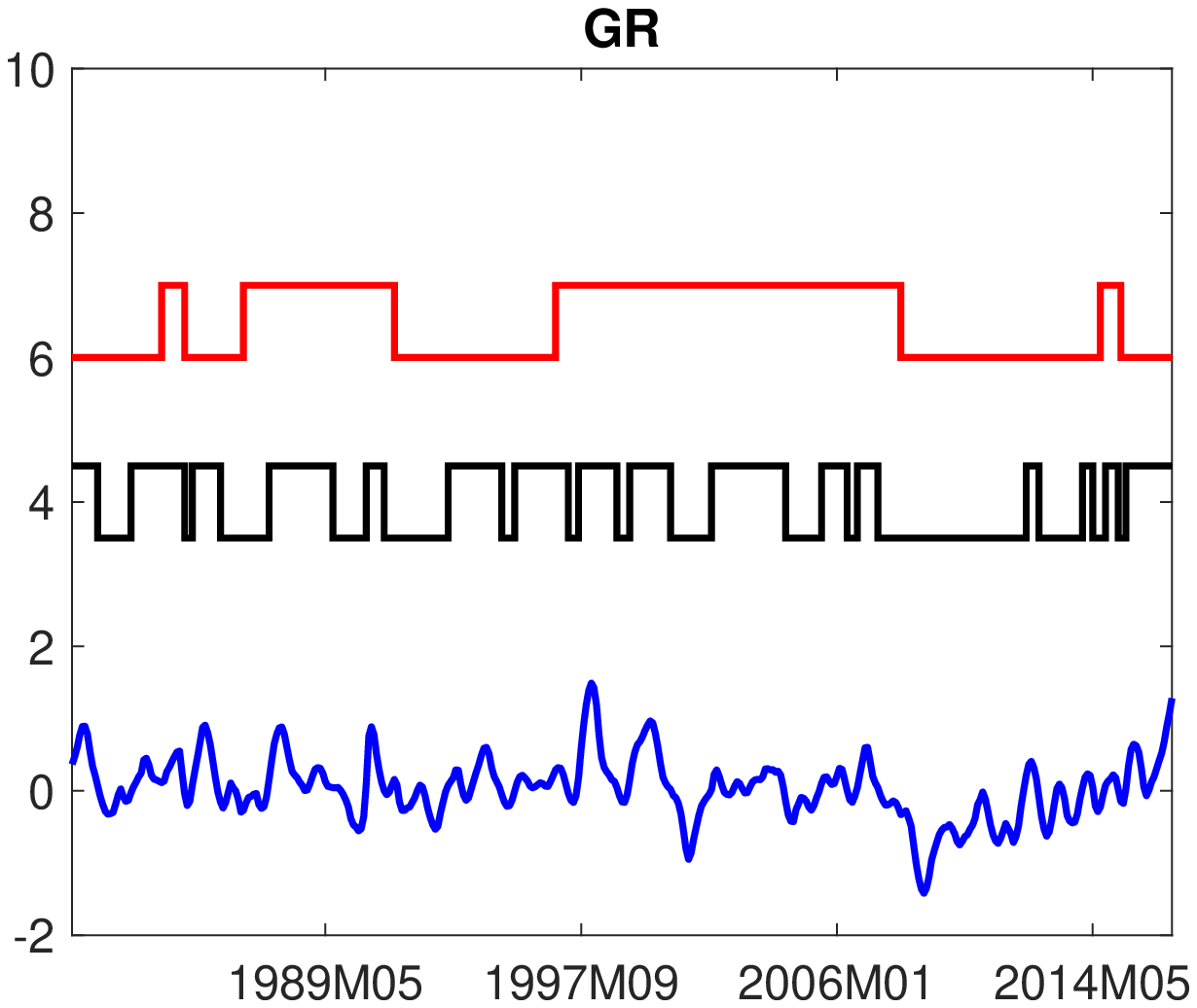}}
  ~
    \subfigure[Temperature only (CSU)]{                
 \includegraphics[width=0.17\textwidth]{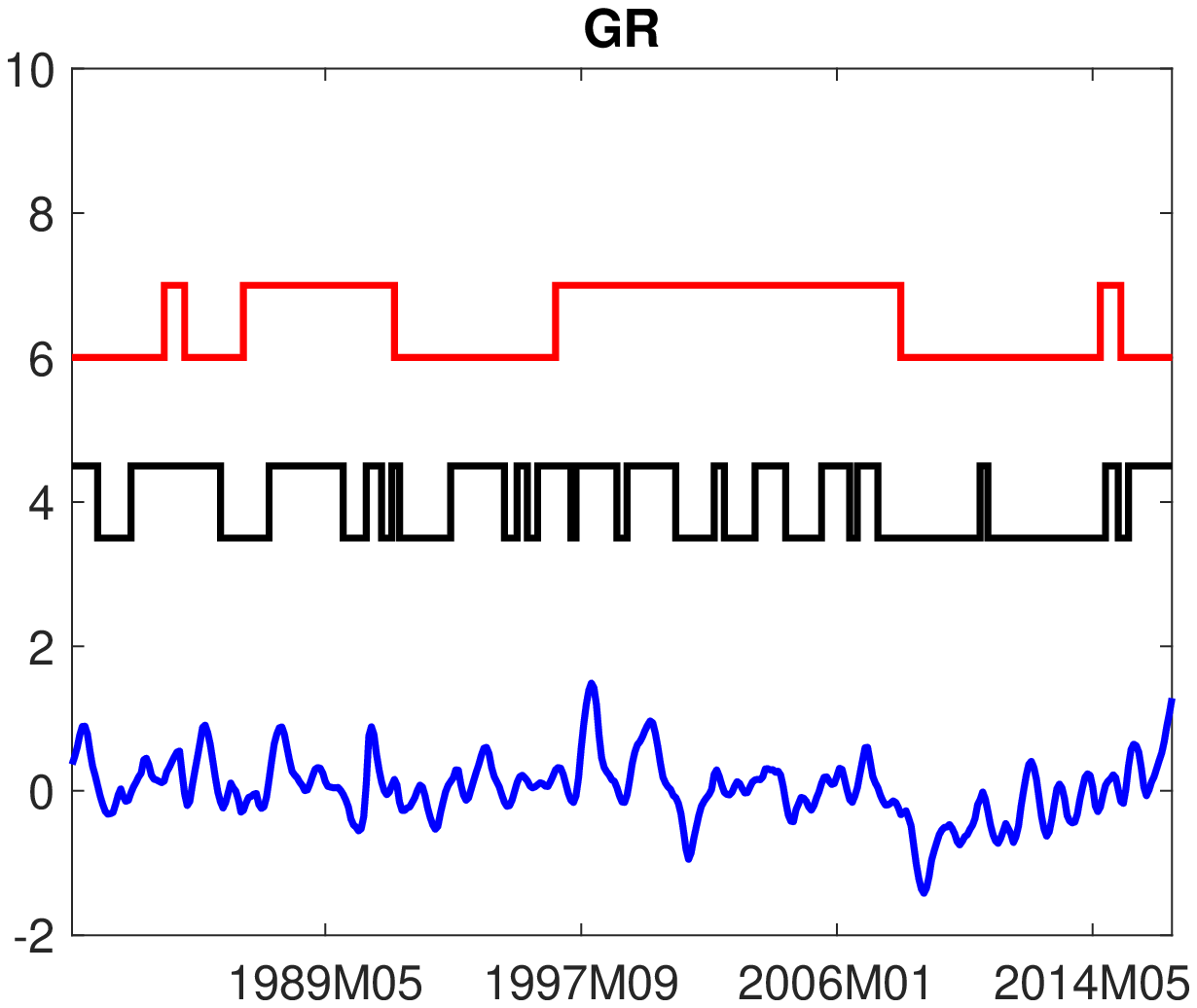}}
 ~  
    \subfigure[Drought only (SPI)]{                
 \includegraphics[width=0.17\textwidth]{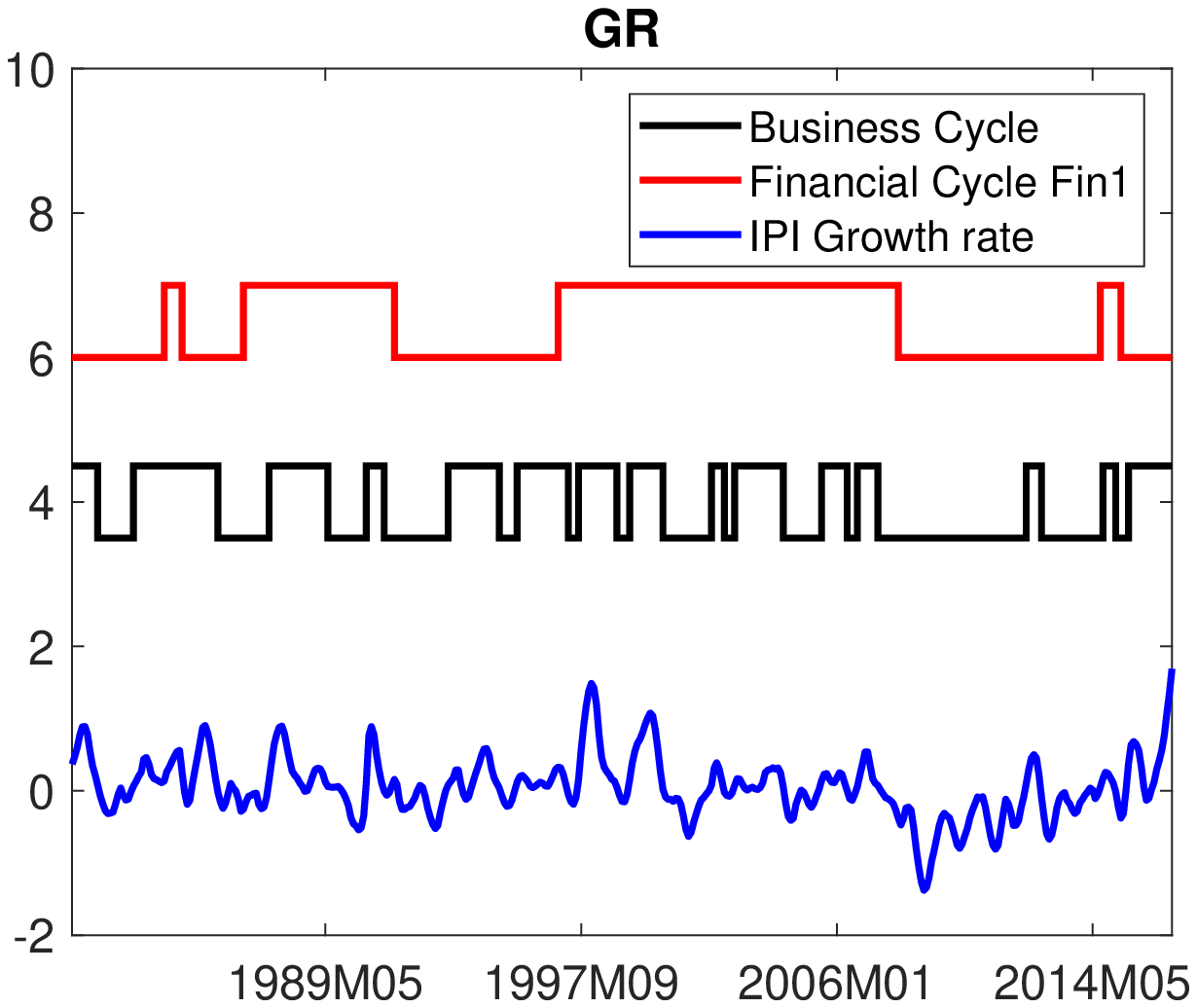}}
 ~ 
    \subfigure[Rainfall only (r20mm)]{                
 \includegraphics[width=0.17\textwidth]{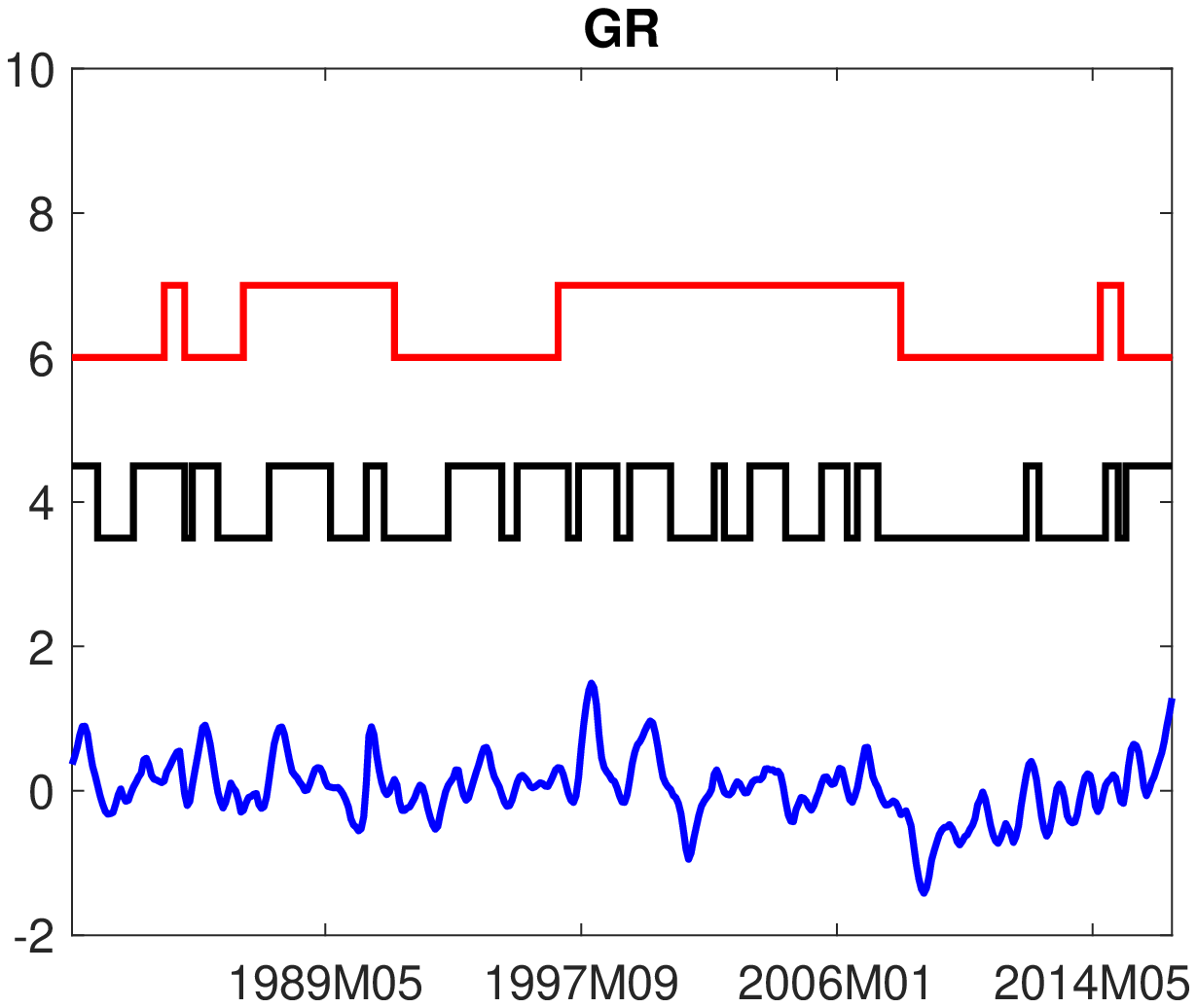}}
 ~ 
    \subfigure{                
 \includegraphics[width=0.17\textwidth]{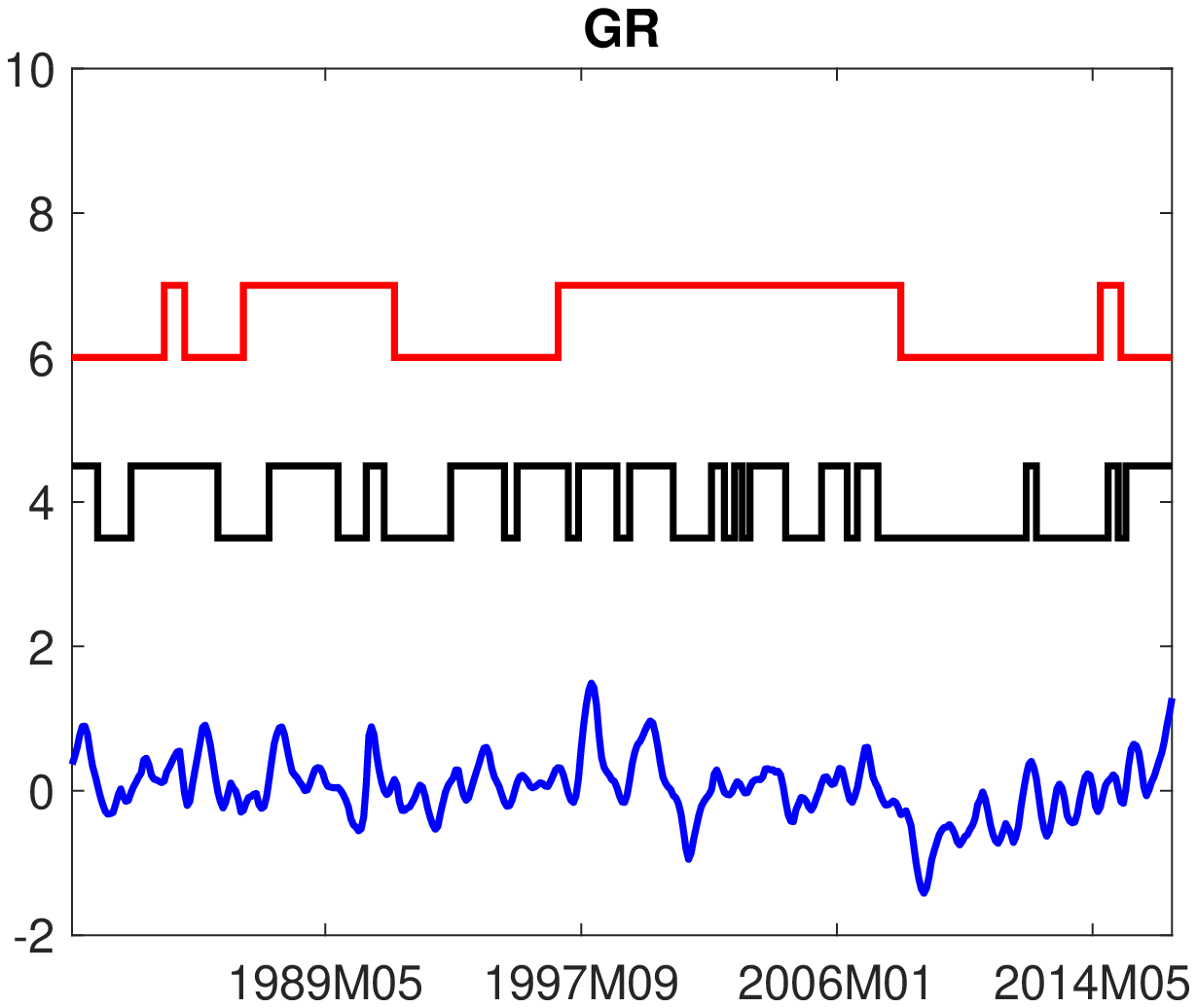}}
 ~ 
    \subfigure{                
 \includegraphics[width=0.17\textwidth]{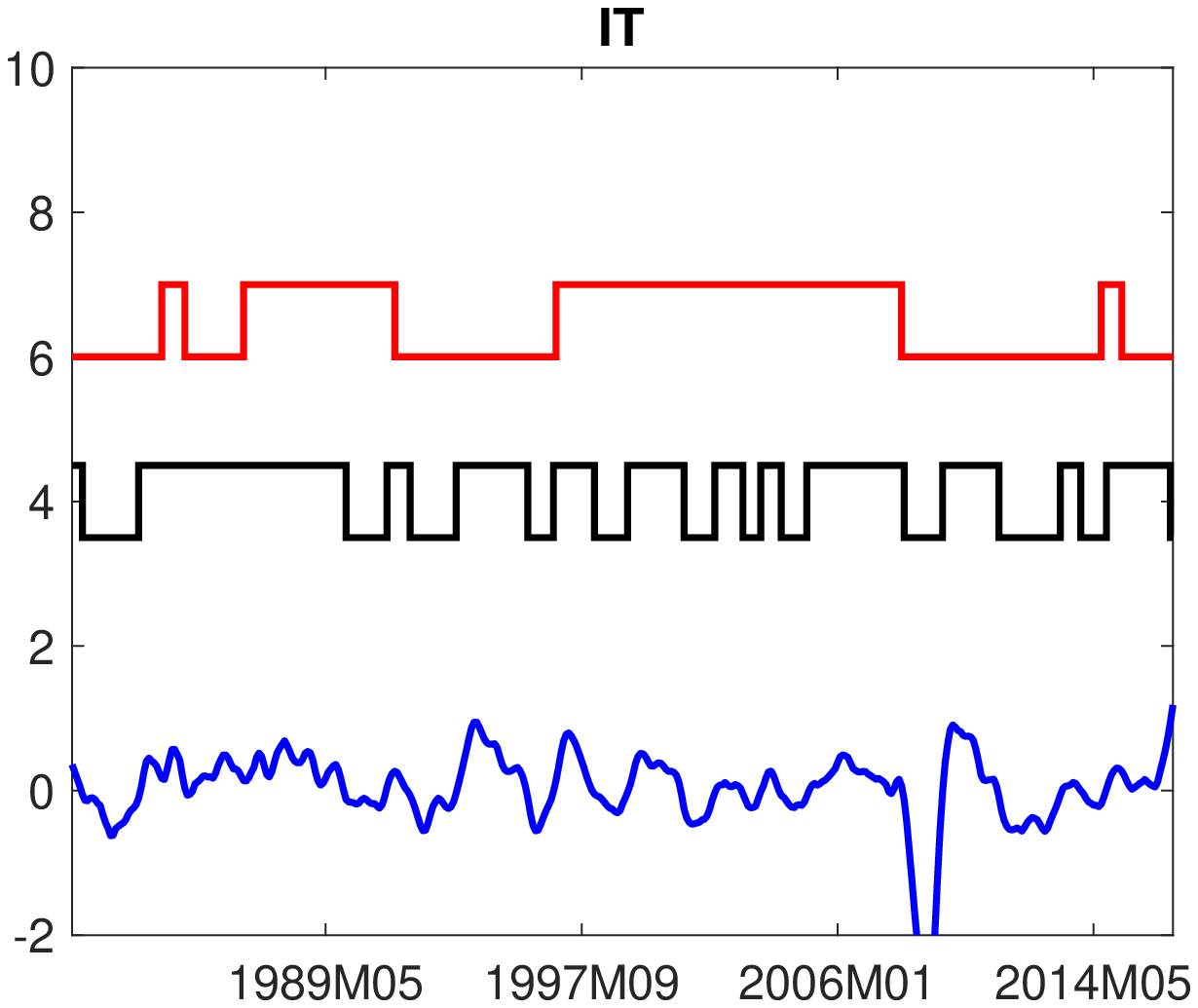}}
  ~
    \subfigure{                
 \includegraphics[width=0.17\textwidth]{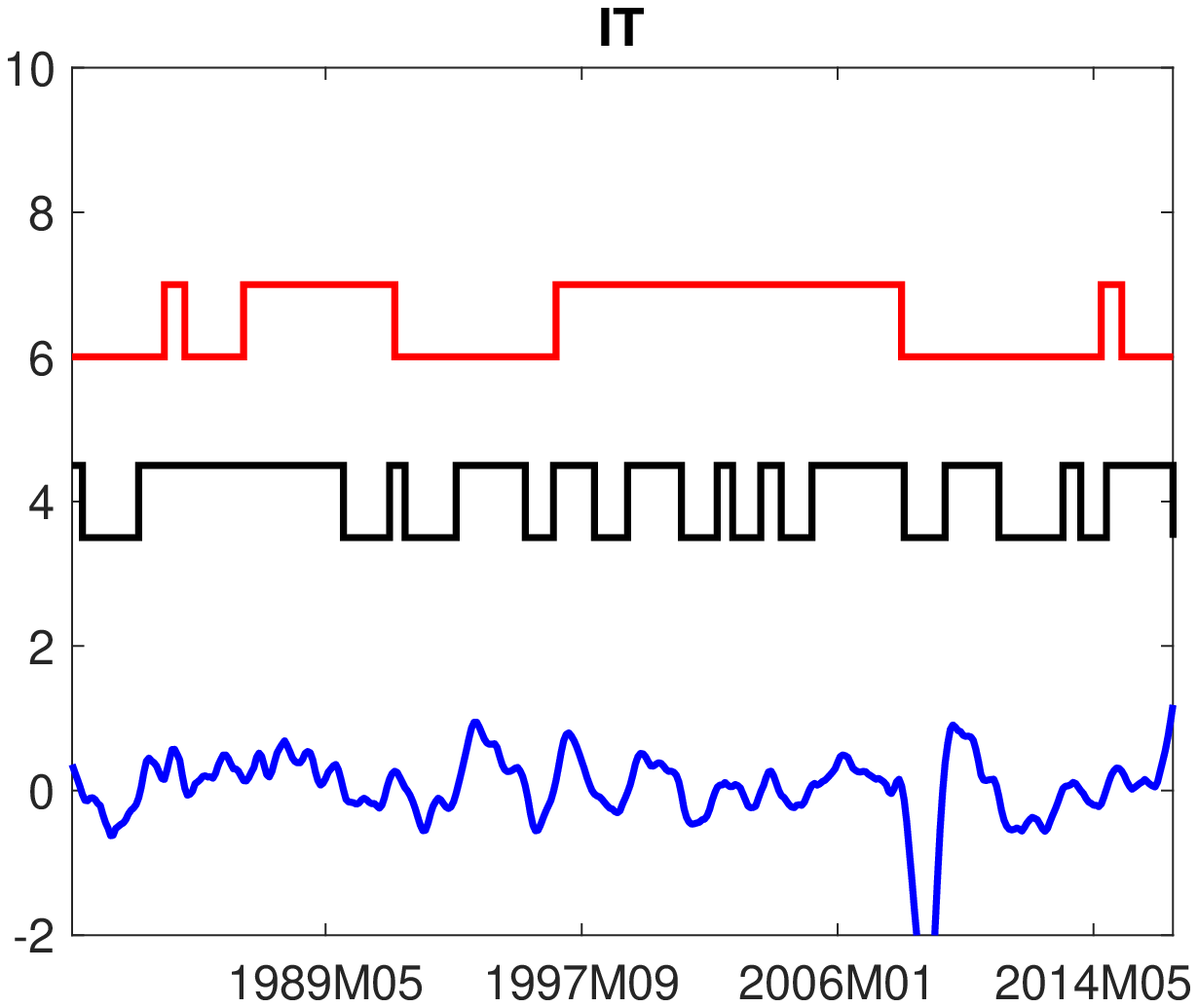}}
 ~  
    \subfigure{                
 \includegraphics[width=0.17\textwidth]{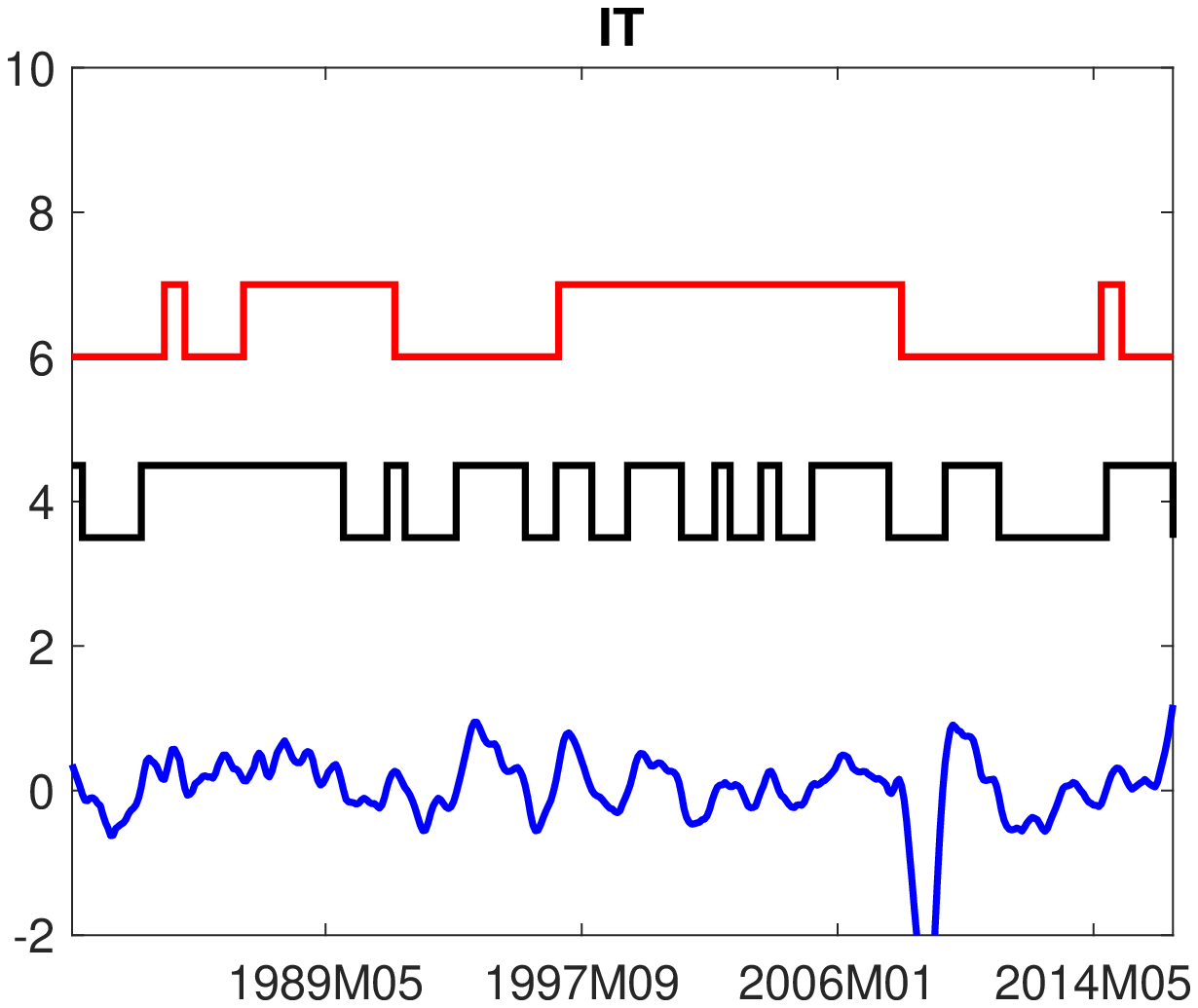}}
 ~ 
    \subfigure{                
 \includegraphics[width=0.17\textwidth]{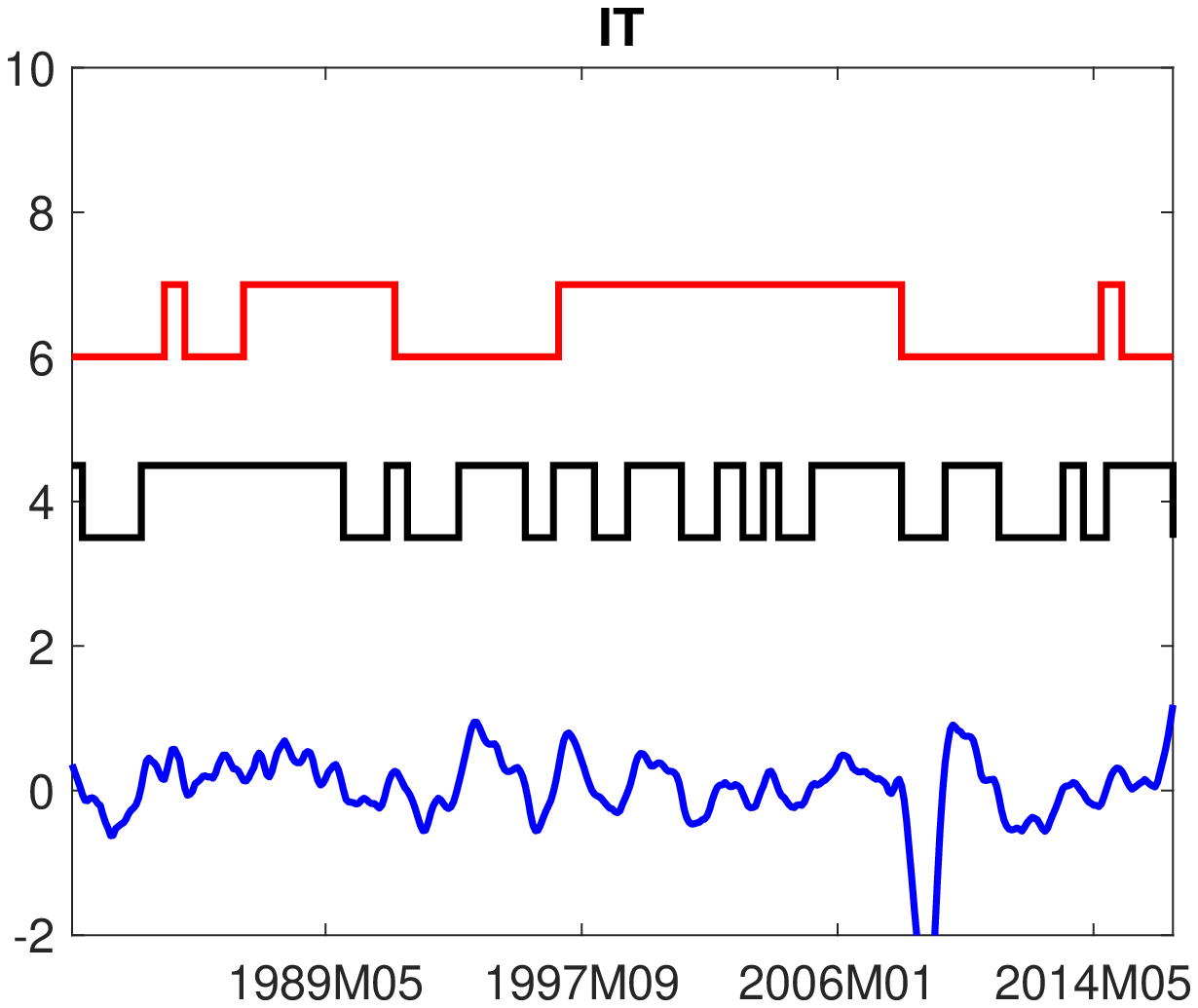}}
 ~ 
    \subfigure{                
 \includegraphics[width=0.17\textwidth]{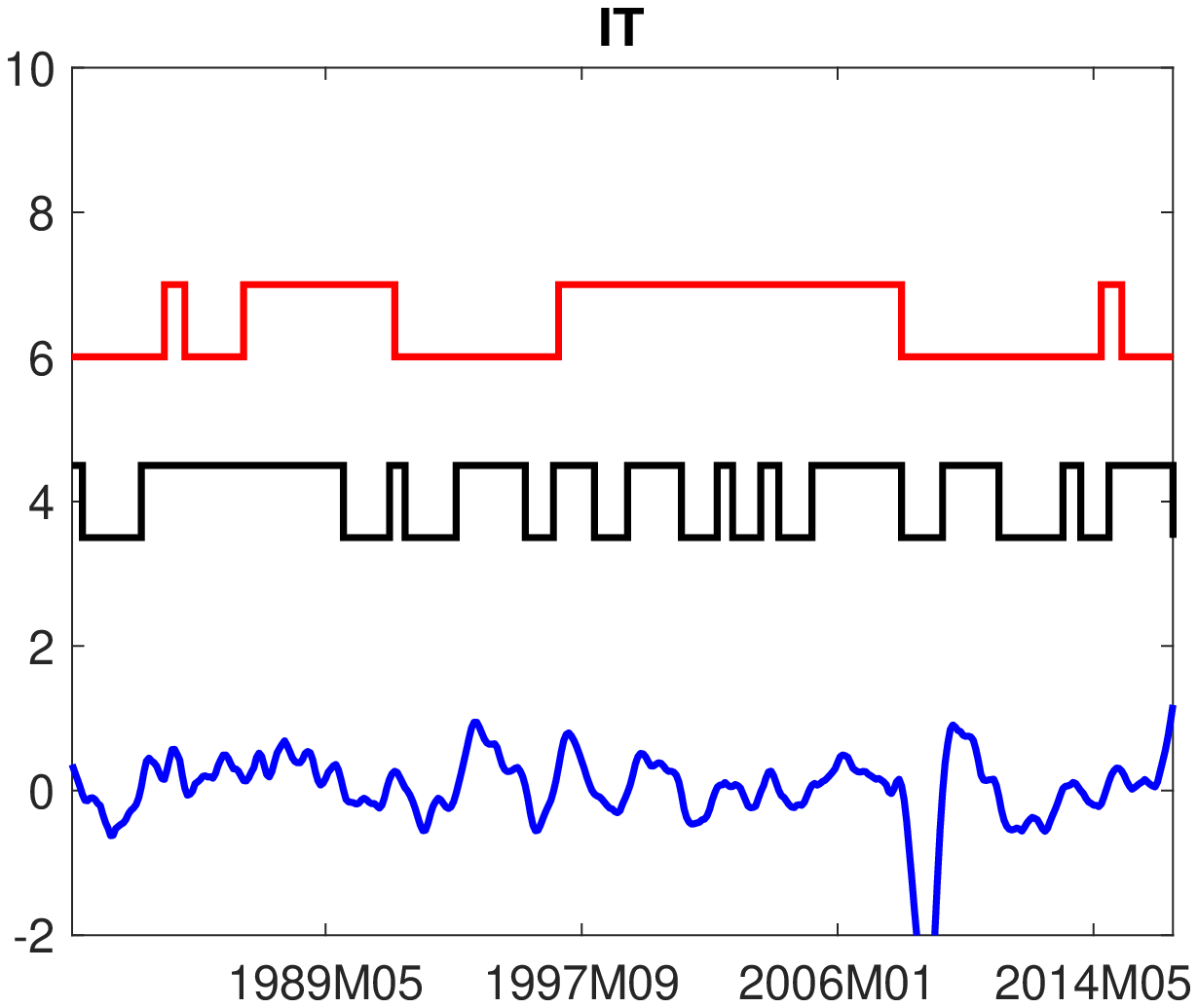}}
 ~ 
    \subfigure{                
 \includegraphics[width=0.17\textwidth]{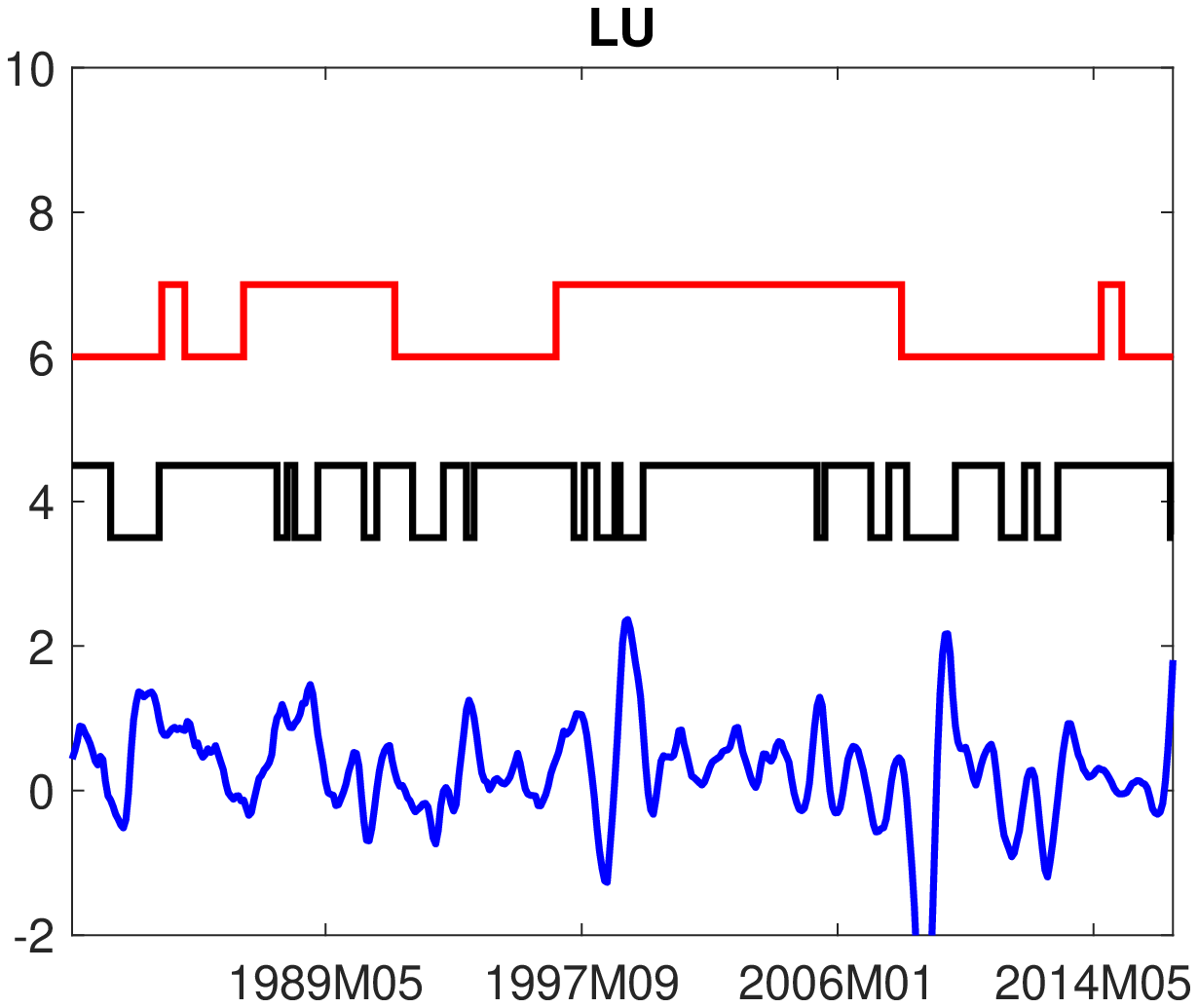}} 
  ~
    \subfigure{                
 \includegraphics[width=0.17\textwidth]{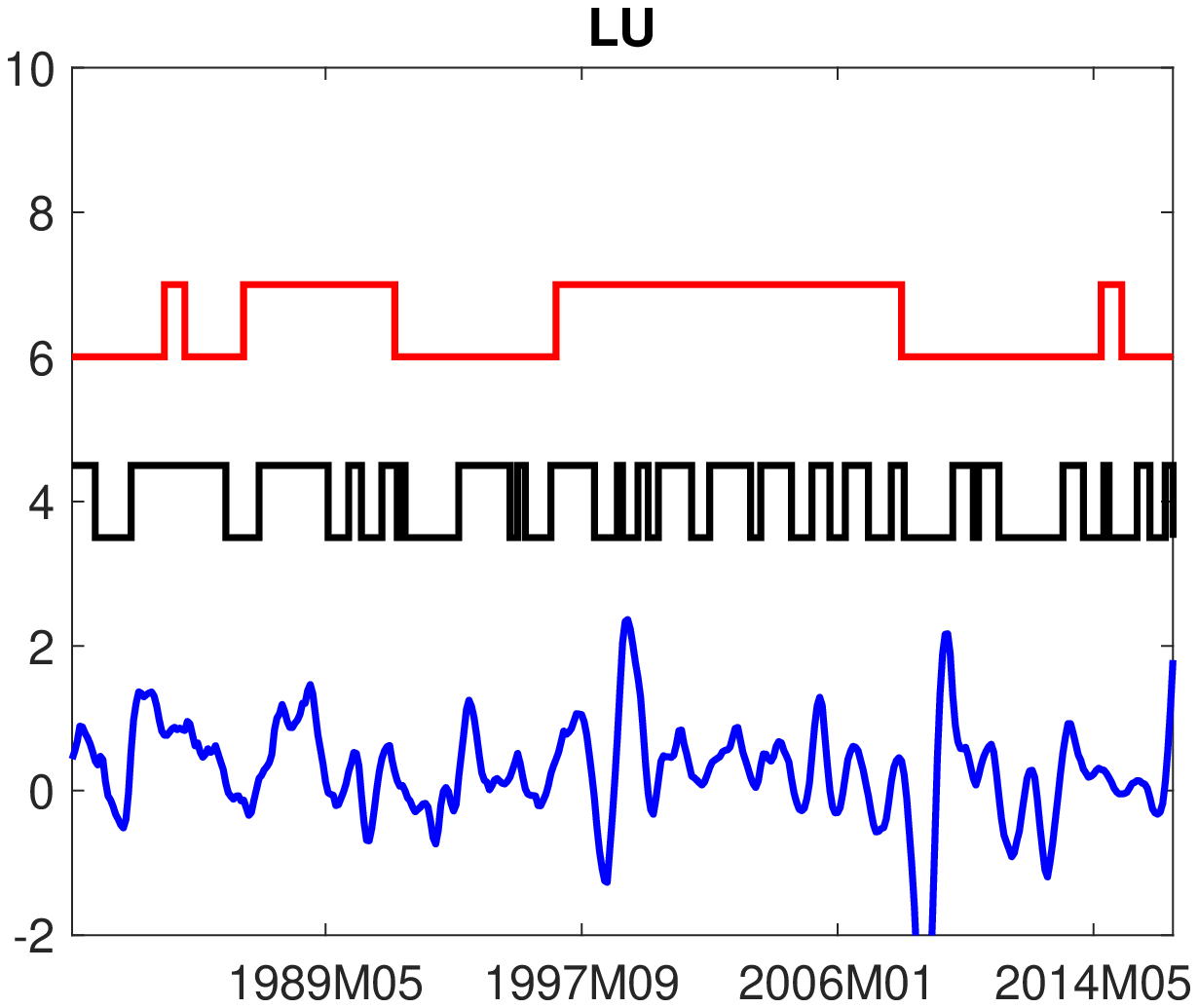}}
 ~  
    \subfigure{                
 \includegraphics[width=0.17\textwidth]{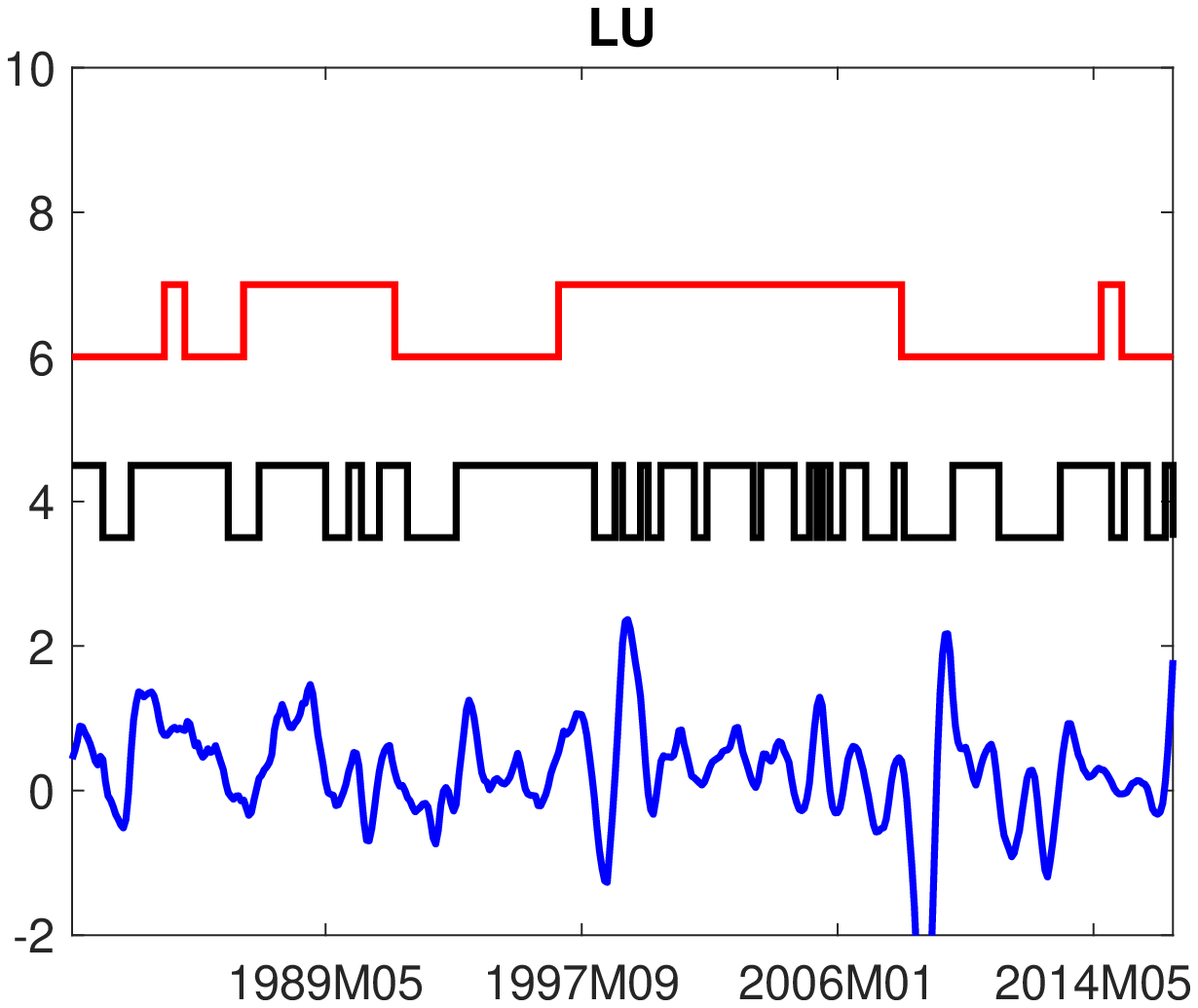}}
 ~ 
    \subfigure{                
 \includegraphics[width=0.17\textwidth]{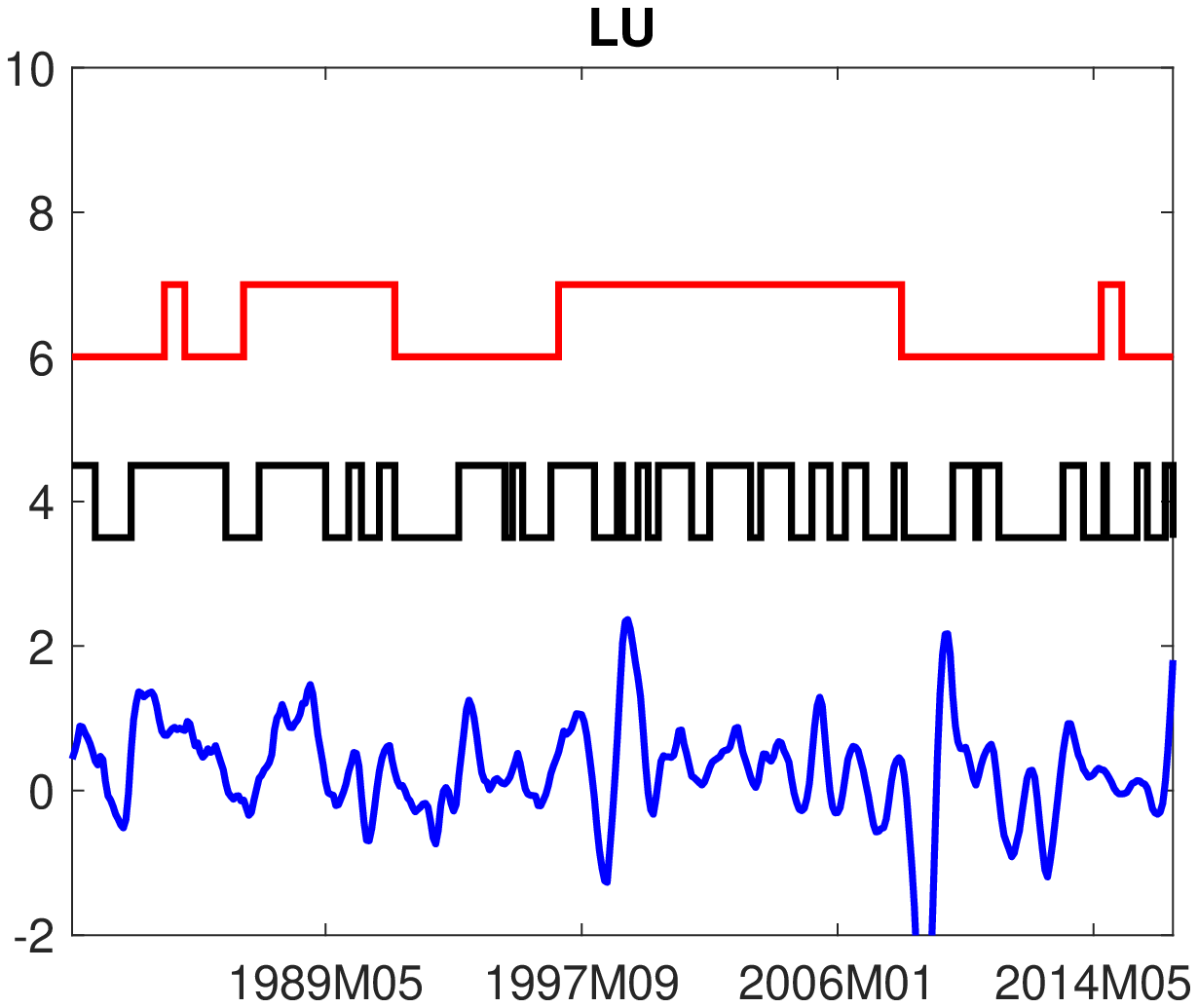}}
 ~ 
    \subfigure{                
 \includegraphics[width=0.17\textwidth]{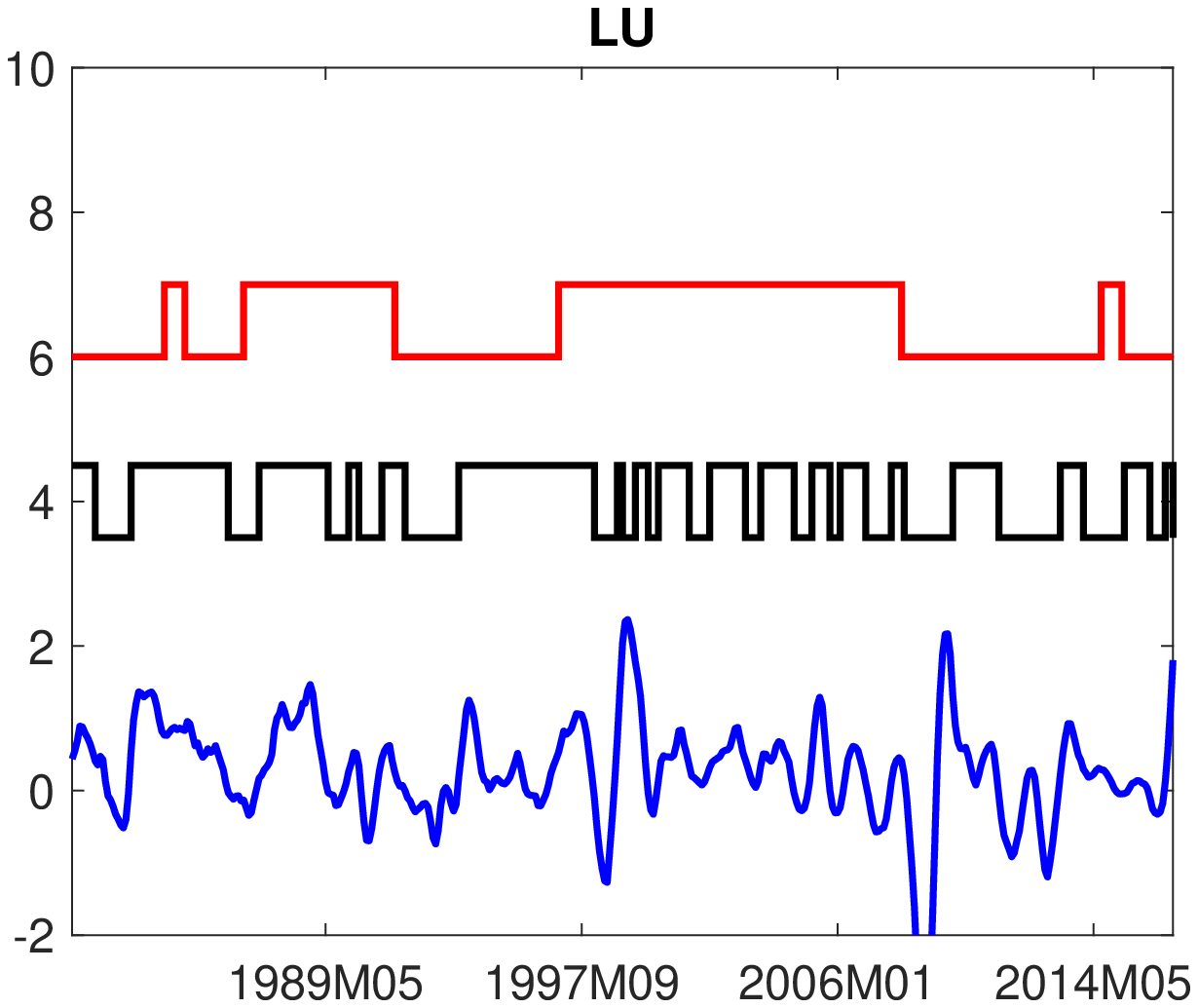}}
 ~ 
    \subfigure{                
 \includegraphics[width=0.17\textwidth]{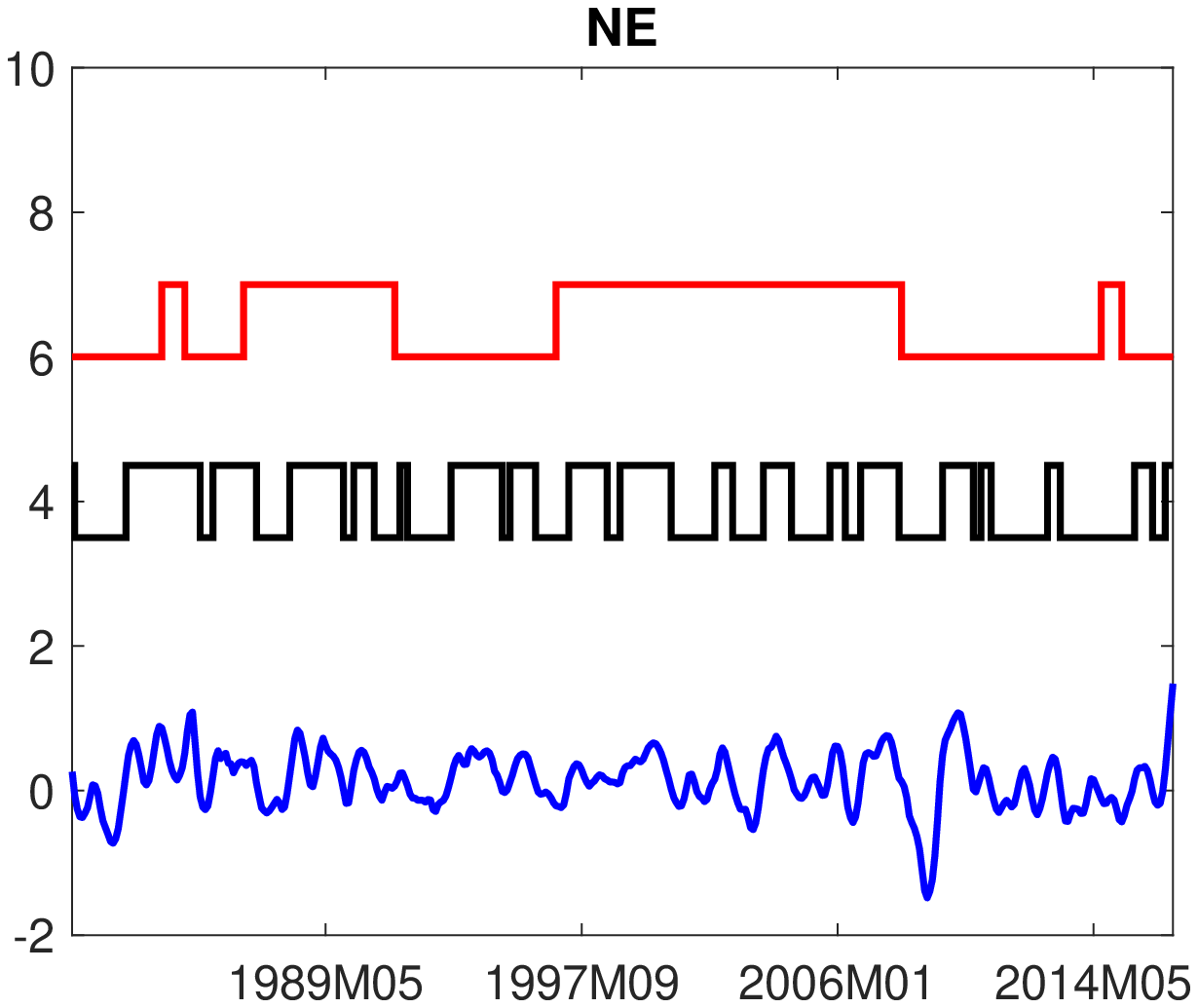}}
  ~
    \subfigure{                
 \includegraphics[width=0.17\textwidth]{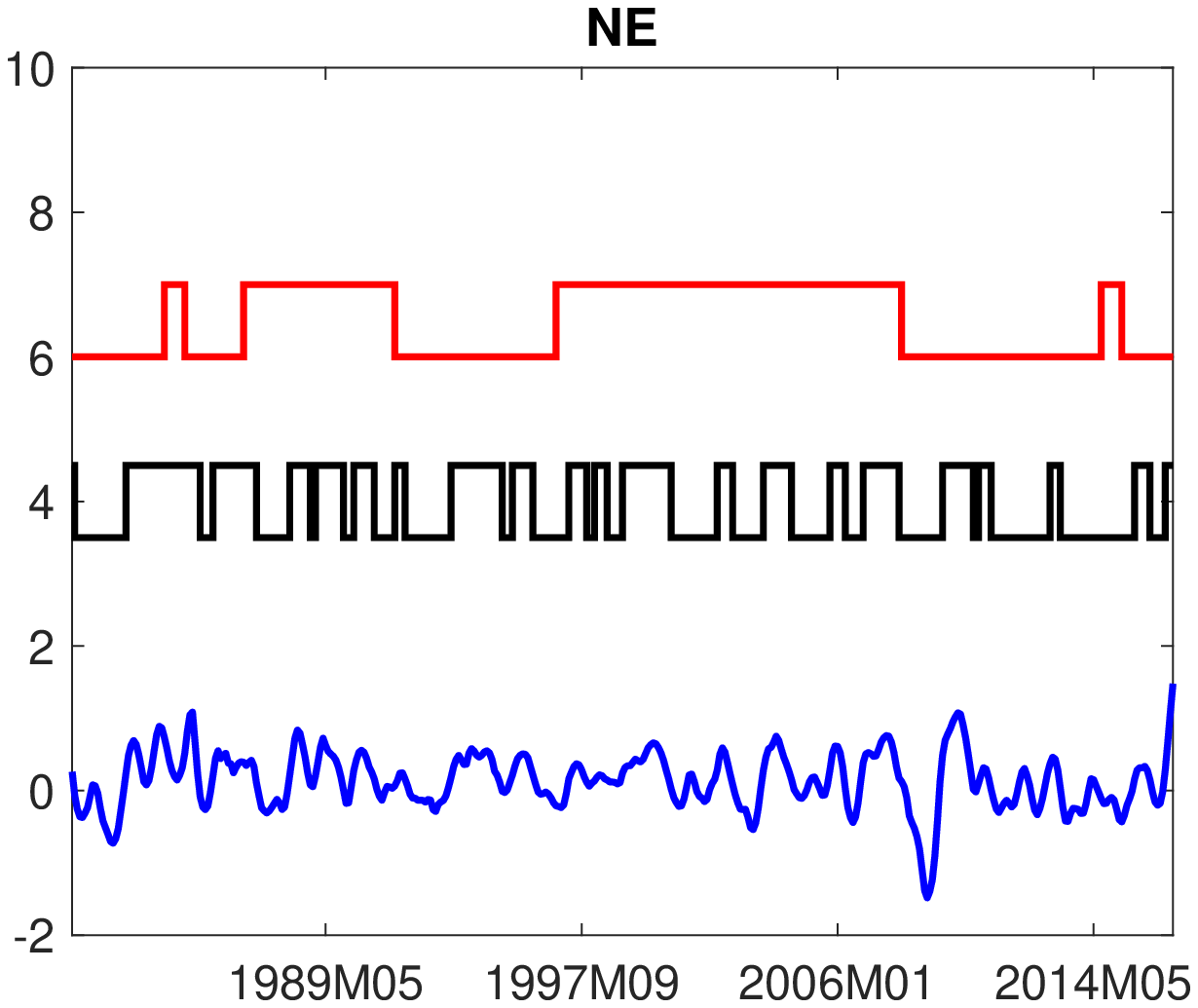}}
 ~  
    \subfigure{                
 \includegraphics[width=0.17\textwidth]{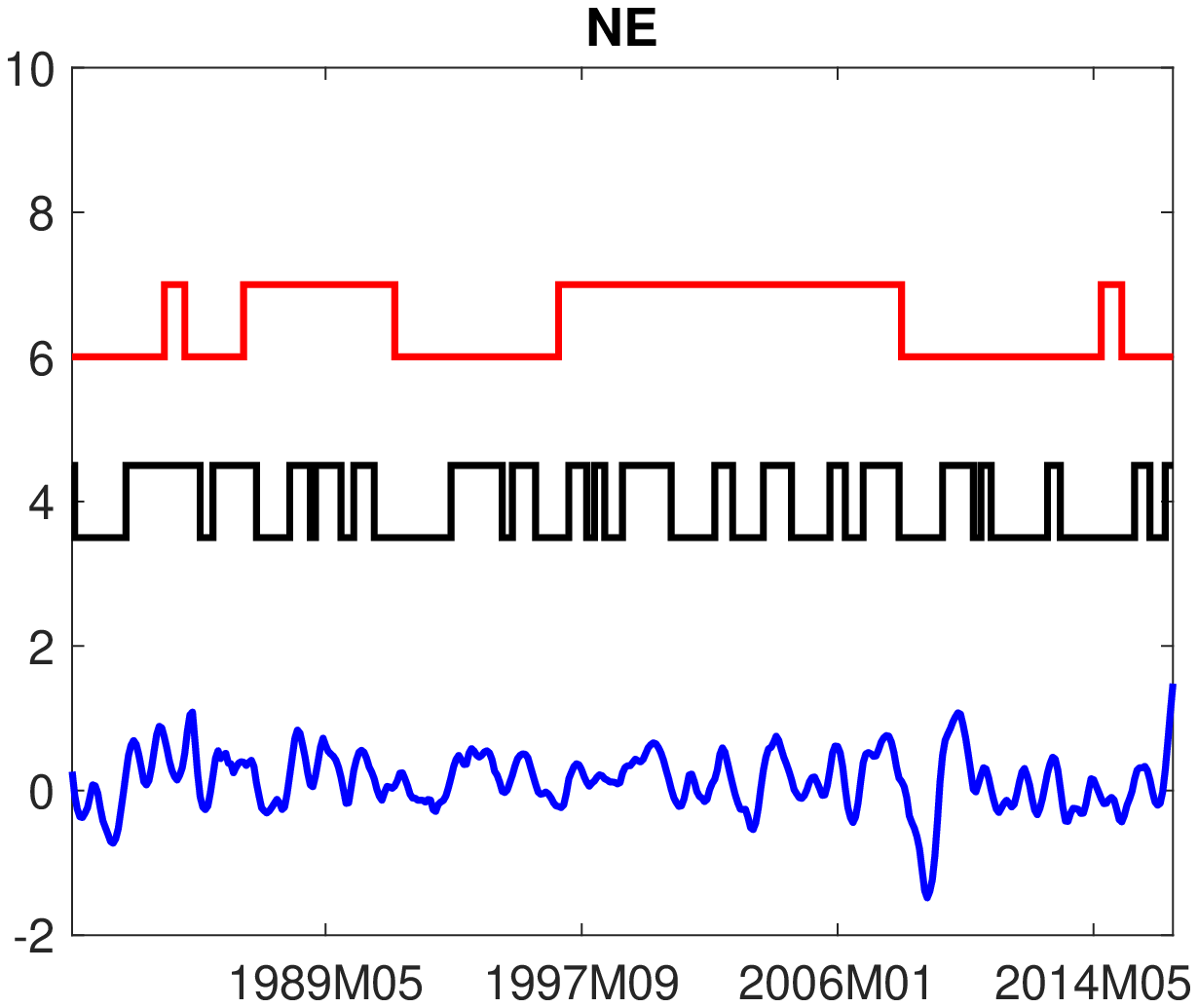}}
 ~ 
    \subfigure{                
 \includegraphics[width=0.17\textwidth]{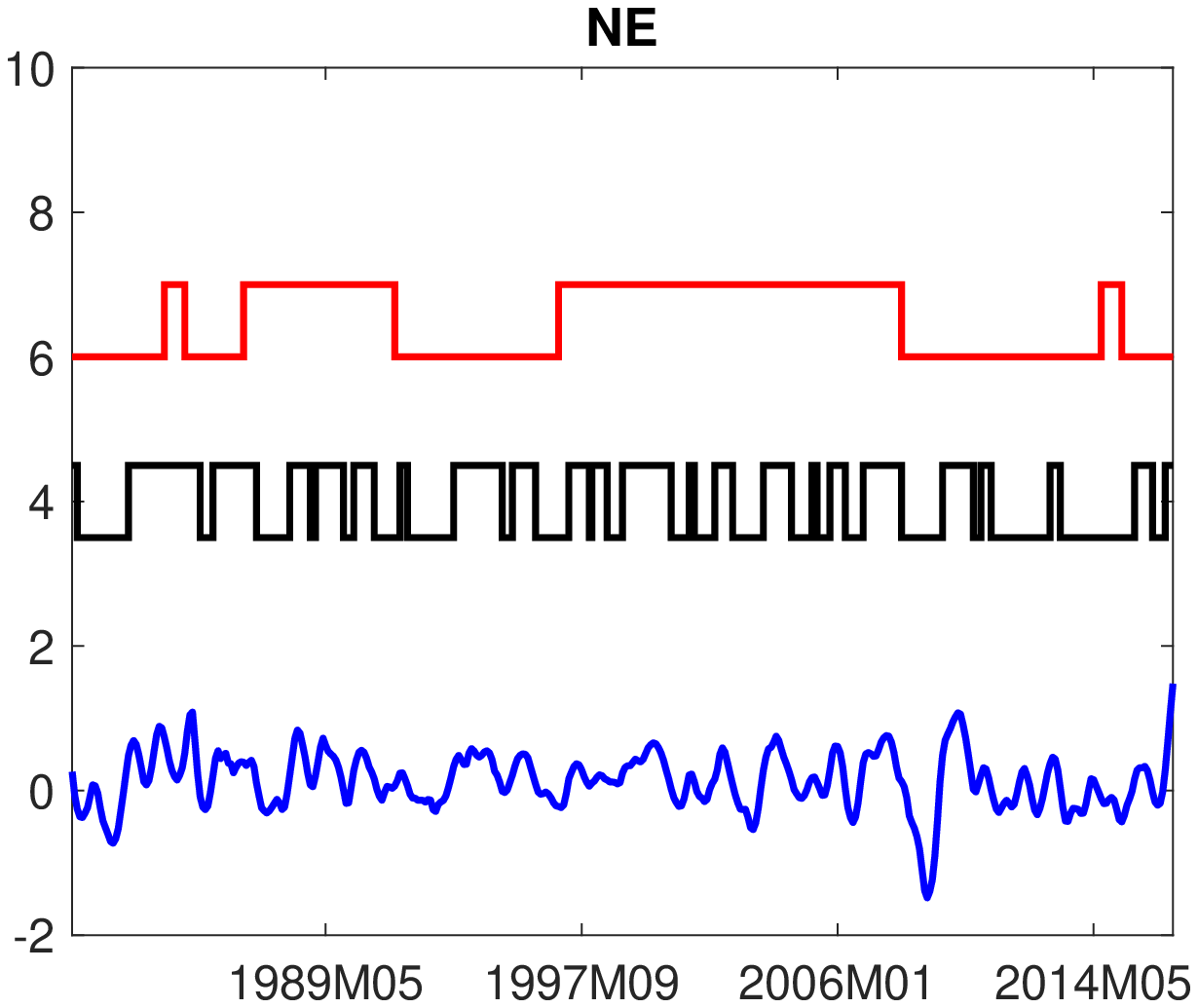}}
 ~ 
    \subfigure{                
 \includegraphics[width=0.17\textwidth]{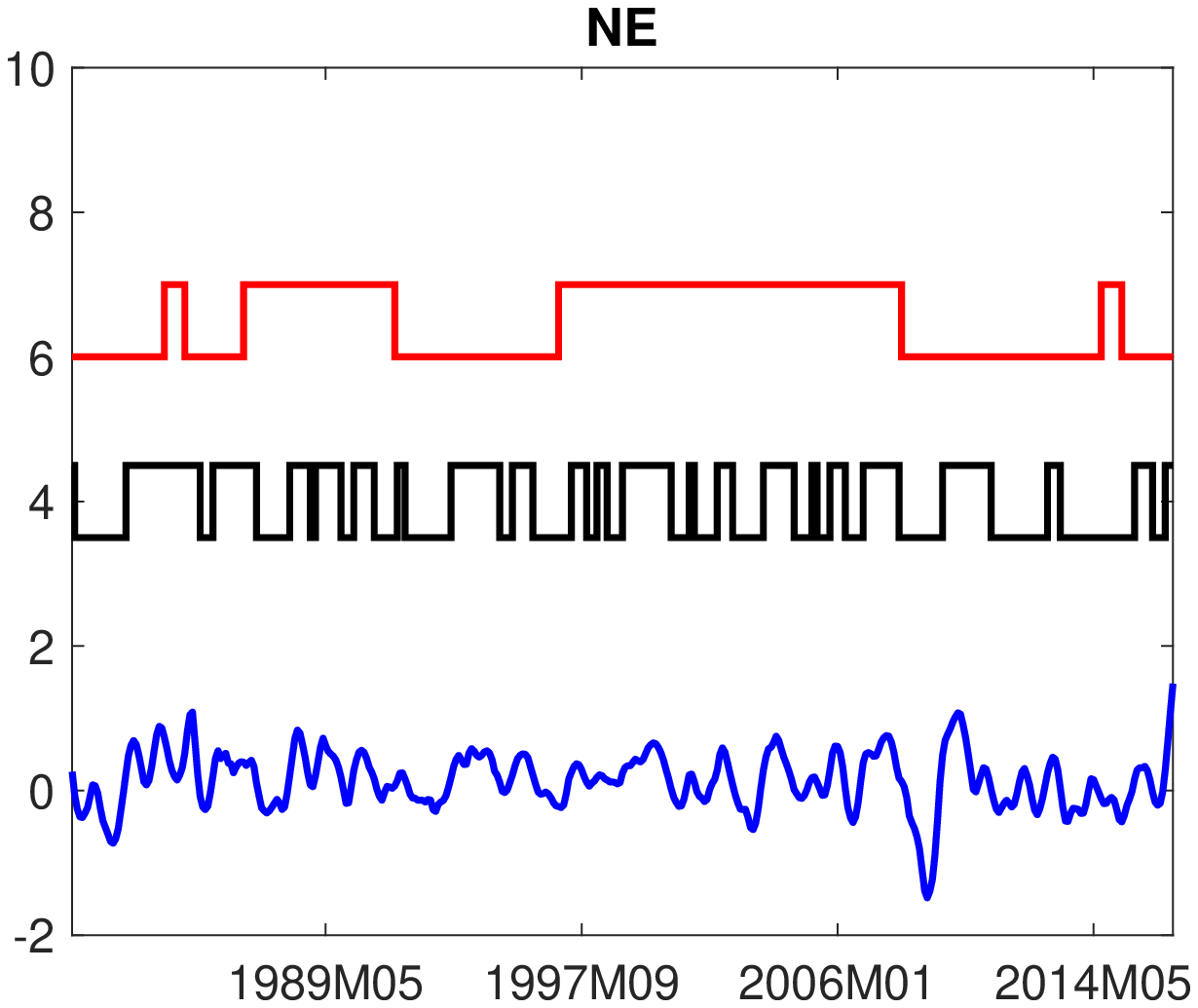}}
 ~ 
    \subfigure{                
 \includegraphics[width=0.17\textwidth]{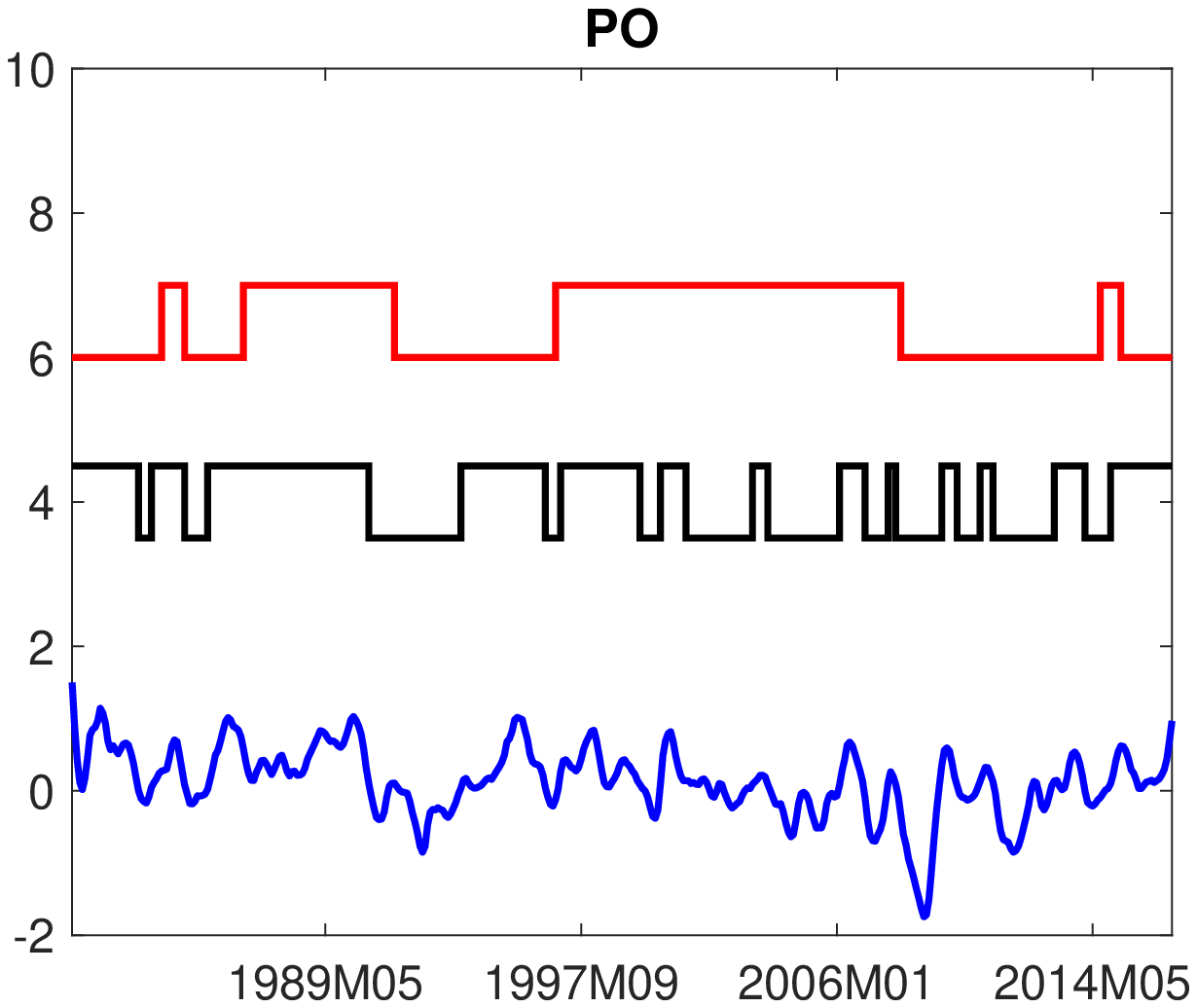}}
  ~
    \subfigure{                
 \includegraphics[width=0.17\textwidth]{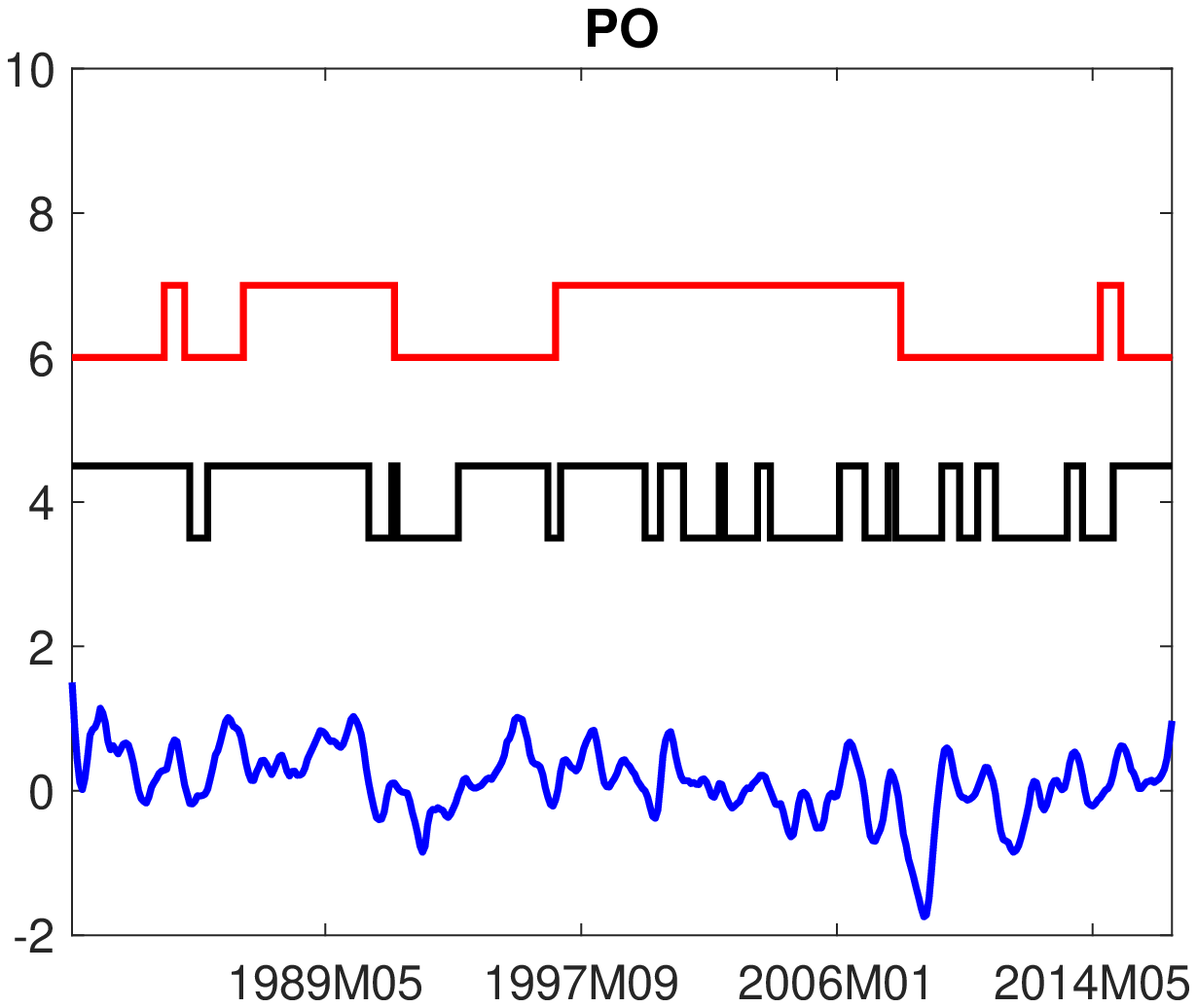}}
 ~  
    \subfigure{                
 \includegraphics[width=0.17\textwidth]{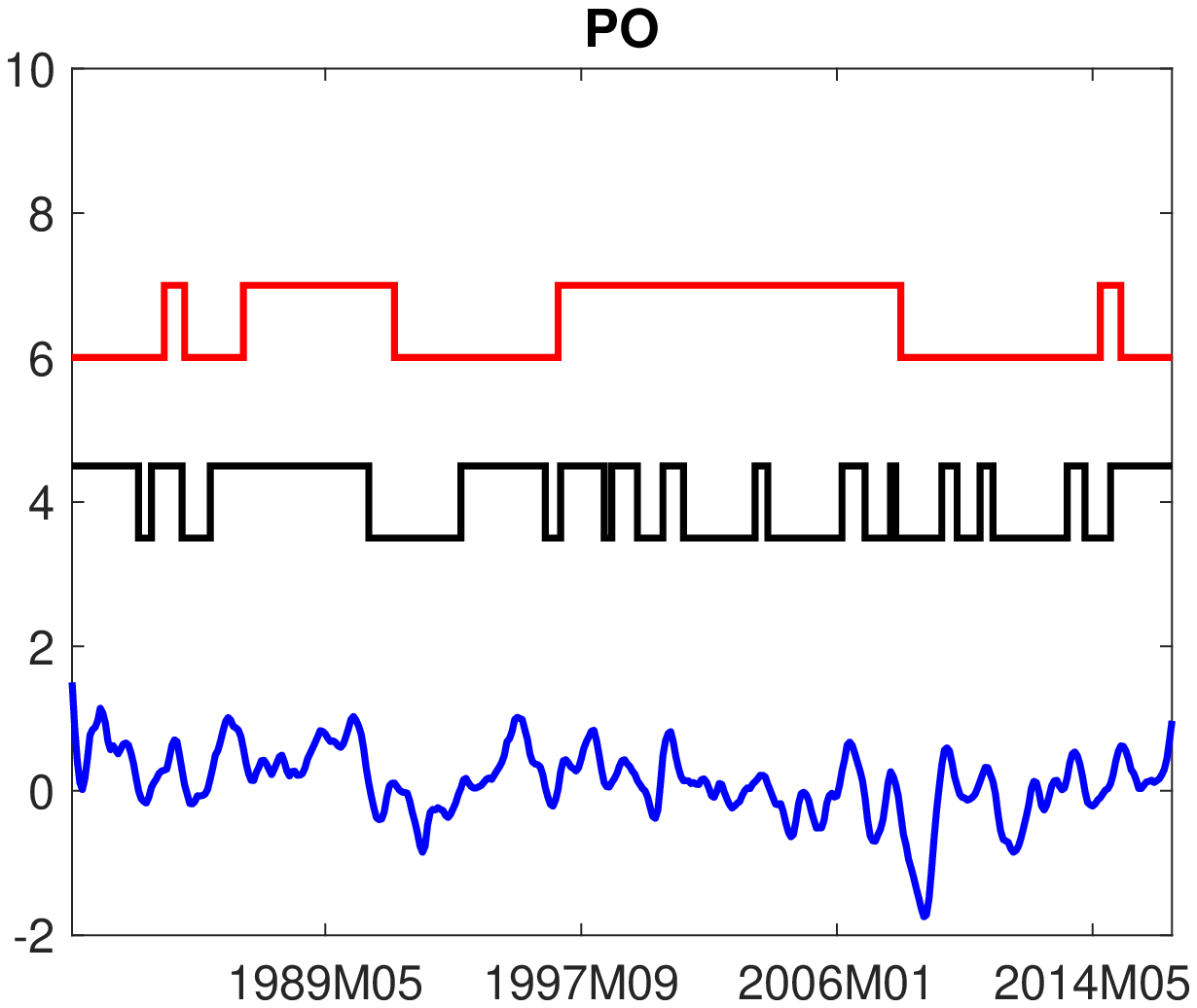}}
 ~ 
    \subfigure{                
 \includegraphics[width=0.17\textwidth]{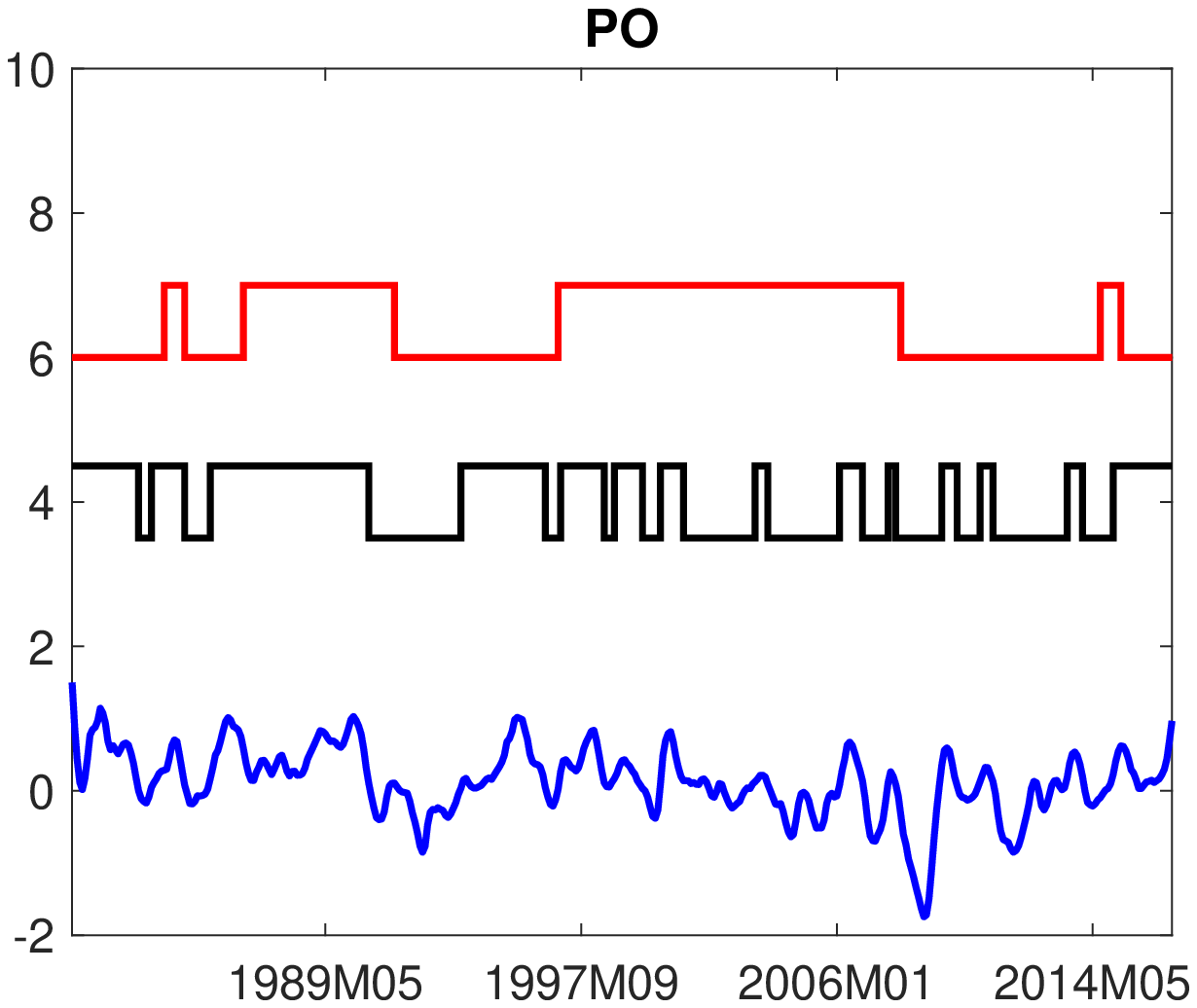}}
 ~ 
    \subfigure{                
 \includegraphics[width=0.17\textwidth]{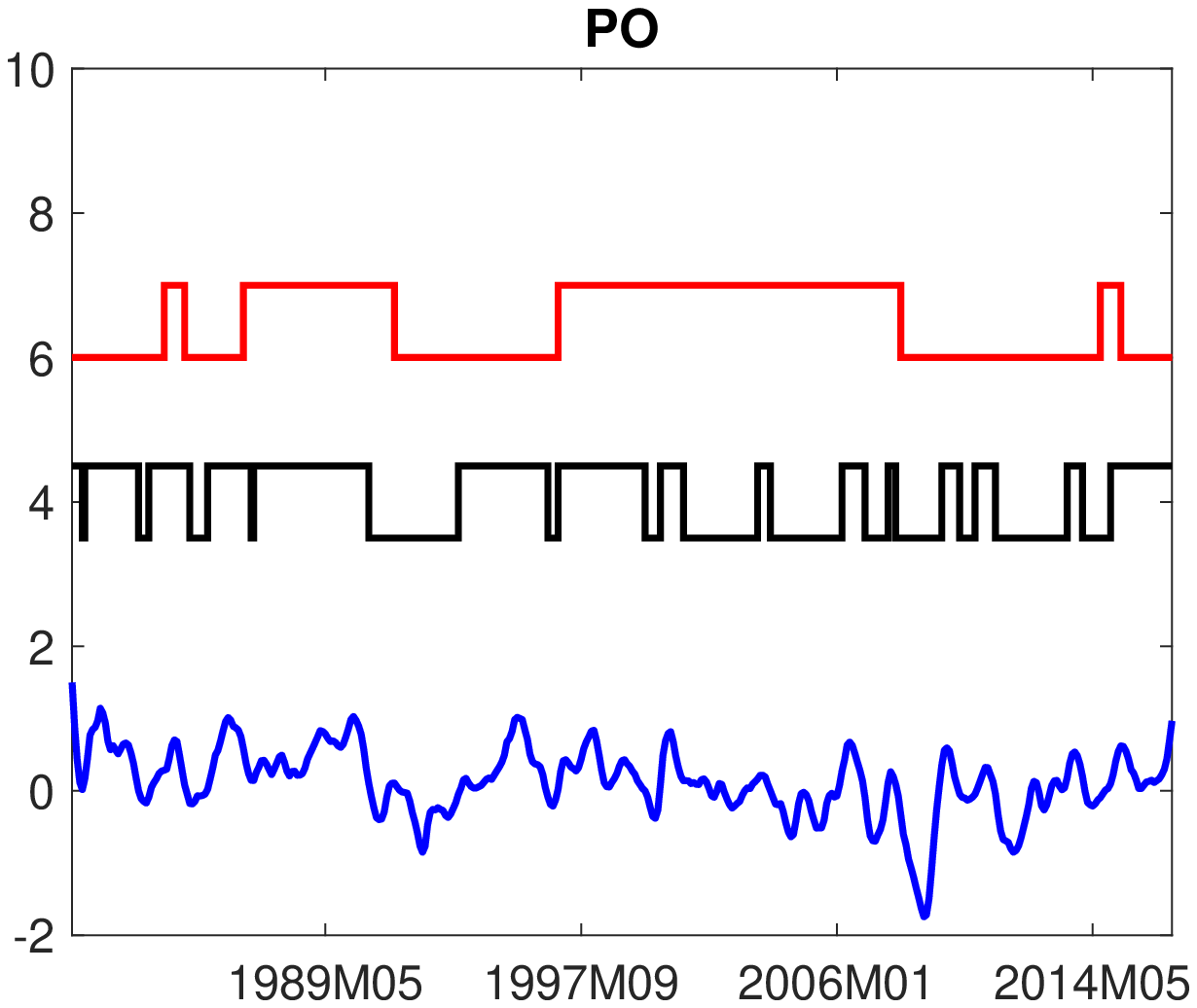}}
 ~ 
    \subfigure{                
 \includegraphics[width=0.17\textwidth]{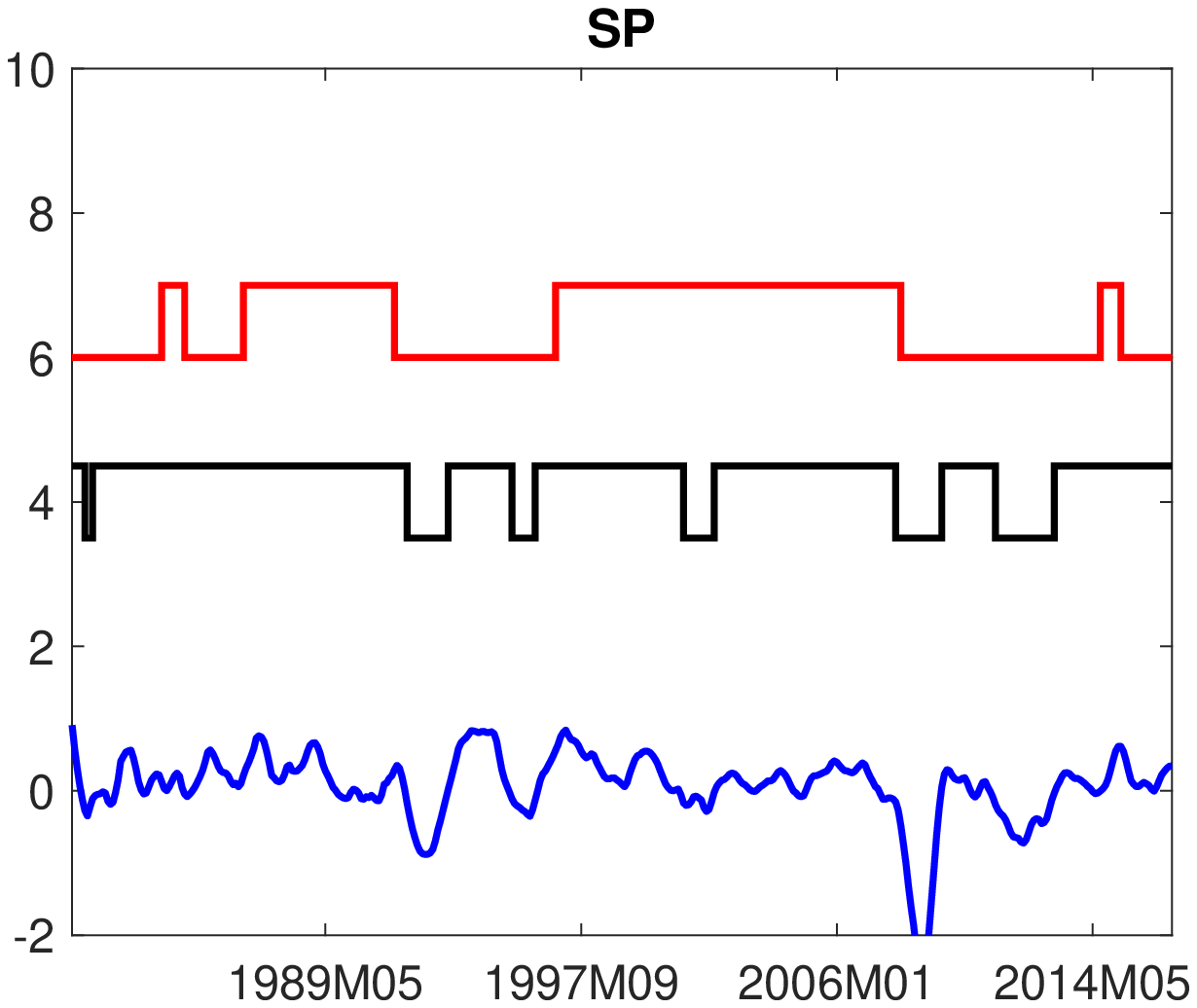}}
  ~
    \subfigure{                
 \includegraphics[width=0.17\textwidth]{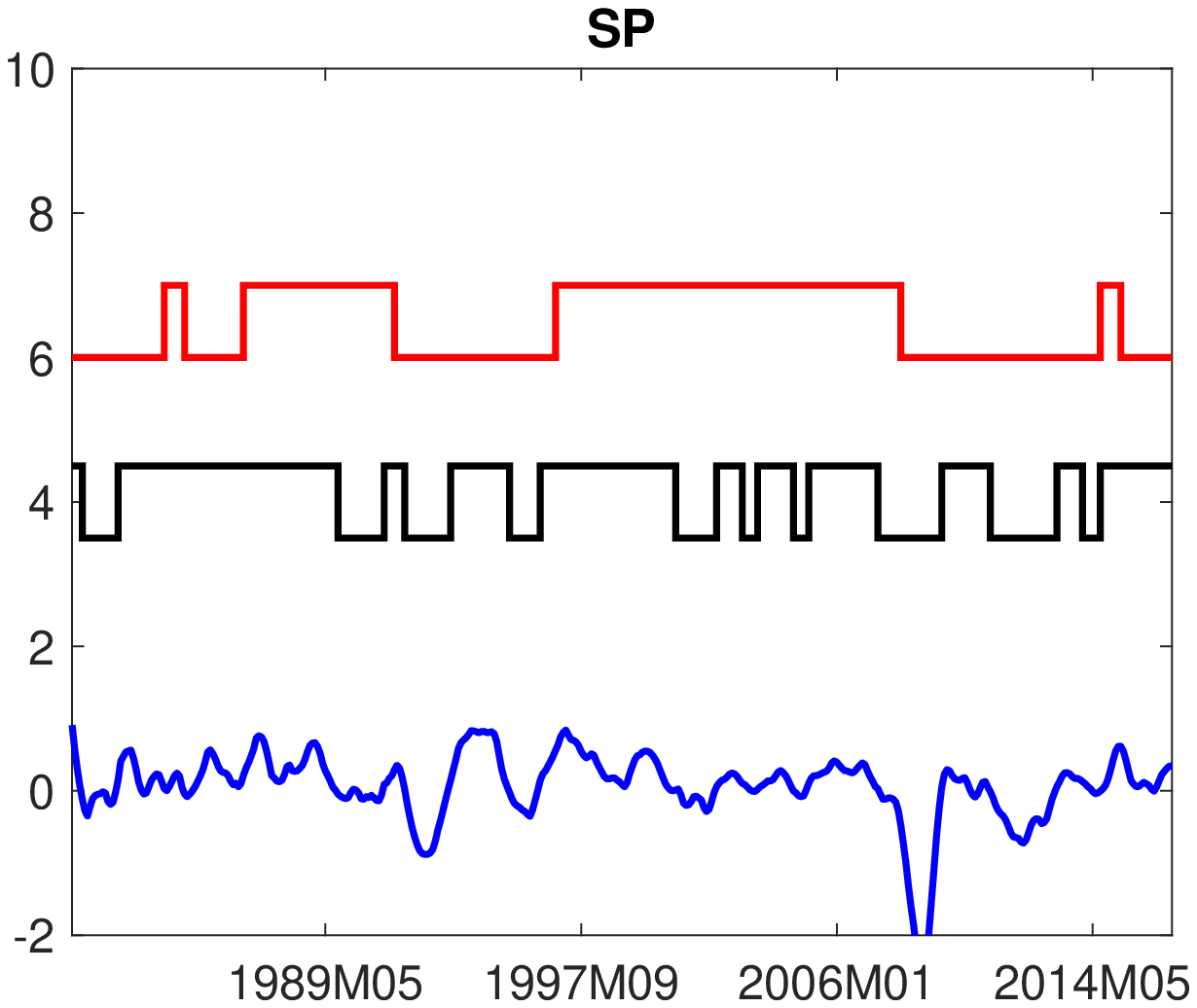}}
 ~  
    \subfigure{                
 \includegraphics[width=0.17\textwidth]{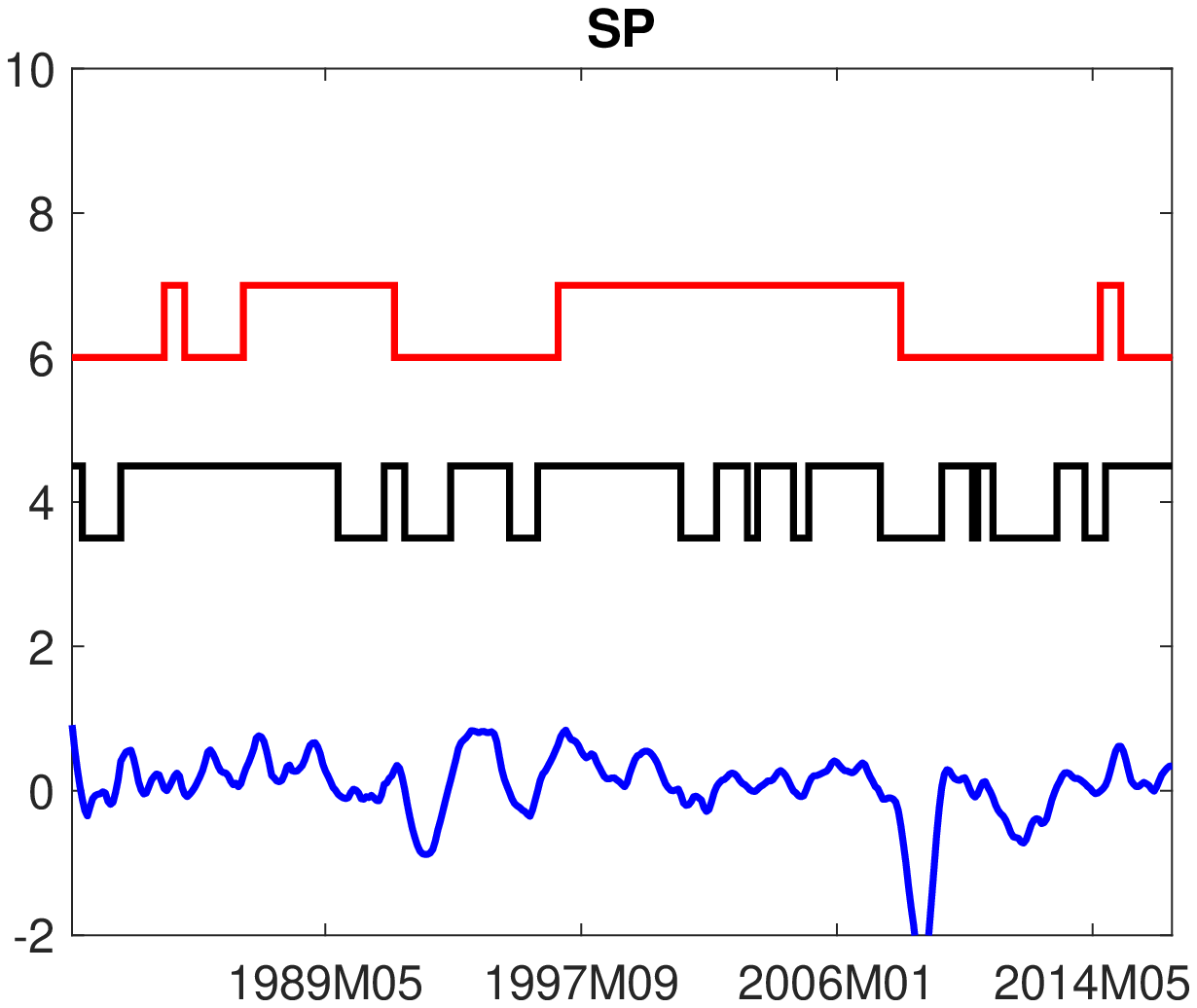}}
 ~ 
    \subfigure{                
 \includegraphics[width=0.17\textwidth]{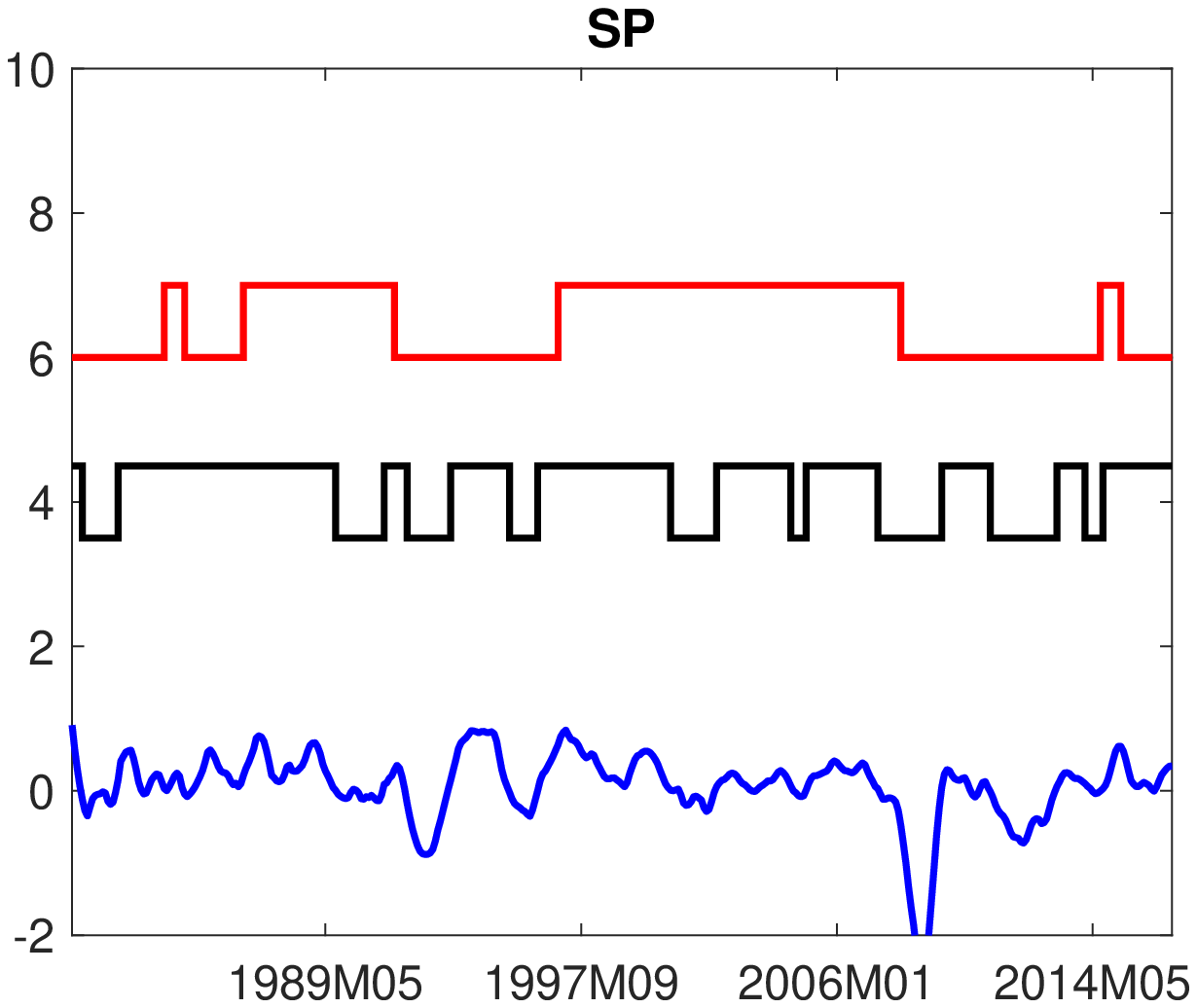}}
 ~ 
    \subfigure{                
 \includegraphics[width=0.17\textwidth]{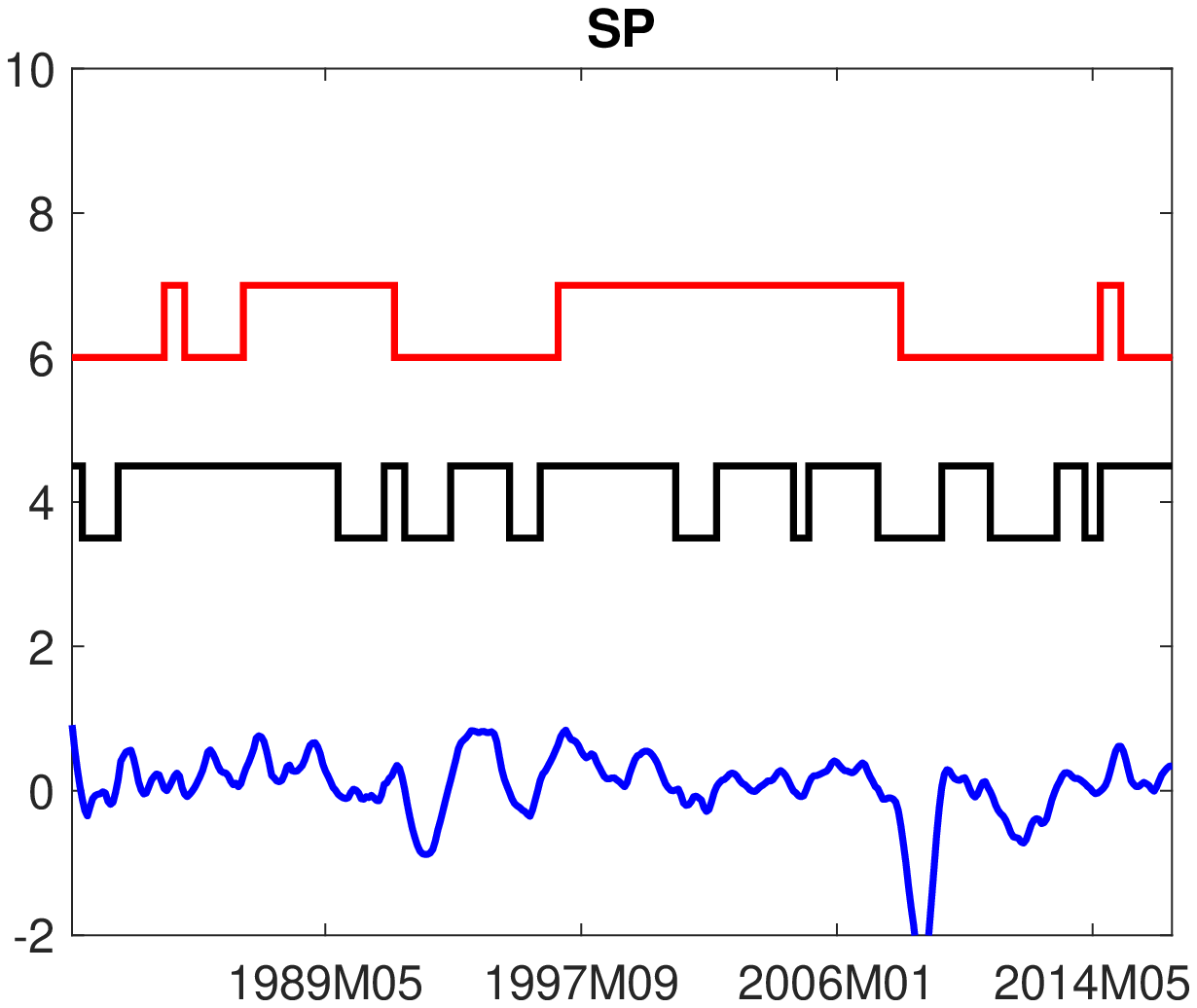}}
 ~ 
    \subfigure{                
 \includegraphics[width=0.17\textwidth]{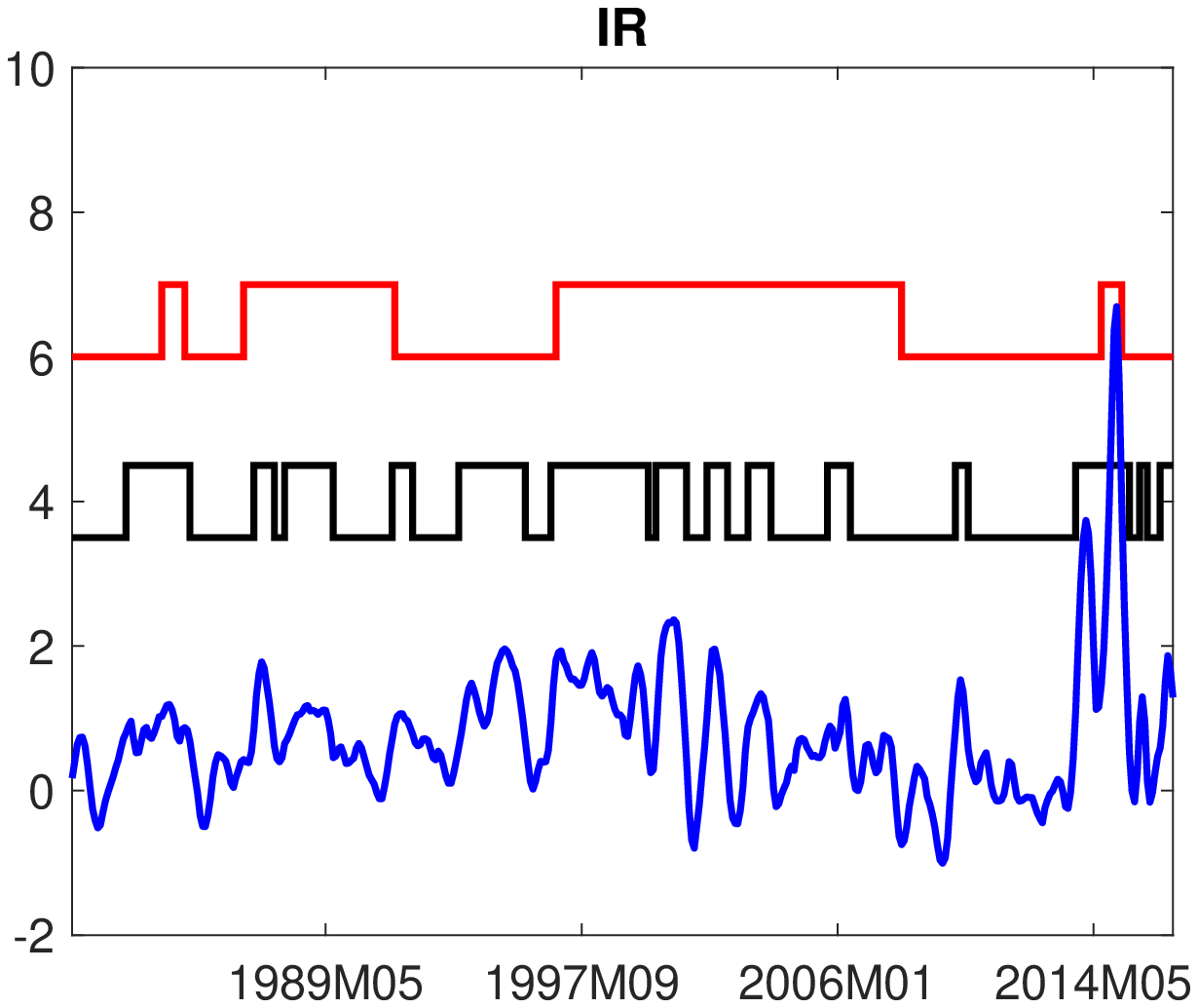}}
  ~
    \subfigure{                
 \includegraphics[width=0.17\textwidth]{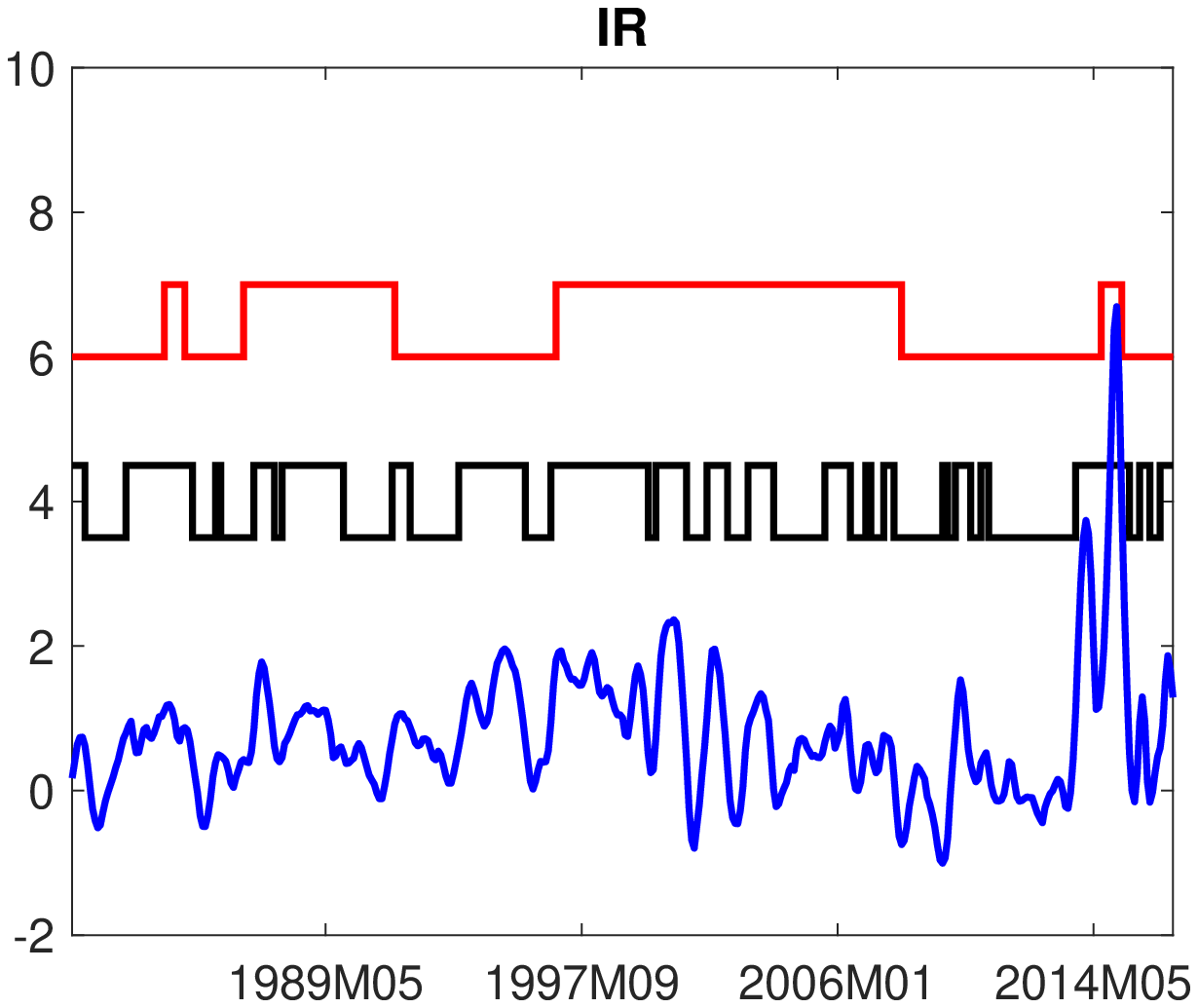}}
 ~  
    \subfigure{                
 \includegraphics[width=0.17\textwidth]{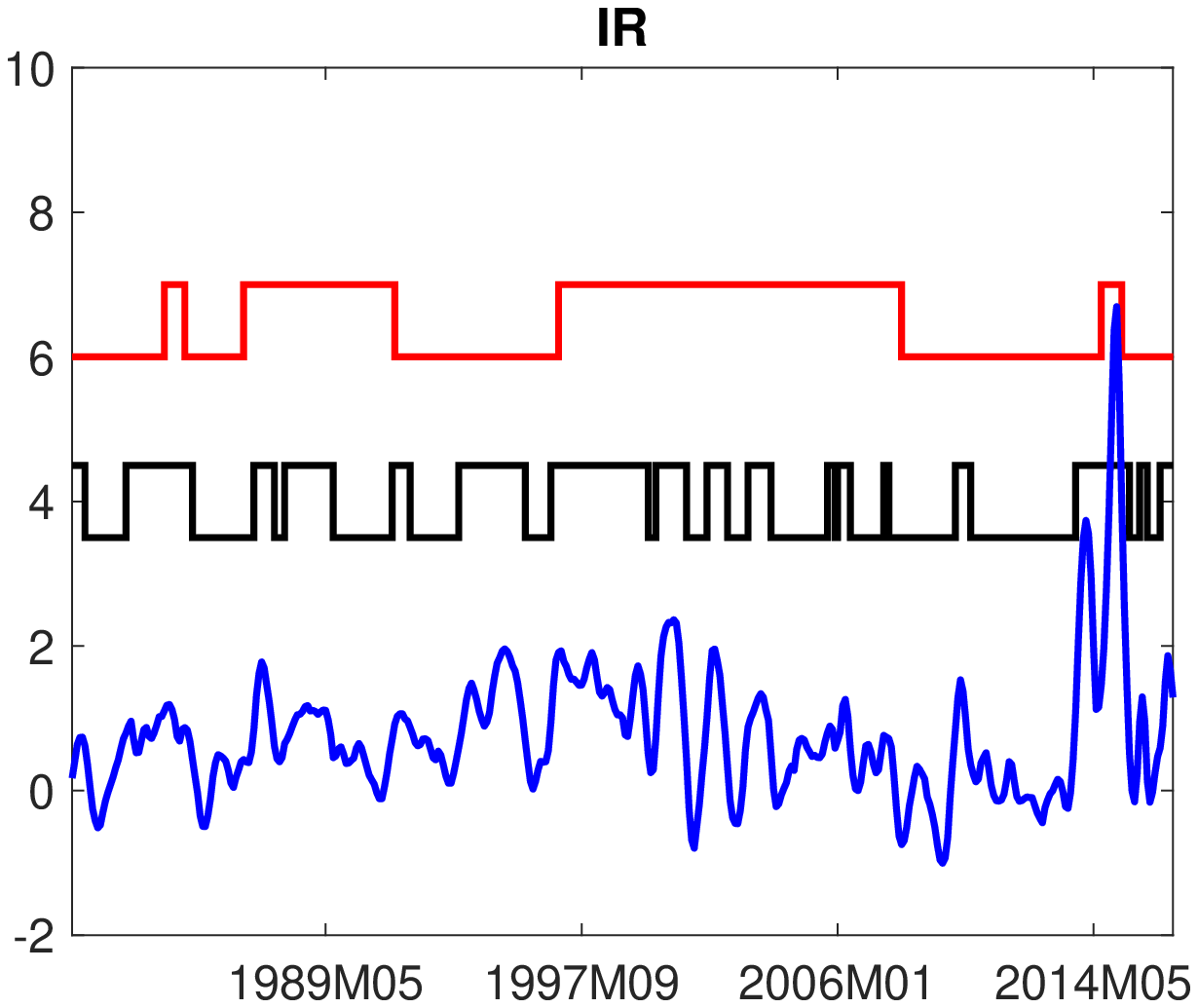}}
 ~ 
    \subfigure{                
 \includegraphics[width=0.17\textwidth]{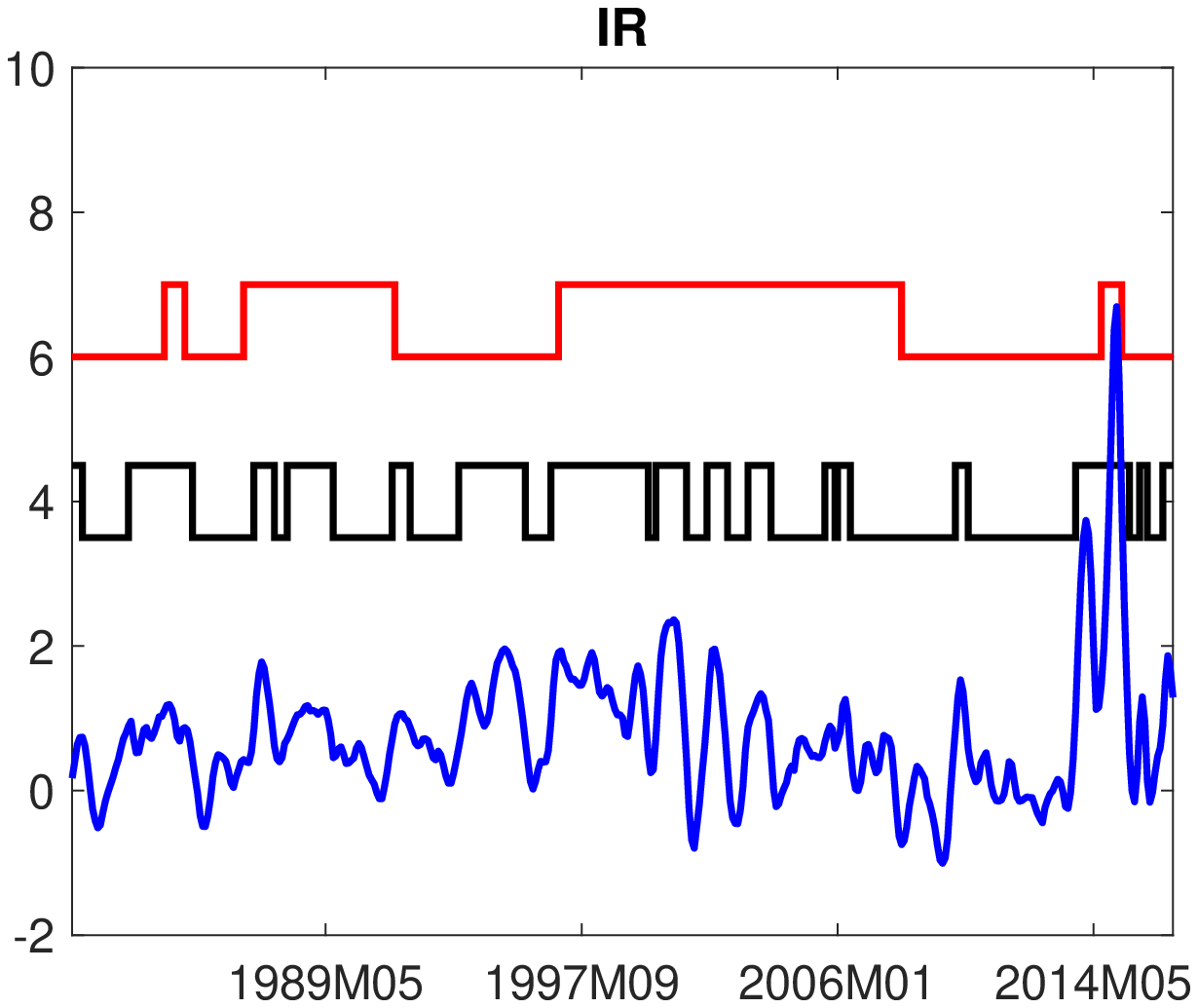}}
 ~ 
    \subfigure{                
 \includegraphics[width=0.17\textwidth]{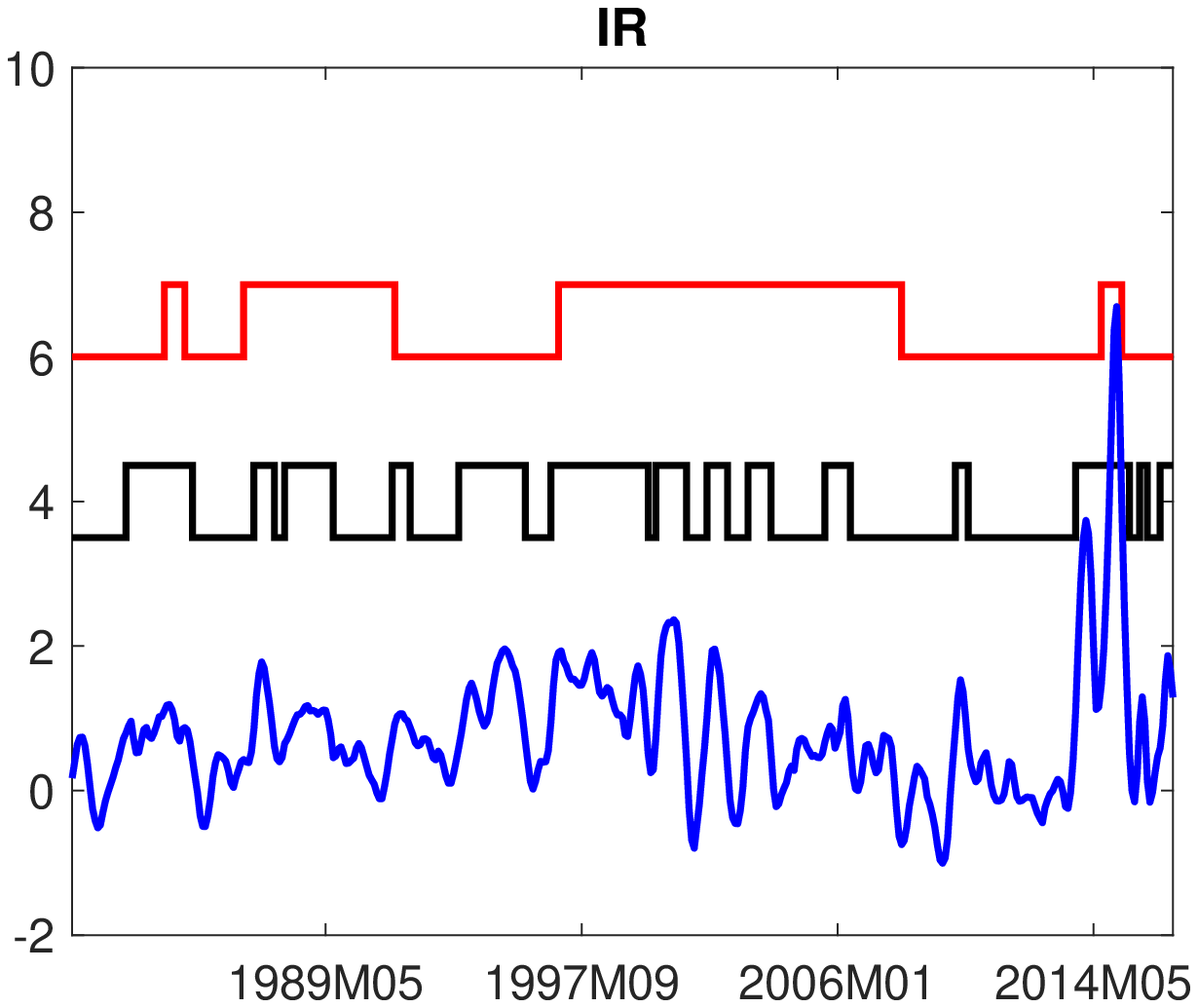}}
 ~ 
 \caption{Estimated cycles for different specifications of the unconstrained PMS models (columns) and countries (rows). The first column represent estimates of the climate conditions independent unconstrained PMS model, the second columns to the fourth column respectively correspond to the estimates of the unconstrained PMS model subject to temperature, drought and rainfall, while the last column represent estimates from unconstrained PMS taking all the climate risk indicators into consideration simultaneously.  In each plot country-specific business cycles (black line), financial cycle (red line) and country specific industrial production (IPI) growth (blue line).}\label{FigCyc1NPooling2}
 \end{figure}

 \begin{figure}[h!]
         \centering
         \subfigure[None ]{                
                 \includegraphics[width=0.17\textwidth]{FigNPooling/FinCy.eps}
                 \label{fig:unconstrainedFinCyApp}}
         ~
         \subfigure[Temperature (CSU) only]{                
                 \includegraphics[width=0.17\textwidth]{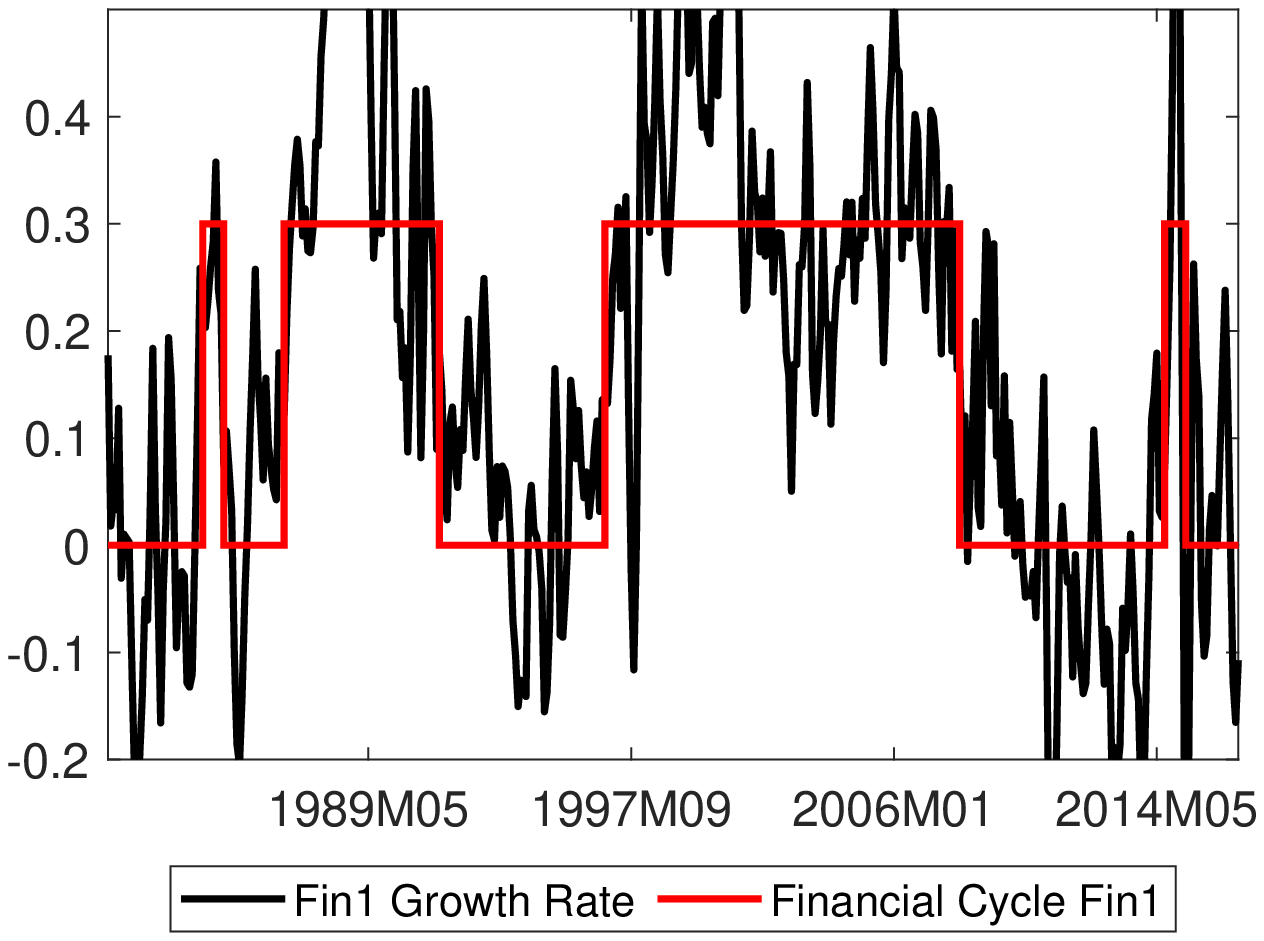}
                 \label{fig:unconstrainedCSUFinCyApp}}
         ~
         \subfigure[Drought (SPI) only]{                
                 \includegraphics[width=0.17\textwidth]{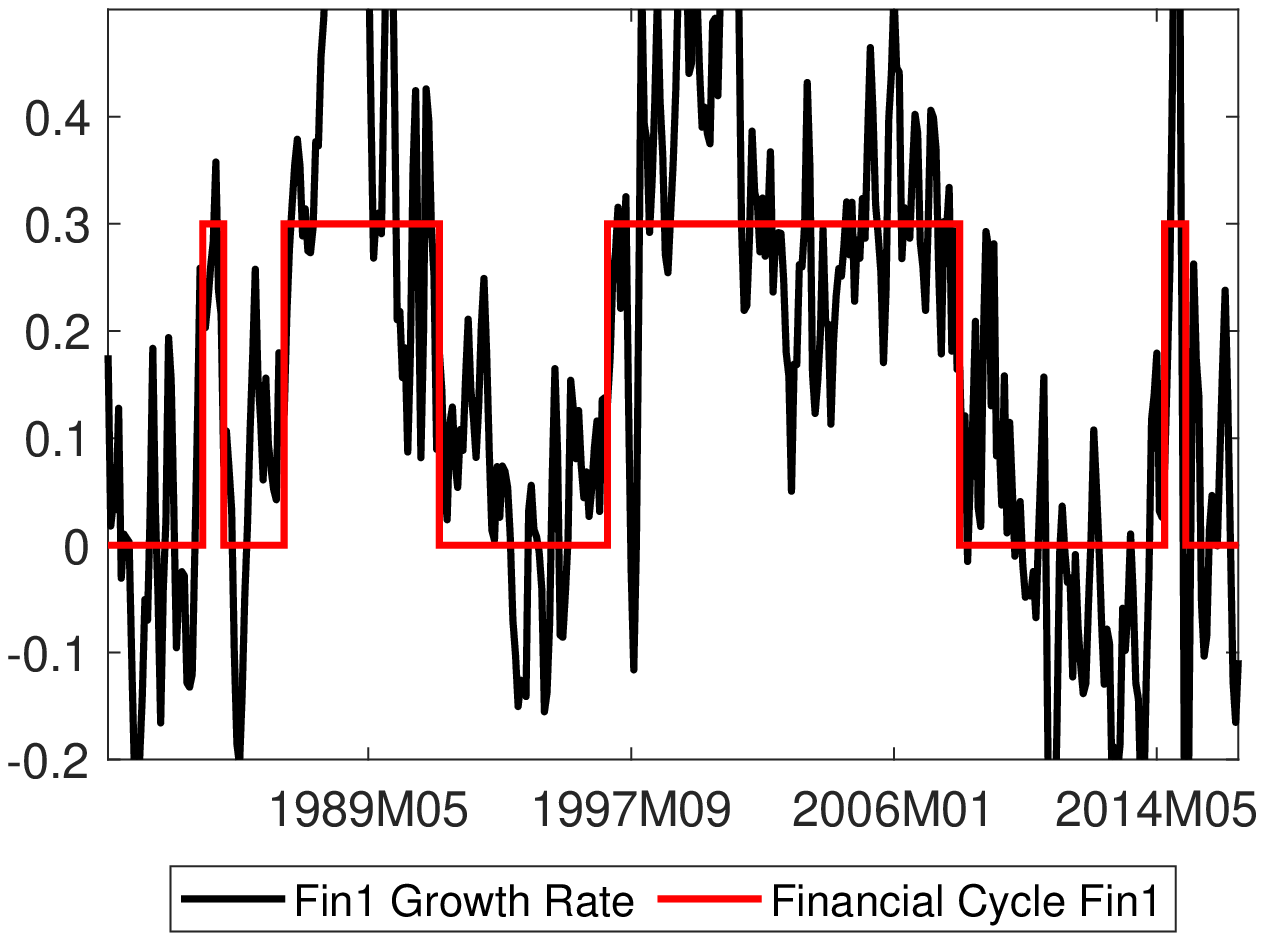}
                 \label{fig:unconstrainedSPIFinCyApp}}
         ~
         \subfigure[Rainfall (r20mm) only]{                
                 \includegraphics[width=0.17\textwidth]{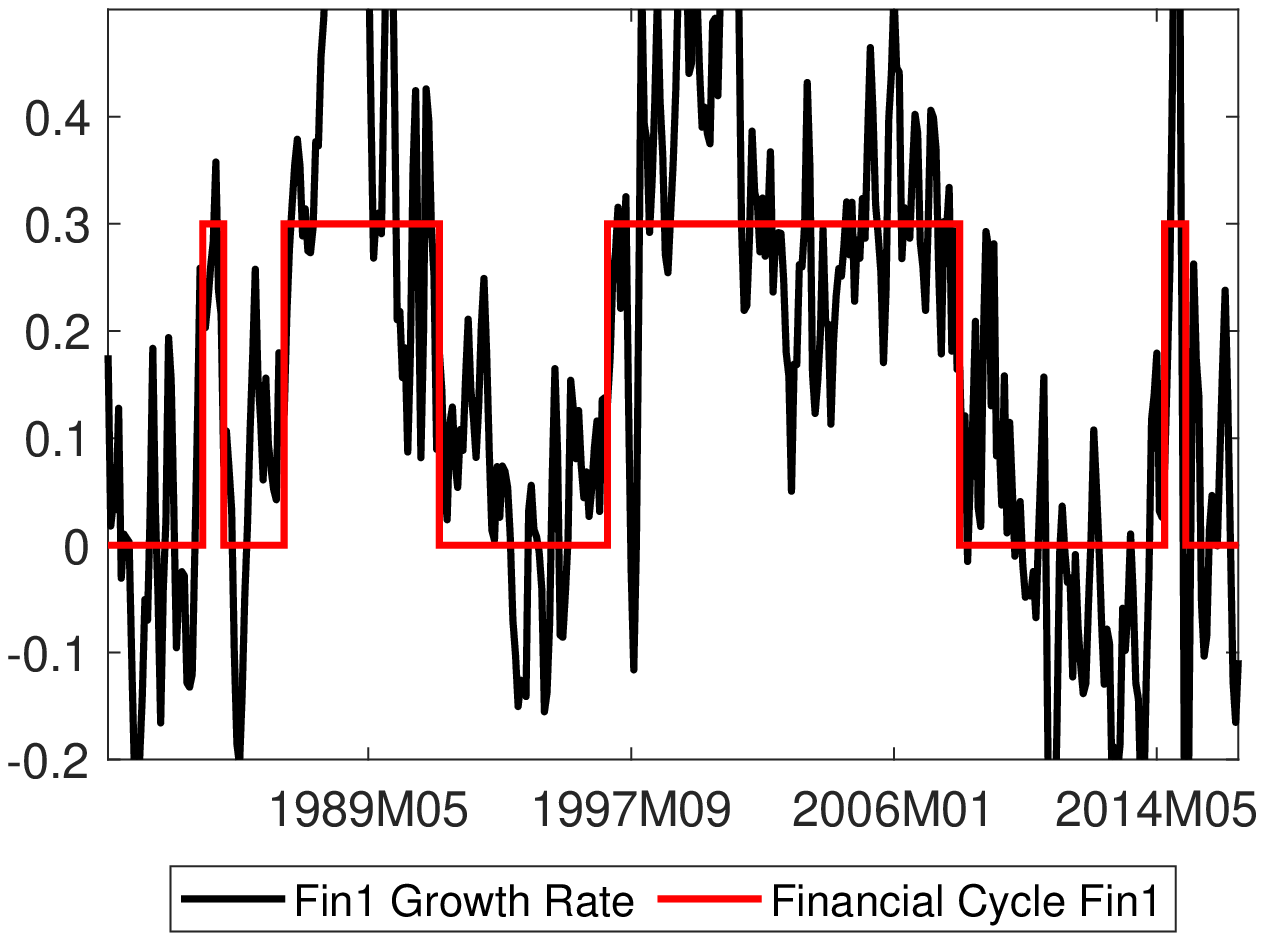}
                 \label{fig:unconstrainedR20MMFinCyApp}}        
         ~
         \subfigure[ALL climate indices]{                
                  \includegraphics[width=0.17\textwidth]{FigNPoolingCSUSPIR20MMCV/FinCy.eps}
                  \label{fig:unconstrainedALLFinCyApp}}        
         \caption{Financial index growth rates (black solid) and the estimated financial cycle $\hat{S}_{x,t}$ for the unrestricted PMS model subject to different climate shock. Each column display the output of the PMS models when it is subjected respectively to no climate condition (column 1), temperature only (column 2), drought only (column 3), rainfall only (column 4) and simultaneously to all the climate change indices  (last column).          
         }
 \label{Fin1}
 \end{figure}
 
 \begin{figure}[h!]
         \centering
         \subfigure[None ]{                
                 \includegraphics[width=0.17\textwidth]{FigNPooling/HidSwitch2.eps}
                 \label{fig:unconstrainedHidSwitchApp}}
         ~
         \subfigure[Temprature (CSU) only]{                
                 \includegraphics[width=0.17\textwidth]{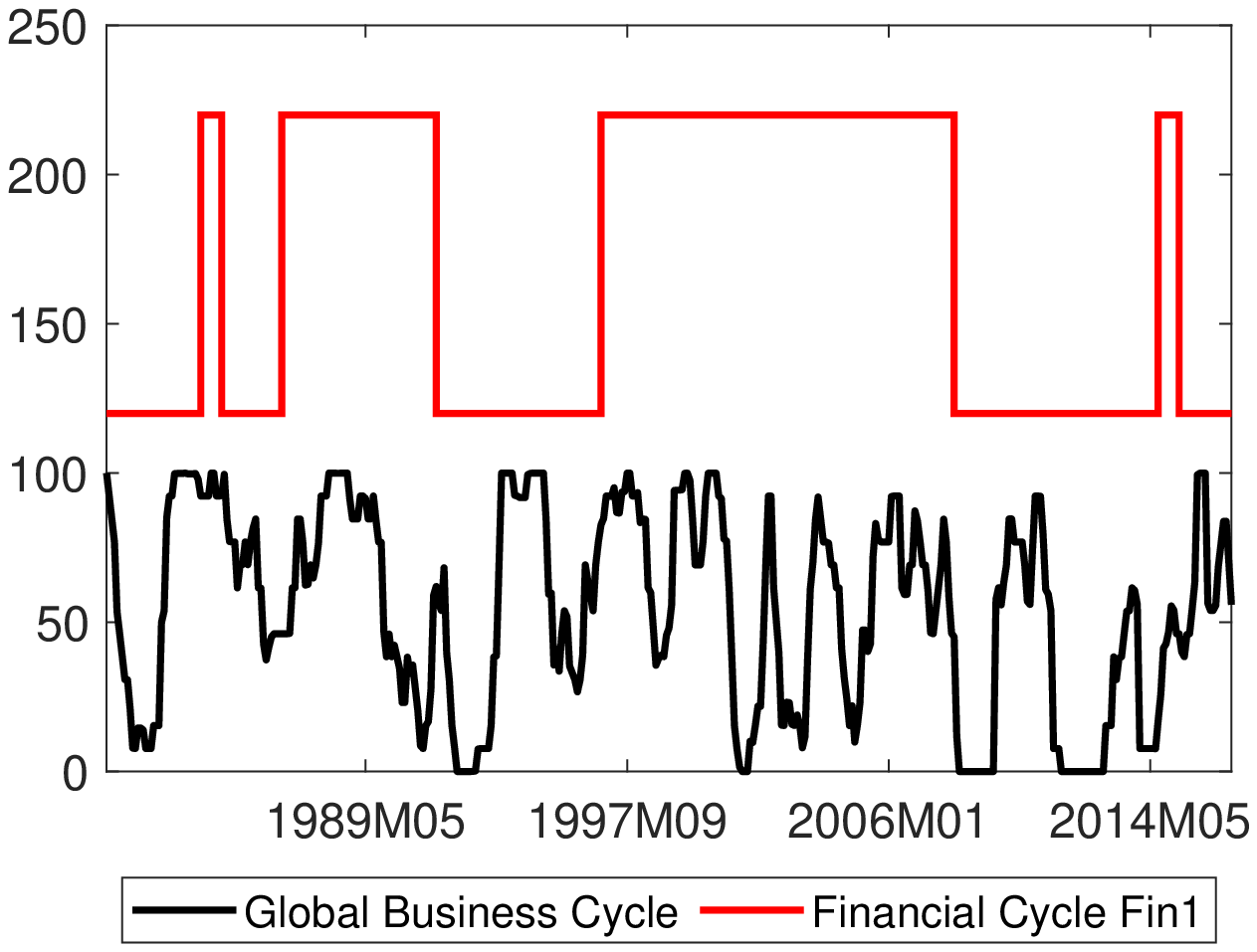}
                 \label{fig:unconstrainedCSUHidSwitchApp}}
         ~
         \subfigure[Drought (SPI) only]{                
                 \includegraphics[width=0.17\textwidth]{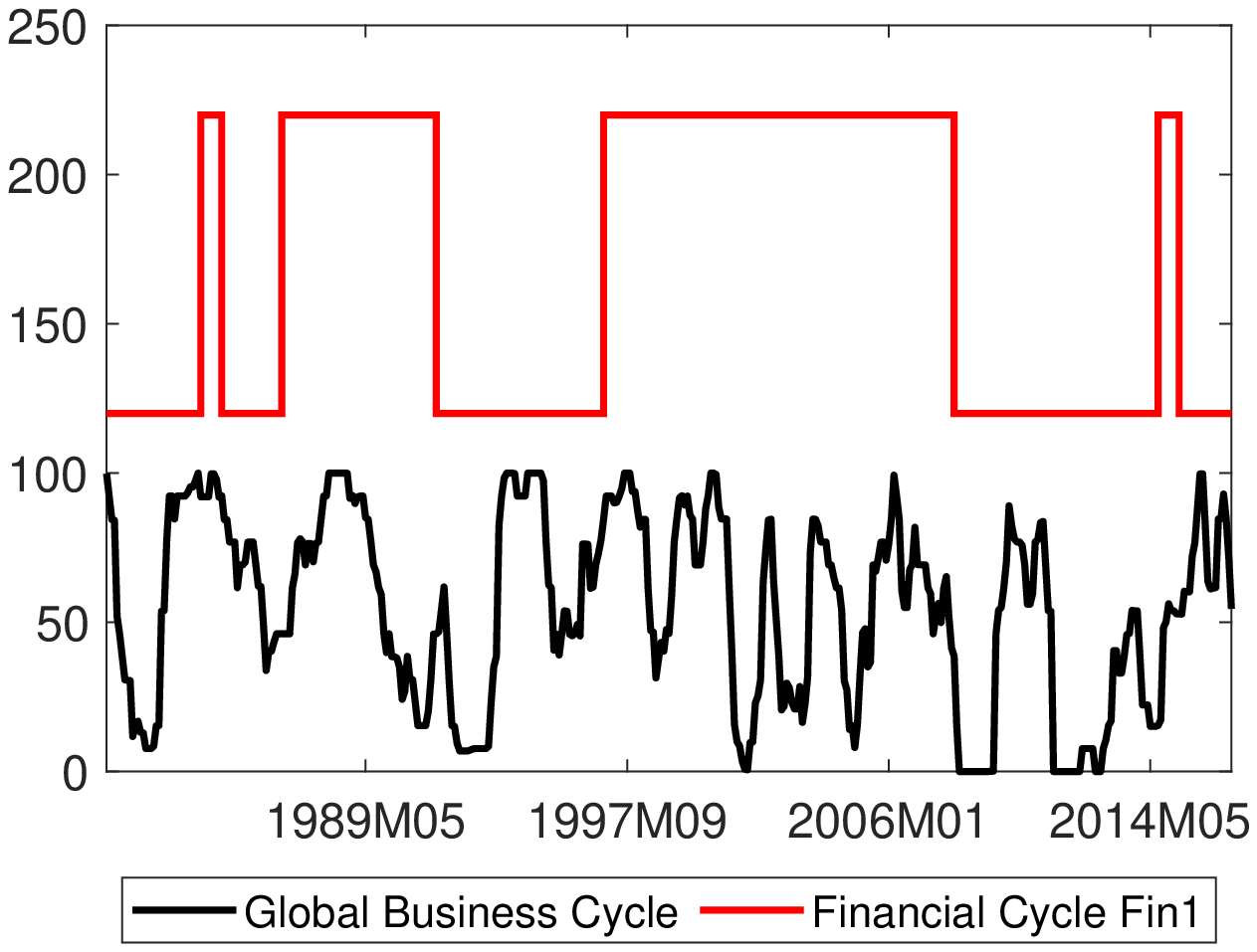}
                 \label{fig:unconstrainedSPIHidSwitchApp}}
         ~
         \subfigure[Rainfall (r20mm) only]{                
                 \includegraphics[width=0.17\textwidth]{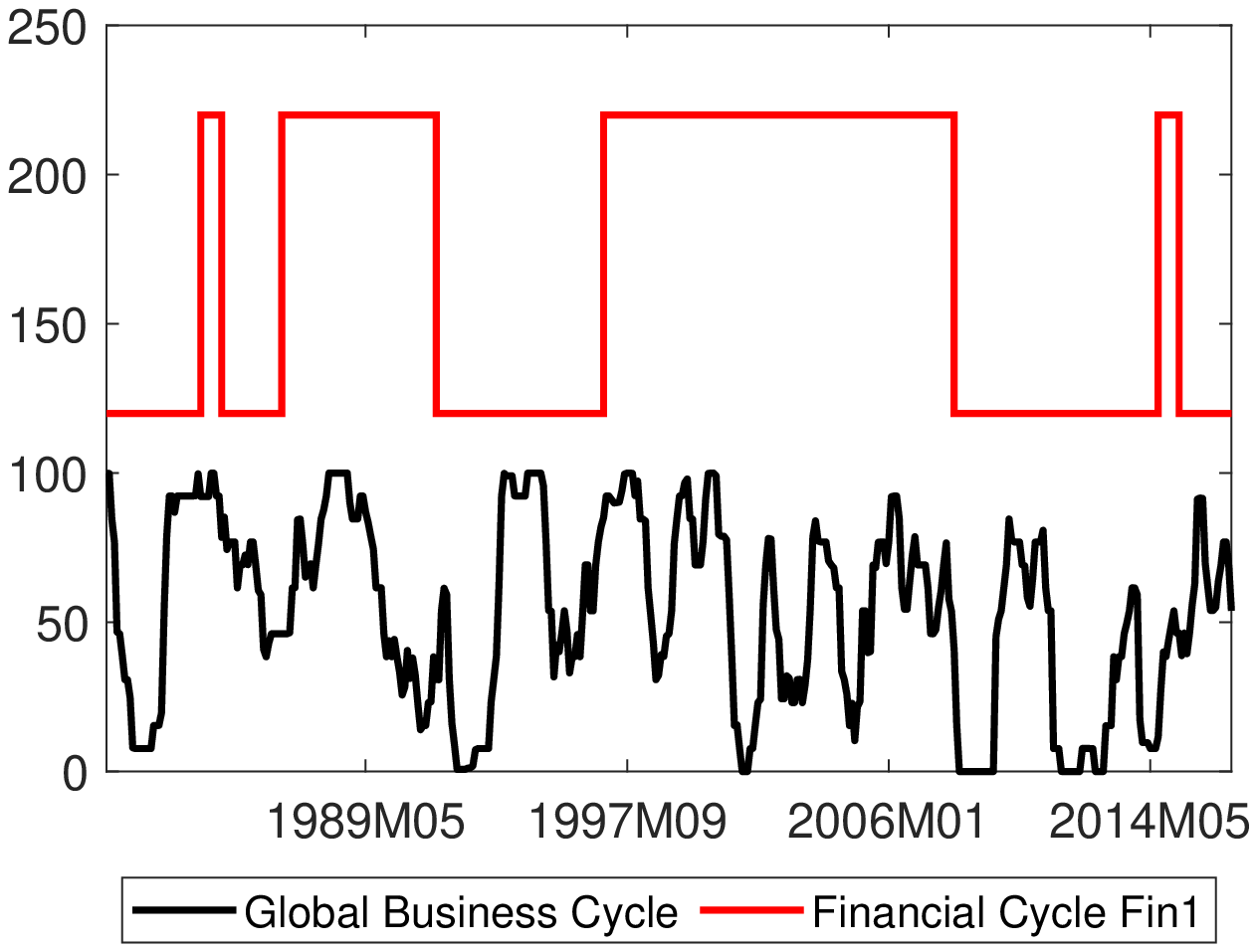}
                 \label{fig:unconstrainedR20MMHidSwitchApp}}        
         ~
         \subfigure[ALL climate indices]{                
                 \includegraphics[width=0.17\textwidth]{FigNPoolingCSUSPIR20MMCV/HidSwitch2.eps}
                 \label{fig:unconstrainedALLHidSwitchApp}}        
         \caption{Estimated global common business cycles ($m_{1}(\hat{S}_{y,t})$, black line) and financial cycle ($\hat{S}_{x,t}$ red line) for the unconstrained PMS model. Column 1 in the figure  display the output of the PMS models in the absence climate conditions, while the others represent output of the PMS models subject to indicators of changes in climate conditions: temperature (CSU); drought (SPI) and rainfall (r20mm). The output of the PMS model when subjected to temperature, drought and rainfall are displayed respectively in Columns 2, 3, and 4, while column 5 displays the PMS output while simultaneously taking all the climate risk indicators into consideration. }\label{GlobalApp}
 \end{figure}
 \newpage
 \clearpage 
 \section{Impact of change in weather conditions}
 \label{Appendix4}
 \begin{figure}[h!]
         \centering
                 \includegraphics[width=0.5\textwidth]{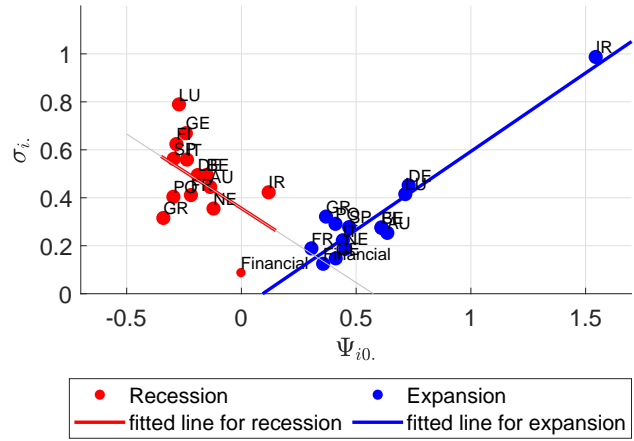}
 		\caption{Estimates of monthly state-level Manufacturing sector indices and Financial index (Fin1) mean growth rates (horizontal axis) and volatilities (vertical axis) with the Bayesian Panel Markov-switching model. The horizontal axis represents the mean growth rates and the vertical axis the volatilities. Sample period: February 1981 to December 2016 (month on month).}\label{MANParam}
 \end{figure}
 
 \begin{figure}[h!]
         \centering
         \subfigure[Temperature (CSU)]{
                 \includegraphics[width=0.4\textwidth]{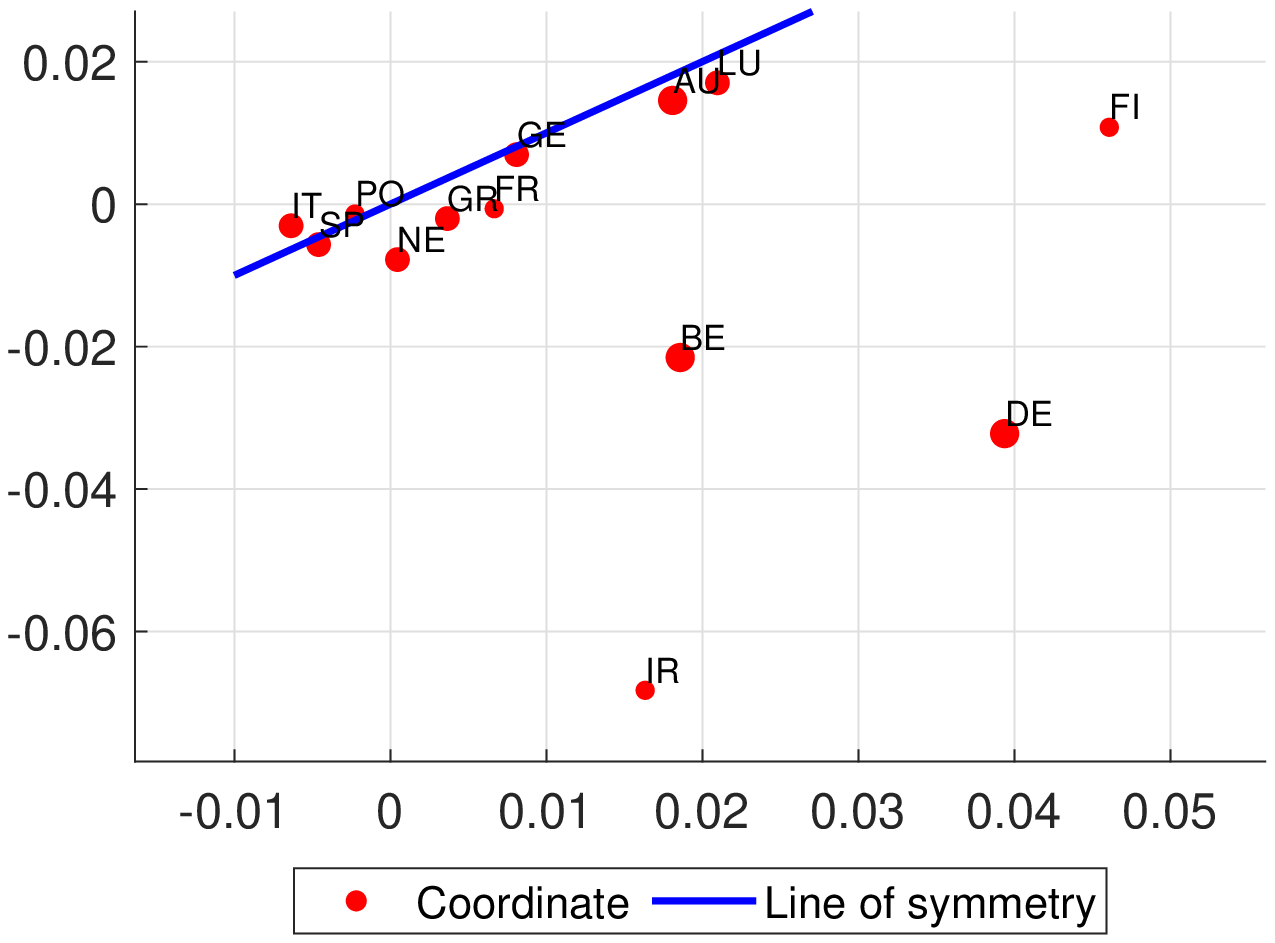}
                 \label{fig:MANunconstrainedCSUAsymCheckALL}}
         ~
         \subfigure[Drought (SPI)]{
                 \includegraphics[width=0.4\textwidth]{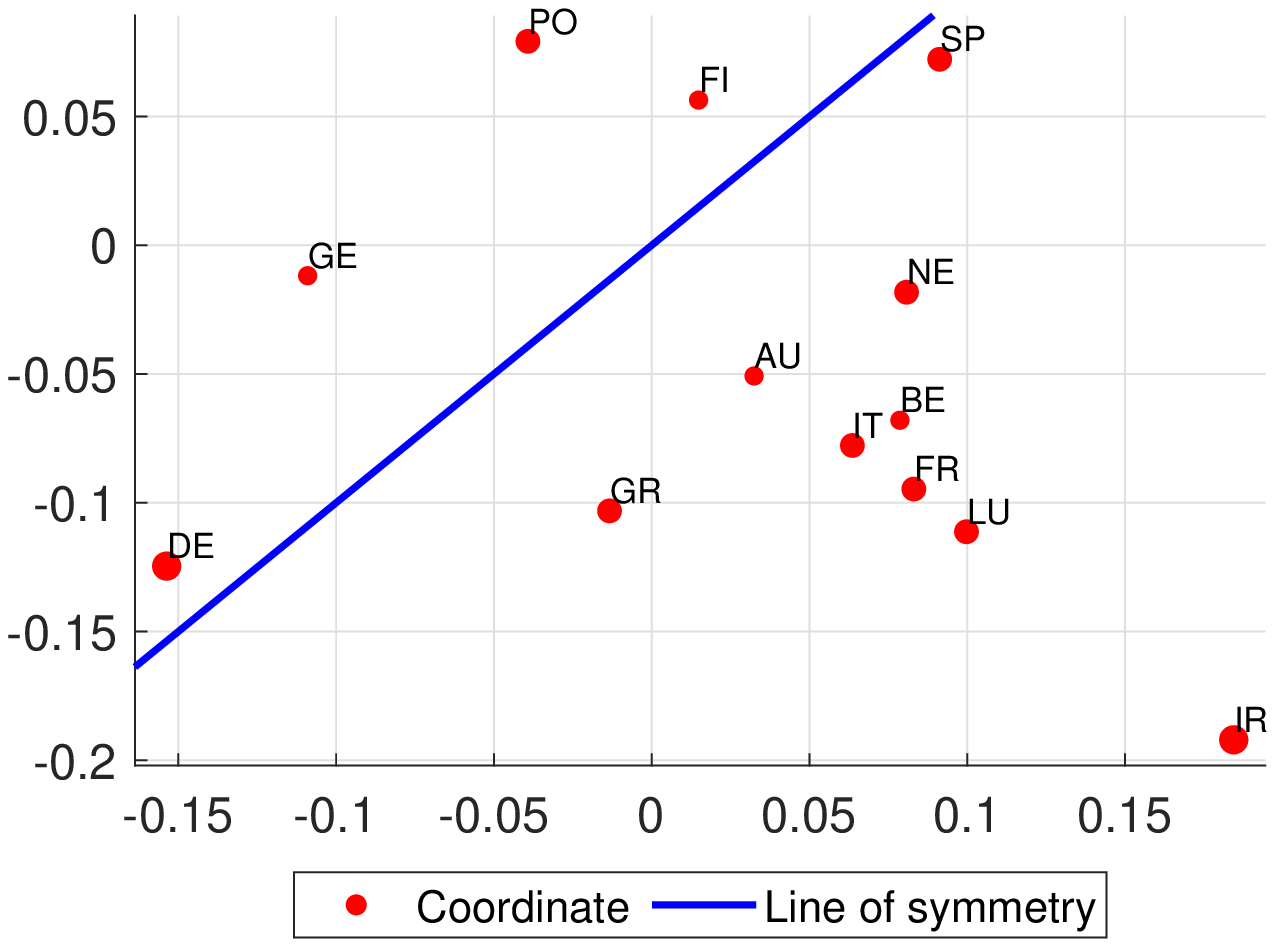}
                 \label{fig:MANunconstrainedSPIAsymCheckALL}}
         ~
         \subfigure[Rainfall (r20mm)]{
                 \includegraphics[width=0.4\textwidth]{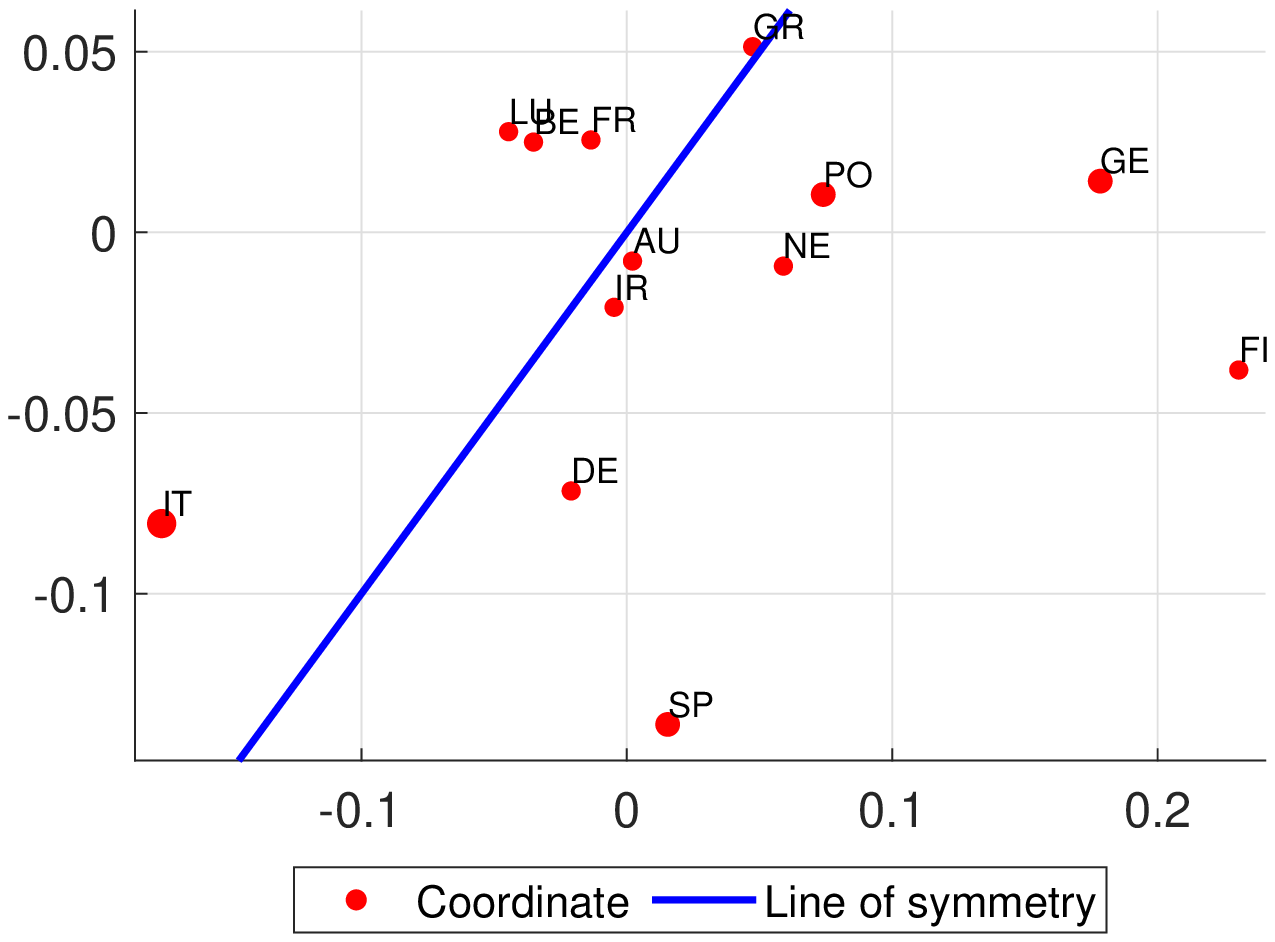}
                 \label{fig:MANunconstrainedR20MMAsymCheckALL}}        
 		\caption{Estimates of the monthly asymmetric effect of the change in the weather conditions on each country's  monthly  Manufacturing sector indices with the Bayesian Panel Markov-switching model. The horizontal and vertical axes respectively, represent parameters capturing the effect of the climate change on IPI during recession and expansion. Sample period: February 1981 to December 2016 (month on month). The climate extreme indices  CSU (panel \ref{fig:MANunconstrainedCSUAsymCheckALL}), SPI (panel \ref{fig:MANunconstrainedSPIAsymCheckALL}) and r20mm (panel \ref{fig:MANunconstrainedR20MMAsymCheckALL}), respectively, captures consecutive hot days , droughts , and number of days with heavy rainfall.
 		}\label{MANAsymCheck}
 \end{figure}
 
 \begin{figure}[h!]
         \centering
         \subfigure[None ]{                
                 \includegraphics[width=0.17\textwidth]{FigNPooling/Parameters.eps}
                 \label{fig:unconstrainedParamApp}}
         ~
         \subfigure[Temperature (CSU) only]{                
                 \includegraphics[width=0.17\textwidth]{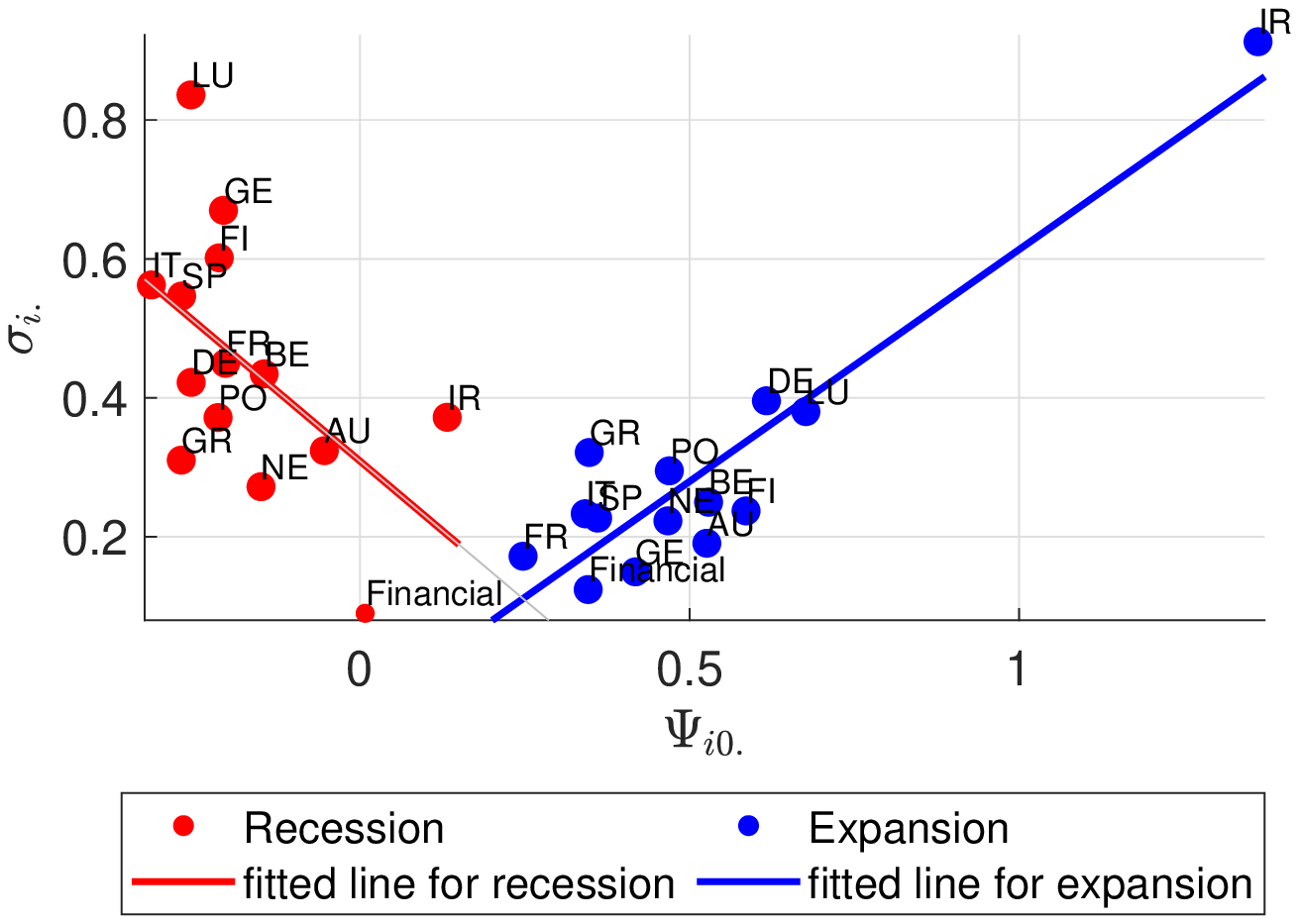}
                 \label{fig:unconstrainedCSUParamApp}}
         ~
         \subfigure[Drought (SPI) only]{                
                 \includegraphics[width=0.17\textwidth]{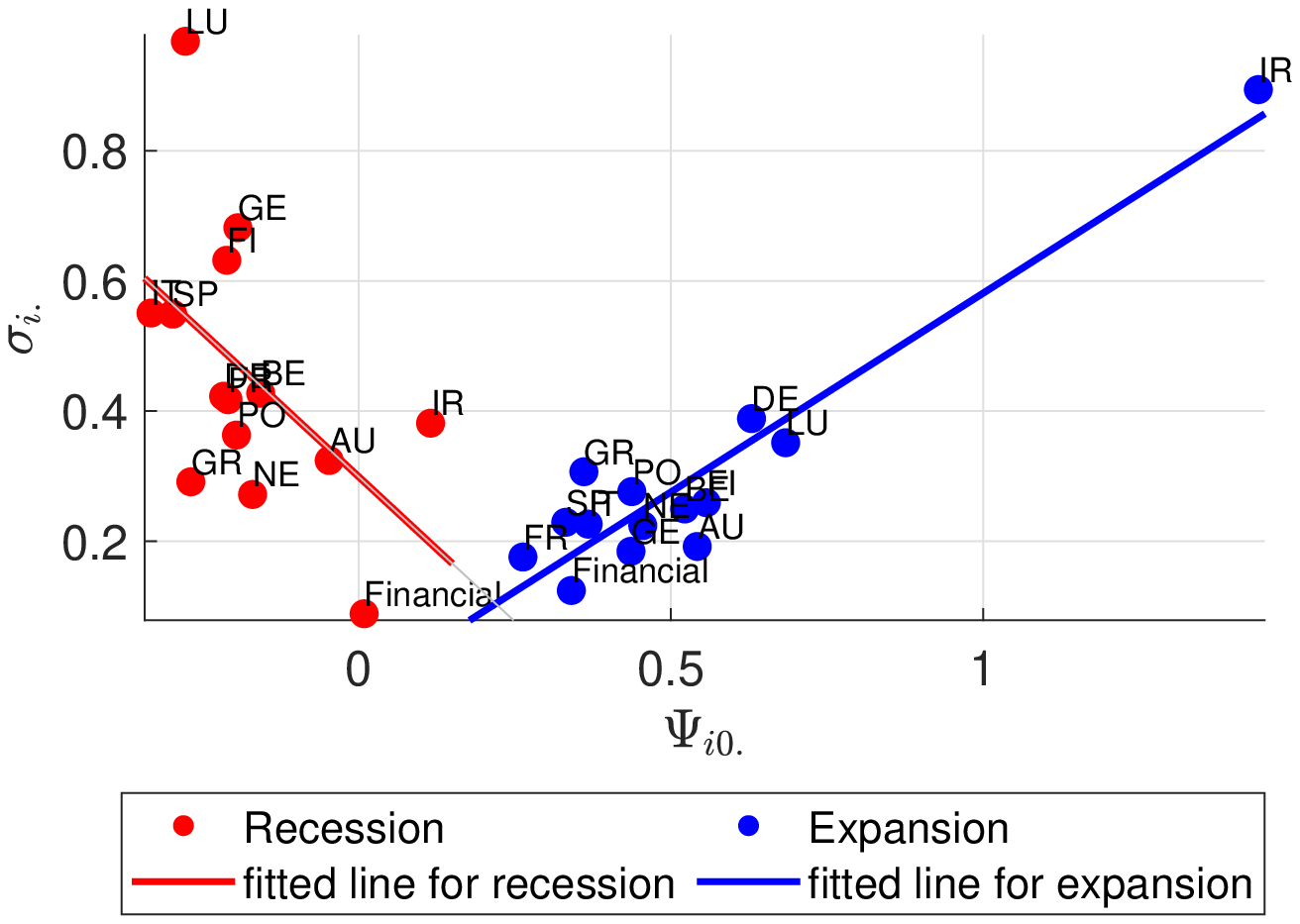}
                 \label{fig:unconstrainedSPIParamApp}}
         ~
         \subfigure[Rainfall (r20mm) only]{                
                 \includegraphics[width=0.17\textwidth]{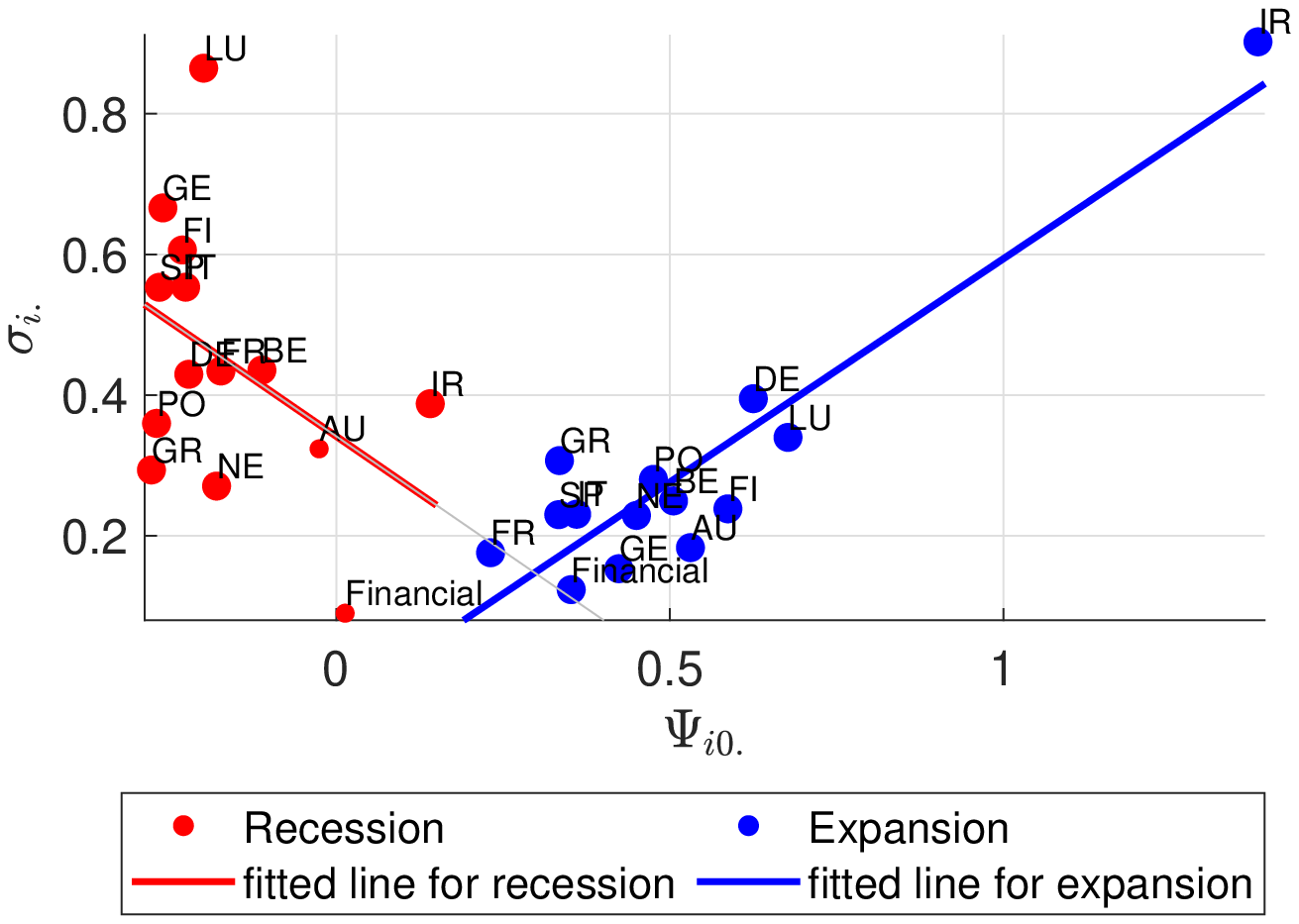}
                 \label{fig:unconstrainedR20MMParamApp}}        
         ~
         \subfigure[ALL climate indices ]{                
                 \includegraphics[width=0.17\textwidth]{FigNPoolingCSUSPIR20MMCV/Parameters.eps}
                 \label{fig:unconstrainedALLParamApp}}                        
 		\caption{Estimates of monthly state-level IPI and Fin mean growth rates (horizontal axis) and volatilities (vertical axis) with the Bayesian Panel Markov-switching model. The horizontal axis represents the mean growth rates and the vertical axis the volatilities. Sample period: February 1981 to December 2016 (month on month). Column 1 of this figure display result obtained from the climate shocks independent PMS models while the others represent the result obtained when the PMS model is subjected to climate shocks: temperature (CSU), drought (SPI) and rainfall ( r20mm). Columns 2, 3, and 4 respectively display result of the PMS model subjected to temperature only, drought and rainfall while column 5 simultaneously considers all the climate change indices in the PMS model. }\label{ParamApp}
 \end{figure}
 
 \begin{figure}[h!]
         \centering
         \subfigure[Temperature (CSU) only]{                
                 \includegraphics[width=0.3\textwidth]{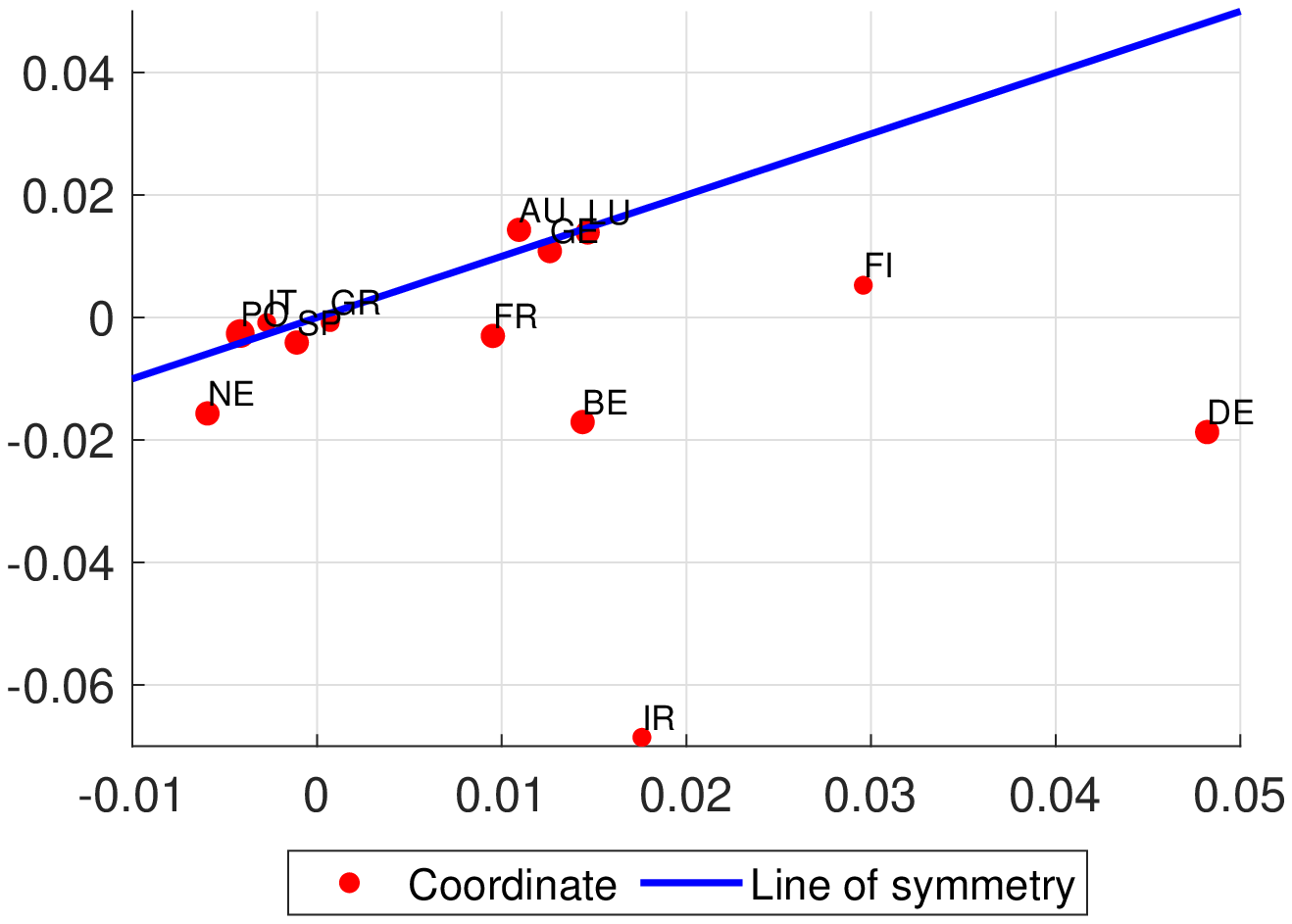}
                 \label{fig:unconstrainedCSUAsymCheckApp}}
         ~
         \subfigure[Drought (SPI) only]{                
                 \includegraphics[width=0.3\textwidth]{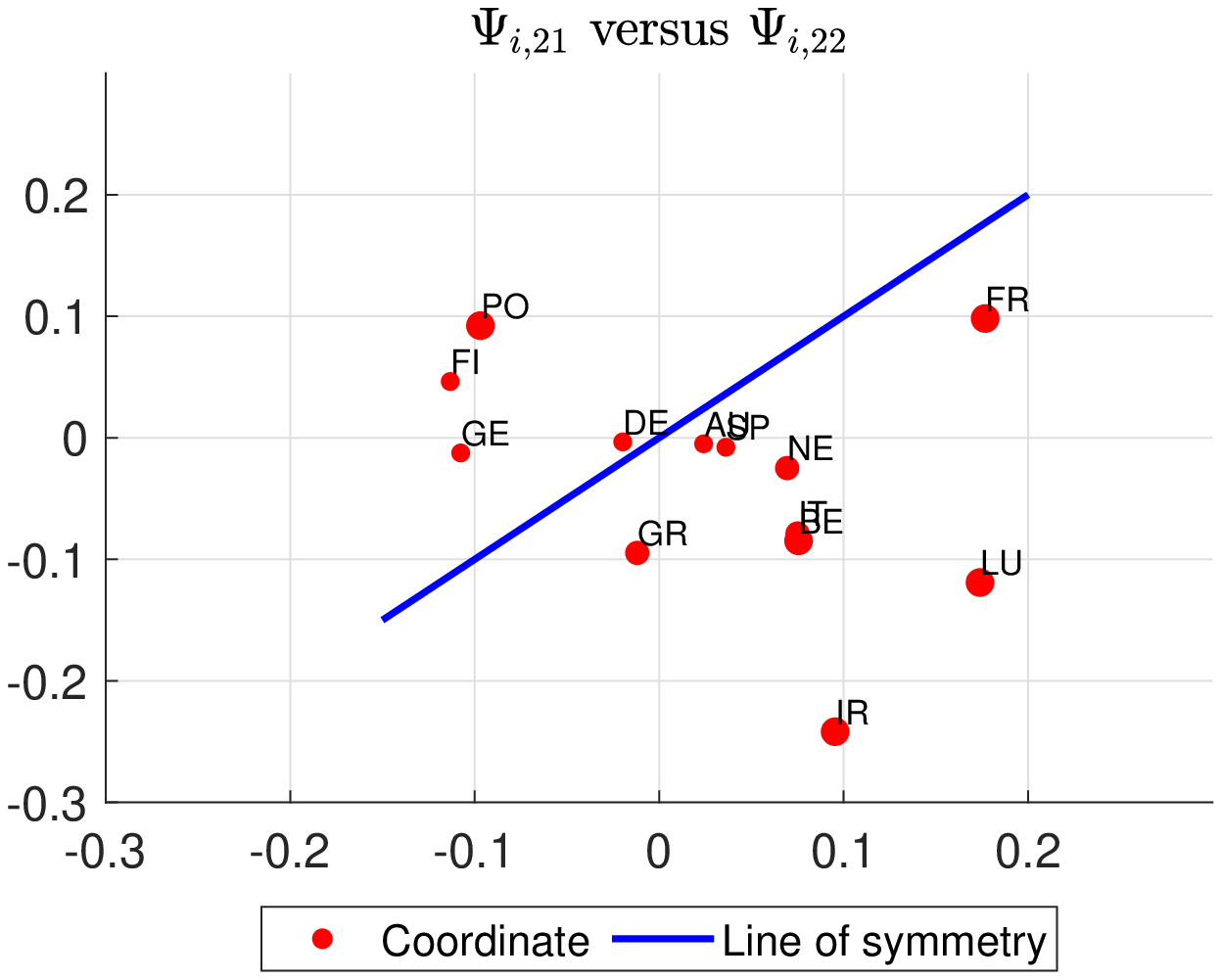}
                 \label{fig:unconstrainedSPIAsymCheckApp}}
         ~
         \subfigure[Rainfall (r20mm) only]{                
                 \includegraphics[width=0.3\textwidth]{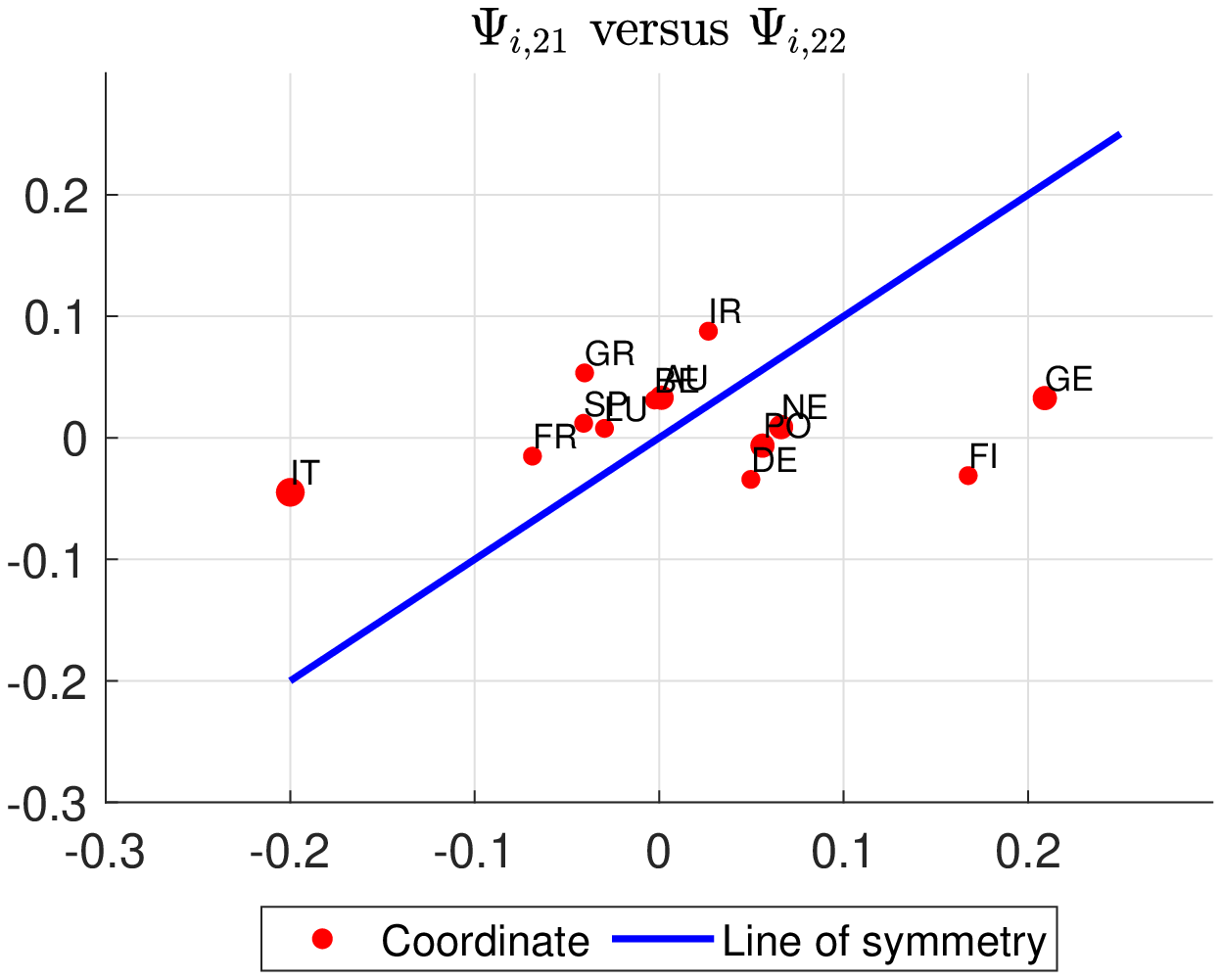}
                 \label{fig:unconstrainedR20MMAsymCheckApp}}        
 		\caption{Estimates of the monthly asymmetric effect of the change in the weather condition on each country's  monthly  IPI with the Bayesian Panel Markov-switching model. The horizontal and vertical axes respectively, represent parameters capturing the effect of the climate change on IPI during recession and expansion. Sample period: February 1981 to December 2016 (month on month). Each column display the result of the PMS model taking into consideration the climate risk indicators (ie. temperature, drought and rainfall) one at a time.
 		}\label{AsymCheckApponly}
 \end{figure}
 
 \begin{figure}[h!]
         \centering
         \subfigure[Temperature (CSU)]{                
                 \includegraphics[width=0.3\textwidth]{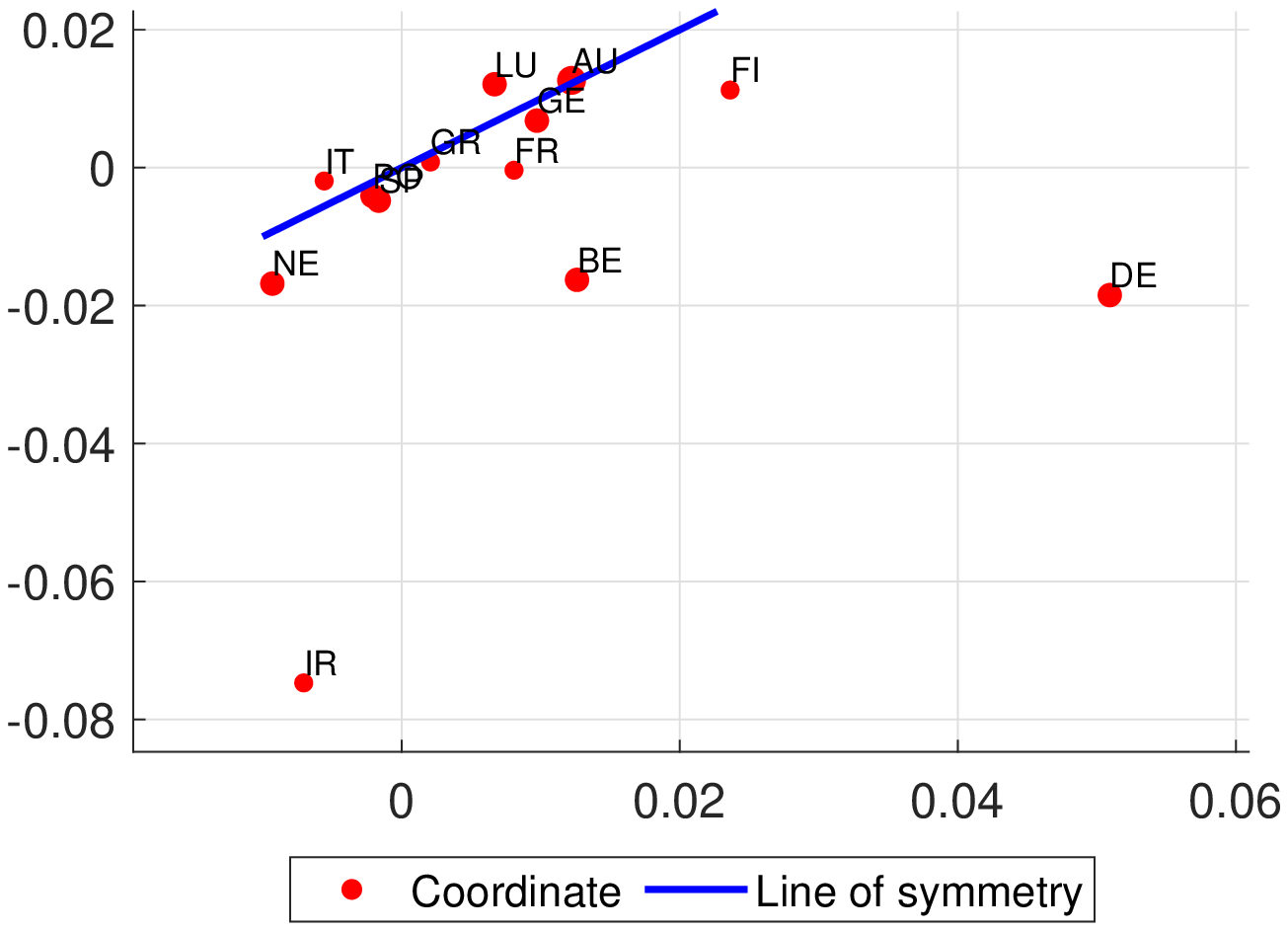}
                 \label{fig:unconstrainedCSUAsymCheckALLApp}}
         ~
         \subfigure[Drought (SPI)]{                
                 \includegraphics[width=0.3\textwidth]{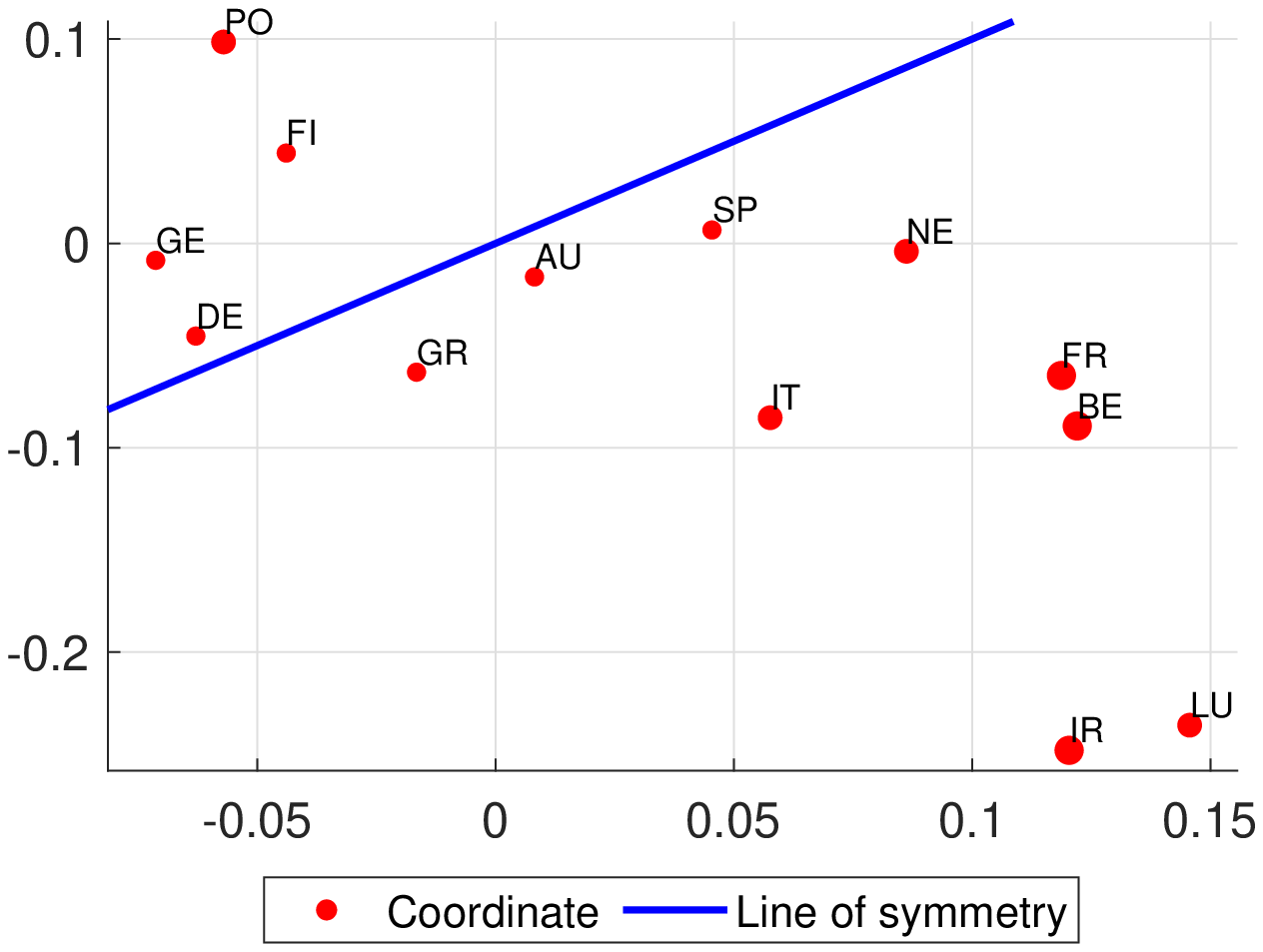}
                 \label{fig:unconstrainedSPIAsymCheckALLApp}}
         ~
         \subfigure[Rainfall (r20mm)]{                
                 \includegraphics[width=0.3\textwidth]{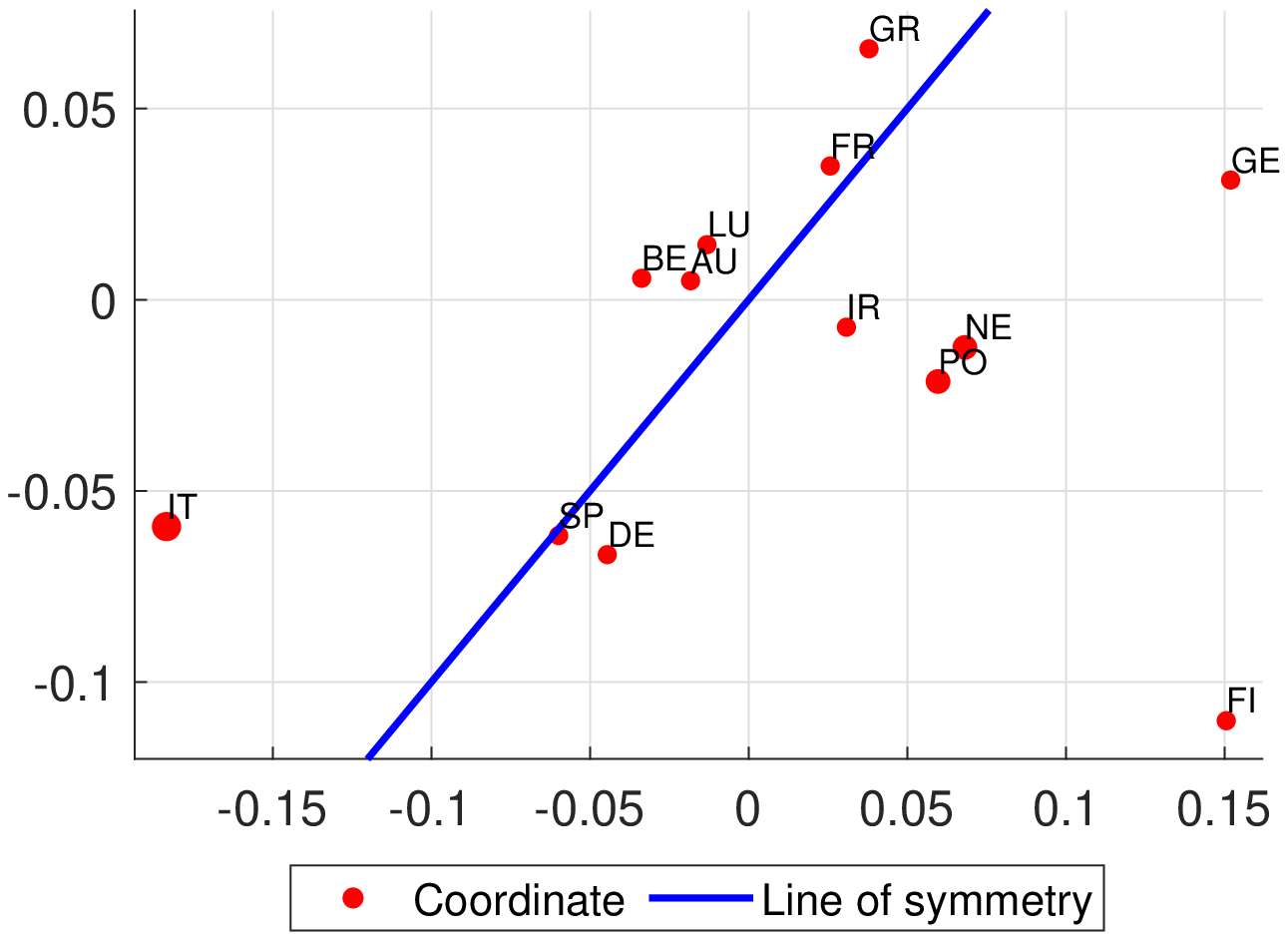}
                 \label{fig:unconstrainedR20MMAsymCheckALLApp}}        
 		\caption{Estimates of the monthly asymmetric effect of the change in the weather condition on each country's  monthly  IPI with the Bayesian Panel Markov-switching model. The horizontal and vertical axes respectively, represent parameters capturing the effect of the climate change on IPI during recession and expansion. Sample period: February 1981 to December 2016 (month on month). The result is this figure take into consideration all the indices of the climate change simultaneously. The columns display the result of the PMS models relative to each of the climate risk indicators (ie. temperature, drought and rainfall) one at a time.
 		}\label{AsymCheckApp}
 \end{figure}
\newpage 
 \clearpage
 \section{Robustness check: Different Transformation of SPI}
 \medskip
 \noindent In this sections we consider 3 alternative transformation of the drought (SPI) index. The threshold for the two indicator transform presented below are set to $a=-1$ and $b=1$ while the percentage growth rate of the IPI is taken as the dependent variable in the business measurement equation.
 \begin{enumerate}
 \item[i.] Transform based on cumulative distribution function (CDF):
 In the first case, the SPI is transformed back to the probability space with the (CDF) of a normal ($\Phi$) to get a continuous indicator. This measure quantifies the intensity of the drought index relative to extreme dryness. 
 \item[2.] Transform based on conditional CDF: In this case, we construct two separate continuous indicators accounting for moderate drought to extreme drought and moderately wet to extremely wet by transforming the SPI back to the probability space with the conditional CDF  of the two tails. This is achieved by doing the following:
 
 \medskip 
 \noindent Let $S$ and $S_{t}$, respectively, denote the random variable representing the SPI and the value of SPI at time $t$. Also, let $d_{t}$ and $w_{t}$, respectively, the conditional cumulative probability that the drought index (SPI) is larger than $S_{t}$ giving that the SPI falls between the extreme dryness and moderate dryness and the conditional cumulative probability that the drought index (SPI) is larger than $S_{t}$ giving that the SPI falls between the moderately wet and extremely wet climatic condition. This measures $d_{t}$ and $w_{t}$ may, respectively, be understood to capture the intensity of the drought condition relative to moderately dry and wet situations.
 $$
 d_t=1-P(S<S_t|S<a)=
 \begin{cases}
 1-(\Phi(S_t)/\Phi(a)), &~{\text{for $S_{t}<a$ }},\\
 0,   &~{\text{otherwise}},
 \end{cases}
 $$ 
 and 
 $$
 w_t=P(S<S_t|S>b)=
 \begin{cases}
 1-(1-\Phi(S_t))/(1-\Phi(b) )  &~{\text{for $S_{t} >b$  }},\\
 0,  &~{\text{otherwise}}.
 \end{cases}
 $$
 \item[3.] Transform based on the two tails of the support of the SPI:  the tails captures the two extremes of the SPI index. Given the thresholds $a$ and $b$ the following segments of the SPI are considered.\\
 
 $w_t=(a-S_t)$ if $S_t<a$ and 0 otherwise and $d_t=(S_t-b)$ if $S_t>b$ and 0 otherwise (you get two indicators).
 \end{enumerate}
 Statistically significant estimates of the parameters of the measurement equations taking each of the above transformation one at a time are reported in Figure \ref{Prop1Map1} to Figure \ref{Prop3Map2}. The impact of both CSU and precipitation remain in these figures are consistent with the result of the model utilizing dummy transformation of the SPI. The models accounting for the two extremes of the SPI index provides further insight into identifying which of the two extremes (dry or wet) of the drought index contributes to the economic growth. As for the SPI index, conditional on a moderate to extreme drought weather condition, increase in the likelihood of the intensity of drought relative to moderately dry
 events seem to support growth in the Netherlands and Belgium irrespective of the state of the economy. Similarly, Denmark and Portugal, respectively, tend to enjoy positive impacts of increasing intensity of drought relative to moderately dry weather conditions, recession and expansion. Finland, Germany, Luxembourg, Italy and Austria however are likely to experience reduction in their growth rates during expansion as a result of increasing intensity in drought. On the other hand, conditional on a moderately to extremely wet weather condition, increase in the cumulative probability of the intensity of wet event seem to provide an enabling environment upon which the negative effect of recession on Portugal and Greece is lessen whereas Belgium, Spain and Finland tend to derive additional growth from this weather condition during expansion. An uneven impact of this weather scenario is experienced by the Netherlands: comparatively to a moderately wet condition, a wetter weather condition seems to ameliorate the adverse effect of recession and reduces the growth rate in expansion.
 
 \begin{figure}[h!]
 \begin{center}
 \begin{tabular}{|c|c|c|}
 \hline
  & Recession & Expansion\\
  \hline
 \rot{Intercept}&
 \includegraphics[scale=0.5]{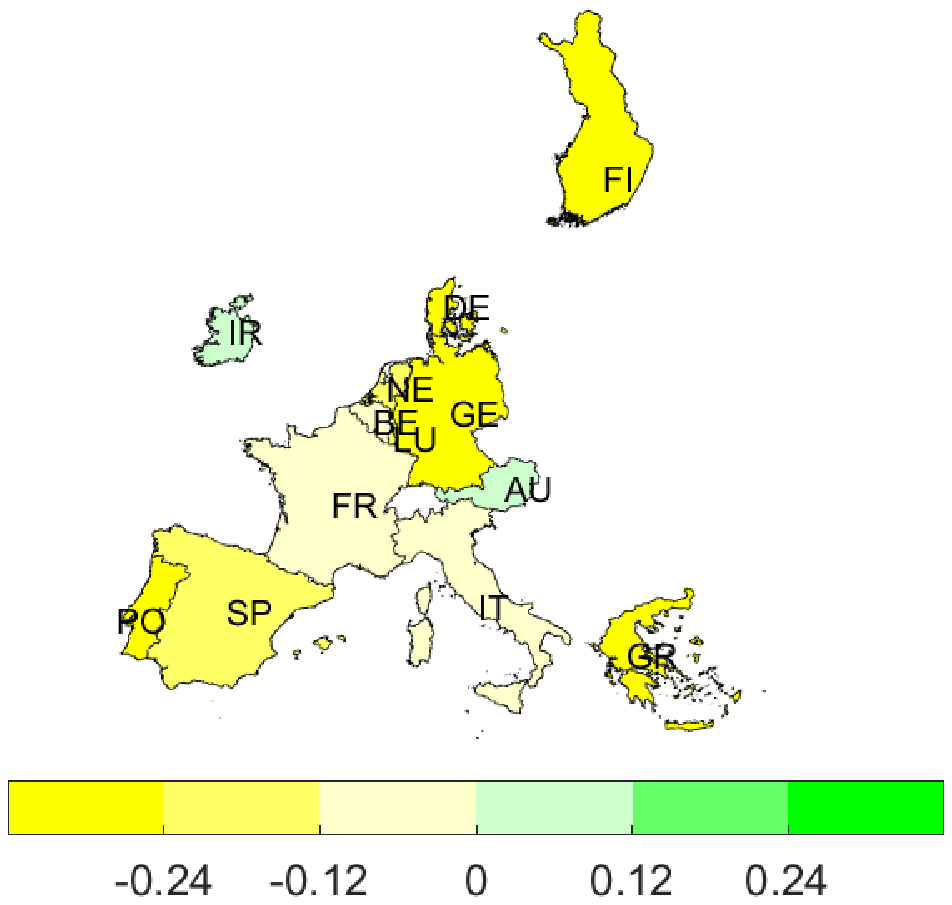}&\includegraphics[scale=0.5]{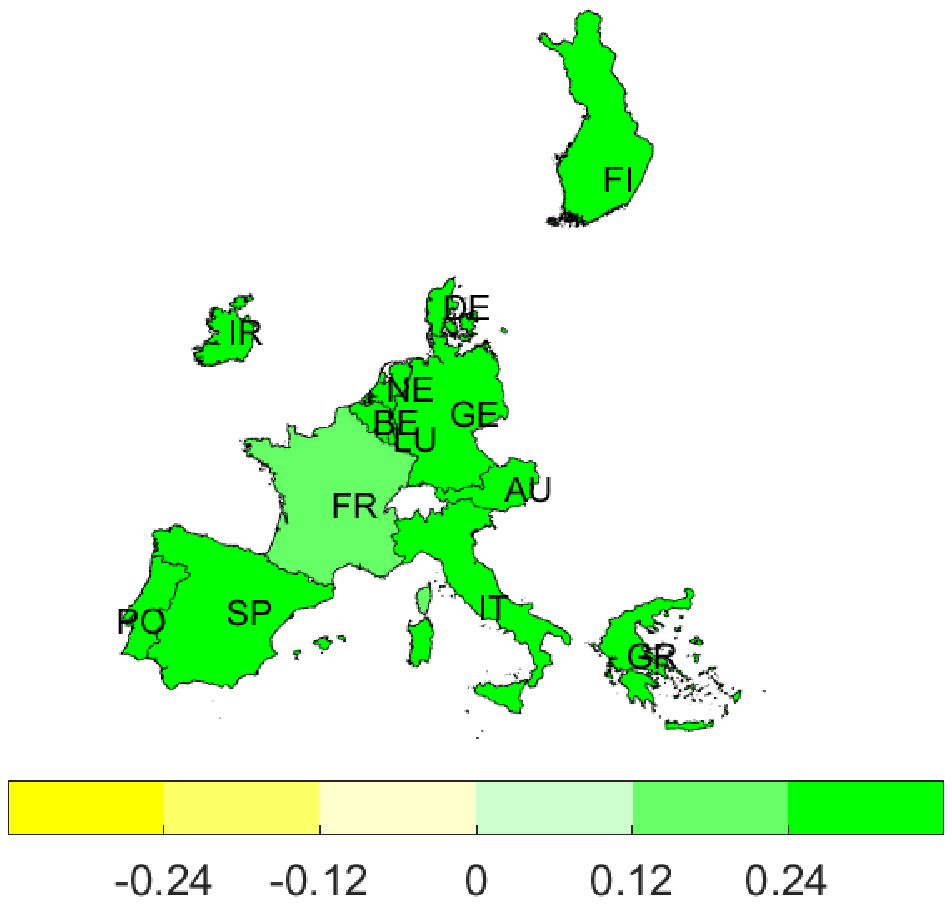}\vspace{5pt}\\
 \hline
 \rot{Temperature (CSU)}&
 \includegraphics[scale=0.5]{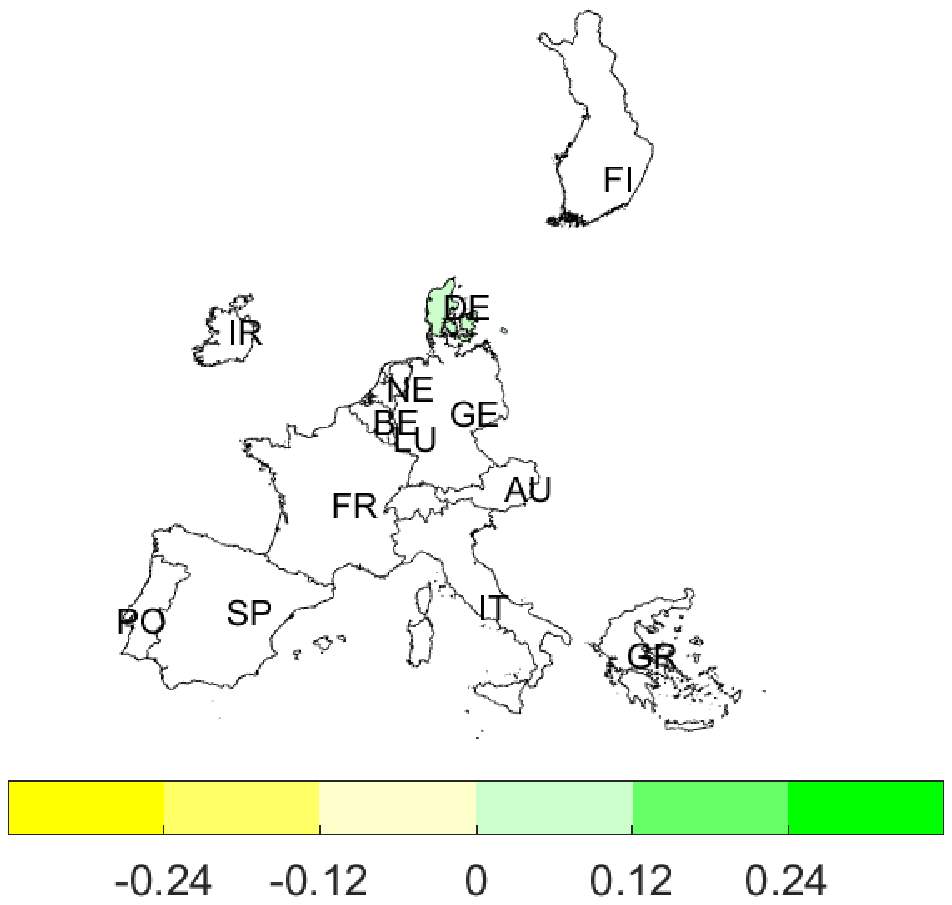}&\includegraphics[scale=0.5]{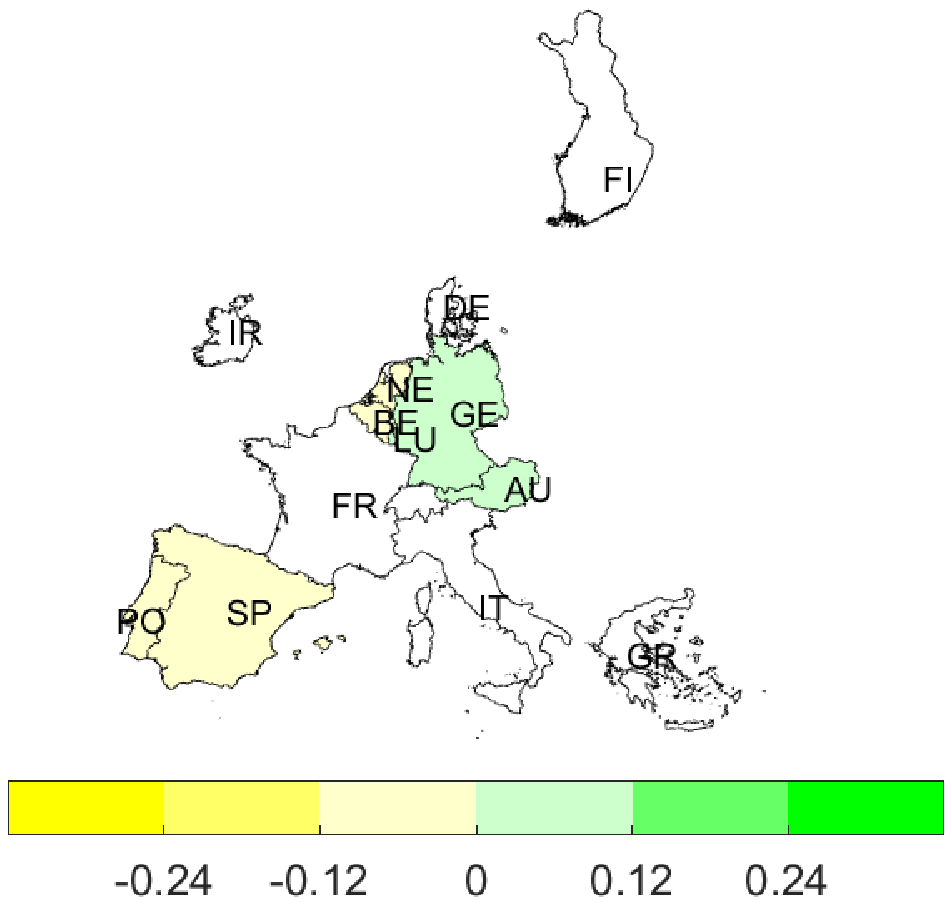}\vspace{5pt}\\
 \hline
 \rot{Drought (SPI)}&
 \includegraphics[scale=0.5]{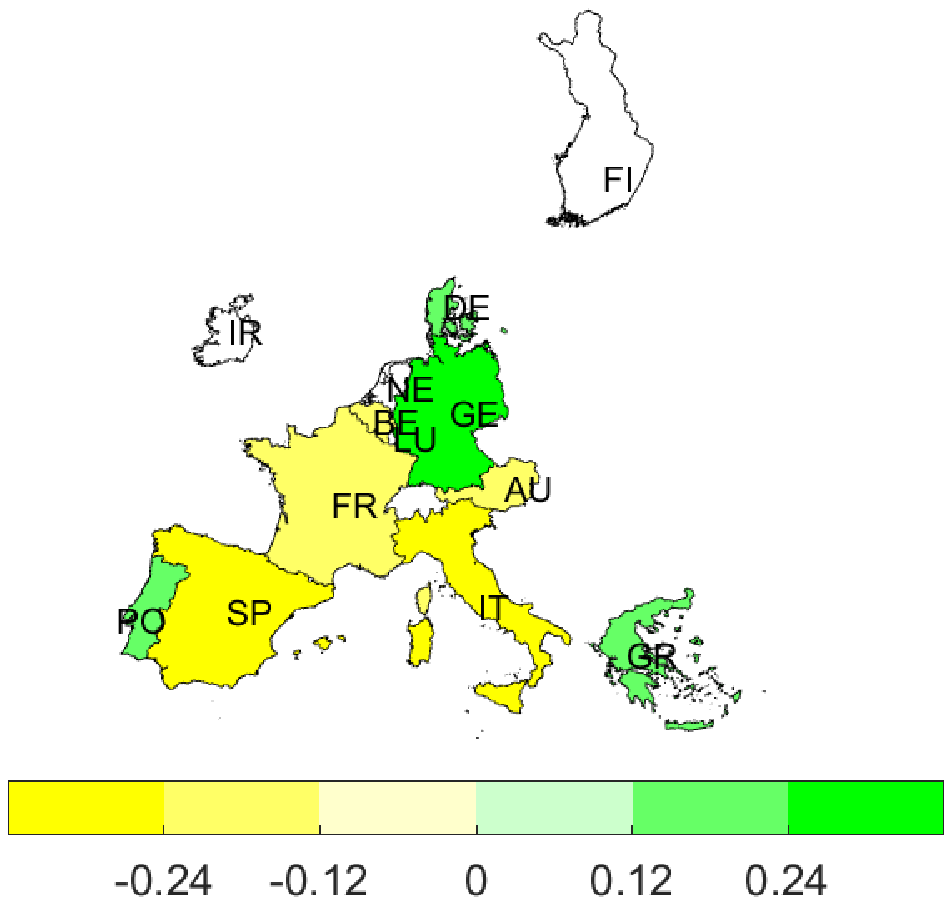}&\includegraphics[scale=0.5]{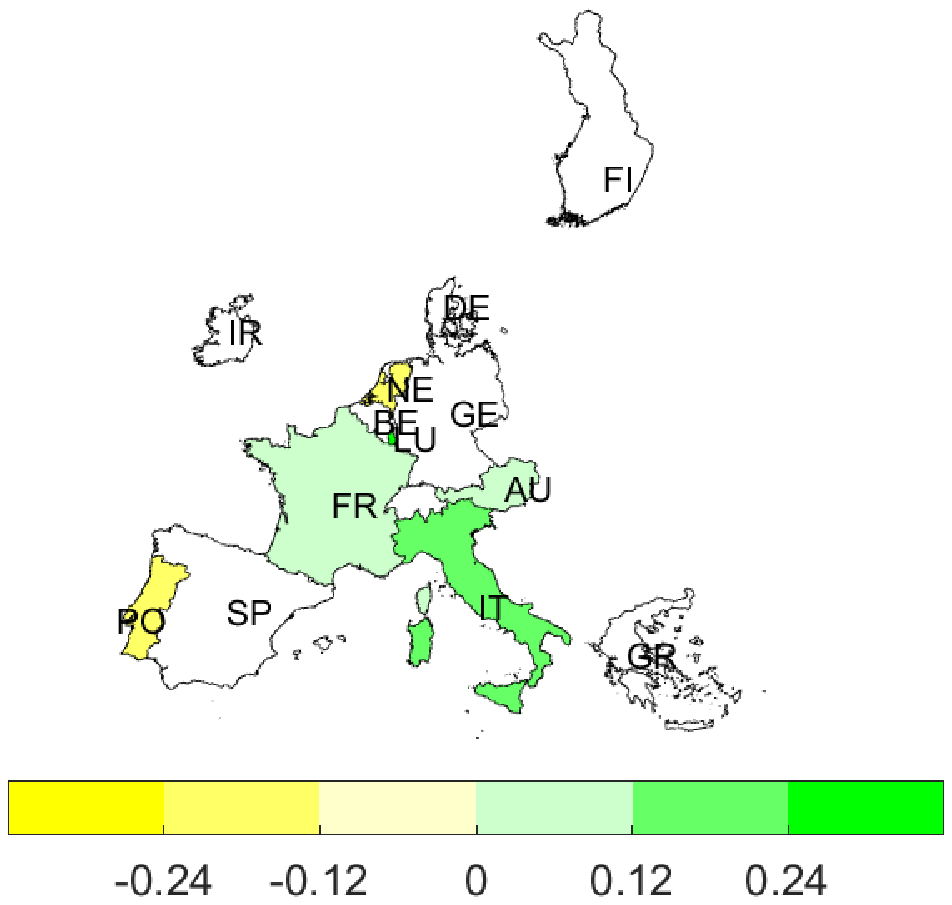}\\
 \hline
 \end{tabular}
 \end{center}
 \caption{Study area with the impact of climate shocks (indicated by different colours) on the growth rate on the represented EU Countries during recession and expansion. White patches suggests that the entity is not responsive (statistically significant) to the shock, while the yellow and green patches respectively denotes negative and positive impact. The measure of the SPI index used here is the associated cumulative distribution of the index with extreme dryness to extremely wet weather condition.}\label{Prop1Map1}
 \end{figure}
 
 \begin{figure}[h!]
 \begin{center}
 \begin{tabular}{|c|c|c|}
 \hline
  & Recession & Expansion\\
 \hline
 \rot{Precipitation (r20mm)}&
 \includegraphics[scale=0.5]{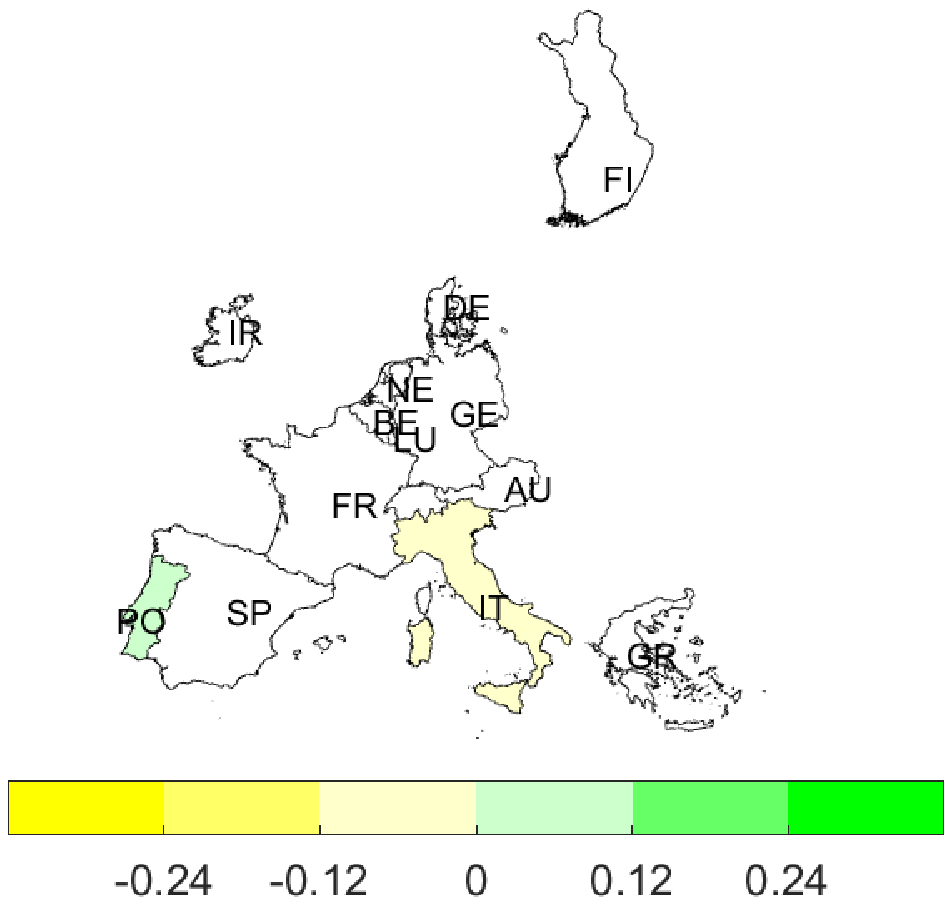}&\includegraphics[scale=0.5]{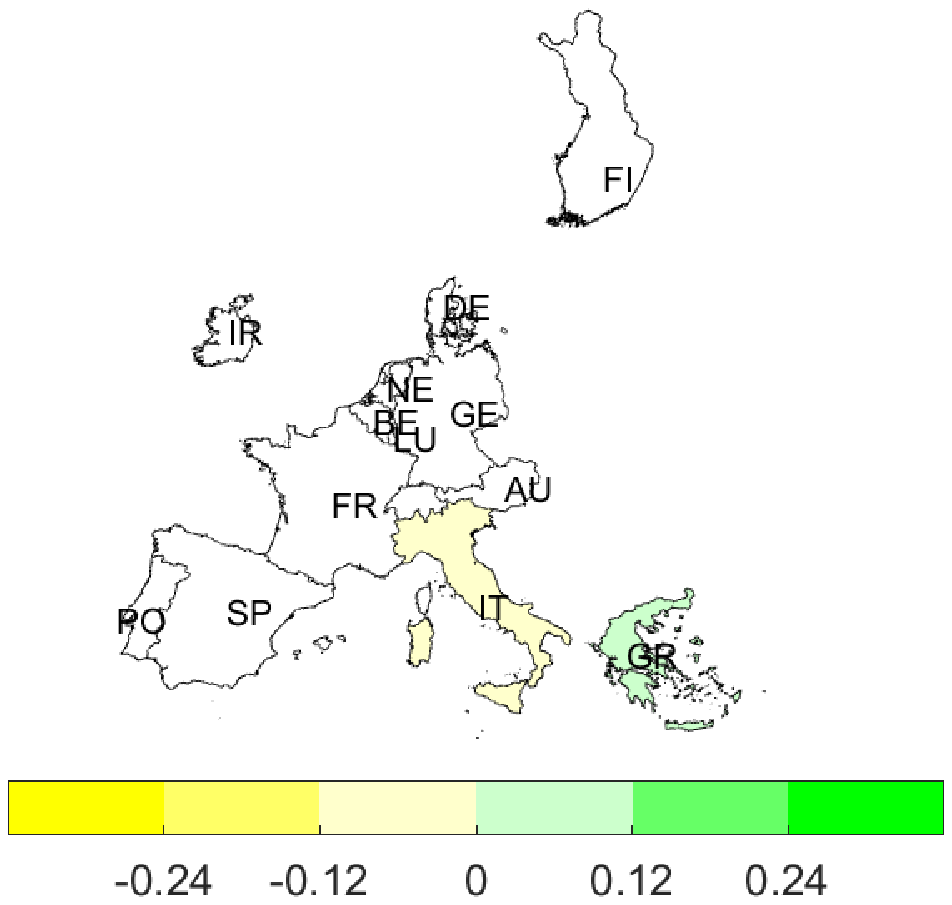}\\
 \hline
 \end{tabular}
 \end{center}
 \caption{Continued from previous page Figure \ref{Prop1Map1}.}\label{Prop1Map2}
 \end{figure}
 
 \begin{figure}[h!]
 \begin{center}
 \begin{tabular}{|c|c|c|}
 \hline
  & Recession & Expansion\\
  \hline
 \rot{Intercept}&
 \includegraphics[scale=0.5]{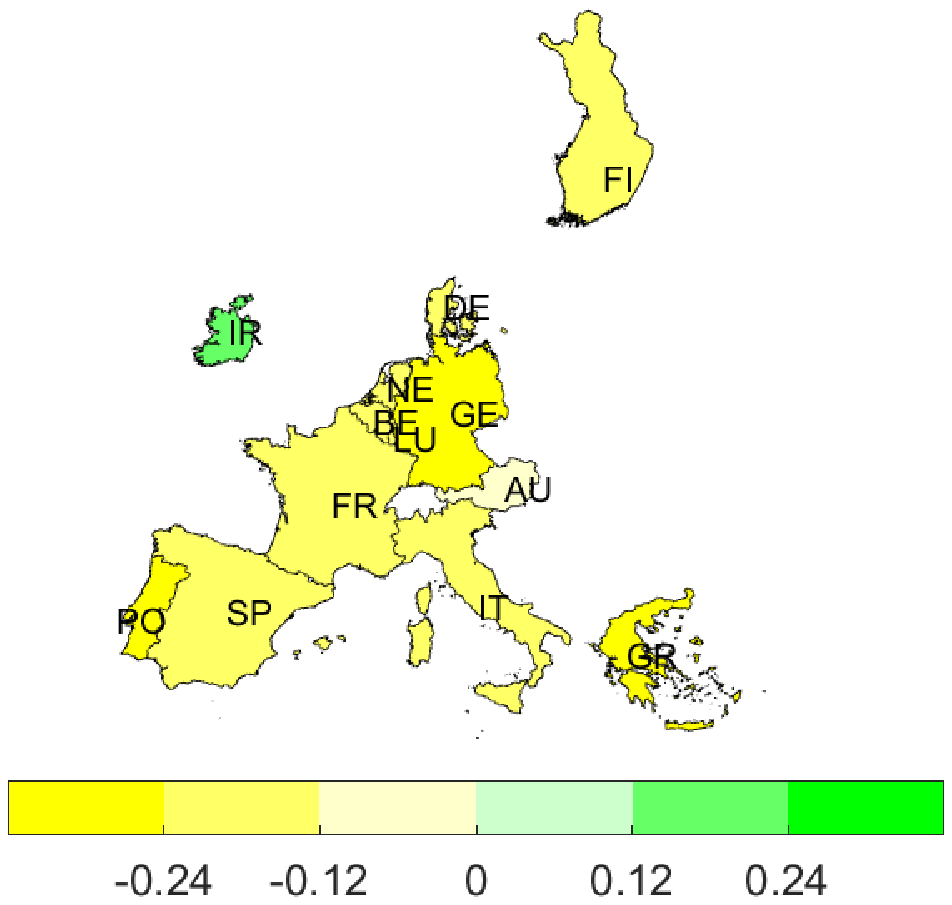}&\includegraphics[scale=0.5]{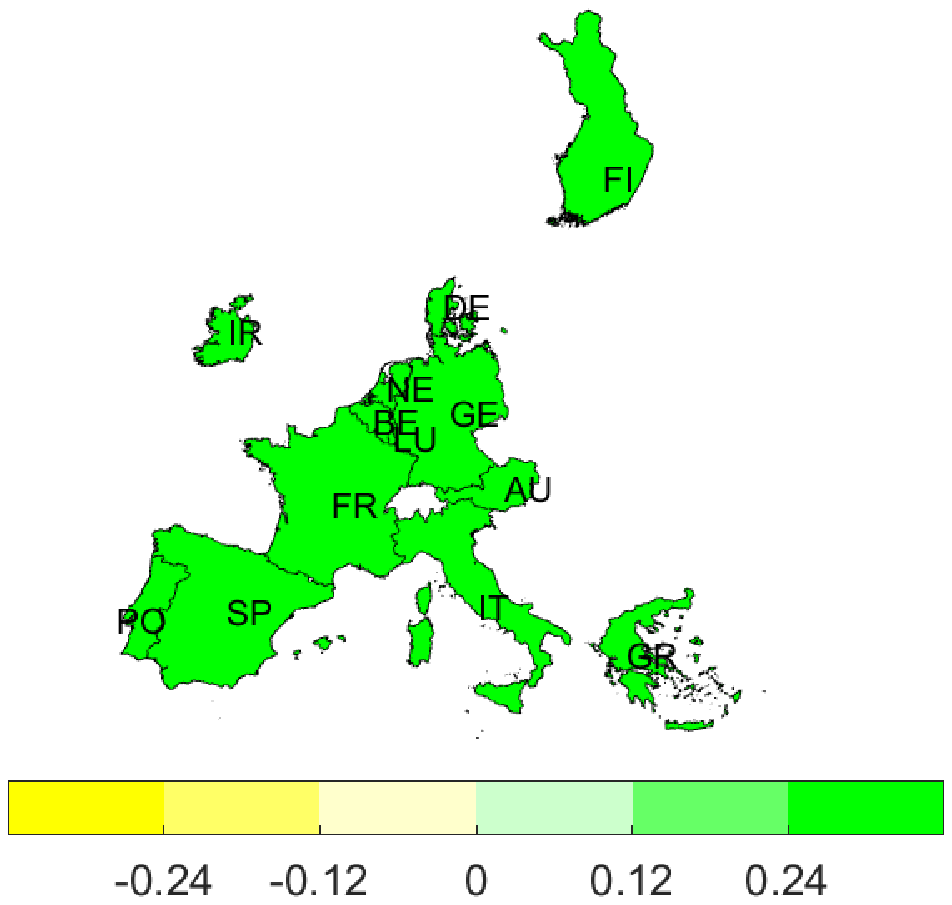}\vspace{5pt}\\
 \hline
 \rot{Temperature (CSU)}&
 \includegraphics[scale=0.5]{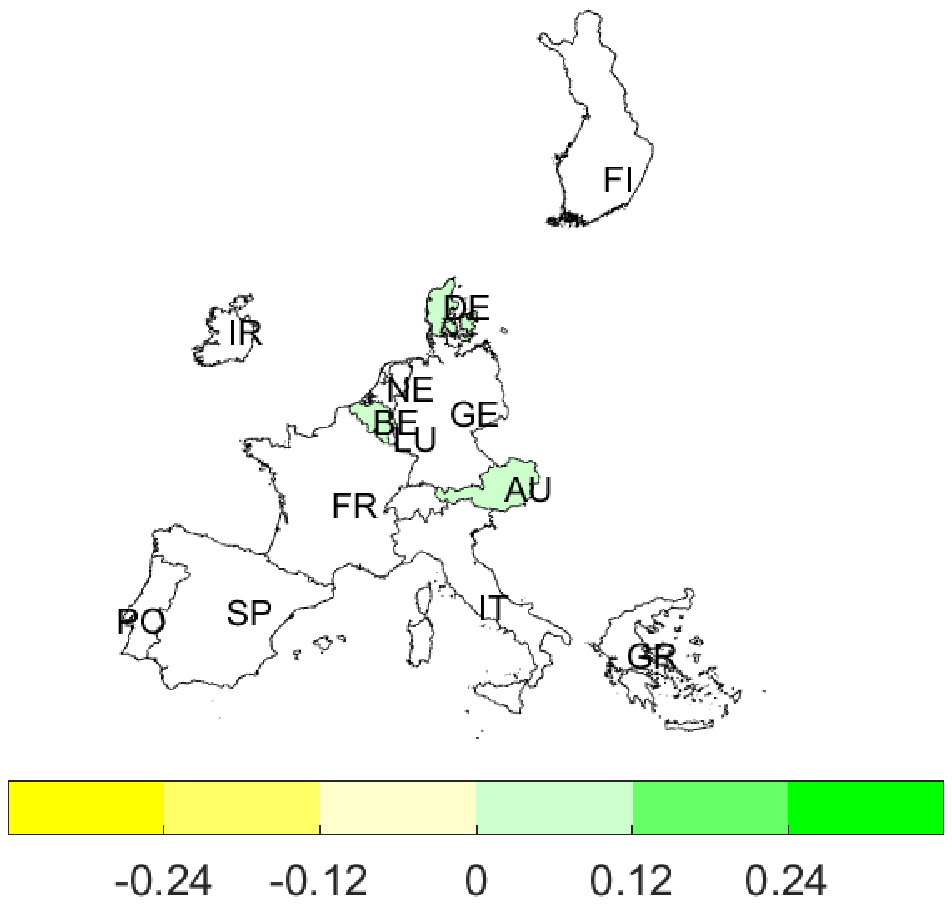}&\includegraphics[scale=0.5]{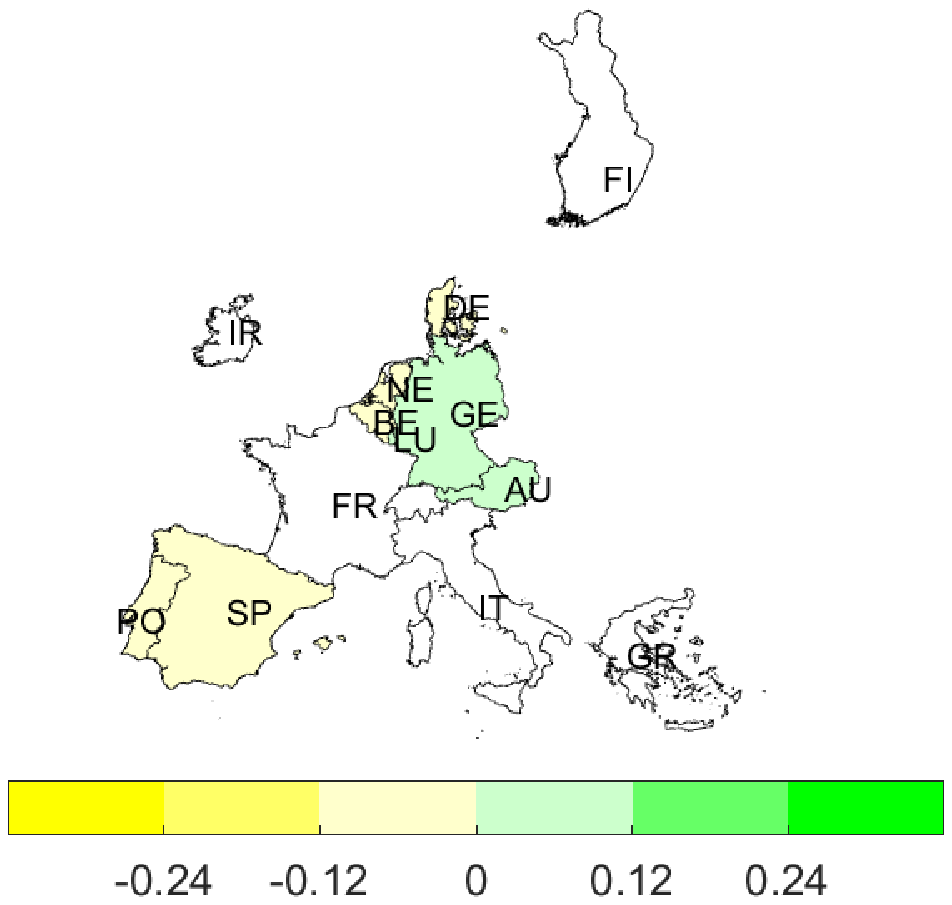}\vspace{5pt}\\
 \hline
 \rot{Drought (SPI dry)}&
 \includegraphics[scale=0.5]{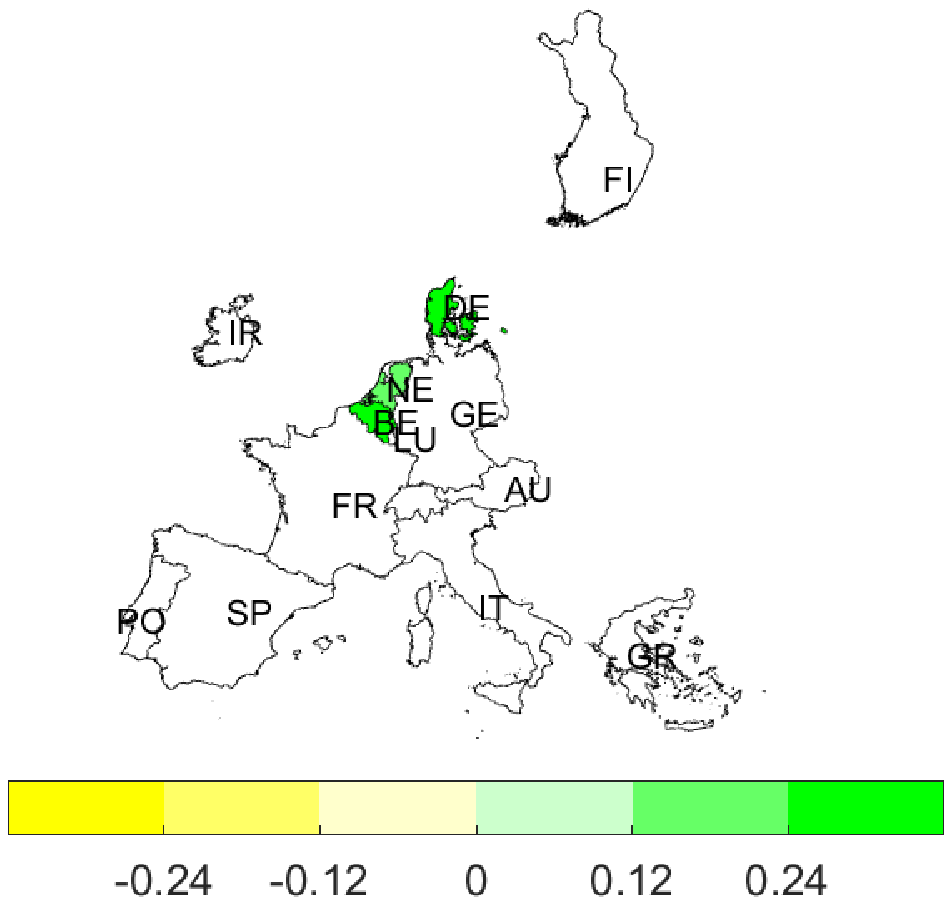}&\includegraphics[scale=0.5]{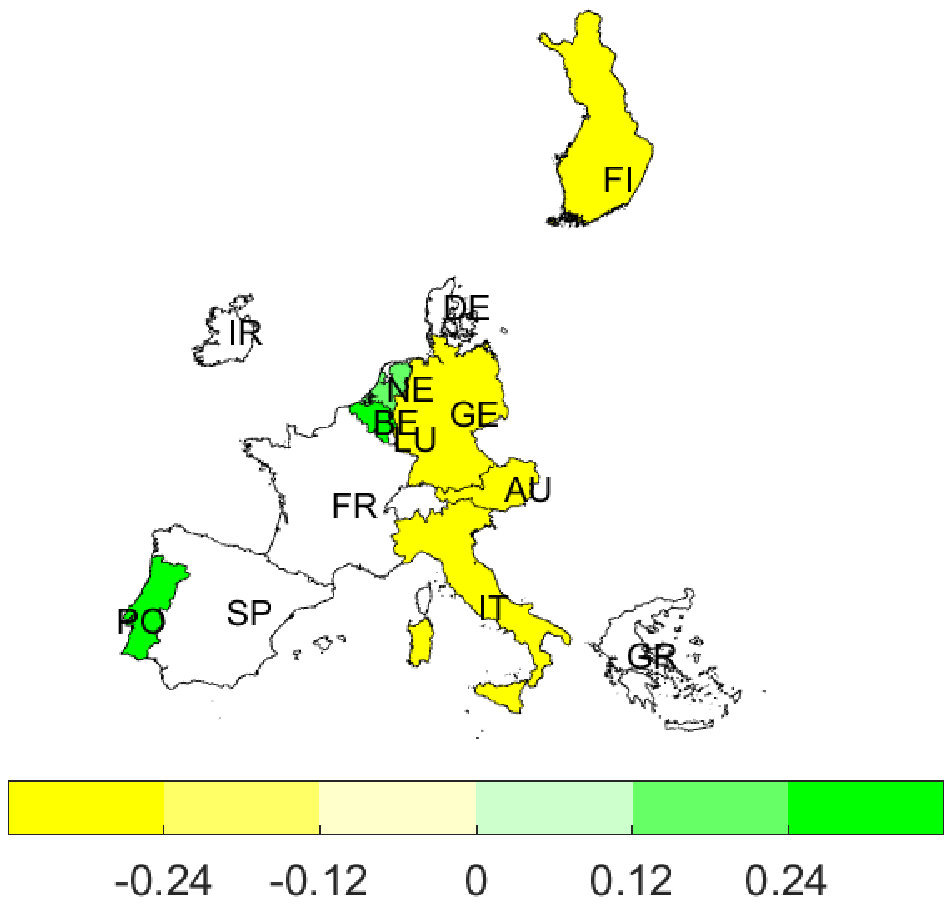}\\
 \hline
 \end{tabular}
 \end{center}
 \caption{Study area with the impact of climate shocks (indicated by different colours) on the growth rate on the represented EU Countries during recession and expansion. White patches suggests that the entity is not responsive (statistically significant) to the shock, while the yellow and green patches respectively denotes negative and positive impact. The measure of drought is constructed based on the conditional cumulative distribution function of the tails of the climate shock index.}\label{Prop2Map1}
 \end{figure}
 
 \begin{figure}[h!]
 \begin{center}
 \begin{tabular}{|c|c|c|}
 \hline
  & Recession & Expansion\\
 \hline
 \rot{Drought (SPI wet)}& 
 \includegraphics[scale=0.5]{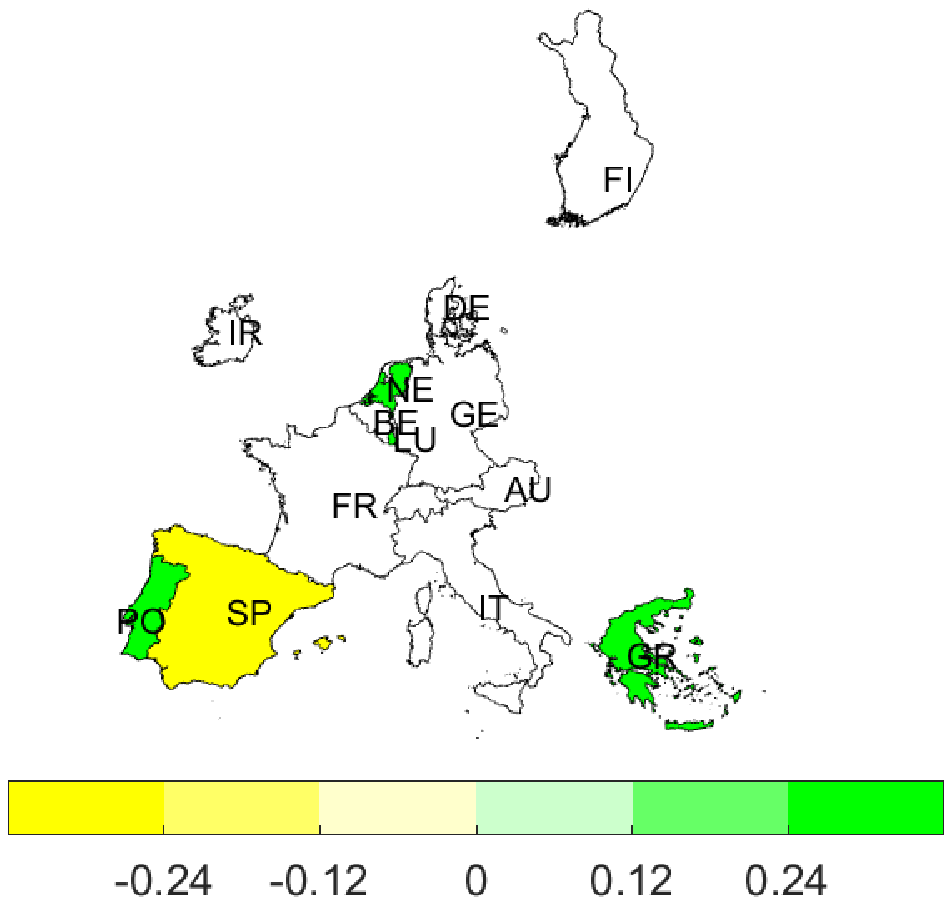}&\includegraphics[scale=0.5]{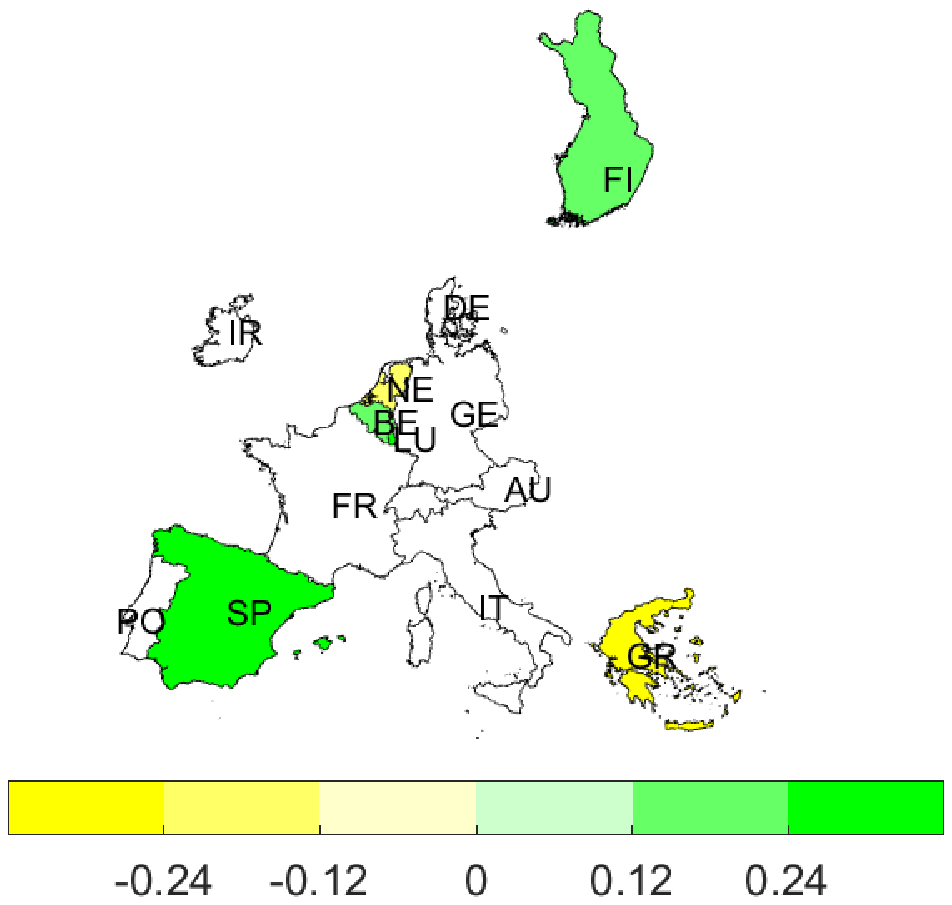}\\
 \hline
 \rot{Precipitation (r20mm)}&
 \includegraphics[scale=0.5]{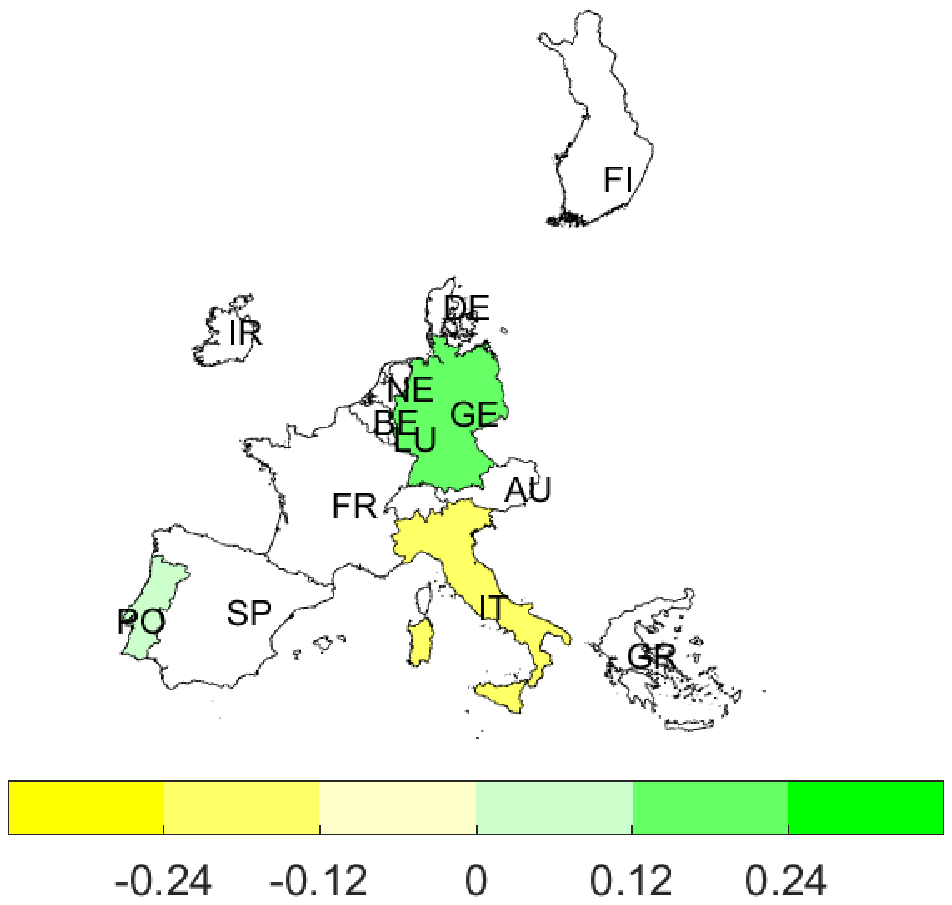}&\includegraphics[scale=0.5]{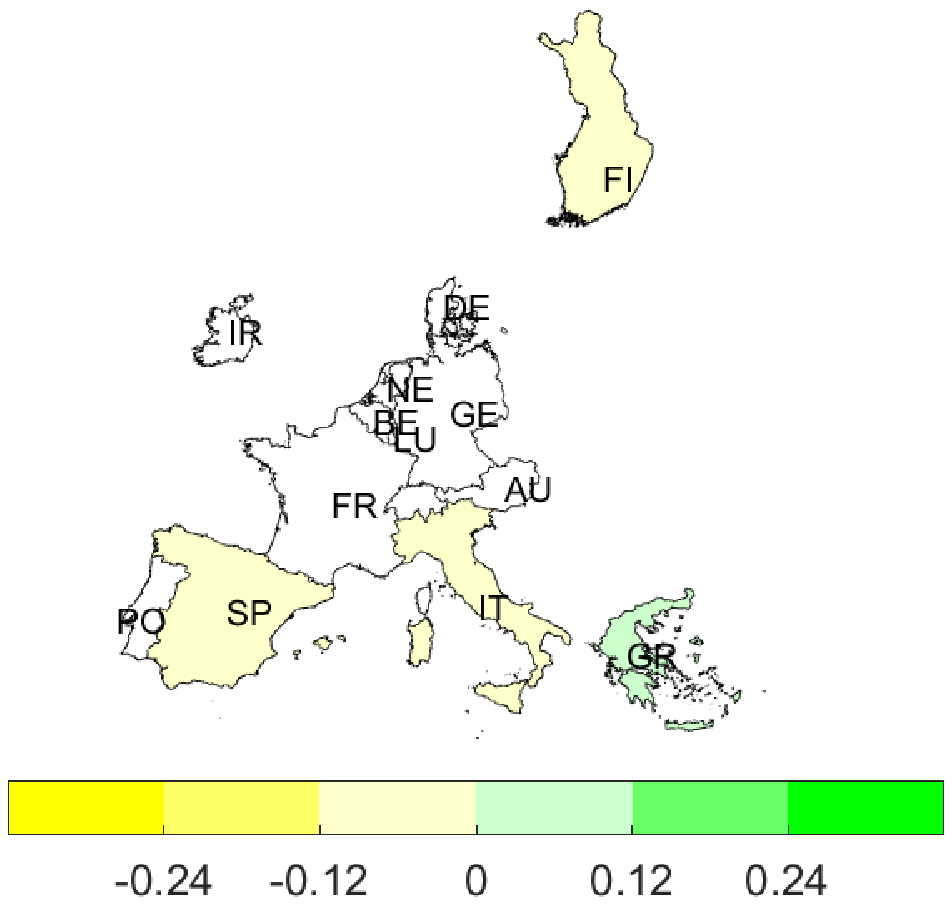}\\
 \hline
 \end{tabular}
 \end{center}
 \caption{Continued from previous page Figure \ref{Prop2Map1} }\label{Prop2Map2}
 \end{figure}
 
 \clearpage 
 \newpage
  The results for this case and case 2 are quite similar. 
 
 \begin{figure}[h!]
 \begin{center}
 \begin{tabular}{|c|c|c|}
 \hline
  & Recession & Expansion\\
  \hline
 \rot{Intercept}&
 \includegraphics[scale=0.5]{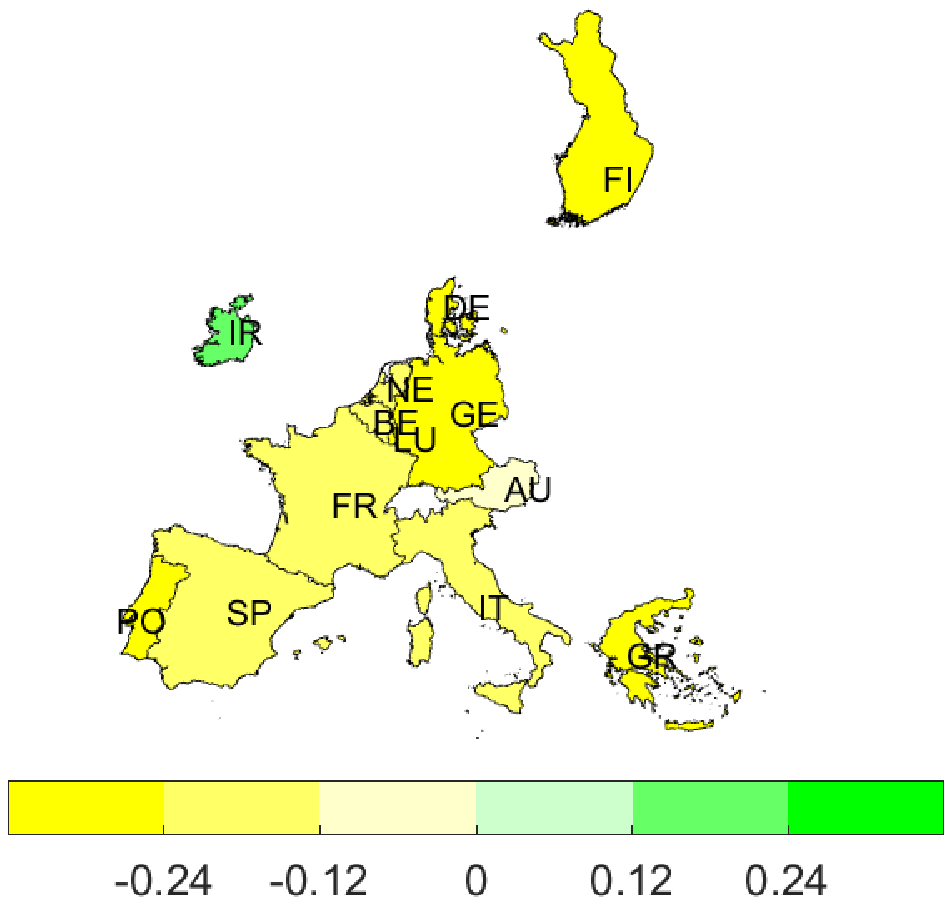}&\includegraphics[scale=0.5]{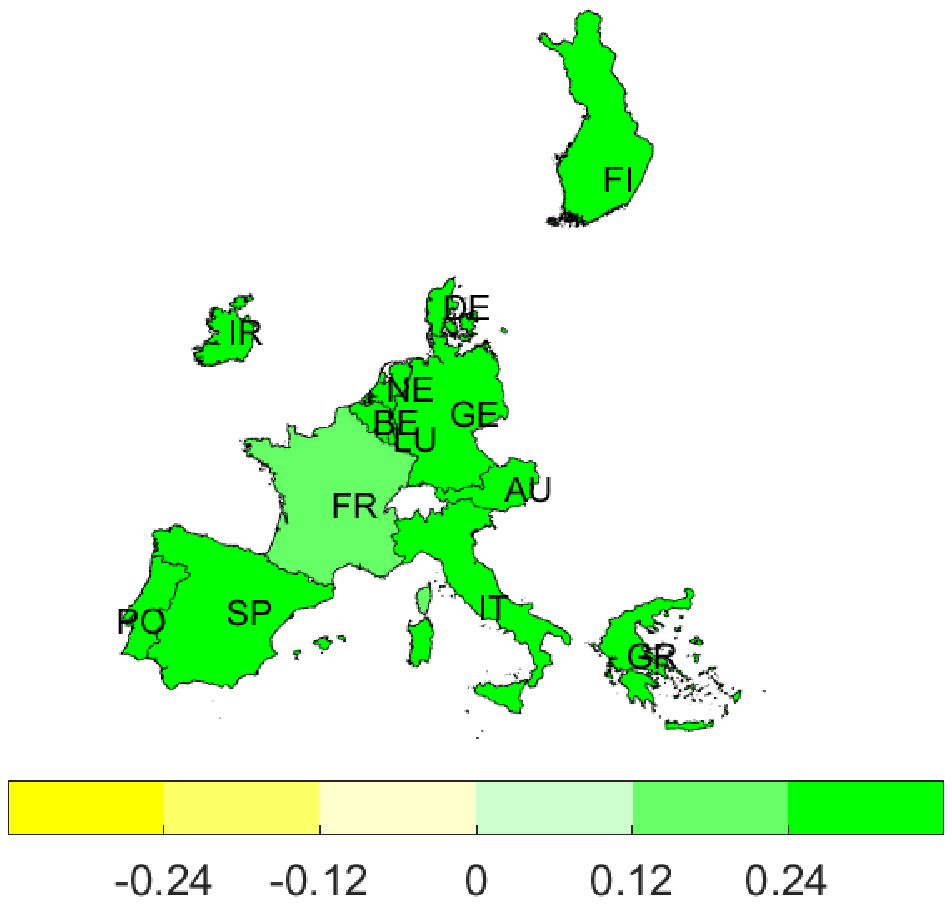}\vspace{5pt}\\
 \hline
 \rot{Temperature (CSU)}&
 \includegraphics[scale=0.5]{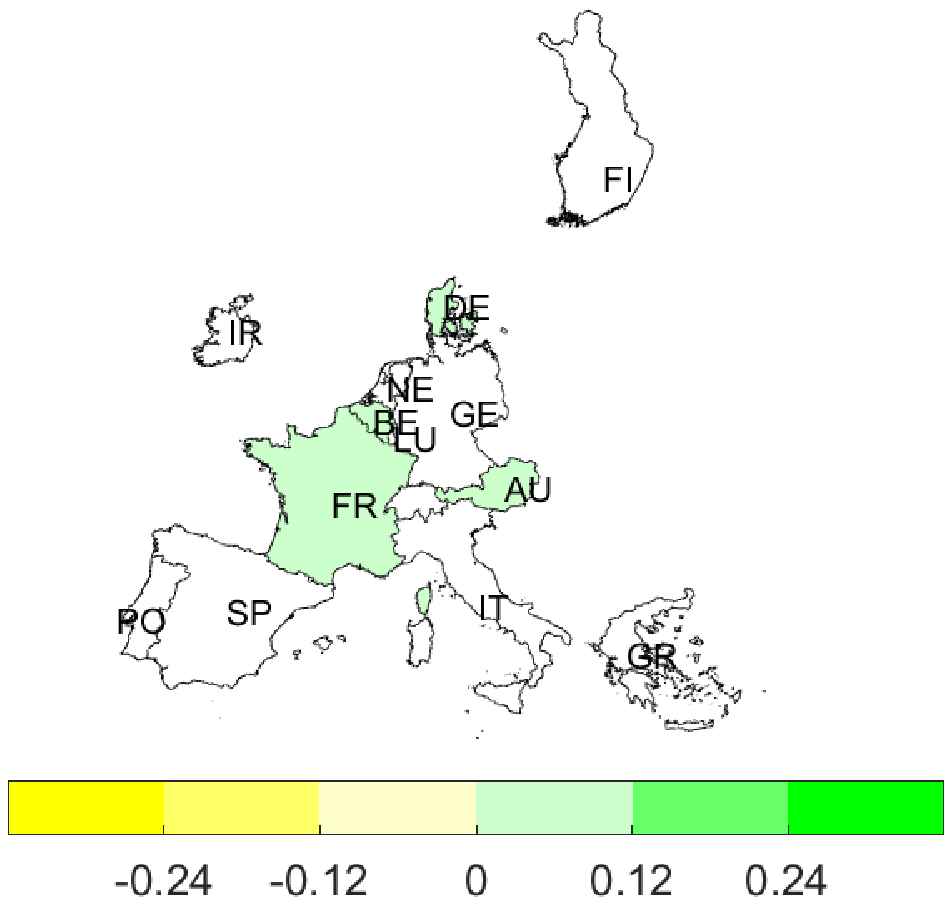}&\includegraphics[scale=0.5]{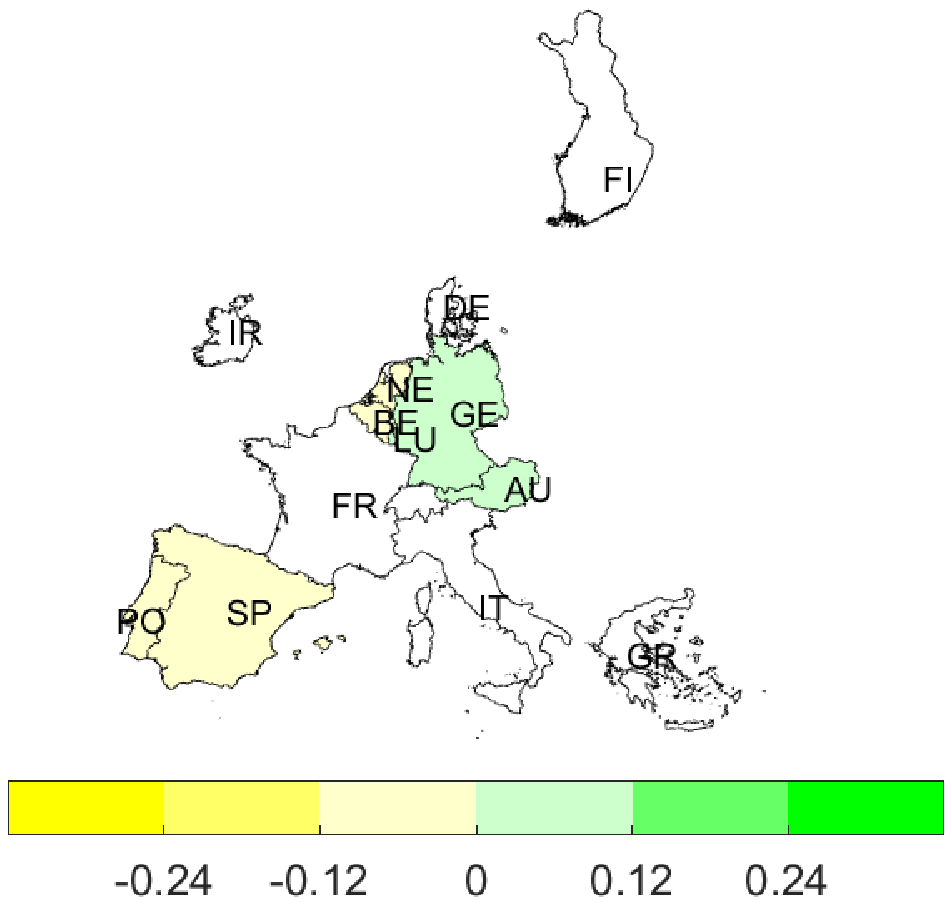}\vspace{5pt}\\
 \hline
 \rot{Drought (SPI dry)}&
 \includegraphics[scale=0.5]{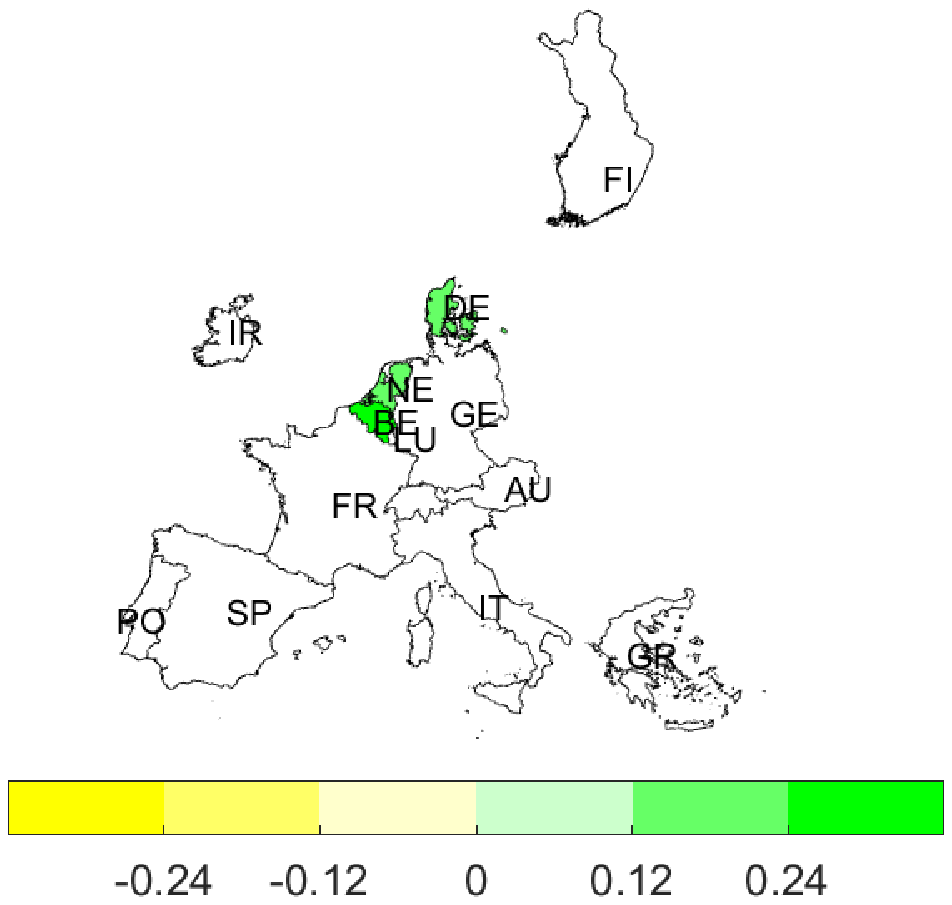}&\includegraphics[scale=0.5]{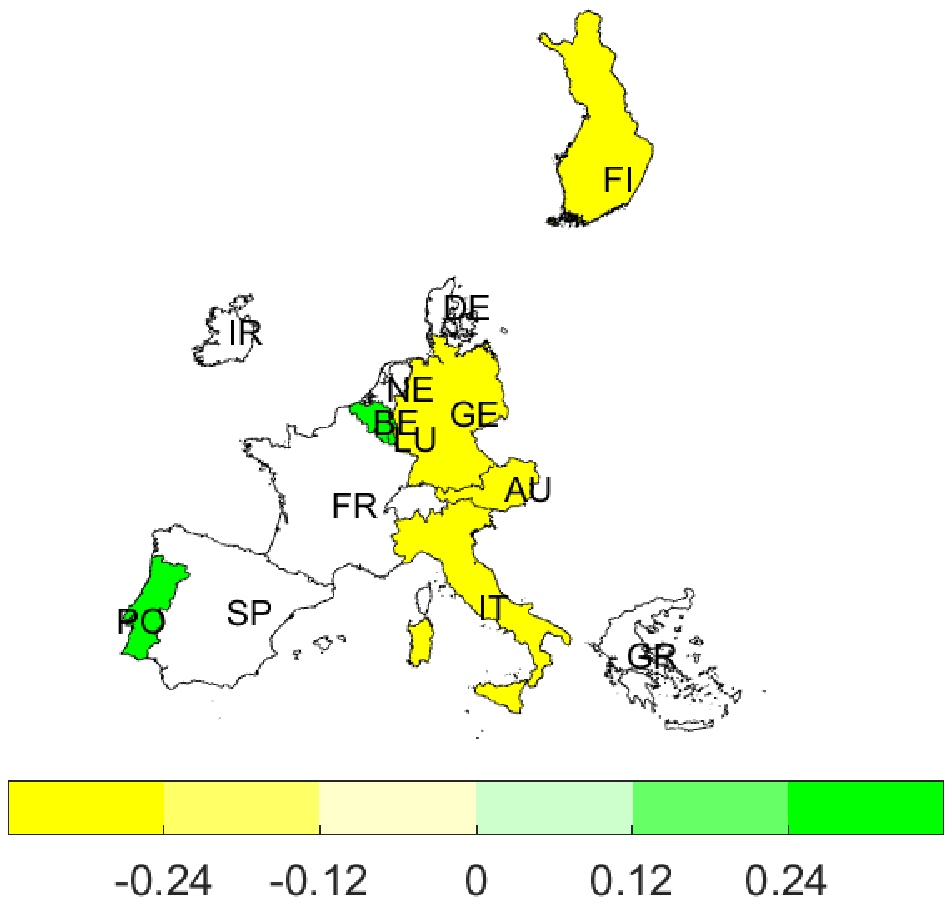}\\
 \hline
 \end{tabular}
 \end{center}
 \caption{Study area with the impact of climate shocks (indicated by different colours) on the growth rate on the represented EU Countries during recession and expansion. White patches suggests that the entity is not responsive (statistically significant) to the shock, while the yellow and green patches respectively denotes negative and positive impact. The measure of drought is constructed based a threshold definition of the tails of the extreme weather index.}\label{Prop3Map1}
 \end{figure}
 
 \begin{figure}[h!]
 \begin{center}
 \begin{tabular}{|c|c|c|}
 \hline
  & Recession & Expansion\\
 \hline
 \rot{Drought (SPI wet)}& 
 \includegraphics[scale=0.5]{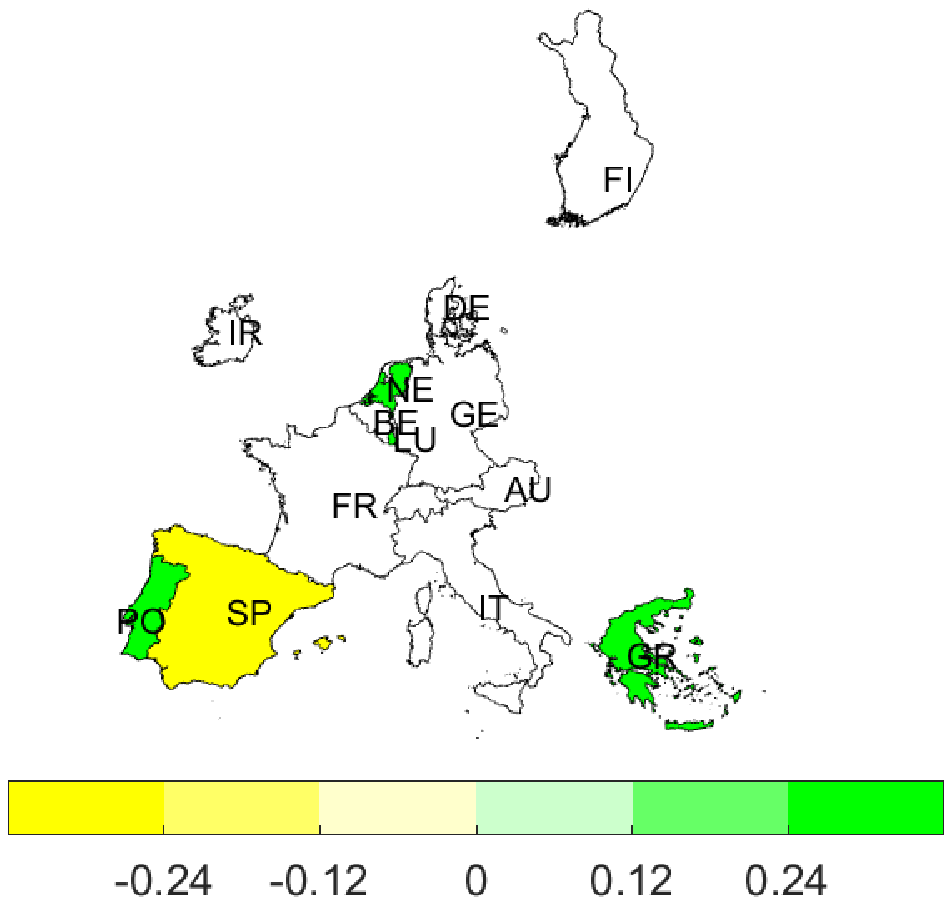}&\includegraphics[scale=0.5]{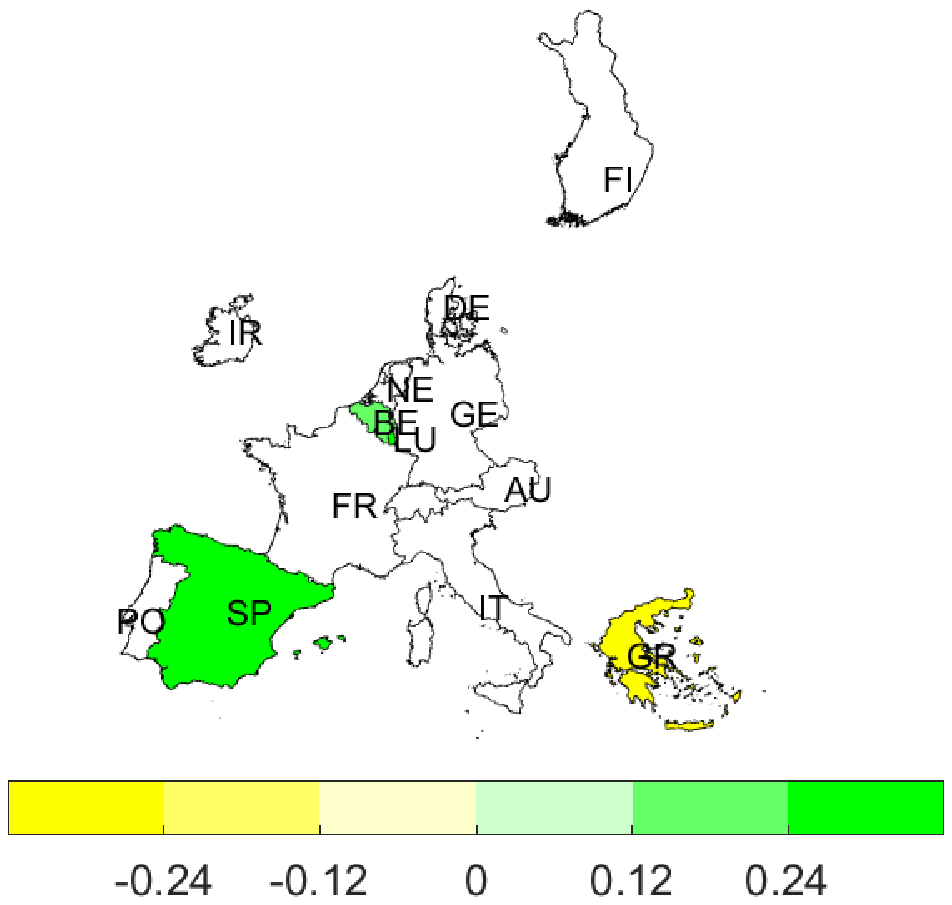}\\
 \hline
 \rot{Precipitation (r20mm)}&
 \includegraphics[scale=0.5]{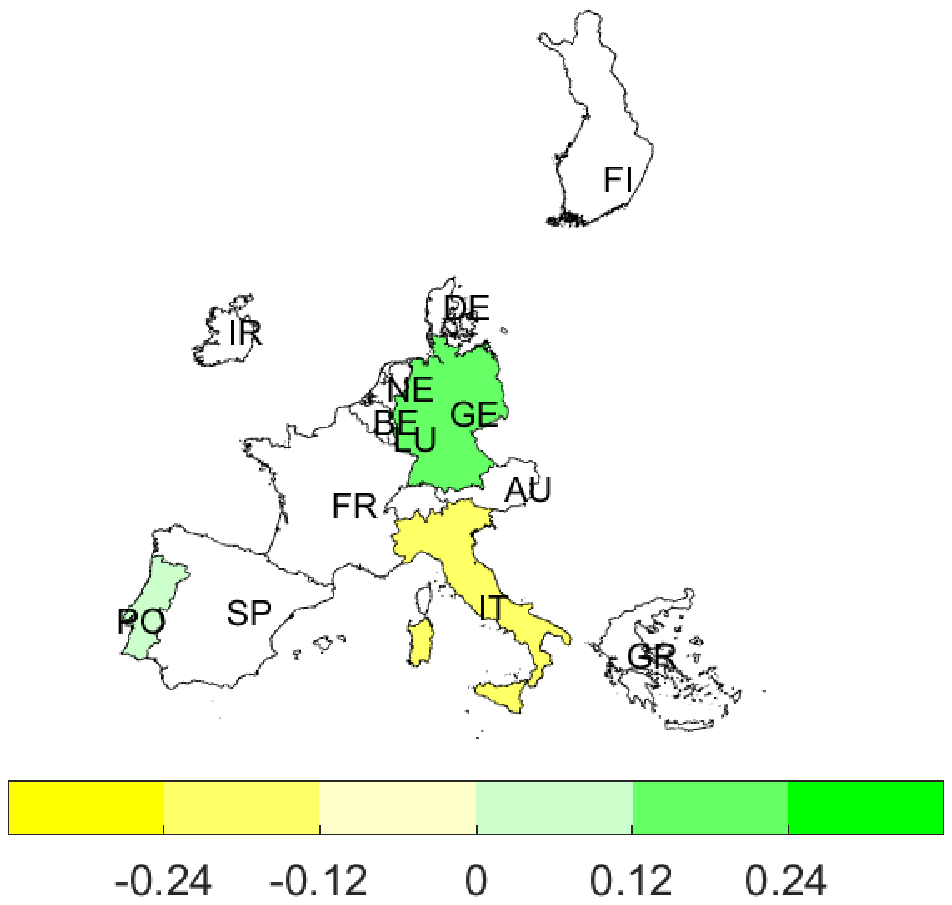}&\includegraphics[scale=0.5]{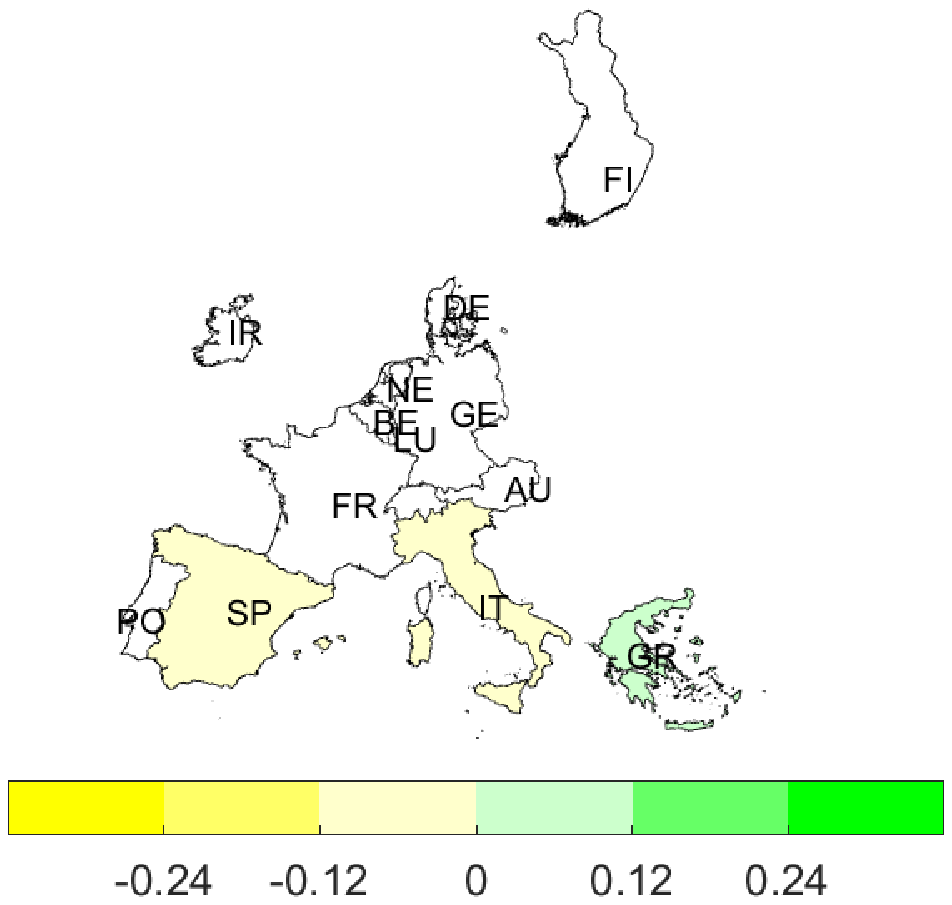}\\
 \hline
 \end{tabular}
 \end{center}
 \caption{Continued from previous page Figure \ref{Prop3Map1}}\label{Prop3Map2}
 \end{figure}
 
 \newpage
 \clearpage
 \section{Posterior distributions}
 \label{Appendix5}
 \begin{figure}[h!]
         \centering
         \subfigure[None]{                
                 \includegraphics[width=0.17\textwidth]{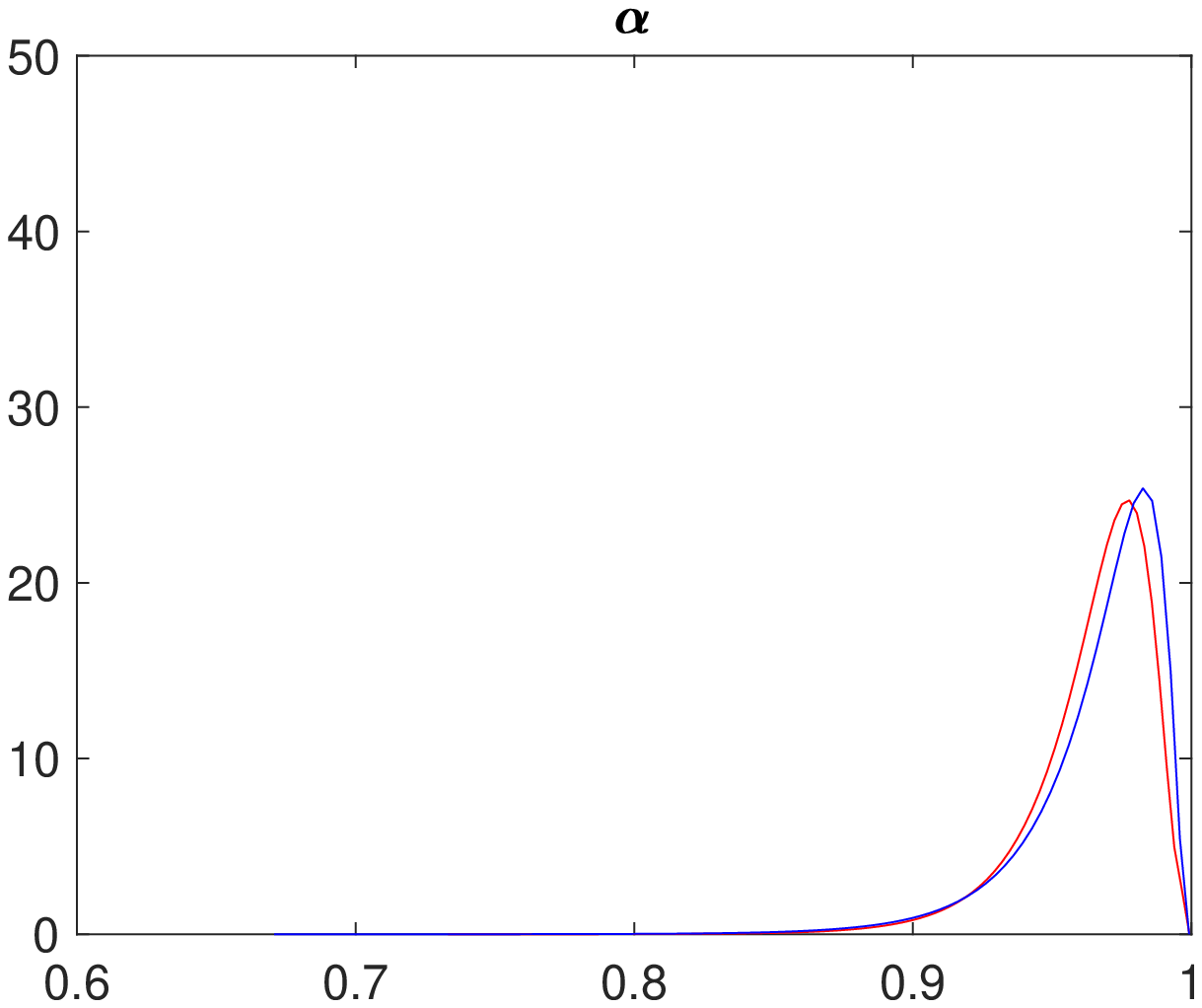}
                 \label{fig:pooledInteractionAlphaApp}}
         ~
         \subfigure[Temperature (CSU) only]{                
                 \includegraphics[width=0.17\textwidth]{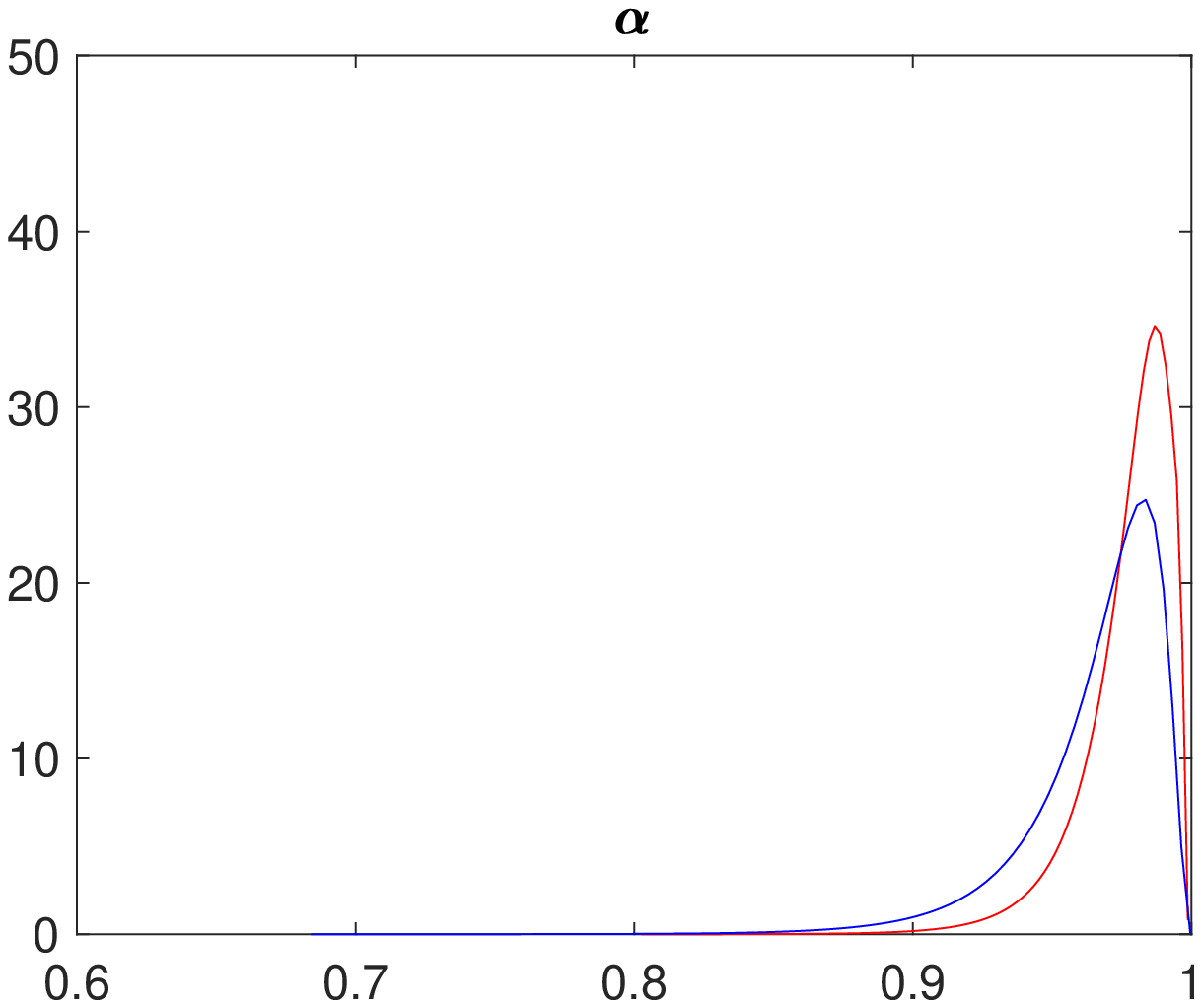}
                 \label{fig:pooledInteractionAlphaCSUApp}}
         ~
         \subfigure[Drought (SPI) only]{
                 \includegraphics[width=0.17\textwidth]{FigPoolingCSUCV/alphaALL.eps}
                 \label{fig:pooledInteractionAlphaSPIApp}}
         ~
         \subfigure[Rainfall r20mm only]{
                 \includegraphics[width=0.17\textwidth]{FigPoolingCSUSPIR20MMCV/alphaALL.eps}
                 \label{fig:pooledInteractionAlphar20mmApp}}               
         ~
         \subfigure[ALL climate indices]{
                 \includegraphics[width=0.17\textwidth]{FigPoolingCSUSPIR20MMCV/alphaALL.eps}
                 \label{fig:pooledInteractionalphaALLApp}}                
         ~        
         \subfigure{                
                 \includegraphics[width=0.17\textwidth]{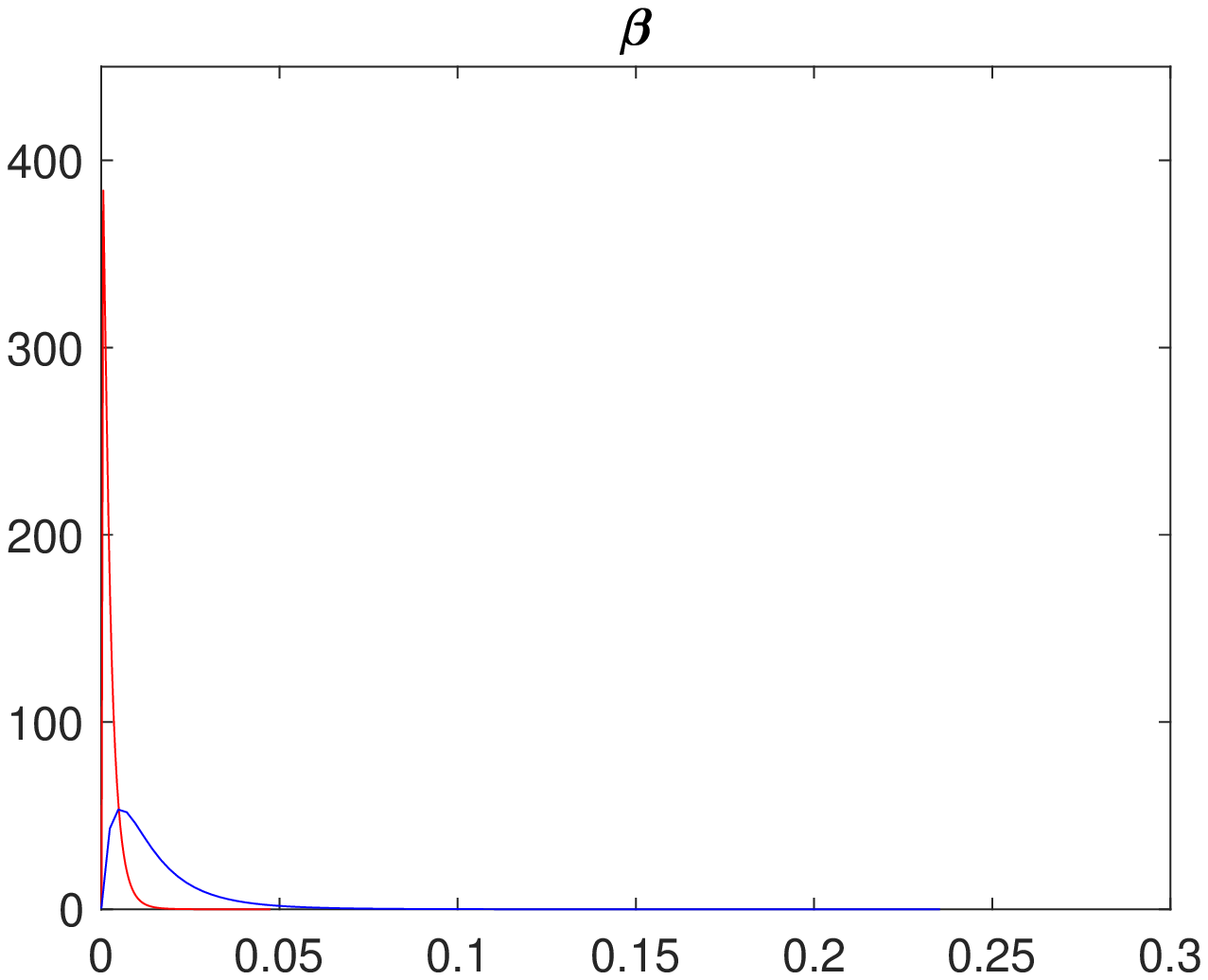}
                 \label{fig:pooledInteractionBetaApp}}
         ~        
         \subfigure{                
                 \includegraphics[width=0.17\textwidth]{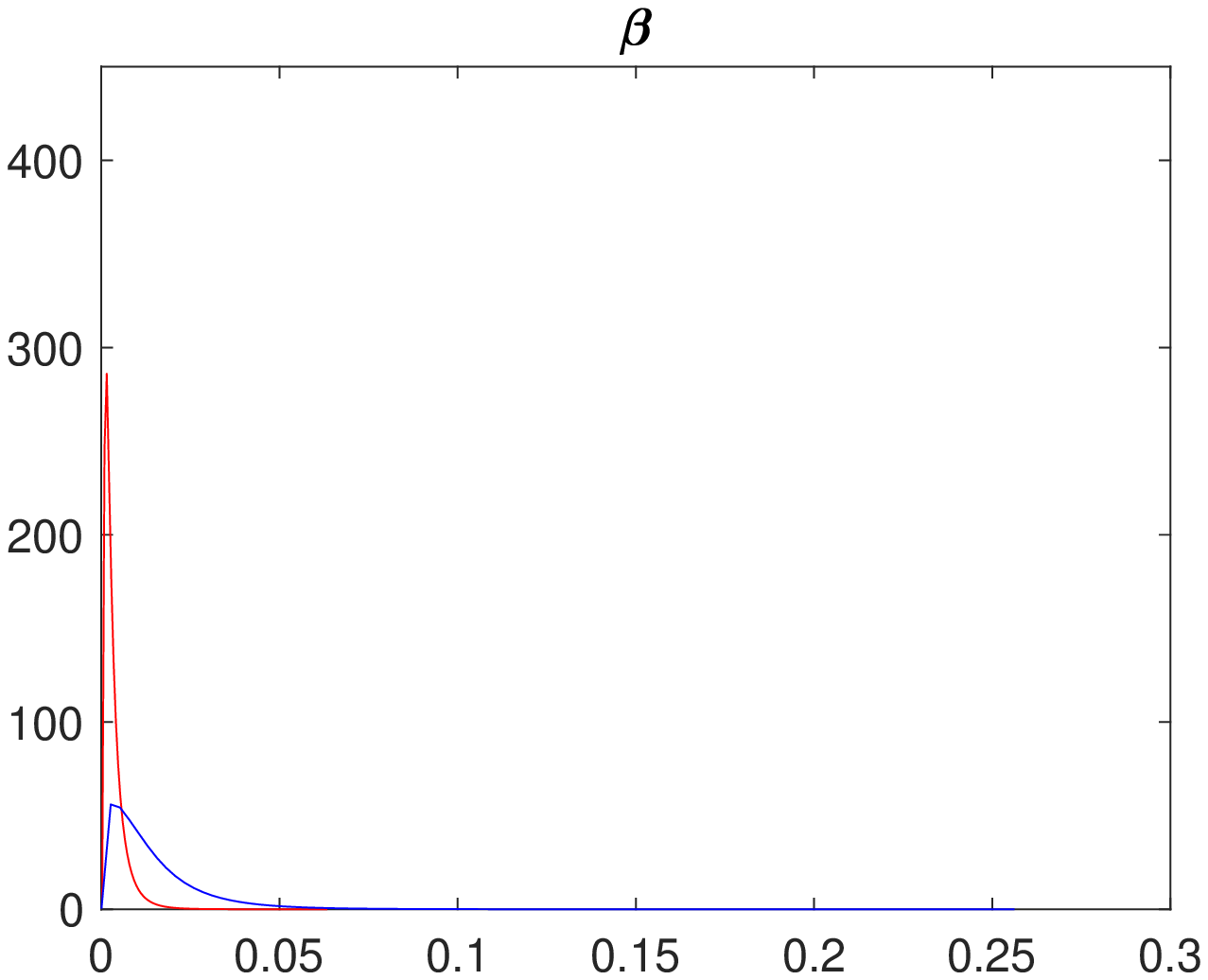}
                 \label{fig:pooledInteractionBetaCSUApp}}
         ~
         \subfigure{
                 \includegraphics[width=0.17\textwidth]{FigPoolingSPICV/betaALL.eps}
                 \label{fig:pooledInteractionBetaSPIApp}}
         ~
         \subfigure{
                 \includegraphics[width=0.17\textwidth]{FigPoolingCSUSPIR20MMCV/betaALL.eps}
                 \label{fig:pooledInteractionBetar20mmApp}}   
         ~
         \subfigure{
                 \includegraphics[width=0.17\textwidth]{FigPoolingCSUSPIR20MMCV/betaALL.eps}
                 \label{fig:pooledInteractionbetaALLApp}}                                             
         ~                
         \subfigure{                
                 \includegraphics[width=0.17\textwidth]{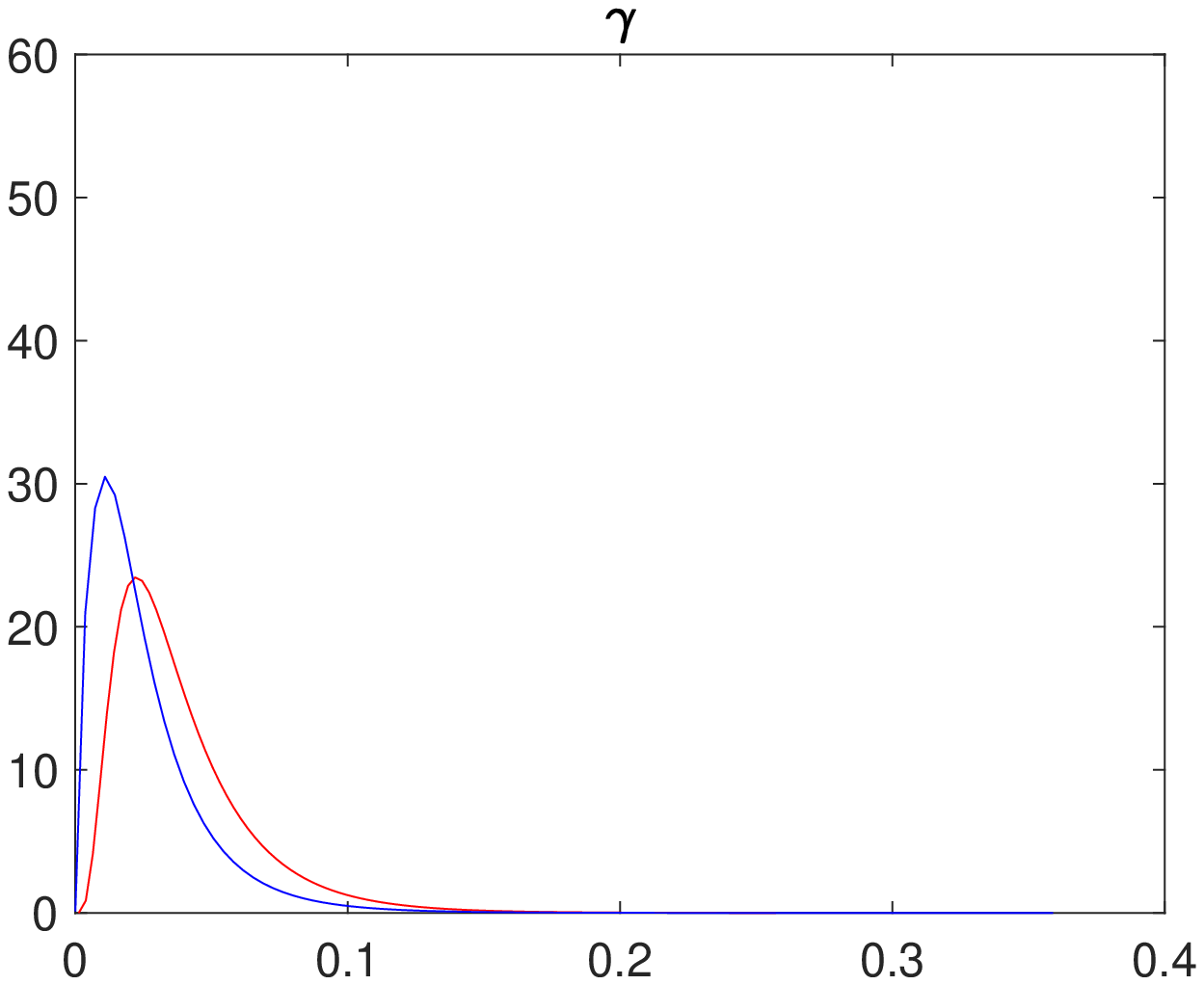}
                 \label{fig:pooledInteractionGammaApp}}              
         ~                
         \subfigure{                
                 \includegraphics[width=0.17\textwidth]{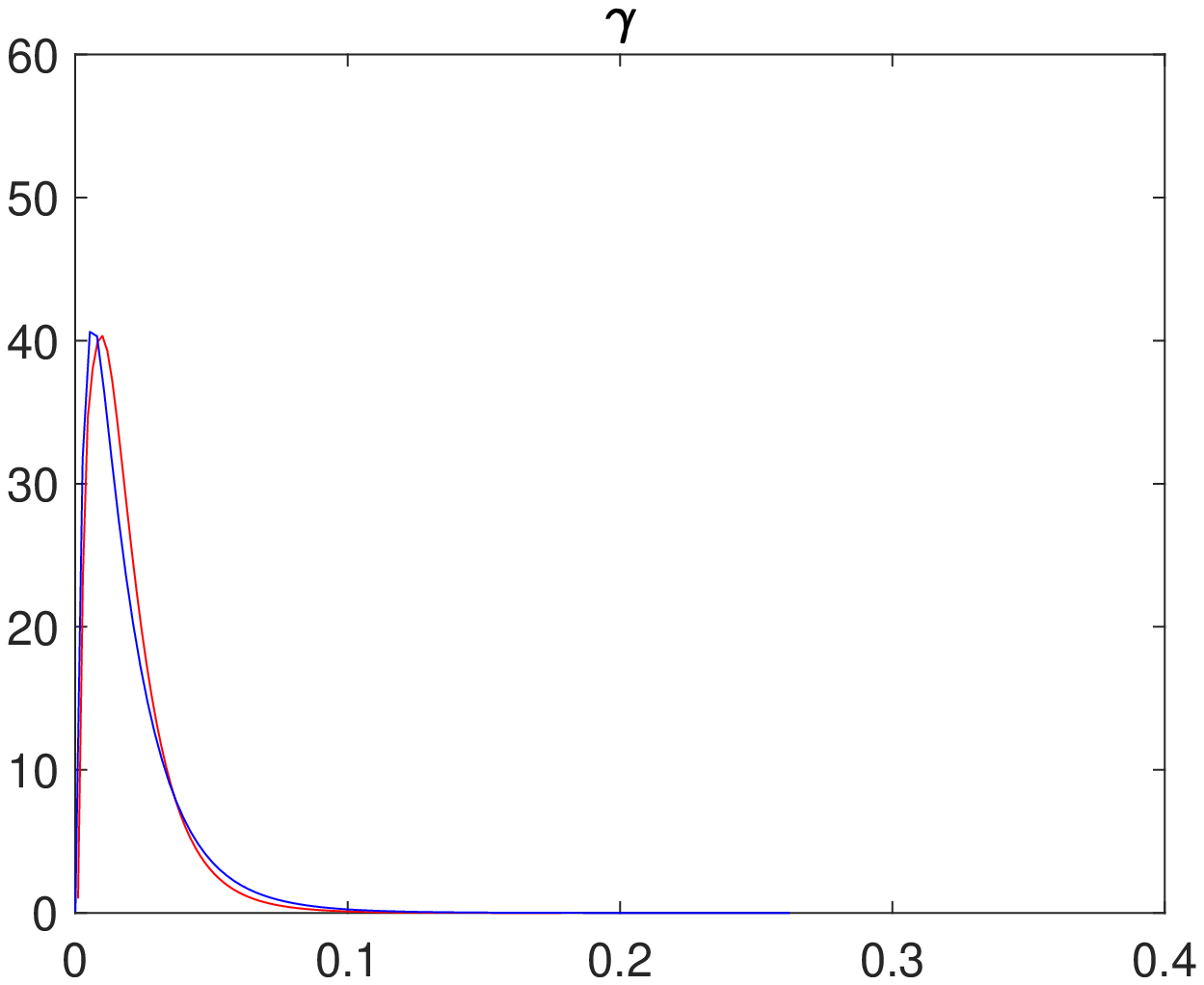}
                 \label{fig:pooledInteractionGammaCSUApp}}              
         ~
         \subfigure{
                 \includegraphics[width=0.17\textwidth]{FigPoolingR20MMCV/gammaALL.eps}
                 \label{fig:pooledInteractionGammaSPIApp}}        
         ~
         \subfigure{
                 \includegraphics[width=0.17\textwidth]{FigPoolingCSUSPIR20MMCV/gammaALL.eps}
                 \label{fig:pooledInteractionGammar20mmApp}}        
         ~
         \subfigure{
                 \includegraphics[width=0.17\textwidth]{FigPoolingCSUSPIR20MMCV/gammaALL.eps}
                 \label{fig:pooledInteractionGammaALLApp}}        
 		\caption{Posterior density for the interaction parameters of both the business cycle (blue line) and the financial cycle (red line) for the pooled PMS models. Row 1, 2, and 3 respectively display the density function of the idiosyncratic parameter ($\alpha$), the local parameter (financial cycle) ($\beta$),  and the global business cycle parameter ($\gamma$). 
 		Column 1 of this figure display result obtained form the climate condition independent PMS models,  columns 2, 3, and 4 respectively display result of the PMS model subjected to temperature only, drought only and rainfall only while column 5 simultaneously considers all the climate change indices in the PMS model.}\label{interactionPooledApp}
 \end{figure}
 
 \begin{figure}[h!]
         \centering
         \subfigure[$\alpha$ under the unconstrained PMS]{                
                 \includegraphics[width=0.17\textwidth]{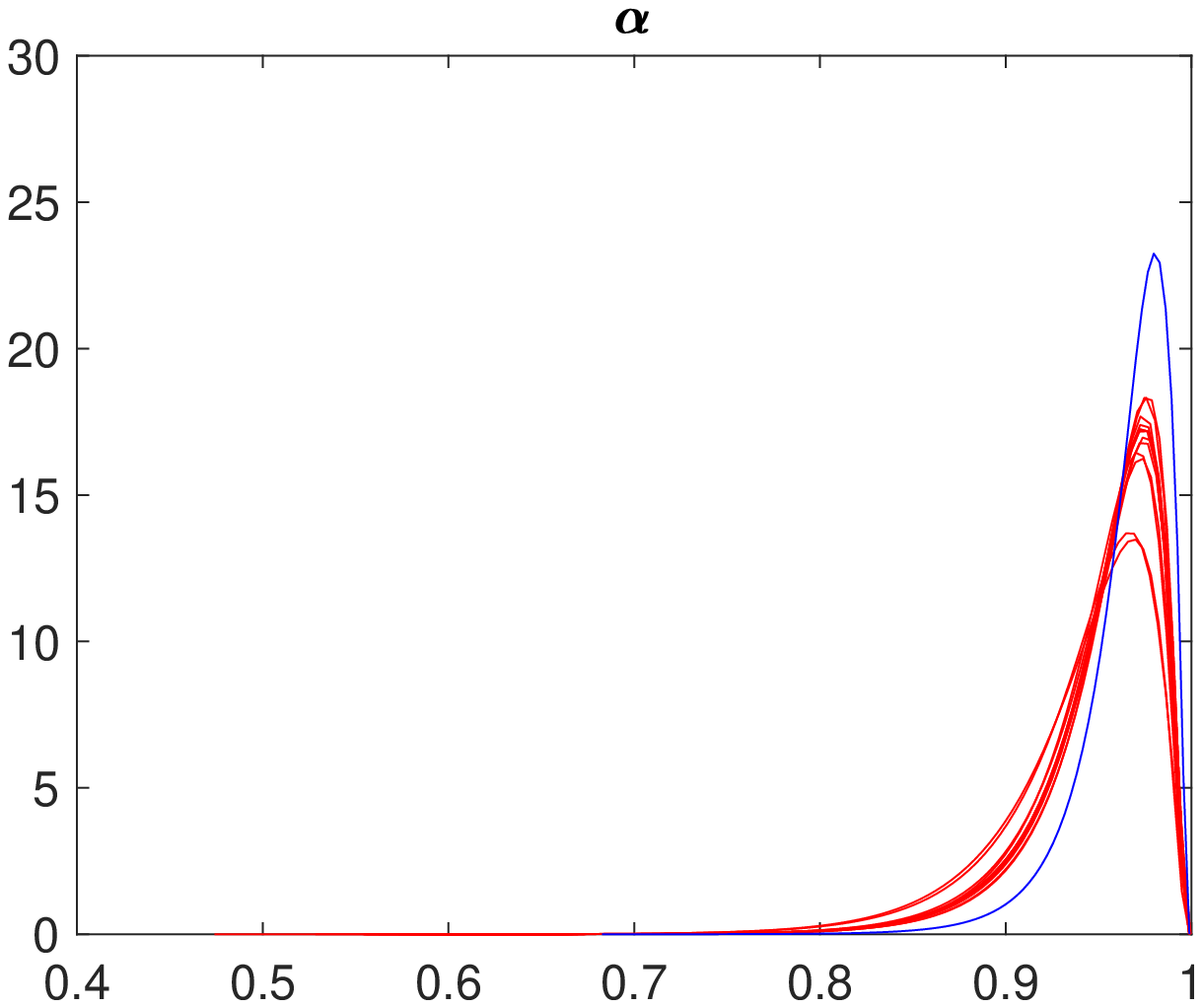}
                 \label{fig:NpooledInteractionAlphaApp}}
         ~
         \subfigure[$\alpha$ under the unconstrained PMS with CSU only]{                
                 \includegraphics[width=0.17\textwidth]{FigNPoolingCSUSPIR20MMCV/alphaALL.eps}
                 \label{fig:NpooledInteractionAlphaCSUApp}}
         ~
         \subfigure[$\alpha$ under the unconstrained PMS with SPI only]{
                 \includegraphics[width=0.17\textwidth]{FigNPoolingCSUCV/alphaALL.eps}
                 \label{fig:NpooledInteractionAlphaSPIApp}}
         ~
         \subfigure[$\alpha$ under the unconstrained PMS with r20mm only]{
                 \includegraphics[width=0.17\textwidth]{FigNPoolingCSUSPIR20MMCV/alphaALL.eps}
                 \label{fig:NpooledInteractionAlphar20mmApp}}
         ~
         \subfigure[$\alpha$ under the unconstrained PMS with ALL climate indices]{
                 \includegraphics[width=0.17\textwidth]{FigNPoolingCSUSPIR20MMCV/alphaALL.eps}
                 \label{fig:NpooledInteractionalphaALLApp}}
         ~        
         \subfigure[$\beta$ under the unconstrained PMS]{                
                 \includegraphics[width=0.17\textwidth]{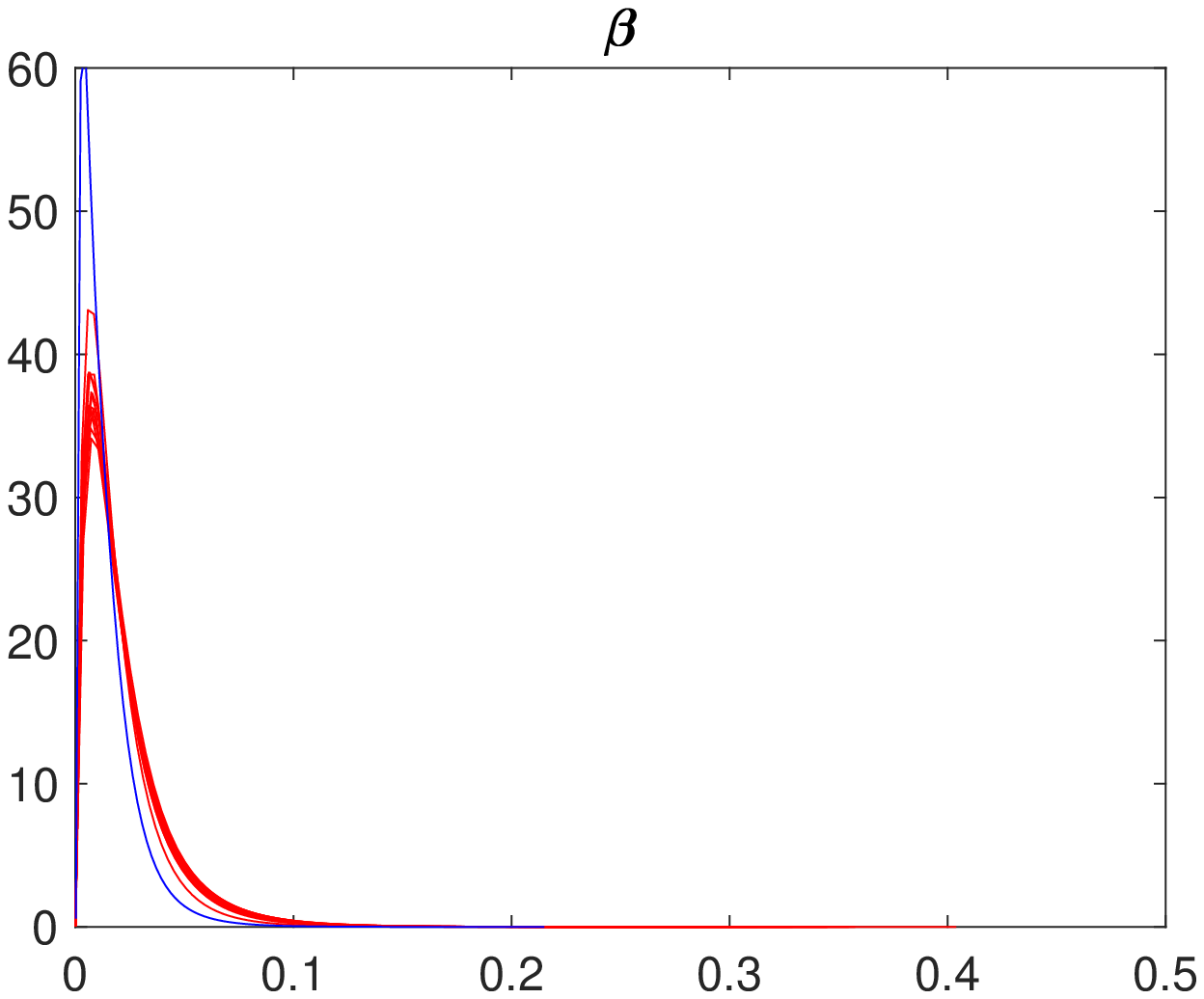}
                 \label{fig:NpooledInteractionBetaApp}}
         ~        
         \subfigure[$\beta$ under the unconstrained PMS with CSU only]{                
                 \includegraphics[width=0.17\textwidth]{FigNPoolingCSUSPIR20MMCV/betaALL.eps}
                 \label{fig:NpooledInteractionBetaCSUApp}}                
         ~
         \subfigure[$\beta$ under the unconstrained PMS wth SPI only]{
                 \includegraphics[width=0.17\textwidth]{FigNPoolingSPICV/betaALL.eps}
                 \label{fig:NpooledInteractionBetaSPIApp}}
         ~
         \subfigure[$\beta$ under the unconstrained PMS with r20mm only]{
                 \includegraphics[width=0.17\textwidth]{FigNPoolingCSUSPIR20MMCV/betaALL.eps}
                 \label{fig:NpooledInteractionBetar20mmApp}}                
         ~
         \subfigure[$\beta$ under the unconstrained PMS with ALL climate indices]{
                 \includegraphics[width=0.17\textwidth]{FigNPoolingCSUSPIR20MMCV/betaALL.eps}
                 \label{fig:NpooledInteractionbetaALLApp}}           
         ~                
         \subfigure[$\gamma$ under the unconstrained PMS]{                
                 \includegraphics[width=0.17\textwidth]{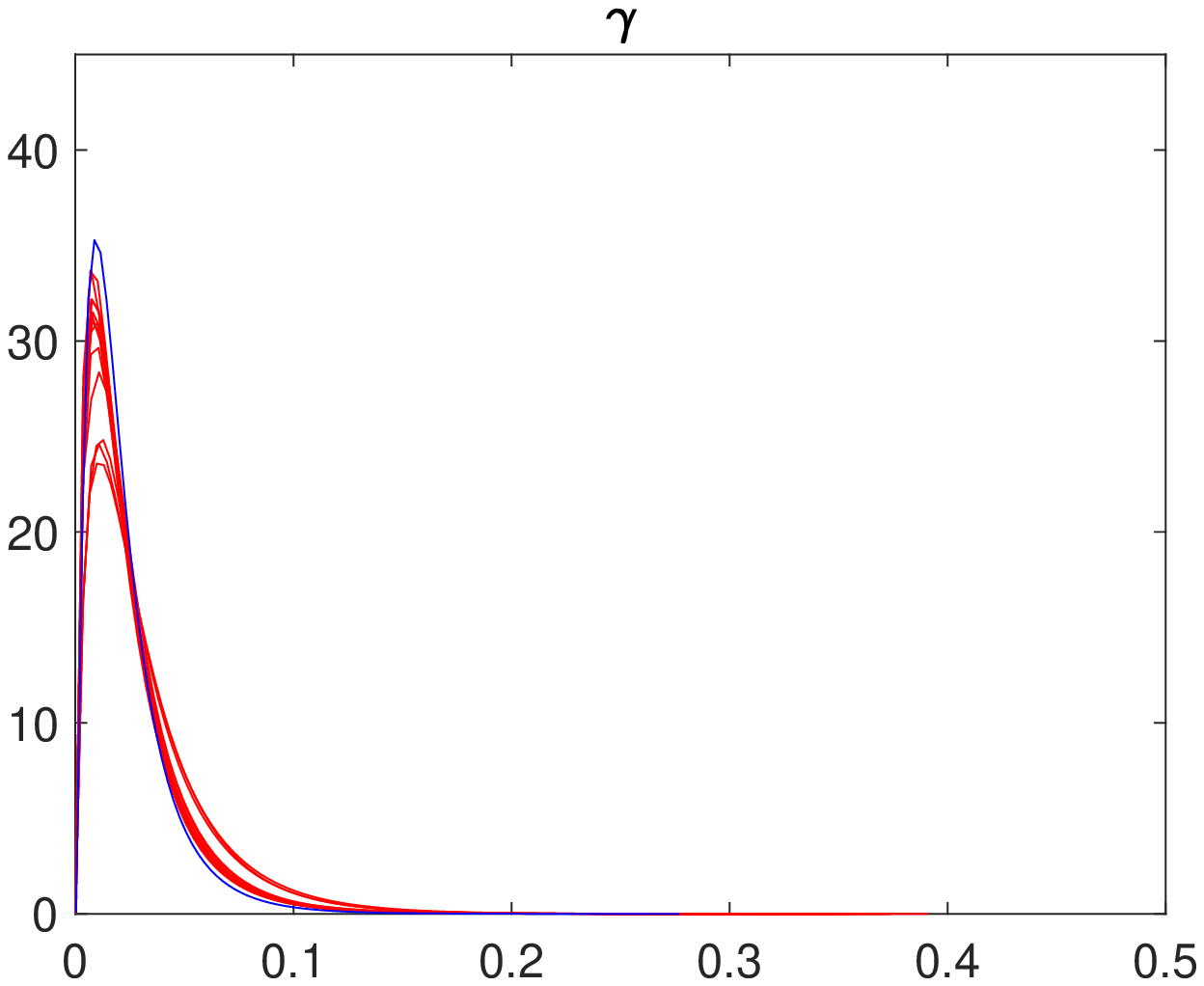}
                 \label{fig:NpooledInteractionGammaApp}}              
         ~                
         \subfigure[$\gamma$ under the unconstrained PMS with CSU only]{                
                 \includegraphics[width=0.17\textwidth]{FigNPoolingCSUSPIR20MMCV/gammaALL.eps}
                 \label{fig:NpooledInteractionGammaCSUApp}}                              
         ~
         \subfigure[$\gamma$ under the unconstrained PMS with SPI only]{
                 \includegraphics[width=0.17\textwidth]{FigNPoolingR20MMCV/gammaALL.eps}
                 \label{fig:NpooledInteractionGammaSPIApp}}                        
         ~
         \subfigure[$\gamma$ under the unconstrained PMS with r20mm only]{
                 \includegraphics[width=0.17\textwidth]{FigNPoolingCSUSPIR20MMCV/gammaALL.eps}
                 \label{fig:NpooledInteractionGammar20mmApp}}        
         ~
         \subfigure[$\gamma$ under the unconstrained PMS with ALl climate indices]{
                 \includegraphics[width=0.17\textwidth]{FigNPoolingCSUSPIR20MMCV/gammaALL.eps}
                 \label{fig:NpooledInteractionGammaALLApp}}        
 		\caption{ Posterior density for the interaction parameters of both the business cycles (red lines) and the financial cycle (blue line) for the unconstrained PMS models. Row 1, 2, and 3 respectively display the density function of the idiosyncratic parameter ($\alpha$), the local parameter (financial cycle) ($\beta$),  and the global business cycle parameter ($\gamma$). Column 1 of this figure display result obtained form the climate condition independent PMS models,  columns 2, 3, and 4 respectively display result of the PMS model subjected to temperature only, drought only and rainfall only while column 5 simultaneously considers all the climate change indices in the PMS model.}\label{interactionNPooledApp}
 \end{figure}
\end{APPENDICES}

\end{document}